\documentclass[oneside,
	10pt, 
	fleqn, 
	a5paper, 
]{LegrandOrangeBook}

\hypersetup{
	pdftitle={Title}, 
	pdfauthor={Author}, 
	pdfsubject={Subject}, 
	pdfkeywords={Keyword1, Keyword2, ...}, 
	pdfcreator={LaTeX}, 
}

\newsavebox{\fminipagebox}
\NewDocumentEnvironment{fminipage}{m O{\fboxsep}}
 {\\\kern#2\noindent\begin{lrbox}{\fminipagebox}
  \begin{minipage}{#1}\ignorespaces}
 {\end{minipage}\end{lrbox}%
  \makebox[#1]{%
    \kern\dimexpr-\fboxsep-\fboxrule\relax
    \fbox{\usebox{\fminipagebox}}%
    \kern\dimexpr-\fboxsep-\fboxrule\relax
  }\\\kern#2
 }

\DeclareUnicodeCharacter{0308}{HERE!HERE!}
\DeclareUnicodeCharacter{0301}{HERE!HERE!}
\definecolor{ocre}{rgb}{0.5, 0.5, 0.5} 
\providecommand{\tightlist}{%
  \setlength{\itemsep}{0pt}\setlength{\parskip}{0pt}}
\chapterimage{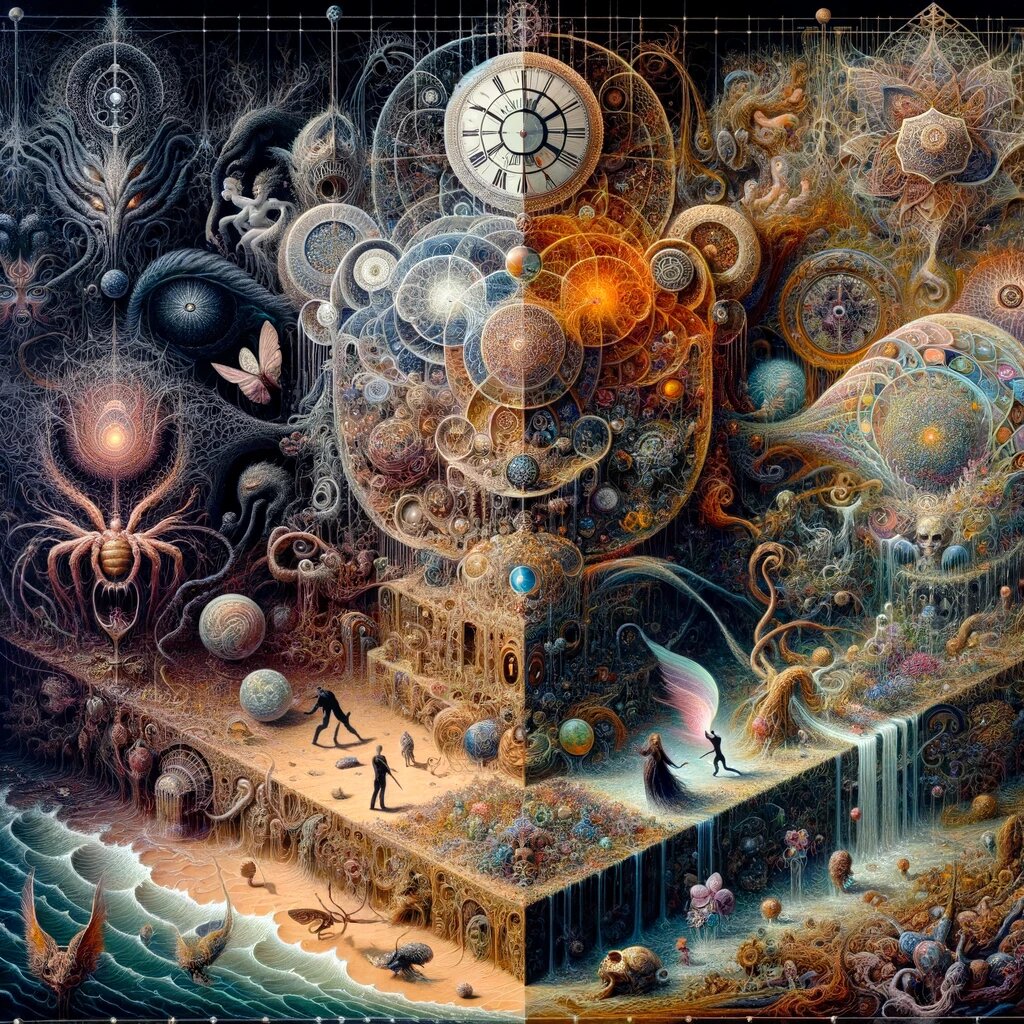} 
\chapterspaceabove{6.5cm} 
\chapterspacebelow{6.75cm} 
\setlistdepth{11}

\begin{document}

\titlepage 
{
    \begin{tikzpicture}[remember picture, overlay]
        \node[anchor=center, inner sep=0pt] at (current page.center) {
            \includegraphics[height=\paperheight, clip, trim=1cm 0 1cm 0]{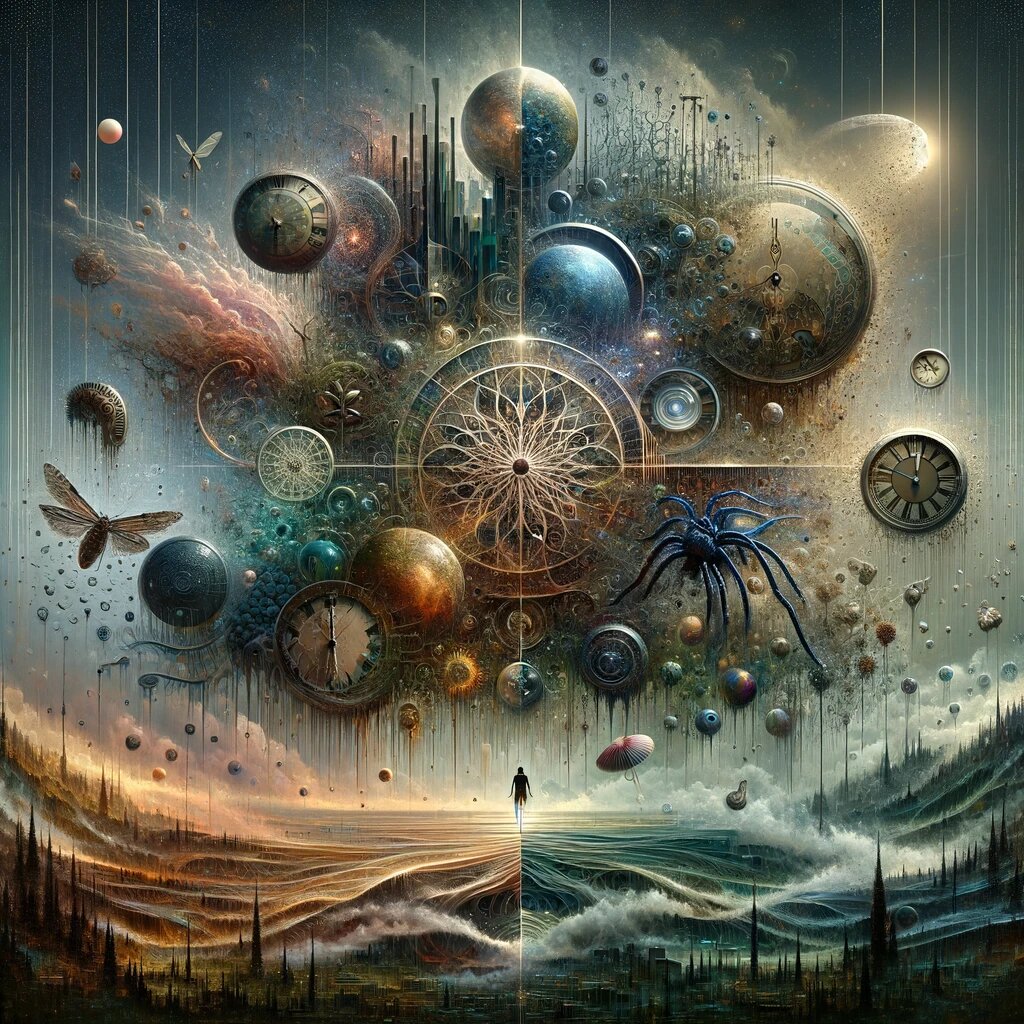}
        };
    \end{tikzpicture}
} 
{ 
    \centering\sffamily 
    {\Huge\bfseries \textbf{Convergence and\\Disruption in\\Digital Society}\par} 
    \vspace{16pt} 
    {\LARGE Money, Secure Communication,\\Digital Objects and Generative AI\\in Spatial Mixed Reality\\}  
    \vspace{24pt} 
    \centering
{
    \hypersetup{urlcolor=black}
    \textbf{\href{https://www.linkedin.com/in/jjohare/}{Dr John O'Hare}}\\
}
     \bigskip
     \bigskip
}

\thispagestyle{empty} 
~\vfill 

\noindent \href{https://creativecommons.org/licenses/by-nc/4.0/}{Attribution-NonCommercial 4.0 International (CC BY-NC 4.0) }\\ 2022 John O'Hare \& Allen Fairchild \& Umran Ali\\ 

\noindent \textsc{Published by john@xrsystems.uk}\\ 

\noindent \textsc{\href{https://github.com/flossverse/origin/blob/draft/Book/metaverseBTC.pdf}{Raw GitHub hyperlink}}\\ 

\noindent 
You are free to:
\begin{itemize}
\item Share — copy and redistribute the material in any medium or format
\item Adapt — remix, transform, and build upon the material
\end{itemize}
The licensor cannot revoke these freedoms as long as you follow the license terms.\\
Under the following terms:
\begin{itemize}
\item Attribution — You must give appropriate credit, provide a link to the license, and indicate if changes were made. You may do so in any reasonable manner, but not in any way that suggests the licensor endorses you or your use.
\item NonCommercial — You may not use the material for commercial purposes.
\item No additional restrictions — You may not apply legal terms or technological measures that legally restrict others from doing anything the license permits.
\end{itemize}
\noindent \textit{First printing, March 2022} 


\pagestyle{empty} 

\tableofcontents 

\listoffigures 

\listoftables 

\pagestyle{fancy} 

\cleardoublepage 


\part{State of the art}


\chapterimage{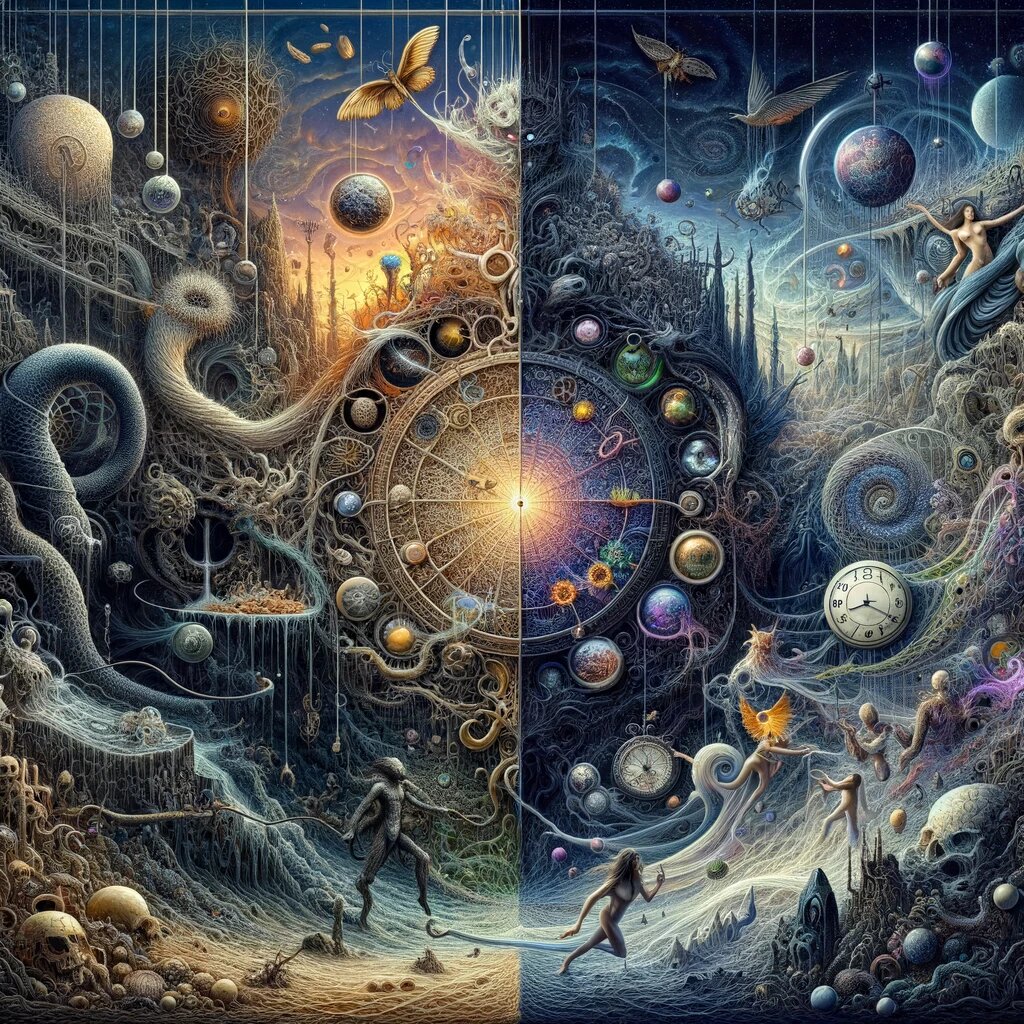} 
\chapterspaceabove{6.75cm} 
\chapterspacebelow{7.25cm} 


\section{Conflict of interest statements}
The authors may own small numbers of the various tokens referenced in the text for experimentation and/or investment purposes. At this time that is only sufficient Ethereum to operate a Web3 wallet, and Bitcoin locked on the Lightning network. No NFTs are owned at this time. There are no financial stakes in the development of any of these ecosystems.
\chapter{Introduction}
\section{Overview}
As human beings, we have always relied on certain social constructs to guide our interactions and transactions with one another. Money and trust are two such constructs that have played a vital role in shaping our societies, and the way we live our lives. However, the digital age has brought with it new challenges that are testing the foundations of these social norms.\par
In a world where we are increasingly connected through the internet and able to communicate with people from all corners of the globe, the concept of money and trust is changing. Gone are the days of the village structure in which we evolved, where personal relationships and face-to-face interactions were ubiquitous. Now, we are faced with the prospect of working and interacting one another, and also with artificial intelligence actors that seem subjectively real, all while navigating the complexities of a global mixed reality.\par
This transition to a more efficient and interconnected world has the potential to bring about great benefits, but it also presents us with an enormous challenge. The chaotic and intangible mix of value, trust, socialisation, generative art, and AI chat actors, is not yet well understood, and it will take time for us to adapt to this new way of living and interacting with one another.\par
We initially wanted to explore exciting new developments in the transmission of value, and trust, in `digital society'. The problem is that each of these topics alone are enormously complex, and the intersections seem to be more so. We have been researching the current state-of-the-art, and the emerging consensus narrative, to try to figure out how the collision of these technologies might serve our virtual production workflows (Figure \ref{fig:vprobot}). As we worked on this research the Cambrian explosion of generative AI added an incredibly important new strand to our investigation.\par
Over the course of a couple of years the focus of the work has developed, and refined. Our tool-kit, as it stands, supports inclusive human creativity and economic exchange, especially for emerging markets and especially perhaps \href{https://www.afrobitcoin.org/}{Africa}. There is a huge proportion of human creativity currently excluded from media production pipelines due to gatekeepers of knowledge, access to identity proofs, and financial infrastructure that is taken for granted in the richer nations. This inclusion will be accomplished for the most part through integration of open source machine learning and AI tools, but this field quite new, and that part of the work is under developed.\par

\begin{figure}
  \centering
   \includegraphics[width=\linewidth]{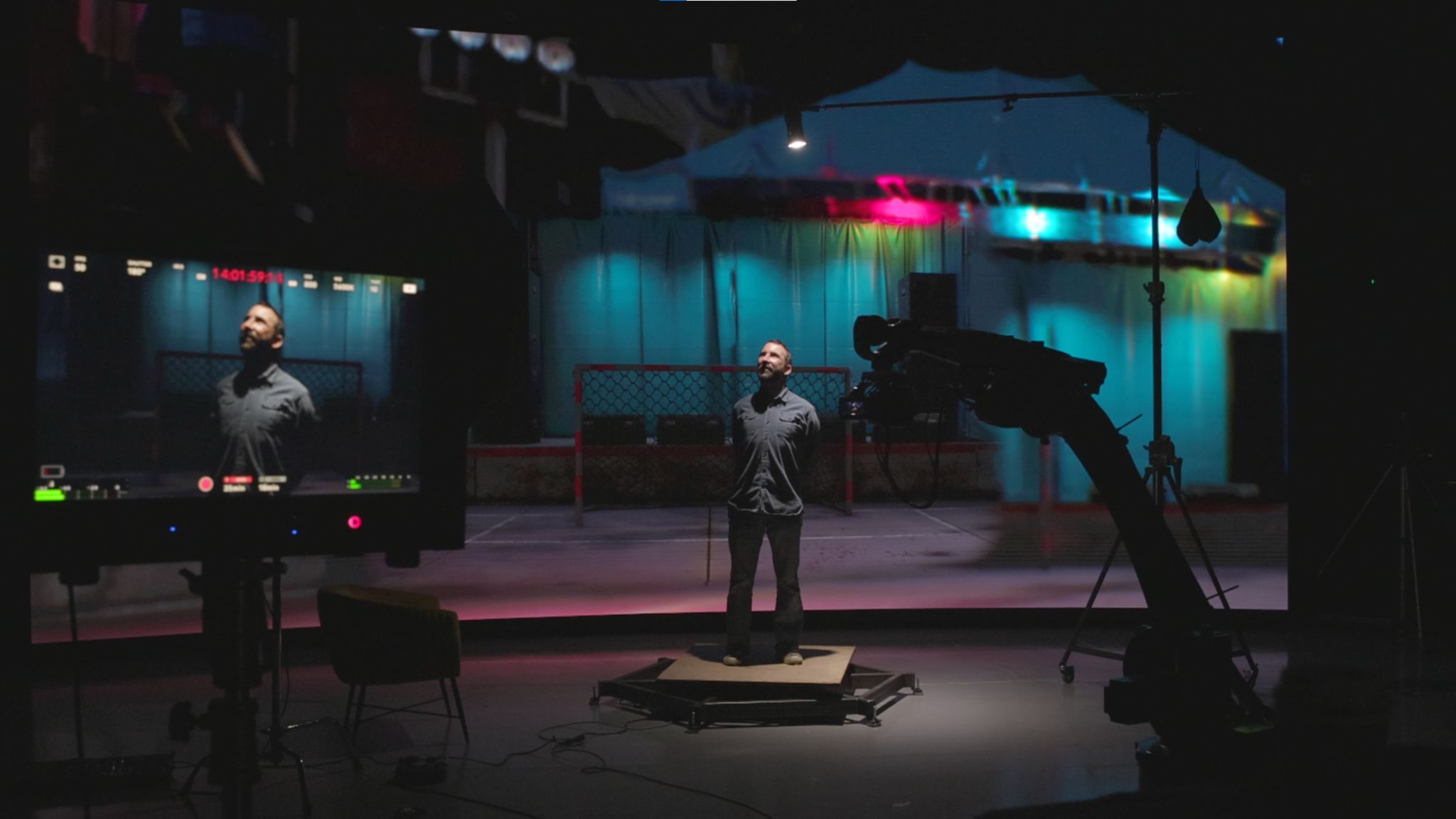}
 \caption{In camera VFX virtual production}
    \label{fig:vprobot}
\end{figure}

\subsection{Why just business to business}
Trust and safety (T\&S) is crucially important for building ethical and inclusive online communities, the landscape is fraught with complex tradeoffs and challenges that make direct involvement difficult for independent researchers and developers like us. As summarized in a recent \href{https://www.atlanticcouncil.org/in-depth-research-reports/report/scaling-trust/}{Atlantic Council report}, T\&S operates in a high-stakes environment driven by commercial incentives that often conflict with safety objectives. Practitioners face threats of trauma, burnout, and retaliation when enforcing policies. The field itself lacks established standards, struggles to quantify impact, and relies heavily on the judgement of private companies.
\begin{itemize}
\item Trust and safety (T\&S) emerged as a field to govern risks in online communities. It became crucial as user-generated content platforms scaled.
\item The emergence of a professionalized T\&S field creates opportunities for collaboration and innovation. But knowledge sharing, tooling, talent pipelines, and metrics need improvement.
T\&S practitioners, especially content moderators, face high risk of trauma. Their wellbeing requires urgent attention.
\item Systemic gaps exist around measuring T\&S value, regulation's impact on incentives, and the role of venture capital. Market failures drive under-investment.
\item Adjacent fields like academia, civil society, and media provide crucial external expertise but lack formal inclusion in T\&S.
\item The gaming industry offers useful insights but has its own major T\&S challenges.
\item Known harms will spread to new technologies, requiring adaptive solutions and proactive design.
\item Systemic gaps exist around measurement, regulation, and investment. Creative initiatives needed to realign incentives.
\item Philanthropy and government can help address systemic gaps through strategic programs and incentives.

\item T\&S must balance competing goals like:
\begin{itemize}
\item Protecting rights vs mitigating harms
\item Efficiency vs accuracy in enforcement
\item Reviewer wellbeing vs review needs
\item Centralized vs decentralized moderation
\item Growth vs safety investments
\item Internal process vs external expertise
\item Reactive enforcement vs proactive design
\end{itemize}

\end{itemize}
As independent researchers focused on decentralized and generative technologies, we believe the greatest value we can provide is advancing capabilities that empower users and diversify governance models. Directly mediating harms involves subjective tradeoffs we are not comfortable imposing on others. Our goal is to expand possibilities for human flourishing, while avoiding direct content moderation that restricts expression based on our own limited perspectives. We aim to enable solutions, not dictate them. As the report advocates, addressing T\&S issues requires a broad coalition encompassing civil society, government, academia, and industry. We can make a positive contribution through technology R\&D, while leaving nuanced governance decisions to more representative institutions. There are many valid ways to build a more trustworthy web; given our background and values, we have chosen to focus on expanding generativity rather than regulating spaces.

\section{Introduction}
This book is a high level view of technologies and their potential within the developing digital society narrative, focusing around the transmission of value within and across immersive global networks, with a further focus on the Bitcoin monetary network.\par
Cybersecurity is top of the list of concerns in the \href{https://digital-strategy.ec.europa.eu/en/policies}{EU digital society strategy}, just ahead of digital inclusion, so we started out with security best practices in mind, and we tried to end the investigation with inclusion. We aimed to support small and especially developing companies in our sector, giving them a foot in the door on a global stage, without their costs spiralling.\par
Fortunately, we discovered a wealth of carefully crafted open source tools which can support this. \href{https://opensource.org/osd}{Open source software} is software that is available with source code that can be modified and distributed by anyone. The model extended into other creative works through things like the \href{https://creativecommons.org/licenses/}{Creative Commons} licenses, which is itself an \href{https://open-stand.org/about-us/principles/}{`open standard'}. We will return to these themes throughout the book, and both this text and the supporting code are open source. This means that anyone can access, use, and modify the source code for their own purposes. Open source software is often developed and maintained by a community of volunteers. This approach to software development allows for more collaboration and innovation, as well as greater transparency and security. Many popular software programs, such as the Linux operating system and the Apache web server, are open source.\par
We have tried to assemble the tools we found, cogently, to deliver an open source kit for experimentation, to curious technically individuals and groups, and we have applied our own security knowledge on top of an already first class set of tools. It’s certainly not production ready, but it's good enough to commit small amounts of money into, and collaboration is welcomed.\par
Whilst researching, it seemed that every door we opened was full of interesting and useful treats. What was supposed to be a short technical paper quickly became a 400-page book, and a deployable virtual machine stack, with a dozen different open source components in it. \par
This book supports the software stack, which supports anyone who thinks this material might be useful. Below is a précis of the chapters of the book, which will hopefully give an insight into what ``this stuff'' is. The reader can decide to download the book and the system ``How To'' guide. All of it can be contributed to on GitHub, all of it will be developed forward, and none of it is really finished yet.\par
Chapter 1 starts with an introduction to the book which is about value transmission, with distributed trust, in global digital society and mixed reality systems. \par
Next is a summary of Web3, as it stands right now. Web3 is a complex term that is cropping up far more in the technical press, so we wanted explain what it might mean. It’s still pretty confusing. There are a bunch of legacy explanations which are Web3.0 (note the `0' there), but these are withering on the vine. Then there’s the new VC funded, super hyped, and potentially useless Web3 incarnations, which again cover a slew of intersecting technologies. Note they dropped the zero to reboot the brand! This doesn’t mean there’s nothing to see here. The astonishing amount of \href{https://mirror.xyz/tr3butor.eth/AlZPMq_syymAoi8M1VVb2xES9Twj1OeetJbEE7EWhiw}{money and developer talent}, and the clear market hunger for things like NFTs (non-fungible tokens) suggest that there’s a future for Web3, it’s just really unclear if this is inherently valuable or just hype.\par
In the next chapter we took a look at blockchain, which is very intersectional with Web3. Even on its own this is a complex emergent set of disciplines. The blockchain chapter was especially interesting to research. It turns out there’s a \textit{lot} of ways to get this technology wrong. Even very appealing options on paper, turn out to have very shaky foundations. There are valuable things here, but given the complexity and scope, we decided to focus on the most promising of the technology stacks; the Bitcoin network.\par
\begin{figure*}[ht]\centering 
	\includegraphics[width=\linewidth]{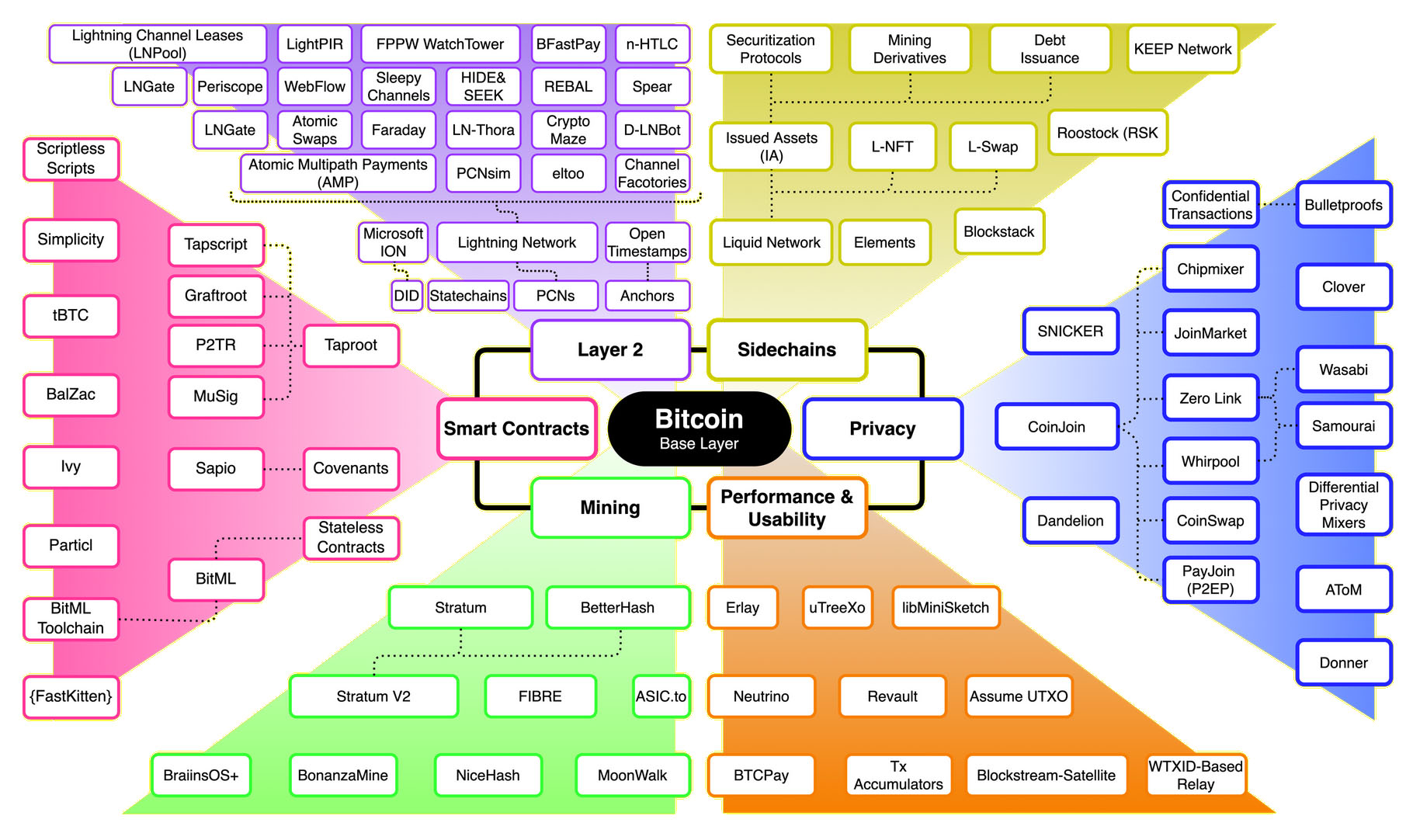}
	\caption{The Open Source Map of The Bitcoin Protocol Ecosystem \href{https://www.ekosys.org/}{maintained by Lucas Nuzzi} of Coinmetrics}
	\label{fig:ecosystemmap}
\end{figure*}

Even Bitcoin isn’t just Bitcoin any more. It’s a swarm of open source tools which can (in theory) accomplish a great many things. These are illustrated well in Figure \ref{fig:ecosystemmap}. 
These newer, ancillary elements to Bitcoin, are emergent right now. Some of them won’t be around until next year, and it’s questionable whether they will even work out. With that said we aren't convinced by the value proposition of of Ethereum, and there’s enough Bitcoin tooling for us to cherry pick useful components. We map those forward into our metaverse product.\par
The next chapter is about Money. In expanding our research on Bitcoin, we found that it’s impossible to think about the tech without opening up a whole line of questions about money itself. This is fine because we set out to look at global value transfer for business. It’s not a trivial subject though, and this section tries to overview why value and Bitcoin are so enmeshed, then what other options there might be in the end (because Bitcoin has kicked off a whole slew of global adoption outside of itself). Chapter five extends this look at global money systems and examines corruption, governance, and opportunities for digital society to affect change and equality.\par
The distributed identity management, and trust chapter follows. Identity management is important for digital society and potentially crucial to metaverse applications which have a value transaction layer. It’s not an easy section to write about, because there’s a lot of research, it’s not our field, and finding the value and the right target audience has been very difficult. It's by no means clear that blockchain is the right tool for this component, and newer cryptographic products are emerging. We select an exciting emergent protocol called Nostr which we believe can help us knit together many elements of digital society.\par 
In chapter 7 we take another look at NFTs. It’s impossible to ignore this stuff now. It’s fundamentally a bit broken, but there are probably use cases, and the money and development attention it’s getting are incredible. We try to navigate our hypothetical virtual production partners through this as best we can. \par 
We’re actually pretty excited about future versions of `digital assets', based around Bitcoin, because that allows us to keep just one software stack, minimising the threat surface. We’ve mapped that forward into the open source tools that we recommend.\par
Chapter 8 is a big one for us as it’s our research area prior to opening up the Bitcoin box(es). Metaverse, or at least one of the current definitions of metaverse, is just social interaction in mixed reality (VR/AR/XR). We’ve been studying that for decades, so this section is more academic and tries to boil down what we think is most important. The choices we made here guided us toward the selection of free and open source metaverse software.\par
We also take a look at the other definitions of `metaverse' which are doing the rounds on the web, try to unpick which is which, and what they are for, and attempt to weave back together the best of both. This ends up looking a bit like the Venn in Figure \ref{fig:landscapevenn}, where we have transmission of provable identity, non-fungible tokens bearing value or data, distributed files, actual money (including micropayments) and a social layer based on our best knowledge about mixed reality. In the end we abandon the word metaverse and settle on `digital society' as our preferred term.\par
Chapter 9 is a work in progress and looks at how the recent explosion in generative art, and machine learning language models might help drive equality and equity of access globally. Past this stage in the book we get into the murky and half developed tail end, where we’re interfacing with our design choices, and the stack which can be deployed into the cloud.\par

\begin{figure*}[ht]\centering 
	\includegraphics[width=\linewidth]{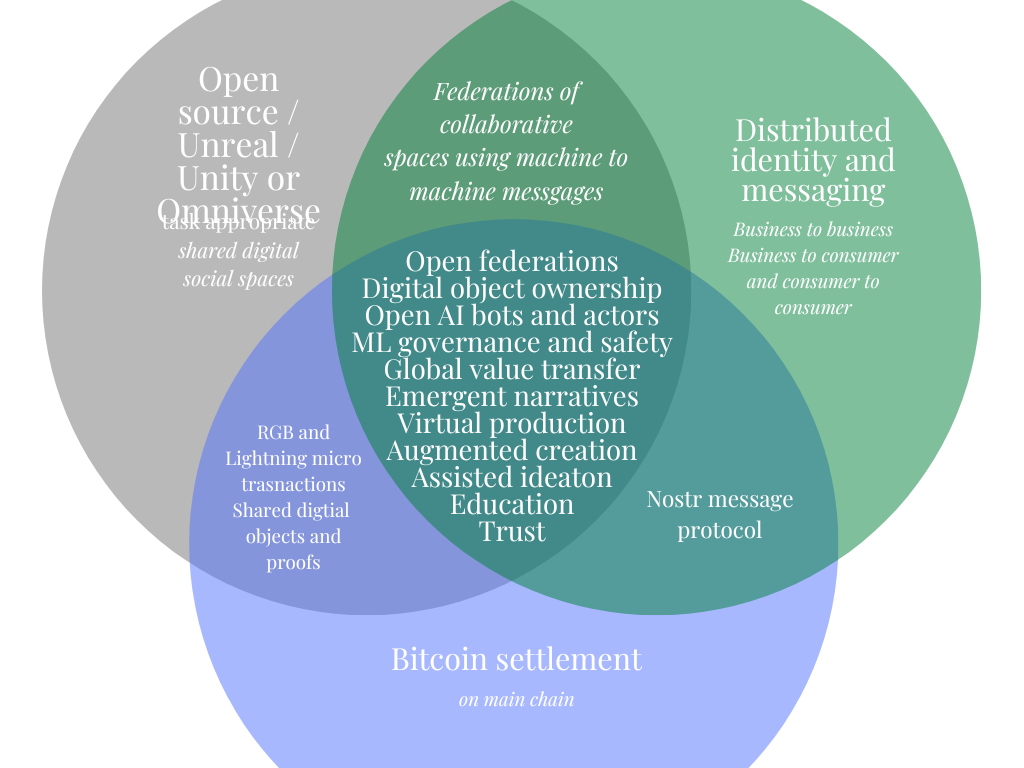}
	\caption{Distributed web, metaverse, and Bitcoin are intersectional technologies.}
	\label{fig:landscapevenn}
\end{figure*}

\subsection{The view of big business and governments}
As adoption of these technologies increases it will be necessary for people, and AI actors, to pass economic value between themselves. These `goods and services' interactions, within the digital and virtual social spaces should be underpinned by a trust system, which scales globally and presents low friction. Current secure international payment rails are poorly suited to such interactions; indeed it is likely with legacy systems, that parties would be forced to leave the metaverse application, and instead navigate their banking applications to exchange value with overseas entities in a secure fashion. This might conceivably take several days.\par
\begin{figure}[ht]\centering 
	\includegraphics[width=0.5\linewidth]{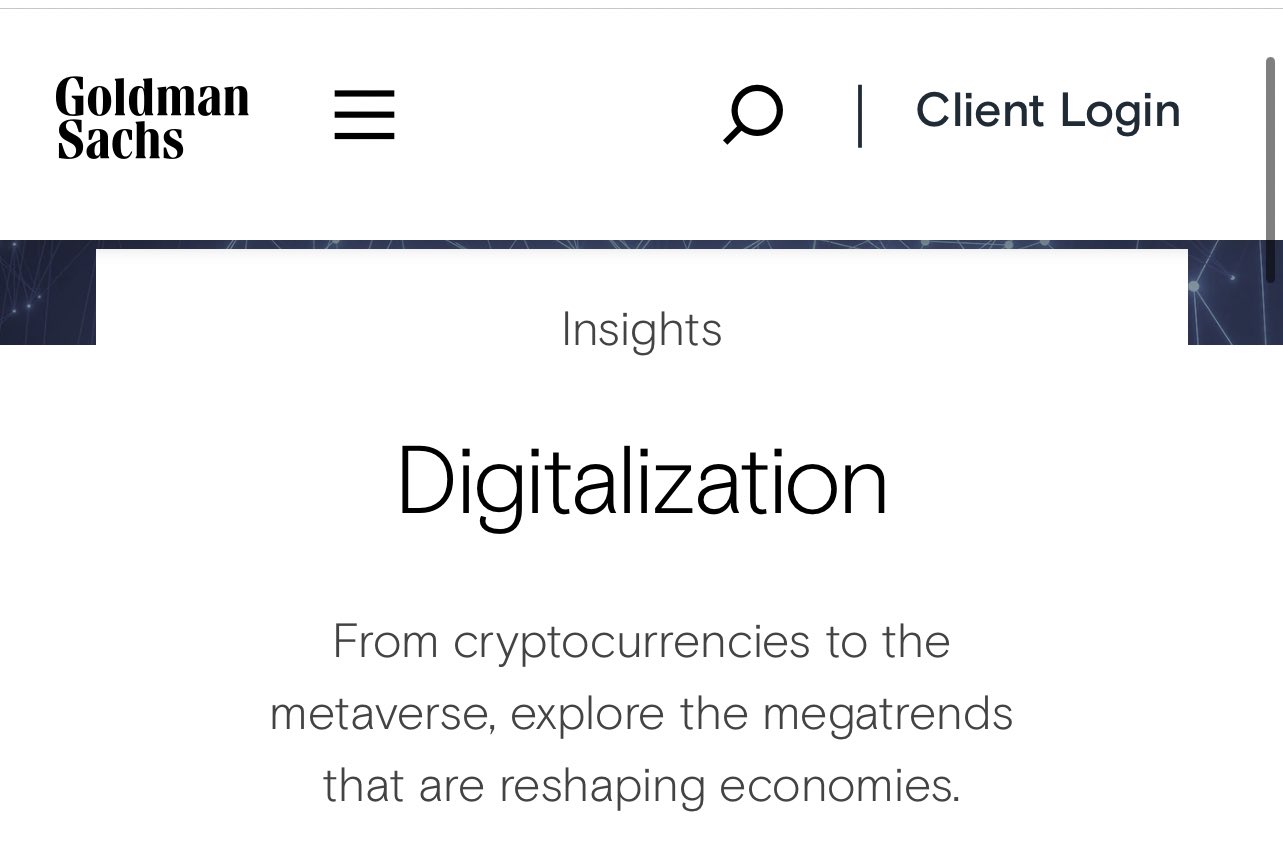}
	\caption{The landing page of global\\financial giant Goldman Sachs shows the hype.}
	\label{fig:goldmanFront}
\end{figure}

 
Fortunately, the whole landscape of money and \href{https://www.omfif.org/futureofpayments2021/}{value transfer is changing}. Huge global financial players are entering the space. HSBC have \href{https://sandboxgame.medium.com/hsbc-to-become-the-first-global-financial-services-provider-to-enter-the-sandbox-c066e4f48163}{just bought} metaverse `land' in The Sandbox, JP Morgan have \href{https://www.forbes.com/sites/ronshevlin/2022/02/16/jpmorgan-opens-a-bank-branch-in-the-metaverse-but-its-not-for-what-you-think-its-for/?sh=2fbd1e90158d}{opened a `lounge'} in another. The worlds largest hedge fund Bridgewater is stepping into \href{https://uk.finance.yahoo.com/news/bitcoin-latest-price-crypto-ray-dalio-bridgewater-investment-fund-ethereum-094946686.html}{acquisition of digital assets}, and the world's largest pension fund manager Blackrock \href{https://blog.coinbase.com/coinbase-selected-by-blackrock-provide-aladdin-clients-access-to-crypto-trading-and-custody-via-b9e7144f313d}{partnered with crypto behemoth Coinbase} and is adding these assets to their management engine (which manages tens of trillions of dollars). America's oldest bank \href{https://www.bnymellon.com/emea/en/about-us/newsroom/press-release/bny-mellon-launches-new-digital-asset-custody-platform-130305.html}{BNY Mellon}, and even the Nasdaq stock exchange are \href{https://www.nasdaq.com/articles/nasdaq-to-launch-institutional-bitcoin-crypto-custody-services\%3A-report}{offering service to institutional clients}, and Fidelity asset management are about to add \href{https://www.wsj.com/articles/fidelity-weighs-bitcoin-trading-on-brokerage-platform-11663008698}{Bitcoin to their pension plans}, and they have asserted their view that Bitcoin is \href{https://www.fidelitydigitalassets.com/sites/default/files/documents/bitcoin-first.pdf}{different, and more meaningful} than the rest of the industry. Fidelity are also offering a \href{}{dedicated metaverse tradable fund}, and considering more direct product offerings through their retail investment engine. \href{https://www.citivelocity.com/citigps/metaverse-and-money/}{Citigroup have a minisite} dedicated to ``Metaverse and Money''. The front page of Goldman Sachs recently says it all (Figure \ref{fig:goldmanFront}).\par
In Gartners \href{https://www.itp.net/emergent-tech/gartner-says-nfts-metaverse-web3-will-expand-immersive-experiences}{2022 hype cycle report} one of their three ``trend themes'' says: \textit{``The future of digital experience is immersive. A collection of emerging technologies supports such experiences through dynamic virtual representations, environments and ecosystems of customers and people, as well as new modes of user engagement. With these technologies, individuals can control their own identities and data and experience virtual ecosystems that can be integrated with digital currencies. These technologies help reach customers in new ways to strengthen or open new revenue streams.
The technologies to watch that deliver evolving and expanding immersive experiences are metaverse, non-fungible tokens (NFTs), super apps and Web3, decentralized identity, digital humans, digital twin of the customer and internal talent marketplaces.''}\par
\begin{figure}[ht]\centering 	\includegraphics[width=0.5\linewidth]{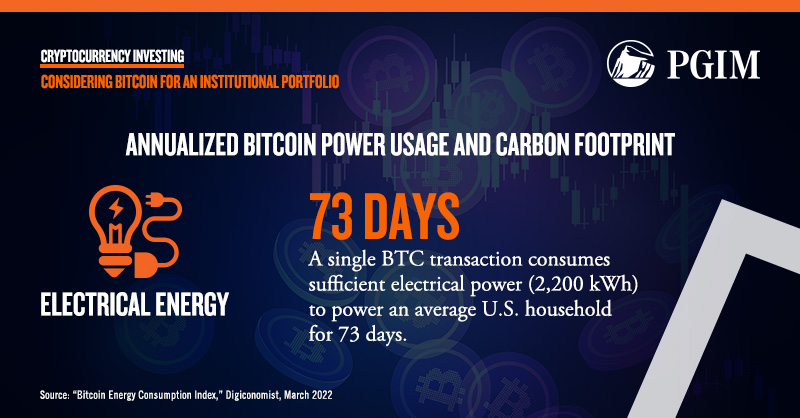}
	\caption{PGIM cite `digiconomist', a prominent critic.}
	\label{fig:pgim}
\end{figure}

Of their recent investments KPMG global said: \textit{``We've invested in a strong cryptoassets practice and we will continue to enhance and build on our capabilities across Decentralized Finance (DeFi), Non-Fungible Tokens (NFTs) and the Metaverse, to name a few''}. This is not to say that all fund managers are so positive. PGIM who manage over a trillion pounds globally have come out very strongly against the technology, with a \href{https://www.pgim.com/megatrends/cryptocurrency-investing/bitcoin?}{slew of reports} to warn off investors (Figure \ref{fig:pgim}).\par
It's possible that for such huge organisations it makes better business sense to take a punt on hype bubbles like this, than to do a proper due diligence with a team of internal staff who understand their business. These endorsements should be taken with a large pinch of salt. As \href{https://newsletter.fintechtakes.com/p/metaverse-branches?s=r}{Alex Johnson says}: \textit{``At some point in the future, it’s possible that the digital worlds being built today will have aggregated sufficient user attention and engagement that financial services companies will need to invest in the metaverse as an acquisition and customer service channel. But we’re not there yet. Until the metaverse is a little less empty, resist the temptation to colonize it with branches and billboards.''}\par
Meanwhile, Meta (ex Facebook) are launching their own \href{https://archive.ph/coyp2}{META Web3 and metaverse} token after abandoning Libre, their global cryptocurrency. Libre became Diem, then was quietly acquired by Silvergate bank, who likely integrated it into their SEN settlement network. Following the collase of Silvergate the bank was sold on without the SEN network, marking an ignominious end to the technology which possibly started the the rush to central bank digital currencies. \href{https://www.coinbase.com/blog/announcing-coinbase-google-cloud}{Google have formed a strategic partnership} with Coinbase, and \href{https://blog.youtube/inside-youtube/innovations-for-2022-at-youtube/}{recently blogged}: \textit{``Web3 also opens up new opportunities for creators. We believe new technologies like blockchain and NFTs can allow creators to build deeper relationships with their fans. Together, they'll be able to collaborate on new projects and make money in ways not previously possible. For example, giving a verifiable way for fans to own unique videos, photos, art, and even experiences from their favourite creators could be a compelling prospect for creators and their audiences. There's a lot to consider in making sure we approach these new technologies responsibly, but we think there's incredible potential as well. Finally, we couldn't have a piece about innovation without touching on the metaverse! We're thinking big about how to make viewing more immersive. ''}\par
It's already the case that the recent bubble of \href{https://www.forbes.com/sites/paultassi/2022/03/10/interest-in-nfts-and-the-metaverse-is-falling-fast/?}{hype is dwindling}, but the enormous investment into teams and startups will potentially bear fruit in the next couple of years, and this perhaps has implications for small and medium-sized enterprises.\par
It's fortunate timing for this book that the UK government has signalled enthusiasm for so called `stablecoins' at the same time that the Bitcoin network is being upgraded to transmit these GBP equivalent tokens around. This gives us a very good idea what it is we can build into our application stack. In the UK the government has stated it's ambition to be a \href{https://www.gov.uk/government/news/government-sets-out-plan-to-make-uk-a-global-cryptoasset-technology-hub}{global cryptoasset technology hub}, and announced, then scrapped plans for the Royal Mint to issue a (novelty) NFT. Fuller, Economic Secretary to the Treasury \href{https://drive.google.com/file/d/19ZYKLeT-ds3TueTpqSM22MUqB4gmN_Pl/view}{said in a speech}: \textit{``We want to become the country of choice for those looking to create, innovate and build in the crypto space [...] By making this country a hospitable place for crypto technologies, we can attract investment, generate new jobs, benefit from tax revenues, create a wave of ground breaking new products and services, and bridge the current position of UK financial services into a new era.''}\par
Their outline plans for \href{https://www.gov.uk/government/news/uk-sets-out-plans-to-regulate-crypto-and-protect-consumers}{`robust regulation'} were published after these seemingly supportive moves, and with the public consultation drawing to a close they have signalled their willingness to differentiate from Europe \href{https://www.cnbc.com/2023/04/18/britain-could-see-crypto-regulation-in-12-months-lawmaker-says.html}{within a year}. Like the assertion by major global businesses it is too early to tell how `sticky' these claims are. Indeed the findings of a recent treasury committee looking at the sector suggest that there is much work to do, with \href{https://committees.parliament.uk/committee/158/treasury-committee/news/175634/treasury-committee-85-of-crypto-firms-failed-to-meet-minimum-standards-according-to-fca/}{85\% of companies} failing to comply with \textit{existing} law. The UK legal system is clear in it's view that all crypto assets \href{https://blockchain.bakermckenzie.com/2020/02/03/uk-court-confirms-bitcoins-status-as-property/}{are `property'}.\par
A Law Commission consultation on ``digital assets'' \href{https://s3-eu-west-2.amazonaws.com/lawcom-prod-storage-11jsxou24uy7q/uploads/2022/07/Digital-Assets-Summary-Paper-Law-Commission-1.pdf}{has proposed a new \textbf{third category} of property}:
\textit{\begin{itemize}
\item it is composed of data represented in an electronic medium, including in the
form of computer code, electronic, digital or analogue signals;
\item it exists independently of persons and exists independently of the legal system;
\item it is rivalrous such that use by one  prejudices the ability of others;
\end{itemize}
}
Consensus seems to be that this is a thorough paper, and demonstrates strong knowledge of digital assets by the authors. \par
\begin{figure}[ht]\centering 
	\includegraphics[width=0.9\linewidth]{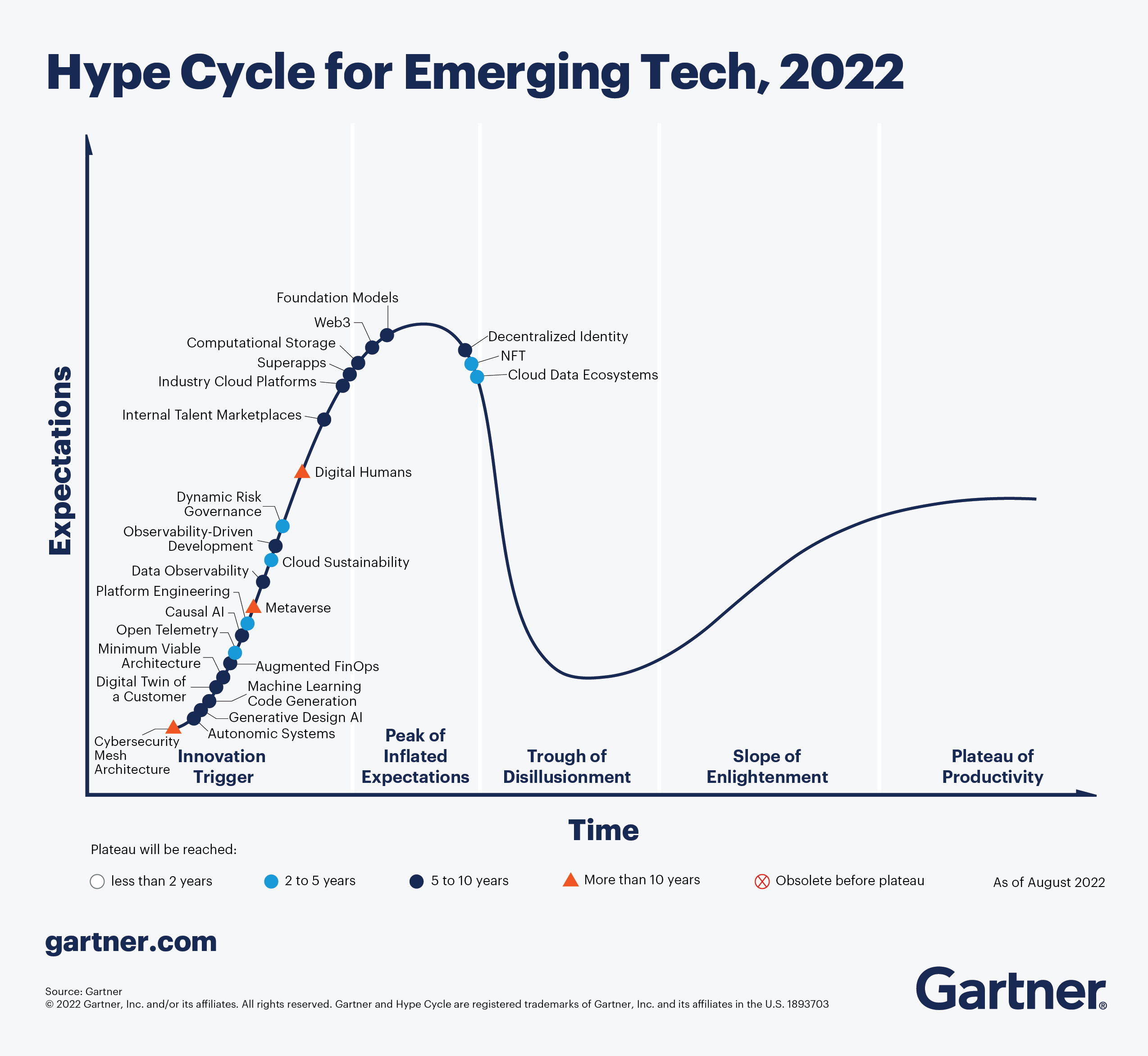}
	\caption{The Gartners Hype Cycle for 2022.}
	\label{fig:gartners}
\end{figure}

Gartner's \href{https://en.wikipedia.org/wiki/Gartner_hype_cycle}{hype cycle} 2022 features \href{https://www.gartner.com/en/articles/what-s-new-in-the-2022-gartner-hype-cycle-for-emerging-technologies}{Web3, distributed identity, NFTs, and Metaverse} and can be seen in Figure \ref{fig:gartners}.\par
\section{Summary}
With all this attention it seems timely to explore the potential of recent technologies, which can address collaborative mixed reality interactions in \textit{business to business} (B2B), \textit{business to customer} (B2C), and the newer C2C (social commerce; \textit{creator to consumer, customer to customer, consumer to consumer\cite{jones2008trust})}. \par 
This book seeks to overview and explain the available open source technologies. It supports an open source \href{https://github.com/flossverse/product}{github repository} which enables SMEs to access these emergent platforms and ecosystems. It aims to build toward a minimum viable product for trust minimised transfer of value within a social immersive space, but also across all internet connected devices.\par
Referencing is in two styles; academic works and books are numeric, while opinion pieces, gray statistics, and pertinent news articles are hyperlinked from the text. This hybrid style yields about twice the citation density of a normal PhD thesis, which is a lot. For this reason the normal blue hyperlink colour was eschewed in favour of a more aesthetic ``gray''. \par

\chapterimage{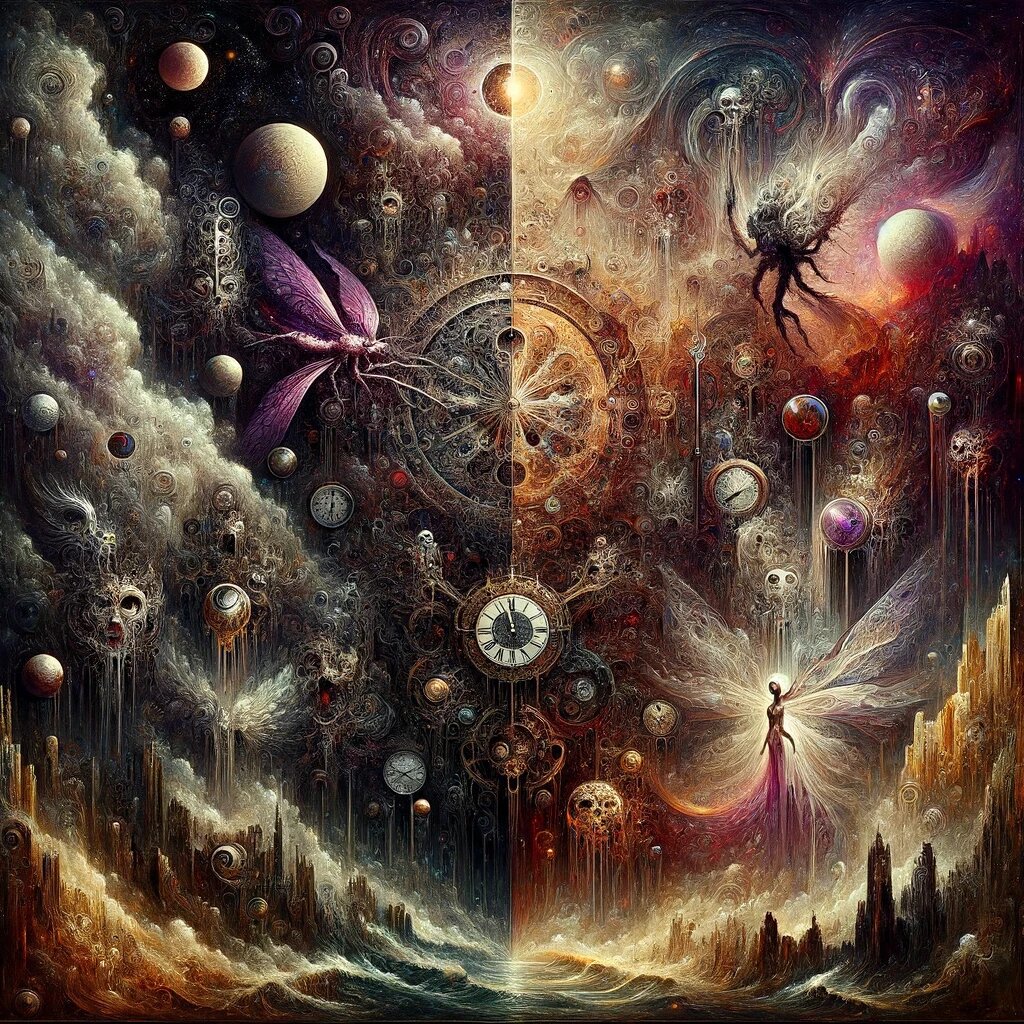}
\chapter{Decentralisation \& The Web}
\section{Semantic web}
The ``semantic web'' definition of Web3.0 has been somewhat overhauled by other innovations in decentralised internet technologies, now evolving toward the slightly different Web3 moniker. Tim Berners Lee (of WWW fame) first mentioned the semantic web in 1999 \cite{semanticWeb}.\par
``I have a dream for the Web [in which computers] become capable of analyzing all the data on the Web – the content, links, and transactions between people and computers. A "Semantic Web", which makes this possible, has yet to emerge, but when it does, the day-to-day mechanisms of trade, bureaucracy and our daily lives will be handled by machines talking to machines. The "intelligent agents" people have touted for ages will finally materialize.''\par
Attention developed around three core themes, ubiquitous availability and searchability of data, intelligent search assistants, and highly available end points such as phones, and `internet of things' devices. This is certainly manifesting in home devices, but few people think of this as a Web3 revolution. 
Since ratification of the standards by the \href{https://www.w3.org/standards/semanticweb/}{World Wide Web (W3C) consortium} it seems that their imperative toward decentralisation has become lost. Instead, it can be seen that Facebook, Amazon, Google, and Apple have a harmful oligopoly on users data \cite{costigan2018world}. This is at odds with Berners-Lee's vision, and he has recently \href{https://thenextweb.com/news/web-inventor-tim-berners-lee-screw-web3-my-decentralized-internet-doesnt-need-blockchain/}{spoken out about this discrepancy}, and attempted to \href{https://www.cnbc.com/2022/11/04/web-inventor-tim-berners-lee-wants-us-to-ignore-web3.html}{refocus the media} onto Web3.0. \par
It is worth taking a look at his software implementation called \href{https://solidproject.org}{Solid}, which is far more mindful of the sovereignty of user data.\par
``Solid is an exciting new project led by Prof. Tim Berners-Lee, inventor of the World Wide Web, taking place at MIT. The project aims to radically change the way Web applications work today, resulting in true data ownership as well as improved privacy. Solid (derived from "social linked data") is a proposed set of conventions and tools for building decentralized social applications based on Linked Data principles. Solid is modular and extensible and it relies as much as possible on existing W3C standards and protocols.'' \par
Excitement around this kind of differentiated trust model, hinted at in ubiquitous availability of data (and implemented explicitly in Solid), has led to exploration of different paths by cryptographers, and this will be described later. For instance, one of the main developers of Solid, \href{https://github.com/melvincarvalho/}{Carvelho}, is now a leading developer and propotent of Nostr, another very interesting option which will be described later. This technology space is prolific, but still comparatively young and small.\par
\section{Spatial web}
\begin{figure}
  \centering
    \includegraphics[width=\linewidth]{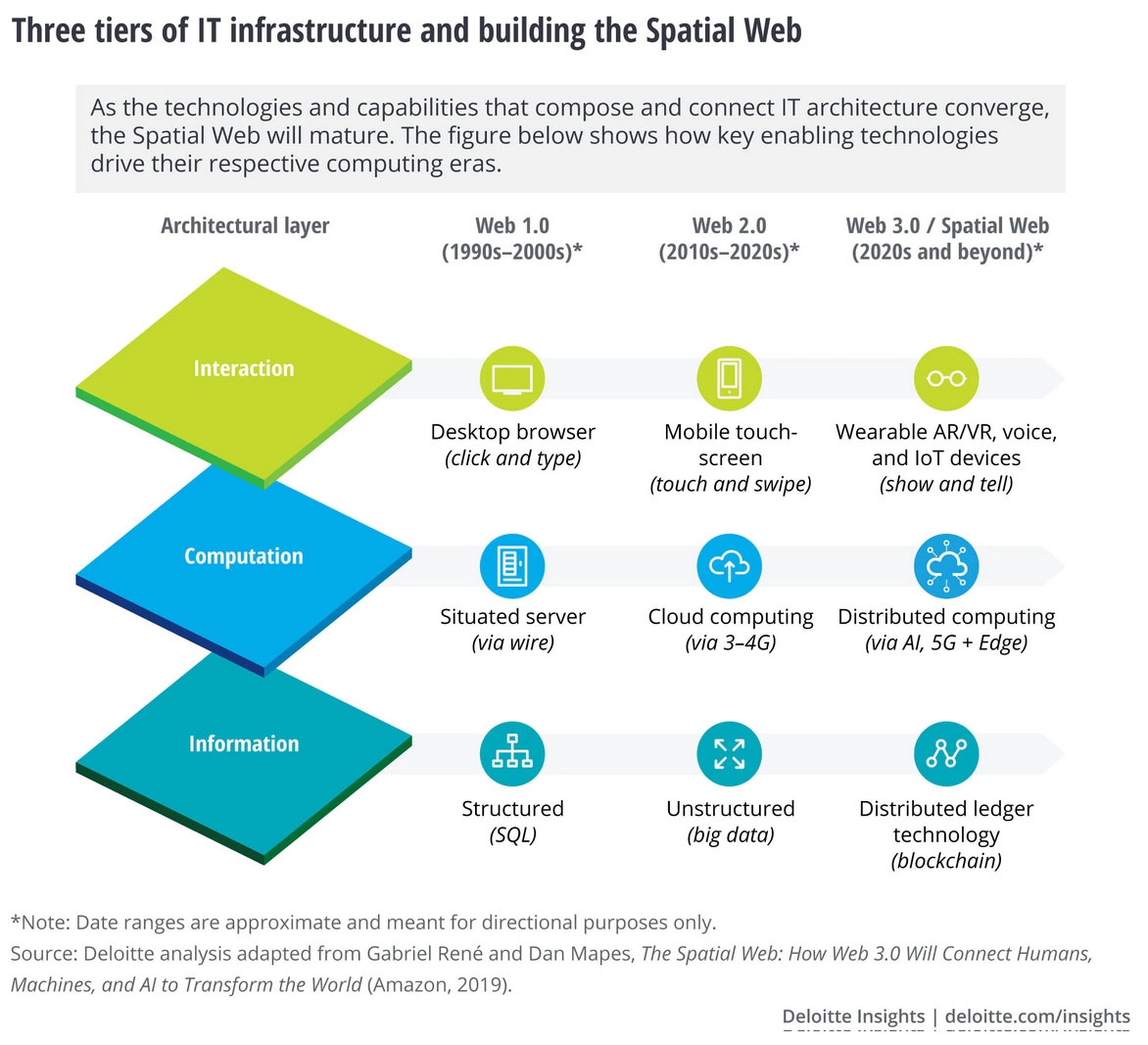}
  \caption{\href{https://www2.deloitte.com/us/en/insights/topics/digital-transformation/web-3-0-technologies-in-business.html}{Deloitte Spatial Web Overview} Reused with permission.}
  \label{fig:deloitteSpatial}
\end{figure}

``The Spatial Web'', a blurring of the boundaries between digital and geospatial physical objects, seems to have developed from the strands in the original W3C scope around devices in the real world. It has been concentrating around AR and VR but is being marketed and amplified with the same references to availability of data (See Figure \ref{fig:deloitteSpatial} from a Deloitte accounting report). This too is finding little traction in practice, though obviously the component technologies continue to enjoy rapid development. Nonetheless, this interpretation of Web3 becomes valuable when examining Metaverse later.\par
\section{Web3}
More recently Web3 is \href{https://trends.google.com/trends/explore?date=all&q=web3}{being touted} as a way to connect content creators directly to content consumers, without centralised companies acting as gatekeepers of the data. It implies that all users have a cryptographic key management system, to which they attach metadata, that they make requirements of peers with whom they communicate, and that they maintain trust `scores' with peers.\par
\begin{figure}
  \centering
    \includegraphics[width=\linewidth]{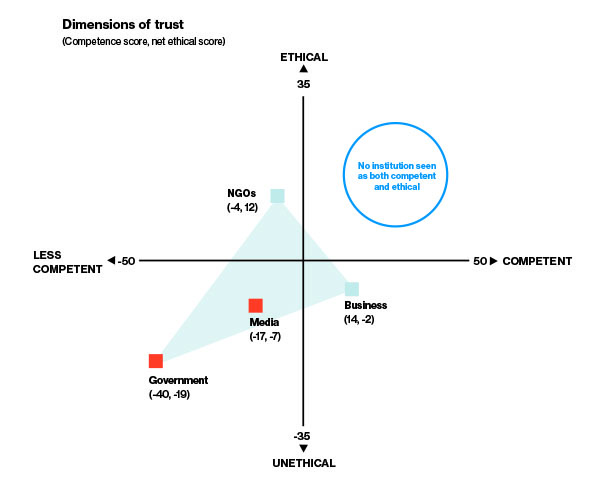}
  \caption{\href{https://www.edelman.com/trust/2020-trust-barometer}{Edelman 2020 trust barometer} [rights requested]}
  \label{fig:trustbarometer}
\end{figure}

It seems likely that this new model is less driven by a market need, and more by the high availability of tools which allow this to happen (the ecosystems described later). Add to this a social response to the \href{https://finance.yahoo.com/news/meta-facebook-worst-company-of-the-year-yahoo-finance-165345819.html}{collapse in trust of companies such as Facebook} and other \href{https://reb00ted.org/tech/20220727-end-of-social-networking/}{social media platforms}\cite{torok2017cascading} (Figure \ref{fig:trustbarometer}). There is perhaps a wish by consumers to pass more of the economic incentive to content creators, without the `rent seeking' layer afforded by businesses, and a healthy dose of mania driven market speculation. \href{https://www.edelman.co.uk/sites/g/files/aatuss301/files/2022-01/2022\%20Edelman\%20Trust\%20Barometer_UK.pdf}{Edelman's latest trust report} is shocking, finding that trust in all institutions has slumped recently to all time lows, and their global survey found that: \textit{``Nearly 6 in 10 say their default tendency is to distrust something until they see evidence it is trustworthy. Another 64\% say it’s now to a point where people are incapable of having constructive and civil debates about issues they disagree on. When distrust is the default – we lack the ability to debate or collaborate.''}

\subsection{Emerging consensus}
The recent hype cycle ignored the legacy definitions described above and instead focusing almost exclusively on Ethereum based peer-to-peer projects. It can be seen that the description is somewhat in the eye of the beholder.\par
It's possible to frame this Ethereum Web3 as a hugely complex and inefficient digital rights management system (DRM). DRM is something that users of the internet are increasingly familiar and comfortable with. It's somewhat debatable whether decentralising this is worthwhile. The thesis of the developers of the technology seems to be that without it, control of `value' will accrete over time, to one or more hegemonic controlling entities. It's a strong argument, but there is a \href{https://moxie.org/2022/01/07/web3-first-impressions.html}{substantial counter argument} emerging that users just don't want this stuff. The nervousness of legislators in the USA to the attempt by Facebook/Meta to enter this peer-to-peer value transmission space is telling in terms of the perception of who is driving Web3.\par
\begin{figure}
  \centering
   \includegraphics[width=\linewidth]{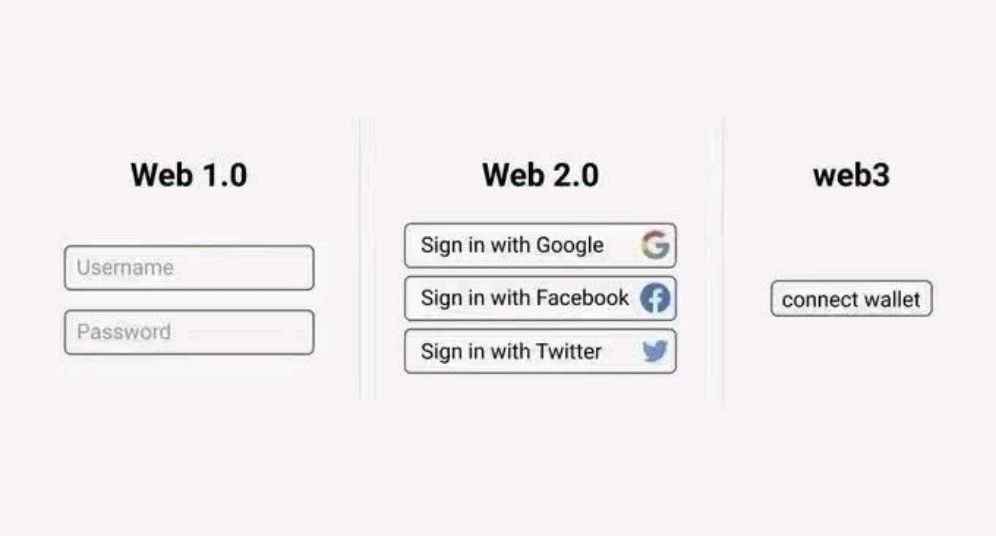}
 \caption{A meme showing differing approached to logging in on a website.}
 \label{fig:web1web2web3}
\end{figure}

Throughout 2022 there was much furore on the internet over what Web3 might be, and who it `serves'. Enthusiasts feel that products such as \href{https://blog.spruceid.com/sign-in-with-ethereum-is-a-game-changer-part-1/}{Sign-In with Ethereum} (EIP-4361) might give users choice over their data sovereignty, and a meme to this effect is seen in Figure \ref{fig:web1web2web3}. In practice though users are expecting to use badly written, buggy, economically vulnerable `crypto' wallets to log into websites. The quality of this wallet software is improving of late with the so called ``wallet wars'' seeing commerce grade offerings from Coinbase and shares platform `Robinhood'. These two companies alone have over 100 million users. It's likely that these wallets will evolve to offer the full spectrum of Web3 functionality. With that said it doesn't seem to make much sense yet on the face of it. There are in fact examples of the technology completely failing at censorship resistance. Popular `Web3' browser extension Metamask and NFT platform Opensea have both \href{https://www.forbes.com/sites/stevenehrlich/2022/03/03/iranian-venezuela-users-abruptly-dropped-from-major-crypto-platforms-as-russian-sanctions-grow/?sh=22bcabc470b0}{recently banned countries} in response to global sanction pressure. This failure to meaningfully decentralise will be explored further in the distributed identity section. \par
\begin{figure}
  \centering
    \includegraphics[width=\linewidth]{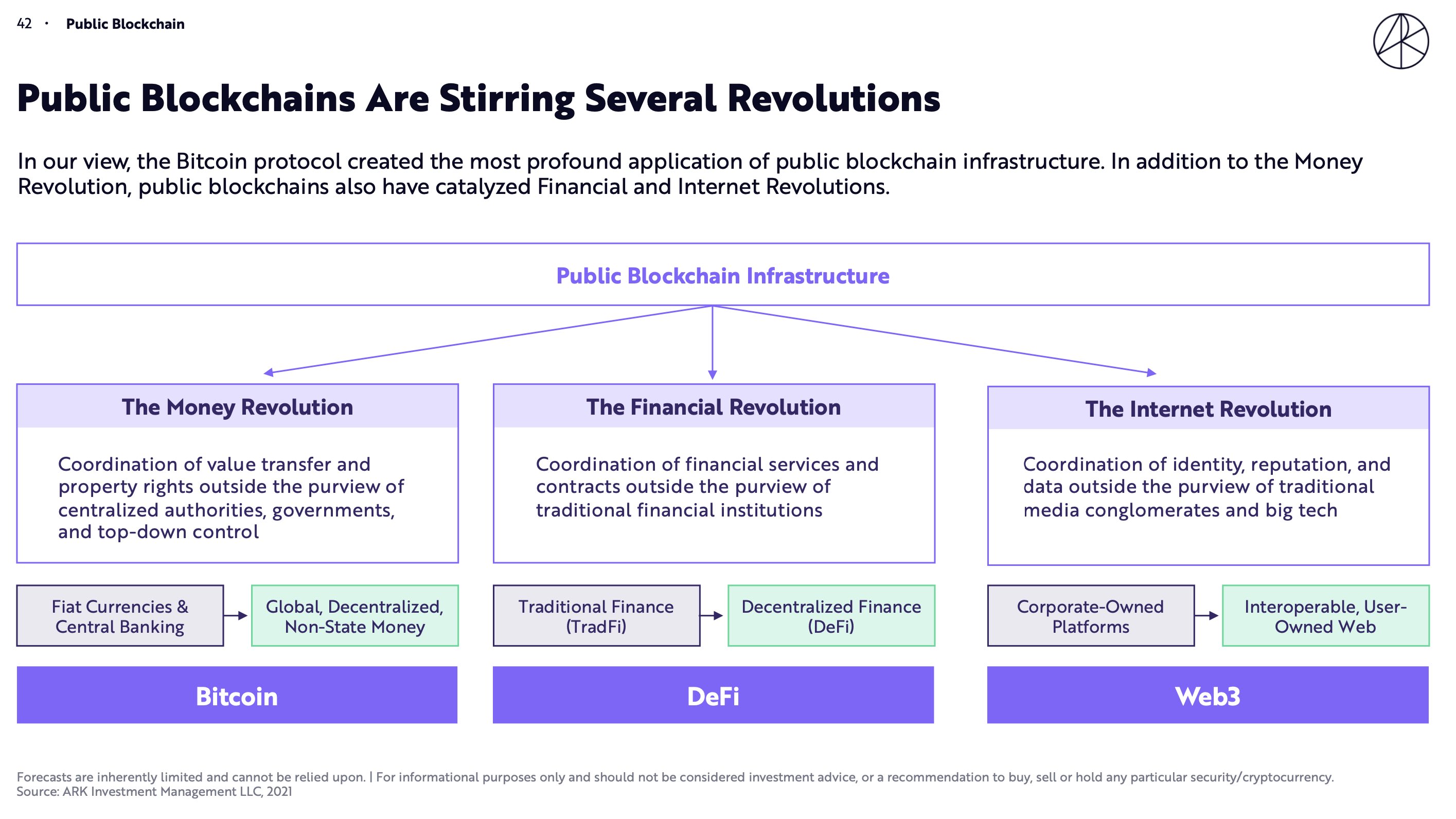}
  \caption{\href{https://twitter.com/wintonARK/status/1486143239753060353}{ARK slide on Web3.} Rights requested}
  \label{fig:ARKWeb3}
\end{figure}

Of their 2022 \href{https://research.ark-invest.com/thank-you-big-ideas-2022?submissionGuid=0937b1ae-9e11-4b46-ae03-6cd8d2f8301b}{`Big Ideas' report}, ARK investment LLC (who manage a \$50B tech investment) \href{https://www.ark-bigideas.com/2022/en/pages/download}{said the following} (Figure \ref{fig:ARKWeb3}), which connects some of the dots already mentioned, and leads us into the next section which is Blockchain and Bitcoin:\par
\textit{``While many (with heavily vested interests) want to define all things blockchain as web3 we believe that web3 is best understood as just 1 of 3 revolutions that the innovation of bitcoin has catalyzed.
\begin{itemize}
\item The Money Revolution
\item The Financial Revolution
\item The Internet Revolution''
\end{itemize}}
This new hyped push for Web3 is being driven by enormous venture capital investment. A16Z are a \href{https://a16z.com/2022/01/07/9b-to-build-the-future/}{major player} in this new landscape and have released their \href{https://a16z.com/2022/01/07/how-to-build-a-better-internet-10-principles-for-world-leaders-shaping-the-future-of-web3/}{ten principles} for emergent Web3. Note here that A16Z are (like so many others) probably a \href{https://twitter.com/coryklippsten/status/1592242420137148416}{house of cards}.
\begin{itemize}
\item Establish a clear vision to foster decentralized digital infrastructure
\item Embrace multi-stakeholder approaches to governance and regulation
\item Create targeted, risk-calibrated oversight regimes for different web3 activities
\item Foster innovation with composability, open source code, and the power of open communities
\item Broaden access to the economic benefits of the innovation economy
\item Unlock the potential of DAOs
\item Deploy web3 to further sustainability goals
\item Embrace the role of well-regulated stablecoins in financial inclusion and innovation
\item Collaborate with other nations to harmonize standards and regulatory frameworks
\item Provide clear, fair tax rules for the reporting of digital assets, and leverage technical solutions for tax compliance
\end{itemize}
This list seems targeted toward the coming regulatory landscape, and could be considered at odds with the original tenants of an organically emergent, decentralised internet. Indeed principles such as `furthering sustainability goals' seem downright incongruous. The community they claim to wish to support here are openly critical of these major institutional players and their motives, with even more pointed criticisms \href{https://www.profgalloway.com/web3/}{coming from outside of the Web3}. This book and lab steer well clear of these companies and their applications.\par
Dante Disparte, chief strategy officer of `Circle' venture capital, said in testimony to a US senate hearing; that Web 1 was `read', Web 2 was `read write', and that Web 3 will `read write own'. The important takeaway here is not so much this oft quoted elevator pitch for Web3, but the fact that legislative bodies now consider this technology a force which they need to be aware of and \href{https://a16z.com/2021/12/17/prediction-for-the-new-year-a-web3-midterm/}{potentially contend with}.\par
Jeremy Allaire, again of Circle', talks about the recent legislative order in the USA as follows:
\textit{``this is a watershed moment for crypto, digital assets, and Web 3, akin to the 1996/1997 whole of government wakeup to the commercial internet. The U.S. seems to be taking on the reality that digital assets represent one of the most significant technologies and infrastructures for the 21st century; it's rewarding to see this from the WH after so many of us have been making the case for 9+ years.''}\par
\begin{figure}
  \centering
    \includegraphics[width=\linewidth]{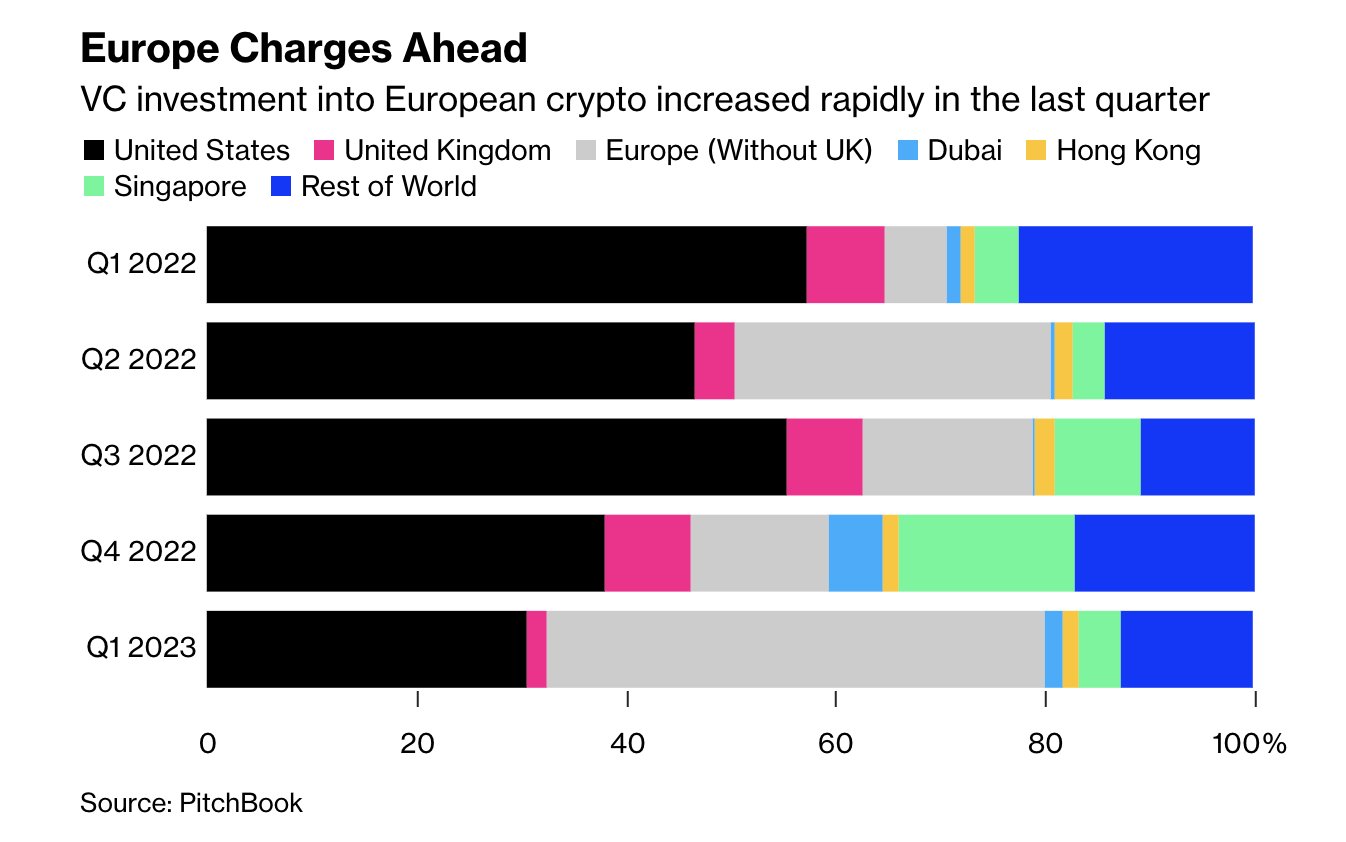}
  \caption{\href{https://twitter.com/paddi_hansen/status/1655883224726241281}{``Regulatory clarity attracts capital \& entrepreneurs from around the world.''}}
  \label{fig:pitchbook}
\end{figure}

We will see in the following chapters that participation in this new Web3 is contingent on owning cryptocurrencies. \href{https://www.finder.com/uk/cryptocurrency-statistics}{It's estimated} that about 6\% of people in the UK own some cryptocurrency, with skews to both younger demographics, and smaller holdings. The legislative landscape in the UK is comparatively strict with \href{https://uk.news.yahoo.com/perverse-impacts-anti-money-laundering-144239343.html}{questionable} ``know your customer / anti money laundering'' (KYC/AML) data collection \href{https://www.gov.uk/guidance/money-laundering-regulations-your-responsibilities}{mandated in law}. Users of UK exchanges must provide a great deal of personal financial information, and undertake to prove that the wallets they are withdrawing to are their own. From the perspective of the UK SME it seems this seriously limits the potential audience for new products. Europe meanwhile has recently voted through even more restrictive regulation, applying the ``\href{https://www.europarl.europa.eu/legislative-train/theme-an-economy-that-works-for-people/file-revision-of-the-regulation-on-transfers-of-funds}{transfer of funds regulation}'' to all transactions coming out of exchanges, enforcing a database of all addresses between companies, and reporting transactions above 1000 Euros to authorities. They have narrowly avoided enforcing KYC on all transfers to private wallets, but have capped transactions at 1000 Euros. The recent \href{https://www.consilium.europa.eu/en/press/press-releases/2022/06/30/digital-finance-agreement-reached-on-european-crypto-assets-regulation-mica/}{``Markets in Crypto Assets (MiCA)} legislation imposes overheads that may make it harder for smaller businesses in the sector to operate within the EU, but is has been cautiously welcomed by established players (Figure \ref{fig:pitchbook}, who have been hungry for clarity. It is certainly far short of the `ban' seen in China, and the regulation be enforcement in the USA.

\begin{itemize}
\item European Parliament approved EU's crypto assets framework, MiCA
\item Enforcement clock starts in June, with 12-18 months for rules to kick in
\item MiCA offers license tailored to crypto asset services and stablecoin issuers
\item Regulation refrains from covering decentralized finance or non-fungible tokens
\item Stablecoin issuer rules boost consumer confidence, potentially increasing institutional comfort
\item Transfer of Funds regulation passed, imposing stronger surveillance and identification requirements for crypto operators
\item Regulations described as world-first and end of Wild West era for crypto assets
\item MiCA represents a crucial step forward for crypto industry, providing comprehensive set of rules
\item Crypto firms must be licensed by the EU and comply with money laundering and terrorism finance safeguards to serve EU customers
\item Concerns about weakened privacy due to reporting standards in the name of customer safety and national security
\item Binance CEO supports MiCA, calling it a pragmatic solution
\item EU's MiCA could become a global template for international companies
\item UK, now outside the EU, is setting similar stablecoin and crypto asset service rules
\end{itemize}
Germany is bringing forward legislation allowing the `tokenisation' of legacy instruments such as stocks, though it's far from clear what the value of this would be, except perhaps lowering risk for custodians. It seems that this EU position has prompted the UK government to seize the potential competitive advantage offered, and there will be more on this later. Japan meanwhile has gone so far as to \href{https://cointelegraph.com/news/japanese-prime-minister-says-gov-t-investment-in-digital-transformation-will-include-metaverse-nfts}{make an announcement} about supporting the technologies at a national level.\par
It's a complex evolving narrative, and clearly contradictions are common. Right now there seems little appeal for stepping into Web3. Into the confusion, this book advances a narrow take, and toolset, which might extract some value from the technologies, while maintaining a low barrier to entry.\par 
\section{Example applications}
It's handy here to get a feel for what this looks like. These aren't things that this book wishes to contribute to, or even have a firm opinion on, they're just representative of current activity in the decentralised web space.
\subsection{Veilid - A Peer-to-Peer Privacy Mesh Project}

Veilid is an open-source, mobile-first, networked application framework for building decentralized apps with networking, distributed data storage, and built-in IP privacy without reliance on external services.

\begin{itemize}
\item \textbf{Platforms}: Runs on Linux, Mac, Windows, Android, iOS, and in browsers via WASM. Bindings available in Rust, Dart, and other languages.
\item \textbf{Protocols}: Supports UDP, TCP, WebSockets. DNS only used briefly during bootstrap.

\item \textbf{Encryption}: Uses Ed25519, XChaCha20, BLAKE3 for end-to-end encryption and authentication.
\item \textbf{Storage}: Distributed hash table for data records close to node keys. Popular data replicated.
\item \textbf{Routing}: Nodes help each other connect. Routing based on node IDs. Private routing over encrypted hops.
\item \textbf{Goals}: Enable decentralized apps without reliance on centralized corporate systems.
\end{itemize}

Key features include strong cryptography, ability to run on a variety of platforms, distributed and replicated data storage, and private routing to provide IP privacy. The decentralized design aims to avoid issues with centralized and corporate controlled systems.
\subsection{Podcasting2.0}
\href{https://medium.com/@everywheretrip/an-introduction-to-podcasting-2-0-3c4f61ea17f4}{Podcasting 2.0} leverages \href{https://www.rssboard.org/rss-specification}{RSS} (the original dissemination system for podcasts) and the Bitcoin Lightning network, to enable so-called `\href{https://www.youtube.com/watch?v=NO1aDZ6L4NQ&t=1123s}{value for value}' broadcasting. Subscribers use one of a variety of apps to stream micro-transactions of Bitcoin directly to the content creators as they listen to the podcast. No intermediate business takes a cut. Some variation on this model exists, such as John Carvalho's crowd funded podcast ``The Biz'' which progressively unlocks more minutes for everyone based on \href{https://thebiz.pro/about#crowdwall}{crowd funded donations}.
\subsection{Crowd funding}
At time of writing a \href{https://www.constitutiondao.com/}{crowd funding initiative} based around a digital decentralised construct called a DAO (explained later in detail) \href{https://www.coindesk.com/business/2021/12/06/daos-and-the-next-crowdfunding-gold-rush/}{managed to raise \$46 million dollars to bid for a copy of the US constitution} at Southerbys auction house. The attempt narrowly failed, but the press \href{https://www.coindesk.com/business/2021/12/09/what-kickstarter-going-decentralized-means-for-web-3/}{heralded this new era of ``Web3'' economic might}. This model might be the only use for DAOs and is likely just a way to avoid regulatory scrutiny. There is more detail on DAOs later.
\subsection{Distributed exchanges}
There are dozens of decentralised exchanges deployed on various blockchains. These platforms allow users to trade back and forth between various tokens (including `normal money' stablecoins) and charge a fee for doing so. They operate within the logic of the smart contracts \cite{szabo1997formalizing}, within the distributed blockchains. This makes them extremely hard to ban, and as a result they operate in a legal grey area. At the extreme end of this is ``distributed apps'' (dApps) and ``Decentralised Finance'' (DeFi) which allows users access to complex financial instruments without legal or privacy constraints. DeFi will be touched on briefly later.\par
This is a huge area, and of only limited interest to the topics expanded in this book. It's perhaps worth noting \href{https://bitcoin-dex.net/about/index.html}{BitcoinDEX}, which runs in JavaScript in a web browser. It is effectively uncensorable, \href{https://bitcoin-dex.net/tokens.js}{auditable by the user}, and has no counter party risk since it operates entirely in the Bitcoin network. It is clearly an early prototype but manages this complex feature without the more expressive logic of more `modern' public blockchains.
\subsection{NFT marketplaces}
NFT markets are far more centralised services which match `owners' of digital assets with potential buyers. The concept is a staple of the more recent interpretation of Web3, even though in practice these seem to be centralised concerns. \href{https://opensea.io/}{Opensea} claims to be the largest decentralised NFT marketplace, but they have the ability to \href{https://thedefiant.io/sad-frogs-delisted-opensea/}{remove listings} in response to legal challenges. This seems to fly in the face of Web3 principles. NFTs are currently a \href{https://tante.cc/2021/12/17/the-third-web/}{deeply flawed} technology but seem likely to persist and will be covered later.
\subsection{Non blockchain webs of trust}
New products like Slashtags and Nostr (covered later) use a web of trust decentralised peer-to-peer (ish) model which assigns metadata and trust scores to `any' data and connection, with a security model rooted in the Bitcoin cryptographic `keys' but crucially not the bitcoin network. This makes it interoperable with bitcoin but not reliant upon it. In principle this allows users to build complex networks of inherited trust bi-directionally with their networks over time. Every connection to a peer can be a new schema, with individual metadata managed by the user. These are new and have low adoption at this time. The user controls the source of the data and can allow them to be used by centralised services. This flips the authentication and data management paradigm of web around, putting the user in charge of their data. This is a familiar concept to the DID/SSI communities (described later) but with significant investment. As Slashtags and Nostr use keys as endpoints they act as a web of naming and routing, bypassing the existing web infrastructure of DNS. It is likely very complex to use in practice and will be revisited later. Slashtags is being paired with the \href{https://hypercore-protocol.org/}{Hypercore protocol} for peer-to-peer data sharing, more specifically the `hole punching' capability of the hypercore system which ensures connections through firewalls\cite{ford2005peer}. The first application by the affiliated Hyperdivision team is an open source peer-to-peer live video streaming app called \href{https://dazaar.com/}{Dazaar}. Once again, it's not clear yet who wants or needs this bit-torrent style service. 
\subsection{Distributed DNS applications} 
\begin{figure}
  \centering
    \includegraphics[width=\linewidth]{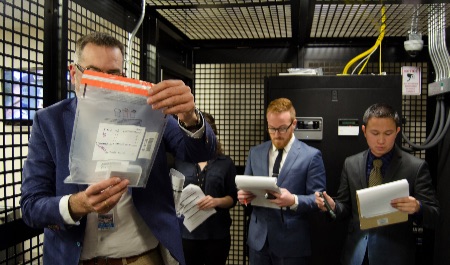}
  \caption{\href{https://www.internetsociety.org/blog/2016/10/watch-live-today-dnssec-root-ksk-ceremony-at-1700-utc/}{DNSSEC ceremony in a faraday cage}}
  \label{fig:dnssec}
\end{figure}

There are many perceived problems with having centralised authorities for overseeing the database which translates between human readable internet names and the underlying machine-readable address notation. The databases which manage this globally are already somewhat distributed, and this distributed trust model is managed through a cryptographic chain of trust called DNSSEC which is capped by a somewhat \href{https://www.iana.org/dnssec/ceremonies}{bizarre key ceremony} seen in Figure \ref{fig:dnssec}. The authority around naming is centralised in ICANN. 
There has been talk for many years about `properly' distributing this database using decentralised/blockchain technologies\cite{karaarslan2018blockchain}. The nature of this problem means that it either moves from control by ICANN, or it does not, and so far it has not, but there are many attempted, and somewhat mature attempts, at this difficult problem. Of these \href{https://www.namecoin.org/}{Namecoin} is the most prominent, and is a fork of Bitcoin. The ubiquity of Bitcoin in such systems is perhaps becoming apparent.
\subsection{Impervious browser}
It might be that the future of Web3 comes in the guise of integrated suites such as the proposed \href{https://newsletter.impervious.ai/impervious-browser-functionality-overview/}{Impervious web browser}. They say that ``without centralized intermediaries'' it features:
\begin{itemize}
\item    Zoom, without Zoom.
\item    Google Docs, without Google.
\item    Medium, without Medium.
\item    WhatsApp, without WhatsApp.
\item    Payments, without banks.
\item    Identity, without the state.
\end{itemize}
This is obviously leading marketing hype, and they're already late for their release deadline, but what they're talking about here is an integration of the components mentioned in this book. If they can get critical mass around this browser then perhaps the Web3 market can be kickstarted. CEO Chase Perkins has \href{https://www.youtube.com/watch?v=2J8v-TMygK8}{recently presented} on this.

\section{Web 4.0}
The EU has released it's \href{https://ec.europa.eu/commission/presscorner/detail/en/ip_23_3718}{positional thinking} on Web 4. This has come pretty much out of nowhere but seems highly relevant to us if it sticks. From the text:
Here is the bullet point list in LaTeX:

\begin{itemize}

\item \textbf{Empowering people and reinforcing skills} to foster awareness, access to trustworthy information and build a talent pool of virtual world specialists. By the end of 2023, the Commission will promote the guiding principles for virtual worlds, put forward by the Citizens' Panel; and will develop guidance for the general public thanks to a ‘Citizen toolbox' by the first quarter of 2024. As specialists on virtual worlds are essential, the Commission will work with Member States to set up a talent pipeline and will support skills development, including for women and girls, through projects funded by the Digital Europe Programme, and for creators of digital content through the Creative Europe programme.

\item \textbf{Business: supporting a European Web 4.0 industrial ecosystem} to scale up excellence and address fragmentation. Currently, there is no EU ecosystem bringing together the different players of the value chain of virtual worlds and Web 4.0. The Commission has proposed a candidate Partnership on Virtual Worlds under Horizon Europe, possibly starting 2025, to foster excellence in research and develop an industrial and technological roadmap for virtual worlds. To foster innovation, the Commission will also support EU creators and media companies to test new creation tools, bring together developers and industrial users, and work with Member States to develop regulatory sandboxes for Web 4.0 and virtual worlds. 

\item \textbf{Government: supporting societal progress and virtual public services} to leverage the opportunities virtual worlds can offer. The EU is already investing in major initiatives, such as Destination Earth (DestinE), Local Digital Twins for smart communities, or the European Digital Twin of the Ocean to allow researchers to advance science, industries to develop precision applications and public authorities to make informed public-policy decisions. The Commission is launching two new public flagships: “CitiVerse”, an immersive urban environment that can be used for city planning and management; and a European Virtual Human Twin, which will replicate the human body to support clinical decisions and personal treatment.

\item \textbf{Shaping global standards for open and interoperable virtual worlds and Web 4.0,} ensuring that they will not be dominated by a few big players. The Commission will engage with internet governance stakeholders around the world and will promote Web 4.0 standards in line with the EU's vision and values.

\end{itemize}
\section{The common thread}
One feature which persists throughout all of these interpretations of Web3 is the need for decentralised trust. According to \href{https://www.coindesk.com/podcasts/the-breakdown-with-nlw/yesterdays-hearing-was-cryptos-most-positive-interaction-with-the-us-government-ever/}{Nathaniel Whittemore}, a journalist for Coindesk, ``The Web3 moniker positions this industry in opposition to big tech''. Alternatively the \href{https://cryptocriticscorner.com/}{many detractors} of the technology think it simply provides avenues for incumbents to experiment with new \href{https://www.cigionline.org/articles/amid-the-hype-over-web3-informed-skepticism-is-critical/}{models of control and monetisation}, \href{https://newsletters.theatlantic.com/galaxy-brain/624cb2ebdc551a00208c1524/crypto-bubble-web3-decentralized-finance/}{increasing systemic risk} at no cost to themselves.\par 
Overall then, perhaps the space is hype, and is certainly \href{https://web3isgoinggreat.com/}{rife with scams}. Fully 24\% of projects in 2022 are \href{https://blog.chainalysis.com/reports/2022-crypto-pump-and-dump-schemes/}{estimated to be built} as `pump and dump' scams. The degree to which it even accomplishes decentralised trust is highly debatable, and meanwhile the limited numbers of Web3 and supporting crypto companies display lamentable cyber security practice themselves, creating \href{https://www.coindesk.com/tag/data-breaches/}{honeypots of personal data} from users of the ecosystem.\par
With that said the component parts necessary to deliver on the promise \textbf{do} exist. If there is to be no central controlling party(s) as in the Web 2 model then nothing can happen without a cryptographically secure underpinning, allowing digital data to be passed around without a prior arrangement.\par
The following chapter will describe how much has been done by computer scientists over the past decades to support that. From this base layer we also get the potential for secure and trust minimised identity management. This nascent field of distributed identity management is explained later. From distributed trust models we can see `trustless' transmission of economic value. The ability to send value from one person to another person or service without a third party. \par
This whole area is `crypto', which is increasingly seeping into the human consciousness, and saw an astonishing \$30B of \href{https://docsend.com/view/nrvsuae85a4dx3jz}{capital investment in 2021} alone. At time of writing the industry is an \href{https://www.coingecko.com/en}{over 1 trillion} dollar market. \par
All the new crypto technologies circling the Web3 narrative are bound tightly together, but there is currently very little meaningful value to be seen.\par
The rest of this book will focus on the trust and value transfer elements of this shift in internet technologies, and attempt to build a case for it's use in decentralised, open source, collaborative mixed reality applications.

\chapterimage{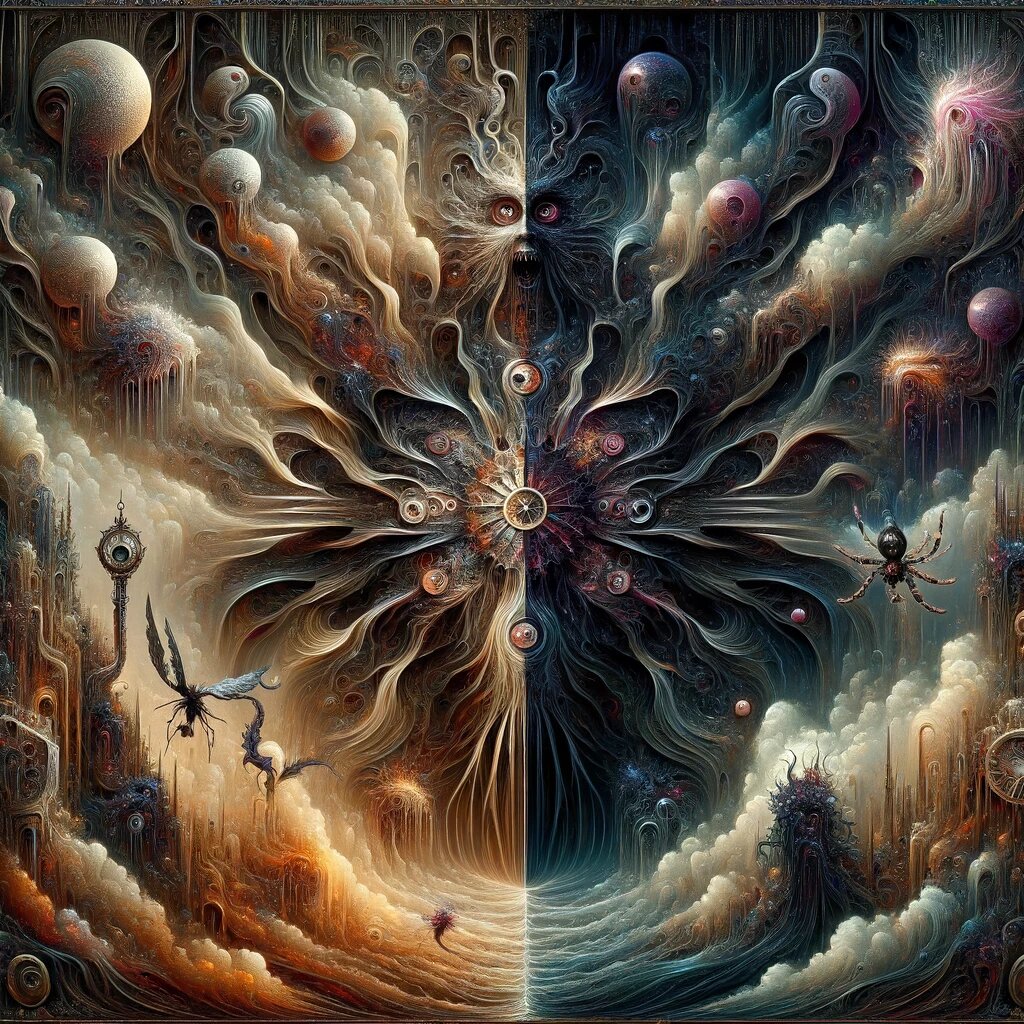}
\chapter{DLT, Blockchain, and Bitcoin}
Distributed ledger technology (DLT) is a data structure distributed across multiple managing stakeholders. A subset of DLT is blockchain, which is a less efficient, immutable data structure with a slightly different trust model. Rauchs et al. of the Cambridge Centre for Alternative Finance provide a detailed taxonomy and conceptual framework \cite{rauchs2018distributed}. It can be seen in their paper that the definitions are somewhat unclear in literature.\par
DLT, and especially blockchain, are rapidly gaining ground in the public imagination, within financial technology companies (FinTech), and in the broader corporate world. \par
The technology and the global legislative response are somewhat immature, and misapplications of both technologies are commonplace. \par
Distributed trust models emerged from cryptography research in the 1970s when Merkle, Diffie, and Hellman at Stanford worked out how to \href{https://medium.com/swlh/understanding-ec-diffie-hellman-9c07be338d4a}{send messages online} without a trusted third party \cite{diffie1976new,merkle1978secure}.\par
Soon after the 1980s saw the emergence of the cypherpunk activist movement, as a reaction to the emerging surveillance state \cite{burnham1983rise, chaum1985security}, a topic which is expanded for this moment in a later chapter. These early computer scientists in the USA saw the intersectionality between information, computation, economics, and personal freedom \cite{lavoie1990prefatory}. Online discussion in the early nineties foresaw the emergence of trans-national digital markets, what would become the WWW \cite{salinCosts, cypherPunkMailList}. The issues of privacy 
 and the exchange of digital value (digital / ecash) 
 were of foremost importance within these discussions 
 and while privacy was within reach thanks to \href{https://www.openpgp.org/about/history/}{``public/private key pairs''}, 
 ecash proved to be a more difficult problem. \par
\begin{figure}
  \centering
    \includegraphics[width=\linewidth]{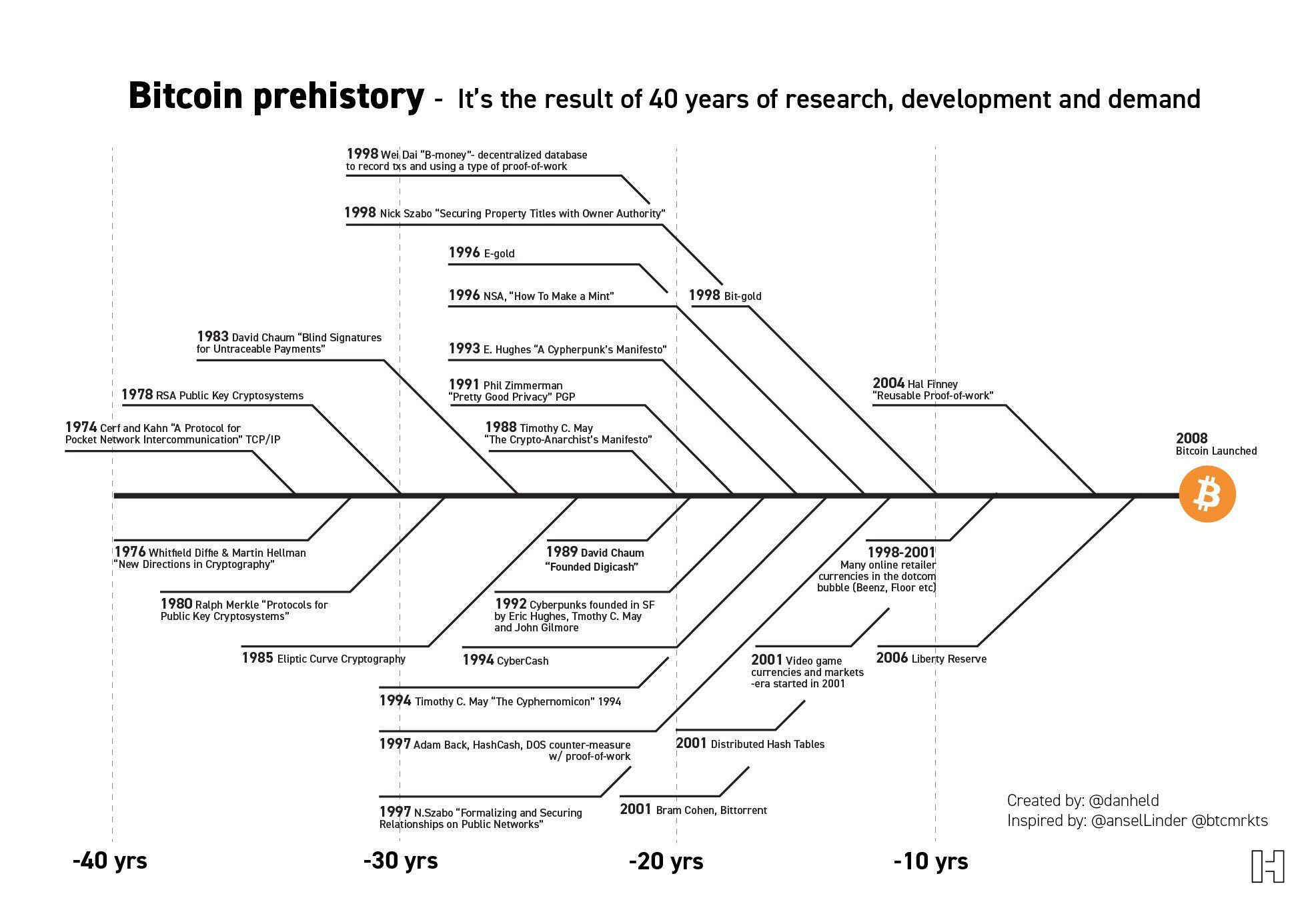}
  \caption{Dan Held: \href{https://www.danheld.com/blog/2019/1/6/planting-bitcoinsoil-34}{Bitcoin prehistory} used with permission.}
  \label{fig:prehistory}
\end{figure}
\begin{figure}
  \centering
    \includegraphics[width=\linewidth]{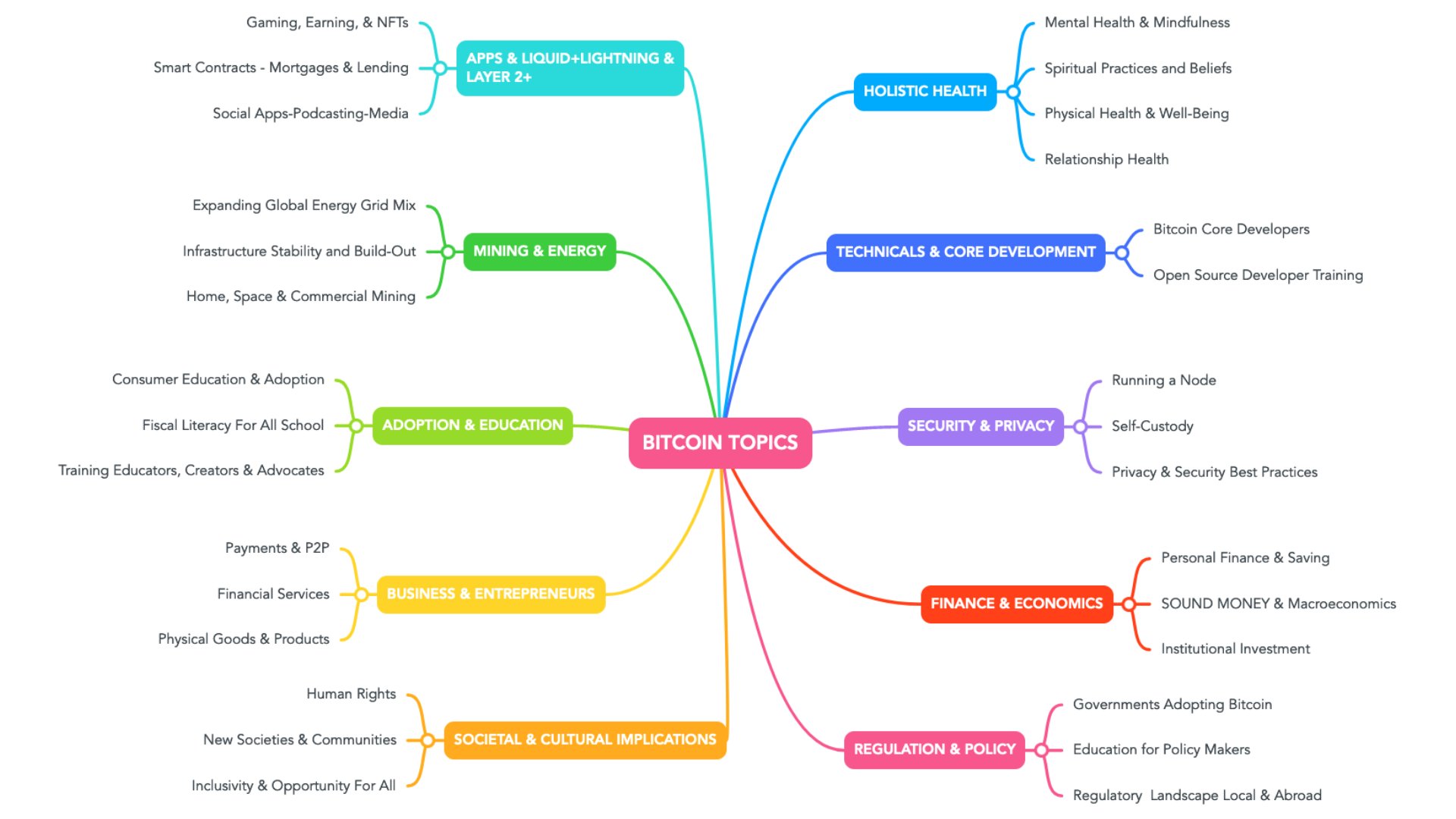}
  \caption{\href{https://twitter.com/djvalerieblove/status/1514703620272394243/photo/1}{Bitcoin Topics} used with permission @djvalerieblove.}
  \label{fig:bitcointopics}
\end{figure}

Adam Back's 1997 `hashcash' \cite{back2002hashcash} paved the way for later work by implementing the concept of what would become `proof of work' \cite{dwork1992pricing, jakobsson1999proofs}. This was built upon by Dai \cite{dai1998b}, Szabo \cite{szabo1997formalizing}, Finney \cite{callas1998openpgp}, and Nakamoto amongst others. In all it took 16 years of collaboration on the mailing lists (and dozens of failed attempts) to attack the problem of trust-minimised, distributed, digital cash. The culmination of these attempts was Bitcoin \cite{Nakamoto2008}. This is illustrated by Dan Held in Figure \ref{fig:prehistory}. This is now a wider ecosystem of technologies and societal challenges (Figure \ref{fig:bitcointopics}). 

\begin{figure*}[ht]\centering 
	\includegraphics[width=\linewidth]{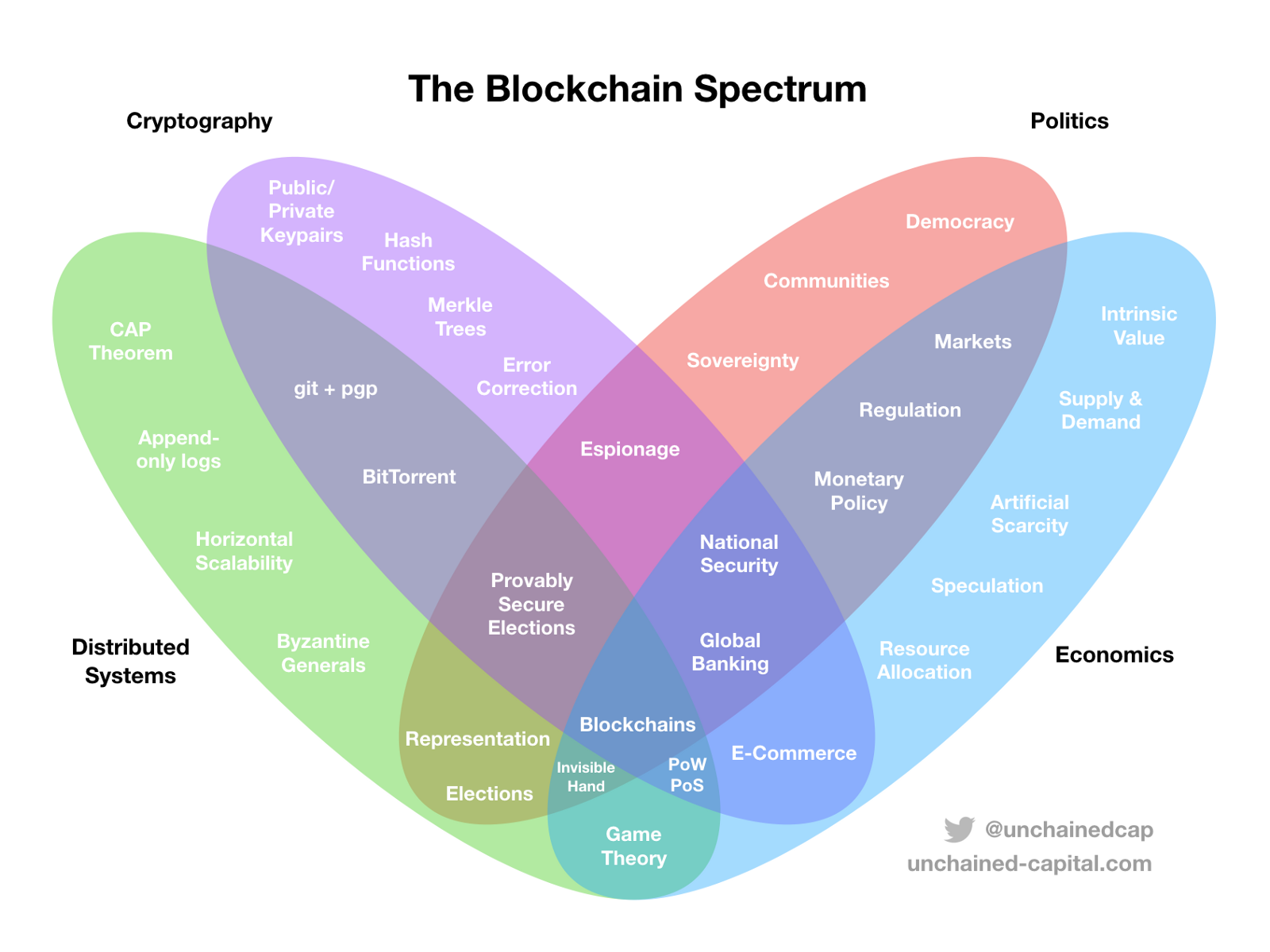}
	\caption{\href{https://unchained.com/blog/blockchain-spectrum/}{Intersecting disciplines}. Reused with permission \href{https://unchained.com/}{Dhruv Bansal}}
	\label{fig:venn}
\end{figure*}

There is enormous complexity and scope, as seen in Figure \ref{fig:venn}, and yet genuinely useful products are elusive.
It is surprisingly hard to pin down a simple explanation for the features which define a blockchain. These ``key takeaway'' \href{https://www.investopedia.com/terms/b/blockchain.asp}{from Investopedia} are a neat summary however.\par
\textit{\begin{itemize} \item Blockchain is a specific type of database. \item It differs from a typical database in the way it stores information; blockchains store data in blocks that are then chained together. \item As new data comes in it is entered into a fresh block. Once the block is \href{https://bits.monospace.live/}{filled with data} it is chained onto the previous block, which makes the data chained together in chronological order. \item Different types of information can be stored on a blockchain but the most common use so far has been as a ledger for transactions. \item In Bitcoin’s case, blockchain is used in a decentralized way so that no single person or group has control—rather, all users collectively retain control. \item Decentralized blockchains are ``append only''. In effect this means that the data entered becomes irreversible over time. For Bitcoin, this means that simple economic transactions are permanently recorded and viewable to anyone. \end{itemize}}
In principle blockchains provide a \textbf{differentiated trust model}. With a properly distributed system a blockchain can be considered ``trust-minimised'', though certainly not risk minimised. This is important for some, but not all people. There is not much emboldening of text within this book. If you start to question the whole reason for this `global technology revolution' then it always comes back to those three words. Put more crispy it's been hiding in plain sight since 20008 as `Magic Internet Money'. Perhaps the lack of a trusted third party, and the potential for instant final settlement will be most important for machine to machine (AI) systems, and that is the primary focus of this book.\par
It can be argued that the whole concept of distributed cryptographic blockchains is \href{https://www.trailofbits.com/reports/Unintended_Centralities_in_Distributed_Ledgers.pdf}{somewhat strained}, as the vast majority of the technology offerings are not distributed, and worse, meaningful distribution may indeed be practically impossible without a trusted third party \cite{kwon2019impossibility}. ``There are many scenarios where \href{https://calpaterson.com/blockchain.html}{traditional databases} should be used instead''\cite{casino2019systematic}.\par
\section{What's this for sorry?}
The proponents of blockchains argue, that in an era when data breaches and corporate financial insolvency intersect with a collapse in trust of institutions, it is perhaps useful to have an alternative model for storage of data, and value. That seems like a lot of effort for a questionable gain. It's far more likely it's simply speculation.\par 
While writing this book the questions of `what is this \textit{really for} and how can it possibly be worth it', came up again and again. In truth it's a very difficult question, without a clear enough answer. It's beyond the scope of this book to figure this out properly, but references to advantages and disadvantages will be made throughout.\par
\begin{figure*}[ht]\centering 
	\includegraphics[width=\linewidth]{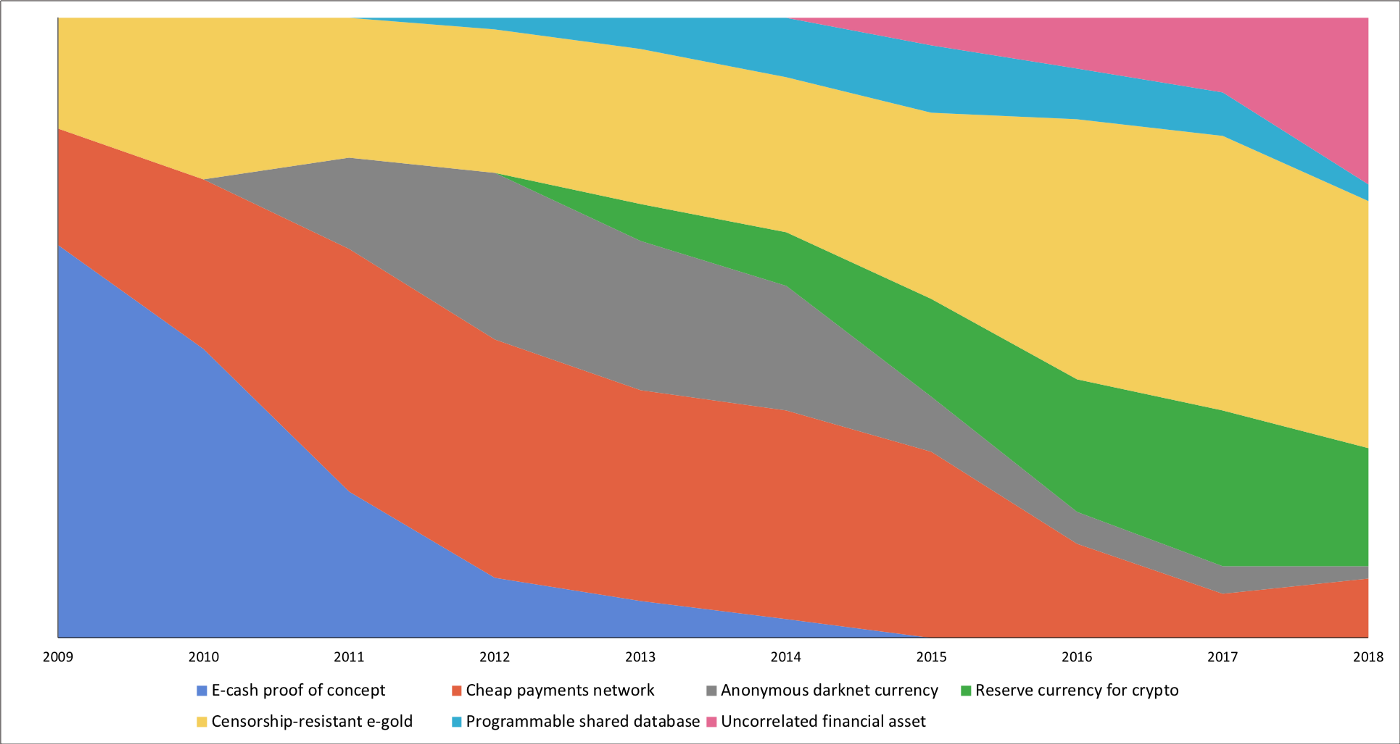}
	\caption{The narrative use of Bitcoin has evolved, by Nic Carter and Hasufly.}
	\label{fig:Evolving}
\end{figure*}

  
It seems that the engineers who created Bitcoin wanted very much to solve a technical problem they saw with money (from their understanding of it), and the transmission of money digitally. As the scale and scope have increased so has the \href{https://medium.com/@nic__carter/visions-of-bitcoin-4b7b7cbcd24c}{narrative evolved} as seen in Figure \ref{fig:Evolving}, but it's never really kept pace with the level of the questions posed. \par
A cost benefit analysis that excludes speculative gains seems to fail for pretty much all of blockchain/DLT. Bitcoin is more subtle as it possibly \textit{can} circumvent the legacy financial systems. This still leaves huge questions. To quote others in the space, is Bitcoin now the iceberg or the life raft? \par
\begin{figure*}[ht]\centering 
	\includegraphics[width=\linewidth]{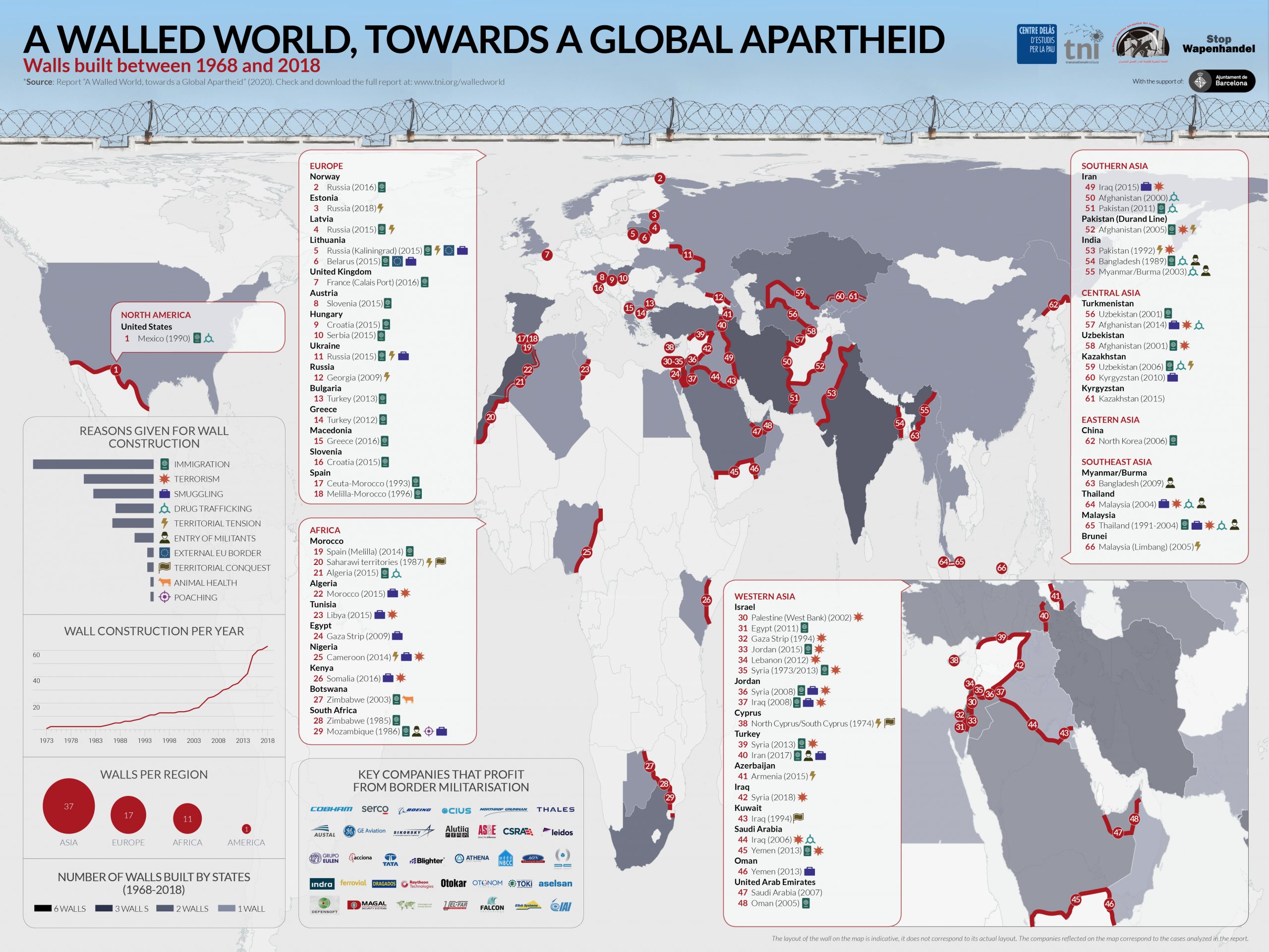}
	\caption{We live in an increasingly \href{https://www.tni.org/en/walledworld}{walled world} (tni, rights requested)}
	\label{fig:walledworld}
\end{figure*}

\begin{figure*}[ht]\centering 
	\includegraphics[width=\linewidth]{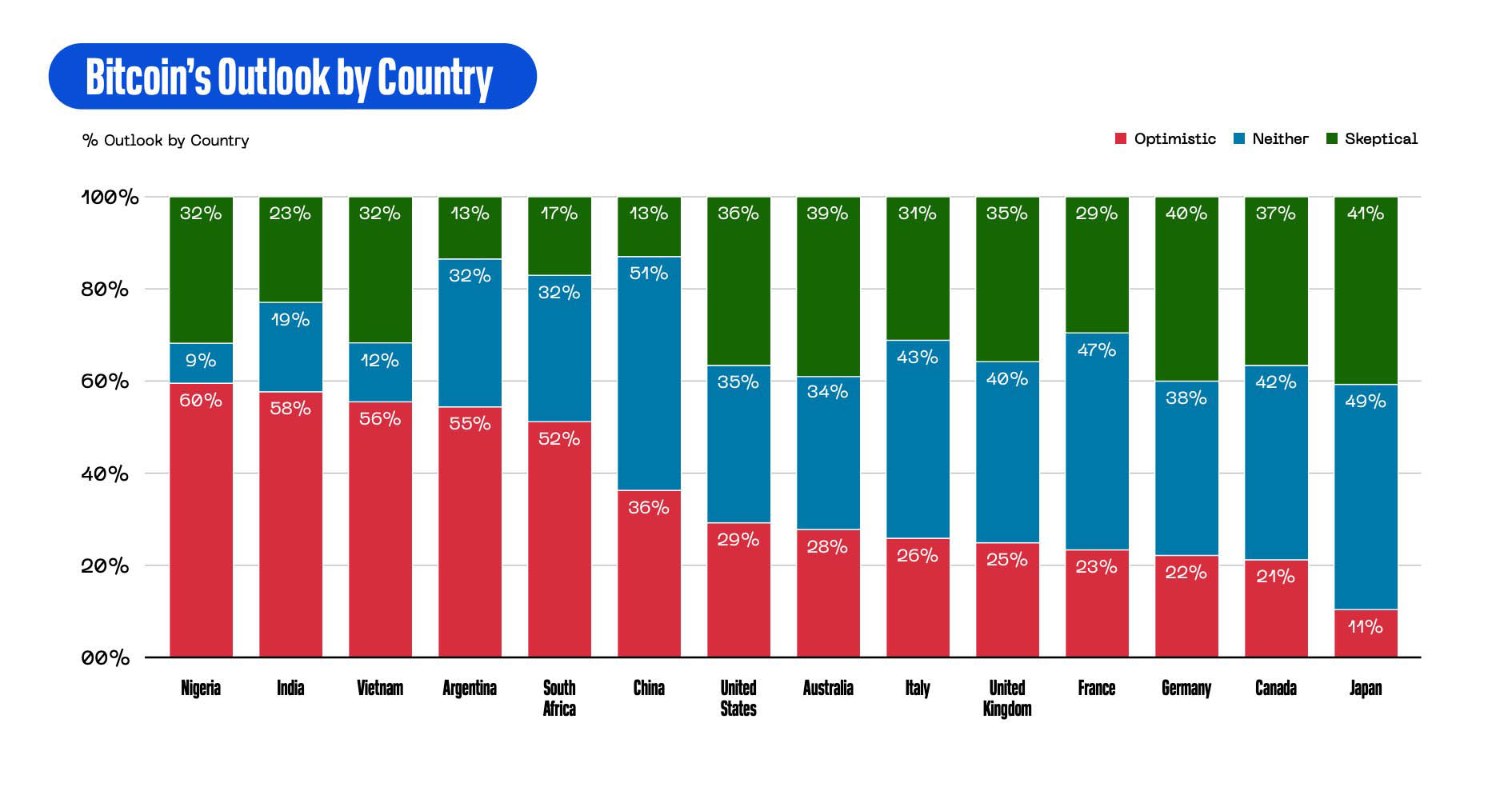}
	\caption{\href{https://twitter.com/gladstein/status/1532054253673406464}{``This new chart from Block is financial privilege visualized.''}}
	\label{fig:optimism}
\end{figure*}

 
For the most developed defence of the technology as it stands in from a Western perspective, in this moment, Gladstein (\href{https://www.financialinclusion.tech/}{and others}) offer a vision for the asset class, in the 87\% of the world he says don't have access to the technology infrastructure benefits enjoyed by the developed west \cite{gladsteincheck2022} (Figure \ref{fig:walledworld}). 
He points to Block and Wakefield Research's report which finds those living under financially oppressive regimes are the most optimistic about the technology as in Figure \ref{fig:optimism}. This argument is suggestive of huge and untapped markets for services which may be accessible to developed nations through telepresence/metaverse interfaces, and which may increase equity of access to opportunity elsewhere. To put some figures against this:
\begin{itemize}
\item Nigeria has the highest number of crypto owners in the world in 2022 with 45\% of its population owning or using cryptocurrency.
\item Thailand occupies the second space with 44\% of its population reported to be using or owning cryptocurrency.
\item Turkey has 40\% of its population owning and using cryptocurrency in 2022, equal to over 33 million people.
\item Argentina occupies the fourth position with an ownership and usage rate of 35\% in 2022, representing almost 16 million people.
\item United Arab Emirates has 34\% of the population owning or using cryptocurrency in 2022, representing almost 10 million people.
\item Philippines is ranked sixth with a 29\% adoption rate.
\end{itemize}
Gladstein's is a carefully developed and well researched book, but is \href{https://bitcoinmagazine.com/culture/imf-world-bank-repress-poor-countries}{written from the western perspective} of (just) Bitcoin `being the raft'. Later in this book we will consider if it might be the iceberg, but this is not the domain expertise we offer in this book. It is crucial to note that Gladstein has vociferous detractors within Africa. It seems entirely possible he's another grifter as suggested by Kimani: \textit{``Gladstein is a charlatan who makes his living by selling the image of a global south that is corrupt, entirely lacking in rational thinking and needing a saviour, like him to swoop in and save us from our floundering selves. He exploits on tired and unproven stereotypes, cherry picks data while ignoring mountains of evidence that disprove him. Because he knows that as the perceived ``morally superior'' ``right thinking'' western superior coming to save, he will mostly go unchallenged. It's a grift, an old grift that many like him have turned into an industry. Where they earn tax free income by selling a delusion and fetish to their western audience who need to think the global south is a failure of the human experience. He is trying to set himself up as some gate keeper and king maker in the Global South. He knows that the next phase of growth is. So he wants to make sure that westerners looking to invest in the global south  see him as some ``expert'' and ask for his unfounded opinions. People like him run global morality extortion rings. How so? Simple: By purporting to know and be the keeper of global south morality, he will use his words to bless or curse your business, well, unless you make a generous donation to his foundation. These are scare tactics employed by charlatans to run tax-evading PR entities, thinly veiled as ``human rights'' organisations. If you are not on his side, he will slander you and your organisation. If you ensure you promote him and his ambitions, he anoints you as the good guy! He is trying to play the role that the Vatican and other corrupt religious organisations played in the 1800. Turning morality into a commodity that can be purchased from his market place: We decide who is good and who is bad and who can do business and who can't. For a ``donation''. He is not the first and he will not be the last. It's a growing industry, driven by shrewd westerners who know that they can sell racial stereotypes back home, but as long as they claim they are the one's helping or saving the coloured peoples from themselves.''} \par
\begin{figure*}[ht]\centering 
	\includegraphics[width=\linewidth]{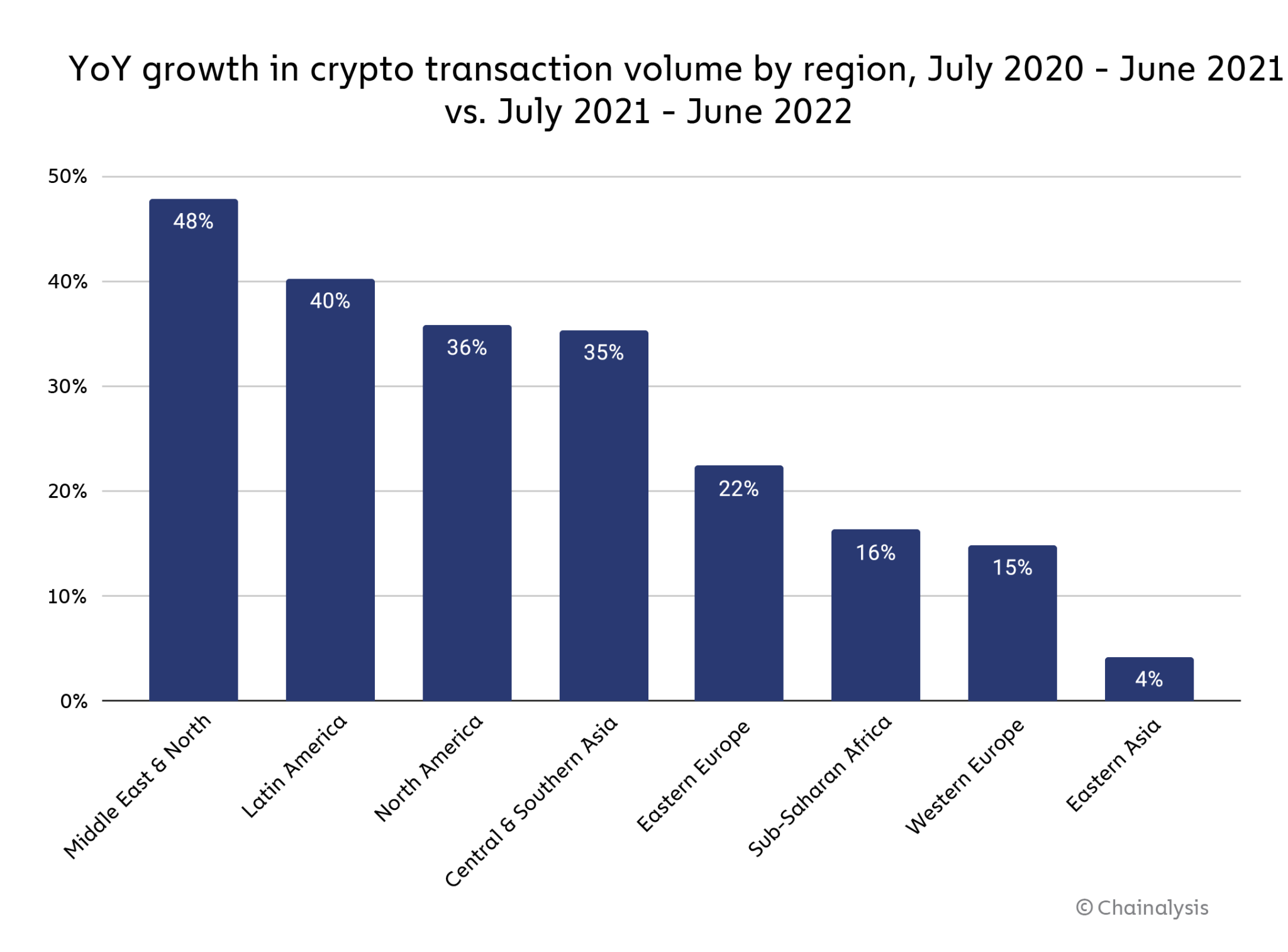}
	\caption{Rapid growth is mainly outside of `Western Markets'}
	\label{fig:grow}
\end{figure*}

\href{https://dailyhodl.com/2022/05/04/crypto-winter-unlikely-as-astonishing-user-growth-dwarfs-internet-adoption-rate-macro-guru-raoul-pal/}{Raoul Pal of RealVision} says: \textit{Crypto adoption is now massively outperforming the internet. It’s been growing at about 165\% a year versus 85\% for the internet for the same period of time now.} According to analytics company Chainalysis; growth is fastest in the Middle east and North Africa (Figure \ref{fig:grow}). \par
\begin{figure*}[ht]\centering 
	\includegraphics[width=\linewidth]{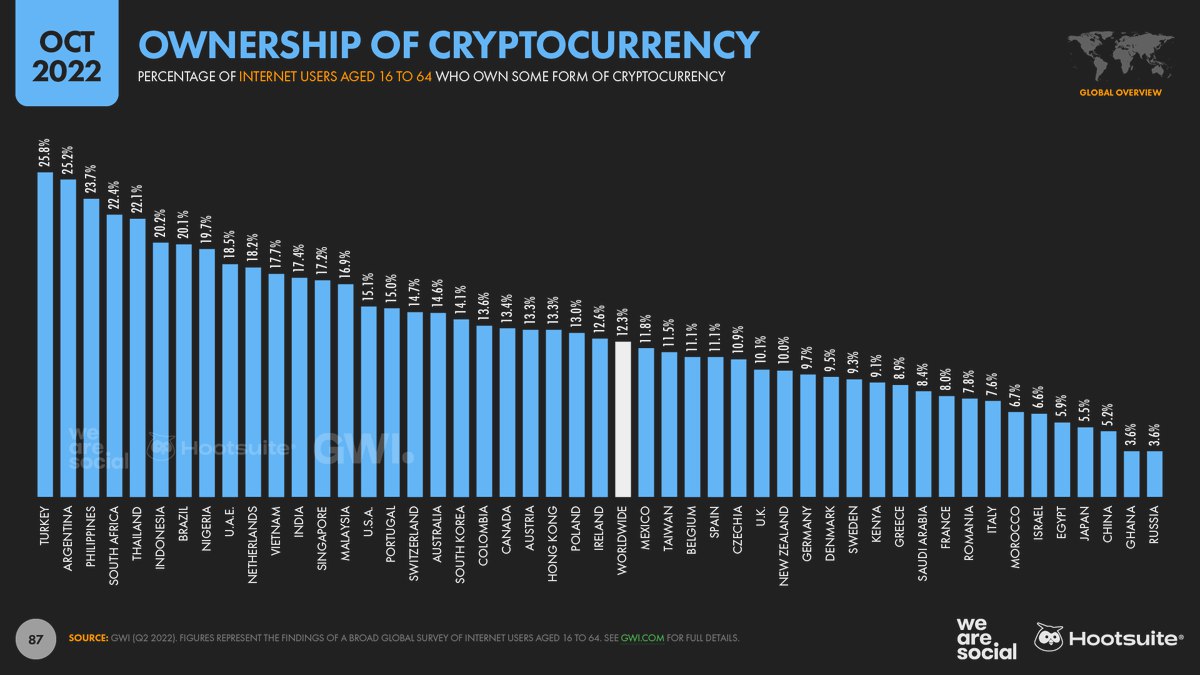}
	\caption{Rapid growth is mainly outside of `Western Markets' - 2}
	\label{fig:ownership}
\end{figure*}

\begin{figure*}[ht]\centering 
	\includegraphics[width=\linewidth]{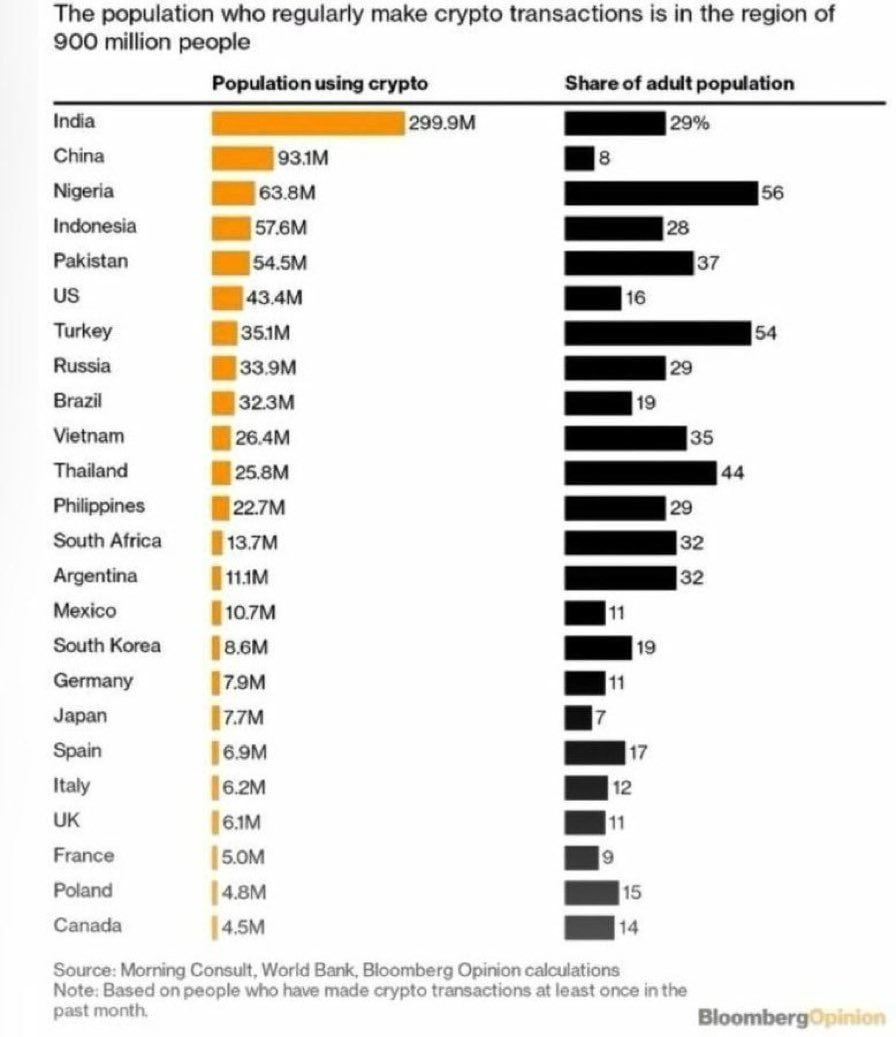}
	\caption{Regular user numbers are surprisingly high.}
	\label{fig:userFigures}
\end{figure*}

\begin{figure*}[ht]\centering 
	\includegraphics[width=\linewidth]{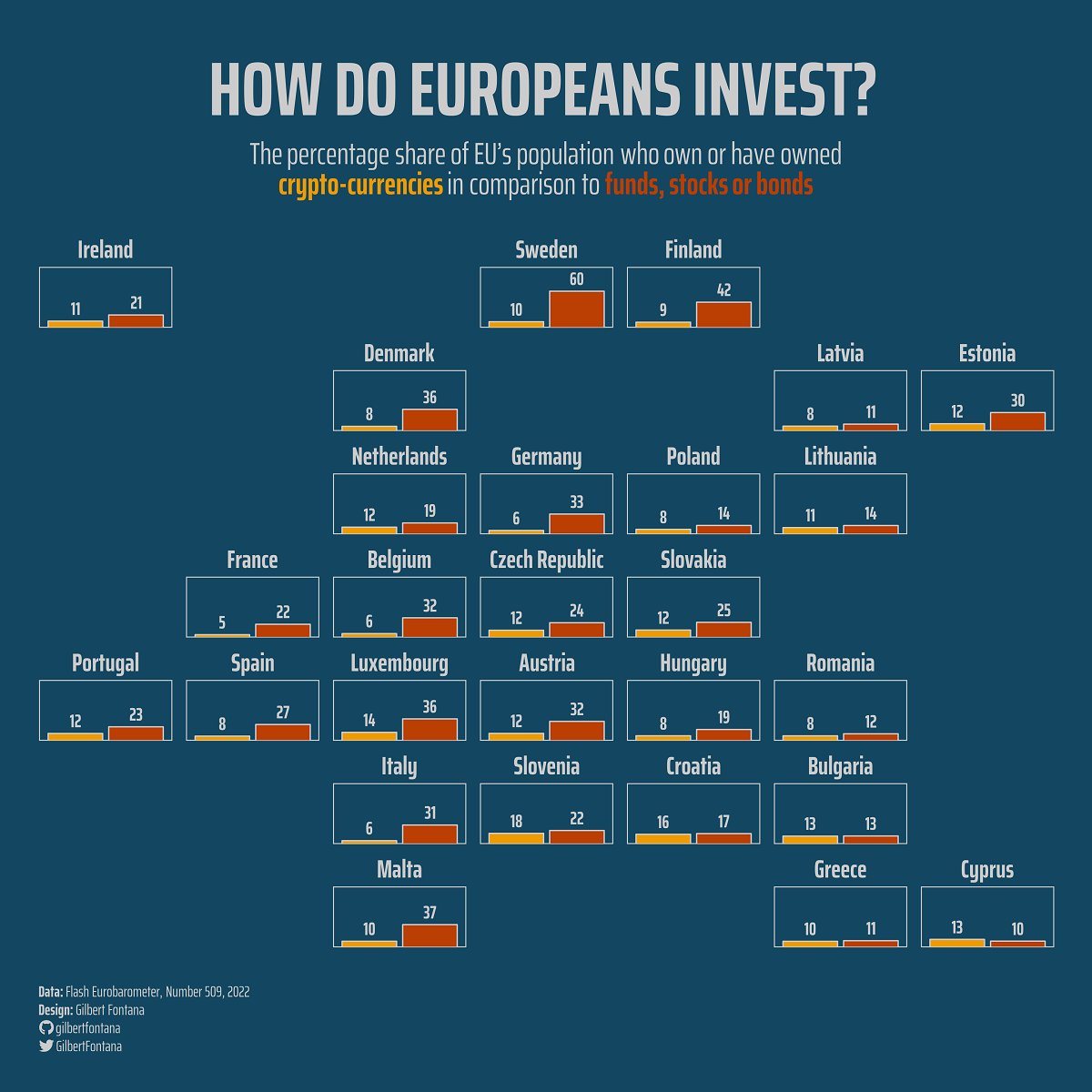}
	\caption{European crypto investment vs funds stocks and bonds from \href{https://github.com/gilbertfontana/DataVisualization}{Fontana} at \href{https://www.visualcapitalist.com/cp/crypto-popularity-in-europeans-union-nations/}{VisualCaptialist}}
	\label{fig:euroInvest}
\end{figure*}

Thanks to a natural fit with strong encryption, and innate resistance to censorship by external parties, these systems do lend themselves well to `borderless' applications, and are somewhat resistant to global regulation (for good or ill). Given the rates of adoption seen in Figures \ref{fig:grow}, \ref{fig:ownership}, \ref{fig:userFigures}, and \ref{fig:euroInvest} it seems that this stuff is coming regardless of their usefulness to the developed world. If we are to take this as a given then we can perhaps logically infer that finding a use case for the technology is important, somewhat irrespective of other arguments. 
\subsection{Machine to machine communication}
A financial instrument which can pass rapidly from computer to computer, AI agent to AI agent, and bounce around `in the cloud' doing real work and activity is where Bitcoin may possibly being to add real global utility. This is such a new field, and better left for the AI section at the end. This seed of an idea provides us a lever to explore for digital society around the asset, and this will be the focus.

\section{A panoply of tech}
Within DLT/blockchain there seem to be as many opinions on the value of the technology as there are implementations. A host of well engineered open source code repositories makes the cost of adoption relatively low. \par
There are thousands of different `chains' and many more tokens which represent value on them. A majority of these are code forks of earlier projects. Most \href{https://99bitcoins.com/deadcoins/}{are defunct} yet still have some residual `value' locked up in them as a function of their `distributed' tokens. \par 
Because the space is comparatively new, subject to \href{https://www.esma.europa.eu/press-news/consultations/call-evidence-dlt-pilot-regime}{scant regulation}, and often open source, it is possible to clone a github, change a few lines of code, and front it with a website in order to create `scams', and this happens frequently \cite{golumbia2020cryptocurrency}.\par
The following sections give an overview of the major strands of the technology. First is Ethereum, mainly to discount it's use for our needs, and move on to more appealing options.

\section{Ethereum}
Ethereum \cite{buterin2013ethereum} is the second most \href{https://www.crypto51.app/}{secure} public blockchain (\href{https://howmanyconfs.com/}{by about 50\%})\cite{sayeed2019assessing}, and second most valuable by \href{https://coinmarketcap.com/}{market capitalisation} (though this comparison is somewhat stretched). It is the natural connection from Web3 to the rest of the book, so it will be considered first.\par
It is touted as `programmable money'. It, unlike bitcoin, is (\href{https://hackernoon.com/turing-completeness-and-the-ethereum-blockchain-c5a93b865c1a}{nearly}) Turing complete \cite{petzold2008annotated}, able to run a \href{https://ethereum.org/en/developers/docs/evm/}{virtual machine} within the distributed network (albeit slowly), and can therefore process complex transactional contracts in the settlement of value. This has given rise to the new field of `distributed finance', or DeFi (described later), alongside many interesting trust-minimised immutable ledger public database ideas. \par
There are trade-offs and problems with Ethereum (Eth/Ether) which currently increase the `participation floor' and make the network far less suitable for entry level business-to-business use. The ledger itself being a computational engine, with write only properties, is enormous. Specialist cloud hardware is required to run a full node (copy of the ledger), and partial nodes are the norm. Many partial nodes are run by one specialist cloud provider (\href{https://consensys.net/blog/news/why-infura-is-the-secret-weapon-of-ethereum-infrastructure/}{Infura}), which has recently been forced to \href{https://finance.yahoo.com/news/metamask-infura-block-certain-areas-173749914.html}{exclude Venezuela} from the network. Network validators are \href{https://mevwatch.info}{refusing to process} addresses on an \href{https://home.treasury.gov/policy-issues/office-of-foreign-assets-control-sanctions-programs-and-information}{OFAC sanction list}. A staggering 58\% runs on \href{https://ethernodes.org/networkType/Hosting}{Amazon AWS servers}. Critics of the project point to these vulnerabilities to outside influence as an existential threat to the aims of the technology. If it can be censured, then what advantage is there over the \href{https://protos.com/consensys-lawsuit-jpmorgan-owns-critical-ethereum-infrastructure/}{founders} simply running a high speed database to the same purpose? \par
This is a function of the so called `scalability trilema' \cite{hafid2020scaling}, in which it seems that only two features from the list of decentralization, scalability or security can be chosen for blockchains \cite{bonneau2015sok}.\par
Moreover the network is \href{https://bitcoinmagazine.com/technical/ethereum-is-coercive-bitcoin-is-not}{centrally controlled} by its creator and the `miners'. There is a strong case to answer that Eth is \href{https://blog.mollywhite.net/blockchains-are-not-what-they-say/}{neither distributed}, nor trustless, and in fact therefore fails to be differentiated from a DLT, undermining some of it's claims. The history of Ethereum is a fascinating case study in human greed. By the time the whitepaper had it's first limited release, Bitcoin (covered next) had already passed \$1000 per token. This led to the creators ambitions for a `fair release' of tokens being voted down by powerful funders, leading to the explosion of similarly structured `pre-mined' coins in the ICO craze, which followed on the Ethereum network. Laura Shin is possibly the most experienced journalist and author in the space and has covered this crazy era in her book `The Cryptopians' \cite{cryptopians}. It's a tough read for the newcomer though, perhaps finish this primer first!\par
\begin{figure}
  \centering
    \includegraphics[width=\linewidth]{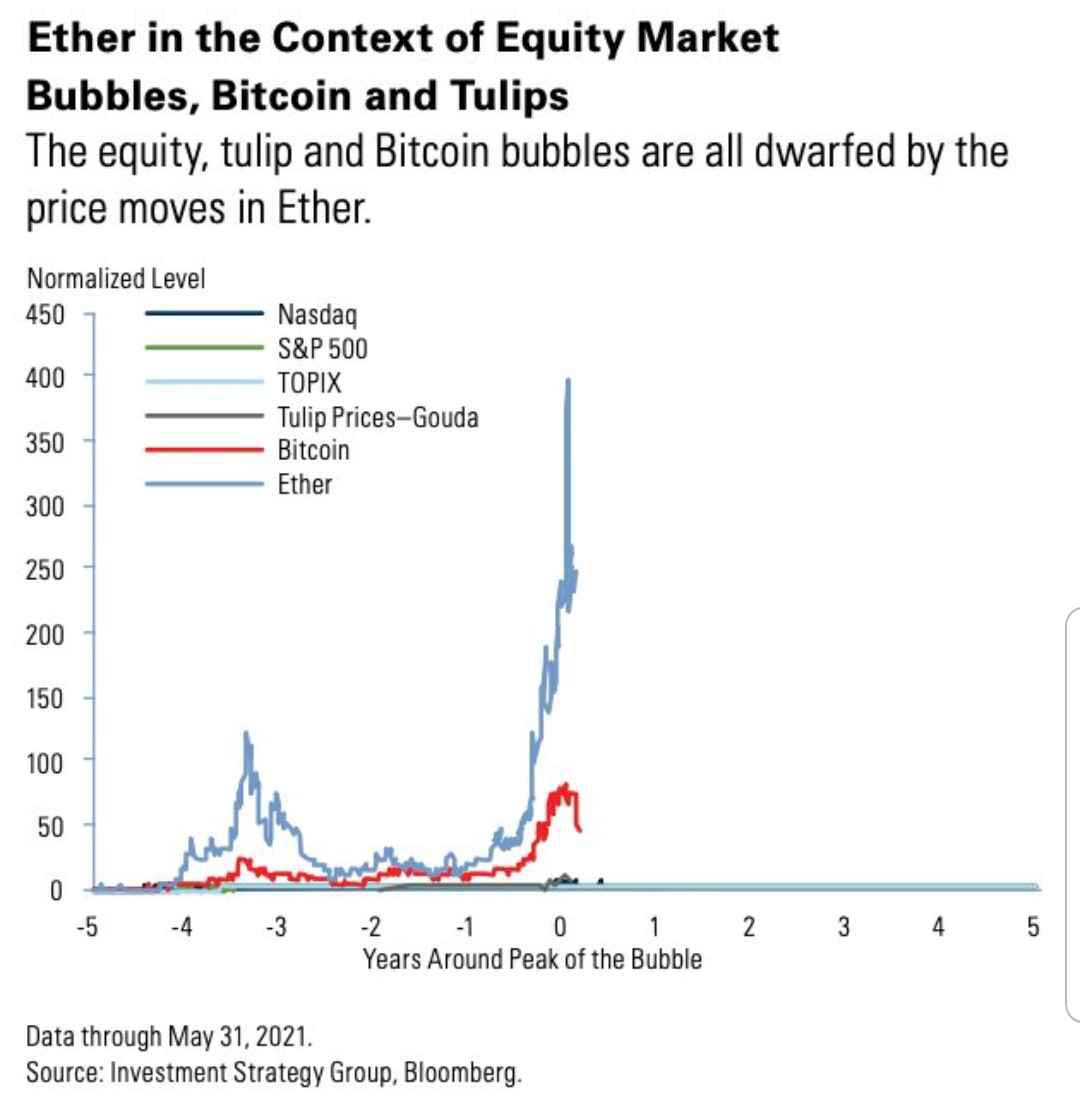}
  \caption{Ethereum is thought to look like a speculative bubble. Rights requested}
    \label{fig:etherbubble}
\end{figure}

With that said there are many talented developers doing interesting work on the platform, and innovation is fast paced. It is entirely normal for technology projects to launch their distributed ledger idea on and within the Ethereum network. These generate tradable `ERC-20' tokens, which can accrue value or demonstrate smart contract utility  (based on the \href{https://soliditylang.org/}{Solidity} programming language). Because the value locked and generated in the Ethereum platform comes not just from the ETH token, but all the ERC technologies built upon it, there are hundreds of billions of pounds `within' the network. All of these projects, and indeed the core technology of Ethereum are subject to exploits and vulnerabilities and tens of billions of pounds have been lost \cite{chen2020survey}. Most of this money is pure market speculation (as is the case across blockchains). Many analysts cannot see this as anything but a speculative bubble, with all the predictable crash yet to come. This can be seen in the context of other bubbles in Figure \ref{fig:etherbubble}. It seems that most of the projects in crypto more generally, but certainly with ETH and the NFTs within it are a new kind of social gambling, where online communities can reinforce groupthink around their speculative choices. This idea that Ethereum is not a commodity, but rather a security, built around promises of returns, is finding recent favour in law. New York attorney general James has alleged that Ether is a security in \href{https://www.docdroid.net/Myyp0yz/kucoin-pdf#page=11}{court proceedings}, which could have enormous consequences, potentially reversing the momentum the asset had been enjoying as a commodity.   Jason Lowery of MIT and US Space Force \href{https://twitter.com/JasonPLowery/status/1572275617344757760}{lays out} a very clear thesis on the difference between Natamoto consensus and most of what followed as part of his PhD \cite{Lowery2023}. His explanation here is proximal to why we focus on Bitcoin, and dismiss `proof of stake' models, though Lowery himself has very serious detractors such as Voskuil who has been working on these problems for a long time \cite{voskuil2020cryptoeconomics} \par
\textit{``The innovation behind PoW is precisely the fact that it \textbf{doesn't} rely exclusively on software (an abstraction) to keep the ledger systemically secure, but instead incorporates real-world physics (watts) to impose real-world physical constraints on people/computers who run it. Stake is an abstraction. It is an imaginary way to describe the complex emergent behaviour of a bunch of general-purpose state machines. The state machines may physically exist, but the way you choose to visualize the complex emergent behaviour of those machines is imaginary. Satoshi didn't couple control authority over ledger to abstract, imaginary things like `stake' or `coin' precisely because these things don't physically exist. If they don't physically exist, they are incapable of imposing real-world physical costs on people seeking control of ledger. The real-world physical cost of controlling the ledger is what keeps control over the ledger decentralized. It is too physically expensive (in watts) to gain and maintain centralized control over the BTC ledger. In proof of stake, there is no physical cost of gaining centralized control. Why? Because stake doesn't physically exist. So all it takes to gain centralized control is majority stake. And once you have it (which, because of math, some combination of people already do), you have it forever.}

With all this said most of the \href{https://www.statista.com/statistics/1266322/nft-user-number/}{couple of million people} who have used NFTs use Ethereum, and if this market of creators and consumers is to be brought into a mixed reality space then they will need a way to bring their objects with them. 
Such is the level of nefarious activity on these networks (within Ethereum) that they have a poor reputation, and are difficult to audit, launch, and maintain. The overriding problem of using a blockchain for utility applications (rather than just as money) is that people can, and will, simply lie for criminal purpose when entering data into the ledger. It is far more likely that Ethereum is simply a speculative bubble than any of the claims for utility being born out. Add to that \href{https://advisor.morganstanley.com/daron.edwards/documents/field/d/da/daron-edwards/Cryptocurrency_201__What_is_Ethereum_.pdf}{Morgan Stanleys recent assertion} that Ethereum is itself threatened by newer contender chains and it's future becomes unclear. The report correctly identifies that ``High transaction fees create scalability problems and threaten user demand. High costs make Ethereum too expensive for small-value transactions.''. It is this high cost of use that most excludes the ERC-20 networks from our consideration.
\subsection{Gas fees}
Ethereum has a significant barrier to entry because of high fees to use the network. The system is Turing complete; able to programatically replicate any other computational system. This includes endless loops in code, so it is trivial to lock up the computational bandwidth of the whole system, in a smart contract commitment, through a web wallet. \par 
To mitigate this existential `denial of service attack' the `gas' system demands that users spend some of their locked up value to operate on the network. In this way a transaction loop would quickly erode the available gas and stop looping. As the popularity of the system has grown, so too have the gas fees. It can \href{https://twitter.com/Blockworks_/status/1521071340517830657}{sometimes cost} over £10,000 to do a single transaction, though it is typically a few tens of pounds. Appallingly if the user pitches their mining fee offer too low, then the money gets spent anyway! \href{https://fees.wtf/#/}{A website} just plucks random Ethereum addresses out of the aether to show you the level of this expense for participants. People can even \href{https://opensea.io/collection/fees-wtf-nft?search[sortAscending]=false&search[sortBy]=PRICE}{buy NFTs} of the worst examples of these, as `tokens', wasting more money. This is a huge problem for potential uses of the network. \par
\subsection{Ether ultra hard money narrative}
\begin{figure}
  \centering
    \includegraphics[width=\linewidth]{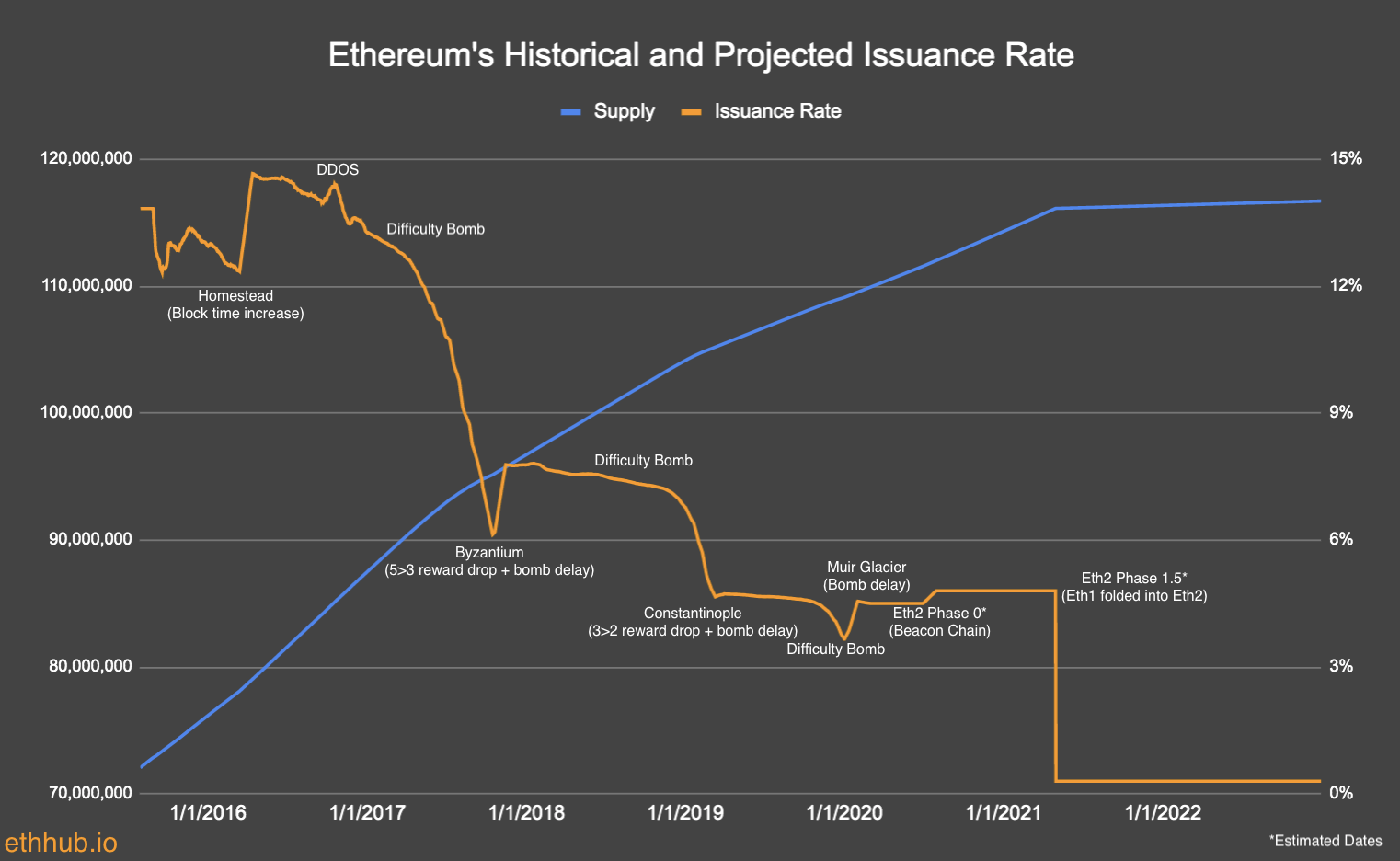}
  \caption{The rate of token generation has changed unpredictably over time. Rights requested}
  \label{fig:ethemission}
\end{figure}

Part of the challenge Ethereum faces is wrapped up with it's complex token emission schedule. This is the rate at which tokens are generated and `burnt' or destroyed in the network. The total supply of tokens is uncertain, and both emission and burn schedules are regularly tinkered with by the project. The changes to the rate at which ETH are generated can be seen in Figure \ref{fig:ethemission}.
In addition, a recent upgrade (EIP-1559) results in tokens now being burnt at a higher rate than they are produced, deliberately leading to a diminishing supply. In theory this increases the value of each ETH on the network at around 1\% per year. It's very complex, with impacts on transaction fees, waiting time, and consensus security, as examined by Liu at al. \cite{liu2022empirical}. Additionally, there is now talk (by \href{https://time.com/6158182/vitalik-buterin-ethereum-profile/}{Butlerin}, the creator of Ethereum) of extending this burn mechanism \href{https://ethresear.ch/t/multidimensional-eip-1559/11651}{further into the network}.\par
Ethereum was designed from the beginning to move to a `proof of stake' model where token holders underpin network consensus through complex automated voting systems based upon their token holding. This is now called \href{https://blog.ethereum.org/2022/01/24/the-great-eth2-renaming/}{Ethereum Consensus Layer}.  This recent `Merge' upgrade has reduced the carbon footprint of the network, a laudable thing, though it seems the GPUs and datacentres have just gone on to be elsewhere. It has not lowered the cost to users nor improved performance. As part of the switching roadmap users were asked to lock up 32ETH tokens each (a substantial allocation of capital). In total there are around 14 million of these tokens, and it is those users who now control the network. This money is likely stuck on the network until at least 2024, a significant delay wen compared to the original promises.\par
This means that proof of stake has problems in that the majority owners `decide' the truth of the chain to a degree, and must by design have the ability to over-ride prior consensus choices. Remember that these users are now trapped in their positions. Four major entities now control the rules of the chain, and have already agreed to censor certain banned addressees. Proof of stake is probably inherently broken \cite{poelstra2015stake}. This has \href{https://notes.ethereum.org/@djrtwo/risks-of-lsd} for malicious actors who have sufficient control of the existing history of the chain, \href{https://twitter.com/MTorgin/status/1521433474820890624}{thought to be} in the region of \$50M \cite{mackinga2022twap}. Like much of the rest of `crypto' the proposed changes will concentrate decisions and economic rewards in the hands of major players, early investors, and incumbents. This is a far cry from the stated aims of the technology. The move to proof of stake has recently earned it the \href{https://www.technologyreview.com/2022/02/23/1044960/proof-of-stake-cryptocurrency/}{MIT breakthrough technology award}, despite not being complete (validators cannot yet sell their voting stakes). It's clearly a technology which is designed to innovate at the expense of predictability. This might work out very well for the platform, but right now the barrier to participation (in gas fees) is so high that we do not intend for Ethereum to be in scope as a method for value transfer within metaverses.\par
\subsection{Inherent Weaknesses}
Ethereum faces a unique dilemma, often overshadowed by its technological capabilities. Unlike Bitcoin (BTC), which has solidified its role as a stable and reliable store of value, Ethereum's value proposition is more complex and, ultimately, paradoxical. The following points elaborate on this conundrum:

\begin{itemize}
    \item \textbf{Lack of Monetary Certainty:} Ethereum's mutable supply schedule and governance model introduce a level of uncertainty not found in Bitcoin. 
    \item \textbf{Equity-like Characteristics:} Ethereum acts more like a share in a semi-decentralized corporation than a straightforward asset, deriving its value from expected future transaction fees.
\end{itemize}

These attributes lead to a value paradox that is two-fold:

\begin{itemize}
    \item \textbf{Fee Dilemma:} High transaction fees, while beneficial for Ethereum's perceived value, deter usage and drive decentralized finance (DeFi) applications to other platforms.
    \item \textbf{Scalability Trap:} Attempts to scale the platform and lower fees would, counterintuitively, reduce Ethereum's intrinsic value by decreasing its future cash flows.
\end{itemize}

This presents a catch-22 situation where Ethereum's value is fundamentally limited by its own economic model. If the asset's value drops significantly, it could undermine the security of the entire platform, making it less reliable for settling large transactions.

In the long run, this creates a feedback loop that could, theoretically, push Ethereum's value towards zero. This issue casts a shadow over Ethereum's long-term viability, presenting a challenge that goes beyond mere technical scalability.

\section{Bitcoin}
The first blockchain was the Bitcoin network \cite{Nakamoto2008}, some two decades after Haber et al. first described the idea \cite{haber1990time}. Prior to Bitcoin these structures were called `timechains' \cite{nakamoto2018}. It can be considered a triple entry book keeping system \cite{ijiri1986framework, faccia2019accounting}, the first of it's kind, integrating a `provable' timestamp with a transaction ledger, solving the ``double spend problem'' \cite{chohan2021double, perez2019double, grunspan2018double}. Some see this as the first major innovation in ledger technology since double entry was codified in Venice in fourteen seventy five\cite{sangster2015earliest}. \par
It was created pseudonomously by an individual or group calling themselves `Satoshi Nakamoto' in 2009, as a direct response to the perceived mishandling of the 2008 global financial crisis \cite{nakamoto2018}, with the stated aim of challenging the status quo, with an \href{https://world.hey.com/dhh/i-was-wrong-we-need-crypto-587ccb03}{uncensorable} technology, to create a money which could not be \href{http://p2pfoundation.ning.com/forum/topics/bitcoin-open-source}{debased by inflation policy}, and outside of the \href{https://www.coindesk.com/layer2/2022/05/04/matt-taibbi-paypals-deplatforming-and-the-case-for-crypto/}{politically captured} fintech incumbents. It's interesting to note that the narrative around the use case for Bitcoin has \href{https://uncommoncore.co/visions-of-bitcoin-how-major-bitcoin-narratives-changed-over-time/}{shifted over it's lifetime}. \par
The \href{https://en.bitcoin.it/wiki/Genesis_block}{``genesis block''} which was hard coded at the beginning of the `chain' contains text from The Times newpaper detailing the second bank bailout.\par 
There will only ever be (\href{https://blog.amberdata.io/why-the-bitcoin-supply-will-never-reach-21-million}{just short of}) 21 million bitcoins issued, of which around 19 million have already been minted, and around 4 million lost forever. This `hard money' absolute scarcity is a strong component of the Bitcoin meme landscape. These are basically arbitrary figures though; a combination of the issuance schedule, and an \href{https://plan99.net/~mike/satoshi-emails/thread1.html}{`educated guess'} by Nakamoto: \cite{nakamoto2018}\par 
\textit{''My choice for the number of coins and distribution schedule was an educated guess.  It was a difficult choice, because once the network is going it's locked in and we're stuck with it.  I wanted to pick something that would make prices similar to existing currencies, but without knowing the future, that's very hard.  I ended up picking something in the middle.  If Bitcoin remains a small niche, it'll be worth less per unit than existing currencies.  If you imagine it being used for some fraction of world commerce, then there's only going to be 21 million coins for the whole world, so it would be worth much more per unit.''}\par
Digital scarcity is incredibly important and is explained well by software engineer Hillibrand in a podcast (this text is paraphased: \textit{``Digital scarcity is an interesting concept that was well explained by German economist Guido H\"ulsmann in his book ``The Ethics of Money Production,'' \cite{hulsmann2008ethics} published in 2007. H\"ulsmann stated that an economic good that is defined entirely in terms of bits and bytes is unlikely ever to be produced spontaneously on a free market, and at the time, he was right. However, the emergence of Bitcoin would soon prove that digital scarcity could indeed be achieved. H\"ulsmann noted that an economic good must be scarce and rivalrous, meaning there is a potential for conflict over who can utilize the resource. For example, air is abundant but still considered scarce as its availability can be limited in specific situations, leading to conflicts over its use. The concept of digital scarcity is built on the idea that information, which is fundamentally not scarce, can be made scarce through specific mechanisms. Bitcoin, for instance, addresses the double-spending problem, where a digital token could be spent more than once, by establishing a decentralized network that prevents the same coin from being used in multiple transactions. Nakamoto devised a system that allows users to establish scarcity and rivalrousness in cyberspace without relying on a single trusted third party. Instead of relying on a central authority, like a government, to determine the validity of transactions, Bitcoin relies on a network of computers known as ``full nodes'' that verify and enforce a set of rules. This decentralized system enables the creation of digital goods that are both scarce and rivalrous, which was previously thought to be impossible.''}\par
In theory there is no \href{https://www.forbes.com/sites/peterizzo/2021/09/29/against-cryptocurrency-the-ethical-argument-for-bitcoin-maximalism/?}{barrier to access}, and \href{https://www.coindesk.com/layer2/2022/02/16/why-bitcoin-is-a-tool-for-social-justice/}{equality of opportunity} to accumulate and save over long periods. This is not true of chains and tokens since, which lock up some of their value for seed investors to cash out later. None of the blockchains since are decentralised in the same way \cite{selvam2021blockchain}. Bitcoin was probably a \href{https://danhedl.medium.com/bitcoins-distribution-was-fair-e2ef7bbbc892}{singular event}.\par
Each Bitcoin can be divided into 100 million satoshis (sats), so anyone buying into Bitcoin can buy a thousandth of a pound, assuming they can find someone willing to transact that with them. \par
Satoshi Nakamoto (the name of the publishing entity) \href{https://bitcoinmagazine.com/technical/what-happened-when-bitcoin-creator-satoshi-nakamoto-disappeared}{disappeared from the forums} forever in 2010. Bitcoin has the marks of cypherpunks and anarcho capitalism. The IMF has recently conceded that the Bitcoin \href{https://blogs.imf.org/2022/01/11/crypto-prices-move-more-in-sync-with-stocks-posing-new-risks/}{poses a risk} to the traditional financial systems, so it could be argued that it is succeeding in this original aim.\par
\begin{figure}
  \centering
    \includegraphics[width=\linewidth]{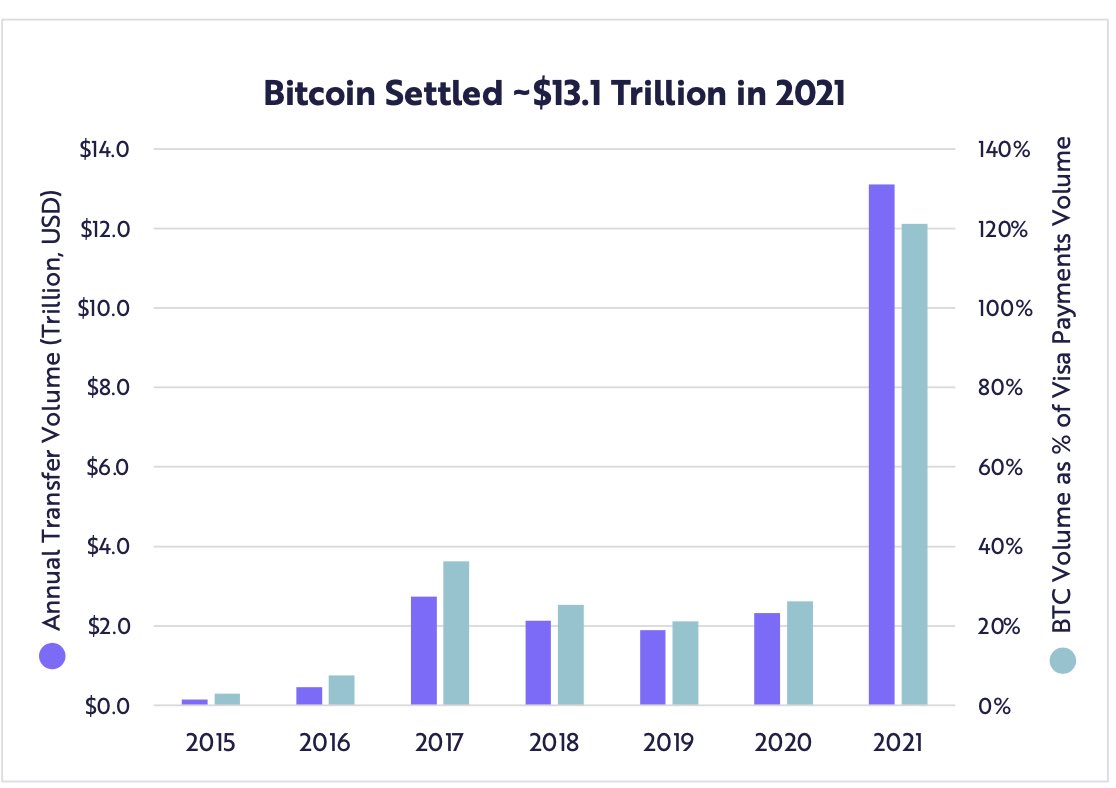}
  \caption{\href{https://twitter.com/glxyresearch/status/1469039427028664320?}{Growth in settlement} value on the Bitcoin network (Forbes).}
  \label{fig:settled2021}
\end{figure}

Although there were some earlier experiments (hashcash, b-money etc), Bitcoin is the first viably decentralised `cryptocurrency'; the network is used to \href{https://www.aier.org/article/why-does-bitcoin-have-value/}{store economic value} because it is judged to be secure and trusted. It is a singular event in that it became established at scale, such that it could be seen to be a fully distributed system, without a controlling entity. This is the differentiated trust model previously mentioned. This relative security is the specific unique selling point of the network. It is many times more secure than all the networks which came after based on a like for like comparison of \href{https://howmanyconfs.com/}{transaction `confirmations'}. This network effect of Bitcoin is a compounding feature, attracting value through the security of the system. It is deliberately more conservative and feature poor, preferring instead to \href{https://bips.xyz/}{add to it's feature set} slowly, preserving the integrity of the value invested in it over the last decade. At time of writing it is a \href{https://fiatmarketcap.com/}{top quartile} largest global currency and has settled over \$13 trillion Dollars in 2021, though Makarov et al. contest this, citing network overheads, and speculation \cite{makarov2021blockchain}. Institution grade `exchange tradable funds' which allow investment in Bitcoin are available throughout the world, and the native asset can be bought by the public easily through apps in all but a handful of countries as seen in Figure \ref{fig:settled2021}. \par
Only around 7 transactions per second can be settled on Bitcoin. The native protocol does not scale well, and this is an inherent trade-off as described by Croman et al. in their positioning paper on public blockchains \cite{croman2016scaling}. Over time, competition for the limited transaction bandwidth drives up the price to use the network. This effectively prices out small transactions, even locking up some value below what is a termed the '\href{https://github.com/bitcoin/bitcoin/blob/v0.10.0rc3/src/primitives/transaction.h#L137}{dust limit}' of unspent transactions too small to ever move again \cite{delgado2018analysis}. \par
Bitcoin has developed quickly, with a \href{https://phemex.com/blogs/crypto-bitcoin-s-curve-adoption-curve}{faster adoption} than even the internet itself. It is already a mature ecosystem, with \href{https://www.fortris.com/}{enterprise grade software} stacks, and is seeing adoption as a \href{https://bitcointreasuries.net/}{corporate treasury asset}. \par
Adoption by civil authorities is increasing, and legislators the world over are being forced to \href{https://www.politico.com/news/2022/01/16/bitcoin-crashes-the-midterms-527126}{adopt a position}. California has an \href{https://www.gov.ca.gov/2022/05/04/governor-newsom-signs-blockchain-executive-order-to-spur-responsible-web3-innovation-grow-jobs-and-protect-consumers/}{explicitly Web 3 and blockchain executive order} to invetigate and support opportunities. Many city treasuries have \href{https://www.bloomberg.com/news/articles/2022-01-14/rio-de-janeiro-wants-to-become-brazil-s-cryptocurrency-capital}{added it} to their balance sheet. Honduras \href{https://www.reuters.com/world/americas/honduras-launches-bitcoin-valley-tourist-town-santa-lucia-2022-07-29/}{has launched} ``Bitcoin Valley'' as a tourist initiative, and the Swiss city of Lugano is launching a \href{https://twitter.com/Stadicus3000/status/1499656424422526977}{huge initiative} alongside Tether. It is already legal tender in the country of El Salvador\cite{oxford2021salvador} and the \href{https://finance.yahoo.com/news/central-african-republic-passes-bill-180910797.html?}{Central African Republic}, and will be soon in \href{https://www.forbes.com/sites/ninabambysheva/2022/04/07/two-new-territories-are-adopting-bitcoin/?sh=7f014ed2499a}{Madeira and Roatán island}. This means it \textit{must} be accepted as a means of payment, with uncertain global political legal consequences \cite{katterbauer2022impact}. 
In places such as Panama it simply has legal status and \textit{can} be accepted without double taxation.\par 
Global asset manager ``Fidelity'' wrote the following in their \href{https://www.fidelitydigitalassets.com/articles/2021-trends-impact}{2021 trends report}. \textit{``We also think there is very high stakes game theory at play here, whereby if Bitcoin adoption increases, the countries that secure some bitcoin today will be better off competitively than their peers. Therefore, even if other countries do not believe in the investment thesis or adoption of bitcoin, they will be forced to acquire some as a form of insurance. In other words, a small cost can be paid today as a hedge compared to a potentially much larger cost years in the future. We therefore wouldn't be surprised to see other sovereign nation states acquire bitcoin in 2022 and perhaps even see a central bank make an acquisition.''}\par
\subsection{The Bitcoin Network Software}
\begin{figure*}[ht]\centering 
	\includegraphics[width=\linewidth]{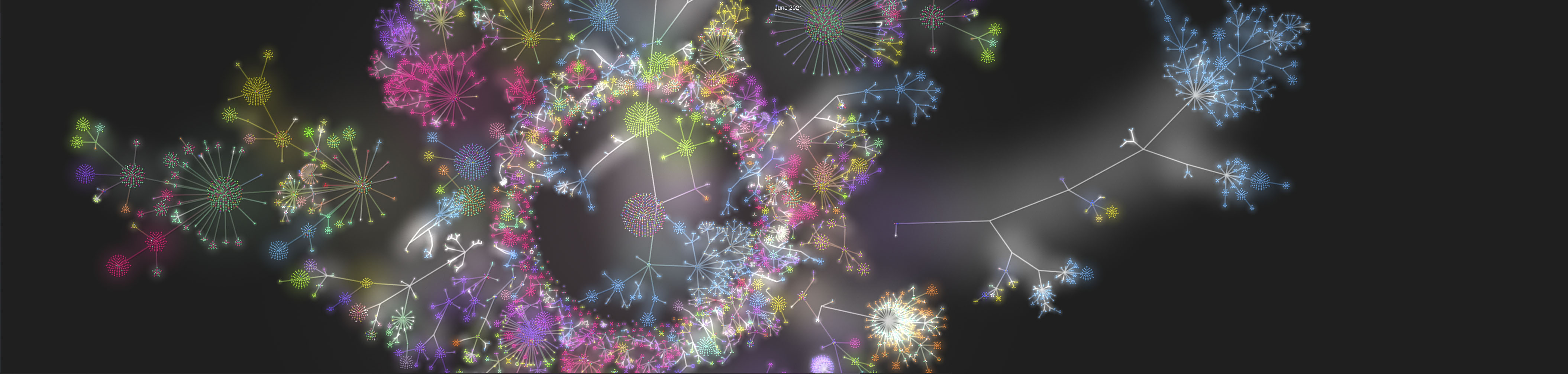}
	\caption{\href{https://github.com/bitpaint/bitcoin-gources}{Bitpaint}: Contributions to the Bitcoin ecosystem. Reused with permission.}
	\label{fig:gource}
\end{figure*}

There isn't a single GitHub which can be considered the final arbiter of the development direction, because it is a distributed community effort with some \href{https://decrypt.co/66740/who-are-the-fastest-growing-developer-communities-in-crypto}{500 developers} out of a wider `crypto' pool of around 9000 contributors (the vast majority are spread across disparate Ethereum and some Solana projects). \href{https://bitcoinops.org/en/newsletters/2021/12/22/}{Development and innovation continues} but there is an emphasis on careful iteration to avoid damage to the network. Visualisation of code commitments to the various open source software repositories can be seen at \href{https://www.youtube.com/channel/UC4DT4qudqogkmbqVAQy8eFg/videos}{Bitpaint youtube channel} and in Figure \ref{fig:gource}.\par
\href{https://github.com/bitcoin/}{Bitcoin core} is the main historical effort (with around a dozen major contributors guiding the direction), but there are alternatives (\href{https://github.com/libbitcoin/libbitcoin-node/wiki}{LibBitcoin in C++}, \href{https://github.com/btcsuite/btcd}{BTCD in Go}, and \href{https://bitcoinj.github.io/getting-started}{BitcoinJ in Java}), and as innovation on layer one slows, attention is shifting to codebases which interact with the base layer asset. Much more on these later.
\subsection{Mining and Energy concerns}
\subsubsection{Mining process overview}
Bitcoin mining is the process of adding public transactions into the ledger, in return for two economic rewards, paid in Bitcoin. These are the mining fee, and the block reward. The transactions which are added into the next `block' of the chain are selected preferentially based on the fee they offer, which is up to the user trying to get their transaction into the chain. This can be within the next 10 minutes (next block), or a gamble out toward 'never' depending how competitive the network is at any time. Miners try to find a sufficiently low result from a cryptographic hash function \cite{rogaway2004cryptographic}(a random process), and upon finding it, they can take their pre-prepared `block' of transactions sourced from their local queue (mempool), and add it into the chain, for confirmation by other miners. In return they take all the fees within that mined block, and whatever the block reward is at the time. When the network started the block reward was 50 Bitcoin, but has \href{https://ma.ttias.be/dissecting-code-bitcoin-halving/}{halved} repeatedly every 210,000 blocks (four years) and now stands at 6.25 BTC. The rate of mining is kept roughly at one block every 10 minutes, by a difficulty adjustment every 2016 blocks (2 weeks). This in a complex interdependent mechanism and is explained very well in \href{https://bitcoinmagazine.com/technical/how-mining-protects-the-bitcoin-network}{this article}. These components are explained in slightly more detail later.\par
\subsubsection{Energy \& policy response}
\begin{figure*}[ht]\centering 
	\includegraphics[width=\linewidth]{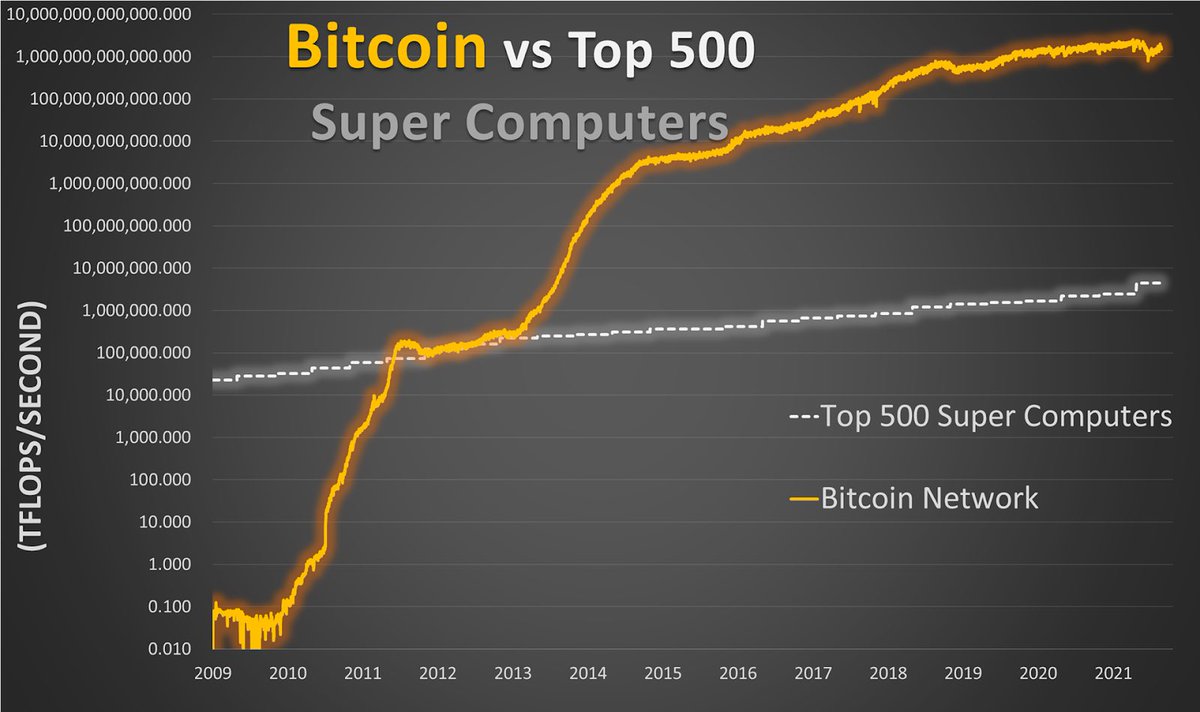}
	\caption{\href{https://twitter.com/blockbain/status/1432824720727105539}{Bitcoin network vs TOP500 supercomputers}}
	\label{fig:top500}
\end{figure*}

Bitcoin uses a staggering amount of energy to secure the blockchain (Figure \ref{fig:top500}), and this \href{https://www.edmundconway.com/bitcoin-money-and-the-planet/}{has climate repercussions}. A simple back of the envelope use of the \href{https://www.iea.org/reports/key-world-energy-statistics-2021/supply}{IEA total energy supply}, and the \href{https://ccaf.io/cbeci/index}{Cambridge Bitcoin energy use} estimate puts the network at \href{https://www.wolframalpha.com/input?i=153+terawatt+hours+as+percentage+of+\%28600+exa+joules+as+terrawatt+hours\%29+}{around 0.1\%} of global energy use.\par
\begin{figure*}[ht]\centering
	\includegraphics[width=\linewidth]{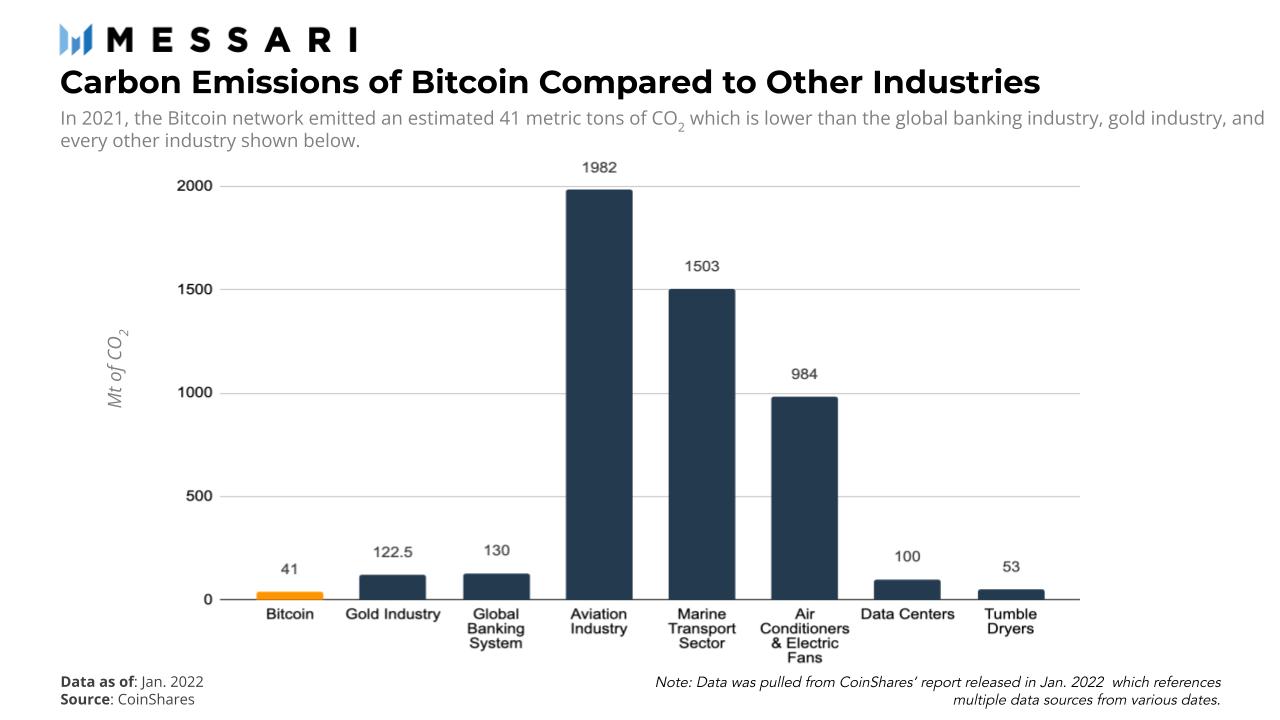}
	\caption{}
	\label{fig:energycompare1}
\end{figure*}
\begin{figure*}[ht]\centering
	\includegraphics[width=\linewidth]{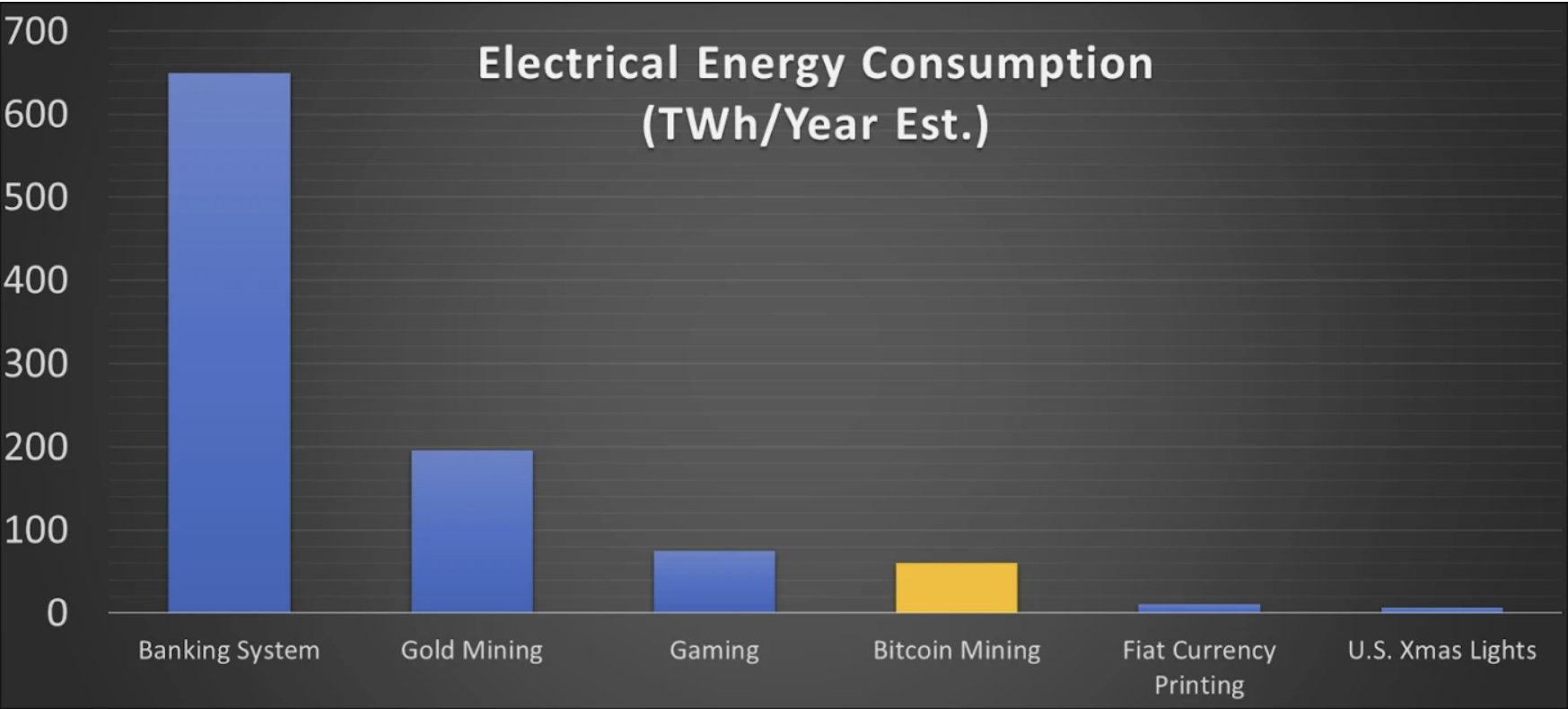}
	\caption{\href{https://bitcoinmagazine.com/business/introducing-cbei-a-new-way-to-measure-bitcoin-network-electrical-consumption}{Bitcoin Magazine}}
	\label{fig:energycompare2}
\end{figure*}
It is an \href{https://www.ruetir.com/2022/03/18/riot-whinstone-the-bitcoin-farm-with-100000-computers-that-uses-excess-energy-from-an-oil-platform-to-mine-cryptocurrencies-ruetir/}{industrial scale} global business with `mining companies' investing \href{https://ir.marathondh.com/news-events/press-releases/detail/1272/marathon-digital-holdings-bitcoin-mining-fleet-to-reach}{hundreds of millions of pounds} at a time in specialist \href{https://en.wikipedia.org/wiki/Application-specific_integrated_circuit}{ASIC} mining hardware and facilities. The latest purpose designed Intel chip \href{https://www.intel.com/content/www/us/en/newsroom/opinion/thoughts-blockchain-custom-compute-group.html#gs.pd9ofu}{touts} both Web3 and metaverse applications. This is ``proof of work'',  and is essential to the technology, and is still thought by some to be the \href{https://www.truthcoin.info/blog/pow-cheapest/}{best available option}. The \href{https://ccaf.io/cbeci/index}{Cambridge Bitcoin Energy Consumption Index} monitors this energy usage. Their \href{https://www.jbs.cam.ac.uk/insight/2022/bitcoin-mining-new-data-reveal-a-surprising-resurgence/}{2022 report} sees American mining leading globally. Even they have had a \href{https://compassmining.io/education/the-worst-of-bitcoin-mining-in-2022/}{terrible time recently} with many companies either failing or looking likely to.\par
\begin{figure}
  \centering
    \includegraphics[width=\linewidth]{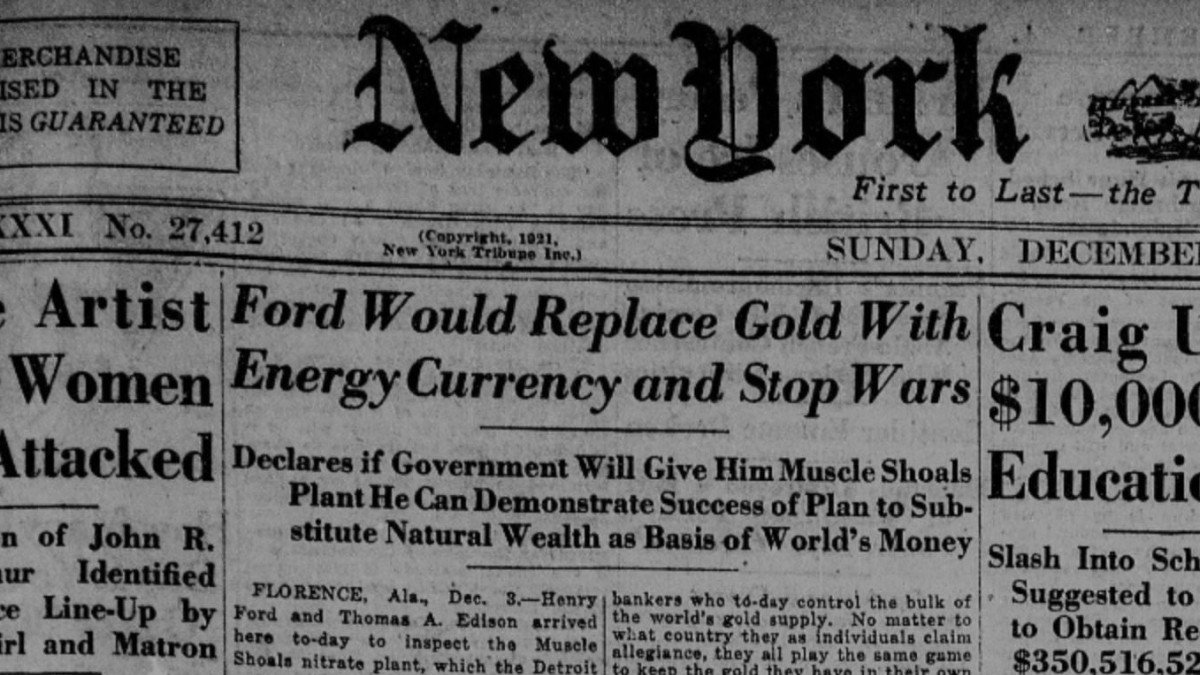}
  \caption{\href{https://www.nytimes.com/1921/12/06/archives/mr-fords-energy-dollar.html}{Intimate tie between energy and money, Henry Ford}}
  \label{fig:energyNYT}
\end{figure}

At the end of 2022 it is thought to be the case that the only profitable miners are the large scale companies who are also providing load balancing services to energy companies. This is unusual in the history of mining, and the situation will likely change over time.
This is not to say that all mining is, or should be, so concentrated. Anyone running the hashing algorithm can \href{https://twitter.com/ckpooldev/status/1485585814419812356}{get lucky} and claim the block reward. PoW ties the value of the `money' component of Bitcoin directly to energy production. This is not a new idea. Henry Ford proposed an intimate tie between energy and money to create a separation of powers from government, as can be seen in Figure \ref{fig:energyNYT}.\
The potential ecological footprint of the network has always been a concern; Hal Finney himself was \href{https://twitter.com/halfin/status/1153096538}{thinking about this issue} with a mature Bitcoin network as early as 2009, and \href{https://satoshi.nakamotoinstitute.org/posts/bitcointalk/threads/167/#35}{a debate} on the Bitcoin mailing lists called the mining process ``thermodynamically perverse''. The most cited negative analysis on the matter by Mora et al sees Bitcoin mining alone warming the planet above 2 degrees \cite{mora2018bitcoin}. \par
\begin{figure}
  \centering
    \includegraphics[width=\linewidth]{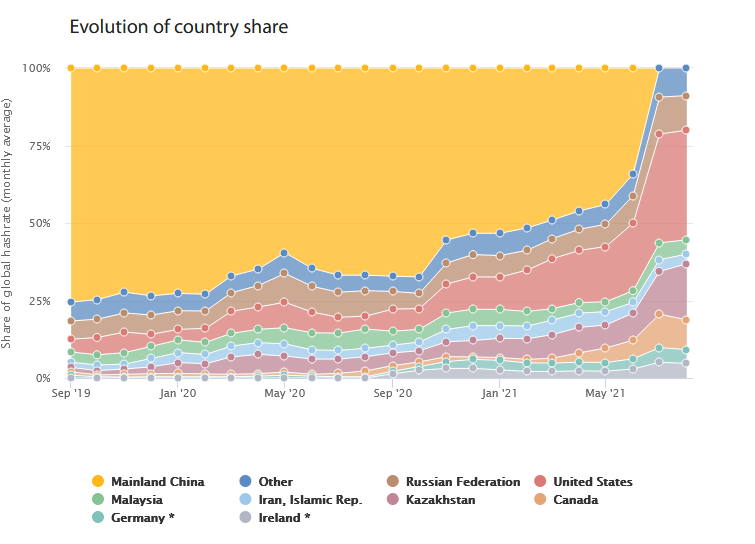}
  \caption{Hash rate \href{https://ccaf.io/cbeci/ining_map}{suddenly migrates} from China [Reuse rights requested]}
  \label{fig:miningshare}
\end{figure}

\href{https://electricmoney.org/}{Proponents of the technology} say that the balance shifted dramatically in 2021 with China outright banning the technology; this has forced the bulk of the energy use \href{https://docs.google.com/spreadsheets/d/1E7489rM7Q62oXwk1f4NUlMvok9noAbpYfTynY2VTyww/edit#gid=0}{toward the USA}, and away from `dirty coal' as seen in Figure \ref{fig:miningshare}. 
Some adherents \href{https://docs.google.com/document/d/1N2N-5jY00cmteoY_puWI9oosM1foa4EQqsO1FFfIFR4/edit}{have proposed mitigations} \cite{cross2021greening}. As a worked example of \href{https://docs.google.com/spreadsheets/d/15e_a-D3x4fv3tglEzFmQ6TLQx0fZe6-iKO9Fc9SyISQ/edit#gid=0}{Cross and Bailey's proposal} a retail investor owning 1 BTC would have to buy around 700 shares of `CleanSpark' mining company (CLSK) to make their \href{https://docs.google.com/spreadsheets/d/1r32T8p_PHTP8S781u7PhPSwehLx2VcJTaJJKesMswD0/edit#gid=0}{holding completely neutral}.  Some more strident voices suggest that \href{https://medium.com/@magusperivallon/a-financial-hail-mary-for-the-climate-an-argument-for-bitcoin-adoption-9c58e707d0}{`ending financialisation' through use of Bitcoin} may be net positive for the environment at a macro level \cite{bitcoinisvenice}. Indeed it may \href{https://www.newsweek.com/bitcoin-mining-americas-most-misunderstood-industry-opinion-1669892}{provide a route} to support \href{https://mobile.twitter.com/DSBatten/status/1514072998881665027}{electrifying everything} through deployment of \href{https://lancium.com/solutions/}{flexible demand load}. This enables a kind of \href{https://medium.com/@theendoftheworldpartyparty/deep-bitcarbonization-c8f483716ff7}{`financial battery'} that can soak up excess capacity from overbuilt renewables (something which needs to be done). Money saving uses like \href{https://www.theguardian.com/technology/2022/feb/09/can-bitcoin-be-sustainable-inside-the-norwegian-mine-that-also-dries-wood}{drying wood}, and even heating greenhouses (\href{https://www.euronews.com/next/2022/12/14/a-bitcoin-miner-and-tulip-grower-team-up-to-reduce-costs}{ironically tulips}) show the use case of the technology as a universal subsidy in heating applications.\par
Some projects are using the financial incentive of Bitcoin to enable trials of new infrastructure. For instance; Bhutan in the Himalayas has been quietly (or secretly!) mining Bitocin for years and plans a \href{https://www.straitstimes.com/business/bhutan-plans-a-500-million-fund-for-crypto-mining-in-the-himalayas}{\$500M fund} to expand specifically Bitcoin mining. Makai Ocean engineering have partnered with Oceanbit Hawaii to trail \href{https://en.wikipedia.org/wiki/Ocean_thermal_energy_conversion}{`ocean thermal energy conversion'} as a possible power source for the Islands. Local subsidy initiatives may begin to \href{https://fortune.com/2022/08/14/bitcoin-has-plunged-but-texas-miners-are-flush-with-profits-thanks-to-an-unusual-arrangement-the-state-is-paying-them-not-to-mine/}{drive this kind of adoption} as seems to be \href{https://braiins.com/blog/bitcoin-mining-the-grid-generators}{happening in Texas}\cite{griffith2021electrify, ercotimpact2021, Menati2022} and \href{https://www.governor.nh.gov/sites/g/files/ehbemt336/files/inline-documents/sonh/cryptocurrencie-report.pdf}{New Hampshire}. Brad Jones, \href{https://www.youtube.com/watch?v=gKnRfDeFgr0}{interim CEO of the Texas grid said}:\par
\textit{``As we get more renewable generation, in particular wind [which] is operating at night ...  we have to find a home for it, otherwise we have to turn the wind down. It’s such a great resource we shouldn’t turn it down. Bitcoin mining or what some call crypto has found a way to come into our markets and take some of that wind in off-peak periods. Then when we get to peak period times they are very quick to remove themselves from the market as prices increases The fact that we can turn down whenever we need the power for other customers is fantastic. We can use that crypto currency to soak up that excess generation when there’s a lot of that, and find a home for more solar and more wind to come to our grid. Then they reduce consumption when we need that power for other customers. So it’s a great balancing act. Most other datacenters [such as] Microsoft or Amazon have other customers to serve every other day, so they can’t just turn off. But these crypto customers can. If the cost of energy gets too high they can remove themselves from the market. They are also helpful if we lose a generator. They can quickly respond to that frequency disruption and allow us to balance our grid.''}\\
Flexible load balancing is entering the \href{https://www.forbes.com/sites/jemmagreen/2023/01/27/why-no-one-saw-the-success-of-demand-response-coming/?}{mainstream news cycle} as and is gaining traction in legislative bodies. This \href{https://www.citadel21.com/bitcoin-is-the-first-global-market-for-electricity-and-will-unleash-renewables}{``global energy market revolution''} is explained by Tabatabai of Modo Energy. Incredibly Bitcoin mining in Texas is now making the grid both more reliable and cheaper for consumers. In the UK we have similar problems because wind power is not evenly distributed, and moving it around is complex and expensive, and the whole system needs smoothing out with \href{https://archy.deberker.com/the-uk-is-wasting-a-lot-of-wind-power/}{gas plants}. \par 
There is growing interest and adoption of so called \href{https://www.bloomberg.com/news/articles/2022-06-01/oman-backs-u-s-firm-mining-crypto-to-cut-natural-gas-flaring}{``stranded energy mining''}  which cannot be effectively transmitted to consumers, and is thereby sold at a huge discount while also \href{https://www.renewableenergyworld.com/wind-power/900mw-wind-farm-to-power-bitcoin-mining-operation/}{developing power capacity}, without the \href{https://batcoinz.com/the-renewable-energy-cannot-happen-without-bitcoin-mining\%ef\%bf\%bc/}{usual constraints} \cite{bastian2021hedging}. One such example is \href{https://gridlesscompute.com/news/}{``Gridless'' in Kenya}, which seeks to harness abundant green energy resources in rural areas with the hope of kick-starting economic growth. This feature of the network first came to popular attention in 2020 when Stevens, CEO of Stoneridge capital included the following text in a letter to shareholders within their \href{https://www.stoneridgefunds.com/documents/AnnualReport.pdf?v=4}{annual report}: \textit{``Bitcoin mining is the only profitable use of energy in human history that does not need to be located near human settlement to operate. The long-term implications of this are world changing and hiding in plain sight.''}\par
In addition to new build it is possible to \href{https://www.curbed.com/2021/07/crypto-currency-mining-old-power-plants.html}{re-purpose historic} infrastructure, and/or \href{https://www.bloomberg.com/news/articles/2022-03-24/exxon-considers-taking-gas-to-bitcoin-pilot-to-four-countries}{reducing the carbon} (or \href{https://batcoinz.com/quantifying-the-potential-impact-of-bitcoin-mining-on-global-methane-emissions-4/}{more interestingly} the methane) of existing and \href{https://www.axios.com/2023/01/27/crypto-mining-advocate-green-abandoned-gas-wells}{abandoned infrastructure}. Both the \href{https://www.weforum.org/videos/this-start-up-catches-waste-methane-to-power-data-centres}{World Economic Forum}, and UK's Department for Energy \href{https://www.contractsfinder.service.gov.uk/Notice/a99aa482-c7d1-4ec4-82fe-5c4b16827a6d}{have expressed interest} in this use case. Adam Wright of \href{https://vespene.energy/}{Vespene Energy} says: \textit{``You could either mine Bitcoin on one small landfill for a year, or you could plant 5 million trees and let them grow for 10 years - both of those are going to have the same environmental impact.''}\par
\begin{figure}
  \centering
    \includegraphics[width=\linewidth]{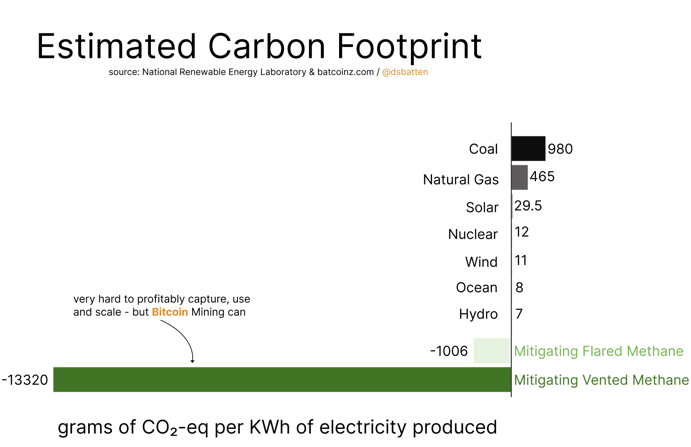}
  \caption{\href{https://twitter.com/DSBatten/status/1566735902617276416}{Climate tech investor Daniel Batten asserts that methane capture could highly impactful}}
  \label{fig:methane}
\end{figure}
Cheikosman, a policy analyst for the World Economic Forum (somewhat surprisingly) \href{https://www.weforum.org/agenda/2022/03/crypto-energy-consumption/?}{wrote} \textit{``Crypto is becoming an essential part of developing a carbon-neutral energy grid and has made it economically viable to invest in, develop and build renewable energy power generation.''} 
This is explored in detail by Ibanez et al \cite{ibanez2023sok}.\par
The most cited example of building capacity before grid connection is El Salvador's `volcano mining' proposal, which is supporting their national power infrastructure plans. Uzbekistan seems to be promoting a \href{https://www.reuters.com/business/finance/uzbekistan-legalises-solar-powered-crypto-mining-2022-05-04/}{similar model} with zero tax provided the Bitcoin mining companies build out their own solar infrastructure. A more poignant example is the \href{https://www.timesunion.com/news/article/Mechanicville-hydro-plant-gets-new-life-16299115.php}{Mechanicville hydro plant in the USA}. The refurbishment of this 123 year old power plant is being funded by Bitcoin mining. This is the \href{https://www.lynalden.com/bitcoin-energy/}{``buyer of last resort''} model first \href{https://squareup.com/us/en/press/bcei-white-paper}{advanced by Square Inc}.\par
Conversely it might be that vertical integration of Bitcoin mining \href{https://bitcoinmagazine.com/business/oil-companies-partner-with-bitcoin-miners}{within legacy fossil fuel stations} gives them a new lease of life. New York State has dealt with this kind of threat by imposing a moratorium on new, fossil fuel powered mining activity. On a global stage something as portable and industrial as Bitcoin mining will have unintended impacts on fragile energy systems, as has happened in \href{https://ceobs.org/environmental-governance-in-frozen-conflicts/}{South Ossetia} and \href{https://restofworld.org/2022/crypto-miners-fleeing-kazakhstan/}{Kazakhstan} (note \href{https://thenewscrypto.com/kazakhstans-crypto-miners-to-acquire-electricity-from-russia/}{Russia has stepped into this} mess). Undeniably the \href{https://time.com/6193004/crypto-climate-impact-facts/}{consensus position} is that it's overall very negative, (with some caveats) and this will \textit{probably} persist. Perhaps though if it's happening anyway, then finding utility of the asset might mitigate the net harm.\par
More pragmatically, Baur and Oll found that \textit{``Bitcoin investments can be less carbon intensive than standard equity investments and thus reduce the total carbon footprint of a portfolio.''}\cite{baur2021bitcoin}. Perhaps of note for the near future is that KPMG whose investment was mentioned in the introduction also matched their position in the space with equivalent  carbon offsets. This may provide an investment and growth model for others.\par
The first US-based \href{https://www.businesswire.com/news/home/20230305005096/en/TeraWulf-Announces-Energization-and-Rapid-Deployment-of-Mining-Operations-at-the-Nautilus-Facility-in-Pennsylvania}{nuclear-powered Bitcoin mining facility} has just opened in Pennsylvania. The facility has been completed by Cumulus Data, a subsidiary of independent power producer Talon Energy. Talon Energy owns the adjacent 2.5 GW Nuclear Power Plant and has been dabbling in Bitcoin mining for some time, opening a zero carbon Bitcoin mining facility in collaboration with Terawolf in August 2021. The new facility will operate with a maximum capacity of 48 MW, drawing on excess power from the nuclear plant. By locating the mining facility on the combined 1200 acre campus, there is no intermediation by legacy electric transmission and distribution utilities, as the mining is directly connected to the power station. Cumulus Data is in the process of building two additional 48 MW facilities and has identified 18 additional Talon Energy sites with potential to host data centres directly connected to electricity generation infrastructure. In general, there is a shift in attitudes towards nuclear in the US. The enormous benefit to this model stems from the cots associated with scaling down atomic power output to match grid requirements. By co-locating in this way the reactor can work at highest efficiency all the time, and can earn money from the generation of Bitcoin when the grid is unable to accept the energy. It is increasingly possible to find excited talk about funding smaller more pragmatic nuclear power plants using the cost benefits of the Bitcoin mining model, though this remains untested.\par
The power commitment to the network is variously projected \href{https://www.nature.com/articles/s41558-018-0321-8}{to increase}, or \href{https://assets.website-files.com/614e11526f6630959fc98679/616df63a27a7ec339f5e6a80_NYDIG-BitcoinNetZero_SML.pdf}{level off over time}. The emission schedule of the code suggests that the energy usage will decrease exponentially over time, and indeed many analysts feel that it has peaked due to a combination of factors. It's one of the maddening unknowns of the technology how this will all pan out. The \href{https://www.forbes.com/sites/martinrivers/2022/04/03/is-bitcoin-really-that-bad-for-the-environment/?sh=6a3203427143}{industry now argues} that economic pressures mean that most of the `hashrate' is \href{https://bitcoinminingcouncil.com/q4-bitcoin-mining-council-survey-confirms-sustainable-power-mix-and-technological-efficiency/}{generated by renewable energy}\cite{blandin20203rd}. As a recent example of this trend Telsa (Elon Musk), Block (Twitters Jack Dorsey), and Blockstream (Adam Back) are teaming up to \href{https://www.cnbc.com/2022/04/08/tesla-block-blockstream-to-mine-bitcoin-off-solar-power-in-texas.html}{mine with solar energy} in Texas.\par 
Paez and Cross \href{https://uploads-ssl.webflow.com/627aa615676bdd1d47ec97d4/62f41b9ce54e014f9869efa7_OSTP.docx.pdf}{prepared a paper} for the White House Office of Science and Technology Policy, submitted through the Bitcoin Policy Institute, which is a growing thinktank for academics and industry leaders. Their summary points echo the assertions made here, but they provide rich additional referencing for those who wish to dig deeper into this:\\
\textit{\begin{itemize}
\item Bitcoin’s value—its economic value and promotion of American values and American
national interests—must frame any discussion of its environmental impact.
\item Bitcoin’s value is inherently tied to its consensus mechanism: proof of work.
\item While bitcoin mining is energy-intensive, its energy use is often overestimated and improperly characterized as a function of transaction volume.
\item Due to bitcoin’s exponentially decreasing schedule of issuance, mining’s actual emissions are likely to peak at under 1\% of global emissions, even if prices rise more than tenfold within the
decade.
\item Mining’s profile as a consumer of energy is unique: extremely cost-sensitive, and invariant
across times and locations.
\item Bitcoin mining, as a buyer of first and last resort, incentivizes the buildout of renewable power production. As a controllable load resource (CLR) bitcoin mining also strengthens the grid, allowing it to reliably function at a high level of renewable penetration.
\item Mining’s energy use is increasingly non-rival, trending towards a diet of renewables andstranded, wasted energy resources such as flared methane.
\end{itemize}}
\par
\begin{figure}
  \centering
    \includegraphics[width=\linewidth]{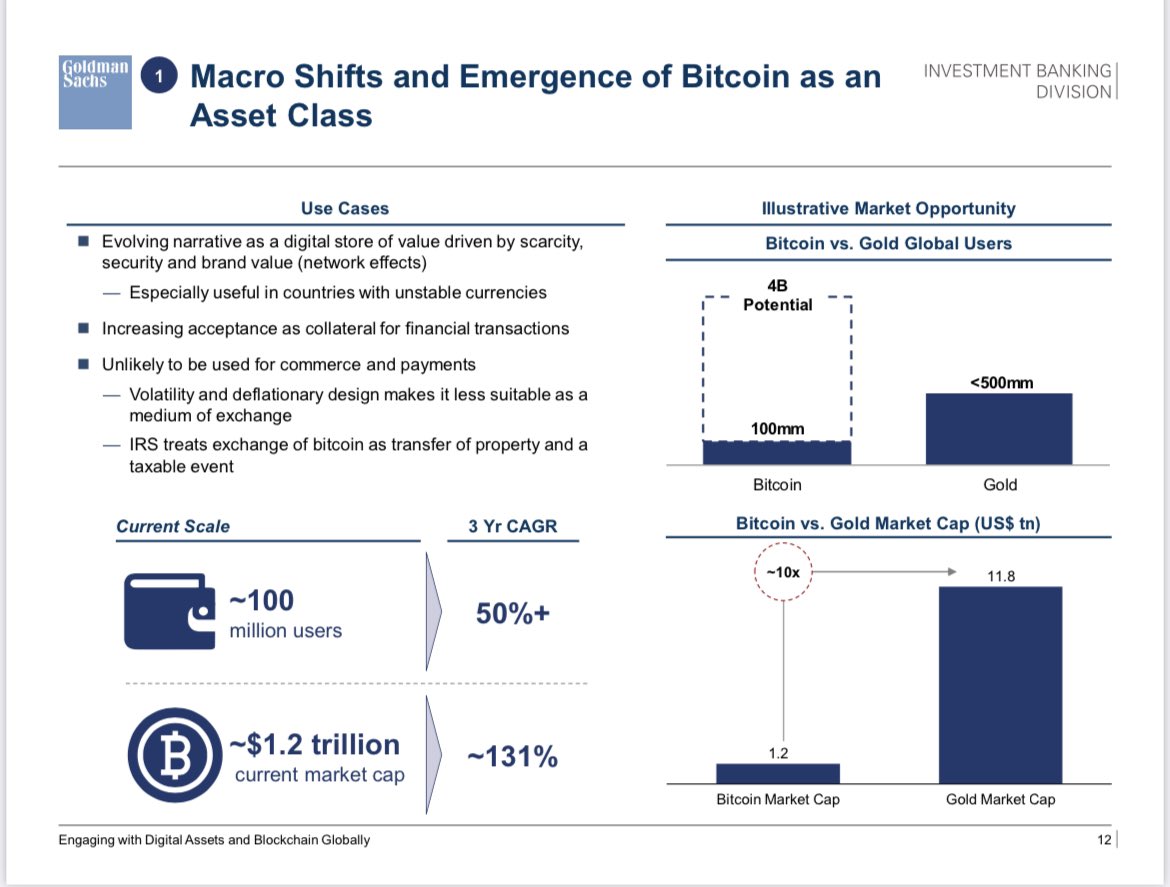}
  \caption{Goldman suggest growth opportunity and potential demonetisation of gold?}
  \label{fig:goldmanVgold}
\end{figure}

\href{https://www.youtube.com/watch?v=6LP8G-oZnEs}{The debate} whether this consumption is `worth' it is \href{https://www.utilitydive.com/news/bitcoin-mining-as-a-grid-resource-its-complicated/617896/}{complex} and \href{https://www.aei.org/technology-and-innovation/no-hearing-on-bitcoins-energy-use-is-complete-without-nic-carter/}{rapidly evolving}. Useful examples of this are:
\begin{itemize}
\item the \href{https://www.zerohedge.com/crypto/questionable-ethics-anti-bitcoin-esg-junk-science}{online pushback} to an academic article by PhD candidate de Vries et al. \cite{de2022revisiting}
\item the assertion that the widely cited Mora er al. paper in Nature \cite{mora2018bitcoin} was based on an \href{https://twitter.com/NateHawaii/status/1460706785216450560}{undergraduate class discussion}, and has had an outsized effect on global policy. 
\item a \href{https://rebrand.ly/v8qq1sx}{paper from the Bitcoin Policy Institute}, 
\item and the industry \href{https://bitcoinminingcouncil.com/wp-content/uploads/2022/05/Bitcoin_Letter_to_the_Environmental_Protection_Agency.pdf}{open letter to the EPA}.\par
\item this well considered \href{https://twitter.com/jyn_urso/status/1508899761319038983}{Twitter thread} by climate scientist Margot Paez.
\end{itemize}
It is somewhat confusing that positive views are coming only from diverse and non-specialist voices in the community, and never the academic community, but the shortcomings they point out in the supposedly considered articles such as Mora et al \cite{mora2018bitcoin} are easily verified. Academia seems \href{https://bitcoinmagazine.com/culture/bitcoin-could-never-be-invented-in-a-university}{poorly positioned} to pivot to this subject, as an ethical bar has to be established before research can commence, and the field is too new to make this an affordable task. This stuff is existentially important to the whole technology. Is a trillion dollar asset which \href{https://www.theheldreport.com/p/bitcoin-vs-gold}{potentially replaces} the money utility of gold, but doesn't need to be stored under guard in vaults (Figure \ref{fig:goldmanVgold}), worth the equivalent power consumption of clothes dryers in North America? Probably not with the current level of adoption, but this is an experiment in replacing global money. If that were to happen then Bitcoin would be around 50 times more efficient than the current system according to Khazzaka \cite{khazzaka2022bitcoin}. To be clear it's not the position of this book that replacing Fiat money is a good idea, but the experiment is being run regardless. This is explored in it's own chapter later.\par
It seems possible that eight value propositions are therefore emerging:
\begin{itemize}
\item Bitcoin the speculative asset (or greater fool bubble \cite{de1990positive}). Nations such as the USA, who own 30\% of the asset have bid up the price of the tokens during a period of very cheap money, and this has led to a high valuation for the tokens, with a commensurately high network security through the hash rate (mining). This could be a speculative bubble, with the asset shifting to one of the other valuations below. There is more on this subject in the money section later.
\item Bitcoin the (human) monetary network, and `emerging market' value transfer mechanism. This will be most useful for Africa (especially Nigeria), India, and South America. There is no sense of the ``value'' of this network at this time, but it's the aspect we need for our collaborative mixed reality application. For this use the price must simply be high enough to ensure that mining viably secures the network. This security floor is unfortunately a `known unknown'. If a global Bitcoin monetary axis evolves (as in the Money chapter later) the network would certainly require a higher rate than currently, suggestive of a higher price of token to ensure mining \cite{Wouters2022}.
\item Bitcoin as an autonomous AI monetary network. In an era where AI actors perform tasks on behalf of humans in digital realms such as cyberspace, these AI actors will require a reliable and efficient means of transaction. AI agents can perform, transact and negotiate, and execute work contracts in near real-time. For this use, the primary requirement is not a high token price, but rather a high level of network security and scalability that can support an enormous volume of transactions. The Lightning Network of Bitcoin might be a starting point but the robustness of the system, against potential AI exploits, is yet to be confirmed. As AI systems become more complex and autonomous, there is an increasing need for decentralized AI governance mechanisms that can prevent the concentration of power and ensure ethical AI development and deployment. Bitcoin can serve as a basis for this, providing a decentralized, transparent, and immutable record of AI decisions and actions. Furthermore, Bitcoin's proof-of-work consensus mechanism could potentially be adapted to enforce AI adherence to agreed-upon rules or norms. In this context, Bitcoin's value extends beyond its token price and into its potential contributions to AI governance and ethics. This is Bitcoin as an AI economy.
\item Bitcoin as a hedge against future quantum computation. It has been argued that the advent of quantum computers could threaten the security of many existing cryptographic systems. Bitcoin's open-source nature allows for the integration of post-quantum cryptographic algorithms, safeguarding it against quantum threats. In this sense, investment in Bitcoin might also be seen as an investment in a future-proof monetary network. This assertion depends on the assumption that Bitcoin's protocol will adapt in time to incorporate such cryptographic advances before quantum computing becomes a real threat to its integrity. The practical implementation of these technologies might see a shift in the network's dynamics - the hash rate, mining cost, and token value.
\item Bitcoin's value in terms of `sunk opportunity cost'. This refers to the value that could have been generated if the resources invested in a particular activity had been utilised elsewhere. In the context of Bitcoin, this includes the investments made in mining equipment, power, facilities, and the hiring of skilled personnel to maintain the operations. The sunk opportunity cost of Bitcoin can be substantial. It can be argued that the value of Bitcoin must take this cost into consideration, as the resources could have been allocated to other productive sectors or investments \cite{chiu2017economics}. Of course, there remains the infamous sunk cost fallacy, which refers to the tendency of individuals or organizations to continue investing in a project or decision based on the amount of resources already spent, rather than evaluating the current and future value of the investment. This indeed tends to lead to a cyclical boom and bust dynamic in the industrial mining communities. The ultimate fallacy would occur if miners or investors continued to invest in mining equipment and operations solely because of the resources that have already been spent on them, and the asset simply crashes to nothing from here. It's a shaky justification because it assumes the future is the same as the past.
\item Bitcoin as a flexible load in power distribution systems, and methane mitigation `asset', and `subsidised heater' for varied applications such as growing and drying. Again there is no price against this, but we can perhaps grossly estimate it at around half the current hash rate if 50\% of the network is currently green energy. This would imply a price for the asset roughly where it is now (ie, not orders of magnitude higher or lower). 
\item The 2023 global bank runs have awoken some companies to the risks of access to cash flows in a potential crisis \cite{jiang2023monetary}. Access to a small cache (in corporate treasury terms) of a highly liquid \& tradable asset could allow continuity of payroll in a `24/7' global context. This could avoid or at least mitigate the panic which ensues in companies when banks are forces to suddenly wind up their operations.  
\item Amusingly \href{https://www.epsilontheory.com/in-praise-of-bitcoin/}{Ben Hunt suggests} in an online article that the true value of Bitcoin can be couched in terms of it's value simply as `art'. He posits that at this time the narrative is simply so seductive and powerful that people (being people) are choosing to value their involvement in the economics of the space as they might a work of art. It's a fascinating idea, and intuitively, probably it's right. 
\end{itemize}
Legislators globally, are \href{https://www.lopp.net/bitcoin-information/legal.html}{starting to codify} their positions on proof of work as a technology (including Bitcoin). US States are variously \href{https://capitol.texas.gov/tlodocs/88R/billtext/html/HC00089I.htm}{supporting} or constricting the technology, according to \href{https://www.ncsl.org/research/financial-services-and-commerce/cryptocurrency-2021-legislation.aspx}{state legislatures}. Notably New York has \href{https://www.nysenate.gov/legislation/bills/2021/A7389}{banned new carbon intensive} mining facilities for 2 years, while rust and farm belt states with energy build-out problems are \href{https://financialpost.com/fp-finance/cryptocurrency/texas-governor-abbott-turns-to-bitcoin-miners-to-bolster-the-grid-and-his-re-election}{providing incentives} and passing legislation to protect \href{https://www.arkleg.state.ar.us/Bills/Detail?id=hb1799}{mining datacenters}. At the federal level the white house has strongly signalled their concerns about the sector \href{https://www.whitehouse.gov/wp-content/uploads/2023/03/ERP-2023.pdf}{in a report}. Many of the points in the report are fair, and true, and reflect things said in this book (which pre-dates the report). It's worth picking out the conclusion of that section verbatim: \textit{``Innovation in financial services brings both risks and opportunities for the broader economy. It can challenge business models and existing industries, but it cannot challenge basic economic principles, such as what makes an asset effective as money and the incentives that give rise to run risk. Although the underlying technologies are a clever solution for the problem of how to execute transactions without a trusted authority, crypto assets currently do not offer widespread economic benefits. They are largely speculative investment vehicles and are not an effective alternative to fiat currency. Also, they are too risky at present to function as payment instruments or to expand financial inclusion. Even so, it is possible that their underlying technology may still find productive uses in the future as companies and governments continue to experiment with DLT. In the meantime, some crypto assets appear to be here to stay, and they continue to cause risks for financial markets, investors, and consumers. Much of the activity in the crypto asset space is covered by existing regulations and regulators are expanding their capabilities to bring a large number of new entities under compliance (SEC 2022). Other parts of the crypto asset space require coordination by various agencies and deliberations about how to address the risks they pose (U.S. Department of the Treasury 2022a). Certain innovations, such as FedNow and a potential U.S. CBDC, could help bring the U.S. financial infrastructure into the digital era in a clear and simple way, without the risks or irrational exuberance brought by crypto assets. Hence, continued investments in the Nation’s financial infrastructure have the potential to offer significant benefits to consumers and businesses, but regulators must apply the lessons that civilization has learned, and thus rely on economic principles, in regulating crypto assets.''}\par
\begin{figure}
  \centering
    \includegraphics[width=\linewidth]{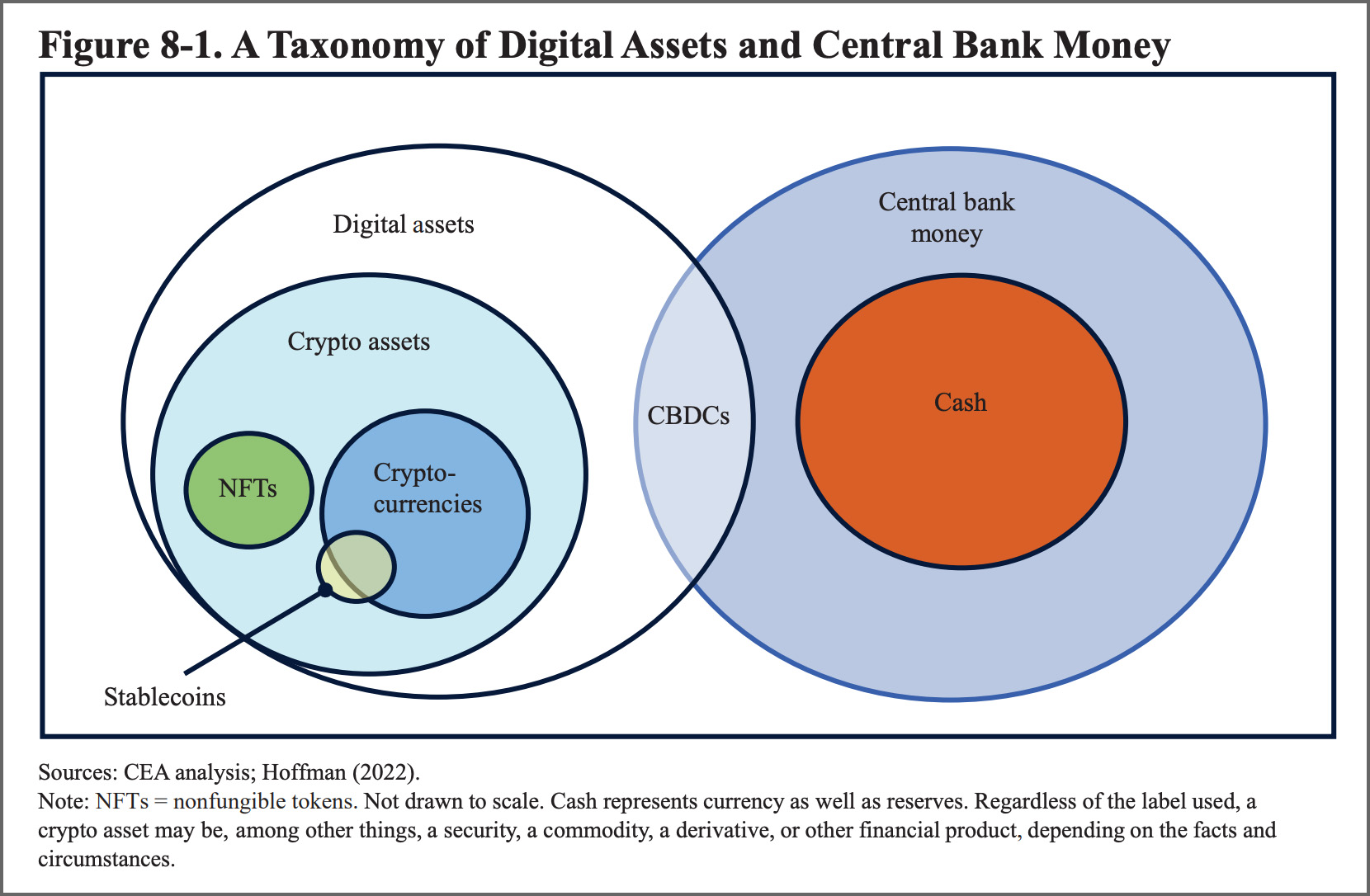}
  \label{fig:PresidentTaxonomy}
\caption{Taxonomy of digital assets Hoffman 2022}
\end{figure}

Reading between the lines suggest that strong regulation is coming. Indeed the \href{https://www.sec.gov/Archives/edgar/data/1679788/000167978823000051/coin-20230322.htm}{SEC is now suing} the major tech company in the space, Coinbase, while closing a bank servicing the sector, and signalling that stablecoins may be unregistered securities in law. The report itself has no `teeth' but is likely a sign of things to come. There is purportedly \$2.4B \href{https://docs.house.gov/meetings/AP/AP23/20230329/115576/HHRG-118-AP23-TTF-GenslerG-20230329.pdf}{entering the regulation ecosystems} to enhance regulatory oversight. In actual fact, because of the nature of the federation of states it is likely that a variety of different approaches in law will be taken across the geography and the sector seems to have responded with a shrug. As an aside the report contains an excellent taxonomy of digital assets from Hoffman (Figure \ref{fig:PresidentTaxonomy}).\par
Conversely the recent ``\href{https://www.whitehouse.gov/ostp/news-updates/2022/09/08/fact-sheet-climate-and-energy-implications-of-crypto-assets-in-the-united-states/}{Climate and energy implications}'' report is parts positive and parts negative about proof of work, and leaves the door open to a legislative clampdown. This is most notable in a \href{https://www.whitehouse.gov/cea/written-materials/2023/05/02/cost-of-cryptomining-dame-tax/}{White House proposal} to tax Bitcoin mining at 30\%, a plan which will destroy much of the US based mining industry over the coming years. Carter provides a \href{https://medium.com/@nic__carter/comments-on-the-white-house-report-on-the-climate-implications-of-crypto-mining-8d65d30ec942}{detailed response} to the tardy scientific analysis in the report. Perhaps most interestingly it notes the potential of methane mitigation as mentioned earlier. It is conceivable that methane mitigation alone could provide a route forward for the technology. The report says:
\textit{``The crypto-asset industry can potentially use stranded methane gas, which is the principal component of natural gas, to generate electricity for mining. Methane gas is produced during natural gas drilling and transmission, and by oil wells, landfills, sewage treatment, and agricultural processes. Methane is a potent GHG that can result in 27 to 30 times the global warming potential of CO2 over a 100-year time frame, and is about 80 times as powerful as CO2 over a 20-year timeframe. Reducing methane emissions can slow near-term climate warming, which is why the Biden-Harris Administration released the U.S. methane emissions reduction action plan in 2021. Venting and flaring methane at oil and natural gas wells wastes 4\% of global methane production. In 2021, venting and flaring methane emitted the equivalent of 400 million metric tons of CO2, representing about 0.7\% of global GHG emissions. This methane is vented or flared, because of the high cost of constructing permanent pipelines or electricity transmission that could transport the methane or its potential electricity generation from remote oil and gas operations to end-users, or because of the high cost of installing equipment on older landfills. Crypto-asset companies are now exploring ways to use electricity generation from vented and flared methane at oil and gas wells and at landfills.
While the EPA and the Department of the Interior have proposed new rules to reduce methane for oil and natural gas operations, crypto-asset mining operations that capture vented methane to produce electricity can yield positive results for the climate, by converting the potent methane to CO2 during combustion. Mining operations that replace existing methane flares would not likely affect CO2 emissions, since this methane would otherwise be flared and converted to CO2. Mining operations, though, could potentially be more reliable and more efficient at converting methane to CO2. While such operations can reduce wasted methane, another option is low-cost recovery of methane using existing vapor capture technologies at oil and gas wells, which can reduce global methane emissions up to 50\% by 2030.''}\par
The EU has just voted to add the whole of `crypto', including PoW, to the EU taxonomy for sustainable activities. This EU wide classification system provides investors with guidance as to the sustainability of a given technology, and can have a meaningful impact on the flows of investment. With that said the report and addition of PoW is not slated until 2025, and it is by no means clear what the analysis will be by that point. Meanwhile they're tightening controls of transactions, on which there will be more detail later. For it's part the European Central Bank has come out \href{}{in favour of strong constraints} on crypto mining. They use the \href{https://medium.com/crescofin/the-reports-of-bitcoin-environmental-damage-are-garbage-5a93d32c2d7}{widely discredited} ``digiconimist'' estimates to assert that mining operations are \href{https://www.ecb.europa.eu/pub/financial-stability/macroprudential-bulletin/html/ecb.mpbu202207_3~d9614ea8e6.en.html}{disproportionately damaging to the environment}.  \par
We have seen that China has cracked down hard on the technology, banning mining and pressuring holders of the assets. They have unwound this somewhat, and based on past experience it seems that they will continue to nuance their position as they seek adoption of their own digital currency. As much as 20\% of all mining activity is now suspected to take place within China.\par
In India there has been confusion for years as more ``local'' law vies with confusing central government signalling. It has variously been banned and unbanned, and is now subject to punitive tax. The central bank of India is \href{http://164.100.24.220/loksabhaquestions/annex/179/AS10.pdf}{strongly in favour} of a complete ban. Ajay Seth, secretary of the Finance Ministry's Department of Economic Affairs recently said \textit{``We have gone through a deep dive consulting with not just the domestic and institutional stakeholders but also organizations like IMF and World Bank.... Simultaneously we are also beginning our work for some sort of a global regulation (to determine) what role India can play... Whatever we do, even if we go to the extreme form, the countries that have chosen to prohibit, they can't succeed unless there is a global consensus''}\par
It feels like a global political response is just around the corner, but reputable voices in the community suggest that it always feels this way. There is more detail on this in Money chapter later in the book.
\subsection{Technical overview}
This section could be far more detailed, but this is pretty complex stuff. Instead, there's plenty of \href{https://github.com/bitcoinbook/bitcoinbook}{books and websites} that do a more thorough job, if the reader is interested. Each subsection will include a good external link where more depth can be found. This whistle stop tour of the main components of the protocol should provide enough grounding, but it's not essential reading for non technical readers.\par
\subsubsection{ECDSA / SHA256 / secp256k1}
These technologies tend to use the same underpinning \href{https://curves.ulfheim.net/}{elliptic curve cryptography}, and it makes sense to unpack this here just once, only in the context of Bitcoin, as this will be the main focus of our attention.\par
Public keys are a huge number used in conjunction with an algorithm to encrypt data. This allows a remote party to interact with an actor on the network whose private keys can decrypt that same data.\par
In Bitcoin the ECDSA algorithm is used on the \href{https://en.bitcoin.it/wiki/Secp256k1}{secp256k1} elliptic function to create a trapdoor. This (essentially) one way mathematical operation was originally the ``discrete log problem'' and part of the research in cryptography by Diffie and Hellman \cite{diffie1976new}. This is what binds the public and private keys in a key pair (the foundation of the whole space).\par 
In their mathematical construct a modulus operator creates an infinite number of possible variations on operations which multiply enormous exponential numbers together, in different orders, to create key pairs. In order to reverse back through the `trapdoor' a probably impossible number of guesses would have to be applied.\par
\begin{figure}
  \centering
    \includegraphics[width=\linewidth]{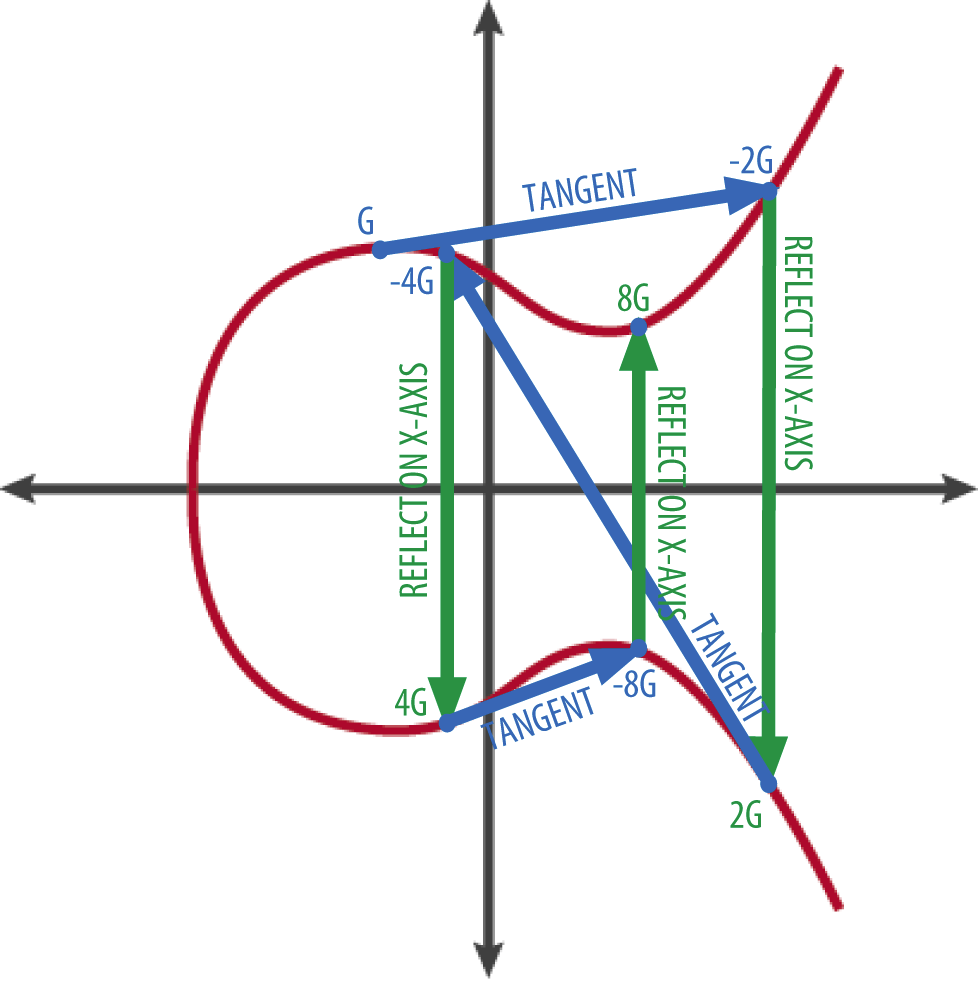}
  \caption{Given a start point on the curve and a number of reflection operations it's trivial to find a number at the end point, but impossible to find the number of \href{https://github.com/bitcoinbook/bitcoinbook/blob/develop/ch04.asciidoc}{hops} from the two end points alone. (CC Mastering Bitcoin second edition)}
  \label{fig:ECDSA}
\end{figure}

Latterly, elliptic curves such as the secp256k1 curve used in Bitcoin have substantially simplified the computation problems. Rather than exponentials used by Diffie Helmman instead a repeated operation is applied to an elliptic curve function, and this itself creates a discrete log problem trapdoor in maths, far more efficiently. Figure \ref{fig:ECDSA} suggests how this works. \par
This makes it easier, faster, and cheaper to provide secure key pairs on basic computational resources. Elliptic curve solutions are not \href{https://safecurves.cr.yp.to/}{`provably' secure} \cite{gayoso2018secure} in the same way as the Diffie-Hellman approach, and the security of this system is very sensitive to the randomness which is applied to the operation. Aficionados of Bitcoin use dice rolls or \href{https://www.hackster.io/news/alex-waltz-s-quantum-random-number-generator-for-bitcoin-uses-radioactive-decay-and-a-raspberry-pi-25a75316220f}{even more exotic} means to add entropy (randomness) when creating keys. This really isn't necessary, the software does this well enough.\par  
ECDSA has already been replaced by the more efficient Schnorr signature method \cite{schnorr1989efficient} which uses the same mathematical curve so is backward compatible. This will take some time for organic adoption, and ECDSA will never be deprecated.\par
\subsubsection{Analysis of the underlying security}
The evidence available paints a picture of strong and improving but not ironclad cryptographic robustness. 
\begin{itemize}
\item Who is involved:
\begin{itemize}
\item Pieter Wuille, Gregory Maxwell, and Andrew Poelstra are some of the most prominent Bitcoin Core developers and cryptographers. Their extensive experience and contributions to Bitcoin speak to their credentials.
\item Jonas Nick is a cryptography researcher who has published analyses of Bitcoin's ECDSA signature scheme.
\item Tim Ruffing, Pedro Moreno-Sanchez, and Yannick Seurin are cryptographers who have published academic papers analyzing and improving Bitcoin's cryptographic constructions.
\item Tadge Dryja is one of the original Lightning developers and an MIT DCI researcher focused on Bitcoin's cryptography.
\item Arvind Narayanan is a professor at Princeton who has published seminal works on Bitcoin and cryptocurrency cryptography.
\item Aviv Zohar at the University of Jerusalem has helped identify \href{https://lists.linuxfoundation.org/pipermail/bitcoin-dev/2023-October/021999.html}{still extant} vulnerabilities in the Lightning network \cite{harris2020flood}.
\item There are likely many other credentialed cryptographers working on Bitcoin, but these are some of the most well-known examples. The common thread is a strong academic background in cryptography.
\end{itemize}
\item Where critical analysis is published:
\begin{itemize}
\item Many analyses have been published at top cryptography and security conferences like IEEE S\&P, ACM CCS, Financial Cryptography, and Real World Crypto. These are generally highly regarded venues.
\item Academic journals like Journal of Cryptography, Ledger, and Journal of Cryptographic Engineering have also published peer-reviewed papers. The impact factors are not as high as broader CS journals however.
\item Bitcoin Improvement Proposals (BIPs) also undergo scrutiny by expert reviewers.
\item There is still room for more peer review in top journals, but publication venue credibility is generally decent.
\end{itemize}
\item Test of time:
\begin{itemize}
\item Bitcoin has been live since 2009 and its core cryptographic protocols have stood the test of time so far, with no major breaks of ECDSA, SHA256, RIPEMD160, etc despite enormous incentives.
\item That being said, 13 years is short in cryptography timescales. Continued analysis over decades is ideal.
\item Some newer proposals like Taproot have undergone less time testing but build on proven constructions.
\end{itemize}
\item Academic acceptance:
\begin{itemize}
\item Acceptance has been increasing steadily, with a growing number of peer-reviewed academic papers, PhD dissertations, and university courses on Bitcoin cryptography.
\item Funding remains limited but is increasing through organizations like Square Crypto that fund open source work.
\item Controversy around Bitcoin's social implications persists but its technical merit and cryptography seem to be gaining more mainstream academic respect.
\end{itemize}
\end{itemize}
\subsubsection{Addresses \& UTXOs}
Ethereum has addresses which transactions flow in and out of. This is synonymous to a bank account number and so makes intuitive sense to users of banks. This is not the case in Bitcoin.\par
Bitcoin is a UTXO model blockchain. UTXO stands for unspent transaction output, and these are `portions' of Bitcoin created and destroyed as value changes hands (through the action of cryptographic keys). They are the basis of the evolving ledger. This process is described well by Rajarshi Maitra in \href{https://medium.com/bitbees/what-the-heck-is-utxo-ca68f2651819}{this post}.\par
\textit{``Every Transaction input consists of a pointer and an unlocking key. The pointer points back to a previous transaction output. And the key is used to unlock the previous output it points to. Every time an output is successfully unlocked by an input, it is marked inside the blockchain database as `spent'. Thus you can think of a transaction as an abstract “action” that defines unlocking some previous outputs, and creating new outputs.\\
These new outputs can again be referred by a new transaction input. A UTXO or `Unspent Transaction Output' is simply all those outputs, which are yet to be unlocked by an input.\\
Once an output is unlocked, imagine they are removed from circulating supply and new outputs take their place. Thus the sum of the value of unlocked outputs will be always equal to the sum of values of newly created outputs (ignoring transaction fees for now) and the total circulating supply of bitcoins remains constant.''}\par
Fresh UTXOs are created as coinbase transactions, rewarded to miners who successfully mine a block. These can be spent to multiple output as normal. This is how the supply increases over time.
\subsubsection{Bitcoin script and miniscript}
A Bitcoin script is a short chunk of code written into each transaction which gives conditions for the next UTXO transfer (spend). Bitcoin script is a programming language invented by Satoshi Nakamoto as part of the Bitcoin system. It's a stack-based language, similar to reverse Polish notation, used to encode transactions and specify the conditions under which a Bitcoin address can be spent. Bitcoin script has 256 op codes, some of which are deprecated or can cause the program to fail.\par
Miniscript is a higher-level language that makes it easier to write robust Bitcoin smart contracts on chain. It smooths out the rough edges of Bitcoin script and makes it more accessible for non-technical users to understand and use. Miniscript provides a more intuitive way to specify spending conditions, making it easier for users to create smart contracts without needing to be an expert in programming languages like Rust or C++. Miniscript takes the basket of 256 op codes in Bitcoin script and simplifies them, making the most commonly used op codes more accessible and usable for average users. The \href{https://bitcoin.sipa.be/miniscript/}{limited scripting language} and the features built into wallets on top, allow for some clever additional options beside receiving and spending. In fact, some of the more innovative features such as discrete log contracts (detailed later) are quite powerful, and can interact with the outside world. Scripts allow spends to be contingent on multiple sets of authorising keys, time locks into the future, or both.\par 
Time locks can be either block height-based or wall time-based, but Miniscript ensures that the user has to choose one or the other within a single Bitcoin script. This is because some of the time lock opcodes like "check lock time verify" (CLTV) and "op sequence verify" change their behavior based on whether the code is four or five bytes long. Miniscript removes these quirks by providing a unified and more intuitive way to write smart contracts. An example of Miniscript's functionality is a decaying multi-sig where a five of five multi-sig can be changed over time to a four or five multi-sig, or a three of five multi-sig, in case one or two keys are lost. This provides the user with more control and flexibility over their money and allows for contingencies in case of loss events. Additionally, Miniscript enables users to have \href{https://bitcoindevkit.org/bdk-cli/playground/}{more control over their funds} by setting rules when money is put into a Bitcoin address, as well as allowing for corporate governance situations. Miniscript can also be used for inheritance planning, where a child's key can be made to activate after a certain number of blocks have passed, creating a "dead man switch" functionality on the chain.\par
Overall, Miniscript enables users to have more control and flexibility over their funds, making Bitcoin smart contracts more robust and secure, but it's important to note that this is new technology and not yet integrated into user wallets.
\subsubsection{Halving}
As mentioned eariler, roughly every four year the `block reward' given to miners halves. This gives the issuances schedule of Bitcoin; \href{http://bashco.github.io/Bitcoin_Monetary_Inflation/}{it's monetary inflation}. This `controlled supply' feature was added to emulate the growth of physical asset stocks through mining. It's exhaustively \href{https://en.bitcoin.it/wiki/Controlled_supply}{explained elsewhere} and is somewhat immaterial to our transactional use case in metaverse applications.
\subsubsection{Difficulty adjustment}
The difficult adjustment (also mentioned earlier) shifts the difficulty of the mining algorithm globally to re-target one new block every 10 minutes. This means that adding a glut of new mining equipment will increase the issuance of Bitcoins, in favour of the new mining entity, for up to 2 weeks, at which point the difficulty increases, the schedule resets, and the advantage to the new miner is diffused. Equally this protects the network against significant loss of global mining hashrate, as happened when China comprehensively banned mining. Again, this is explained in \href{https://en.bitcoin.it/wiki/Difficulty}{more detail} elsewhere.
\subsubsection{Bitcoin nodes }
The Bitcoin network can be considered a triumvirate of economic actors, each with different incentives. These are:
\begin{itemize}
\item Holders and users of the tokens, including exchanges and market makers, who make money speculating, \href{https://en.wikipedia.org/wiki/Arbitrage}{arbitraging}, and providing liquidity into the network. Increasingly this is also real `money' users of BTC, earning and spending in pools of circular economic activity. Perversely Bitcoin as a money is the fringe use case at this time.
\item Miners, who profit from creation of new UTXOs, and receive payments for adding transactions to the chain. In return they secure the network by validating the other miners blocks according the rules enforced by the node operators.
\item Node operators, \href{https://www.truthcoin.info/blog/measuring-decentralization/}{who enforce the consensus} rule-set, which the miners must abide by in order to propagate new transaction into the network. In return node operators optimise their trust minimisation, and help protect the network from changes which might undermine their speculation and use of the tokens \cite{blocksizewars}.
\end{itemize}
There are currently around \href{https://bitnodes.io/}{15,000 Bitcoin nodes} distributed across the world. 
Since IT engineer \href{https://stadicus.com/}{Stadicus} released his \href{https://raspibolt.org/backstory.html}{Raspibolt guide} in 2017 there has been an explosion of small scale Bitcoin and Lightning node operators. 
Around thirty thousand Raspberry Pi Lightning nodes (which are also by definition Bitcoin nodes) run one of a big selection of \href{https://github.com/bavarianledger/bitcoin-nodes}{open source distributions}. We will build toward our own throughout the book.
\subsubsection{Wallets, seeds, keys and BIP39}
In all the cryptographic systems described in this book everything is derived from a private key. This is an enormous number, and the input to the trapdoor function described earlier. As usual, it's beyond the scope of this book to `rehash' the detail. Prof Bill Buchanan OBE has a \href{https://medium.com/asecuritysite-when-bob-met-alice/can-i-derive-the-private-key-from-the-public-key-ba3609256ec}{great post} on the basic version of this process.\par
In modern wallets, private keys (and so too their public keys), and addresses, are generated hierarchically. This is all part of \href{https://github.com/bitcoin/bips/blob/master/bip-0032.mediawiki}{BIP-0032}. It starts with a single \href{https://www.wolframalpha.com/input?i=2\%5E512}{monstrously large} number of up to 512 bits. From this are crafted Hierarchical Deterministic (HD) wallets, which use `derivation paths' to make a tree of public/private key pairs, all seeded from this first number. This means that knowing the initial number, and the derivation path applied to it (just another number), wallets can search down the tree of derivations and find all the possible addresses. In this way a whole group of active addresses belonging to an entity can be held conveniently in one huge number (a concatenation of the input and path). This is the seed. Seeds are even more conveniently abstracted into a mnemonic called a seed phrase. Anyone interacting with these systems will see a 12 word (128 bits of entropy which is considered to be \href{https://twitter.com/adam3us/status/1433375602808066049}{`enough'}) or 24 word (256 bit) seed phrase. That phrase accesses the whole of the assets stored by that entity in the blockchain under it. A master key. These seeds can be \href{https://vault12.com/securemycrypto/cryptocurrency-security-how-to/dice-crypto-recovery-seed/}{generated by hand} with dice, remember it's just a huge number and the onward cryptography at play here.\par
An address in Bitcoin is derived from the public/private key pair. Again this is a one way hash function. The public/private keys can't be found from the address. Addresses are really only a thing in wallets. They contain the element necessary to interact with the UTXOs. Many UTXOs can reside under an address, in that they just share the same keys. Wallets and nodes can monitor the blockchain to see transactions that `belong' to addresses owned by the wallet, then they can perform unlocking of those funds to move them, through operations on the UTXOs via keys.\par
\subsubsection{HD wallet encoding - ideas}
\begin{figure}
  \centering
    \includegraphics[width=\linewidth]{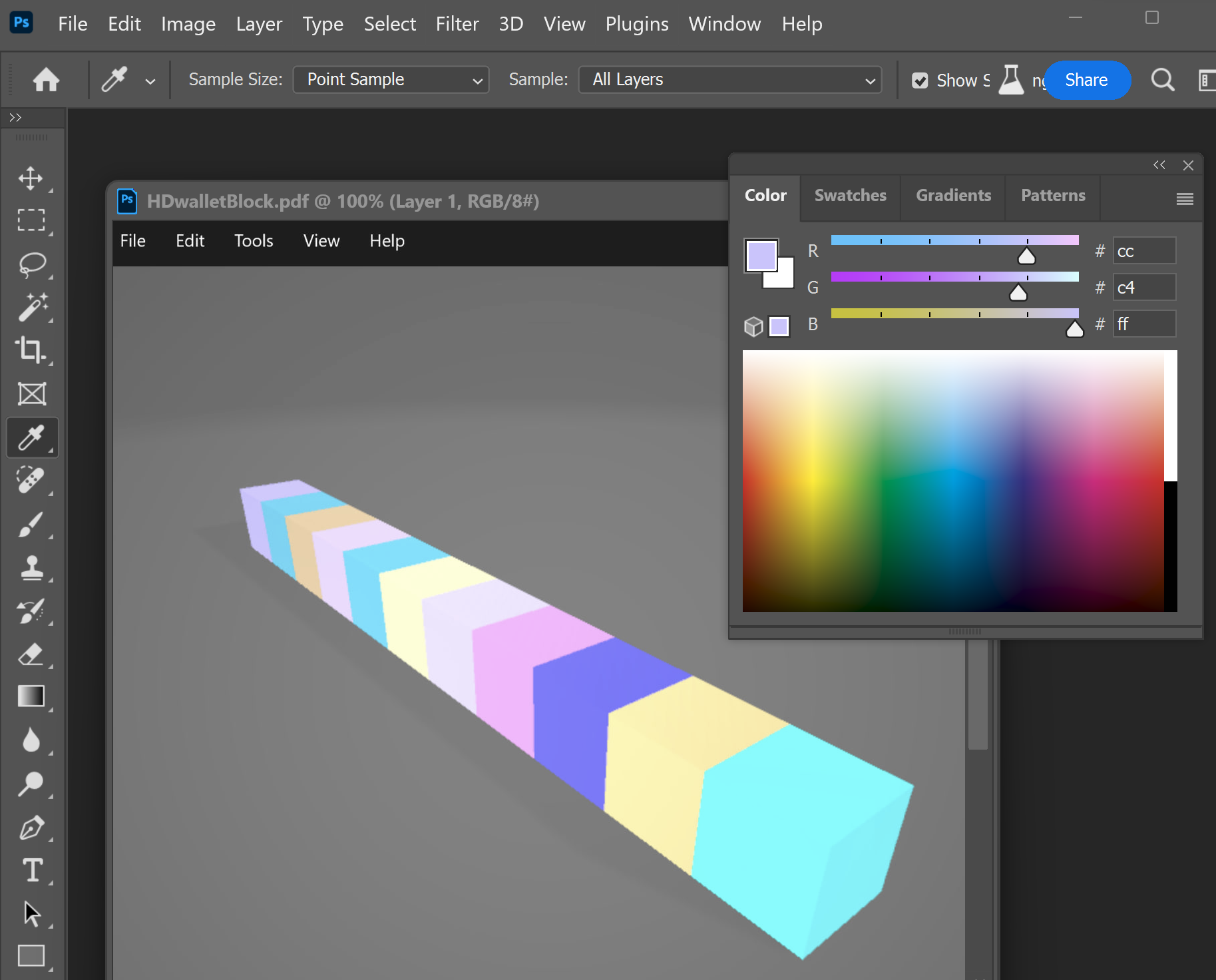}
  \caption{This is the nostr pubkey for flossverse, encoded into the far larger HD wallet space (hence the muted colours) and then displayed as blocks.}
  \label{fig:HDwalletBlock}
\end{figure}

The BIP39 standard is designed to create a human-readable and easily portable format for Bitcoin and other cryptographic wallet seeds. By representing these seeds as a series of colors in a 3D model, we add a new dimension of portability, visual appeal, and potential applications. 
\begin{itemize}
    \item \textbf{Easter Egg Hunts in Social Metaverses}: The color-based representation of BIP39 seeds opens up opportunities for creative and engaging experiences in metaverse environments. For instance, mnemonic seeds could be hidden within digital artifacts in the form of color sequences. These could be used as treasure hunts or easter eggs, potentially carrying real-world value in the form of Bitcoin or other cryptocurrencies. Recovery would mean some kind of sampling as in Fig \ref{fig:HDwalletBlock} and software.
    \item \textbf{Provenance Encoding in Digital Art}: The mnemonic color-coding could be embedded within digital art pieces, effectively encoding the provenance of the artwork directly into its visual representation. This could add an extra layer of security and uniqueness to the art piece, and also serve as a novel way of proving ownership or creatorship.
    \item \textbf{Steganographic Transfer of Funds}: By incorporating the color-encoded BIP39 seeds into various aspects of a metaverse or digital environment, they can serve as a form of steganography. This allows for the transfer of funds or sensitive information covertly within the visual and experiential components of the environment.
    \item \textbf{Gamification of Cryptographic Keys}: Cryptographic keys are typically represented as long, random strings of characters that are hard to remember and not very user-friendly. Representing these keys as a sequence of colors could make them more approachable and memorable. This could also introduce an aspect of gamification into the world of identity and value, possibly increasing their appeal to a broader audience.
    \item \textbf{Embedding in Physical Objects}: 3D printing technologies could be used to create physical representations of these 3D models (assuming some variance of the colour as the materials age), embedding BIP39 seeds into tangible, physical objects. These could serve as novel physical wallets, gift items, or physical tokens representing digital assets.
\end{itemize}
While this encoding scheme opens up numerous creative opportunities, it is important to be aware of potential security implications. The use of mnemonic seeds in this way should be done with care and an understanding of the risks involved.\par
The GitHub repository for this book \href{https://github.com/flossverse/bip39Geom}{has some example code} playing around with this idea further, generating a color-based and a three-dimensional graphical representation of nostr addresses. Each mnemonic word is mapped to a unique color and a 3D model is created, in which each word of the mnemonic is represented by a cube of the corresponding color arranged in a circle.
\begin{itemize}
    \item \textbf{BIP39Colors}: A class that contains a list of BIP39 words and methods to convert a hexadecimal seed to RGB colors and mnemonic words.
    \item \textbf{hex\_to\_rgb}: A static method within the BIP39Colors class that converts a hexadecimal color to an RGB color.
    \item \textbf{seedToColors}: A static method within the BIP39Colors class that converts a seed (a string of hexadecimal characters) into colors and mnemonic words. This function first converts the seed into bytes, then divides these bytes into groups and converts each group into a position in the word list and a color.
    \item \textbf{generate\_3d\_model}: A function that takes a list of colors as input, and generates a 3D model with cubes of these colors arranged in a circular formation.
    \item \textbf{main}: The main function of the script, which takes a 64-character hexadecimal string (a nostr ID) as an argument, converts it to colors and mnemonic words, and generates the 3D model.
\end{itemize}

The output of the program is a GLB file, which is a binary file format representation of 3D models saved in the GL Transmission Format (glTF). The file "model.glb" will be created in the same directory as the script.

\subsubsection{Custody}
The topic of `custody' of Bitcoin (addresses,UTXOs) can be confusing at first. This is another area where there's a lot of detail available, but not all of it is appropriate because increased complexity increases risk. Broadly though it's important to remember that ownership of a UTXO is passed around using signing keys, which are functions of wallets. Wallets themselves don't contain Bitcoin, they contain keys. The simplest approach is a software wallet. This is an application on a device, which stores the private keys, and manages signing of transactions which go onto the blockchain. It's beyond the scope of this book to review or suggest software in detail, but \href{https://bluewallet.io/}{Bluewallet} on mobile devices, and \href{https://sparrowwallet.com/}{Sparrow Wallet} on desktop devices provide rich basic functionality if a reader wishes to get started immediately. Note that these software wallets send your extended public key (the path of those keys) to the wallet providers server, for the monitoring of the blockchain to happen on it's behalf. They're updated by the software vendor, not the blockchain direct. To get this to `privacy best practice' commensurate with the aim of this book it's necessary to run a full node as detailed above, and connect the wallet software to that on a secure or local connection.\par
So called \href{https://unchained.com/blog/best-bitcoin-hardware-wallets/}{hardware wallets} should perhaps be termed signing devices. A \href{https://www.reddit.com/r/Bitcoin/comments/z27jg8/comment/ixfj0w4/?}{reddit user} simplifies this concept very well: \textit{``Your hardware wallet is a safe that holds a key. Your bitcoin is in a mailbox that anyone can look at or put more bitcoin into, but nobody can take the bitcoin out unless they have the key stored in your safe. The 24 word seed phrase are the instructions needed to cut a new key.''}. Rather than store Bitcoin they store the private key in a more secure way, in a device which interacts with a computer or phone. More recently the hardware and associated software of such devices are building in \href{https://content.trezor.io/coinjoin}{much needed privacy technology}. Remember that all transactions on the Bitcoin blockchain are public. Interestingly these privacy technologies are subject to chain surveillance under the terms of their agreements, and companies like \href{https://www.wired.com/story/bitcoin-fog-roman-sterlingov-blockchain-analysis/}{Chainalysis rightly draw criticism}.\par
\begin{figure}
  \centering
    \includegraphics[width=\linewidth]{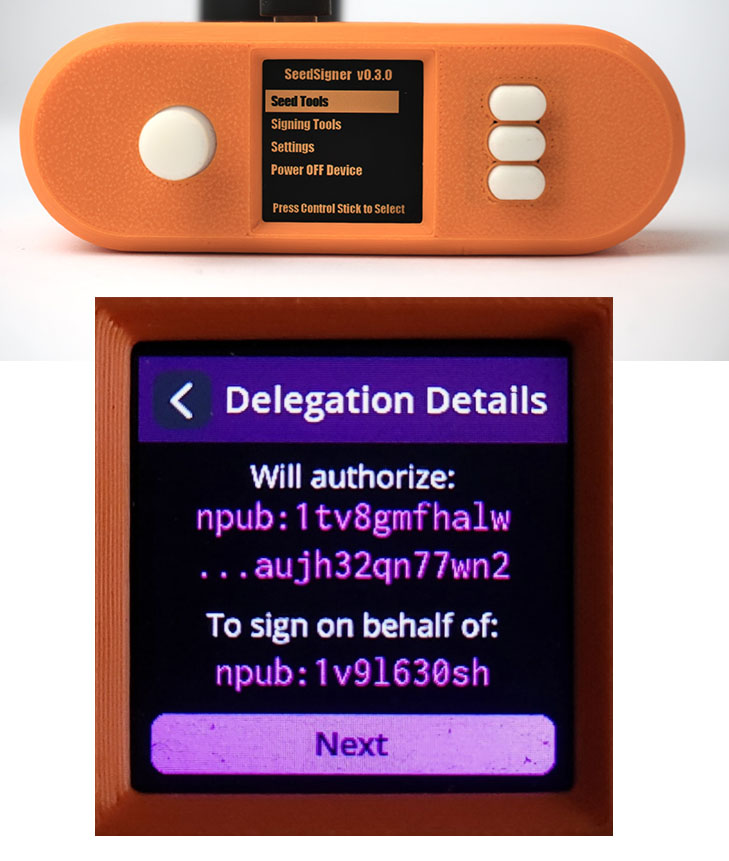}
  \caption{Seedsigner is an inexpensive open source project which scans the master seed in from a QR code to enable signing. One device can run a quorum based wallet (multisig) and manage Nostr identity.}
    \label{fig:seedsigner}
\end{figure}

 
We prefer hardware signing devices which can scan in the seed each time themselves as is the case with opensource \href{https://seedsigner.com/}{``seedsigner''}, which also supports Nostr delegation (Figure \ref{fig:seedsigner}). This makes the device perfect for management of all of our proposed metaverse keys.\par

For higher security it's possible to combine hardware and software wallets (signing devices) to provide a quorum of signatures required to move funds. More exotic still are \href{https://fedimint.org/}{proposals like ``Fedimint''} which allows groups such as families or villages to leverage their personal trust to co-manage Bitcoin. What is not/rarely secure is leaving Bitcoin with a custodian such as an exchange as they simply issue you with an IOU and may abscond. In building toward a proposal for a product in this book it would be simple for us to build a metaverse which users simply paid to use. This is the norm up to now. Representative money would flow around in the metaverse and be changed back like game money at some point. This is not what we wish to promote, so everything will be a variation on ``self-custody'', minimising third party trust for users.
\subsection{Upgrade roadmap}
\subsubsection{Taproot}
`Taproot' is the most recent upgrade to the Bitcoin network. It was first \href{https://lists.linuxfoundation.org/pipermail/bitcoin-dev/2018-January/015614.html}{described in 2018} on bitcoin-dev mailing list, and become \href{https://github.com/bitcoin/bips/blob/master/bip-0341.mediawiki}{BIP-0341} in 2019. It brings improved scripting, smart contract capability, privacy, and Schnorr signatures \cite{schnorr1989efficient}, which are a maximally efficient signature verification method. The network will always support older address types. It is rare to get such a large update to the network, and deployment and upgrade was carefully managed over several months under BIP-0008. Uptake will be slow as wallet manufacturers and exchanges add the feature. It can be considered an \href{https://transactionfee.info/charts/transactions-spending-taproot/}{upgrade in progress (0.3\%)}. Aaron van Wirdum, a journalist and educator in the space describes Taproot in detail in \href{https://bitcoinmagazine.com/technical/taproot-coming-what-it-and-how-it-will-benefit-bitcoin}{an article}.\par
\subsubsection{AnyPrevOut}
\href{https://anyprevout.xyz}{BIP-0118}, is a ``\href{https://en.bitcoin.it/wiki/Softfork}{soft-fork} that allows a transaction to be signed without reference to any specific previous output''. It enables ``Eltoo, a protocol that fulfils Satoshi's vision for nSequence''\par
This is Lightning Network upgrade technology in the main. The Eltoo \href{https://blockstream.com/eltoo.pdf}{whitepaper} or this more \href{https://fiatjaf.alhur.es/ffdfe772.html}{readable explanation} from developer fiatjaf go into detail.\par 
\subsubsection{CheckTemplateVerify}
\href{https://utxos.org/}{BIP-0119} is ``a simple proposal to power the next wave of Bitcoin adoption and applications. The underlying technology is carefully engineered to be simple to understand, easy to use, and safe to deploy''. At it's most basic it is a constructed set of output hashes, creating a Bitcoin address, which if used, can only be spent under certain defined conditions. This is a feature called `covenants'. It enables a feature called `vaults' which provides \href{https://github.com/jamesob/simple-ctv-vault/blob/7dd6c4ca25debb2140cdefb79b302c65d1b24937/README.md}{additional safety features} for custodians. There is currently \href{https://blog.bitmex.com/op_ctv-summer-softfork-shenanigans/}{some debate about the activation process}, and the feeling is that it won't happen (soon).
\subsubsection{Blind merge mining}
BIP-0301 allows `other' chains transactions to be mined into Bitcoin blocks, and for miners to take the fees for those different chains, without any additional work or thoughts by the miners. This is also a prerequisite for Drivechains (mentioned later), and improve Spacechains. In a way this can offer other chains the security model of the Bitcoin network, while increasing fees to miners, which might be increasingly important as the block subsidy falls. This is pretty fringe knowledge \href{https://bitcointalk.org/index.php?topic=1790.msg28696#msg28696}{originally proposed} by Satoshi, but has been refined since and is best explained by \href{https://www.youtube.com/watch?v=xweFaw69EyA}{Paul Sztorc elsewhere}. It is likely an important upgrade in light of the \href{https://www.truthcoin.info/blog/security-budget/}{security budget} of Bitcoin.
\subsubsection{Simplicity scripting language}
\href{https://blockstream.com/simplicity.pdf}{Simplicity} is a proposed contract scripting language which is \href{https://coq.inria.fr/}{`formally provable'}. This would provide a radical upgrade to confidence in smart contract creation. It is \href{https://github.com/ElementsProject/simplicity/blob/pdf/Simplicity-TR.pdf}{work in progress}, and looks to be incredibly difficult to develop in, despite the name. It is more akin to \href{https://en.wikipedia.org/wiki/Assembly_language}{assembly language}. Development has recently slowed, and the proposal requires a soft fork to Bitcoin. The main reason to think it stands a chance of completion vs other \href{https://lists.linuxfoundation.org/pipermail/bitcoin-dev/2022-March/020036.html}{similar proposals} is the powerful backing of \href{https://blockstream.com/}{Blockstream}, one of the main drivers of the Bitcoin ecosystem, run by Adam Back (potential co-creator of Bitcoin). 
\subsubsection{Tail emission}
It is conceivable though unlikely that Bitcoin will choose to remove the 21 million coin hard cap in the end. This would potentially result in a stable and predictable supply, compensating for lost coins, and reinvigorating the miner block reward. The Bitcoin narrative is \textbf{so} invested in the `hard money' thesis that is seems such a hard fork would be contentious at least, and possibly existentially damaging. Peter Todd, long time Bitcoin Core contributor things the idea has merit \href{https://petertodd.org/2022/surprisingly-tail-emission-is-not-inflationary}{and has described it in a blog post}.
\subsubsection{Ossification}
The Bitcoin code is aiming toward so called \href{https://en.wikipedia.org/wiki/Protocol_ossification}{``ossification''}. The complete cessation of development of the feature set. This would provide higher confidence in the protocol moving forward, as long term investors would be somewhat assured that the parameters of the technology would not change, and potentially pressure on the developers would reduce. There's a push to get some or all of the features described above in over the next few year before this happens. As ever this is a controversial topic within the development community. Notably Paul Sztorc, inventor of Drivechain \href{https://www.truthcoin.info/blog/sc-vision/}{feels strongly} that cessation of innovation is a fundamental mistake, while also agreeing that ossification is necessary.
\section{Extending the BTC ecosystem }
The following section are by no means an exhaustive view of development on the Bitcoin network, but it does highlight some potentially useful ideas for supporting collaborative mixed reality interactions, in a useful timeframe.
\subsection{Keet by holepunch}
Tether and Bitfinex have released \href{https://keet.io/}{Keet messenger} which allows peer to peer video calling and file sharing. It will be BTC and Tether enabled which allows transmission of value in a trust minimised fashion. Non custodial Lightning is coming to the product soon. It looks like an incredibly strong and interesting product suite is emerging here. If possible we would like to integrate this open source platform with our metaverse. It is built upon the same \href{https://tether.to/en/tether-bitfinex-and-hypercore-launch-holepunch-a-platform-for-building-fully-encrypted-peer-to-peer-applications/}{Hypercore} ``holepunch'' technology used by Synonym.
\subsection{Block \& SpiralBTC}
Block (formally the payment processor ``Square'' is now an umbrella company for several smaller 'building block' companies, all of which are major players in the space. Block itself is now part of the \href{https://www.w3.org/Consortium/Member/List}{W3C web consortium}, so they will be driving a new era of standards in distributed identity and value transfer. Like much of the industry lately they have either a \href{https://hindenburgresearch.com/block/}{cloud over their reputation}, or else are subject to a targetted opportunistic attack.\par
SpiralBTC, formally `Square Crypto' (a subsidiary of Square) is funding development in Bitcoin and Lightning. Their main internal product is the \href{https://spiral.xyz/blog/what-were-building-lightning-development-kit/}{Lightning Development Kit} (LDK). This promising open source library and API will allow developers to add lightning functionality to apps and wallets. It is a useful contender for our metaverse applications. They also fund external open source development.\par
\subsection{BTCPayServer}
BTCPayServer is one of the recipients of a Spiral grant. It is a self hosted Bitcoin and Lightning payment processor system which allows merchants, online, and physical stores and businesses to integrate Bitcoin into their accounting systems. It might seem that if one were to use Bitcoin then a simple address published on a website might be enough, but this is far from privacy best practice. Using a single address creates a data point which allows external observers to tie all interactions with a given point of sale to all of the customers, and onward to all of their other transactions through the public ledger. Since we seek to employ cyber security best practice will avoid \href{https://en.bitcoin.it/wiki/Address_reuse}{the issues with address reuse}. Each Bitcoin address should be used just once. This is fine as there's essentially an \href{https://privacypros.io/btc-faq/how-many-btc-addresses}{unlimited number} of addresses.\par
In a metaverse application there is no website to interact with, but fortunately BTCPayServer is completely open source and extensible, has a strong support community, \href{https://docs.btcpayserver.org/API/Greenfield/v1/#operation/Invoices_CreateInvoice}{and an API} which could be integrated with a virtual world application. 
BTCPayServer supports the \href{https://docs.btcpayserver.org/LightningNetwork/}{main three} distributions of Lightning but would potentially need extending in order to work with newer technology like RGB or Omnibolt.
\section{Lightning (Layer 2)}
Lightning was a 2016 proposal by Poon and Dryja \cite{poon2016bitcoin}, and is a method for networks of channels of Bitcoin between parties, which can transfer value. The main public network is a community driven liquidity pool which enables scaling and speed improvements for the Bitcoin network. It makes Bitcoin more like money \cite{divakaruni2022lightning}. As with Bitcoin base chain there are multiple standards and approaches, but within Lightning these are not necessarily cross compatible with one another, resulting in several Lightning networks. This is to our advantage as innovation is possible within these smaller networks. It is mainly `powered' by \href{https://plebnet.wiki/wiki/Main_Page}{thousands of volunteers} who invest in hardware and lock up their Bitcoin in their nodes, to facilitate peer-to-peer transactions. Zebka et al. found that although the network is ``fairly decentralised'' it is more recently skewing to larger more established nodes \cite{zabka2022short}. Though this is a grassroots technology the nature of the design means it can likely be trusted for small scale commercial applications.\par
The following text is from \href{https://medium.com/@johncantrell97?p=5cc72f2c664}{John Cantrell}, an engineer who works on Lightning.\par

\textit{``The Lightning Network is a p2p network of payment channels. A payment channel is a contract between two people where they commit funds using a single onchain tx.  Once the funds are committed they can make an unlimited amount of instant \& free payments over the channel.
You can think of it as a tab where each person tracks how much money they are owed.  Each time a payment is made over the channel both parties update their record of how much money each person has.  These updates all happen off-chain and only the parties involved know about them. When it`s time to settle up the two parties can take the final balances of the channel and create a channel closing transaction that will be broadcast on chain.  This closing transaction sends each party the final amounts they are owed. This means for the cost of two on-chain transactions (the opening and closing of the channel) two parties can transact an unlimited number of times and the overall cost of each transaction approaches zero with every additional transaction they make over the channel. Payment channels are a great solution for two parties to transact quickly and cheaply but what if we want to be able to send money to anyone in the world quickly and cheaply?  This is where the Lightning Network comes into play, it`s a p2p network of these payment channels. This means if Alice has a payment channel with Bob and Bob has a channel with Charlie that Alice can send a payment to Charlie with Bob`s help. This idea can be extended such that you can route a payment over an arbitrary number of channels until you can reach the entire world. Routing a payment over multiple channels uses a specific contract called a Hash Time Locked Contract (HTLC).  It introduces the ability for Bob and any other nodes you route through to charge a small fee.  These fees are typically orders of magnitude smaller than onchain fees. This all sounds great but what if someone tries to cheat? I thought the whole point of Bitcoin was that we no longer had to trust anyone and it sure sounds like there must be some trust in our channel partners to use the Lightning Network? The contracts used in Lightning are built to prevent fraud while requiring no trust.  There is a built-in penalty mechanism where if someone tries to cheat and is caught then they lose all of their money.  This does mean you need to be monitoring the chain for fraud attempts.''}

\begin{figure}
  \centering
    \includegraphics[width=\linewidth]{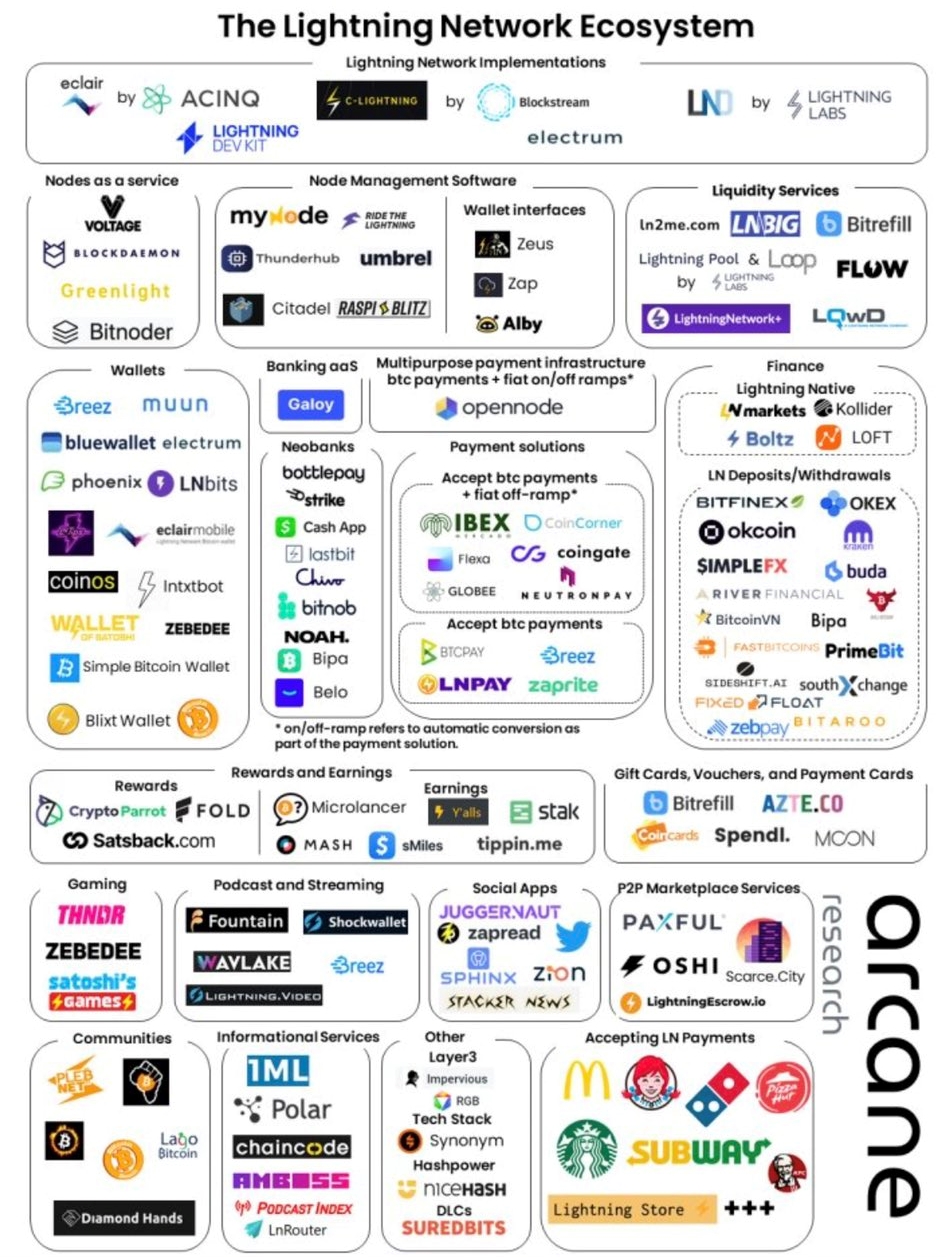}
  \caption{\href{https://www.research.arcane.no/the-state-of-lightning}{Arcane research lightning adoption overview}.}
  \label{fig:lightningAdoption}
\end{figure}

Lightning is a key scaling innovation in the bitcoin network at this time. It is seeing rapid development and adoption (Figure \ref{fig:lightningAdoption}). The popular payment app ``Cash App'' integrates the technology allowing lightning interactions for their 40M users, and `Lightning Strike' services the USA, El Salvador, \href{https://www.bloomberg.com/news/articles/2022-12-06/nigeria-limits-cash-transactions-to-push-enaira-and-other-payments}{large parts of Africa}, and Argentina with zero exchange and transmission fees.
It allows for unbound scaling of transactions (millions of transations per second compared for instance to around 45,000 TPS in the VISA settlement network). Transaction costs are incredibly low, and the transaction speed virtually instantaneous.\par
The most popular lightning software is \href{https://github.com/lightningnetwork/lnd#readme}{LND} from Lightning Labs or \href{https://github.com/ElementsProject/lightning}{C-Lightning} from Blockstream. The software can be run on top of any Bitcoin full node, in a browser extension with a limited node, in a mobile app as a client or a server, or a hybrid such as the Greenlight server \href{https://medium.com/breez-technology/get-ready-for-a-fresh-breez-multiple-apps-one-node-optimal-ux-519c4daf2536}{used by Breez wallet}. Different trust implications flow from these choices.
\subsection{Lightning service providers}
The model of the `LSP' has been refined since it's first introduction by \href{https://medium.com/breez-technology/introducing-lightning-service-providers-fe9fb1665d5f}{Breez in 2019}. At their core they provide the following services, at some expense to the concept of decentalisation of the network.
\begin{itemize}
\item Payment Channel Management: LSPs may offer users the ability to open and close payment channels on the Lightning Network, as well as manage the routing of payments through these channels.
\item Liquidity Provision: LSPs may provide liquidity to users on the Lightning Network, allowing them to make and receive payments even when they do not have sufficient funds in their own payment channels.
\item Node Hosting: LSPs may host Lightning Network nodes on behalf of users, providing them with access to the network without requiring them to maintain their own node.
\item Payment Processing: LSPs may offer payment processing services, allowing merchants to accept Lightning Network payments from customers.
\item Wallet Integration: LSPs may integrate Lightning Network functionality into Bitcoin wallets, making it easier for users to access the network and make payments.
\end{itemize}
There are multiple companies experimenting this space now, and it's unclear how useful the ideas are for our use cases at this time. It's notable that slashtags (mentioned later as a potential for digital assets) are themselves an LSP, and that breez have now \href{https://medium.com/breez-technology/the-breez-open-lsp-model-scaling-lightning-by-sharing-roi-with-3rd-party-lsps-e2ef6e31562e}{introduced a profit sharing model} to assist the adoption.
\subsection{Micropayments}
\begin{figure}
  \centering
    \includegraphics[width=\linewidth*\real{0.5}]{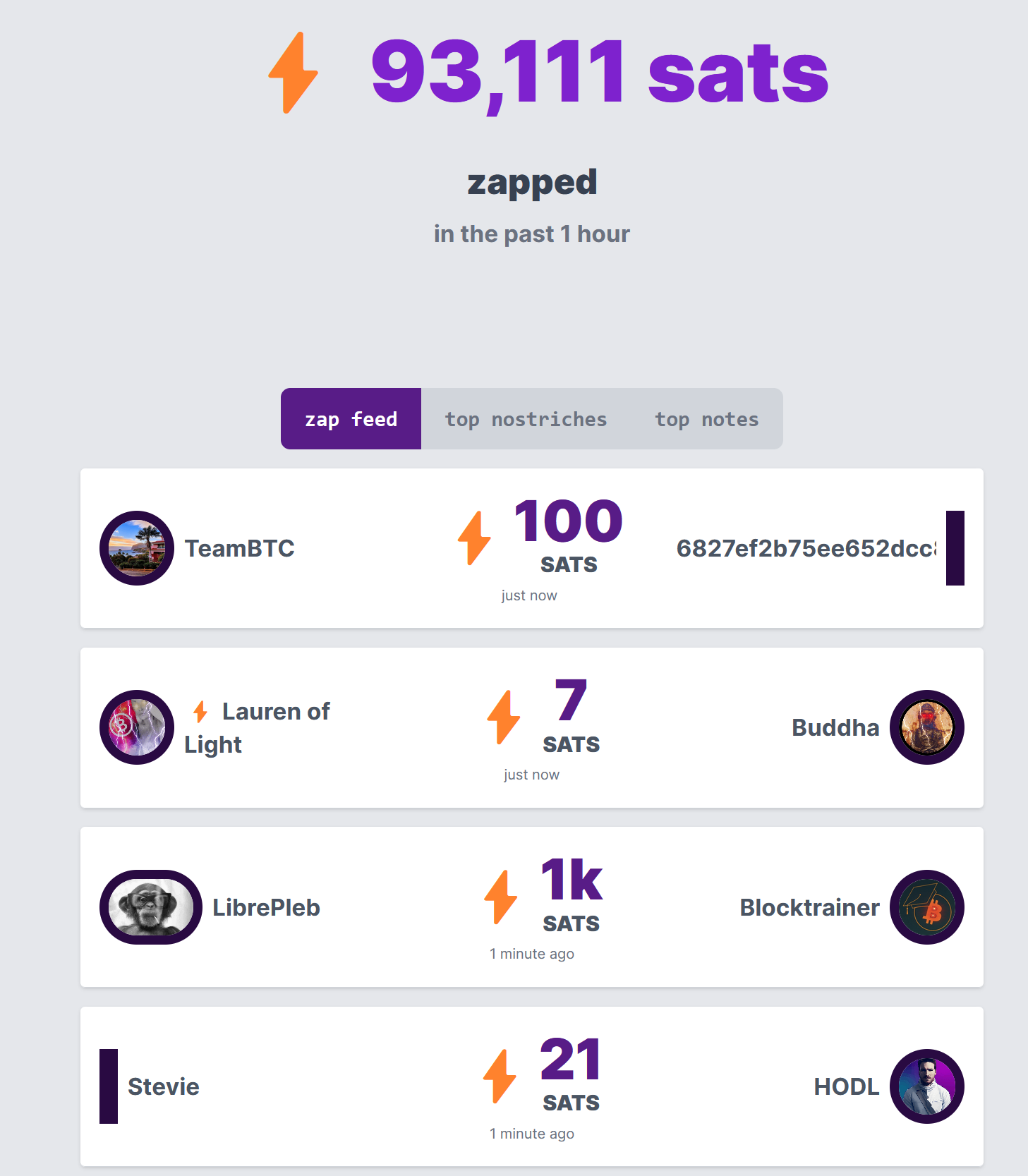}
  \caption{Within weeks of launch thousands of people are pinging micropayments to one another}
  \label{fig:nostrzaps}
\end{figure}

Possibly the most important affordance of the Lightning network is the concept of micropayments, and streaming micropayments. It is very simple to transfer even \href{https://satsymbol.com/}{one satoshi} on Lightning, which is one hundred millionth of a bitcoin, and a small fraction of a penny. This can be a single payment, for a very small goods or service, or a recurring payment on any cadence. This enables streaming payments for any service, or for remittance, or remuneration. These use cases likely have enormous consequences which are just beginning to be explored. Nostr users are seemingly have an enormous amount of fun sending one another tiny amounts of money for free, in response to good posts on the new social media and blog platforms designed around the protocol. This can be seen in Figure \ref{fig:nostrzaps} and nostr will be described in more detail later. Integration of this capability into metaverse applications will be explored later.

\subsection{BOLT12 and recurring payments}
\begin{figure}
  \centering
    \includegraphics[width=\linewidth*\real{0.5}]{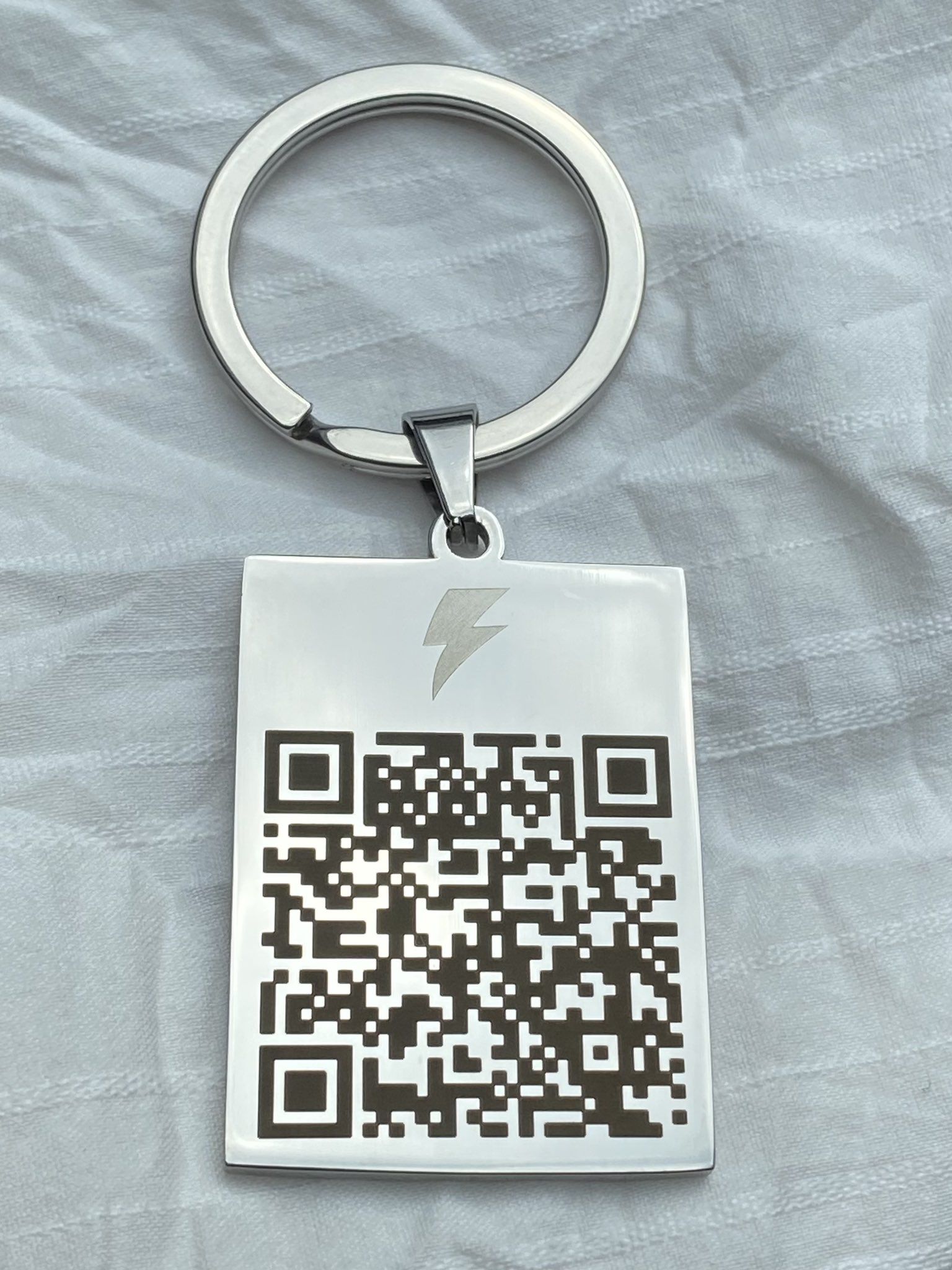}
  \caption{\href{https://twitter.com/SeedMint21/status/1518934554840600579}{A key fob with a Bolt12 QR code}}
  \label{fig:bolt12keyring}
\end{figure}

\href{https://bolt12.org/}{BOLT12} is a new and developing 'standard' which simplifies and extends the capability of the network for recurring payments, but can negotiate single payments too. The example keyring QR code seen in Figure \ref{fig:bolt12keyring} can be scanned to send single or recurring payments securely and anonymously to the holder.
\subsection{Cashu and Fedimint}
\subsubsection{Cashu}
Cashu, an implementation of the e-cash mechanism, is an electronic cash system that allows users to hold and transfer digital representations of money on their devices. In this system, e-cash is a piece of electronic data that represents a certain amount of money, such as satoshis in the case of Bitcoin. Users can send e-cash to others through various encrypted channels like Telegram or email.\par
The concept of e-cash relies on blind signatures, which ensure that transactions cannot be correlated to specific users, providing a high level of privacy. When a user wants to send e-cash, they send the data representing the money to another user. The receiving user then sends the e-cash to a server (or mint) to be recycled. This process ensures that the e-cash cannot be double-spent. However, the server cannot correlate the incoming e-cash to any specific user or transaction.\par
Cashu, as a protocol, enables the implementation of this e-cash mechanism and facilitates interoperability among different Cashu wallets. The system is designed to be private, with no user accounts or wallets associated with any central authority. The e-cash is stored in the user's browser data, and can be backed up using a seed phrase, similar to traditional cryptocurrency wallets.\par
In its current state, Cashu is in its early stages of development and is primarily intended for experimentation. However, it has the potential to enable a wide range of applications, such as streaming services, where users could pay for content on a per-use basis without the need for an account, and without the service provider being able to track their usage. This would allow for greater privacy and flexibility in how people interact with online services.
\subsubsection{Fedimint}
From the \href{https://www.fedi.xyz/blog/introducing-fedi-the-global-bitcoin-adoption-technology}{blog post} on the Fedi App website; Fedimint is:
\begin{itemize}
\item a form of community Bitcoin custody,
\item utilising federations (a byzantine fault tolerant multi-sig wallet technology similar to Blockstream's Liquid network),
\item run collectively by groups of trusted community members we call “guardians”,
\item for and on behalf of their communities,
\item with privacy through Chaumian e-cash,
\item and with close integration with the Lightning Network
\end{itemize}
Obi Nwosu sees Fedimint as the third vital pillar of the Bitcoin ecosystem. If Bitcoin is secure decentralised money, and Lightning is decentralised payments, then he says \href{https://bitcoinmagazine.com/technical/fediment-evolution-of-bitcoin-custody}{Fedimint is decentralised custody} of the Bitcoin asset. The excitement in the community is such that this protocol is included in our metaverse stack later. With Fediment a clade of users within the metaverse would have near perfect transactional privacy within their group inside the metaverse \cite{chaum1985security}. This could be a potentially huge group of users, and could include AI actors in the scene. Transactions with the outside world could be through lightning as already planned.
\subsection{LNBits}
\begin{figure}
  \centering
    \includegraphics[width=\linewidth*\real{0.8}]{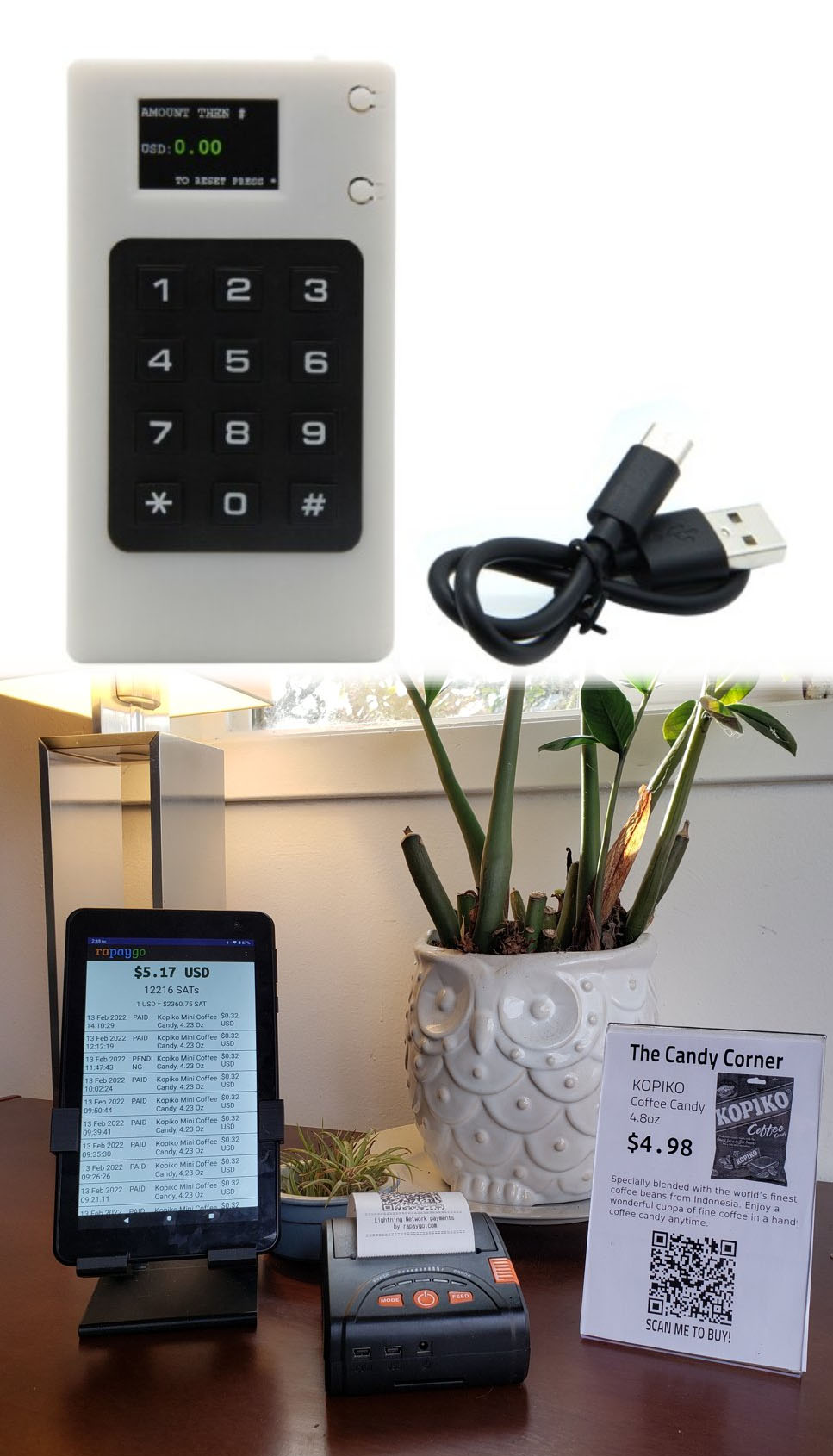}
  \caption{Two of the many \href{https://rapaygo.com/}{prebuilt} and \href{https://github.com/arcbtc/LNURLPoS}{kit} options for Lightning `point of sale'}
  \label{fig:LnBitsPoS}
\end{figure}

LNBits is an open source, extensible, Lightning `source' management suite. It is self hosted, and can connect to a variety of Lightning wallets, further abstracting the liquidity to provide additional functionality to network users. Remember that all of these tools run without a third party, on a £200 setup, hosted at home or within a business. The best way to explore this is to describe \textit{some} of the plugins. 
\begin{itemize}
\item ``\href{https://github.com/lnbits/lnbits-legend#lnbits-v03-beta-free-and-open-source-lightning-network-walletaccounts-system}{Accounts System}; Create multiple accounts/wallets. Run for yourself, friends/family, or the whole world!''
\item \href{https://github.com/lnbits/lnbits-legend/tree/quart/lnbits/extensions/events#events}{Events plugin} allows QR code tickets to be created for an event, and for payments to be taken for the tickets.
\item \href{https://github.com/lnbits/lnbits-legend/tree/quart/lnbits/extensions/jukebox#jukebox}{Jukebox} creates a Spotify based jukebox which can be deployed online or in physical locations.
\item \href{https://github.com/lnbits/lnbits-legend/tree/quart/lnbits/extensions/livestream#dj-livestream}{Livestream} provides an interface for online live DJ sets to receive real-time Lightning tips, which can be split automatically in real-time with the music producer.
\item \href{https://github.com/lnbits/lnbits-legend/tree/quart/lnbits/extensions/tpos#tpos}{TPoS}, \href{https://github.com/arcbtc/LNURLPoS#lnurlpos}{LNURLPoS} \& \href{https://github.com/lnbits/lnbits-legend/tree/quart/lnbits/extensions/watchonly#watch-only-wallet}{OfflineShop} support online \href{https://rapaygo.com/}{and offline} point of sale (Figure \ref{fig:LnBitsPoS}).
\item \href{https://github.com/lnbits/lnbits-legend/tree/quart/lnbits/extensions/paywall#paywall}{Paywall} creates web access control for content. 
\item \href{https://github.com/LightningTipBot/LightningTipBot#lightningtipbot-}{LightningTipBot} is a custodial Lightning wallet and tip handling bot within the popular on Telegram instant messenger service.
\end{itemize}
Together these plugins are incredibly useful primitives which are likely to be translatable to a multi party collaborative mixed reality application. A proposal for building a more specific plugin along these lines is detailed later.\par
\textbf{LnBits is capable of backing every object in a metaverse scene as an economic actor, with a key which is compatible with Nostr. This makes it the best choice and it will likely form the core of the proposed metaverse stack.}

\subsection{Mutiny web wallet}
Mutiny Wallet, a new self-custodial lightning wallet, has launched in open beta. It is web-based so requires no app download.
Key features include just-in-time channels via Voltage, separating on-chain and Lightning balances, encrypted remote backups, and Nostr wallet connections for social tips and subscriptions.\par
Mutiny aims to make lightning accessible to the billions of internet users worldwide and provide features not possible on most custodial wallets due to app store restrictions. It is our chosen user wallet for our design.
\subsection{Daric}
Lightning isn't the only solution to layer 2, as evidenced by Daric \cite{mirzaei2022daric}. This is a complex and technical proposal which claims to improve upon Lightning.
\subsection{Ark}
Ark is a privacy-focused off-chain protocol for bitcoin transactions. Here's how it compares to existing systems:

Ark vs. Chaumian eCash: Both ensure transaction anonymity, but unlike eCash, Ark's transactions are backed by real bitcoins, making them immune to theft or inflation by the service providers.

Ark vs. Lightning: Similar to the Lightning network, Ark is a liquidity network but without liquidity constraints or direct link between sender and receiver. It uses significantly less on-chain footprint as it lacks concepts of opening and closing channels.

Ark vs. On-chain: Ark is similar to on-chain wallets in terms of UX, receiving payments asynchronously without introducing liquidity constraints. However, users must "refresh" their coins regularly to avoid service providers sweeping the funds.

Ark vs. Validity Rollups: Ark doesn't require on-chain data per transaction, providing higher throughput. Both need a soft-fork on the base layer, but Ark doesn't need a soft-fork for the interactive version.

Ark service providers can, in theory, double-spend their transactions in mempool. However, it's counterproductive as recipients can use incoming zero-conf vtxos to pay lightning invoices. If a double-spend occurs, a future Ark extension can allow users to claim their previously redeemed vtxos. This provides an inbound liquidity-like tradeoff without protocol compromise.
\section{Liquid federation (layer 2)}
Liquid is an implementation on Blockstream \href{https://elementsproject.org/}{Elements}, and is itself part of the open source development contribution of Blockstream, the company started by Adam Back (of hashcash fame) and nearly a dozen other early cypherpunks and luminaries.\par 
The Liquid side chain network, and it's own attendant Lightning layer 2, is a fork of Bitcoin with different network parameters. In liquid the user of the network `pegs' into the Bitcoin network, swapping tokens out from BTC to L-BTC (this can of course mean very small subunits of 1 Bitcoin). Once tokens have been `locked' and swapped to Liquid the different network parameters used in the fork allow a different trust/performance trade-off. Liquid is fast on the L1 chain, cheaper to use at this time, and more private. The consensus achieved on this side chain network is faster because it is a far smaller group of node operators. The next block to be written to the side chain is chosen by a node operated by a member of a federation of dozens of major contributors to the Bitcoin technology space. These `trusted' nodes all check one another's security and network operations, meaning that the network is as secure as the aggregate of the trust placed in half of the membership at any one time. There are
\href{https://bitcoinmagazine.com/business/bitcoin-liquid-network-gains-six-new-federation-members}{still dozens} of major companies, development teams, and individual actors, with significant reputational investment.\par
``Federation members contribute to the Liquid Network's security, gain voting rights in the board election and membership process, and provide valuable input on the development of new features. Members also benefit from the ability to perform a peg-out without a third party, allowing their users to convert between L-BTC and BTC seamlessly within their platform.''\par
Crucially for our purposes here Liquid allows tokenised asset transfer. Anyone \href{https://docs.blockstream.com/liquid/developer-guide/developer-guide-index.html#issued-assets}{can issue} an asset on Liquid. Such transfers of assets may be orders of magnitude cheaper than on chain Bitcoin transactions, but still potentially orders of magnitude more expensive than a simple Lightning transaction of value on the Bitcoin network. \par 
Blockstream plan to add arbitrary (user generated) token support to their `Core Lightning' implementation at some point. This would be a very strong choice for specific use cases within an economically enabled metaverse application. When participants wish to `cash out' of the Liquid network they must do this through one of the federation members who activate the other side of the `two-way peg', dispensing the equivalent amount of Bitcoin. This is transparently handled through Blockstream's ``green wallet''.\par
All of this has the advantage of a far lower energy footprint compared to the main chain, but it's not quite ready with a full suite of affordances. \par
The Liquid network is being used as the underlying asset for a novel new global financial product. El Salvador are working with Blockstream to issue a nation state backed bond. 
\section{Bitcoin Layer 3}
Increasingly important features of modern blockchain implementations are programmability through smart contracts, and issuance of arbitrary tokens. Assigning a transaction to represent another thing like an economic unit, energy unit, or real world object, and operating on those abstractions within the chain logic. Chief among these use cases are stablecoins such as Tether, which are pegged to national currencies and described in the next section. Bitcoin has always supported very limited contracts called scripts, and stablecoin issuance has existed in Bitcoin since 
\href{https://www.omnilayer.org/}{Omni Layer}. Omni was the first issuer of Tether, but more recently these important features have passed to other layer one chains. This year is likely to see the \href{https://www.hiro.so/blog/bitcoin-ecosystem-a-guide-to-programming-languages-for-bitcoin-smart-contracts}{resurgence of this capability} on Bitcoin, which of course benefits from a better security model. Once again, there is a stong assertion by some that \href{https://lists.linuxfoundation.org/pipermail/bitcoin-dev/2022-April/020227.html}{this isn't even possible}. The debate is complex and unresolved.\par 
In order to properly understand the use of Bitcoin based technologies in metaverse applications it is necessary to examine what these newer `layer 3' ideas might bring. 
\subsection{LNP/BP and RGB}
\href{https://giacomozucco.com/layers-before-bitcoin}{LNP/BP} is a non profit standards organisation in Switzerland which contributes to open source development of Bitcoin layer 3 solutions into the Lightning protocol, and Bitcoin protocol (LNP/BP). One of the core product developments within their work is the \href{https://www.rgb.tech/}{`RGB' protocol}, which is somewhat of a meaningless name, evolved from ``coloured coins'' which were an early tokenised asset system on the Bitcoin network. RGB represents red, green, and blue. The proposal is built upon research by \href{https://petertodd.org/2016/commitments-and-single-use-seals}{Todd} and \href{https://giacomozucco.com/#intro}{Zucco}. RGB is regarded as arcane Bitcoin technology, even within the already rarefied Bitcoin developer communities. Zucco provides the \href{https://bitcoinmagazine.com/culture/video-interview-giacomo-zucco-rgb-tokens-built-bitcoin}{following explanation}: \par
\textit{``When I want to send you a bitcoin, I will sign the transaction, I will give the transaction only to you, you will be the only one verifying, and then we’ll take a commitment to this transaction and that I will give only the commitment to miners. Miners will basically build a blockchain of commitments, but without the actual validation part. That will be only left to you. And when you want to send the assets to somebody else, you will pass your signature, plus my signature, plus the previous signature, and so on.''}\par
This is non-intuitive explanation of Todds `single-use-seals', applied to Bitcoin, with the purpose of underpinning arbitrary asset transfer secured by the Bitcoin network. In this model the transacting parties are the exclusive holders of the information about what the object they are transferring actually represents. This primitive can (and has) been expanded by the LNP/BP group into a concept called `client side validation'. 
It's appropriate to explain this concept several times from different perspectives, because this is potentially a profoundly useful technology for metaverse applications.\par
\begin{itemize}
\item A promise is made to spend a multi output transaction in the future. This establishes the RGB relationships between the parties.
\item One of the pubkeys to be spent to is known by both parties.
\item The second output is unknown and is a combination of the hash of the state, and schema, from the operation which has been performed.
\item When the UTXO is spent the second spends pubkey can be processed against the shared data blob to validate the shared state in a two party consensus  (sort this out, it's nonsense).
\item This is now tethered to the main chain. Some tokens from the issuance have gone to the recipiant, and the remainder have gone back to the issuer. More tokens can be issued in the same way from this pool. 
\item A token schema in the blob will show the agreed issuance and the history back to the genesis for the token holder. 
\item The data blob contains the schema which is the key to RGB functions and the bulk of the work and innovation. 
\item Each issuance must be verified on chain by the receiving party. 
\end{itemize} 
This leverages the single-use-seal concept to add in smart contracts, and more advanced concepts to Bitcoin. Crucially, this is not conceptually the same as the highly expressive `layer one' chains which offer this functionality within their chain logic. In those systems there is a globally available shared consensus of `state'. In the LNP/BP technologies the state data is owned, controlled, and stored by the transacting parties. Bitcoin provides the crytographic external proof of a state change in the event of a proof being required. This is an elegant solution in that it takes up virtually no space on the blockchain, is private by design, and is extensible to layer 2 protocols like Lightning.\par
This expanding ecosystem of client side verified proposals is as follows:
\begin{itemize}
\item RGB smart contracts
\item RGB assets are fungible tokens on Bitcoin L1 and L2, and non fungible Bitcoin L1 (and somewhat on L2).
\item Bifrost is an \href{https://github.com/LNP-BP/presentations/blob/master/Presentation slides/Bifrost.pdf}{extension} to the Lightning protocol, with it's own Rust based node implementation, and backwards compatibility with other nodes in the network. This means it can transparently participate in normal Lightning routing behaviour with other peers. Crucially however it can also negotiate passing the additional data for token transfer between two or more contiguous Bifrost enabled parties. This can be considered an additional network liquidity problem on top of Lightning, and is the essence of the ``Layer 3'' moniker associated with LNP/BP. It will require a great number of such nodes to successfully launch token transfer on Lightning. As a byproduct of it's more `protocol' minded design decisions Bifrost can also act as a generic peer-to-peer data network, enabling features like Storm file storage and Prometheus.
\item \href{https://www.aluvm.org/}{AluVM} is a RISC based virtual machine (programmable strictly in assembly) which can execute Turing complete complex logic, but only outputs a boolean result which is compliant with the rest of the client side validation system. In this way a true or false can be returned into Bitcoin based logic, but be arbitrarily complex within the execution by the contract parties.
\item Contractum is the proposed smart contract language which will compile the RGB20 contracts within AluVM (or other client side VMs) to provide accessible layer 3 smart contracts on Bitcoin. It is a very early proposal at this stage.
\item  Internet2: ``Tor/noise-protocol Internet apps based on Lightning secure messaging
\item Storm is a lightly specified escrow-based bitcoin data storage layer compliant with Lightning through Bifrost.
\item Prometheus is a lightly specified multiparty high-load computing framework.
\end{itemize}
Really, any compute problem can be considered applicable to client side validation. In simplest terms a conventional computational problem is solved, and the cryptographically verifiable proof of this action, is made available to the stakeholders, on the Bitcoin ledger.\par 
Less prosaically, at this stage of the project the more imminent proposed affordances of LNP/BP are described in `schema' \href{https://github.com/LNP-BP/LNPBPs}{on the project github}. The most interesting to the technically minded layperson are:
\begin{itemize}
\item \href{https://github.com/LNP-BP/LNPBPs/blob/master/lnpbp-0020.md}{RGB20} fungible assets. This could be stablecoins like dollar or pounds representation. Bitfinex exchange \href{https://github.com/RGB-Tools/rgb-lightning-sample}{have code} which already works with RGB to transmit Tether stablecoins on testnet. This is a huge application area for Bitcoin, and similar to Omni, which will also be covered next.
\item \href{https://github.com/LNP-BP/LNPBPs/blob/master/lnpbp-0021.md}{RGB21} for nonfungible tokens and ownership rights. In principle BiFrost allows these to be transferred over a the Lightning network, significantly lowering the barrier to entry for this whole technology. DIBA \href{https://diba.io/}{have this technology working} on testnet.
\item \href{https://github.com/LNP-BP/LNPBPs/issues/29}{RGB22} may provide a route to identity proofs. This is covered in detail later.
\end{itemize}
Federico Tenga is CEO of `Chainside' and an educator and consultant in the space. He has written an up-to-date \href{https://medium.com/@FedericoTenga/understanding-rgb-protocol-7dc7819d3059}{``primer''}, which is still extremely complex for the uninitiated, but does capture how the RGB token transfer system works. That medium article also touches on Taro, which is next.
\subsubsection{DIBA and Bitcoin's Unique Digital Assets}
\textit{DIBA} is a pioneering digital asset marketplace, powered by the \href{https://www.rgb.tech/}{RGB Smart Contract Protocol}. It permits the creation and direct transaction of Unique Digital Assets (UDAs), akin to Non-Fungible Tokens (NFTs), on Bitcoin without the necessity of other tokens. UDAs are special digital assets linked to a Bitcoin UTXO (unspent transaction output). These assets embody distinctive attributes like ownership, transferability, and divisibility, and remain under the full control and ownership of their creators.\par
Through DIBA, users can explore, purchase, and sell a vast array of UDAs, taking advantage of the robustness and permanence of the Bitcoin blockchain. DIBA's innovation extends to the integration of a Lightning layer 2 solution, aiming to facilitate faster and more affordable transactions.\par
Assets minted via DIBA are bound to Bitcoin's base layer with an on-chain UTXO. They are subsequently stored on the Arweave permaweb alongside a cryptographic hash, which, combined with a digital signature, can validate their authenticity. The RGB Smart Contract Protocol executes UDA transactions via BitMask, a wallet engineered by the DIBA Team.
\subsubsection{BitMask}
\textit{BitMask} is a browser extension wallet birthed by DIBA, intended for decentralized applications on Bitcoin. It grants access to Bitcoin Finance, UDAs, and more, utilizing the RGB protocol. It delivers comprehensive financial autonomy with its taproot-enabled Bitcoin and Lightning Network wallet, establishing it as a gateway to DeepWeb3 on Bitcoin. More details can be found on Bitmask.app.\par
\subsubsection{DIBA's Launch and Marketplace Timelines}
DIBA has initiated an Open Marketplace for Beta testing as of April 2022, available on Bitcoin testnet. The full launch on Bitcoin mainnet is expected to occur in the second or third quarter of 2022. Submissions for the DIBA Curated Marketplace are currently open, allowing interested individuals to apply as an Artist or a Curator.\par
These developments represent a substantial expansion of the capabilities within the Bitcoin network and further attest to the potential of the LNP/BP's work. Notably, DIBA \href{https://diba.io/}{has this technology working} on testnet. The extensive application areas for Bitcoin, such as the transmission of Tether stablecoins on testnet by Bitfinex exchange \href{https://github.com/RGB-Tools/rgb-lightning-sample}{via RGB}, emphasize the potential impact of these advancements.
\subsection{Taro / Taproot Assets}
Taproot Assets is the new name for Lightning Labs `Taro', a new \href{https://lightning.engineering/posts/2022-4-5-taro-launch/}{initiative} to allow assets to transmit on the Lightning network. It is more similar to RGB above than Omnibolt below. \href{https://docs.lightning.engineering/the-lightning-network/taproot-assets}{They say}: \textit{``Taproot Assets is a new Taproot-powered protocol for issuing assets on the bitcoin blockchain. Taproot Assets (formerly Taro) is a new Taproot-powered protocol for issuing assets on the bitcoin blockchain that can be transferred over the Lightning Network for instant, high volume, low fee transactions.''}\par
The project has clearly been \href{https://github.com/roasbeef/bips/tree/bip-taro}{under development} by the lead developer at Lightning Labs for some years and seems both \href{https://lightninglabs.substack.com/p/bitcoinizing-the-dollar-and-the-world?s=r}{capable} and mature, though they are obviously following the model of `co-opting' open source ideas (from RGB) to garner venture capital funding. They \href{https://github.com/bitcoin/bips/pull/1298/commits/4daba8c373c777defb48136795382803c137502c}{credit RGB} in the github. More will doubtless be added to this section and it seems a contender for our metaverse purposes, being less broadly ambitious than RGB upon which it's based, but perhaps more focused and implemented. The key feature of Taro seems to be that only the first and last hop in a multi-hop lightning transaction need to support Taro, because of external data validation databases called ``universes''. This is an advance on the RGB proposal. 
The technical specs are now on the \href{https://docs.lightning.engineering/the-lightning-network/taro}{lightning labs web pages}, and \href{https://lightning.engineering/posts/2022-9-28-taro-launch/}{code has been released}. The beta programme uses testnet. There are concerns that large amounts of synthetic dollars on the protocol could be used to create `incentive' for one Bitcoin hard fork or another, under the control of Tether. 
\subsection{ZeroSync}
The recent zerosync paper offers a tantalising glimpse of a highly compressed, performant, and private take on Bitcoin. It potentially addresses Bitcoin's scalability challenges using advanced cryptographic techniques (SNARKs). It compresses the entire Bitcoin blockchain, enabling instant verification and various innovative applications like faster full nodes, trustless light clients, improved privacy, and secure cross-chain bridges. Additionally, zkCoins is a protocol that enhances privacy and throughput of arbitrary tokens, which could lead to more private and scalable digital assets. 
\subsection{Spacechains}
Spacechains is a \href{https://medium.com/@RubenSomsen/21-million-bitcoins-to-rule-all-sidechains-the-perpetual-one-way-peg-96cb2f8ac302}{proposal} by Ruben Somsen. It is a way to provide the functionality of any conceivable blockchain, by making it a sidechain to Bitcoin. \par
Like RGB described earlier it's a single use seal, but which can be closed by the highest bidder.\par
In a spacechain the Bitcoin tokens are destroyed in order to provably create the new spacechains tokens at a 1:1 value. These new tokens only have worth moving forward within the new chain ecosystem they represent, as they cannot be changed back. They nontheless have the same security guarantees as the bticoin main chain, though with a radically reduced ecological footprint (x1000?), and higher performance. Each `block' in the new chain is a single bitcoin transaction. The high level features are:\par
\begin{itemize}
\item Outsource mining to BTC with only a single tx per block on the main chain.
\item One way peg, Bitcoin is burnt to create spacechain tokens.
\item Allows permissionless chain creation, without a speculative asset.
\item Fee bidding BMM is space efficient and incentive compatible. Miners just take the highest fees as normal.
\item Paul Sztorc raised the idea
\item It's best with a soft fork but possible without
\end{itemize}
The concept is \href{https://vimeo.com/703246895/d89aba6e56}{explained fully} in a recent presentation at Advancing Bitcoin conference.\par 
Developer Fiatjaf, a stalwart of the the lightning developer community has a basic Spacechains based \href{https://github.com/nbd-wtf/soma}{asset trading system} which can be run already called Soma, though it is limited to Signet, one of the local bitcoin testnets, modified with AnyPrevOut described elsewhere in the book.

\subsection{Statechains, drivechain, softchains} 
There are many \href{https://gist.github.com/RubenSomsen/96505e99dc061d6af6b757ff74434e70}{proposals for layer 2 scaling solutions} for the bitcoin network. Ruben Somsen \href{https://gist.github.com/RubenSomsen/c9f0a92493e06b0e29acced61ca9f49a}{describes Softchains, Stateschains, and Spacechains}, while  \href{https://www.drivechain.info/literature/index.html}{Drivechain is described} by the author Paul Sztorc on the project web pages and is split across \href{https://github.com/bitcoin/bips/blob/master/bip-0300.mediawiki}{BIP-0300} for drivechain and \href{https://github.com/bitcoin/bips/blob/master/bip-0301.mediawiki}{BIP-0301} for a ``blind merge mining'', a soft fork which it's unlikely to get. They are all hypothetical with the exception of sidechains.  

\section{Risks and mitigations}
Looking across the whole sector, this paragraph from the Bank of International Settlement (BIS) \href{https://www.bis.org/publ/arpdf/ar2022e3.htm}{sums everything up}: \par
\textit{``...it is now becoming clear that crypto and DeFi have deeper structural limitations that prevent them from achieving the levels of efficiency, stability or integrity required for an adequate monetary system. In particular, the crypto universe lacks a nominal anchor, which it tries to import, imperfectly, through stablecoins. It is also prone to fragmentation, and its applications cannot scale without compromising security, as shown by their congestion and exorbitant fees. Activity in this parallel system is, instead, sustained by the influx of speculative coin holders. Finally, there are serious concerns about the role of unregulated intermediaries in the system. As they are deep-seated, these structural shortcomings are unlikely to be amenable to technical fixes alone. This is because they reflect the inherent limitations of a decentralised system built on permissionless blockchains.''}\par
This might seem like reason enough to  stop here and wait for proper digital currency (expanded later), but Bitcoin is here now, is likely unstoppable in, and with mitigations in place might have uses if developed properly. Perhaps surprising the same BIS is \href{https://www.bis.org/press/p221216.htm}{allowing up to 2\%} of bank reserves to be held in crypto assets, including Bitcoin, \href{https://www.bis.org/bcbs/publ/d533.pdf}{according to their June 2022 Basel Committee on Banking Supervision report}, though the BIS chief believe the \href{https://www.bloomberg.com/news/articles/2023-02-22/crypto-has-lost-battle-against-fiat-currency-bis-chief-agustin-carstens-says}{``battle'' against crypto} has already been won following the turmoil described in the next section. \par
Lightning is still considered to be experimental and not completely battle tested. There have been various attacks and a major double spend attack may be possible \cite{https://doi.org/10.48550/arxiv.2208.01908}, but there have been no major problems in the years it's been running with careful design choices and cybersecurity best practice it it likely a production ready component of our planning.
\subsection{Sociopaths \textbf{everywhere}}
In the wake of the \href{https://www.bloomberg.com/opinion/articles/2022-11-14/ftx-s-balance-sheet-was-bad}{rampant crime spree} by Sam Bankman-Freid and his top teams at Alameda research and the Bahamas registered exchange `FTX' the whole industry has suffered, and will continue to suffer, seismic shocks. There is a chance the sector will never recover, and that we have already seen the top of the hype bubble. Fortunately this doesn't diminish our use cases for these technologies, as we were never planning to speculate with the asset, but rather use the network.
\subsection{Digital assets}
For digital assets more generally it is useful to look at the recent \href{https://www.whitehouse.gov/briefing-room/presidential-actions/2022/03/09/executive-order-on-ensuring-responsible-development-of-digital-assets/}{``whole government executive order''} signed by President Biden early in 2022. It was mainly framed in terms of ``responsible innovation, and leadership'' in the new space. The resulting, ``Comprehensive Framework for Responsible Development of Digital Assets'' is a product of multi agency collaboration and can be seen as 9 reports and a summary document, and has been long anticipated. The summary itself is neither particularly comprehensive nor a framework, and mainly serves to identifies high level risks, aspirations, and challenges, and strongly hints toward eventual development of a ``digital dollar'' (CBDC, expanded later). \par
The risks section of the original executive order shows how legislators are framing this, so it's useful to break down here.\par
\begin{itemize}
\item Consumer and business protections. This is likely to pertain to custodians and is much needed. Misselling is rife. Security presents a challenge.  
\item Systemic risk, and market integrity are a concern. The legislators clearly worry about contagion risks from the sector.
\item Illicit finance (criminality and sanction busting etc) are a concern, but not particularly front and centre\cite{moser2013inquiry}. Criminality in 2021 was a mere 0.15\% of transactions according to Chainalysis, but this number varies year to year. There are claims that Iran have begun official overseas buying with cryptocurrencies, but again, the \href{https://finbold.com/iran-makes-the-first-ever-import-of-goods-using-cryptocurrency-worth-millions/}{numbers are small}. One of the better sections of the work is the US treasury department's recently published `National Risk Assessments for Money Laundering, Terrorist Financing, and Proliferation Financing'. This is a comprehensive report and speaks to careful research across the space. It is broken into \href{https://home.treasury.gov/news/press-releases/jy0619}{three parts}. Perhaps surprisingly, while they do see activity in these areas, they do not rate the risk as very significant. Cash remains the main problem for illicit funding. There is some talk that the nature of public blockchain analysis allows greater oversight of these tools and that this is to the advantage of government and civil enforcement agencies.
\item Highlighting the need for international coordination suggests they are mindful of jurisdictional arbitrage. 
The partial regulatory capture of these technologies, where activity flows to globally more lenient legislative regimes, continues to be a concern. Many of the centralised exchanges for instance are located in tax havens such as Malta. As the world catches up with these products it is likely that this will be smoothed out.
\item Climate goals, diversity, equality and inclusion are mentioned. It seems that the ``environment'' aspect of ESG is more important then ``social'' and ``governance'' at this time.
\item Privacy and human rights are mentioned.
\item Energy policy is highlighted, including grid management and reliability, energy efficiency incentives and standards, and sources of energy supply.
\end{itemize}
The \href{https://www.whitehouse.gov/briefing-room/statements-releases/2022/09/16/fact-sheet-white-house-releases-first-ever-comprehensive-framework-for-responsible-development-of-digital-assets/}{latest summary report} resulting from the above guidance actually adds little tangible meat to the bones. This possibly reflects the complexity of these issues. The recommendations seem to be broadly as follows, and are really a copy/paste of the executive order.
\begin{itemize}
\item Carry on doing research into central bank digital currencies, but there's no particular rush.
\item Support development of better instant payment methods both at home and globally. 
\item Ensure consumer and systemic protections.
\item More monitoring, civil and criminal prosecutions.
\item Issue more rules and clarity in response to risks (this is actually likely net positive as rules are currently unclear).
\item Improve global reporting on users (KYC/AML).
\end{itemize}
The government rhetoric to date in the USA can be seen to be converging on an understanding of the technology, at different rates in different parts of government. One thing that seems to shine through is their own perception of their global leadership on legislation on these matters. They seems to assume that what they decide will guide the world, and this may be true through their KYC/AML pressures.\par
A recent proposed \href{https://bitcoinmagazine.com/business/heres-whats-in-senator-lummis-bitcoin-bill}{bi-partisan bill in the USA} will likely help inform global law, though it is unlikely to pass itself. It encourages the use of Bitcoin as a medium of exchange by applying a tax exemption on transactions of less than \$200. The issue of whether an asset is a commodity (a raw material thing) or a security (a promise) is left to a couple of major government agencies to unpick, with corresponding reporting requirements. Crucially for this book these nascent bills all regard both Bitcoin and Ethereum as sufficiently decentralised to \href{https://www.coincenter.org/a-new-senate-bill-focuses-on-cryptocurrency-exchanges-heres-what-developers-and-users-should-keep-an-eye-on/}{qualify as commodities}, meaning they would enjoy more lenient oversight. Far more likely to pass is the \href{https://www.agriculture.senate.gov/imo/media/doc/crypto_one-pager1.pdf}{proposed DCCPA bill} which has senior lawmaker support and would see commodities in the space regulated in such a way that trading of it could be halted in the USA. In this line of policy, exchanges will be required to do far more reporting, and would be penalised for trading against their customers. DOAs and DeFi are the big potential losers. In a maddening twist the Office of Government Ethics in the USA has banned anyone who owns digital assets from working on the legislation. This is an exceptional move and likely to result in poorly crafted laws in the first instance.\par
The most recent and troubling example is the US ban on any Ethereum assets which have been through a ``mixer service'' \href{https://www.coincenter.org/u-s-treasury-sanction-of-privacy-tools-places-sweeping-restrictions-on-all-americans/}{that obfuscates history}. This is a huge constraint on the code and smart contract itself, not just sanctions against individuals. It has \href{https://hoffmang9.github.io/free-speech/the-history-code-is-free-speech.html}{`free speech'} and constitutional implications \cite{anderson2002free}. More such actions and \href{https://www.dw.com/en/dutch-investigators-say-developer-of-tornado-cash-arrested/a-62793823}{arrests of developers} are feared. It has led to Circle (who issue the USDC stablecoin) blacklisting every \href{https://home.treasury.gov/policy-issues/financial-sanctions/recent-actions/20220808}{address sanctioned by the US government}. Centrally issued digital assets are obviously neither uncensorable nor permissionless. This intersects (again) with the whole question of what decentralisation means and how effective it can be in it's stated goal of circumventing global policies.
\subsection{Bitcoin specifically}
\noindent In addition it's useful for this document to focus more on the technical challenges to the Bitcoin network.\par
\begin{itemize}
\item The block reward is reduced every 4 years (epochs). This means a portion of the mining reward is trending to zero, and nobody knows what effect this will have on the incentives for \href{https://www.truthcoin.info/blog/security-budget-ii-mm/}{securing the network} through proof of work \cite{carlsten2016instability}. It is increasingly \href{https://cryptostackers.substack.com/p/bitcoin-is-not-a-store-of-value?sd=pf&s=r}{being discussed} as the major eventual problem for the network.
\item Stablecoins are a vital transitional technology (described later) but do not meaningfully exist yet on the Bitcoin network. This may change.
\item Bitcoin lacks privacy by design. All transactions are publicly viewable. This is a major drag to the concept of BTC as a money. Upgrade of the network is possible, and has indeed been achieved for a Bitcoin fork called Litecoin \cite{fuchsbauer2019aggregate}. 
\item The Lightning network (described later) has terrible UX design at this time. 
\item The basic `usability' of the network is still poor in the main. Any problems which users experience demand a steep learning curve and risk loss of funds. There is obviously no technical support number people can call. 
\item Only around one billion unspent transactions can be generated a year on the network. This means that it might become impossible for everyone on the planet to have their own Bitcoin address (with it's associated underpinning UTXO).  
\item Chip manufacture is concentrated in only a few companies and countries, as identified by \href{https://www.btcpolicy.org/authors/matthew-pines}{Matthew Pines}. 
\item Potential constraints on monetary policy flexibility.
\item Future protocol changes.
\item Unanticipated effects on the domestic and international energy system.
\item Vulnerability to adversary attacks are \href{https://braiins.com/blog/bitcoin-mining-attacks-explained}{widely studied}\cite{apostolaki2016hijacking, apostolaki2017hijacking, johnson2014game, stinner2022proof}, and still pretty much completely speculative because of the complex nature of the attack surface.
\item Mining tends toward economy of scale concentration. Many are already on their \href{https://bitcoinfibre.org/}{own specialised network} to connect to one another.
\item Future hard forks. There will doubtless be pressure to fork the code to add inflation, or ESG mitigations, or to fix the UNIX clock issue in 2106. Each fork is a risk.
\item Other unknown, unanticipated risks given Bitcoin’s limited 13-year history.
\item There is a ``non-zero'' chance that Bitcoin is a complex government intelligence agency construct, \href{https://en.wikipedia.org/wiki/Crypto_AG}{much like Crpto AG was} toward the end of the last century \cite{dymydiuk2020rubicon}. 
\end{itemize}

%

\chapterimage{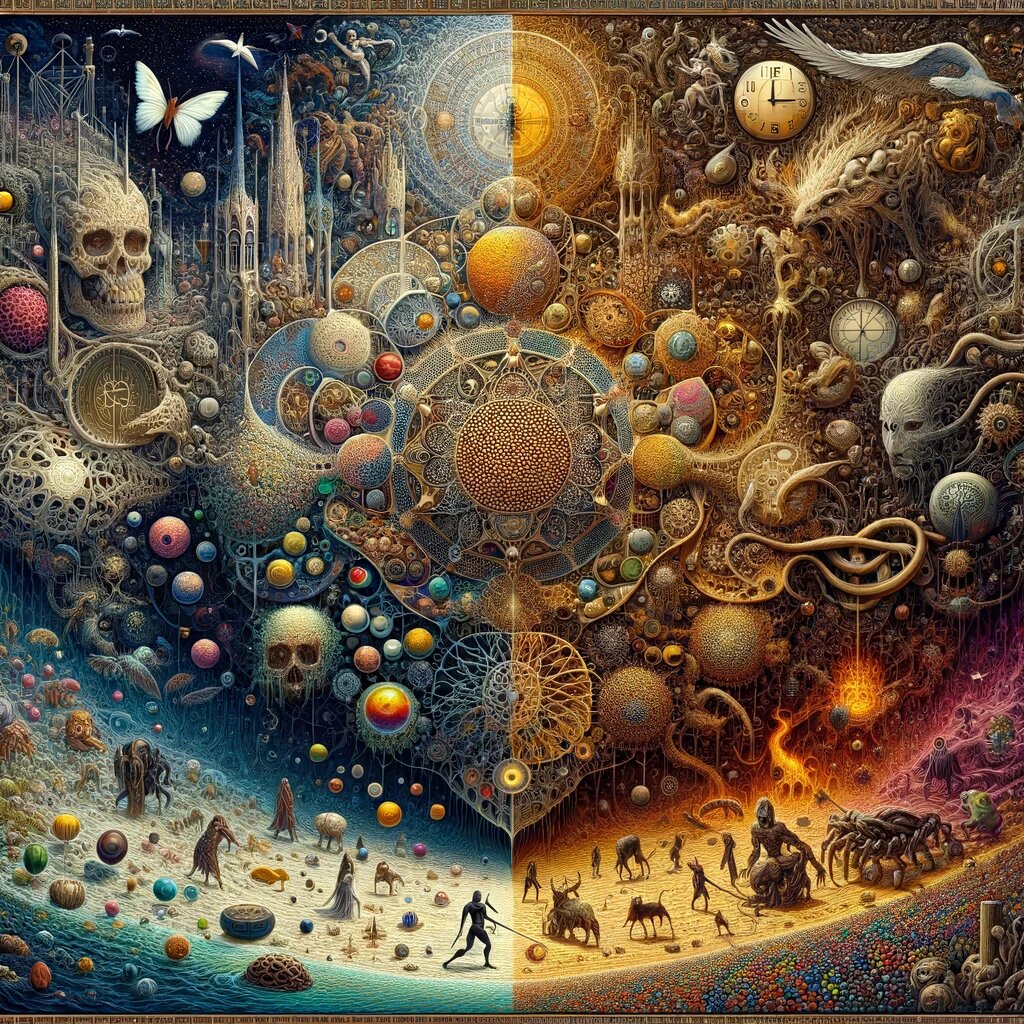}
\chapter{Money in the real world}
It is necessary here to briefly examine what money actually is in the world outside of metaverses, so we can understand it in the context of a virtual global space. In the previous section Bitcoin can be viewed in a couple of different lights. As a self custody digital bearer asset it can be viewed as `property', like gold, i.e. not a liability on someone else's asset sheet. Indeed this has long been one of the assertions of the community and it finds favour in law, \href{https://www.regulationasia.com/shanghai-court-says-bitcoin-is-protected-by-law-as-virtual-property/}{possible most ironically in China} which of course banned mining. `Money' though is a far more \href{https://www.bankofengland.co.uk/knowledgebank/what-is-money}{slippery concept} to grasp. It seems very likely that Bitcoin is evolving as a ``base money'', and it's important to define that, but there are many other kinds of money within the online world which can potentially transfer value within virtual social spaces.

\section{Defining money}
Money is an economic good, that is generally accepted as a medium of exchange. This simple and specific description doesn't do justice to the complexity of everything that humans consider to be money. Even the Encyclopaedia Britannica strays from this immediately in their definition:\par
\textit{``money, a commodity accepted by general consent as a medium of economic exchange. It is the medium in which prices and values are expressed; as currency, it circulates anonymously from person to person and country to country, thus facilitating trade, and it is the principal measure of wealth.''}. \\
In which it can be seen that the principle measure of wealth might not be money at all, but rather property, credit, etc. So are these things money? Is a promise on a ledger money? The assertion at the top of this section is challenged by different schools of economic thinking. Global debt is around an order of magnitude larger than base money, and most wealth is stored in illiquid land/built environment (some \$300T), and yet the system seems to work fine. The debt theory of money offered by anthropologist David Graeber suggests that money is an abstraction of barter, and thereby `credit', but credit clearly pre-dates money, and needs no barter, commodity, intermediary nor underlying asset \cite{homer1996history}. This suggests that money is something slightly different.\par
Money seems to have evolved for two principle purposes; trade outside of a village context, and inheritance \cite{szabo2002shelling}. In doing this it somewhat replaced and augmenting `credit', which as said above, was a promise between parties based on future actions, and likely as old as rudimentary language itself. The anonymous Heavyside blog \href{https://heaviside.substack.com/p/the-forgotten-fourth-function-of}{powerfully argues} that it is the relative stability of money over time which creates a less discussed composite feature; that of `confidence' in being able to defer labour into money, basically credit again.\par
Money can be divided into two categories, which are fungible (interchangeable) from the point of view of the users. Base money is `commodity' money which is backed by assets, or tangible physical (or digital) goods through the actions of a central bank ledger, and is around \$30-\$40T. Everything else is `fiduciary media' \cite{selgin1996defense}.\par
All fiduciary money is credit but not all credit is fiduciary money. Nobody knows the extent of the global supply of fiduciary media. It encapsulates all the new digital money platforms like PayPal, gift cards, offshore accounts and all manner of other vehicles, and is thought to be many \href{https://www.bis.org/publ/qtrpdf/r_qt2212h.pdf}{tens of trillions of pounds}\cite{borio2017fx}. This somewhat muddies the waters since money that is backed by `something' blends away into money which cannot reasonably be assayed. This in turn undermines the assertion that money is backed. It seems that a combination of available raw materials and labour, central banks and their associated political structures \cite{barsky1987fisher}, and global markets drive the value of money up and down relative to ``stuff'' in the shops. This manifests as `inflation', which is `possibly' the effect of not pegging money to an asset such as silver, or gold as in the past \cite{hall2009inflation}. While the gross drivers of inflation seems to be accepted and understood, nobody \href{https://www.dailymail.co.uk/news/article-10966165/Jerome-Powell-admits-understand-better-little-understand-inflation.html}{seems very sure} how the \href{https://www.bloomberg.com/opinion/articles/2022-08-19/this-economy-is-proving-too-complicated-for-economists}{various aspects interact}. Dickson White wrote in 1914 about hyperinflation in France due to excessive money printing, and this driver and causal link persists as a primary hypothesis for inflation and hyperinflation, a cautionary tale especially today after the huge fiscal responses to the 2008 global financial crisis and COVID \cite{white1914fiat}. It may be that central banks actually have no decent response to global monetary pressures and are overdue a paradigm shift, as \href{https://www.ft.com/content/2d79d153-fffa-4441-b79f-0a808a51108f}{explained by Daniela Gabor} (Professor of economics and macrofinance at UWE Bristol):\\ 
\textit{``...last stage of a central banking paradigm, when it implodes under the contradictions of its class politics? Under the financial capitalism supercycle of the past decades, inflation-targeting central banks have been outposts of (financial) capital in the state, guardians of a distributional status-quo that destroyed workers’ collective power while building safety nets for shadow banking.\\
The limits of this institutional arrangement that concentrates (pricing) power and profit in (a few) corporate hands are now plain to see. If the climate and geopolitical of 2022 are omens of Isabel Schnabel’s Great Volatility that most central banks and pundits expect for the near future, then macro-financial stability requires new framework for co-ordination between central banks and Treasuries that can support a state more willing to, and capable of, disciplining capital.\\
But such a framework would threaten the privileged position that central banks have had in the macro-financial architecture and in our macroeconomic models. The history of central banking teaches us that policy paradigms die when they cannot offer a useful framework for stabilising macroeconomic conditions, but never at the hands of central bankers themselves.''}\par
All this makes it \href{https://www.lynalden.com/what-is-money/}{hard to find} a universally accepted and explicable definition of money. The best approach may be to look at the properties of a thing which is asserted to be a money. In his book `A history of money', Glyn Davies identifies ``cognisability, utility,  portability, divisibility, indestructibility, stability of value, and homogeneity'' \cite{davies2010history}.\par
Stroukal examines Bitcoins' likely value as a money from an Austrian economics perspective and identifies ``portability, storability, divisibility, recognizability, homogeneity and scarcity'' \cite{stroukal2018can}.\par
A helpfully brief and useful \href{http://money.visualcapitalist.com/infographic-the-properties-of-money/}{web page by Desjardins from 2015} describes some properties and explains them in layman's terms below:
\begin{itemize}
\item Divisible: Can be divided into smaller units of value.
\item Fungible: One unit is viewed as interchangeable with another.
\item Portable: Individuals can carry money with them and transfer it to others.
\item Durable: An item must be able to withstand being used repeatedly.
\item Acceptable: Everyone must be able to use the money for transactions.
\item Uniform: All versions of the same denomination must have the same purchasing power.
\item Limited in Supply: The supply of money in circulation ensures values remain relatively constant.
\end{itemize}

\subsection{The origins of money}
Lyn Alden has written an excellent book which leads through the history and mechanics of money as a technology \cite{Alden2023}
\begin{itemize}
\item \textbf{Origins and Early Forms}
\begin{itemize}
\item Money originally emerged to solve issues with barter and the double coincidence of wants
\item Early forms of money included commodities like shells, cocoa, salt, furs, feathers
\item These served as money due to properties like portability, divisibility, durability, fungibility
\item Social credit also played a role, enabling delayed settlement between known parties
\end{itemize}
\item \textbf{Precious Metals as Money}
\begin{itemize}
\item As societies advanced, precious metals like gold and silver emerged as the dominant monies
\item They survived debasement from more technologically advanced societies
\item This was due to their scarcity and difficulty to produce more even with modern techniques
\end{itemize}
\item \textbf{Layers Added to Enhance Metals as Money}
\begin{itemize}
\item Coinage added verifiability of weight and purity to raw metals
\item This involved blending metals with authority of issuing institutions
\item Legal tender laws mandated acceptance of certain coinages
\end{itemize}
\item \textbf{Emergence of Paper Money and Banking}
\begin{itemize}
\item Paper money initially emerged to enhance metals for long distance trade
\item Evolved from bilateral credit channels to broadcast systems of bank notes
\item Allowed easier transfer and divisibility without physically moving metals
\end{itemize}
\item \textbf{Credit Theory vs Commodity Theory of Money}
\begin{itemize}
\item Credit theory sees money as shared ledger, its value comes from authority
\item Commodity theory sees money emerge naturally due to properties of commodities
\item Debates center around role of state and intrinsic value of money
\end{itemize}
\item \textbf{Flaws of State Controlled Money}
\begin{itemize}
\item State controlled money allows non-transparent taxation via inflation/debasement
\item Credit theory underestimates degradation of state controlled monetary systems
\item Most state currencies have experienced high inflation or hyperinflation over time
\item Incumbent currencies survive due to lack of convenient alternatives, not soundness
\end{itemize}

\end{itemize}
She argues that as societies advanced, precious metals like gold and silver became the dominant monies. Unlike other commodities, precious metals maintained their value even when more technologically advanced societies attempted to debase them by mixing in other metals. This durability was due to the metals' scarcity and the difficulty of acquiring more even with modern mining techniques.\par

To enhance the use of precious metals as money, layers were added on top. The creation of coinage allowed the weight and purity of raw metals to be verified. Coins also blended the intrinsic value of metals with the authority of the institutions issuing them. Legal tender laws mandated the acceptance of certain coinages for debt repayment.\par

The emergence of paper money and banking provided another layer of enhancement. Paper banknotes initially facilitated long distance trade by avoiding the need to physically move heavy metals. This evolved from bilateral credit channels between specific parties to broadcast systems where banknotes could circulate among many holders. Banknotes increased transferability and divisibility of money without moving the underlying metals.

Debates arose between the credit and commodity theories of money. The credit theory views money as a shared ledger, with value derived from the authority of the issuer. In contrast, the commodity theory sees money emerge naturally due to the properties of the underlying commodity. Disagreements still center on the role of the state and whether money requires intrinsic value.

Alden thinks that over time, flaws became apparent in state controlled monetary regimes. The ability to non-transparently tax through inflation and currency debasement led to the degradation of monetary systems. Most state currencies have experienced high inflation or hyperinflation. They survive due to lack of alternatives rather than soundness. Commodity-based monies constrained state overreach and lasted longer before breaking down.

\subsection{Understanding money creation}

There are two main types of money in our current system: financial money and real economy money.
Financial money refers to bank reserves, which are created by central banks through quantitative easing (QE). The central bank buys bonds from banks and credits their reserve accounts with new digital bank reserves.
Bank reserves are an asset for commercial banks. Reserves allow banks to settle transactions with each other and meet liquidity requirements set by regulators.
Importantly, bank reserves do not directly translate into increased lending or stimulus for the real economy. There is no direct channel for reserves to enter the broader economy. The amount of reserves does not drive bank lending.
Real economy money refers to money that households and businesses can use for transactions. This includes physical currency and bank deposits.
Real economy money is created through government deficits and private sector credit expansion.
\subsubsection{Government deficits drive money creation and inflation}
When the government spends more than it taxes, it is creating net new financial assets in the economy. Government deficits add net financial wealth to the private sector.
This increases private sector deposits and spending capacity. When people then spend this new money, it stimulates aggregate demand and economic activity.
Several empirical examples demonstrate how increasing government deficits leads to higher GDP growth and inflation by injecting more money into the real economy.
Conversely, austerity policies that reduce deficits, like higher taxes or less spending, destroy private sector financial assets and reduce economic activity.
Therefore, government deficits and surpluses have a much more direct impact on the real economy compared to central bank operations that alter the supply of bank reserves.\par
In their paper for the 2023 central bank meeting at Jackson Hole Eichengreen and Arslanalp \cite{Eichengreen2023} argue that the 2008 global financial crisis and COVID-19 pandemic have caused public debt levels to balloon to unprecedented heights across advanced, emerging, and developing economies, and that contrary to the calls of major financial institutions, high public debts are unlikely to meaningfully decline in the foreseeable future.\par 
They say this is because the conventional options for debt reduction like running large primary budget surpluses, relying on higher growth rates, or using inflation to erode real debt burdens are politically and economically infeasible today (though AI perhaps offers a slim productivity `out'). At the same time, changes in the global financial system like the rise of private creditors have made coordinated debt restructuring more challenging. As a result, the world will have to learn to live with persistently high public debts. This may be manageable for major advanced countries like the US that benefit from structural demand for their safe assets. But it poses greater risks for emerging and developing economies that lack this advantage. Creative solutions like GDP-indexed bonds, credit enhancements, and legal reforms are needed to facilitate sustainable debt restructuring for weaker countries weighed down by debt overhangs. Overall, the shift from bank to bond financing and the changing composition of creditors have reduced options for unwinding high public debts accumulated due to recent crises. It's a mess.
\subsubsection{Private credit drives money creation and asset inflation}

Private credit creation through bank lending also increases the money supply by allowing households and businesses to purchase assets they couldn't otherwise afford.
When banks create new loans, they are simultaneously creating new purchasing power in the form of deposits for the borrower, allowing asset purchases with new credit.
This increases broader money supply and spending capacity, but also creates an offsetting debt liability owed back to the bank.
Rapid private credit growth risks fueling asset bubbles and financial instability if debts can't be repaid. This primarily benefits those who already own assets.
Private credit growth is more disciplined by market forces compared to unchecked government deficits, but still risks inflating asset prices.
\subsubsection{The risks of the current system and alternatives}
The current elastic credit money system aims to prevent recessions by constantly expanding credit. However, this artificial stability leads to financial instability long-term.
Alternatives like Bitcoin have a firm supply anchor and cannot rapidly expand the money supply. This prevents runaway credit growth and provides monetary discipline.
However, Bitcoin and hard money standards also provide less flexibility to respond to economic crises by expanding credit. There are tradeoffs between flexibility and discipline.
Our current monetary system relies heavily on expanding real economy purchasing power through government deficits and private credit in order to drive economic growth. However, this constant elasticity promotes financial instability and inequality over the long-run, mainly because of shorter term political incentives. We will see that potential alternatives like Bitcoin offer more stability through monetary discipline, but sacrifice flexibility. It's likely that trading off a known flawed system for an unknown replacement is far too risky, but with sufficient adoption there may be a `flight to safety'. Bitcoin represents a serious risk if it compounds the worst elements and outcomes of a mishandled cyclical credit based system.
\subsection{Global currency interactions}
The legacy moniker ``third world'' came from a division of the world along economic lines \cite{tomlinson2003third}. At the time this was the petrodollar / neo-institutional hegemony \cite{caballero2008financial, spiro2019hidden}, vs the economic superpower of the soviet block, and then `the rest'; unaligned economic powers.\par
This old framework has fallen away with the associated terminology, but it's useful to look at what money `is' from a global viewpoint, because all money is effectively trust in the liability held by some defined counter party.\par
Right now the dollar system is still predominant, but it seems likely that there are new axes forming, especially around the \href{https://www.wsj.com/articles/saudi-arabia-considers-accepting-yuan-instead-of-dollars-for-chinese-oil-sales-11647351541}{Chinese Yuan}. It's clear that central banks have been aware of this potential transition away from a global dollar / energy system. The Dollar has potentially suffered from the radical expansion of the money supply over the last 70 years or so under the private ``Eurodollar'' system \cite{grewal2020struggling}. Macro markets commentator Peccatiello \href{https://themacrocompass.substack.com/p/usd-hidden-debt#details}{describes this} as follows: \textit{``Our monetary and credit system is USD-centric: the lion share of international debt, trade invoices, asset classes and FX volume is settled or denominated in US Dollars. Funnily enough though, direct access to \$ liquidity is only available to entities located in the United States but in a credit-based system the rest of the world also has an incentive to leverage in US Dollars to boost or enhance their global business models. That means European banks, Brazilian corporates or Japanese insurance companies which want to do global business will most likely get exposure to \$-denominated assets and liabilities (\$ debt) despite being domiciled outside the United States.''}\par

 Some policy makers have been looking back to the great economist John Maynard Keynes' ideas for a neutral basket of assets as a global synthetic hedgemonic currency \cite{carney2019growing, piffaretti2009reshaping} which would almost certainly consist partly of gold \cite{stoeferle2018gold}. Gold as a utilitarian commodity trades at a premium because of it's history as a money, and like Bitcoin, there are \href{https://www.newyorker.com/magazine/2023/02/27/the-dystopian-underworld-of-south-africas-illegal-gold-mines}{serious consequences} to it's perceived value to humans. \par
Use of the dollar system has recently been shown more and more to be contingent on adherence to US defined political principles. This is evidenced most starkly by the seizure of Russian central bank \href{https://twitter.com/RussianEmbassy/status/1504530573527760909}{foreign reserves}, a new and untried projection of monetary power. Counter intuitively this allowed Russia to demand sale of it's natural resources in their native Ruble, rapidly increasing the buying power of their currency. It seems that the \href{https://mronline.org/2022/04/16/russias-sergey-glazyev-introduces-the-new-global-financial-system/}{`currency wars'} are accelerating. Putin (who to be clear, is a dictator and aggressor) \href{https://finance.yahoo.com/news/russia-calls-payment-system-based-135512758.html}{recently said} \textit{``The technology of digital currencies and blockchains can be used to create a new system of international settlements that will be much more convenient, absolutely safe for its users and, most importantly, will not depend on banks or interference by third countries'' }\par  
The Chinese Yuan/Renminbi is potentially stepping in where the petrodollar is now waning \cite{mathews2018china}. The effects of this expansion of economic influence by China, through a potential petro-Yuan, and the belt and road initiative \cite{huang2016understanding}, are not yet felt, but the lines are fairly clearly defined and may be felt over the coming decades. The Euro system is potentially even less stable because of recent energy supply pressures, and \href{https://www.fitchratings.com/research/sovereigns/energy-crisis-increases-fiscal-risks-to-western-europe-sovereigns-23-09-2022}{internal tensions} in the bond markets. Though it seems to be less `weaponised' \cite{hudson2021destiny}, it comes with it's own restrictions for use, especially through the International Monetary Fund (IMF). They are opposed to global fragmentation and multi-polarity, seeing is as disproportionately impacting emerging economies. They say in ther \href{https://www.imf.org/en/Publications/WEO/Issues/2023/04/11/world-economic-outlook-april-2023?cid=bl-com-spring2023flagships-WEOEA2023001}{2023 outlook report} that the rise of geoeconomic fragmentation could cause shifts in foreign direct investment (FDI), hitting emerging economies the hardest. They feel that policymakers and companies are focusing on making supply chains more resilient by moving production closer to home or to trusted countries. As a result, FDI flows are becoming more concentrated within blocs of aligned countries. It is likely true that emerging market and developing economies are more vulnerable to FDI relocation, as they rely more on flows from geopolitically distant countries, though this could be viewed as a reduction in economic imperialism. Such economies may face reduced access to capital and technological advancements. It is into this gap that our work presenting AI collaborative tooling wishes to step.\par
To give context to this it is useful to paraphrase Whittemore's \href{https://www.youtube.com/watch?v=LOqQSKbfRu4}{podcast} which gave a high level view of Gladsteins \href{https://bitcoinmagazine.com/culture/imf-world-bank-repress-poor-countries}{critique of the IMF}: \textit{``The terms of the most recent IMF loans to Argentina; one that was just finalized this year was that the country's leadership had to try, as part of their agreement, to discourage citizens from engaging in the use of cryptocurrencies. The most recent deal was a 45 billion dollar deal which is a restructuring of that 57 billion program that Alex mentioned. The provision in question was called `strengthening Financial resilience', and says `to further Safeguard Financial stability we are taking important to discourage the use of cryptocurrencies with a view to preventing money laundering informality and disintermediation'. They explicitly do not want citizens of that country to disintermediate. They want them to have to go through the system that the IMF is ``restructuring'', meanwhile inflation this
year is around 72 percent. Last year it was 48 the year before 42 the year before that 53 percent clearly something is not working. It's not surprising to me then that Argentina is an absolute hotbed for people who are involved in
Bitcoin''} \par
The new `third world' who are excluded from the Dollar and/or Yuan poles of the global economy might drift toward the `basket of assets' discussed by Keynes and Carney above. As mentioned this will certainly have a component of gold, and likely other commodity assets such as rare metals. This is described at length by Hudson\cite{hudson2021destiny}. For our purposes here it's also possible that there would be a small `hedge' allocation of Bitcoin or \href{https://www.independent.co.uk/tech/bitcoin-el-salvador-crypto-btc-b2079881.html}{even a global axis} of `unaligned' nations using the asset \cite{hendrickson2021value, ferranti2022hedging}. Block and Wakefield research \href{https://block.xyz/2022/btc-report.pdf}{found that in developed nations} Bitcoin is treated as in investment, while in less wealthy demographics there is interest in the utility. This is evidenced in the early nation state adoption seen and described to date, and the game theory incentive explained by Fidelity in the introduction. It's too early to tell if this `unaligned money' could constitute a global economic pole, but it's interesting that some commentators are now even discussing this, and that \href{https://docs.google.com/document/d/1Ynl5bbdTqev-wbTAWQoeWdh1cJVf3ortuSjre9K9wGQ/edit}{carbon neutrality research} is being undertaken specifically for this application.
\subsubsection{Central Banks}
\begin{enumerate}

\item Central banks were established to be lenders of last resort, providing liquidity to commercial banks during financial crises to prevent bank runs and systemic crises. This remains a core function.

\item Over time, many central banks have expanded their role as lender of last resort beyond just commercial banks to also support non-bank financial entities that face liquidity shortages in crises. Central banks have effectively become backstops for the broader financial system.

\item Central banks control short-term interest rates through policy tools like adjusting benchmark rates (e.g. fed funds rate), reserve requirements, open market operations, etc. This allows them to influence longer-term rates and overall financial conditions.

\item Central banks engage in quantitative easing and asset purchase programs to lower longer-term rates. They buy financial assets like government bonds and mortgages to inject liquidity and expand the money supply.

\item As a result of asset purchases and liquidity programs, most major central banks have dramatically expanded their balance sheets and the monetary base since the 2008 financial crisis.

\item Central banks earn income on assets purchased but also pay interest on reserves. Most remit profits back to national treasuries/governments after covering expenses. Some now face losses.

\item While politically independent, central banks face pressure from politicians and the public. They have mandates like inflation targeting, financial stability, employment, etc. that shape policy.

\item Central bank policies like QE and low rates for long periods are criticized for enabling fiscal deficits and debt levels to rise and inflating asset bubbles. But also defended as supporting growth.

\item Extraordinary central bank actions during crises like COVID-19 have fueled high inflation worldwide. They face challenges normalizing policy and credibility issues.

\item As lenders of last resort with balance sheet expansion power, central banks have uniquely influential roles in national and global finance. Their policies have major economic and political impacts.

\end{enumerate}

\section{International money transfer networks}
Transferring money from one financial jurisdiction to another is itself a global marketplace which has accreted over the entire course of human history. It's far less useful here to discuss the mythos of salt and seashells as a mechanisms of international remittance and taxation \cite{gainsford2017salt, goldberg2005famous}. Suffice it to say that there are dozens, if not hundreds, of cross border payment companies who make their business from taking a percentage cut of an international money transfer. There are also hundreds if not thousands of banks who offer this service as part of their core business portfolio. This section looks at some of the major players, and their mechanism, to contextualise the more recent shifts brought about by technology.
\subsection{Swift, ISO 20022, and correspondence banking}
Society for Worldwide Interbank Financial Communiactions (SWIFT) was initially formed in 1973 between 239 banks across 15 countries. They needed a way to improve handling of cross border payments. It is now the global \href{https://www.swift.com/standards}{standard} for financial message exchange in over 200 countries, and has recently found itself under a fresh spotlight, during the invasion of Ukraine. The system handles around 40 million short, secure, code transmissions a day, which represent crucial data about a transaction and the parties involved. It is used by both banks and major financial institutions to speed up settlement between themselves, on behalf of the clients and customers. It replaced the Telex (wire transfer) system. The new incoming standard to replace SWIFT is \href{https://www.swift.com/standards/iso-20022}{ISO20022} is a complex and data rich arrangement. The SWIFT consortium are promoting this new standard to their 11,000 plus global user base. A group of `crytocurrencies' are heavily involved in the ISO20022 standard, and there's been experimentation with private permissioned distributed ledger technologies. It's somewhat unclear what value they bring, and possible that the relationship of these public ledgers to international bank to bank messaging is a marketing distraction. The Bank Of England is \href{https://www.bankofengland.co.uk/payment-and-settlement/rtgs-renewal-programme/consultation-on-a-new-messaging-standard-for-uk-payments-iso20022}{transitioning to the system} in June 2023. Note that SWIFT, ISO20022, and the associated tokens within crypto are all themselves products which have a business model. They are all intermediaries which will demand a mediating fee somewhere. All of this proposed functionality could be replaced by central bank digital currencies, which will be discussed later in the section.
\subsection{FEDNOW}
Seemingly in direct response to the pressures of cryptocurrencies The USA is launching \href{https://www.federalreserve.gov/paymentsystems/fednow_about.htm}{FEDNOW}. This section will get revised.
\subsection{SPFS and BRICS}
While media outlets like the Financial Times are \href{https://www.ft.com/content/f8f3b2cd-6690-4f26-b81e-e972751c8799}{seemingly concerned} about the proposal for a BRICS based currency, and a multi-polar economic world (as we have suggested), \href{https://twitter.com/robfnunn/status/1641743274997055490}{Nunn opines} Brazil's reliance on China for inward investment and the impact of US foreign policy. He highlights that Brazil has no choice but to trade with China, who sets the rules. Nunn also points out the reluctance of Brazilians to hold Chinese treasuries. He emphasizes the misunderstanding of international currency usage and states that the Euro-Dollar system, supported by currencies like the Pound and Yen, dominates the market. Nunn argues that the possibility of the US dollar losing reserve currency status is sensationalist nonsense. Meanwhile, chief foreign policy advisor in Brazil has said: \textit{``I think the two countries can also have an important role in building a more multipolar world, in which power is less centralized and there is no hegemony. I think this is a very important aspect in which China and Brazil can play important roles.''}
\subsection{VISA and Mastercard}
Both major credit card companies are building out their ``crypto'' capabilities. Mastercard have \href{https://finance.yahoo.com/news/mastercard-crypto-secure-200559003.html}{launched a back end platform} to mitigate fraud when buying digital products with their cards. VISA have announced a ``\href{https://investor.visa.com/news/news-details/2021/Visa-Introduces-Crypto-Advisory-Services-to-Help-Partners-Navigate-a-New-Era-of-Money-Movement/default.aspx}{crypto business to business support unit}''. They have also \href{https://usa.visa.com/solutions/crypto/auto-payments-for-self-custodial-wallets.html}{published a white paper} to allow users to improve their experience.
\subsection{Money transfer operators}

\href{https://www.toptal.com/finance/market-research-analysts/international-money-transfer}{International Money Transfer Operators analysis}

western union etc, moneygram, transferwise,
\subsection{Digital disruptive fintech}
It seems that the neobank providers of digital banking apps are likely to converge with native digital asset ``wallets''. This is also the thesis advanced by the Ark intestments Big Ideas paper.\par
\begin{figure}
  \centering
    \includegraphics[width=\linewidth]{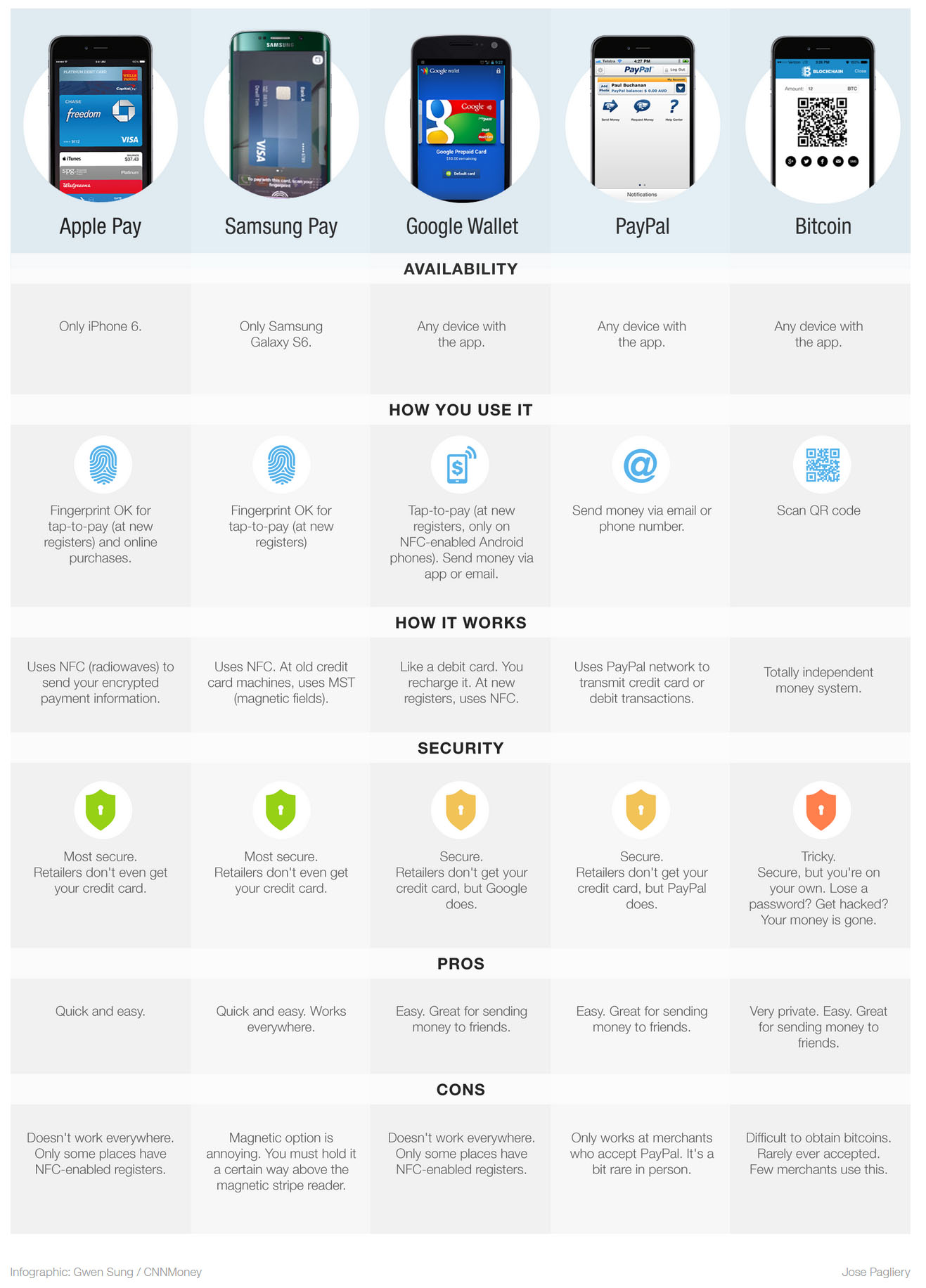}
  \caption{Comparison of mobile based payment systems}
  \label{fig:CNNmobile}
\end{figure}

CNN have a \href{https://money.cnn.com/infographic/technology/mobile-payment-comparison/index.html}{useful primer} of the most prevalent mobile digital payment methods. This can be seen in Figure \ref{fig:CNNmobile}.
This comparison makes it pretty clear that Bitcoin is not ready as a personal mobile payment system. That's not to say that there isn't a place for the underlying technology in global payment processing. 
The most interesting example of this is Strike, a product in the international fintech arena. It is a `global' money transmitter which uses bank connections in local currencies, but a private version of the Lightning network with settlement on the Bitcoin main chain. In practice users connect the app to their bank and can send money to the bank connected Strike app of another user instantly, and without a fee. This is a far better product than those previously available. In principle it's open API allows many more applications to be integrated into the Strike back end. Twitter already uses this for international tipping (and remittance). It seems that this is a perfect contender for supporting transactions in open metaverse applications, and that may be true, but Strike is currently only available in three countries (USA, El Salvador, Argentina).\par
Paypal, xoom, Strike, servicing smaller payments, cashapp, venmo, revulot, 
Paypal especially is noteworthy for their recent Orwellian gaffe suggesting in their terms and conditions that they would be able to fine users \$2500 for ``disseminating informational''. They \href{https://www.yahoo.com/video/paypal-policy-permits-company-fine-143946902.html}{quickly walked this back} but this kind of private fintech action is highly suggestive of a need for uncensorable money such as Bitcoin.\par 
Apple has recently introduced a high-yield Savings account with an impressive 4.15\,\% APY, far surpassing the national average of 0.35\,\% APY. This development represents another milestone for the tech giant as it progresses towards potentially becoming the world's largest bank. The Apple Savings account, established through a collaboration with Goldman Sachs, offers numerous benefits, including an interest rate more than ten times the national average, zero fees, no minimum deposit or balance requirements, and an efficient, user-friendly interface. Additionally, it provides FDIC insurance for balances up to \$250,000. Although the 4.15\,\% APY is lower than returns from money market funds and 1\,\% below the 3-Month Treasury Bill Yield of 5.2\,\%, most people may not be aware of these alternatives. The key factors driving demand for Apple's offering are convenience and the strong brand trust it enjoys.
With products such as Apple Pay, Apple Cash, mPOS, Apple Card, Apple Pay Later, and now Apple Savings, the company is strategically constructing an Apple Finance empire poised to disrupt the traditional financial services landscape.

\subsection{Stablecoins}
Stablecoins are `crypto like' instruments which are `pegged' at a 1:1 ratio with nationally issued Fiat currencies. In fact they usually correspond to units of privately issued  debt underwritten by a variety of different assets. This is (depending on the issuing company's model) a \href{https://www.americanbanker.com/opinion/ststablecoins-are-backed-by-reserves-give-us-a-break}{far more risky} unit of money than the nominal currency that they represent, but they offer significant utility. They allow the user to self custody the cryptographic bearer instrument representing the money themselves, as with blockchain. This may afford the user less friction in that they can transmit the instrument through the newer financial rails which are emerging. Once again, this is likely a product most useful to \href{https://www.cigionline.org/articles/the-future-of-fintech-is-unfolding-in-africa/?}{emerging markets}, those living under oppressive regimes, currencies \href{https://www.bloomberg.com/news/articles/2022-07-03/argentines-seek-hedging-in-crypto-after-economy-minister-resigns}{suffering from high inflation}, and countries who rely on the dollar as their currency, and within digitally native metaverse applications. These are \textit{enormous} global uses though. The use in the west is prominently for `traders' on exchanges at this time. /par
The caveat of such products is that such `units' of money can be frozen by the issuer, and they are subject to the third party risk of the issuer defaulting on the underlying instrument, instantly wiping out the value.\par
Klages-Mundt et al. wrote a paper in 2020, which explains the details of the different mechanisms and risks.\par
The following text paraphrases Spencer noon of on-chain analytics company ``OurNetwork'', who provides an \href{https://twitter.com/spencernoon/status/1524752048121466883}{useful summary} of the paper.
\textit{There are two major classes of stablecoins:
\begin{itemize}
\item Custodial: entrusted by off-chain collateral assets like fiat dollars that sit in a bank. Requires trust in third party.
\item Non-custodial (aka decentralized): fully on-chain and backed by smart contracts \& economics. No trusted parties.
\end{itemize}
In custodial stablecoins, custodians hold a combination of assets (currencies, bonds, commodities, etc.) off-chain, allowing issuers (possibly the same entity) to offer digital tokens of an reserve asset. The top 2 custodial stablecoins today are USDT and USDC.
There are 3 types of custodial stablecoins.
\begin{itemize}
\item Reserve Fund: 100\% reserve ratio. Each stablecoin is backed by a unit of the reserve asset held by the custodian. A useful example of this the \href{https://www.americanbanker.com/news/bank-stablecoin-consortium-usdf-gets-a-ceo-grows-to-9-members}{USDF banking consortium}.
\item Fractional Reserve Fund: The stablecoin is backed by a mix of both reserve assets and other capital assets.
\item Central Bank Digital Currency (CBDC): A digital form of central bank money that is widely available to the general public. CBDCs are in their nascency as today only 9 countries/territories have launched them, many of them small.
\end{itemize}
Custodial stablecoins have three major risks:
\begin{itemize}
\item Counterparty Risk (fraud, theft, govt seizure, etc.)
\item Censorship Risk (operations blocked by regulators, etc.)
\item Economic Risk (off-chain assets go down in value)
\end{itemize}
Each can result in the stablecoin value going to zero.}\par
It's worth taking a look at these tokens individually, to get a feel for the trade-offs, and figure out how they might be useful for us in our proposed metaverse applications. It's important to know that these tokenised dollars and/or other currencies are issued on top of the public blockchains we have been detailing throughout. Which tokens are on what blockchains is constantly evolving, so it's not really worth enumerating specifics. In a metaverse application it would be necessary to manage both the underlying public blockchain and the stablecoin issued on top of it, making the interaction with the global financial system perversely more not less complex. In the following list of a few of the major coins, the first hyperlink is the whitepaper if it's available.
\begin{itemize}
\item \href{https://f.hubspotusercontent30.net/hubfs/9304636/PDF/centre-whitepaper.pdf}{USDC} is a dollar backed coin issued by a consortium of major players in the space, most notably Circle, and Coinbase. It's has a better transparency record than tether but is still not backed 1:1 by actual dollars in reserve. It may or may not be a fractional reserve asset. It's  well positioned to take advantage of regulatory changes in the USA, and seems to be quietly lobbying to be the choice of a government endorsed digital dollar, at least a significant part of a central bank digital currency initiative. It's too early to tell how this will work out, but it has \href{https://www.forbes.com/sites/ninabambysheva/2022/04/13/blackrocks-newest-investment-paves-the-way-for-digital-assets-on-wall-street/?}{substantial `legacy finance backing'}. It is the only stablecoin to increase slightly in value (depegging upward) in the wake of the UST implosion. This `flight to quality' shows the advantage of the work that CENTRE put into regulatory compliance. It runs on Ethereum, Algorand, Solana, Stellar, Tron, Hedera, Avalanche and Flow blockchains. At this time USDC may be \href{https://twitter.com/Excellion/status/1567472488589963264}{under speculative attack} by Chinese exchange Binance, in favour of their own offering BUSD, and is losing market share. 
\item Binance USD is the dollar equivalent token from global crypto exchange behemoth Binance. It's released in partnership with Paxos, who have a strong record for compliance, and transparency. Paxos also offer USDP. Both these stablecoins claim to be 100\% backed by dollars, or US treasuries. They are regulated under the more restrictive New York state financial services and have a monthly \href{https://paxos.com/attestations/}{attestation report}.
\item \href{https://makerdao.com/en/whitepaper#abstract}{MakerDAO Dai} is an Ethereum based stablecoin and one of the older offerings. It's been `governed' by a DAO since 2014. `Excess collateral', above the value of the dai-dollars to be minted, is voted upon before being committed to the systems' cryptographic `vaults' as a backing for the currency. These dai can then be used across the Ethereum network. Despite the problems with DAOs, and the problems with Ethereum, DAI is well liked by its community of users and has a healthy billion dollars of issuance. They may be \href{https://thedefiant.io/tornado-impact-makerdao-dai}{dangerously exposed} to the new crackdown in the USA, and there is \href{https://twitter.com/bantg/status/1557733094899138560}{internal talk} of pro-actively abandoning DAI altogether.
\item \href{https://trueusd.com/pdf/TUSD_WhitePaper.pdf}{TrueUSD} claims to be fully backed by US dollars, held in escrow. It runs on the Ethereum blockchain. They have attestation reports \href{https://real-time-attest.trustexplorer.io/truecurrencies}{available on demand} and claim fully insured deposits. It's not quite that simple in that a portion of the backing is `cash equivalents'.
\item \href{https://www.gemini.com/static/dollar/gemini-dollar-whitepaper.pdf}{Gemini GUSD} claim reserves are ``held and maintained at State Street Bank and Trust Company and within a money market fund managed by Goldman Sachs Asset Management, invested only in U.S. Treasury obligations.'' which seems pretty clear.
\item \href{https://assets.website-files.com/611153e7af981472d8da199c/618b02d13e938ae1f8ad1e45_Terra_White_paper.pdf}{TerraUSD} (UST) \textbf{was} a newer and more experimental stablecoin, and one of a set of currency representations within the network. It worked in concert with the LUNA token on the Cosmos blockchain in order to keep it's dollar stability. It was not backed in the same way as the other tokens, instead relying on an arbitrage mechanism using LUNA. In essence the protocol paid users to destroy LUNA and mint UST when the price was above one dollar, and vice versa. This theoretically maintained the dollar peg. There was much concern that this model of \href{https://mirror.xyz/damsondao.eth/OVeBrmrfcWm7uKLlA2Q4W1XTVkFU3cMKfNWhgf7mQuM}{`algorithmic stable coin'} is unstable \cite{clements2021built}. The developers of the Terra tried to address this concern by \href{https://etherscan.io/address/0xad41bd1cf3fd753017ef5c0da8df31a3074ea1ea}{buying enormous amounts} of Bitcoin, which they quickly had to employ to address UST drifting downward from \$1. This failed to address the `great depegging', with LUNA crashing to essentially zero, destroying some \$50B of capital. It will now likely act as a cautionary tale to other institutions considering Bitcoin as a `reserve asset'. An \href{https://github.com/GMCyberFoundry/Metaverse/blob/b06547bf290392d2ff02e5142dae7386d888a9de/Book/04_money.tex#L186}{earlier version of this book} highlighted the specific variation of the risk which quickly manifested.
\item \href{https://tether.to/en/whitepaper/}{Tether} is the largest of the stablecoins, with some \$70B in circulation, and the third largest `crypto'. This has been a meteoric rise, attracting the ire and scrutiny of \href{https://www.cftc.gov/PressRoom/PressReleases/8450-21}{regulators} and \href{https://www.bloomberg.com/news/features/2021-10-07/crypto-mystery-where-s-the-69-billion-backing-the-stablecoin-tether}{investigators}. There was considerable doubt that Tether had sufficient assets backing their synthetic dollars, but the market seems not to mind. Recently however they have transitioned to being backed by US treasury bills, a perfect asset for this use case. It's resilience against `bank runs' was tested in May 2022 when \$9B was redeemed directly for dollars in a few days following the UST crash (more on this later). They are \href{https://tether.to/en/tether-to-launch-gbpt-tether-tokens-pegged-to-the-british-pound-sterling/}{shortly to launch} a GBP version for the UK. It's an important technology for this metaverse conversation because of intersections with Bitcoin through the Lightning network. Tether might actually provide everything needed. It's only as safe as the trust invested in the central issuer though, and the leadership and history of the company \href{https://www.wsj.com/articles/tether-ownership-and-company-weaknesses-revealed-in-documents-11675363340}{are questionable}. It's notable and somewhat ironic that it's perhaps  better and more transparently backed than most banks, and probably all novel fiat fintech products. We can employ the asset through the Taro technology described earlier but we would rather use something with higher regulatory assurances.
\end{itemize}  .
\subsubsection{The evolving US position}
In most regards the legislative front line is happening in the USA. Treasury Secretary Yellen responded to the collapse of Terra/UST \href{https://www.youtube.com/watch?v=kU0xYBRfgvU}{saying that}: \textit{``A comprehensive regulatory framework for US dollar stablecoins is needed''}. She also said that the stablecoin market is too small to pose systemic risk at this time. This is clearly an evolving situation, but the incredible consumer exposure to these risky products is likely to elicit a swift and significant response, and the timing seems right for intervention. The markets suggest that USDC will be the eventual winner.\par Koning meanwhile has looked into the different \href{http://jpkoning.blogspot.com/2021/08/stablecoin-regulatory-strategies.html}{regulatory approaches} used by various stablecoins.\par
\begin{itemize}
\item The highly regulated New York state financial framework (Paxos, Gemini)
\item Piggyback off of a (Nevada) state-chartered trust [TrueUSD, HUSD]
\item Get dozens of money transmitter licenses [USDC]
\item Stay offshore [Tether]
\end{itemize}

\href{https://www.americanbanker.com/news/toomey-unveils-stablecoin-bill-granting-occ-authority-for-payments-charter}{Proposed legislation} specific to the concept of stablecoins has been advanced by Sen Toomey. There are many provisions in the bill, mostly pertaining to convertibility and the ever present problem of attestation of the `backing' of these products. Mention has already been made of the major bill advanced by Sen. Lummis and Gillibrand. This bill also includes significant provision around stablecoins. Lummis said \textit{``Stablecoins will have to be either FDIC insured or more than 100\% backed by hard assets.''}. This is good news for this section of the digital asssets space.\par
Crucially there is also more clarity on privacy. This is a huge threat from digital money systems, and the USA is likely to lead. Remember though that none of this is yet law.\par 
Valkenburg, the lead researcher of a US think tank in digital assets \href{https://twitter.com/valkenburgh/status/1511783339065237521}{says the following}: \textit{``Stablecoin TRUST Act, is a discussion draft mostly about stablecoins, but it also has important privacy protections for crypto users broadly: it puts real limits on warrantless surveillance by narrowing what info can be collected from third parties. Last summer we fought a provision in the infrastructure bill that damaged the privacy of crypto users by expanding the broker definition (who needs to report information about transactions to the IRS) \& crypto 6050I reporting (reports on business transactions over \$10,000). The winter before we fought and successfully delayed a rushed proposal from the outgoing Trump administration to mandate that exchanges collect information about persons who are not their customers, who hold crypto at addresses in wallets they control directly. the Stablecoin TRUST Act would stop these encroachments, constrain the treasury from collecting any nonpublic information unless they get a search warrant or collect only information voluntarily provided to an exchange by a customer and for a legitimate business purpose. If “voluntarily provided for a legitimate business purpose” sounds familiar to you, that’s b/c it's the constitutional standard articulated by the Court in Carpenter describing LIMITED circumstances where warrantless searches of customer data are ok.It’s the standard we’ve advocated must also limit warrantless data collection at crypto exchanges. If exchanges must collect information about non-customers, that information is, by definition, not voluntarily provided for a legitimate business purpose.''}\par

The ongoing battle for control over emerging stablecoins by the CFTC and the SEC \href{
https://www.reuters.com/legal/transactional/presidents-working-group-report-calls-stablecoin-regulation-2021-12-02/}{seems to be pushing} the American government into legislation. They have \href{https://docs.house.gov/meetings/BA/BA21/20230419/115753/BILLS-118pih-Toproviderequirementsforpaymentstablecoinissuersresearchonadigitaldollarandforotherpurposes.pdf}{published a draft bill} and there have been some congressional hearings over the matter. At this time the bill is nascent, and there are as yet no firm decisions, though as seems typical in the USA there are hardening opinions along political lines.

  



\subsubsection{Paypal - The mainstream stablecoin}
Paypal accomplish what Libra did not, and have launched a dollar stablecoin into the aggressive regulatory landscape in the USA.

\begin{itemize}
\item PayPal is launching a new ERC-20 stablecoin called PayPal USD (PYUSD) pegged 1:1 to the US dollar and issued by Paxos
\item It will be compatible with the Ethereum ecosystem and can be transferred between PayPal and Ethereum wallets
\item The stablecoin will support P2P payments, PayPal checkout integration, and convertibility to other cryptocurrencies
\item PayPal's massive reach could drive significant crypto adoption if users take up the stablecoin
\item Regulatory comfort with PayPal's stablecoin shows preference for tradfi over non-compliant crypto firms. This embrace by lawmakers signals a shift to encourage crypto innovation from compliant US firms, not "shady" crypto natives
\item Likely pressures Congress to finalize clear stablecoin regulation to enable innovation
\item Fits growing trend of tradfi firms like BlackRock entering crypto as regulation tightens
\end{itemize}

\subsubsection{The evolving European and UK position}
Societe Generale is a leading European financial services group, based in France. Founded in 1864, it provides a wide range of services, including retail banking, corporate and investment banking, asset management, insurance, and financial solutions for both individual and institutional clients. The bank operates globally, with a strong presence in Europe, Africa, and the Middle East, as well as a growing presence in the Americas and Asia-Pacific regions. Societe Generale is recognized as one of the largest banks in Europe. They have announced a \href{https://www.sgforge.com/societe-generale-forge-launches-coinvertible-the-first-institutional-stablecoin-deployed-on-a-public-blockchain/}{Euro based stablecoin} initiative on the Ethereum blockchain, which has been met with howls of derision from the crypto and Bitcoin communities, since every transaction needs to be manually approved by the banking groups. In addition there is code in the contracts allowing them (or any party with access) to remotely \href{https://etherscan.io/address/0xf7790914dc335b20aa19d7c9c9171e14e278a134#code}{`burn' or revoke the money} from a wallet. This is ``decentralisation theatre''. The stablecoin is available only for institutional clients only, `aiming to bridge the gap between traditional capital markets and the digital assets ecosystem'. It is likely that this project is too clunky and experimental to ever see adoption.\par
\begin{figure}
  \centering
    \includegraphics[width=\linewidth]{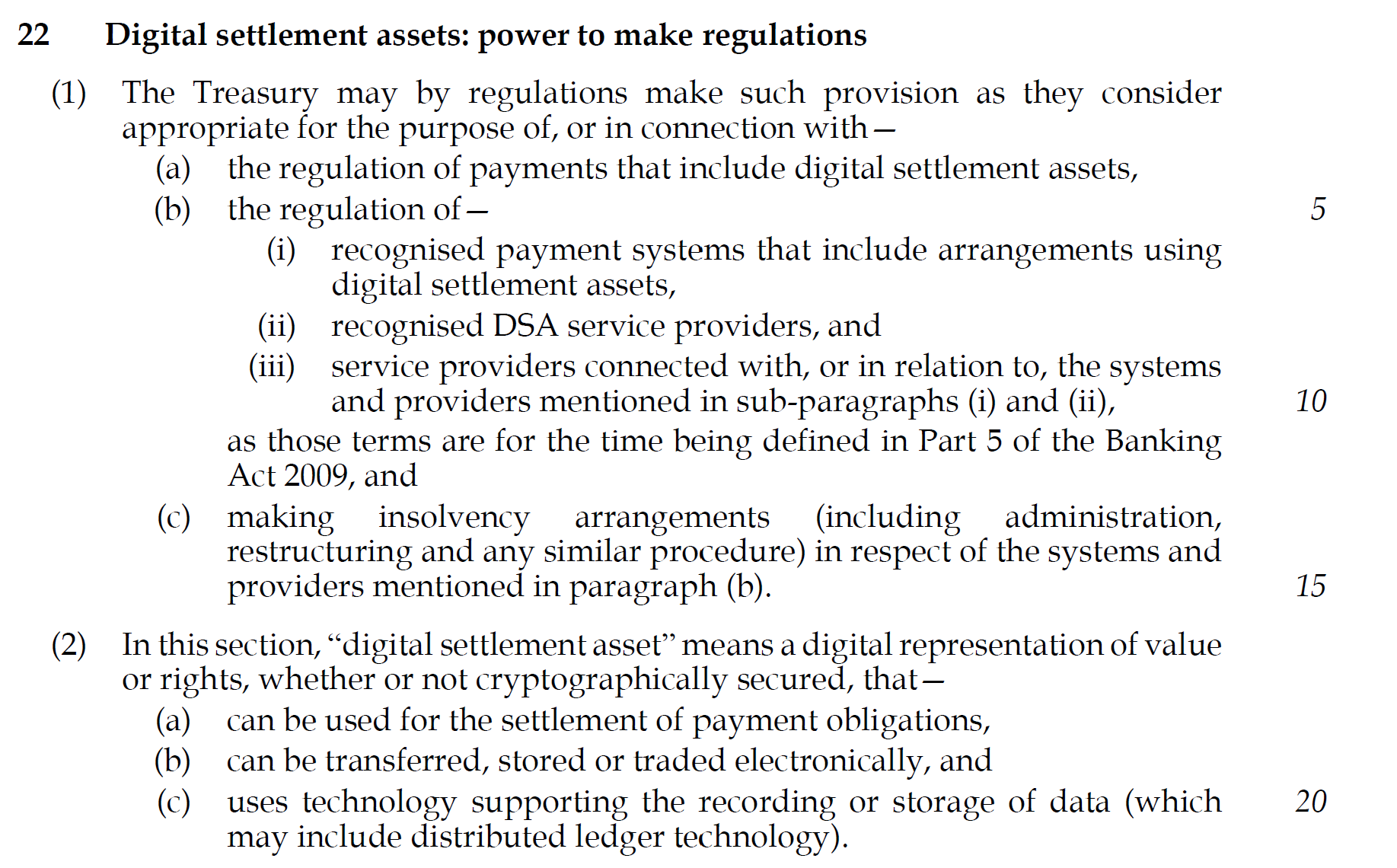}
  \caption{The UK signs into law regulation of digital representatives of value}
  \label{fig:ukdigitalbill}
\end{figure}

 
As mentioned briefly in the introduction the UK has recently \href{https://www.gov.uk/government/news/government-sets-out-plan-to-make-uk-a-global-cryptoasset-technology-hub}{signalled an enthusiasm} for stablecoins as ``means of payment''. This is a stark reversal of their previous legislative momentum is possibly a response to the \href{https://www.coindesk.com/policy/2022/05/11/eu-commission-favors-ban-on-large-scale-stablecoins-document-shows/}{tightening of rhetoric} in Europe around such assets. The \href{https://publications.parliament.uk/pa/bills/cbill/58-03/0146/220146.pdf}{Financial Services and Markets Bill.} became law in July 2022. An excerpt pertaining to stablecoins can be seen in Figure \ref{fig:ukdigitalbill}. \par
The U.K. Financial Conduct Authority’s chief executive, Nikhil Rathi, outlined the FCA’s regulatory goals at the Peterson Institute for International Economics: \textit{``The U.S. and U.K. will deepen ties on crypto-asset regulation and market developments — including in relation to stablecoins and the exploration of central bank digital currencies.''} \par
The timing seems right to explore the use of stablecoins in metaverse applications up the list of choices. 

\subsubsection{Stables in metaverse applications}
It makes a \textbf{lot} of sense to consider stablecoin transfer as the money in metaverses. USDC is furthest along this possible adoption curve. Their partnership with global payment provider Stripe has \href{https://stripe.com/blog/expanding-global-payouts-with-crypto}{enabled global dollar transfer} within Twitter for users of their `Connect' platform. This leverages the Polygon chain (mentioned in the blockchain chapter). Many digital wallets can be connected from the user end, with Metamask potentially being the easiest to integrate. This has also been mentioned in the book. The downside of this for our open platform is that none of these elements are particularly open, or distributed, and the users of the platform will still need to use an exchange to get the USDC to spend. This approach makes it easier for the vendors and product providers in the metaverse applications to accept USDC, but everything else is actually harder.

\section{Central bank digital currencies}
If 2022 was the year of the stablecoin then 2023 is likely to be the year of the central bank digital currency (CBDC). CBDCs would likely not exist without the 2019 catalyst of \href{https://www.thetimes.co.uk/article/facebooks-libra-cryptocurrency-project-ends-in-failure-cxvnnc3kx}{Facebook Libre} crypto currency project, which is now \href{https://fortune.com/2022/07/01/meta-novi-crypto-payments-wallet-end-september-2022/}{cancelled} and defunct, \href{https://www.theguardian.com/world/2021/jul/09/currency-and-control-why-china-wants-to-undermine-bitcoin}{pressure exerted on central banks} by the concept of Bitcoin, and the stablecoins which emerged from the technology.\par
\begin{figure}
  \centering
    \includegraphics[width=\linewidth]{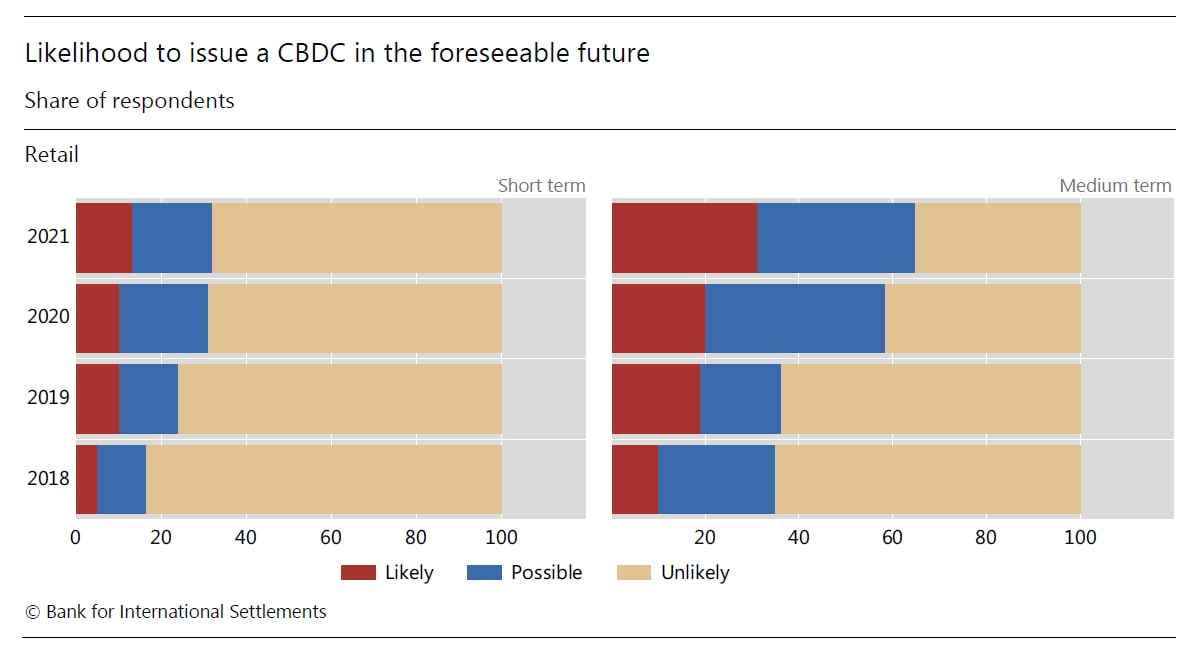}
  \caption{More than half of central banks \href{https://www.bis.org/publ/bppdf/bispap125.htm}{surveyed by the BIS} said they saw issuance of a CBDC as possible.}
  \label{fig:CBDClikely}
\end{figure}

It now seems plausible that the world is moving toward a plurality of national and private digital currencies. Figure \ref{fig:CBDClikely} from the Bank for International Settlement, shows the growing acceptance within central banks. Their 2022 annual economic report dedicates \href{https://www.bis.org/publ/arpdf/ar2022e3.pdf}{a 42 page chapter} to the subject. Hyun Song Shin, head of research at BIS said \textit{``Our broad conclusion is captured in the motto, ‘Anything that crypto can do, CBDCs can do better.``}\par
Bank of America analysts Shah and Moss think that CBDC's \href{}{are `inevitable'} by 2030, and believe that in the meantime stablecoins will fill what they perceive to be this market gap.\par
This text from the \href{https://voxeu.org/article/benefits-central-bank-digital-currency}{thinktank VoxEU} highlights the pressure on not to be \href{https://himes.house.gov/u-s-central-bank-digital-currency}{`left behind'}: \textit{``Given the rapid pace of innovations in payments technology and the proliferation of virtual currencies such as bitcoin and ethereum, it might not be prudent for central banks to be passive in their approach to CBDC. If the central bank does not produce any form of digital currency, there is a risk that it loses monetary control, with greater potential for severe economic downturns. With this in mind, central banks are moving expeditiously when they consider the adoption of CBDC.''} The Atlantic Council \href{https://www.atlanticcouncil.org/cbdctracker/}{have a website} which tracks global adoption.\par\par
CBDCs are wholly digital representations of national currencies, and as such are centralised database entries, endorsed and potentially issued by national governments. The \href{https://www.federalreserve.gov/publications/files/money-and-payments-20220120.pdf}{USA's whitepaper} shows the approach. This thinking seems to have emerged in part from the `Digital Dollar Project', an Accenture funded think tank founded by ex CFTC chairman Giancarlo \cite{giancarlo2021cryptodad}. Curiously only \href{https://www.sanddollar.bs/about}{The Bahamas} seem to have a successful implementation, but it is a rapidly evolving space, and many nations are now scrambling to \href{https://twitter.com/GobiernoMX/status/1476376240873517061}{catch up}. A \href{https://www.linkedin.com/feed/update/urn:li:activity:6980330210030145536/}{post on the LinkedIn page} of the Bank of International Settlements highlights a research project between 20 Asian banks which settles tens of millions of dollars using CBDC tooling.\par
The following text is taken from the March 2021 Biden government ``executive order'' on digital assets, and defines the current global legislative position well.\\
\textit{``Sec. 4.  Policy and Actions Related to United States Central Bank Digital Currencies.  (a)  The policy of my Administration on a United States CBDC is as follows:\\
(i) Sovereign money is at the core of a well-functioning financial system, macroeconomic stabilization policies, and economic growth.  My Administration places the highest urgency on research and development efforts into the potential design and deployment options of a United States CBDC.  These efforts should include assessments of possible benefits and risks for consumers, investors, and businesses; financial stability and systemic risk; payment systems; national security; the ability to exercise human rights; financial inclusion and equity; and the actions required to launch a United States CBDC if doing so is deemed to be in the national interest.\\
(ii)   My Administration sees merit in showcasing United States leadership and participation in international fora related to CBDCs and in multi‑country conversations and pilot projects involving CBDCs.  Any future dollar payment system should be designed in a way that is consistent with United States priorities (as outlined in section 4(a)(i) of this order) and democratic values, including privacy protections, and that ensures the global financial system has appropriate transparency, connectivity, and platform and architecture interoperability or transferability, as appropriate.\\
(iii)  A United States CBDC may have the potential to support efficient and low-cost transactions, particularly for cross‑border funds transfers and payments, and to foster greater access to the financial system, with fewer of the risks posed by private sector-administered digital assets.  A United States CBDC that is interoperable with CBDCs issued by other monetary authorities could facilitate faster and lower-cost cross-border payments and potentially boost economic growth, support the continued centrality of the United States within the international financial system, and help to protect the unique role that the dollar plays in global finance.  There are also, however, potential risks and downsides to consider.  We should prioritize timely assessments of potential benefits and risks under various designs to ensure that the United States remains a leader in the international financial system.''}\par

In traditional nation state currencies the central banks \href{https://www.bankofengland.co.uk/markets/bank-of-england-market-operations-guide}{control the amount} of currency in circulation by issuing debt to private banks, which is then loaned out to individuals \cite{wang2021central}. The debt is `destroyed' on the balance sheet to remove currency through the reverse mechanism. They also facilitate government debt \cite{filardo2012central}, and work (theoretically) outside of political control to adjust interest rates, in order to manage growth and flows of money. \par
It is somewhat surprising that Powell, chair of the US Federal Reserve has \href{https://www.federalreserve.gov/newsevents/speech/powell20220617a.htm}{recently said} \textit{``Rapid changes are taking place in the global monetary system that may affect the international role of the dollar. A US central bank digital currency is being examined to help the US dollar's international standing.''}. This is a rapid evolution of the narrative, with implications. It seems unlikely that the world would sacrifice the traditional banking system in favour of centrally controlled money, but many things which cannot be done with traditional nation state money systems are possible with CBDCs, because they \href{https://voxeu.org/article/benefits-central-bank-digital-currency}{remove the middleman} of private banking between the end user and the policy makers. 
\begin{itemize}
\item Negative interest rates are possible, such that all of the money can lose purchasing power over time, and at a rate dictated by policy. This ``removal of the lower bound'' has been discussed by economists over the last couple of decades as interest rate mechanisms have waned in efficacy. It is not possible in the current system, and instead money must be added through \href{https://www.bankofengland.co.uk/monetary-policy/quantitative-easing}{quantitative easing}, which disproportionately benefits some though Cantillon effects \cite{cantillon1756essai, bordo1983some}.  
\item Ubiquitous basic income is possible in that money can be issued directly from government to all approved citizens, transferring spending power directly from the government to the people. This also implies efficiency savings for social support mechanisms.
\item Asset freezing and confiscation are trivial if CBDCs can replace paper cash money completely, as a bearer asset. Criminals and global `bad actors' could have their assets temporarily or permanently removed, centrally, by suspending the transferability of the digital tokens.
\item Targeted bailouts for vital institutions and industries are possible directly from central government policy makers. Currently private banks must be incentivised to make cheap loans available to sectors which require targeted assistance.
\item Financial surveillance of every user is possible. In this way a `panopticon of money' can be enacted, and spending rulesets can be applied. For instance, social support money might only be spendable on food, and child support only on goods and services to support childcare. This is a very dystopian set of ideas. Eswar Prasad says ``In authoritarian societies, central bank money in digital form could become an additional instrument of government control over citizens rather than just a convenient, safe, and stable medium of exchange\cite{prasad2021future}.'' This is possibly \href{https://twitter.com/WallStreetSilv/status/1581378124452753408}{already happening} in China through integration of outstanding debt data with the social credit system.  
\item It's a virtually cost free medium of exchange, since there is no physical instrument which must be shipped, guarded, counted, assayed, and securely destroyed.
\item The counterfeiting risk is significantly reduced because of secure cryptographic underpinnings rather than paper or plastic anti counterfeiting technologies.
\item Global reach and control is instantly possible for the issuer. This is a big problem especially for a reserve currency such as the dollar. Two thirds of \$100 bills are \href{https://www.federalreserve.gov/pubs/ifdp/2012/1058/default.htm}{thought to} reside outside of the USA.
\item System level quantitative easing and credit subsidies are made far simpler and less wasteful when centrally dictated.
\item Transfer of liability and risk to the holder globally reduces the management costs for global deposits of a currency. 
\item It may be possible to automate the stability of a currency through continuous adjustment of the `peg' through algorithms or AI.
\end{itemize}

The UK had been signalled that it is not interested in developing a CBDC stating that it seemed to be a \href{https://committees.parliament.uk/publications/8443/documents/85604/default/}{solution in search of a problem}, with the Lords economic affairs committee saying:\textit{``The introduction of a UK CBDC would have far-reaching consequences for households, businesses, and the monetary system for decades to come and may pose significant risks depending on how it is designed. These risks include state surveillance of people’s spending choices, financial instability as people convert bank deposits to CBDC during periods of economic stress, an increase in central bank power without sufficient scrutiny, and the creation of a centralised point of failure that would be a target for hostile nation state or criminal actors.''}\par
Since those initial statements however it seems that the previously mentioned ``fear of missing out'' has forced legislators hand. The UK Treasury and the Bank of England are now exploring the possibility of launching a retail central bank digital currency (which they desperately hope will not end up called `Britcoin'), judging that ``it is likely a digital pound will be needed in the future''. They have released a \href{https://www.gov.uk/government/news/hm-treasury-and-bank-of-england-consider-plans-for-a-digital-pound}{consultation paper} inviting public comment. The consultation is aimed at informing the decision on whether to build the infrastructure for a digital pound, with a pilot test not expected before 2025.\par
One of the key points raised in the proposal is the potential to cap citizens' CBDC holdings, with a range suggested between \textsterling 10,000 to \textsterling 20,000, to strike a balance between managing risks and supporting the usability of the digital pound. This limit would allow most UK wage earners to receive their salary in the form of a CBDC but would still allow for competition with commercial banks. The digital pound would not offer interest, enabling banks to offer competitive deposit accounts.\par
They are once again clear about the risks, highlighting that banks currently use deposits as a cheap source of funding for loans, and without that flow, they may become more reliant on expensive wholesale markets, driving up borrowing costs for users. In a severe scenario, a bank run could undermine the capital base for the commercial banking system. \par
The Bank of England has stated that it will not implement central bank initiated programmable functions, but instead provide the necessary infrastructure for the private sector to implement such features with user consent. The digital pound is intended to have at least the same level of privacy as a bank account and users would be able to make choices about data use. This is scant comfort, as such features are intrinsic to the technology, and we have seen time and again that if legislative and economic bodies are given a hammer, they will eventually find a nail to hit. Carlo, the director of `The Big Brother watch' \href{https://twitter.com/silkiecarlo/status/1622912736102346752}{pointed to} 2021 comments from John cunliff of the bank of England when he said ``There’s a whole range of things that programmable money could do like giving the children pocket money but programming the money so it couldn’t be used for sweets''. Again, this raises the potential for government stimulus that has to be spent within a certain time or it disappears. As an interesting side note here it's thought that up to 14\% of American stimulus cheques went to buying crypto, most notably in less well off families \cite{bertomeu2023uncle}. It's easy to imagine that a CBDC would be barred from such a thing.\par
\begin{figure}
  \centering
    \includegraphics[width=0.5\linewidth]{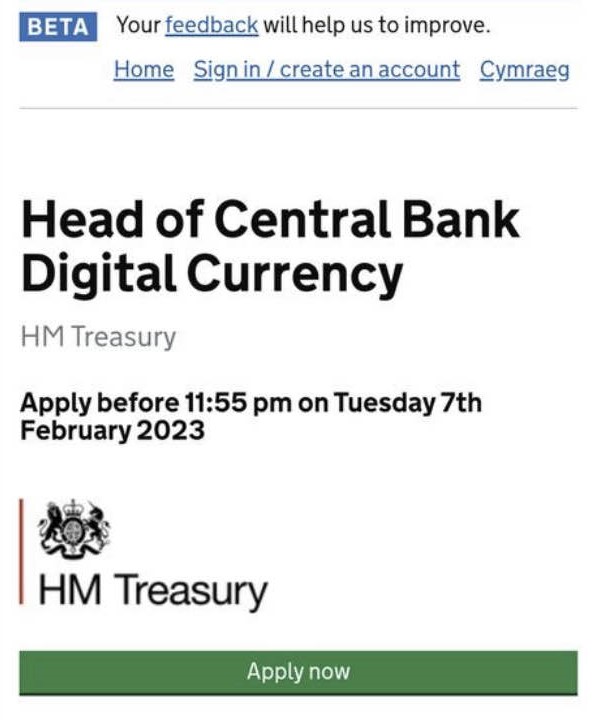}
  \caption{The UK is clearly making moves to staff a new department for CBDC.}
  \label{fig:treasury}
\end{figure}

 
The main motivation for issuing a CBDC is the assumption that there is demand for a safe and stable way to use money online. A digital pound, issued and backed by the Bank of England, could be a trusted, accessible, and easy-to-use form of payment. The infrastructure would allow firms to design innovative and user-friendly services. The civil service is hiring a ``Head of CBDC'' as seen in Figure \ref{fig:treasury}.

Meanwhile in Europe, ECB President \href{https://www.ecb.europa.eu/press/pressconf/2022/html/ecb.is220310~1bc8c1b1ca.en.html#qa}{Christine Legarde} said: \textit{``On your question concerning CBDC, you know my views on CBDC and you know that I have pushed that project. Fabio Panetta is working hard on that together with members in the entire Eurosystem with the high-level taskforce that is working really hard on moving forward. But in a way, I am really pleased that attention is now focussed on the role that cryptos can play and the role that Central Bank Digital Currency can have when they are implemented. We have a schedule, as you know. The Governing Council decided back in October '21 to launch a two-year investigation phase, and it is at the end of that investigation phase that the decision will definitely be made to launch the CBDCs and to make it a reality. We can't go wrong with that project. I am confident that we will move ahead, but that's going to be a decision of the Governing Council. I think it's an imperative to respond to what the Europeans expect, and I think we have to be a little bit ahead of the curve if we can on that front. If we can accelerate the work, I hope we can accelerate the work. I will certainly support that and I was delighted to see that in the United States there was an executive order by President Biden to actually expect similar effort and focus and progress on CBDC, cryptos. I think that it will take all the goodwill of those who want to support sovereignty, who want to make sure that monetary policy can be transmitted properly using our currency, will endeavour.''}\par
She has expanded on these points saying in a video interview that the digital euro will be decided in October 2023. If passed, the current paradigm of cash spending and transfers will become even more restrictive. Lagarde justified the move by saying that she did not want the EU to be `dependent on the currency of an unfriendly country' or a friendly currency activated by a private corporate entity'. She identified Meta, Google, and Amazon.\par
India has expressed far more interest in the technology, and of course their addressable market is huge! They have published a  `\href{https://twitter.com/RBI/status/1578329048446828544?}{concept note}' in which they assert that a digital Rupee would be faster, cheaper, and easier to maintain. The key difference in India's situation is the large areas of the rural population where mobile internet is more patchy. In such situations a cash equivalent stablecoin token with cash finality which can be transferred between mobile phone wallets \textit{without} an internet connection is a huge boon. It seems very likely that India is moving to react to the innovation threat posed by cryptocurrencies to their own cash infrastructure. They are \href{https://www.reuters.com/article/idUSKBN2RQ0WO}{piloting the technology} already. Similarly there seems to be a strong, and \href{https://www.bloomberg.com/news/articles/2022-12-06/nigeria-limits-cash-transactions-to-push-enaira-and-other-payments}{predictably illiberal} push for transition to digital money in Nigeria. Again this is an enormous number of people, and it is hard not to be suspicious of future abuse of the system by governments.\par
In the USA this text from Congressman Tom Emmer shows how complex and interesting this debate is becoming.\textit{``Today, I introduced a bill prohibiting the Fed from issuing a central bank digital currency directly to individuals. Here’s why it matters: As other countries, like China, develop CBDCs that fundamentally omit the benefits and protections of cash, it is more important than ever to ensure the United States’ digital currency policy protects financial privacy, maintains the dollar’s dominance, and cultivates innovation.\\
CBDCs that fail to adhere to these three basic principles could enable an entity like the Federal Reserve to mobilize itself into a retail bank, collect personally identifiable information on users, and track their transactions indefinitely.\\
Not only does this CBDC model raise ``single point of failure'' issues, leaving Americans’ financial information vulnerable to attack, but it could be used as a surveillance tool that Americans should never be forced to tolerate from their own government.\\
Requiring users to open an account at the Fed to access a United States CBDC would put the Fed on an insidious path akin to China’s digital authoritarianism.\\
Any CBDC implemented by the Fed must be open, permissionless, and private. This means that any digital dollar must be accessible to all, transact on a blockchain that is transparent to all, and maintain the privacy elements of cash.\\
In order to maintain the dollar’s status as the world’s reserve currency in a digital age, it is important that the United States lead with a posture that prioritizes innovation and does not aim to compete with the private sector.\\
Simply put, we must prioritize blockchain technology with American characteristics, rather than mimic China’s digital authoritarianism out of fear.''}\par
Most analysts now seem to think that there is little appetite to replace established 'Western' cash with CBDCs. Most significantly such products would need the support of retail banks, and it is not in their interest to service such a product. Their business model relies on using retail deposits for providing loans, and it is these deposits, not cash itself that would be the most addressable market for a CBDC. Banks don't want people to self custody money. In addition it exposes the whole banking system to a higher risk of bank runs. Such a self custody, interest bearing, central government backed asset would have significantly less counterparty risk than even bank deposits, and at times of high systemic stress it seems likely that money would flow to where it's thought safest, exposing the retail banks to runs. Fabio Panetta of the ECB said: \textit{``If we give access to a means of payment, which is relatively limited, there are no transaction costs because you only need to have a smartphone. There will be risks that people could use this possibility to move, for example, their deposits of other banks or their money out of financial intermediates.''} - All of the proposed solutions to these problems such as caps and negative interest penalties seem poorly thought through.
Held and Smolenski present a \href{https://www.btcpolicy.org/articles/why-the-u-s-should-reject-central-bank-digital-currencies}{detailed and rigorous negative critique} of the dystopian ramifications of the technology. In their conclusion they point out that: \textit{``Central bank digital currencies (CBDCs) represent an extension of state control over economic life. CBDCs provide governments with direct access to every transaction in that currency conducted by any individual anywhere in the world. As governments worldwide routinely share data with one another, individual transaction data will quickly become known to any government in a datasharing arrangement. Given the frequency with which government databases are compromised, this arrangement virtually ensures that anyone’s transaction data will eventually become available for global perusal.''}\par

\subsection{Global Centralised Ledgers}

Beyond even national CBDCs it is now possible to find discussion around weaving these together at a supranational level. Indeed it seems that competition is starting to emerge. The Bank for International Settlements (BIS) and the International Monetary Fund (IMF) have both presented plans to deploy global ledgers to support programmable Central Bank Digital Currencies. 
\subsubsection{BIS proposal}
The BIS proposed the concept of a Unified Electronic Ledger, which would combine Central Bank Digital Currencies, tokenized money, and assets on a single platform. This ledger would enable smart contract functionality similar to Ethereum on a global scale. The BIS emphasized the benefits of integrating different types of money and assets on a unified ledger, such as reducing delays, uncertainties, and trade financing costs. They also highlighted the importance of policy harmonization across jurisdictions for the success of such a system.

\subsubsection{The IMF's Proposal}
The IMF proposed a global CBDC platform that would facilitate cross-border CBDC settlement. The stated aim is to enhance interoperability, efficiency, and safety in cross-border payments, while allowing individual nations to maintain capital controls and limits on the flow of funds. The IMF envision a permissioned (closed, but distributed) ledger, perhaps controlled by a platform operator (in the manner of SWIFT), to ensure unique ownership descriptions and prevent double spending. They emphasized the need for maintaining capital controls during national financial crises.

\subsubsection{The Bank of England's Experiment}
The Bank of England, in collaboration with the BIS Innovation Hub, conducted a field test of CBDC technology known as Project Rosalind. The test explored various CBDC use cases, including offline payments, retail transactions, and micropayments. The test focused on a centralized ledger hosted by the Bank of England and involved the development of API functionalities for different scenarios. The BIS considered these experiments informative for the ongoing discussions on CBDCs.

\subsection{Risks and mitigations}
The introduction of CBDCs could have a significant impact on both individuals and the existing private financial sector.\par
For individuals, the risks of CBDCs include:
\begin{itemize}
\item Privacy concerns: CBDCs could potentially be used to track and monitor individuals' financial transactions, raising concerns about privacy and government surveillance.
\item Lack of anonymity: Unlike cash, CBDCs could be easily traced and linked to individuals, which may compromise their financial privacy and security.
\item Cybersecurity risks: CBDCs could be vulnerable to cyberattacks, which could lead to the loss of funds and personal information.
\end{itemize}
For the existing private financial sector, the risks of CBDCs include:
\begin{itemize}
\item Competition: CBDCs could potentially compete with private sector financial institutions, which could lead to a decline in the use of traditional financial services.
\item Disruption: CBDCs could disrupt existing financial systems and business models, which could lead to a decline in profits and revenue for private sector financial institutions.
\item Regulation: The introduction of CBDCs could lead to increased regulation of the private financial sector, which could increase compliance costs and reduce profitability.
\item CBDCs could have implications on monetary policy, financial stability, and international relations. For example, it could change the way central banks conduct monetary policy, and it could also impact the global financial system and the role of the US dollar as a global reserve currency. 
\item CBDCs could also bring about significant changes in the global payment system, which could have major implications for the financial industry, and for the private sector as well as for the central banks.
\item A single global ledger, especially one controlled by large international bodies like the BIS and IMF, could potentially lead to an unhealthy centralization of power. This could exacerbate existing imbalances in global financial control and further marginalize countries with less political and economic power.
\item A centralized ledger system would necessitate a high degree of technological dependence, which could leave countries vulnerable in the event of technological failure or cyber attacks.
\item A global ledger system would require a high level of standardization. This could limit the ability of individual nations to adapt their financial systems to local conditions and needs, potentially leading to a ``one-size-fits-all'' approach that might not be appropriate for all contexts.
\item There are concerns that the shift to digital currencies could leave behind those without access to necessarytechnology, contributing to financial exclusion rather than mitigating it.
\end{itemize} 

In conclusion, the risks associated with CBDCs are significant and multi-faceted. It is (hopefully) more likely that a blend of stablecoins, private bank issued digital currency (with a yield incentive) and perhaps some limited CBDC, alongside the new contender Bitcoin, will present a new landscape of user choice. Different models of trust, insurance, yields, acceptability, and potentially privacy, will emerge. \par
Clearly a global, stable, wholly digital bearer asset in a native currency would ostensibly be the ideal integration for money in a metaverse application, but the whole concept seems deeply `wrong', and it is likely that a transition to such a technology would be complex and painful. Either way, it is certainly not ready for consideration now.
It's important that central banks and governments carefully consider these risks before introducing CBDCs, but it's not clear this is happening. It is conceivable that by working closely with the private sector policy makers could minimize any negative impacts, ensuring that any new regulations are designed in a way that protects the rights and privacy of individuals, while also promoting financial stability and economic growth. Gerard, an incredibly staunch critic of \textit{all} things crypto \href{https://davidgerard.co.uk/blockchain/2023/02/28/news-blockchain-mep-eva-kaili-corruption-arrest-nigerian-cbdc-rewrite-bbc-blockchain-misadventures/}{points to woeful adoption and manifest corruption} in the attempts so far. We are not particularly hopeful either.

\subsubsection{The Role of Tokenisation}
Tokenisation represents a paradigm shift from traditional cryptocurrencies. The concept was introduced and popularised by the wider cypto movement, and it's somewhat absurd claims around `tokenising everything'. After this fab died down post the `initial coin offering' craze of 2018 attention shifted elsewhere. Curiously however the `Office of the Comptroller of the Currency' and the BIS have been focusing on resolving settlement issues within financial systems. It deviates from the blockchain dependency, (correctly) and simply offers a more streamlined approach to financial transactions. This innovation will notably be explored in the OCC's tokenisation symposium held on February 8th 2024, with an aspiration of integrating different types of money and assets on a unified platform. The symposium, a public event featuring keynotes from prominent figures in the financial world, will highlight the burgeoning interest in tokenisation (\href{https://www.occ.gov/news-issuances/news-releases/2023/nr-occ-2023-18.html}{OCC Tokenization Symposium Details}).

\subsubsection{Implications and Potential Risks}
While tokenisation presents significant potential for improving transaction efficiency and reducing risk, it is not without its challenges. A key concern is the impact on the traditional financial sector and the regulatory complexities it introduces. The integration of diverse forms of digital assets on a unified platform necessitates robust regulatory frameworks to ensure stability and prevent misuse.\par
In truth this, like the global push toward central bank digital currency, seems inspired by but asymptotic to the concept of cryptocurrencies. They are important technologies to consider as digital society tooling evolved, but they remain curiously far behind the retail technologies which spawned them. As the banking sector evolves with technological advancements, the role of tokenisation and its interaction with existing financial systems become increasingly crucial. The potential for a more efficient, secure, and integrated global financial system is evident, yet the path to achieving this is laden with regulatory, technical, and ethical challenges. The success of tokenisation initiatives will largely depend on the collaborative efforts of regulatory bodies, financial institutions, and technology experts to navigate these challenges effectively.

\section{Bitcoin as a money}
Nwosu, cofounder of Coinfloor exchange in the UK, and cofounder of the aforementioned Fedimint and says that a digital money needs the following four characteristics:
\begin{itemize}
\item that it be technically mature. 
\item it should have strong community support and network effect. We have seen that this is more simply a feature of money itself.
\item that there should be regulatory clarity around the asset, a feature which even Bitcoin currently struggles with.
\item it should demonstrate a core use case of `store of value' which sounds simple enough, but again is contestable because of the volatility of Bitcoin.
\end{itemize}
\subsection{Spending it}
Since this book seeks to examine transfer of value within a purely digital environment it is necessary to ask the question of whether Bitcoin is money. This short \href{https://bitcoin-zar.blogspot.com/2018/07/duality-excerpt-by-satoshi-nakomoto.html}{`story'}, purportedly written by Nakamoto, is a fabulous look at the money values of the technology, irrespective if it's provenance. In it is the following text: \textit{``Here, for once, was this idea that you could generate your own form of money. That's the primary and sole reason, is because it was related to this thing called money. It wasn't about the proficiency of the code or the novelty, it was because it had to do with money. It centered around money. That is something people cared about. After all, plenty of projects on Sourceforge at the time were just as well coded, well maintained, if not better, by teams, and even if someone else had created the blockchain before me, had it been used for something else beyond currency, it probably would not have had much of an outcome.}\par
Again, irrespective of the author here, this point seems to ring true. The memetic power of Bitcoin is in it's proximity to `money', and the potential of the separation of money from the state. \par
It is beyond argument that the Bitcoin network is a rugged message passing protocol which achieves a high degree of consensus about the entries on it's distributed database.\par Ascribing monetary value to those database entries is a social consensus problem, and this itself is a contested topic. The most useful `hot take' here is that Bitcoin behaves most like \href{https://twitter.com/saylor/status/1395788419301773312}{a `property'}, while it's network behaves far more like a monetary network which is created and supported by the \href{https://saito.tech/an-response-to-paul-krugman-from-a-keynesian-bitcoiner/}{value of the Bitcoin tokens}. \par
Jack Mallers, of Strike \href{https://www.youtube.com/watch?v=jb-45m9f76I}{presentation to the IMF} identified the following challenges which he claims are solved by the bitcoin monetary network.
\begin{itemize}
\item Speed
\item Limited transparency and dependability
\item High cost
\item Lack of interoperability
\item Limited Coverage
\item Limited accessibility
\end{itemize}
He further identifies the attributes of the ideal global money. 
\begin{itemize}
\item Uncensorable
\item Unfreezable
\item Permissionless
\item Borderless
\item Liquid
\item Digital
\end{itemize}
Mallers has recently announced USA focused partnerships which leverage his Strike product to enable spending Bitcoin, through Lightning, as Dollars in much of the \href{https://www.ncr.com/point-of-sale-pos-systems}{point of sale} infrastructure in the USA. This is a huge advance as it immediately enables the vendors both online and at physical locations to either save 3\% costs for card processors, or else pass this on as a discount. Crucially for `Bitcoin as a money' it also allows the vendors to receive the payment \textbf{as} Bitcoin, not Dollars. A possible further and highly significant feature is that it might now be possible to divest of Bitcoin in the USA, buying goods, without a capital gains tax implication. Mallers claims to have legislative backing for this product, but the devil will likely be in the detail. The likely mechanism for this product is that the EPOS partner sends a Lighting request to Strike, which liquidates some of their Bitcoin holding to a dollar denominated stablecoin, but in a tax free jurisdiction such as El Salvador. This stablecoin will then be sent to the EPOS handing partner such as NCR. Stablecoin to Dollar transactions in the USA are much murkier and likely don't cost anything for these companies. This agent will then authorise the Dollar denominated sale to the American digital till. Crucially nobody has a US capital gains tax exposure in this chain, and all of the settlements were near free, and instantaneous, with `cash finality' for everyone except the EPOS company. They are likely actually exposed to a small risk here because uptake will be very low level. The novelty opportunity will likely cover any potential exposure to stablecoin collapse. This is a radical upgrade on the normal flow of divesting Bitcoin for American users. \par
Using this open product to spend Bitcoin as Bitcoin to vendors might be available through Shopify globally. Again, it's too new to be sure. Promisingly a \href{https://www2.deloitte.com/content/dam/Deloitte/us/Documents/technology/us-cons-merchant-getting-ready-for-crypto.pdf}{Deloitte study} has found that 93\% of businesses accepting Bitcoin have seen revenue and brand perception improve, and 75\% of USA sales execs plan to accept digital assets at some point in the next 2 years. This ambition in the US markets is likely to benefit from the proposed \$200 tax exempt law for purchasing goods and services with Bitcoin.\par
Of these recent developments in Lightning \href{https://twitter.com/LynAldenContact/status/1512188883101966351}{Lyn Alden says}: \textit{Some people naturally dismiss [strike] because they don't want to spend their BTC; they want to save it. However, the more places that accepted BTC at point of sale (on-chain or Lightning or otherwise), the more permissionless the whole network is. This is because, if all you can do with BTC is convert it back into fiat on a major exchange, then it's easy to isolate it, effectively blacklist addresses, etc. But if you can directly spend it on goods and services across companies and jurisdictions, it's harder to isolate. There are now plenty of vendors that make this easy for merchants to implement, and the merchant can still receive dollars if they want (rather than BTC), or can decide their \% split. Since it's an open network, anyone can build on it, globally. And then when you add fiat-to-BTC-to-fiat payments over Lightning, it gets even more interesting because it doesn't necessarily need to be a taxable event. Lightning wallets with a BTC balance and a USD/stablecoin balance. Lower fees than Visa and others.}\par
\subsubsection{African adoption}
Africa has one of the most fragmented banking, payment, and currency systems in the world, which makes simple financial tasks like paying a bill, sending money, or accepting money extremely difficult. Over half of Africa does not have access to a bank account, so people hold and save everything in cash, which is often stolen and loses value due to inflation. It is also difficult to get money in and out of many African countries because only about 40\% of people have active internet access, and must rely on financial institutions. Bitcoin is being used in Africa as an alternative form of money that resolves these issues. It is being taught in education centers in underdeveloped areas, giving children the opportunity to learn about and use Bitcoin as a way to access financial services that have been unavailable to them for generations. However, the rest of the country may have difficulty implementing the use of Bitcoin due to issues such as a lack of electricity and internet access, as well as government policies that centralize power.
\subsubsection{Bitcoin based FIAT}
More interestingly for metaverse applications Mallers has opened this section of the company to interact with the public Lightning network, allowing people with a self hosted wallet or node to pay directly for goods across America, settling immediately in Dollars, using their Bitcoin, at zero cost. \textbf{This opens the possibility to buy from US based (Dollar denominated) metaverse stores, using the capabilities of the stack assembled at the end of the book}. The implications globally are unclear at this time.\par
\href{https://stablesats.com/}{Stablesats} is another approach which uses exclusively lightning bitcoin but makes the value stable against the US dollar using an algorithm. This is a very interesting option and will be explored in detail at some point.
\subsection{Saving with it} 
The Bitcoin community believes that \href{https://svetski.medium.com/why-bitcoin-not-shitcoin-6cc826f4fa52}{Bitcoin is the ultimate money}, a \href{https://www.coindesk.com/business/2022/01/07/jpmorgan-sees-more-crypto-adoption-in-2022-debates-bitcoins-status-as-store-of-value/}{`store of value'}, chance to \href{https://www.forbes.com/sites/leeorshimron/2020/06/30/bitcoin-is-the-separation-of-money-and-state/?sh=49294a8356db}{separate money from state}, increase \href{https://www.washingtonpost.com/national/locked-out-of-traditional-financial-industry-more-people-of-color-are-turning-to-cryptocurrency/2021/12/01/a21df3fa-37fe-11ec-9bc4-86107e7b0ab1_story.html}{equality of opportunity} and \href{https://iai.tv/articles/the-rich-get-richer-the-poor-get-bitcoin-auid-1766}{ubiquity of access}, while others view it as \href{https://www.cnbc.com/2021/06/22/a-third-of-investors-think-bitcoin-is-rat-poison-jpmorgan-survey-says.html}{`rat poison'}, or a \href{https://jacobinmag.com/2022/01/cryptocurrency-scam-blockchain-bitcoin-economy-decentralization}{fraudulent Ponzi scheme} \cite{ponzi2021alden}. A notable exclusion from the negative rhetoric is Fidelity, the global investment manager, who have always been positive and have \href{https://www.fidelitydigitalassets.com/articles/bitcoin-first?sf253214177=1}{recently said}: 
\textit{``Bitcoin is best understood as a monetary good, and one of the primary investment theses for bitcoin is as the store of value asset in an increasingly digital world.''}\par
The following paraphrases Eric Yakes, author of \href{https://yakes.io/book/}{`The 7th Property'}. Again, this is an Austrian economics perspective, and like much economic theory the underlying premise \href{https://medium.datadriveninvestor.com/do-you-understand-the-austrian-vs-keynesian-economic-debate-2f4b152c6a6b}{is contested}\cite{maurel2012keynesian}: \textit{``Paper became money because it was superior to gold in terms of divisibility and portability BUT it lacked scarcity. People reasoned that we could benefit from the greater divisibility/portability of paper money as long as it was redeemable in a form of money that was scarce. This is when money needed to be ``backed'' by something. \\
Since we changed money to paper money that wasn't scarce, it needed to be backed by something that was. Since the repeal of the gold standard, politicians have retarded the meaning of the word because our money is no longer backed by something scarce.\\
So, what is bitcoin backed by? Nothing.\\
Sound money, like gold, isn’t ``backed''.
Only money that lacks inherent monetary properties must be backed by another money that maintains those properties. The idea that our base layer money needs to be backed by something is thinking from the era of paper money. Bitcoin does not require backing, it has inherent monetary properties superior to any other form of money that has ever existed.''}\par
\begin{figure}
  \centering
    \includegraphics[width=\linewidth]{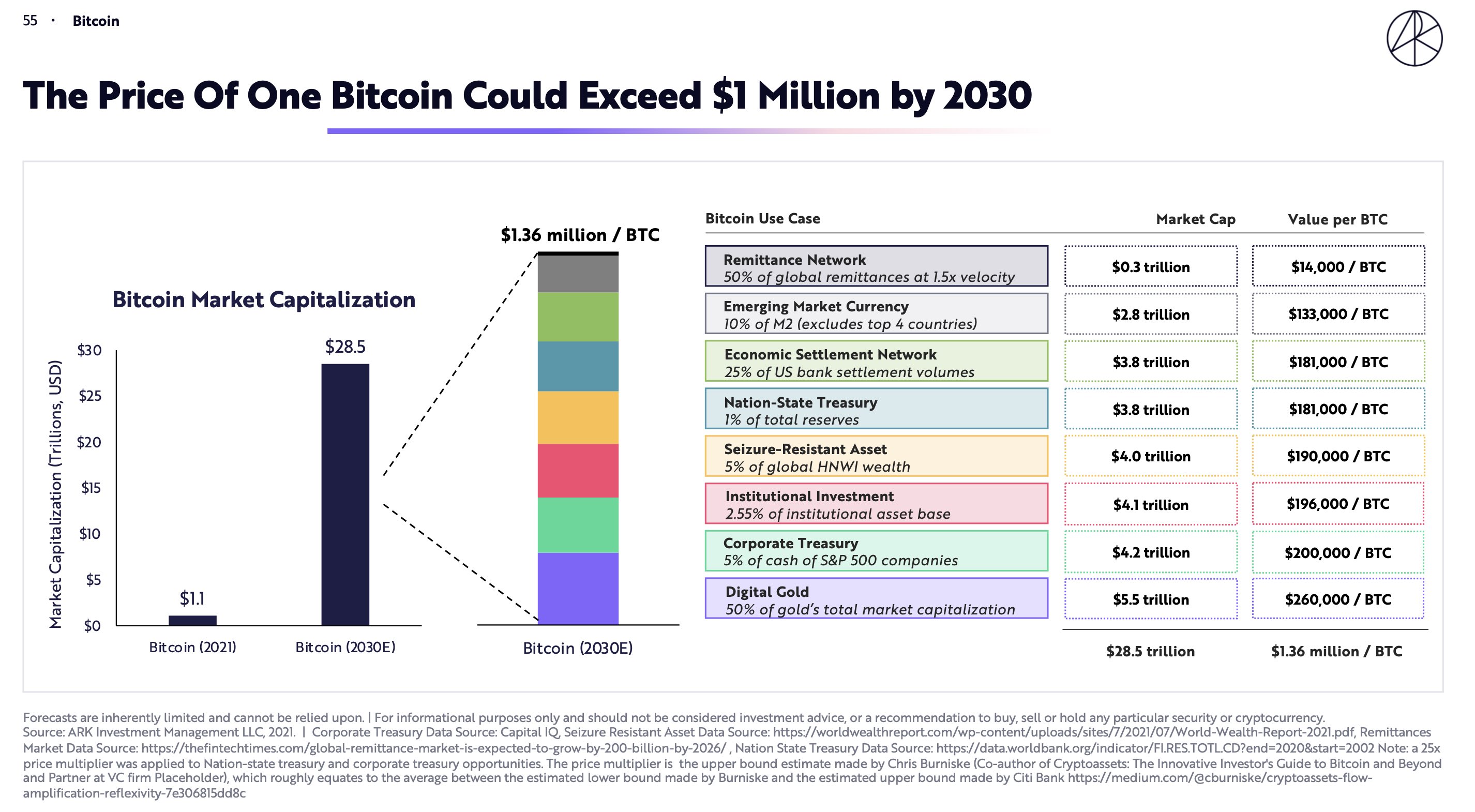}
  \caption{Potential market exposure to Bitcoin as a money}
  \label{fig:BitcoinShareOfMoney}
\end{figure}

The 2022 ARK Big Ideas report again provides some useful market insight. They posit that demand for the money features of Bitcoin could drive the price of the capped supply tokens to around 1M pounds per Bitcoin as in Figure \ref{fig:BitcoinShareOfMoney}. Take this with the usual pinch of salt, as Ark have been performing notably badly lately with their predictions.
Perhaps more than any of these takes, it is worth considering the current public perception of the technology as a money and store of value. This \href{https://twitter.com/saquon/status/1480738426236375041}{twitter thread} from professional sportsman Saquon Barkley, to his half million followers on the platform, captures the mood. He is one of a handful of athletes now being \href{https://www.buybitcoinworldwide.com/athletes/}{paid directly} in Bitcoin.\par
\textit{``I want my career earnings to last generations. The average NFL career is 3 years and inflation is real. Saving and preserving money over time is hard, no matter who you are.
In today’s world: How do we save? This is why I believe in bitcoin. Almost all professional athletes make the majority of their career earnings in their 20s. With a lack of education, inaccessible tools, and inflation, a sad yet common reality is many enter bankruptcy later on. We can do better. We need to improve financial literacy. Bitcoin is a proven, safe, global, and open system that allows anyone to save money. It is the most accessible asset we’ve ever seen.''}\par
This ubiquity of access is what probably most distinguishes Bitcoin. Previously it could be argued that only the most wealthy could access the `means' to store their labour without loss of value over time (through inflation). To be clear, inflation is an important part of the money system, somewhat within the control of the central banks, and approximate to taxation. It applies equally to all holders of the money supply. Asserting that money should be replaced by a `hard asset' such as Bitcoin, in the place of the more controllable utility of money, is likely both a fantasy, and wrong minded. This conflation of money and property is a confusion caused by Bitcoin's proximity to money, and it's `money like' network, and is extremely commonplace.\par
These narrative takes are all rooted in the popular idea that Bitcoin is a `hedge against inflation'; an increasingly fragile take, as the price plummets with global markets. The Bitcoin community seems somewhat confused about the nature of money, which is predictable because we can see in these sections that money is pretty confusing. Money is the fluid, elastic \cite{cagan1958demand}, and thin `working credit' layer on top of historical human production, which provides transaction convenience, and tools for credit. Value is effectively swapped in and out of this layer through the actions of central banks, controlling inflation into acceptable margins. Simplistically this is done through manipulation of interest rates (the easiness of credit), quantitative easing (buying of assets) and quantitative tightening (selling of assets). To give a quick high level view of central bank activity we can use the PESTLE framework:
\begin{itemize}
\item Political:
\begin{itemize}
\item Government policies and regulations can impact central banks by influencing monetary policies.
\item Political stability and government credibility can impact the confidence in central banks and currency.
\end{itemize}
\item Economic:
\begin{itemize}
\item Economic growth and inflation rates influence central bank decisions on monetary policy.
\item The level of debt and balance of payments can impact a central bank's ability to control monetary policy.
\end{itemize}
\item Sociocultural:
\begin{itemize}
\item Consumer behavior and demographics can influence demand for money and credit.
\item Attitudes towards saving and borrowing can impact the economy and central bank decisions.
\end{itemize}
\item Technological:
\begin{itemize}
\item Technological advancements can change the way money is distributed and managed, such as the increasing use of digital currencies.
\item Technology can also influence the accuracy of economic data, affecting the decisions of central banks.
\end{itemize}
\item Legal:
\begin{itemize}
\item Central banks are subject to laws and regulations, such as those related to banking and finance.
\item Changes in laws and regulations can impact central bank policies and operations.
\end{itemize}
\item Environmental:
\begin{itemize}
\item Environmental factors, such as natural disasters, can affect the economy and central bank decisions.
\item The focus on sustainability and reducing carbon emissions can impact the decisions of central banks.
\end{itemize}
\end{itemize}
Fiat money is primarily \textit{not} a long term store of value, as Austrian economists perhaps believe it should be. This function is left to assets. The Austrian thesis of `hard money' (which cannot be `debased' by government action) seems somewhat naive when one considers that if credit exists anywhere in the world (ie, the creation of paper money through loans) then this would be used to buy up a hard money asset in the long run, causing a scarcity crisis. This is what happened to gold in the middle of the last century. \par 
Fundamentally, Bitcoin isn't money (in the traditional sense) because it's not an IOU, which money certainly is. It's a bearer instrument, novel asset class, with money like properties, as identified above. As said again and again it functions most like a `property' which can be invested in by anyone, with all the attendant risks of that property class to the holder. Lyn Alden says it sits \href{https://www.lynalden.com/what-is-money/}{somewhere between} a saving tool, and an investment, acting as ``programmable commodity money''.\par 
\href{https://andrewmbailey.com/}{Andrew M. Bailey} says \textit{``in an ideal world where governments honour the rights of citizens, they don't spy, they don't prohibit transactions, they manage a sound money supply, and they make sound decisions, the value of bitcoin is very low; we're just not in an ideal world''}\par
Another potentially important differentiating affordance is censorship resistance. There's really nothing else like it for that one feature. With that said Bitcoin is only a viable `money like thing' when viewed in the layers described in this book, \href{https://giacomozucco.com/layers-before-bitcoin}{and elsewhere}\cite{Bhatia2021}. The base chain layer is an apex secure store of value. Whatever layer 2 ultimately emerges is the transactional layer which could replace day to day cash money, while the hypothetical layer 3 might be useful for complex financial mechanisms and contracts operating automatically, and also provides the opportunity for using the security model of the chain to support other digital assets, including government currencies through stablecoins. All these things have a natural home in borderless social spaces.
\section{Risks (money, not technical)}
Special thanks to economist Tim Millar for help with this section.
\subsection{Risks to Bitcoin the money}
\subsubsection{Geopolitics}
It can be seen that following the invasion of Ukraine by Russia, that sanctions of various kinds were applied to the Russian economy. One of these was the previously dicussed Swift international settlement network. Another whole catagory was the removal of support by private businesses domiciled outside of Russia and Ukraine, and pertinent here is that VISA, Mastercard, Paypal, and Western Union all removed support for their product rails. This means that while some cards and services still work, and will likely work again through Chinese proxies in the coming months, considerable disruption will be felt by Russian companies and individuals. This is not to say that this disruption is necessarily wrong, but it is clear now that all of these global financial transfer products and services are contingent on political factors. The same might be true of CBDC products if they gain traction globally. There is certainly no reason why all money within a physically delineated border could not be blocked or cancelled. This is not as true for Bitcoin at this time. \par 
However, with enough political will it is technically plausible to incentivise miners with additional payments to exclude transactions from geolocated wallets. This would be mitigated by Tor, and in a global anonymous network it is very likely that a miner could be found at a higher price for inclusion in the next block. \par
We have already seen much negative political positioning related to the energy concerns in an earlier chapter. There are similar noises coming from policy makers with regard to the money utility of the technology. The United Nations \href{https://unctad.org/system/files/official-document/presspb2022d8_en.pdf}{have made the following recommendations}:
\textit{``Developing countries may have less room to manoeuvre, yet the regulation of cryptocurrencies is possible. The following policies, among others, have the potential to curb the further spread of the risks of cryptocurrencies and stablecoins:
\begin{itemize}
\item Ensuring comprehensive financial regulation, through the following actions:
\begin{itemize}
\item Require the mandatory registration of crypto-exchanges and digital wallets and make the use of cryptocurrencies less attractive, for example by charging entry fees for crypto-exchanges and digital wallets and/or imposing financial transaction taxes on cryptocurrency trading;
\item Ban regulated financial institutions from holding stablecoins and cryptocurrencies or offering related products to clients;
\item Regulate decentralized finance (such finance may, in fact, not be fully decentralized, given its central management and ownership, which form an entry point for regulation);
\end{itemize}
\item Restricting or prohibiting the advertisement of crypto-exchanges and digital wallets in public spaces and on social media. This new type of virtual, and often disguised, advertisement requires policymakers to expand the scope of regulation beyond traditional media. This is an urgent need in terms of consumer protection in countries with low levels of financial literacy, as even limited exposure to cryptocurrencies may lead to significant losses;
\item Creating a public payment system to serve as a public good, such as a central bank digital currency. In the light of the regulatory and technological complexity of central bank digital currencies and the urgent need to provide safe, reliable and affordable payment systems, authorities could also examine other possibilities, including fast retail payment systems.
\end{itemize}
}
This is tough talk. We have seen that the IMF is willing to make their loans contingent on such regulation, and are increasingly \href{https://www.imf.org/en/News/Articles/2023/02/23/pr2351-imf-executive-board-discusses-elements-of-effective-policies-for-crypto-assets}{talking about banning} the technology. This global response to the technology is a significant headwind, but like the internet itself, it's very hard to actually stop these products being used.
\subsubsection{Capture by traditional finance}
As the popularity of Bitcoin continues to grow, traditional financial market incumbents have begun to take notice. In an effort to assert their dominance and protect their interests, these incumbents have turned to regulation and acquisition as means of capturing the growing markets. This is most clear in the 'alt coin' space where traditional banks have leveraged their knowledge and marketing to transfer money from retail investors into their own venture capital operations. This is not to say that Bitcoin is immune from these harms.\par
One way that traditional financial market incumbents have sought to capture the bitcoin market is through the use of regulatory frameworks. By working with government agencies (as described in previous chapters), to develop and implement regulations governing the use and trade of cryptocurrencies, these incumbents are able to limit competition and control the flow of capital into and out of the markets. They are also able to ``print paper bitcoin'', running a fractional reserve operation, as happened in the FTX/Alameda fiasco.\par
We have already described how, in the United States, the Securities and Exchange Commission (SEC) has implemented regulations governing the issuance and trading of bitcoin-based securities. These regulations, which require issuers of bitcoin-based securities to register with the SEC and comply with a variety of reporting and disclosure requirements, have effectively made it difficult for small and independent players to enter the market. \par
Another way that traditional financial market incumbents have sought to capture the bitcoin market is through the use of partnerships and acquisitions. As the newer companies stumble and fail as a result of poor risk management and over-leverage it seems that Wall Street incumbents like Goldman Sax are \href{https://www.reuters.com/technology/goldman-sachs-hunt-bargain-crypto-firms-after-ftx-fiasco-2022-12-06/}{taking advantage of the opportunity} at structural scale. By acquiring existing crypto companies, these incumbents are able to gain access to the technology, expertise, and customer base of these companies, giving them a significant advantage over their competitors.\par
For example, in 2017, the Chicago Mercantile Exchange (CME) partnered with the CBOE to launch bitcoin futures trading. This partnership allowed the CME and CBOE to tap into the growing market for bitcoin derivatives, while also providing a means for traditional financial market participants to gain exposure to bitcoin without having to hold the underlying asset. This is a crucial risk to the emerging technology as ownership of the underlying asset (self custody) was supposed to be the whole point of the technology. Ben Hunt of epsilon theory recently said: \textit{``..if you don't see that the crypto quote-unquote industry has become just as blindingly corrupt as the traditional Financial Services industry it was supposed to replace well you're just not paying attention what made Bitcoin special is nearly lost and what remains is a false and constructed narrative that exists in service to Wall Street in Washington rather than in resistance; the Bitcoin narrative must be renewed and that will change everything''}
\subsubsection{Liquidity Lottery}
Because holders of BTC are disincentived to sell the asset (assuming future gains) it is likely vulnerable to something \href{https://twitter.com/UrbanKaoboy/status/1526311908709502977}{Kao called the `liquidity lottery'}. This is a supply/demand mismatch which he thinks could spell the end of the asset class in time. Macro analyst group \href{https://doomberg.substack.com/}{`Doomberg'} believe that this mispricing of the asset is the significant risk, and point out that if Bitcoin is approached within the framework of government controlled Fiat, then there is no \href{https://en.wiktionary.org/wiki/there_is_no_there_there}{`there there'}. Bitcoin does not generate more fiat money within it's ecosystem (as say an energy extraction company would), and as such is very suggestive of the features of a Ponzi. They have recently softened on this view, and are now clear to separate Bitcoin from the wider `crypto' world, which they remain convinced are simply scams, wash trading magic beans without any productivity. The value is dependent on finding the `greater fool' mentioned near the start of the book. \href{https://doomberg.substack.com/p/dollars-ex-machina}{Doomberg assert} that the price of the asset has been inflated by manipulation in the unregulated stablecoin markets (specifically Tether), and in the event of a `run for the exits' there would be a serious repricing. This seems entirely possible, and perhaps even likely, below an unknown threshold of confidence. They are now asserting that if the manipulation and mispricing could be `washed out' of Bitcoin then it would present an investment opportunity, and they estimate that price at around \$3000. We think that the combination of global speed of exchange of value, generative AI, and bots which leverage the network to create value within the ecosystem of the network, that this thesis does not stand true, but there is no way to know for sure at this time.
\subsubsection{Manipulation of price or the network}
Bitcoin is still young and illiquid enough to be highly manipulable. Imagine for instance if a major organisation or nation state wished to accumulate a significant amount of the asset, but would prefer a lower price. \par
There is an unknown level of exposure to risk from centralised mining. If a few of the major mining pools were simultaneously infiltrated by a nation state actor then it might be possible to engineer a `deep re-org' of a large transaction. This would be dealt with quickly and almost certainly be a transient attack, but the damage to the narrative might be substantial. The proposed solution to this known vulnerability is called \href{https://braiins.com/stratum-v2}{`Stratum V2'} in which the transaction in the blocks would be organised by pool miners or their delegates, with an increase in efficiency as a driving incentive. A similar vulnerability exists in the centralisation at the level of internet service providers \cite{apostolaki2017hijacking}. This or some other flaw might lead to a selling cascade. Nobody knows just how vulnerable to selling cascades Bitcoin might be against a really serious challenge by an empowered actor, but it's already high volatility is suggestive of risk. 
\subsubsection{Rehypothecation}
\begin{figure}
  \centering
    \includegraphics[width=0.7\linewidth]{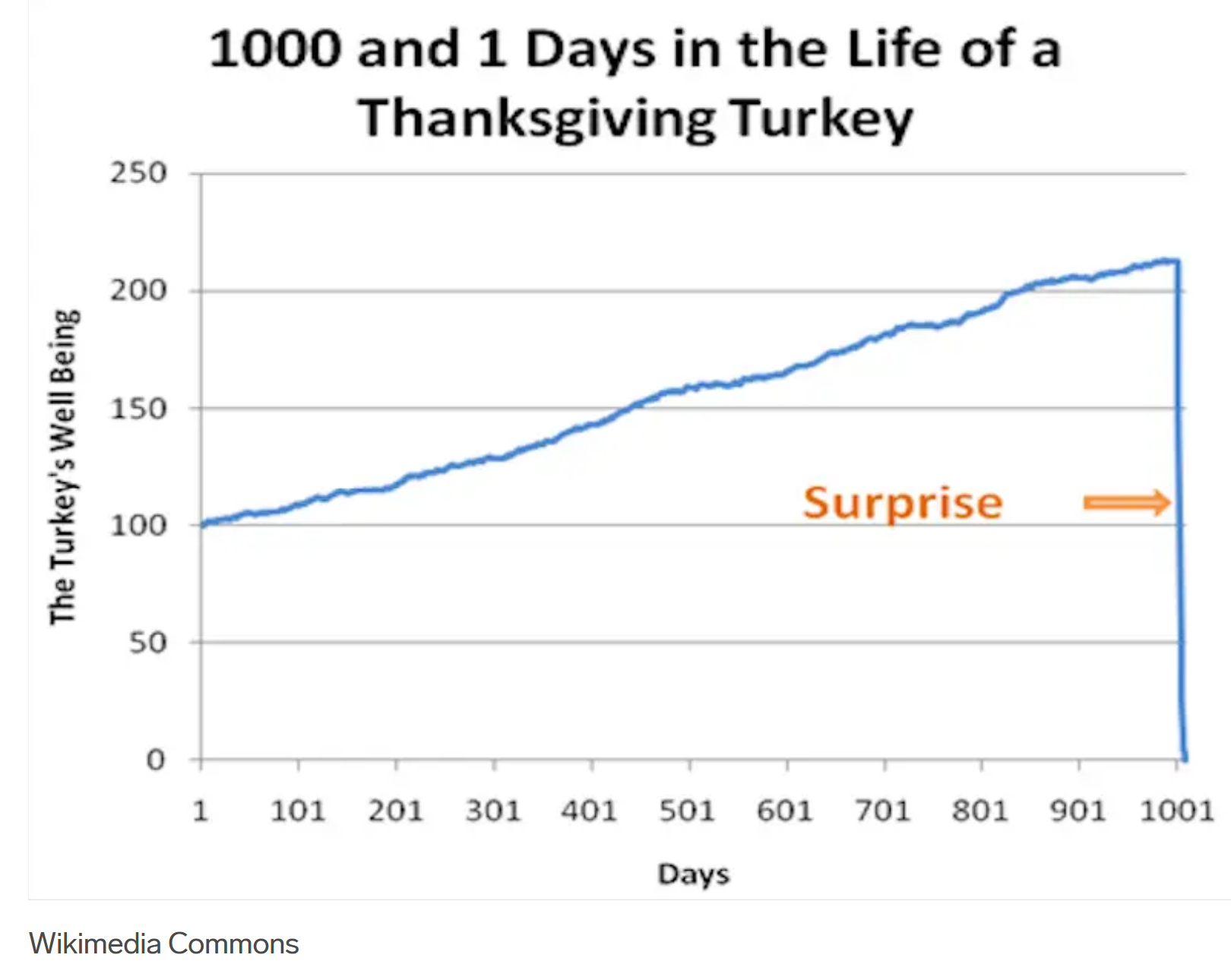}
  \caption{Nassim Taleb's Turkey Problem}
  \label{fig:talebturkey}
\end{figure}

It's vulnerable to rehypothecation (paper bitcoin managed by centralised entities running a fractional reserve). It seems that Figure \ref{fig:talebturkey} by  Nassim Taleb is a cautionary tale \cite{taleb2012antifragile}.
\subsubsection{Scale}
Scalability is always going to be a problem for Bitcoin, for all the reasons discussed in the blockchain chapter. There is no ``ready to go'' solution (except perhaps federations) that could onboard the whole world at this time because of the limited number of available UTXOs.\par 
Finally, a lack of fungibility, and privacy by default in Bitcoin, trends towards blacklists and over time this could seriously compromise the use of the asset. 
\subsubsection{Centralisation of the money over time}
In a medium term future it's possible to imagine a smart enough autonomous AI or ML actor managing to accrue Bitcoin through fast and smart `decisions'. This could unreasonably centralise the asset, and it would be impossible to claw this situation back. These constructs would last for the lifetime of the chain unless constrained by timelock multisigs for instance. 
\subsection{Bitcoin externalities}
This section is the risks that Bitcoin poses to external money systems, but it's worth pointing out that a risk to wider society is clearly \textit{also} a risk to Bitcoin itself.
\subsubsection{Inherent volatility}
\begin{figure}
  \centering
    \includegraphics[width=\linewidth]{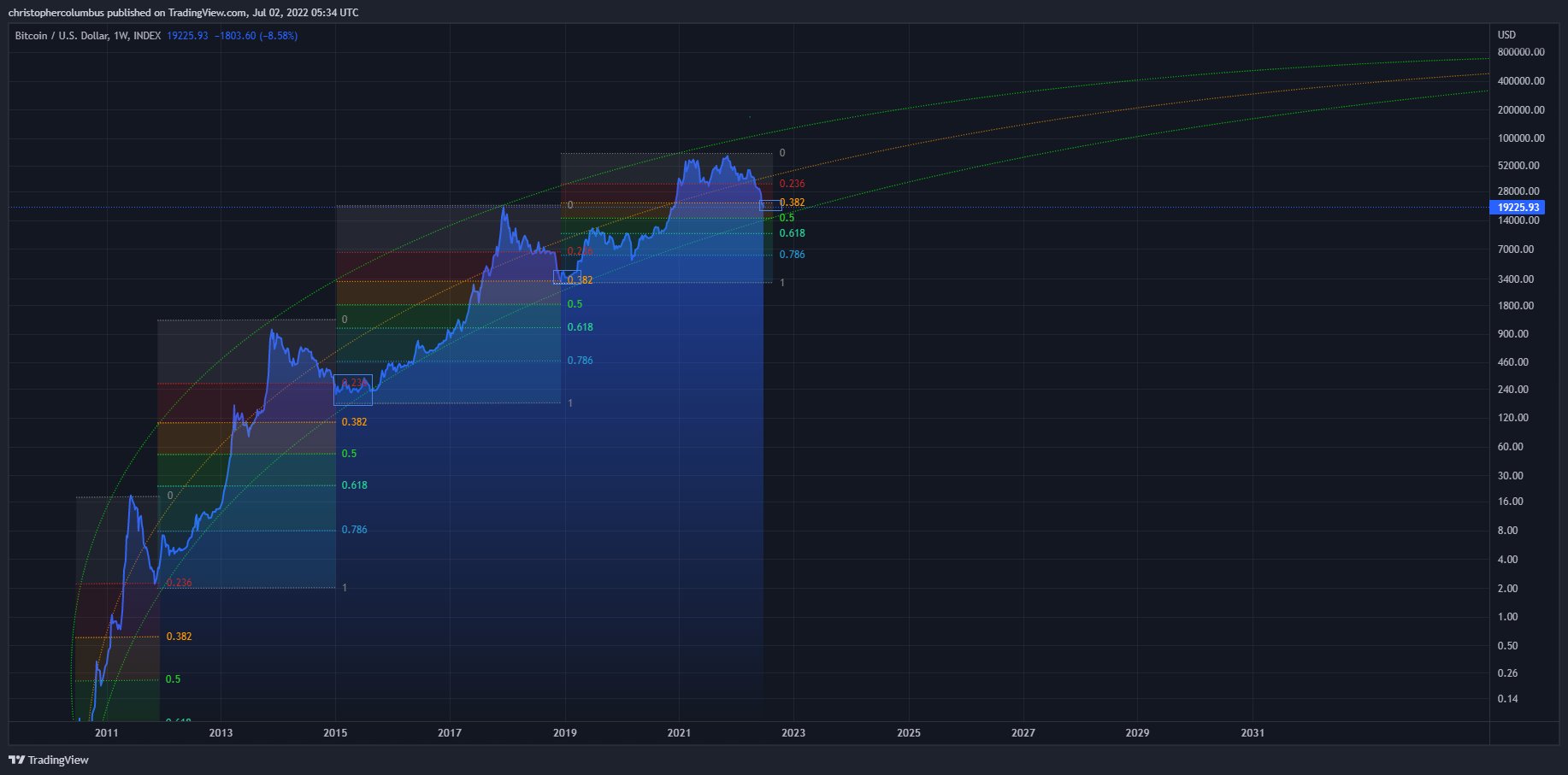}
  \caption{\href{https://davethewave.substack.com/p/cycle-theory-revisited?s=r}{Cycle theory revisited blog post} [Image used with permission]}
  \label{fig:davethewave}
\end{figure}

One of the better public analysts of the asset, \href{https://twitter.com/davthewave/status/1072441941390974982/photo/1}{sees the price} eventually fluctuating somewhere between ~\$700k and ~\$300k.  Figure \ref{fig:davethewave}. This is not how a money is supposed to work.  

Neither though is it the endless \href{https://stephanlivera.com/episode/147/}{``number go up''} that speculators have been promised. The aims of the project have a cognitive dissonance right at the core. The volatility trends toward:
\subsubsection{Unfair distribution}
\begin{figure}
  \centering
    \includegraphics[width=\linewidth]{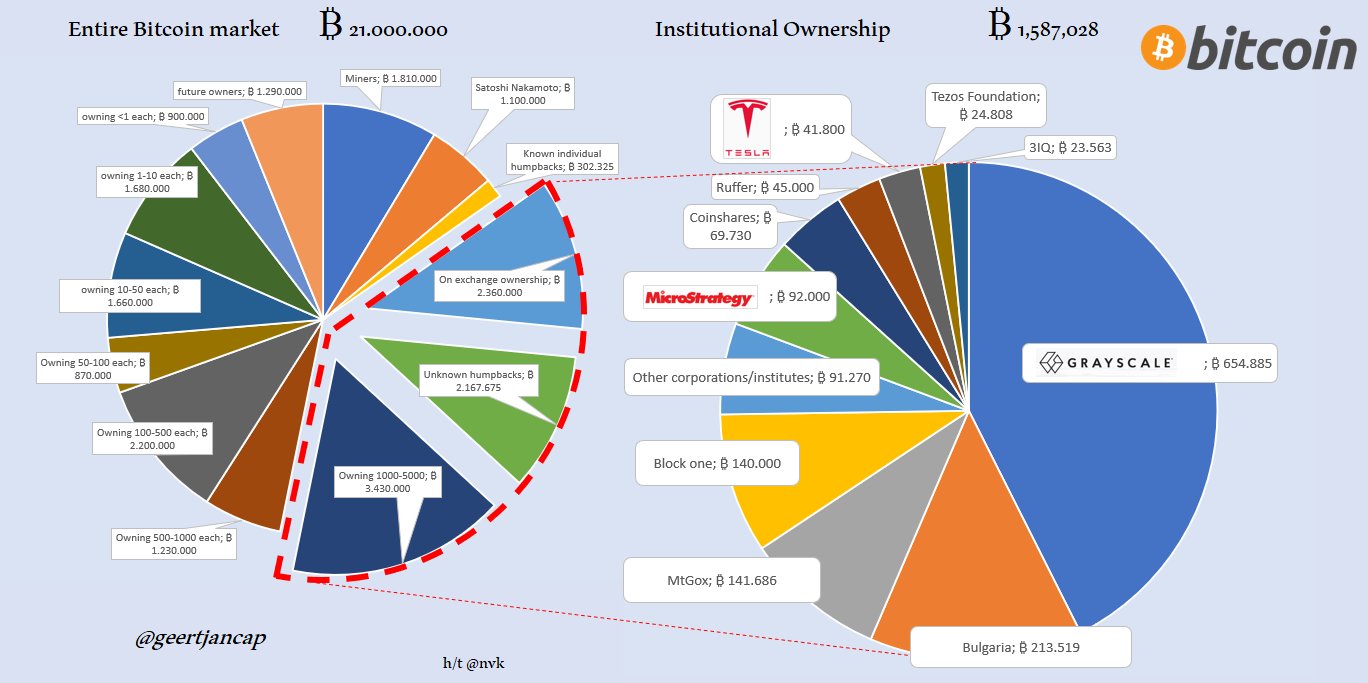}
  \caption{\href{https://twitter.com/Geertjancap/status/1380972132990136322/photo/1}{Bitcoin distribution is skewed to s few early holders, but it likely \textbf{is} fair.} [Image used with permission]}
  \label{fig:btcdistribution}
\end{figure}

By design the distribution of Bitcoin is likely `fair`, in that everyone has been able to access and secure the asset long term without prejudice. Figure \ref{fig:btcdistribution} from Twitter user @Geertjancap shows the distribution in 2021. Whether this is judged to be fair if the asset jumps to 10 times it's current value, minting a new class of hyper rich holders, is another matter. 
This pressure to emulate the early winners leads to:
\subsubsection{Endless HODL}
\begin{figure}
  \centering
    \includegraphics[width=0.5\linewidth]{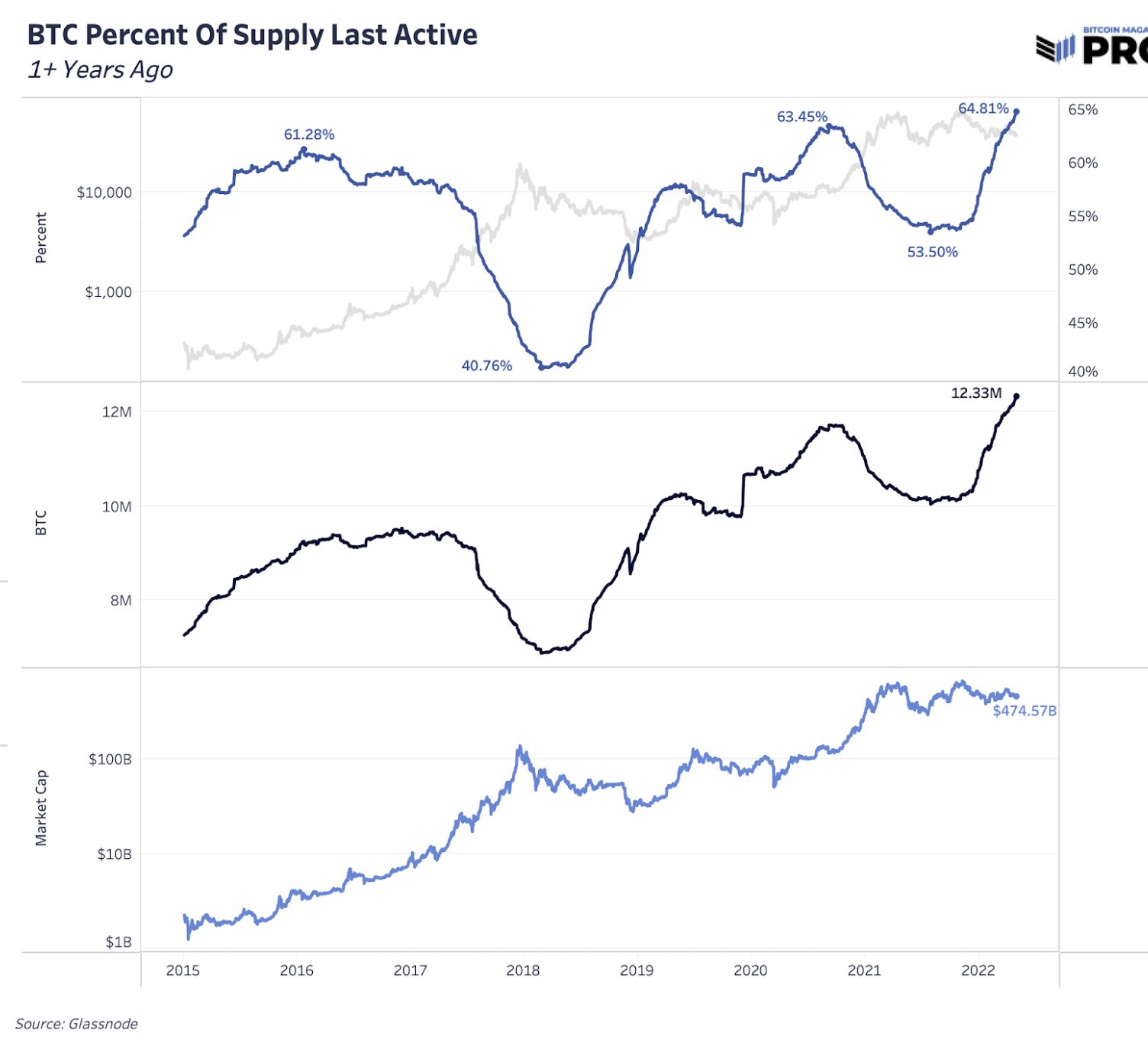}
  \caption{Supply of bitcoin that \href{https://en.macromicro.me/charts/32355/bitcoin-supply-last-active-1plus-years-ago}{hasn't moved} for over 1 year}
  \label{fig:notselling}
\end{figure}

It's possible that there's a problem with people not wanting to sell the asset, because they are predisposed to a particular fervour promoted within the community. This can be seen in the \href{https://en.macromicro.me/charts/32355/bitcoin-supply-last-active-1plus-years-ago}{glassnode data}, where the black line in Figure \ref{fig:notselling} shows that the asset held for more than a year (illiquid) has increased over the years.
There's real recalcitrance about using the asset as a money, which potentially negatively impacts the security model \cite{Wouters2022} and leads to:
\subsubsection{Reduction of funding source / liquidity in legacy finance}
In the current financial system remuneration for labour performed in the workforce is loaned into the money system, where it's put to work providing liquidity for creation of more opportunity. This system actually works pretty well. The more of this deferred labour that's taken out of the legacy system, the less work can be done with what remains. This isn't to say that Bitcoin will cause a liquidity crisis, but there is possibly a cost if the current trend continues. This isn't as bad as:
\subsubsection{Bitcoin collapse system shock}
In the event of an existential collapse of the Bitcoin network the erasure of so much capital would certainly have a contagion effect on the whole global financial system. It's hard to imagine what such an event could be, this being the nature of ``black swans''. One cited example is the unravelling of cryptography by quantum computing. Some conspiracy theorists in the past have even speculated that Bitcoin is itself a canary in the coal mine, engineered by the NSA to warn about emergent quantum computing somewhere in the world. It's all pretty silly because without cryptography Bitcoin would be the least of humanities problems. The risk of `something' does exist though. The same anti-fragile feature can't be said about the technologies around Bitcoin, which gives us:
\subsubsection{Stablecoin collapse system shock}
This is much more likely. Stablecoins are under regulated, centralised, under collateralised, ponzo like structures, which could quite clearly fall apart at any point. The contagion effects of this are unclear as they're not yet too significant. They're a risk nontheless, and may be an indicator of:
\subsubsection{Tech for techs sake yielding unexpected outcomes}
The whole question of what Bitcoin addresses, whether it's been properly thought about, what the end goals are, and what the risks are is significant. It's a computer science and engineering solutions gone completely wild. It's clearly got benefits and there's clearly human appetite for this technology, but it's probably running ahead of the knowledge base around it. This is most exemplified in:
\subsubsection{No agreed measurable end goal}
 Bitcoin is a game theoretic juggernaut, where success of the network breeds more success for the network. The was obviously a great design choice for the computer scientists trying to solve the problem of a secure, and scalable, electronic cash, which couldn't be confiscated. Ironically for a global consensus mechanism it seems that nobody wants to discuss what constitutes a successful end point to this, and especially not what `successful' endpoints for the game theory which have calamitous negative repercussions for wider society look like \cite{warren2023bitcoin}.  This might have implications for:
\subsubsection{National security / actual warfare}
There's some national security implications for Bitcoin which are discussed both in the \href{https://twitter.com/JasonPLowery/status/1512775981693648897?}{fringes} and the \href{https://www.coindesk.com/layer2/2022/04/04/why-bitcoin-mining-is-a-matter-of-national-security/}{sector media}. Essentially, the industrial mining complexes which are more commonplace now, are easily identifiable targets, and provide nations with both some leverage over the global network, and a considerable source of income. The IMF correctly identifies these facilities as a way for nation states to \href{https://www.imf.org/en/Publications/GFSR/Issues/2022/04/19/global-financial-stability-report-april-2022}{monetise their energy reserves} without the need for foreign markets, opening the door to sanction avoidance. In the case of smaller and developing nation states who are perhaps subject to financial penalties on the global stage for whatever reason, these facilities start to look like legitimate targets for cyber and conventional warfare. Lowry explain the potential strategic importance of Bitcoin in Softwar \cite{Lowery2023}, though to be clear his motives are unclear and his thesis is neither peer reviewed nor publicly accessible. This `weaponisation' of a neutral technology is already manifest in:
\subsubsection{Bitcoin as a culture war foil}
Bitcoin's online community skews very hard toward right wing libertarianism. This isn't to say there are no other voices, but they are certainly outnumbered. This imbalance is almost certainly a product of the ESG concerns around the technology. There has been a notable increase in diversity of thought since the evolution of the energy narrative, but it persists. This leads to a paucity of voices in policy making circles, and in the USA a strong delineation between policy makers along party lines. This kind of thing tends to be self reinforcing, and it seems very possible that the global liberal left will swing mainly against the technology, while the neoliberal right will be attracted more to it. As tensions increase so it seems does the online rhetoric. Even scientists now seem to agree that Bitcoin investors are calculating psychopaths \cite{martin2022dark}. This leads to:
\subsubsection{Self reinforcing monocultures}
There are some powerful `pockets' of fringe thinking within the \href{https://pourteaux.substack.com/p/bitcoin-culture-burn-it-to-the-ground?}{vocal, online, Bitcoin communities}. The \href{https://www.forbes.com/sites/peterizzo/2022/07/04/bitcoin-maximalism-is-dead-long-live-bitcoin-maximalism/?}{most palatable} of these are figures like \href{https://www.saylor.org/about/}{Michael Saylor}, Elon Musk and Jack Dorsey, but there's whole subcultural intersections around antivax, anti-woke, anti cancel culture, and fad diets. These are the so-called ``\href{https://blog.lopp.net/history-of-bitcoin-maximalism/}{toxic maximalists}''. There are a disproportionate number of adherents of the failed global ``neocon'' experiment \cite{va2010neoconservatism}, and not a few outright bigots. Lopp's list linked above is an amusing roundup:
\begin{itemize}
\item Carnivory
\item Laser eyes
\item Anti-woke
\item Weightlifting
\item Tradwife culture
\item Climate change denial
\item Overt Christian moralizing
\item ``Have fun staying poor'' retorts
\item Rejection of seed oils and sunscreen
\item Vaccine conspiracies, alt-health cure-alls
\item Contrarianism for the sake of contrarianism
\item Political populism and support of strongmen
\item ``Fiat'' criticism of contemporary art and architecture
\end{itemize}
As Lopp himself points out, it's not that these things are necessarily wrong or bad, but more that adherence to the set became a purity test for the whole space. It might seem that this isn't terribly important, but Bitcoin viewed though the lens of these of these communities looks pretty strange to the newcomer. The early adopters are just using their wealth to leave the battlefield behind using:
\subsubsection{Jurisdictional / legislative  arbitrage}
The reach of Bitcoin and it's ability to undercut the global money systems, delivering savings for those with a first mover advantage, and the current paucity of agreed legislation has set up an interesting and rare condition. Bitcoin encourages something called jurisdictional arbitrage; the race to take advantage of the variance in national approaches to the asset class. This section could perhaps be explored as a list of opportunities, but from the viewpoint of our SME business use case it's far more likely that these destabilising `features' are risks: 
\begin{itemize}
\item \textbf{Difference in `crypto' profit models}. Countries and jurisdictions can apply different charges for use of trading platforms and capital gains tax enjoys huge variance. Some countries are now competing to offer zero tax as a way to attract valuable tech mind share. 
\item \textbf{Income tax} is harder to monitor in a truly international context. This is variously pitched around the world.  It's hard to monitor this stuff and tax at source like with company employees wages, because it's basically designed to be hard to monitor. This results in:
\item \textbf{Passport perks}. Countries are already selling residence and company rights against Bitcoin marketing. There's a lot of new ways to buy passports and citizenship based on `inclusion' in this community now. It's a terrible look. The early adopters can live international jetsetter lifestyles and ca benefit from:
\item \textbf{Business subsidies} such as those appearing in Switzerland, \href{https://davisclute.medium.com/visiting-a-startup-city-in-honduras-73d9c026ee6d}{Hondoras}, El Salvador, Africa etc. This means a new divide is emerging since some countries are in instead applying:
\item \textbf{KYC/AML} rules which make onboarding into this technology harder. Currently there's a trend toward globally capturing information about people buying these assets, but it's effectively tech warfare now with engineers, rapidly producing tools to circumvent slow and varied legislation. The best example of this remains El Salvador, where Bitcoin is legal tender, and has perhaps kickstarted:
\item \textbf{Bond issuances}. El Salvador are having a \href{https://www.ft.com/content/4fa63c8c-51f5-4512-b522-76dd75e62916}{faltering start} to their promised bond issuance. It might be that all of this is a harbinger of the rise of: 
\item \textbf{The Network State} is a proposal by Srinivasan \cite{Srinivasan2022}. His is a transhumanist thesis which he describes: \textit{``The fundamental concept behind the network state is to assemble a digital community and organize it to crowdfund physical territory. But that territory is not in one place — it’s spread around the world, fully decentralized, hooked together by the internet for a common cause, much like Google’s offices or Bitcoin’s miners. And because every citizen has opted in, it’s a model for 100\% democracy rather than the minimum threshold of consent modeled by 51\% democracies.''}
\end{itemize}
\subsubsection{Hyperbitcoinization}
All of the above starts to look like convergence on something the crypto community regularly describes to itself within it's internal media. Hyperbitcoinization is a term coined in 2014 by Daniel Krawisz \cite{krawisz2014hyperbitcoinization}. It is the hypothetical rise of Bitcoin to become the global reserve currency, and the demonetisation of all other store of value assets. This seems unlikely but is hinted at in a game theoretic analysis of both Bitcoin and current macro economics. Again, Bitcoin is a likely very poor replacement for money. The ability to monetise assets through banks, backed by law and contracts (the debt based system), is a highly refined human concept, while Bitcoin is a fusion of Austrian economics, and a computer science project. The hyperbitcoinization idea finds it's ultimate expression in Svalholm's ``Everything Divided by 21 Million'', a hypothetical re-accounting of all human production into the Bitcoin ledger \cite{booth2022bitcoin}.\par
Nobody is sure what a \href{https://fredblog.stlouisfed.org/2022/07/inflation-and-deflation-with-a-fixed-money-supply/}{regular deflationary cycle} might do to global supply chains. Malherbe et al. point out the inherent unsuitability of a deflationary asset such as Bitcoin as the global reserve currency \cite{malherbe2019cryptocurrencies} and feel that perhaps other cryptocurrencies might be more suitable for adoption by governments.  Interestingly this is the only paper to reference `Duality' (the only thing purportedly written by Satoshi Nakamoto after they left the project). \par
Writer and activist Cory Doctorow is \href{https://onezero.medium.com/the-byzantine-premium-8411521db843}{not a fan of Bitcoin}. He provides an excellent summary of what he sees as the \href{https://doctorow.medium.com/finance-caused-the-fall-of-rome-fd091fa02973}{basic societal mistake} of the libertarian ideals around strong property rights and hard money. In a hyperbitcoinised world where debt law would be enforced by distributed code, it might be far harder to prevent the ``fall of Rome'' scenario he describes. It is notable that he is also \href{https://pluralistic.net/2023/03/09/autocomplete-worshippers/#the-real-ai-was-the-corporations-that-we-fought-along-the-way}{strongly opposed} to the current hype in AI and it's possible this is just his stock in trade.\par  
Fulgur Ventures (a venture capital firm) provide a \href{https://medium.com/@fulgur.ventures/the-roads-to-hyperbitcoinization-part-1-27dc84d0e5e5}{blog post series} about the route this might take. It's important to note that Budish suggested that the usefulness of Bitcoin (and blockchain) cannot exceed the cost to attack it. The is highly suggestive that hyperbitcoinisation is impossible \cite{budish2018economic}. It's beyond the scope of this book to look at the implications of all this. 

\section{Is DeFi any use?}
DeFi is decentralised finance, and might only exist because of partial regulatory capture of Bitcoin. If peer-to-peer Bitcoin secured yield and loans etc were allowed then it seems unlikely that the less secure and more convoluted DeFi products would have found a footing. DeFi  has been commonplace over the last couple years, growing from \href{https://a16zcrypto.com/state-of-crypto-report-a16z-2022/}{essentially zero to \$100B} over the last two or three. It enables trading of value, loans, and interest (yield) without onerous KYC. If Bitcoin's ethos is to develop at a slow and well checked rate, and Ethereum's ethos is to move fast and break things, then DeFi could best be described as throwing mud and hoping some sticks. A counter to this comes from Ross Stevens, head of NYDig \href{https://nydig.com/on-impossible-things-before-breakfast}{who says} \textit{``The concept of decentralized finance is powerful, noble, and worthy of a lifetime of focused effort.''}. This may be true in principle, but certainly isn't the case as things stand.\par
According to a recent JPMorgan industry insider report, around 40\% of the locked value on the Ethereum network is DeFi products. It is characterised by rapid innovation, huge yields for early adopters, incredibly high risk, and a culture of speculation which leads to products being discarded and/or forked into something else in the pursuit of returns. Ethereum also allows miners of the blockchain to cheat the system \cite{piet2022extracting}.\par 
Much of the space is now using arcane gamification of traditional financial tools, combined with memes, to promote what are essentially pyramid schemes. Scams are very commonplace. Loss of funds though code errors are perhaps even more prevalent.\par
The Bank for International Settlements have the stated aim of supporting central banks monetary and financial stability. Their \href{https://www.bis.org/publ/qtrpdf/r_qt2112b.pdf}{2021 report on DeFi} noted the following key problems.
\begin{itemize}
\item ..a ``decentralisation illusion'' in DeFi due to the inescapable need for centralised governance and the tendency of blockchain consensus mechanisms to concentrate power. DeFi`s inherent governance structures are the natural entry points for public policy.
\item DeFi’s vulnerabilities are severe because of high leverage, liquidity mismatches, built-in interconnectedness and the lack of shock-absorbing capacity.
\end{itemize}
These are two excellent and likely true points. European Parliament Vice President \href{https://cointelegraph.com/news/wef-2022-most-defi-protocols-aren-t-really-decentralized-says-european-parliament-vp?}{Eva Kaili made this same point} at the World Economic Forum, so clearly regulators are aware of the lack of meaningful distribution in DeFi. In addition access to DeFi is `usually' through web.0 centralised portals (websites) which are just as vulnerable to legal takedown orders as any other centralised technology. Given who the major investment players seem to be in this `new' financial landscape it seems very likely that regulatory capture is coming. The seemingly unironic trend towards CeDeFi (\href{https://www.nasdaq.com/articles/cedefi-what-it-is-and-why-it-matters}{centralised decentralised finance}) illustrates this.\par
With this said, it is notable that in the wake of the FTX debacle and unwinding of counter party risk across the whole extended ecosystem, it is DeFi which seems to have fared best, suggesting there might be a viable product here in the end. Circle and DeFi infrastructure lab Uniswap have recently published a paper which asserts that use of the technology could de-risk foreign exchange markets \cite{Adams2023}. It feels regrettably close to the endless broken promises of `blockchain for remittance' which have circulated for a decade \cite{sood2019implementation, bechtel2022future}. They estimate that it may be possible to cut the costs of cross border remittances by 80\% This is a big claim and time will tell.\par 
There are more recent DeFi on Bitcoin contenders, but these are vulnerable to the \href{https://bisq.community/t/trading-halted-until-v1-3-0-hotfix/9208}{same attacks} and problems in the main. \par 
There is likely no use for this technology for small and medium sized companies on the international stage, at least until the proposed Forex integrations appear. It is far more likely that reputation would be damaged. It's possible to \href{https://www.coindesk.com/layer2/2022/07/20/the-credit-crunch-is-not-the-end-of-crypto-lending/}{get loans} (by extension business loans) out of such systems at relatively low risks. The best `distributed' example of this is probably \href{https://lend.hodlhodl.com/}{Lend, at HODLHODL}, which is a peer-to-peer loan marketplace. \href{https://atomic.finance/blog/a-laypersons-guide-to-discreet-log-contracts-atomic-yield-series-part-3/}{Atomic Finance} leverages \href{https://adiabat.github.io/dlc.pdf}{discrete log contracts} amongst other more edge uses of Bitcoin, to provide financial services without custody of the users' Bitcoin. It is possible to make the argument that between hodlhodl loans, taro asset issuance, boltz exchange, and lightning escrow that all of the ``classes'' of DeFi smart contract can be serviced already by Bitcoin alone, but this tech is fringe at best.\par
Many more custodial options exist for loans (CASA, Nexo, Ledn, Abra etc). These might not really fit the definition of DeFi at all. Many of these centralised DeFi companies (CeDeFi) have imploded in the wake of the Terra/Luna collapse since they were generating yield from one another and ultimately Terra. The maxim seems to be that if you don't know how the system is monetised then you are likely the product. As mentioned, DeFi itself weathered the recent market turmoil comparatively well and it's possible that as these products evolve they may be useful to companies who have Bitcoin and stablecoins on their balance sheet long term. Dan Held maintains an \href{https://docs.google.com/spreadsheets/d/1ZoapTCl76wahFMeNISSx9UdC3QBx-zC_jY4Le1H5Sdg/htmlview#}{online spreadsheet} which compares these products.\par

\chapterimage{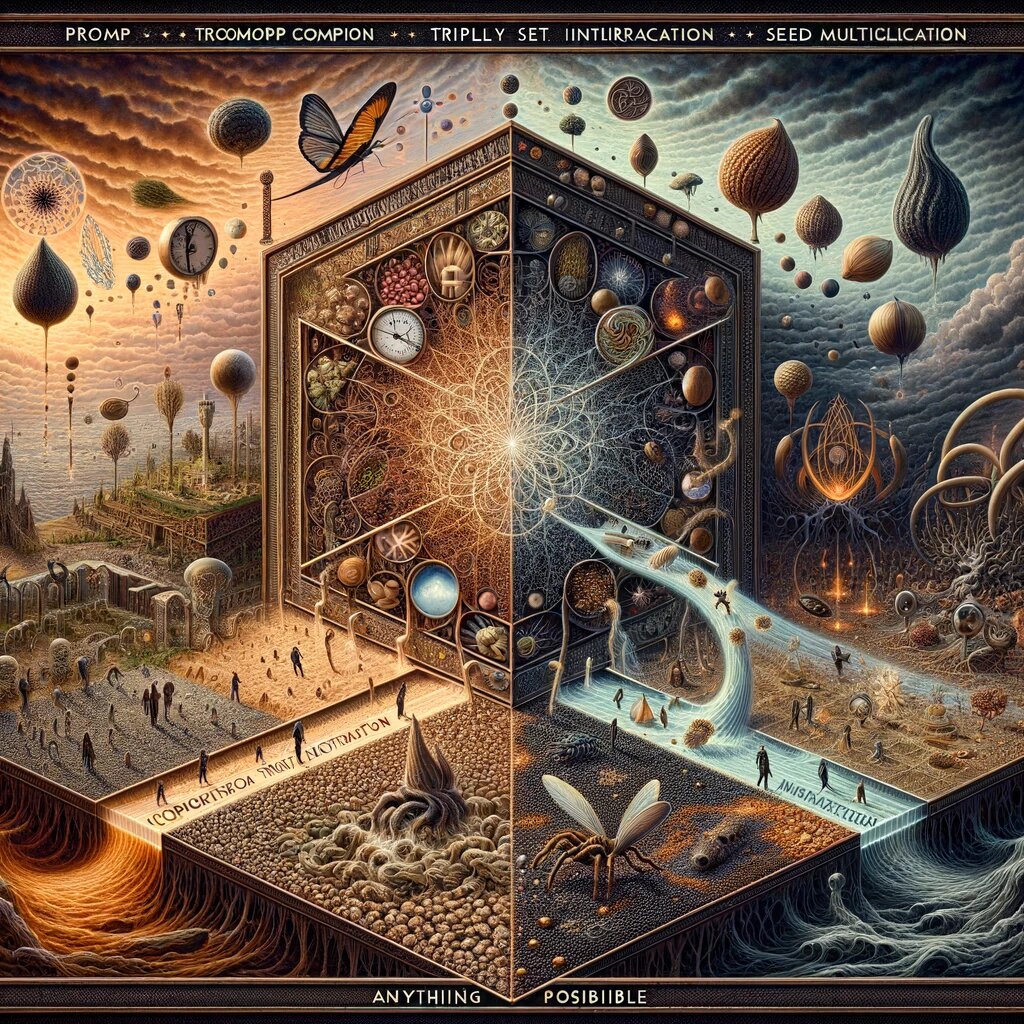}
In his latest book, Runciman, professor of Politics at Cambridge University, traces contemporary anxieties about artificial intelligence back centuries to the origins of the modern state and corporation. There are interesting an striking parallels between the apparatus of state, and the emergent field of AI. \par
In the 17th century, Thomas Hobbes described the ideal state as a kind of "automaton" - a human-made machine that could provide stability and security beyond fickle, emotional human politics. Later, the invention of the limited liability corporation allowed artificial entities to take on previously unthinkable risks and debts. Runciman argues that states and corporations function essentially as ``robots'' - artificial, human-made creations constructed to make decisions and take actions. Much like our worries about AI today, these entities were designed to take over certain tasks and responsibilities from human hands.\par
States and corporations have acquired immense, sometimes unchecked power, persisting and protecting themselves as emergent features of their creation. They remain fundamentally inhuman; they do not think, feel, or have a conscience as individual humans do. Runciman suggests that the story of the modern world is the story of handing over decision-making and control to these robots, AI and systemic alike.\par
Today, our prosperity, health, and safety depend deeply on these state and corporate machines. Yet Runciman warns they could also lead to catastrophe if we fail to maintain control. Their vast powers, from mass surveillance to nuclear weapons, remind us of their inhuman, robotic nature. Runciman argues we must focus not just on regulating new AI, but on democratizing and improving oversight over existing state and corporate "robots" we rely upon daily. More transparency, public input, and innovation in governance is needed to retain human agency. Though we created them, these powerful machines can take on a life of their own.\par
This chapter attempts to speak to these issues, and the wider global need to reassert control over the human condition. We will start by looking at global economics, then talk briefly about the global approaches to AI which are emerging, before exploring AI in detail in it's own chapter.\par
Malone, an ex central banking analyst now working in crypto, links across the last two chapters of blockchain, and money, \href{https://twitter.com/brendanpmalone/status/1628067806984937472}{in a Twitter thread}. He believes that policymakers should focus on the underlying problems in the financial system, rather than just focusing on crypto. He has a lot of appreciation for US policymakers worrying about risk in the financial system. Crypto gets  attention because it's an easy target, but Malone believes that the real problems are so much bigger. According to Malone, people want to hold USD money to store value and make payments. Most are familiar with cash and bank deposits, but there's actually a spectrum of assets of varying quality that act like money, as we saw in the previous chapter. These include Euro dollars, repo, commercial paper, and more. This is what people are talking about when they reference the shadow banking system - money moving around the financial system outside of traditional banks, primarily in non-banks. As an aside, the name `euro dollar' predates the Euro currency, and has nothing to do with it. The origins of the eurodollar market can be attributed to the Cold War in the 1950s. At that time, the Soviet Union and its Eastern European allies began depositing their US dollar holdings in European banks, primarily in London, to avoid the risk of their assets being frozen by the US government. These dollar-denominated deposits held outside the United States became known as eurodollars. Malone notes that some amount of shadow banking activity is good because it allows the money supply to be more reactive and expand and contract with economic activity, which helps fuel economic growth. However, the regulatory and political apparatus and the underlying systems weren't really designed for a system this large, opaque, and multi-dimensional. This was seen in 2008 and 2009, which was as much about shadow banking and financial plumbing as it was about subprime housing and complex derivatives. The same was seen in 2020 with COVID-19.\par
In times of crisis, people want to be able to freely convert whatever they are holding into something safer on the spectrum. Sadly, sometimes market liquidity isn't there, so the central banks and come to save the day, and this kicks the can down the road. The core issue is that people an institutions want to store capital in places they can't access due to technical, institutional, or geopolitical reasons. Sovereigns hold US treasuries, hedge funds and HRTs use repo, and we have seen that the crypto and Bitcoin economies have stablecoins.\par
Since 2008 and 2009, the Treasury Market has gotten significantly larger, more fragile, and more complex. Banks have even more restrictions on creating deposits, and the demand for safe assets keeps skyrocketing. On top of that, the geopolitical landscape has changed dramatically, with US sanctions and seizure of Russian USD assets. Malone notes that crypto is a response to these underlying problems. Although it is not perfect, it is getting better as people learn from past experiences and begin to build regulatory clarity. This issue of regulatory clarity leads us into this section of the book, which looks are implicit or explicit corruption of governance.\
As a uueful example; The New York Magazine article provided an in-depth interview with Gary Gensler, the head of the Securities and Exchange Commission, in which he shared his thoughts on the cryptocurrency industry. One of the key takeaways was his belief that all cryptocurrencies, except for Bitcoin, should be considered securities, as they involve relying on the work of others to give them value. Gensler is an ex banker, and an ambitious politician, with his eyes on bigger prizes. He openly courted the attention of the now disgraced top team at FTX which failed so spectacularly. His assertions have sparked controversy, as it raises questions about the feasibility of registering all tokens as securities, given the unique challenges posed by open-source protocols and the changing nature of blockchain technologies. Critics argue that Gensler's stance could harm innovation and capital formation, as companies and entrepreneurs may struggle to comply with onerous regulations or abandon their projects altogether. The current system simply doesn't fit this new self forming marketplace, and his implication seems to be that the legal end game here is the destruction of the invested capital, because of non compliance. This has led to frustration and concern among crypto advocates and investors, who worry about the impact of such policies on the industry's growth and development.\par
The discourse should be on the much more fundamental questions of the monetary system and fragility of past assumptions and their ability to predict what comes next. Even as these conversations happen, however, the Bitcoin and stable coin builders will keep building because they are not going to sit around and wait for solutions to be presented to them.
\section{Global politics \& digital society}
\subsection{Inequality as the driving force}
\subsubsection{Inequality on the Rise}

In Britain inequality has returned to levels not seen since the 1930s. After steadily rising between 1600 to 1913, Britain's wealth as a share of the global total peaked and then began falling until the end of the 1970s [ref required]. During this time, Britain became one of Europe's most equal countries, even without the support of its Empire [ref needed]. Some argue this relative equality enabled Britain's economic growth and international standing to keep pace with its European neighbours, despite the loss of imperial power [ref needed]. During this period there was much upheaval in global monetary systems. More recently we have seen that trust has diminished, and inequality has risen, with social media perhaps acting as an accelerator. 

\subsection{The Social Cost of Inequality}

Four decades later, the social impacts of rising inequality are becoming clear. Of the 14 million people living in poverty in Britain today, most are in working families [ref needed]. Upward mobility is declining, as the continued dominance of the privately educated elite in top jobs hinders meritocracy [The Gender Wage Gap Among University Vice Chancellors in the UK 
2022] . The lack of affordable housing and regulation in the rental market has led to increasing homelessness [ref needed]. And with the super-rich able to avoid taxes, the burden falls more heavily on lower income groups [ref needed].

\subsection{When Inequality Declines, Life Improves}

However, in societies that prioritize equality, life improves for all citizens. Infant mortality falls, lifespans lengthen, and population health increases [dorling, finland, ref]. Access to education rises, enabling greater social mobility [The Parenthood Effect on Gender Inequality 
2013 ]. With reduced poverty and homelessness, there is less crime and violence [ref needed].

\subsection{Tackling Inequality}

Dorling [oxford, reference] Tackling inequality requires recognizing that excessive wealth concentration is detrimental to social cohesion and national prosperity. A modicum of inequality may be inevitable, but the widening chasm between rich and poor in Britain has passed sustainable limits. With common purpose and political will, a more equitable path is possible. As inequality lessened for decades before, supportive policies enabled the rise of a thriving middle class [The Persistence in Gendering: Work-Family Policy in Britain since Beveridge 
]. By pursuing greater fairness once more, Britain can regain its balance.

\subsection{Anacyclosis}
It's interesting in the current global political moment to look briefly at Anacyclosis. This is a political theory attributed to the ancient Greek historian Polybius, which posits that political systems evolve in a cyclical manner. The theory is based on the observation that governments tend to progress through six stages, each corresponding to a specific form of governance: monarchy, tyranny, aristocracy, oligarchy, democracy, and ochlocracy (mob rule). These stages are organized into three pairs, with each pair consisting of a 'good' form of governance and its corresponding 'bad' form.\par
\begin{itemize}
\item Monarchy (benign) -> Tyranny (corrupt): Monarchy is the rule by a single individual, such as a king or queen, who is considered to be a wise and benevolent ruler. However, as the monarchy endures, there is a risk that the ruler becomes corrupted or that a less competent or tyrannical successor takes over. This leads to tyranny, the degenerate form of monarchy, where the ruler becomes oppressive and self-serving.
\item Aristocracy (benign) -> Oligarchy (corrupt): To counter the tyranny, a group of nobles or elites may overthrow the tyrant and establish an aristocracy, which is the rule by a select group of individuals who are considered wise and virtuous. Over time, the aristocracy may become more focused on their own interests and power, leading to an oligarchy. This is the degenerate form of aristocracy, where a small group of elites control the government for their own benefit.
\item Democracy (benign) -> Ochlocracy (corrupt): The populace, dissatisfied with the oppressive rule of the oligarchs, may rise up and establish a democracy, which is the rule by the majority of the people through voting and participation in the political process. Democracy has the potential to create a fair and representative system of governance. However, as the democratic process becomes more susceptible to demagoguery, populism, and factionalism, it can devolve into ochlocracy or mob rule, where the government is influenced or controlled by unruly masses.
\end{itemize}
According to Polybius, these stages form a continuous cycle, as one form of governance gives way to another, and each form eventually becomes corrupted and degenerates into its corresponding 'bad' form. The theory of anacyclosis suggests that political systems are inherently unstable, with each form of governance containing the seeds of its own destruction. 
\subsection{The World Economic Forum}
The World Economic Forum (WEF) is a non-governmental organization founded in 1971 by Klaus Schwab. It is well known for its annual meeting in Davos, Switzerland, where world leaders, CEOs, and various stakeholders gather to discuss global issues and potential solutions. Although the WEF does not have direct control over policymaking, its influence on global policy arises from its role as a platform for dialogue and idea exchange, as well as its ability to bring together influential individuals.\par
As unelected technocrats, the WEF's impact on global policy can be observed through these aspects:
\begin{itemize}
\item Convening power: The WEF's Davos meeting is a high-profile event that attracts prominent political figures, business executives, and other influential individuals. This ability to assemble people allows the WEF to initiate conversations on global issues, create networks, and establish connections among key players. These interactions can lead to ideas and initiatives that might eventually shape global policy.
\item Knowledge sharing and thought leadership: The WEF produces a range of publications, reports, and research that provide insights into various global challenges. By disseminating this knowledge, the WEF contributes to the broader understanding of complex issues and helps to inform policymaking by governments, businesses, and other organizations.
\item Agenda-setting: Through its conferences and publications, the WEF identifies and highlights emerging trends, risks, and opportunities, which can help to set the agenda for global policy discussions. By bringing attention to specific issues, the WEF can indirectly influence the priorities of governments and other decision-makers.
\item Public-private cooperation: The WEF actively promotes collaboration between the public and private sectors in addressing global challenges. By fostering partnerships and facilitating dialogue between these sectors, the WEF can help drive the development and implementation of policies that require cooperation between governments, businesses, and civil society.
\end{itemize}
Despite its influence, critics argue that the WEF's position as unelected technocrats raises concerns about the organization's legitimacy and accountability. They contend that the WEF's ability to shape global policy without being directly answerable to citizens can undermine democratic processes and result in policies that prioritize the interests of elites over the broader public. However, others argue that the WEF's role in facilitating dialogue and collaboration is essential for tackling complex global challenges that require coordinated action across sectors and borders.\par
Interesting for us the WEF recently released its annual \href{https://www3.weforum.org/docs/WEF_The_Global_Risks_Report_2022.pdf}{Global Risks Report}, which highlights various threats and challenges facing the world today, and which intersect with all of the narratives in this book. The report discusses issues related to cybersecurity, public trust, and social cohesion, and underscores the importance of a comprehensive approach to addressing these challenges.\par
The WEF's founder, Klaus Schwab, has previously argued for a ``great reset'' in society and the economy, which involves revamping various aspects of our lives, from education to social contracts and working conditions. This reset would require the construction of new foundations for economic and social systems.\par 
The WEF Global Risks Report 2022 focuses on five main categories, which are also part of their ``Great Narrative for Humankind'' initiative:
\begin{itemize}
\item Economy
\item Environment
\item Geopolitics
\item Society
\item Technology
\end{itemize}
The report emphasizes that the erosion of social cohesion has been a significant global issue since the start of the COVID-19 crisis. In addressing these challenges, the WEF suggests that public-private collaborations are necessary to ensure effective decision-making and to safeguard the future of humanity.\par
The report also highlights the increasing digital dependency that intensifies cyberthreats, as the WEF has long warned of the potential for a significant cyber pandemic. The rapid spread of a cyber attack with ``COVID-like characteristics'' could potentially cause more damage than any biological virus.\par
The WEF Global Risks Report 2022 delves further into the potential consequences of a cyber pandemic. In a section titled ``Shocks to Reflect Upon'' the report explores the possibility of a wide-ranging and costly attack that could lead to cascading failures in systemically important businesses and disrupt services, ultimately undermining digital transformation efforts made in recent year.\par
The report also emphasizes the need for governments to address cyberthreats and warns that without mitigation, the escalation of cyberwarfare and the disruption of societies could result in a loss of trust in governments' ability to act as digital stewards.\par
To better understand the risks associated with technology, the WEF report explores the concept of the fourth industrial revolution, which Schwab believes will lead to the fusion of our physical, biological, and digital identities. This fusion will be facilitated by technologies such as artificial intelligence, internet of things-enabled devices, edge computing, blockchain, and 5G. You can see they're examining similar things to this book.\par 
As explaining in this work, these technologies present numerous opportunities for businesses and societies, they also expose users to elevated and more pernicious forms of digital and cyber risk. The report also discusses the potential emergence of the metaverse, which could create new vulnerabilities for malicious actors by increasing the number of entry points for malware and data breaches, again a central theme of this text.\par
In light of these risks, the WEF report suggests that users will need to navigate security vulnerabilities inherent in complex technologies characterized by decentralization and a lack of structured guardrails or sophisticated onboarding infrastructure.\par
The report also touches on the issue of digital identity as we do. They view digital identity is a crucial component of accessing products, services, and information in a digital world, but again, this raises concerns about privacy, security, and the potential for misuse.\par
\begin{figure}
  \centering
    \includegraphics[width=0.9\linewidth]{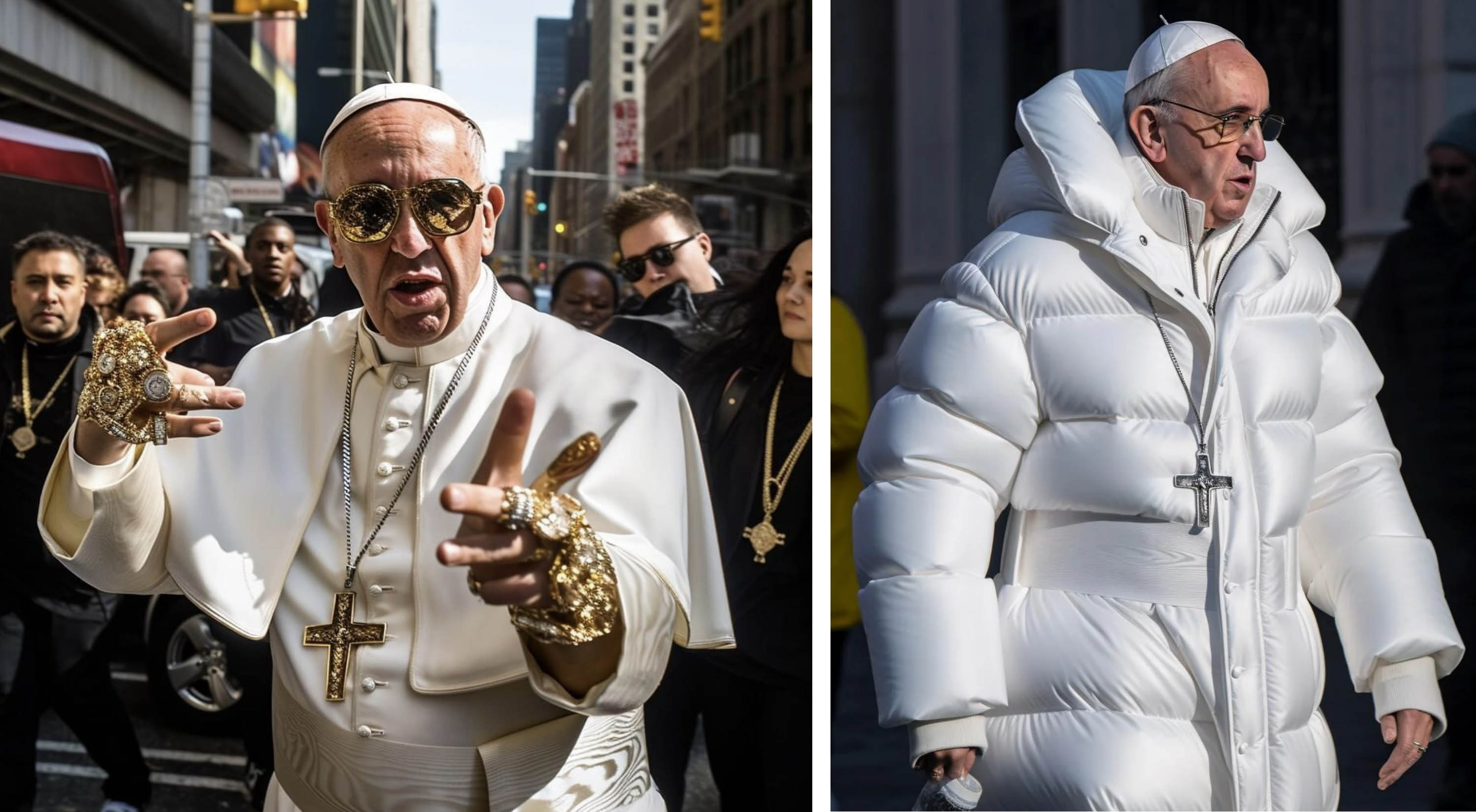}
  \caption{Midjourney 5 fake images of The Pope Francis which are \href{https://www.reddit.com/r/midjourney/comments/120vhdc/the_pope_drip/}{circulating as memes} and show the power and the danger of the technology even at this early stage.}
  \label{fig:pope}
\end{figure}

Finally, the WEF Global Risks Report 2022 addresses the issue of public trust, noting that the growth of deepfakes and disinformation-for-hire can deepen mistrust between societies, businesses, and governments. We can already see this starting to happen as Musk's defence lawyers \href{https://www.theguardian.com/technology/2023/apr/27/elon-musks-statements-could-be-deepfakes-tesla-defence-lawyers-tell-court}{point to possible deepfake use} around video comments he is alleged to have made with regard to Tesla's software safety. To rebuild trust and social cohesion, the report calls for leaders to adopt new models, look long term, renew cooperation, and act systemically. Quite what they think they can do in the face of images like the recent memes of the Pope is unclear \ref{fig:pope} .\par 
It's absolutely crucial to note that the WEF is a powerful organisation, with global sway over policy, and is an enormous concentration of power in the hands of unelected technocrats. The authors are very sceptical of the WEF, but this report highlights what both technocrats and policy makers are thinking.

\subsection{Money and The State}
It seems a pretty reasonable that the best `systemic' approach is a separation between major centralising forces such as state, church, and money. In practice we can see that globally, this isn't the case, with bad hotspots of high corruption where all three meld together into kleptocratic dictatorships, or theocracies. For our purposes in the UK it's useful to look at the concept of `austerity'.\par
Austerity is a term used to describe a set of economic policies that aim to reduce government spending and debt, often through cuts to public services and welfare programs. The concept of austerity has its origins in the 1920s, following the end of World War I and the economic crisis that ensued. In the wake of the war, many Western European countries were struggling with high levels of debt and inflation. In response, governments began implementing policies to reduce spending and balance their budgets.\par
We have seen in the previous chapter that the concept of inflation itself is complex, and somewhat argued about still. Globally, on aggregate, the efficiencies of increasing technology are thought to be deflationary to the tune of between 3 and 5 percent annually, though this may radically spike up in the era of AI which will be covered later. This is counter to the current need for inflation to maintain debt repayments at a national level. Central banks manipulate interest rates to control inflation, aiming to keep it at sustainable levels. This process is necessary because as national debt and deficits grow, governments need inflation to prevent these debts from spiraling out of control. Higher inflation results in higher nominal GDP, which in turn increases the tax base, providing governments with the revenue needed to pay down debt. To achieve this. The natural progression of humanity inherently deflationary, which forces central banks to print more money and further manipulate the monetary system in order to generate the desired inflation. This can be seen as a hidden tax on citizens, as it devalues their money over time. The negative effects of this system are disproportionately felt by lower-income groups. As inflation rises, the cost of living increases, and many households struggle to make ends meet. This has led to a situation where households need multiple incomes to maintain their standard of living, forcing individuals to work longer hours and take on multiple jobs. As a result, people have less free time and energy to engage in rewarding activities or spend time with their families. This need for constant economic growth, as measured by GDP, has led to an environment where individuals are pushed to be more productive at the expense of their well-being. This has resulted in a society where many people are overworked and struggling to keep up with the rising cost of living. Booth discussed this at length in his book `The Price of Tomorrow'. His is a rare thesis based around the ideas that technology is deflationary, that the marginal cost of goods trends over zero over time, and that the current system of debt and inflation are inherently unsustainable in the face of exponential technology improvements and automation. We discuss the concept of inflation and deflation, and both their risks throughout the book, but Booth has been very clear on this for many years. He thinks the current global monetary system ill-suited to handle the challenges and opportunities presented by deflation. He suggests that embracing deflation is the key to unlocking a prosperous and sustainable future. The book delves into the implications of deflation on various aspects of society, including wealth distribution, job markets, and the role of governments in shaping economic policies. \cite{booth2020price}.\par
In the 1920s, Keynes was one of the first to argue against the austerity measures which seem part of the cyclical playbook around debt and inflation. He argued that that cutting government spending during a recession would only worsen the economic downturn. Instead, he advocated for increased government spending to stimulate economic growth and reduce unemployment. Despite this, many governments continued to implement austerity policies throughout the 1920s and 1930s.\par
In the post-World War II period, the rise of the welfare state and the adoption of Keynesian economic policies led to a shift away from austerity in many countries. However, in the 1970s, a new economic crisis led to a resurgence of austerity policies, particularly in the United States and United Kingdom. In the 1980s, the rise of neoliberalism and the influence of economists such as Milton Friedman led to further cuts to government spending and the rolling back of the welfare state.\par 
Today, the concept of austerity continues to shape economic policy, particularly in the wake of the 2008 financial crisis. Many governments, particularly in Europe, have implemented austerity measures in response to the crisis, leading to cuts to public services and welfare programs. The effectiveness of these policies remains a contentious issue, with some arguing that they have helped to reduce debt and stabilize economies, while others argue that they have led to increased inequality and hindered economic growth. Looking around at the state of the world, and the widening gap between the rich and the poor, it is possible to have some sympathy with those who see patterns in the bahaviour of political leaders and the controllers of Western capital and global resources. The system seems engineered to reward a few. It is possible to view `austerity' as a means of political control of economic levers, in order to de-democratise populations. This mantra of `do more, consume less' has perhaps become a defacto methodology to constrain popular ideas, diverting capital back into the hands of incumbents, land owners, and the politically and economically motivated \cite{mattei2022capital}. It seems that the controlling nexus of this political framework globally is the concept of the central bank, unelected technocrats whose tenures span across political administrations. Again, this can be traced back to the 1920's. Hawtrey’s 1925 ``Currency \& Public Administration'' asserts that a central bank should \textit{``Never explain; never regret; never apologise.''}, and speaks glowingly of the selfish market \cite{hawtrey1925currency}. This economic model is referred to as Dirigisme and feels increasingly the global norm \cite{balassa2013theory}.  We can perhaps here see the divergent point at which the lionization of the market began. Again, to be clear, the authors are not economists, but it does seem that in a global digital society there is room to explore more equitable models of global value, governance, and trust.\par
Remember that these centrally planned national and global actions provide liquidity to the private banking sector. Like the digital money analogues discussed earlier in the book private banks operate fractional reserve banking. This is a banking system where banks hold only a fraction of the deposits they receive as reserves, while the rest is lent out to customers. This means that the money supply in an economy can be increased through the lending activities of banks (itself a complex inflationary force which devalues money over time, feeding back into the policy directives of the central banks. The fractional reserve system is useful for capital creation in times of growth, but relies on the confidence of the depositors. Historical examples of bank runs which threatened systemic risk or caused failures of the banking system include:
\begin{itemize}
\item The Bank of United States crisis in the 1930s: This was the largest bank failure in American history and was a result of a bank run caused by rumors of financial mismanagement.
\item The Savings and Loan crisis of the 1980s: This was a result of a large number of failed savings and loan associations in the United States, which were caused by a combination of factors including poor management, risky lending practices, and a decline in real estate values.
\item The Nordic banking crisis of the 1990s: This crisis was caused by a combination of factors including a real estate bubble, high levels of debt, and a lack of regulation. It resulted in the collapse of several major banks in Sweden, Finland, and Norway, and had a significant impact on the economies of the region.
\item The Bank of Japan crisis in the late 1990s: This crisis was caused by a combination of factors including a real estate bubble, high levels of debt, and a lack of regulation. It resulted in the collapse of several major banks and had a significant impact on the Japanese economy.
\item The Asian Financial Crisis of 1997: This crisis was triggered by a devaluation of the Thai baht and quickly spread throughout the region, causing a number of major banks to fail. The crisis was largely a result of a lack of transparency and poor regulation in the banking industry.
\item The 2008 financial crisis in Iceland: This crisis was caused by the collapse of the country's three largest banks, which had been engaging in risky lending practices and had accumulated large amounts of debt. The crisis had a devastating impact on the Icelandic economy and resulted in a severe recession.
\item The Global Financial Crisis of 2007-2009: This was a result of a widespread failure of the global banking system, caused by a combination of factors including the housing market collapse, risky lending practices, and a lack of regulation.
\item The collapse of Banco Popular in Spain in 2017: This was one of the largest bank failures in European history, and was caused by a combination of factors including a large amount of bad debt and a declining real estate market.
\item There were many bank runs on smaller rural banks in China during 2022. The financial conditions of Chinese banks are somewhat reminiscent of the 2008 American landscape.
\end{itemize}
In response to the Global Financial Crisis, many measures have been taken to shore up the banking system, including the creation of new regulatory bodies, the implementation of new regulations, such as the Dodd-Frank Wall Street Reform and Consumer Protection Act, which increased the regulatory oversight of the banking industry. The introduction of stress testing for banks, to ensure that they have enough capital to withstand financial shocks, globally, has radically deleveraged banks from around 1:40 fractional reserve, to around 1:10.\par 
There is increased political pressure to regulate the banking industry and prevent another financial crisis. However, there is also political opposition to excessive regulation, as some argue that it may stifle economic growth. There are concerns about rising levels of debt and the potential for another financial crisis.\par 
It's interesting that Brett, a former FDIC regulator \href{https://blog.orchid.com/exfdic-regulator-on-trust-and-the-battle-of-the-social-media-videos/}{believes that} the 2008 US bank run was sparked by youtube posts of queues forming at banks. He says those that formed the initial lines carried memories of the great depression, but that once youtube started showing the footage more broadly the contagion struck. In the world of instant messaging media today we can perhaps see how this might happen again. More recently, the 2023 `wobble' in global banking caused by the collapse of America's 5th largest \href{https://theconversation.com/why-svb-and-signature-bank-failed-so-fast-and-the-us-banking-crisis-isnt-over-yet-201737}{bank SVB} has precipitated strong intervention by the federal government, who have opted to `backstop' investor deposits. In the midst of this potential crisis it it notable that TikTok (now arguably the world's \href{https://blog.cloudflare.com/popular-domains-year-in-review-2021/}{most popular search engine}) is carrying millions of hashtag references to \href{https://www.tiktok.com/tag/bankrun?lang=en}{bankruns}. Senator Kelly in the USA \href{https://public.substack.com/p/exclusive-senator-mark-kelly-called}{allegedly inquired} about the potential for limiting such references on social media, and a UK minister is \href{https://news.sky.com/story/tiktok-ban-minister-asks-national-cyber-security-centre-to-look-into-safety-of-app-12833371}{asking for security services} to examine the risks of the Chinese application. The perhaps reflects concern about algorithmically driven geopolitically motivated threats to the banking system.\par
There is a growing awareness of the role of banks in the economy, and a growing desire for greater transparency and accountability. There is also a growing mistrust of banks, particularly in light of the Global Financial Crisis. As we have seen, the advent of new technologies, such as blockchain CBDC, and fintech, is changing the way that banks operate and interact with customers. This presents both opportunities and challenges for the banking industry. As a final controversial aside, there is \href{https://apnews.com/article/signature-bank-fdic-barney-frank-silicon-valley-6ad86262d9945675a42d735b66ace4f2}{industry suspicion} that the collapse of SVB has been used as cover to close the final US bank servicing crypto, effectively decapitating the banking rails of the industry, and forcing it overseas. Were it not for the credibility of the people making these claims, this would seem pretty wild, but the prevailing winds are surely blowing against the disruptive potential of a money system which is beyond the control of legislators.
\subsection{Surveillance Capitalism}
Surveillance capitalism is a term coined by Harvard Business School professor Shoshana Zuboff to describe the business model of using data collected from individuals to target advertising and influence behavior. The concept of surveillance capitalism emerged in the late 20th and early 21st centuries with the rise of technology companies that specialize in gathering and analyzing personal data.\par
The history of surveillance capitalism can be traced back to the early days of the internet. In the 1990s, companies such as DoubleClick and Omniture began collecting data on internet users' browsing habits in order to target advertising. As the internet grew in popularity, these companies were able to gather an increasing amount of data on individuals, allowing them to more effectively target advertising and increase profits.\par
The advent of smart phones and mobile technology in the 2000s further expanded the reach of surveillance capitalism. With the widespread adoption of smart phones and mobile apps, companies were able to collect even more data on individuals, including location data and information about their physical activity. This data was used to target advertising and influence behavior, leading to the rise of companies such as Google and Facebook, which have become dominant players in the digital advertising market.\par
The use of data collected from individuals to influence behaviour has also been used to influence political campaigns. In the 2016 US presidential election, Cambridge Analytica, a data analytics firm, used data collected from Facebook users to influence voter behaviour. The firm used the data to target advertising and create psychological profiles of individuals, allowing them to more effectively influence voter behaviour.\par
The business model of surveillance capitalism has been widely criticized for its ethical implications. Critics argue that the collection and use of personal data without consent is a violation of individuals' privacy and that the use of data to influence behaviour is manipulative and unethical. In recent years, there have been calls for greater regulation of the tech industry to address these concerns.\par 
Surveillance capitalism has led to significant compliance overheads for companies that collect and use personal data. There are a number of laws and regulations that have been put in place to protect individuals' privacy, such as the General Data Protection Regulation (GDPR) in the European Union and the California Consumer Privacy Act (CCPA) in California, USA. These laws require companies to obtain consent from individuals before collecting and using their data, and to provide individuals with the right to access, correct, and delete their data.\par 
Complying with these laws can be costly and time-consuming for companies. They may need to hire additional staff to handle data privacy compliance, and may also need to invest in new technology to manage and protect personal data. In addition, companies are at risk of significant fines if they fail to comply with these laws.\par
In terms of who profits from surveillance capitalism, the primary beneficiaries are technology companies such as Google and Facebook, which have become dominant players in the digital advertising market. These companies collect and analyse large amounts of personal data, which they use to target advertising and influence behavior. This allows them to generate significant profits from advertising revenue.\par
On the other hand, those who suffer the most negative impact from surveillance capitalism are individuals, whose personal data is collected and used without their consent. They are also at risk of their privacy being violated, and their personal data being misused. Additionally, the collection and use of personal data can lead to the manipulation of individuals' behaviour and decision-making, which can have negative consequences for their lives and society at large.\par
Moreover, the business model of surveillance capitalism has also been criticized for creating a power imbalance between companies and individuals. Companies have access to vast amounts of personal data, which they can use to influence behavior and make decisions that affect individuals' lives. This can lead to a lack of privacy and autonomy for individuals, and can also lead to discrimination and bias in decision-making.\par 
This is collectively an erosion of the demarcation between data, state surveillance, banking, and political leadership globally.\par
The term "surveillance state" refers to a state in which government agencies have the power to collect and analyze large amounts of personal data, often without the consent of individuals. The rise of surveillance capitalism has led to concerns about the potential for the creation of a surveillance state, as government agencies may use the data collected by companies for surveillance purposes.\par 
There have been instances where government agencies have used data collected by companies for surveillance purposes. For example, in the United States, the National Security Agency (NSA) has been accused of using data collected by companies such as Google and Facebook for surveillance purposes. The agency's PRISM program, which was revealed by Edward Snowden in 2013, was designed to collect and analyze data from internet companies in order to identify and track individuals. Europe is \href{https://www.patrick-breyer.de/en/posts/chat-control/}{clear about it's intentions} to mandate their complete access to all encrypted personal communications in forthcoming legislation.\par
The use of data collected by companies for surveillance purposes can have significant implications for individuals' privacy and civil liberties. It can also lead to a lack of transparency and accountability, as government agencies may use the data without the knowledge or consent of individuals. In addition, the use of data for surveillance purposes can lead to discrimination and bias in decision-making, as well as a chilling effect on free speech and the exercise of other rights.\par
Akten has been \href{https://memoakten.medium.com/all-watched-over-by-machines-of-loving-grace-8c2464aa6fda}{talking about} the phase transition from digital surveillance to pernicious corporate AI in terms of a modern `religion' for many years \cite{bayer2023artificial}. He feels that despite public awareness of privacy invasion, there has been no significant outcry or unanimous demand for privacy. Instead, most individuals seem to find comfort in the belief that a higher force is watching and protecting the virtuous, while punishing wrongdoers. The concept of a `digital deity' emerge from his thinking in this context, reflecting the role that religion and traditional gods have played in providing ethical frameworks, security, discipline, power, and other societal functions. More recently O'Gieblyn has been drawing the same conclusions \cite{o2021god}, explicitly linking religiosity to the imperative to `create a godhead' simply because it can be done, not pausing to discuss if it should be. Rosenberg calls this `a threat to Epistemic Agency' \cite{rosenbergmanipulation}More recently the Harari, author of Sapiens \cite{harari2014sapiens} \href{https://forumlive.frontiersin.org/agenda/speakers/2977577}{said of AI}: \textit{``For thousands of years, prophets and poets and politicians have used language and storytelling in order to manipulate and to control people and to reshape society. Now AI is likely to be able to do it. And once it can... it doesn't need to send killer robots to shoot us. It can get humans to pull the trigger. We need to act quickly before AI gets out of our control. Drug companies cannot sell people new medicines without first subjecting these products to rigorous safety checks.''} (AI will be discussed in detail in a later chapter).\par
Klein at the New York Times has been \href{https://www.nytimes.com/2023/03/12/opinion/chatbots-artificial-intelligence-future-weirdness.html}{writing against this point} for some time. His well articulated fear, is that the current model where three major Western companies, with similar highly competitive capitalist origins and values, should certainly not be in charge of racing to monetise the most compelling and innately unknowable chat bot experience. As societies shift towards materialism and technological dependence, traditional gods lose their relevance, and the need for a new form of overseer arises. This digital deity, existing within the realm of technology and the cloud, perhaps represents an adaptation of primal human belief systems. This will be explored further in the AI/ML chapter later.\par 
In conclusion, the rise of surveillance capitalism has led to concerns about the potential for the creation of a surveillance state, or worse, a new kind of omnipresent digital culturla authority. Corporations and government agencies may use the data collected by companies for surveillance purposes. This can have significant implications for individuals' privacy and civil liberties. It's important for laws and regulations to be in place to safeguard citizens' rights and privacy in regards to the use of data by government agencies, and to hold them accountable for any misuse of data, and yet it seems the reality of the situation in `post Snowden' seems far from that.\par 
Surveillance Capitalism. As a quick round-up of this area, which is best researched elsewhere:
\begin{itemize}
\item The global digital advertising market is expected to reach \$335 billion by 2023.
\item In 2020, Google and Facebook accounted for 60\% of the global digital advertising market.
\item The data brokerage industry, which includes companies that collect and sell personal data, is estimated to be worth \$200 billion.
\item In 2020, Google and Facebook were reported to have data on over 4 billion active users.
\item As of 2021, the number of data breaches reported worldwide has grown from 4.1 billion in 2018 to 4.9 billion in 2020.
\item In 2013, it was revealed that the US National Security Agency (NSA) had been collecting the phone records of millions of Americans under its PRISM program.
\item In 2013, Edward Snowden leaked classified documents that revealed the scale of the NSA's surveillance programs.
\item In the US, the Foreign Intelligence Surveillance Act (FISA) allows the government to conduct surveillance on non-US citizens outside the US without a warrant.
\item The UK's Investigatory Powers Act 2016, also known as the "snooper's charter," gives government agencies wide-ranging powers to collect and analyze personal data.
\item In 2021, it was reported that the Chinese government has been collecting and analyzing the data of its citizens through a system of "social credit" scores, which are used to monitor and control individuals' behaviour.
\item Surveillance capitalism refers to the business model of collecting and analyzing personal data for the purpose of targeted advertising and other forms of monetization.
\item A recent study by the Center for Digital Democracy found that the top 100 global digital media companies are projected to generate over \$1 trillion in revenue by 2020, much of which is derived from surveillance-based advertising.
\item The number of surveillance cameras in use worldwide is estimated to be over 1 billion, with the majority located in China.
\item A 2018 study by Comparitech found that the average person in the UK is captured on CCTV cameras over 300 times per day.
\item According to a report by the American Civil Liberties Union (ACLU), the FBI has access to over 640 million photographs for facial recognition searches, including driver's license and passport photos.
\item The U.S. government's use of surveillance technologies, such as drones and mass data collection, has been a subject of ongoing controversy and debate.
\item Some experts warn that the increasing use of surveillance technologies by governments and private companies could lead to the erosion of privacy rights and the creation of a "surveillance state."
\item In the USA senate hearing following the collapse of FTX Rep. Jesus Garcia described bitcoin and crypto as an industry that operates outside of the law and relies on hype, implying that the communities that have adopted bitcoin are ill-informed and vulnerable.
\item Bitcoin has been adopted by a variety of communities worldwide, particularly in countries such as Vietnam, the Philippines, Ukraine, India, Pakistan, Brazil, Thailand, Russia, and China.
\item There is an outsized level of adoption among Black Americans in the United States. This trend is not a result of targeted advertising by companies such as FTX, but rather a response to a legacy financial system that has limited individuals' potential.
\item Marginalized early adopters of bitcoin still constitute a minority in their communities, but the worldwide adoption trend among these groups is on the rise.
\item The solutions that outsiders build in bitcoin will ultimately be the source of the technology's promised revolution. Adoption in Africa and possibly India seems likely to be capable of driving this.
\item The paradigm shift will come from those who bring local, real-world focused use cases to their communities, separating bitcoin from the empty hype of speculation.
\item Marginalized communities will lead the industry's recovery and redefine the purpose of bitcoin in the future.
\end{itemize} 

Much of the following text is paraphrased from the work of Guy Turner of `The Coin Bureau', and Lawyer and academic Eden Moglen, and needs more work because of it's critical importance to the book.\par 
The adoption of printing by Europeans in the 15th century led to concerns around access to printed material. The right to read and the right to publish were central subjects in the struggle for freedom of thought for most of the last half millennium. The basic concern was for the right to read in private and to think, speak, and act based on a free and uncensored will. The primary antagonist for freedom of thought at the beginning of this struggle was the universal Catholic Church, an institution aimed at controlling thought in the European world through weekly surveillance of individuals, censorship of all reading material, and the ability to predict and punish unorthodox thought. In early modern Europe, the tools available for thought control were limited, but they were effective. For hundreds of years, the struggle centered around the book as a mass-manufactured article in Western culture, and whether individuals could print, possess, traffic, read, or teach from books without the permission or control of an entity empowered to punish thought. By the end of the 17th century, censorship of written material in Europe began to break down in waves throughout the European world, and the book became an article of subversive commerce, undermining the control of thought.\par
Currently, a new phase in human history is beginning as we are building a single extraneous digital nervous system, that will connect every human mind. Within two generations, every single human being will be connected to this network, in which all thoughts, plans, dreams, and actions will flow as nervous impulses. The fate of freedom of thought and human freedom as a whole will depend upon the organization of this network. Our current generation is the last in which human brains will be formed without contact with this network, and from now on, every human brain will be formed from early life in direct connection to the network, with input from generative AI/ML systems. This possibly results in humanity becoming a super organism of a sort, where each of us is but a neuron in the brain. Unfortunately, this generation has been raised to be consumers of media, which is now consuming us.\par Anonymous reading is being determined against. Efforts discussed throughout the book to ensure privacy, from Zimmerman and the cypherpunks onward, have been met with resistance from government efforts to monitor and control information flow. The outcome of the organization of this network, and the freedom it allows, is currently being decided by this generation.\par
It is not solely the ease of surveillance, nor solely the permanence of data, that is concerning, it is the relentless nature of living after the ``end of forgetting''. Today's encrypted traffic, which is used with relative security, will eventually be decrypted as more data becomes available for crypto analysis. This means that security protocols will need to be constantly updated and redone. Furthermore, no information is ever truly lost, and every piece of information can be retained and eventually linked to other information. This is the rationale behind government officials who argue that a robust social graph of the United States is needed. The primary form of data collection that should be of most concern is media that is used to spy on us, such as books that watch us read them and search boxes that report our searches to unknown parties. There is a lot of discussion about data coming out of Meta/Facebook, but the true threat is code going in. For the past 15 years, enterprise computing has been adding a layer of analytics on top of data warehouses, which is known as business intelligence. This allows for the vast amount of data in a company's possession to be analyzed and used to answer questions the company did not know it had. The real threat of Facebook is the business intelligence layer on top of the Facebook data warehouse, which contains the behaviour of nearly a billion people. Intelligence agencies from around the world want to access this layer in order to find specific classes of people, such as potential agents, sources, and individuals that can be influenced or tortured. The goal is to run code within Facebook to extract this information, instead of obtaining data from Facebook, which would be dead data once extracted. Facebook wants to be a media company and control the web, but the reality is the true value of Facebook is the information and behavior of it's users, and the ability to mine that data.
Distributed internet protocols are important in the context of government overreach into digital society and people's private lives because they provide a level of decentralization and resilience that can help protect against censorship and surveillance.\par
For example, if a government were to attempt to censor or block access to a centralized internet service, it could potentially do so with relative ease. However, if that same service were distributed across a network of nodes, it would be much more difficult for the government to effectively censor or block access to it.\par
Another advantage of distributed protocols is that they are typically more resilient to attacks or failures. If one node in the network goes offline or is compromised, the others can continue to operate, ensuring that the service remains available. This can be especially important in situations where the internet is being used for critical communication, such as during a natural disaster or political crisis.\par
In addition to their benefits for censorship resistance and resilience, distributed protocols can also help protect people's privacy. Because they do not rely on centralized servers or infrastructure, they can be more difficult for governments or other entities to monitor or track. This can be especially important in countries where government surveillance is prevalent or where individuals may be at risk of persecution for their online activities.\par 
There are a number of distributed protocols that have been developed specifically to address issues of censorship and privacy, and these will be covered in more detail later.\par
It is important to note that distributed protocols are not a silver bullet for censorship or privacy concerns. They can be vulnerable to certain types of attacks, such as those that target the nodes of the network, and they may not always be practical for certain types of applications. However, they do provide an important tool for those seeking to protect their freedom of expression and privacy online. They offer a valuable tool for those seeking to protect their freedom of expression and privacy online, and they will likely continue to play a critical role in the future of the internet.\par
In recent years, several countries have proposed or passed bills that would result in unprecedented levels of online censorship. One such example is Canada's Bill C-11, also known as the Online Streaming Act. This bill was first proposed in November 2020 as Bill C-10, but failed to pass due to its controversial provisions. It was reintroduced in February 2021 as Bill C-11 and was approved by the Canadian House of Commons, the first step in the process of becoming law. If passed, the bill would give the Canadian Radio, Television and Telecommunications Commission (CRTC) the power to decide what content Canadians can view on YouTube and other social media platforms. The CRTC would also have the power to dictate what content creators can produce, with a focus on promoting "Canadian content." Additionally, the bill would require certain broadcasters to contribute to the Canada Media Fund, which is used to fund mainstream media in Canada. The bill is currently being considered by the Canadian Senate, which will vote on it in February. If passed, it will then be debated by the Canadian Parliament. Tech companies such as YouTube have reportedly failed to convince the Senate to exclude user-generated content from the bill, indicating a high likelihood of it becoming law. The potential impact on the internet and free expression in Canada is significant, as the bill would give the government significant control over online content and restrict the ability of individuals to share their views and perspectives.\par
In a similar vein the forthcoming RESTRICT act in the USA gives huge powers without oversight to a single branch of the US government.
\begin{itemize}
\item The bill is called the ``Restricting the Emergence of Security Threats that Risk Information and Communications Technology Act'' 
\item It was initially thought to be about banning TikTok due to its connections to the Chinese government and the data it collects on its users.
\item The RESTRICT Act has very little to do with banning TikTok and instead grants the US Secretary of Commerce significant powers to determine which entities are foreign adversaries and what technology poses a risk to national security.
\item The bill defines critical infrastructure broadly, which means it could apply to almost anything the government deems necessary.
Lobbyists will be allowed to advise the Secretary of Commerce on which products and services should be labeled as foreign adversaries, potentially leading to monopolies.
\item  Fines and jail time for interacting with foreign adversaries or posing a risk to national security could reach up to \$1 million, 20 years in prison, and asset seizures.
\item The bill aims to crack down on VPNs (Virtual Private Networks), which provide privacy and access to foreign websites.
\item There is no oversight for the actions taken by the Secretary of Commerce under this act, and neither Congress nor the courts can request information on these decisions.
\end{itemize}

The European Union (EU) has separated its online censorship efforts into two separate bills: the Digital Markets Act and the Digital Services Act. These bills were introduced in December 2020 and are part of the EU's Digital Services package, which aims to be completed by 2030. The Digital Services package is the second phase of the EU's digital agenda, which is being enforced through regulation in the public sector and through ESG investing in the private sector. Both the Digital Markets Act and the Digital Services Act were passed in spring 2022 and went into force in autumn 2022, but will not be enforced until later this year and early next year, depending on the size of the relevant entity. The Digital Markets Act aims to increase the EU's competitiveness in the tech space by imposing massive fines on "gatekeepers," or companies that maintain monopolies by giving preference to their own products and services. This could open the door to innovation in cryptocurrency in the EU, but also requires gatekeepers to provide detailed data about the individuals and institutions using their products and services to the EU. The Digital Services Act, on the other hand, aims to regulate the content that is available online, including user-generated content. It does this by requiring companies to remove illegal content within one hour of it being reported and by imposing fines for non-compliance. The act also requires companies to implement measures to protect users from illegal content and from "other forms of harm," which is defined broadly and could include a wide range of content. The EU is also in the process of passing the Artificial Intelligence Regulation Act, which will be discussed later this year and is reportedly the first of its kind. All five bills in the EU's Digital Services package are regulations, meaning they will override the national laws of EU countries. The potential impact on the internet and free expression in the EU is significant, as the Digital Services Act would give the government significant control over online content and restrict the ability of individuals to share their views and perspectives.\par
In the United States, two significant documents related to online censorship are the Kids Online Safety Act and the Supreme Court case Gonzalez v. Google. The Kids Online Safety Act was introduced in February 2021 and is expected to pass later this year due to bipartisan support. The act requires online services to collect Know Your Customer (KYC) information to ensure that they are not showing harmful content to minors. It also gives the Federal Trade Commission (FTC) the power to decide when children have been made unsafe online and allows parents to sue tech companies if their children have been harmed online. The act has received criticism from both sides of the political spectrum and entities outside of Congress, as it is seen as giving too much power to the government to regulate online content and could lead to increased censorship by tech companies.\par
The Supreme Court case Gonzalez v. Google involves the question of whether Google's algorithmic recommendations supported terrorism and contributed to the 2015 terrorist attacks in Paris. The case has been picked up by the Supreme Court after being passed up by various courts of appeal. It is being heard alongside another case, Twitter v. Tumne, involving the role of Twitter's algorithms in a terrorist attack in Istanbul. There are two potential outcomes for the case. If the Supreme Court sides with Gonzalez, it could increase the liability of social media companies under Section 230 of the Communications Decency Act, which allows them to moderate content to a limited extent without violating the First Amendment. Alternatively, the Supreme Court could declare Section 230 unconstitutional, which would make online censorship illegal but also hinder the use of algorithms on the internet. The ideal outcome, in theory, would be for the Supreme Court to side with Google and for Congress to change Section 230. However, giving Congress the power to change the law could lead to increased censorship and the potential for abuse of power.\par
In the UK forthcoming legislation will see tech company leaders liable for \href{https://www.independent.co.uk/news/uk/politics/bill-mps-iain-duncan-smith-molly-russell-rishi-sunak-b2263353.html}{prison sentences} if they fail in their duty to protect minors. This will doubtless lead to both stringent universal requirements for identity proof (KYC), and significantly muted and controlled content on the platforms.\par
Our research focuses on business to business use cases for distributed technologies, and will provide mechanisms for verifying who is communicating with whom, to avoid falling foul of these swinging global infringements on privacy.\par
It is the opinion of this book that information should be free \cite{swartz2008guerilla}
\section{Government over-reach through bureaucracy}
As an contextual example of the soft power which political apparatus uses to influence emergent human behaviour and their markets it is useful to look again to the USA. In 2013, the Obama Administration, faced with a divided Congress, resorted to using the banking system as a means to implement policy through non-traditional channels. This effort, known as Operation Choke Point, was a continuation of their success in cutting off the offshore online poker industry from banking services. Initially, the crackdown was aimed at the payday lending industry, but it soon expanded to include gun sales and adult entertainment, and eventually up to 30 different industries.\par
The rationale behind Operation Choke Point was to target banks that facilitated fraud, as indicated by a high ratio of fraud and disputes. However, the operation soon evolved into a redlining of industries based on nothing more than the perceived risk of reputational harm. Financial institutions were investigated without any evidence of losses. Throughout the entire operation, there was no new legislation or written guidance issued. Banks were simply warned of increased regulatory scrutiny if they did not comply.\par
Major banks continue to deny services to industries such as firearms and fossil fuels, and they continue to assign higher risk ratings to industries that may face government criticism, even in the absence of any official guidance. This utilisation of the financial system as a means of driving change is seen by some as a legitimate, if not ideal, mechanism; as just one more type of market actor. Regardless of one's political perspective, it is important to consider the moral hazard of bypassing traditional political channels and using bureaucratic mechanisms as a means of affecting change in the free market. It is important to consider how the power of these tactics might be used in the future by opposing political groups. For example, supporters of Operation Choke Point who were in favour of increased financial pressure on the oil and gas industry may not feel the same if the same techniques were applied to organizations like Planned Parenthood. From this perspective, the tactics used by Operation Choke Point can be seen as undemocratic, regardless of who is deploying them. Bringing this back to our study of new financial tooling in crypto we can look to recent events:

\begin{itemize}
\item January: Some banks start to wind down activity in the crypto industry
\item January 21st: Binance announces its banking partner, Signature Bank, refuses to process Swift payments for less than \$100,000
\item January 27th: Federal Reserve denies Custodia Bank's application to access Federal Reserve System
\item January 27th: Federal Reserve denies Custodia Bank's application for a master account
\item January 27th: Federal Reserve releases statement discouraging banks from holding crypto assets or issuing stable coins
\item January 27th: National Economic Council issues policy statement discouraging banks from transacting with crypto assets or maintaining exposure to crypto depositors
\item February 2nd: DOJ announces investigation into Silvergate Bank over dealings with FTX and Alameda Research
\item February 6th: Binance announces suspension of USD bank transfers to and from offshore exchange
\item February 8th: Binance announces search for another banking partner
\item February 7th: Fed's policy statement enters Federal Register as a final rule
\item Two outstanding applications for National Trust Bank licenses from Anchorage and Paxos likely to be rejected by the OCC
\item Banking services becoming increasingly difficult for crypto firms, some startups will likely now not make the attempt
\end{itemize}
It seems that in the absence of democratic the SEC is attempting to use their tools to control and centralise the `ramps' into and out of digital assets, and the rules around holding them for investors. The SEC has proposed a new rule that would require registered investment advisors to use qualified custodians for all assets, including cryptocurrencies. The intention behind this proposal is to improve investor protection by mandating that custodians hold customer assets in segregated and identifiable accounts. However, critics argue that this proposal would limit the number of qualified cryptocustodians and deter investment advisors from advising their clients on crypto. The few banks with the necessary technical capabilities and regulatory approvals will have a monopoly on crypto custodial services, while exchanges without a banking license or trust bank will likely lose out. The proposal assumes that crypto assets are securities without going through a process to determine that. The outcome of the proposal will depend on the stringency of the SEC's qualified custodian registrations. The proposal is currently in a 60-day public comment period before the Commissioners hold another vote on whether to pass the rule.\par
Caitlyn Long explains that the proposed rule would not necessarily kill crypto custody, but would be a move against State Charter trust companies. She points out the big issue with the proposal, which is the requirement for custodians to indemnify for negligence, recklessness, or willful misconduct. This would apply to all asset classes, including commodities and crypto, which could kill the custody business broadly. The SEC proposal would apply the custody rule to all asset classes, including commodities and crypto, which is okay, but the SEC also wants custodians to indemnify the full asset value for losses in which the custodian played any role, even for physical assets like oil, cattle, and wheat. This would upset long-standing insurance terms and could cause huge pushback from the banking, Wall Street, commodities, and crypto industries. Sarah Brennan believes that the proposal represents continued governmental efforts at denial of service attacks on crypto, and that the SEC's approach only seeks to chill digital asset markets. She and the Republicans on the House Financial Services Committee are urging stakeholders to submit public comments on the proposed amendments to ensure the custody rule for investment advisors is modernized appropriately. The U.S. Internal Revenue Service plans to hire nearly 30k new staff and technology over the next two years, spending \$80 billion to improve tax enforcement, much of it focussing on crypto markets. It might be that the industry follows the prevailing winds \href{https://noelleacheson.substack.com/p/weekly-feb-25-2023}{and pivots to the East}. As usual, none of this particularly impacts our use case and thesis.
\section{Global monetary policy}
The term ``don't fight the Fed'' has been used in trading circles for many years. Owing to the pre-eminent role of the dollar in global markets actions of the political and central banking bodies which impact the dollar always have global reach. It is worth knowing that these decisions are usually contested, and worse, the power of the decision makers seems rooted in their narrative impact. It's a pretty terrible system given the impact on billions of lives. The Federal Reserve System, which is comprised of a Board of Governors, 12 regional banks, and an Open Market Committee, is a privately-owned central banking system in the United States. The member banks of each Federal Reserve Bank vote on the majority of the Reserve Bank's directors and the directors vote on members to serve on the Open Market Committee, which determines monetary policy. The president of the New York Federal Reserve Bank is traditionally given the vice chairmanship of the Open Market Committee and is a permanent committee member. This means that private banks are the key determinants in the composition of the Open Market Committee, which regulates the entire economy. The Federal Reserve is an independent agency and its monetary policy decisions do not have to be approved by the President or anyone else in the executive or legislative branches of government. The Fed's profits are returned to the Treasury each year, but the member banks' shares of the Fed earn them a 6\% dividend. The 2008 financial crisis and subsequent bailouts exposed the fundamental conflicts of interest at the heart of the Federal Reserve System, where the very banks that caused the crisis were the recipients of the trillions of dollars in bailout money. These conflicts of interest were baked into the Federal Reserve Act over 100 years ago and are a structural feature of the institution. The concentration of power within this group is staggering. 
\section{Opportunities in Africa}
\subsection{Gridless}
In the course of researching this book we see most opportunity for change in Africa. As an example the company `Gridless' began by examining different energy sources in Africa and exploring opportunities for larger energy generation and grid-connected energy. However, they found that the real benefit of gridless energy was in providing energy to places that were not well connected and did not have a good grid. They contacted mini-grid providers all over East and Southern Africa to learn about their problems. A mini-grid is defined as a project that generates energy under 2 megawatts, often under 1 megawatt. They discovered that these providers had to overbuild for the community, resulting in stranded energy. The company found a way to utilize this stranded energy by placing Bitcoin miners on it and paying the mini-grid providers for it. They tested this method and found it to be successful. Additionally, they implemented a system to automate and remotely turn off the power during periods of high usage to make the grid more efficient and sustainable. This solution provided a win-win-win situation for the company, the mini-grid providers, and the communities they served.\par
The company utilizes Bitcoin miners to create space for other activities and to increase access to affordable energy for communities and small businesses. As energy usage increases in the community, the company decreases their usage of miners and moves them to other locations. This is outlined in their contracts with partners. The company is currently testing this method and has encountered some challenges, such as losing internet connection at one of their sites and poor rainfall affecting the amount of water flowing into turbines. They have found that building a lean operation with flexible and adaptable staff is crucial, as well as creating processes and systems to manage variables. The company also faces unique environmental factors such as lightning strikes, which require them to turn off their operations temporarily.\par
Gridless suggest that those who are critical of opportunities like this often come from a place of privilege and do not understand the consequences of their actions in places like Africa where access to electricity and other resources is limited. They argue that these critics, who are often from the West, have blinders on and cannot see the impact of their actions on a global scale. They suggest that more people need to travel and have diverse experiences in order to change their perspective on Bitcoin and its potential to support human flourishing in underprivileged areas. They also mention that gridless plans may become a case study for the positive impact of Bitcoin mining on economic opportunities, particularly in rural Africa.
\subsection{Machankura}
Mobile phone users in Nigeria, Tanzania, South Africa, Kenya and five other African countries can now \href{https://www.forbes.com/sites/digital-assets/2023/03/15/how-africans-are-using-bitcoin-without-internet-access/?sh=434df18b7428}{send and receive bitcoin} without a smartphone or Internet connection. Just a basic feature phone and text code will suffice, thanks to a digital wallet from software developer Ngako. No internet connection and low power handsets means using SMS and the Lightning network, with the phones SIM acting as the wallet private keys.
\section{El Salvador as a case study}
El Salvador became the first country in the world to adopt Bitcoin as legal tender. El Salvador's adoption of Bitcoin was a historic moment in the world of Bitcoin and was met with a mix of excitement and scepticism. On June 9, 2021, the country's Legislative Assembly approved a bill introduced by President Nayib Bukele to make Bitcoin a legal tender alongside the US dollar, which has been used as the country's official currency since 2001.\par
President Bukele, who has been a vocal proponent of Bitcoin, stated that the adoption of Bitcoin was a way to promote financial inclusion and stability in the country, where more than 70\% of the population is unbanked or underbanked. In a tweet, he stated, ``Bitcoin will have the same value as the US dollar. We will support both. They will have the same power of purchase and will be accepted in the same way.''\par
The move was met with a lot of media attention and reaction, with some praising it as a bold and innovative step, while others raised concerns about the volatility of Bitcoin and their potential impact on the economy. President Nayib Bukele himself has faced criticism for his handling of political power and some of his actions have raised concerns about the potential for abuses of power. In 2021, President Bukele faced widespread criticism for his handling of the legislative process and his use of the military to secure the Legislative Assembly building during a political standoff with lawmakers. This led to allegations of intimidation and a violation of democratic norms, and raised concerns about his willingness to use force to achieve his political goals. Additionally, President Bukele has faced criticism for his use of social media to communicate with the public and his tendency to bypass traditional media outlets, which has raised concerns about the potential for censorship and the manipulation of information. With that said he seems much loved in the country, and the previously appalling safety statistics of the nation have radically improved.\par
In addition to the adoption of Bitcoin as legal tender, El Salvador has also proposed the issuance of a Bitcoin-backed bond to finance various public works projects and promote the use of Bitcoin. The bond would be denominated in Bitcoin and would allow investors to directly participate in the country's development while also supporting the growth and adoption of Bitcoin.\par
Another ambitious project that has been proposed by President Bukele and his administration is the creation of ``Bitcoin City'', a new city that would mine Bitcoin at the base of a dormant Volcano, and offer considerable tax benefits to holders. The city would serve as a hub for innovation and a showcase for the potential of Bitcoin, and would offer a wide range of services, including housing, healthcare, education, and entertainment.\par
There has been a significant increase in the adoption of Bitcoin in El Salvador, and apparently increased inward investment to the country. Many businesses, both small and large, have started accepting Bitcoin as a form of payment, and there has been a growing interest in Bitcoin among the general population. Additionally, the government has been actively promoting the use of Bitcoin through various initiatives. There have also been efforts to educate the public about Bitcoin and its potential benefits, including increased financial security and reduced transaction fees compared to traditional banking systems.\par
Overall, the adoption of Bitcoin in El Salvador has been positive, far outstripping the number of people in the country with traditional bank accounts, and has the potential to greatly impact the country's economy and financial sector. However, it is important to note that there are still challenges to overcome, such as regulatory and infrastructure limitations, as well as ongoing concerns about the volatility and stability of Bitcoin.\par 
Somewhat surprisingly the IMF have de-escalated their previously highly critical assessment of the move, toward a more \href{https://www.imf.org/en/News/Articles/2023/02/10/el-salvador-staff-concluding-statement-of-the-2023-article-iv-mission}{concerned and conciliatory tone}:
\textit{``Bitcoin’s risks should be addressed. While risks have not materialized due to the limited Bitcoin use so far—as suggested by survey and remittances data—its use could grow given its legal tender status and new legislative reforms to encourage the use of crypto assets, including tokenized bonds (Digital Assets Law). In this context, underlying risks to financial integrity and stability, fiscal sustainability, and consumer protection persist, and the recommendations of the 2021 Article IV remain valid. Greater transparency over the government's transactions in Bitcoin and the financial situation of the state-owned Bitcoin-wallet (Chivo) remains essential, especially to assess the underlying fiscal contingencies and counterparty risks.''}\par
In terms of economic impact, it is still too early to determine the full effects of the adoption of Bitcoin in El Salvador. However, it is expected to have a positive impact on financial inclusion and stability, as well as reducing the reliance on traditional banking systems. The use of Bitcoin has the potential to lower transaction fees and increase financial security, which could be particularly beneficial for those who do not have access to traditional banking services.\par
Overall, the adoption of Bitcoin in El Salvador marks a significant step forward in the mainstream acceptance and adoption of Bitcoin and has the potential to set a precedent for other countries to follow. However, it is important to monitor the situation and assess the long-term impacts on the economy and financial sector.

The swift rise of digital walled gardens, moving towards a less transparent internet, reveals both a need for user data protection and a corporate push for greater control and profit. Tech giants like Google, Reddit, and Twitter are increasingly controlling their platforms, adjusting data flows for revenue growth. Google's new privacy policy, which allows data collection for AI model training, increases public concerns over user consent and privacy rights. The wide-ranging language of the policy gives Google considerable power in using user-generated content, fueling debates on data usage ethics. Simultaneously, the social web's shift towards an entertainment-focused business model prioritizes revenue over human connection. Platforms target ad revenue through vertically scrolling videos, risking reduced content diversity and creating echo chambers.\par 
Entertainment unions like the International Alliance of Theatrical Stage Employees (IATSE) are grappling with AI's impact on employment. Their approach includes research, collaboration, education, political advocacy, organizing, and collective bargaining to protect members' interests, including upskilling initiatives. Upskilling is gaining industry attention. Companies like Tata Consultancy highlight the need to equip engineers with AI skills. Recognizing AI technologies' potential, they invest in reskilling programs to stay competitive and effectively use AI tools.\par
The rise of generative AI and the declining open web raises concerns about maintaining digital commons and encouraging diverse perspectives. AI-generated content could overshadow human contributions, making meaningful information harder to find and increasing misinformation risks. This situation highlights the need for balance between AI-generated and human-generated content.\par

Data control battles between platforms and users fuel debates on data ownership and profit sharing. Users demand more control over their data use and potentially a share in the resulting profits. This issue emphasizes the need for transparent data policies and fair user compensation models.\par

The influence of AI on the job market and the future of work is a significant concern. As AI technology progresses, the need for upskilling and reskilling programs grows to ensure workers can adapt to changing job requirements. Collaboration between industries, governments, and educational institutions is essential to address AI-induced disruptions and ensure a smooth workforce transition.\par

Finally, the importance of AI ethics and governance grows as AI technologies become more prevalent. The development and deployment of AI systems require ethical frameworks, transparency, and accountability. Collaboration between AI researchers, policymakers, and ethicists is critical to address potential risks and societal implications of AI technology.

\section{Artificial Intelligence in a global context}

This currently borrows heavily from \href{https://www.youtube.com/watch?v=5clOHBo8HP8}{the AI breakdown podcast}, is an AI generated placeholder, and needs considerably more more.

\subsection{Perception of AI and Society}
The examination of AI's implications on societal structures should undoubtedly receive the necessary attention. Soros's language and perception of reality seem particularly interesting, especially in the era of AI. He emphasizes his belief in reality and its importance in providing moral guidance, a concept that seems increasingly challenged in the age of AI.

\subsection{AI, Propaganda, and Authoritarianism}
In an opinion piece for The Hill by Bill Drexel and Caleb Withers, titled "Generative AI could be an authoritarian breakthrough in brainwashing," the authors argue that the concern isn't just external attempts to influence U.S. elections, but the impact on the populations within authoritarian countries. They posit that foreign disinformation efforts by Chinese and Russian entities are only the tip of the iceberg, with Beijing and Moscow disseminating massive amounts of propaganda to their own populations. The authors also cite instances of AI-enabled propaganda and misinformation campaigns, both in the context of undermining democracies and consolidating control within authoritarian states.

\subsection{Increased Surveillance Through AI}
Another critical concern around AI and authoritarianism is the potential for increased surveillance. With the integration of AI and data scraping techniques, governments can employ extensive teams to facilitate unprecedented levels of surveillance, compromising privacy. Such concerns are raised in the works of authors like Daniel Oberhaus, who posits that authoritarian regimes may have an advantage in AI due to their willingness to exploit data, such as advanced facial recognition data, in ways that open societies might not.

\subsection{Worker Surveillance and Remote Work}
Furthermore, the issue of worker surveillance, especially with the rise of remote work regimes, has garnered the attention of various entities, including the White House. This is due to concerns over automated systems that employers are using to monitor their remote workers, highlighting a less benign context of surveillance.

\subsection{AI and Ideology}
One way AI might foster authoritarianism is by supporting the ideology of closed societies or authoritarian regimes, such as China. These societies may leverage their global influence to disseminate their particular AI model, aligning it with their motivations and goals. The Carnegie Endowment for International Peace points out that for most countries, AI technology is viewed as an economic development factor that determines their standing in the global technology race, rather than as an ideological preference.

\subsection{AI and Central Planning}
Another concern is the fear that AI will make centrally planned economies seem viable, where past attempts failed due to the lack of data. This idea was discussed in a conversation between Peter Thiel and Reed Hoffman hosted by Neil Ferguson at Stanford in 2018. Thiel posited that AI appears to favor centralization, an aspect that supports the principles of central planning.

\subsection{Uncontrolled AGI Creation}
On the other hand, some suggest that capitalist competition could result in the creation of AGI that cannot be controlled. Dr. Jeffrey Hinton, a vocal advocate of this view, argues that AI's potential to disrupt business models could drive companies to recklessly pursue advancements in AI to stay competitive. This could lead to increased state power as people become more reliant on the state in an AI-dominated economy, potentially resulting in increased authoritarianism.

\subsection{AI Promoting Freedom}
However, AI could also promote freedom in several ways. For instance, AI tools like Altana have been used to identify goods made using forced labor, helping companies make informed supply chain decisions. AI could also serve as a new interface for disseminating information, such as a chatbot that aids detainees in requesting legal assistance.

\subsection{AI, Integrity, and Accessibility}
Yet, for AI to achieve its full potential in promoting freedom, the integrity of the information it disseminates must be uncompromised, and its accessibility must be ensured despite potential firewalls.

\subsection{AI's Impact on Societal Organization}
Given these diverse viewpoints, it seems that the potential of AI to either aid authoritarianism or promote freedom is yet to be fully explored. However, the inherent ability of democracies to encourage disagreement and diverse perspectives may serve as a counterbalance to the potential of AI for authoritarian control. Moreover, AI's capacity as a catalytic force in societal organization should not be underestimated. The increasing discourse around AI and its implications for labor and technology usage suggests that AI technology is reshaping the world in ways that were unimaginable just a few years ago. Its capabilities in data analysis, decision making, and automation are transforming industries and redefining the scope of what's possible.

\subsection{Democratization of AI Technology}
An argument often made in favor of democratization of AI technology is that it should be made open-source and freely available, thus creating a challenging framework for global political incumbents. This perspective is grounded on the belief that technology - and its underlying power - must be accessible to everyone to mitigate the risks of misuse and ensure fair benefits distribution.

\subsection{Open-source AI and Innovation}
Open-source AI can be a vehicle for widespread innovation. It can spur creativity, leading to breakthroughs in various sectors, from healthcare and education to energy and transportation. Open-source technologies facilitate collaboration, accelerate the pace of research, and democratize access, enabling researchers and developers across the globe to contribute to the expansion of AI's capabilities. It opens the possibility for rapid iteration and innovation, reducing the likelihood that a few powerful entities monopolize control over these transformative technologies.

\subsection{Open-source AI and Global Politics}
However, as beneficial as open-source AI may appear, the complexity of global politics can make the transition challenging. A landscape where AI technologies are open-source and freely available brings about potential dilemmas in various areas including national security, economic competitiveness, intellectual property rights, and data privacy.

\subsection{National Security and Open-source AI}
To start, national security is a primary concern. AI has a myriad of applications in defense and security sectors, many of which could potentially be exploited by adversarial entities. As such, unrestricted access to AI technologies could pose a risk to nations' security. Nevertheless, it is crucial to note that security risks also stem from concentrated AI power. A handful of nations or corporations owning the majority of AI developments may lead to destabilization, power imbalance, and heightened global tensions.

\subsection{Economic Competitiveness and Open-source AI}
Economic competitiveness is another intricate aspect. Countries and corporations are engaged in a fiercely competitive race to advance in AI technologies, recognizing the economic gains and strategic advantages tied to AI leadership. Open-source AI might challenge this dynamic, disrupting traditional models of competition. However, it could also create an environment of shared growth, leading to a more balanced global AI landscape.

\subsection{Intellectual Property Rights and Open-source AI}
Intellectual property rights form another complex dimension in the discussion. Open-source AI challenges traditional notions of ownership and patents, potentially undermining the incentives for companies and individuals to invest in AI research and development. Balancing the need for innovation with the necessity to protect inventors' rights becomes critical in an open-source framework.

\subsection{Data Privacy and Open-source AI}
Data privacy is a further point of contention. Open-source AI, coupled with increasingly ubiquitous data collection methods, raises concerns about individuals' privacy. However, it also provides an opportunity to develop robust, decentralized, and transparent AI systems that respect user privacy.

\subsection{A New Social Contract for AI}
Thus, navigating the intersection of AI and global politics necessitates careful consideration. It requires establishing a new social contract for AI—one that respects human rights, promotes equitable economic growth, and protects national security.

\subsection{Conclusion}
In conclusion, making AI open-source and freely available represents a shift from the status quo, with both promising potentials and daunting challenges. A global AI framework that upholds democratic principles and values, promotes shared prosperity, and safeguards security and privacy is the aspiration. To achieve this, an inclusive and multidimensional discourse is essential, involving governments, corporations, civil society, academia, and individual citizens. It is through this collective effort that AI's true potential can be harnessed for the global good.

There is skepticism the idea of artificial general intelligence (AGI) leading to superintelligent machines that threaten humanity in the near future. This supposed risk of AGI is described as a "red herring" - an unfounded fear. The reasons given are:

\begin{itemize}
\item We do not have a clear understanding or definition of general intelligence or consciousness.
\item Current AI like large language models are limited in scope. They are good at statistical pattern matching in language, not generally intelligent.
\item The hypothesis that intelligence and consciousness emerge simply from increasing computational power is unproven. There are likely other components we don't understand.
\end{itemize}

The real risk is perhaps government control and regulation of AI development and applications, justified by arguing it is needed for safety and responsible AI. This could impose limits on acceptable speech and thought. Centralised entities could become gatekeepers for how people access and interpret information about the world. Mandating allowable language could narrow ideas and speech to fit an official narrative. Fears of AGI, even if exaggerated, open the door for regulators and bureaucrats to intervene in the name of safety. The risk is not AGI itself but the government control that hype about it enables.

There is speculation that AI will automate many white collar cognitive jobs, similar to how industrial machinery automated manual labor. This may "chase humans up the value stack" as lower value work is handled by AI, freeing people to focus on higher value creative activities. 
\chapterimage{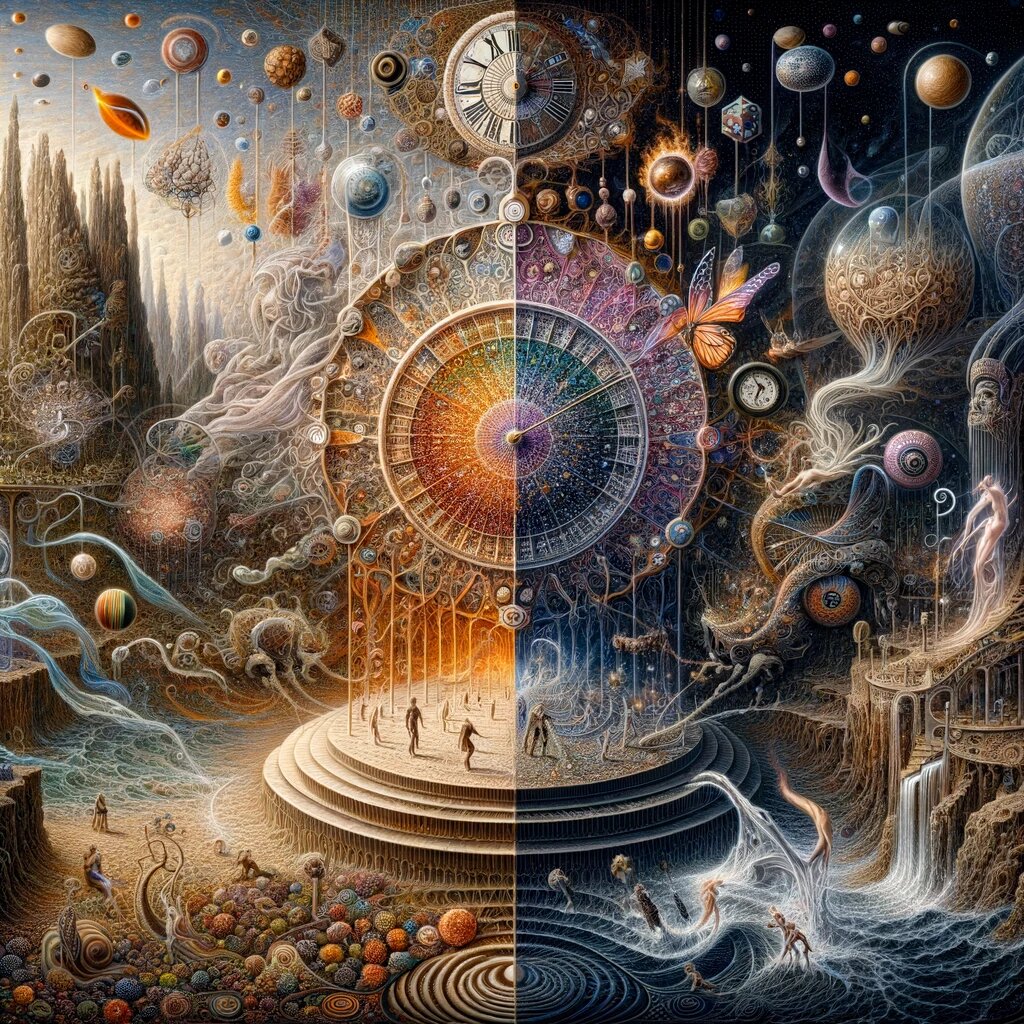}
\chapter{Distributed Identity \& Trust}
For distributed Web3, and by extension metaverse applications to flourish it is necessary to solve the identification problem \cite{king1966fisher}. Without a \href{https://joshgans.medium.com/web3-isnt-going-to-work-without-identification-6aa776d674}{solution to this} bots, scammers, and AI actors will reduce usefulness and usability of and already quite arcane user experience.\par  
This chapter is an oddity because most of traditional DID/SSI isn't really fit for purpose. Distributed self sovereign identity has a great elevator pitch though. Individuals should be empowered through technology to manage their own data, without manipulation or exploitation by centralised corporate behemoths. In practice it's a staggeringly complex proposition which increases risk to the individual, decreases convenience, and despite much work, does not even make much sense in it's own terms. Webs of trust are viable so this means Nostr, \href{https://github.com/project-bitmark/marking/wiki#marking}{Marking}, or Slashtags which will be discussed, but are early products. 
\section{Applications of DID/SSI}
Some of the likely, and discussed applications for DID/SSI are the more inherently private and personally valuable sets of data an individual might generate throughout their life. The theory is that subsets of such data could then be digitally revealed by the individual when required, and that cryptographic verification built into the system would guarantee the veracity of the data to the receiving party. It is also possible to make use of ``zero-knowledge proof'' such that assertions can be made about about the contents of the data without revealing the data itself. A good example of this an age verification challenge, where a threshold age could be asserted without necessarily revealing the date of birth. 
Other keystone uses of the technology are:
\begin{itemize}
\item health documents history
\item qualifications and certifications
\item financial record and relationships with those of others
\item contacts, connections to other people and their appropriate data, including things like shared and personal calendars
\end{itemize}  
It's also possible to extend this key management ethos to all login credentials, and all data currently stored on centralised servers. This is the tension discussed in the chapter about Web3. Proponents think that using something like a DID/SSI stack to manage encryption, decryption and access to data within cloud services gives the user the best of all worlds. They see simply logging in with a cryptographic wallet, and using that same public/private key pair to manage the data beyond as some kind of panacea. This is very complex stuff though, and it seems very likely they just haven't thought this through enough.
\section{Classic DID/SSI}
\begin{figure}
\includegraphics[width=\linewidth]{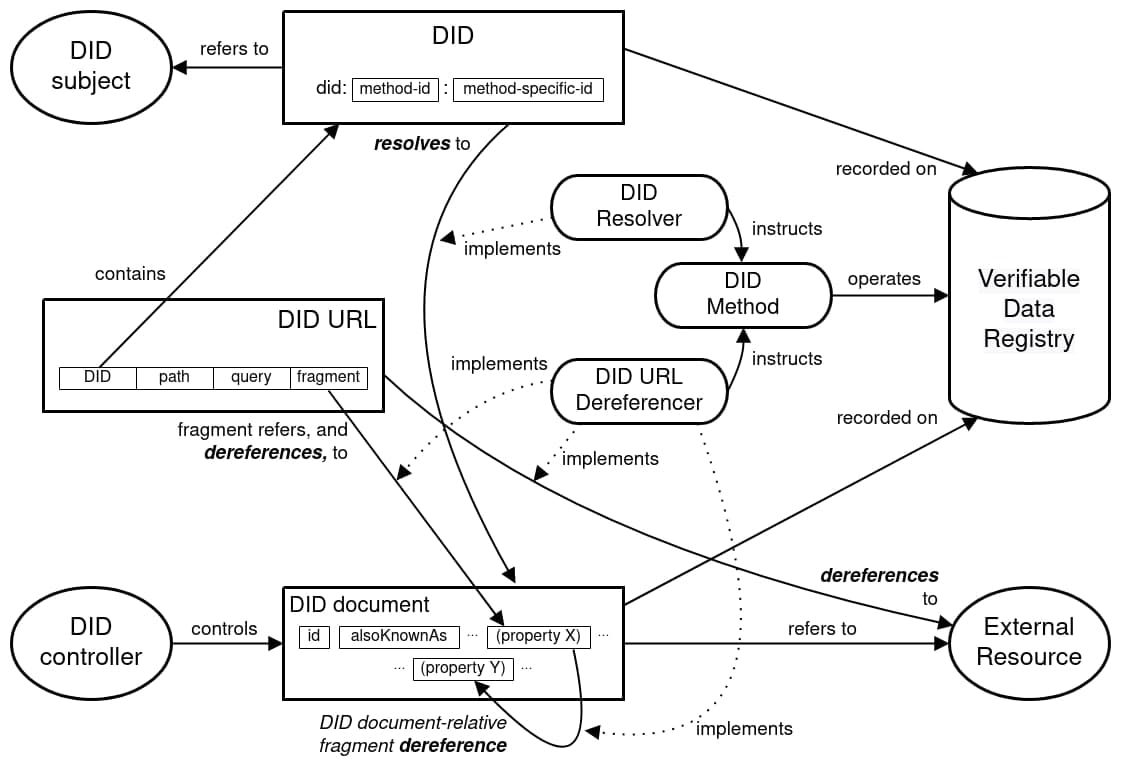}
  \caption{Part of the DID SSI specs}
  \label{fig:DID}
\end{figure}

Distributed identity / self sovereign identity has been extensively researched for decades, with hundreds of peer reviewed papers, and extensive support from the \href{https://www.w3.org/TR/did-core/}{world wide web consortium}. The academic field now seems quite ossified and has settled on a couple of hundred `schema' which they feel underpin the next layer of development. It is a \href{https://medium.com/decentralized-identity/overview-of-decentralized-identity-standards-f82efd9ab6c7}{complex field}, and the language and diagrams are arcane and self referential as seen in Figure \ref{fig:DID}.
. 
Moreover the minimal implementation of such proposed systems hints at a \href{https://www.w3.org/community/perma-id/}{federated model} of \href{https://github.com/w3c/vc-data-model/issues/947#issuecomment-1276186406}{centralised/federated `truth'} to enable persistence of identifiers over time.\\
The major failing of the DID/SSI work to date is a lack of meaningful use cases with incentives for adoption. This is clearly explained by Lockwood \cite{lockwood2021exploring} who proposes that the pathway to adoption of `classic' DID/SSI requires an incentive over and above the current identity management on the web. Being distributed is not enough. Especially in the light of questionable assurances of this even being true.\\
Perhaps most concerning is this \href{https://lists.w3.org/Archives/Public/public-credentials/2022Mar/thread.html}{recent exchange} on the mailing lists. Here, two long standing developers of DID say the following:\\
\textit{``Not a single entity I know that's doing production deployments has actually vetted did:ion and found it to be production capable. This goes for every DLT-based DID Method out there - even the one we're working on. I am highly sceptical of anyone that says that any DID Method is ready for production usage at present.\\
Agreed — as one of the proponents of DLTs (in particular permissionless public ones) none are mature enough yet for production.''}.
It seems then that we can rule out use of these technologies?

\subsubsection{DID principles}
The core principles of distributed identity are that there should be persistent identifiers, like real world documents which assert identity, but with extended use cases. These should be permanent, and resolvable everywhere, forever. Underpinning this is cryptographically verifiable and decentralised data, managed by the user, or their trusted proxy. As primitives this makes them lifetime digital assets, that are portable, and unconfiscatable, with no required reliance on a trusted third party. By this stage in the book you should be familiar with these concepts, but application of this fundamental mindset to all personal data and digital interactions is a bigger reach even than money and value.
\subsubsection{What's in a DID document?}
All classic DID is underpinned by a DID document what bootstrap the services it's connected to. It is made up of one or more public keys. The documents can make use of services such as timestamps, cryptographic signatures, proofs, delegations, and authorisations. They should contain the minimum amount of information to accomplish the specific task required of them.
\section{Federated social media trust}
This section about newer technologies is perhaps best \href{https://www.getrevue.co/profile/jackjack/issues/a-native-internet-protocol-for-social-media-1503112?via=twitter-card&client=DesktopWeb&element=issue-card}{summarised by Jack Dorsey}, ex CEO of twitter, paraphrased here:\\
\textit{``I'll start with the
principles I've come to believe based on everything I've learned and experienced through my past actions as a Twitter co-founder and Lead:
\begin{itemize}
\item Social media must be resilient to corporate and government control.
\item Only the original author may remove content they produce.
\item Moderation is best implemented by algorithmic choice.
\end{itemize} 
The biggest mistake I made was
continuing to invest in building tools
for us to manage the public conversation versus building tools for the people using Twitter to easily manage it for themselves this burdened the company with too much power and opened us to significant outside pressure such as advertising budgets. I generally think companies have become far too powerful. The only way I know of to truly live up to these three principles is a free and open protocol for social media that is not owned by a single company or group of companies and is resilient to corporate and government influence the problem today is that we have companies who own both the protocol and discovery of content which ultimately puts one person in charge of what's available and seen or not this is by definition a single point of failure no matter how great the person and over time will fracture the public conversation and may lead to more control by governments and
corporations around the world.}\par

The following technologies were selected for this book long before Dorsey wrote those words, but they \textit{are} the technologies in which he is investing his time and money to further those 3 principles.
Keybase is an interesting example of how proofs ont he internet can lean upon one another to provide a corpus of trusts. It provides a model of \href{https://book.keybase.io/account#proofs}{importing proofs} from various social media sites. This allows importing of reputation into new ecosystems.
\subsection{Lightning}
It is possible to log into a website using only Lighting, as in \href{https://stacker.news/login?callbackUrl=https://stacker.news/}{Stacker News}. 
\subsection{Web5, Bluesky, \& Microsoft ION}
Promisingly Jack Dorsey's company TBD is working on a project \href{https://developer.tbd.website/projects/web5/}{called ``Web5''}. Details are scant but the promise is decentralised and/or self hosted data and identity running on Bitcoin, without recourse to a new token. \textit{``Components include decentralized identifiers (DIDs), decentralized web node (DWNs), self-sovereign identity service (SSIS) and a self-sovereign identity software development kit (ssi-sdk)''}.\par
Web5 leverages the ION identity stack. All this looks to be exactly what our metaverse system requires, but the complexity is likely to be quite high as it is to be built on existing DID/SSI research which is pretty complex and perhaps has problems.\par 
They readily admit they \href{https://atproto.com/guides/identity}{do not have a working solution} at this time: \textit{``At present, none of the DID methods meet our standards fully. Many existing DID networks are permissionless blockchains which achieve the above goals but with relatively poor latency (ION takes roughly 20 minutes for commitment finality). Therefore we have chosen to support did-web and a temporary method we've created called did-placeholder. We expect this situation to evolve as new solutions emerge.''}
\subsubsection{ION} 
While working at Microsoft on ION Daniel Buchner (now working at Square) or Henry Tsai \href{https://github.com/decentralized-identity/ion/blob/master/docs/Q-and-A.md}{said the following}, which is worth quoting verbatim:\par
``While ledger-based consensus systems, on the surface, would seem to provide the same general features as one another, there are a few key differences that make some more suitable for critical applications, like the decentralized identifiers of human beings. Some of these considerations and features are:
\begin{itemize}
\item The system must be open and permissionless, not a cabal of authorities who can exclude and remove participants.
\item The system must be well-tested, and proven secure against attack over a long enough duration to be confident in.
\item The system must produce a singular, independently verifiable record that is as immutable as possible, so that reversing the record the system produces is infeasible.
\item The system must be widely deployed, with nodes that span the globe, to ensure the record is persisted.
\item The system must be self-incentivized, so that nodes continue to operate, process, and secure the record over time. The value from operation must come from the system directly, because outside incentive reliance is itself a vector for attack.
\item The cost to attack the system through any game theoretically available means must be high enough that it is infeasible to attempt, and even if an ultra-capitalized attacker did, it would require a weaponized mobilization of force and resources that would be obvious, with options for mitigation.\par

The outcome:

\item Number 1 eliminates private and permissioned ledgers
\item Number 2 eliminates just about all other ledgers and blockchains, simply because they are inadequately tested
\item For the metrics detailed in 3-6, Bitcoin is so far beyond all other options, it isn't even close - Bitcoin is the most secure option by an absurdly large margin.''
\end{itemize}

On the surface then it might seem that the choice is Bitcoin again, and indeed that the open source Microsoft ION stack is a natural choice, but it's complex to run, the interactions with the blockchain have a cost implication which can't be surmounted without every user owning some Bitcoin, and as we have seen, there is no formal validation of this system. In addition (in the current implementation) an identity proof does not need to be published to be valid, just timestamped. In this way an identity can be stolen and used years later to claim later chains of proof. It seems that it might be somewhat useful `at scale' and is worth additional monitoring and investigation, especially given it's integration into TBD - Web5.
\subsection{Slashtags}
Slashtags is a distributed identity open method being developed by Bitfinex and Tether under the Synonym suite. It's origins date back to 2011 and was initially seeded through academia, and government innovation grants to build on the concepts of BitTorrent, and later \href{https://dat-ecosystem.org/timeline.html}{DAT}. This eventually became the Hypercore protocol, with an additional rebranding to Holepunch in 2021. It is essentially this system, a mobile app UX, and Bitcoin integration which forms the Synonym/Slashtags stack. There is a lot of historical investment, new focus, and promising product design in the Synonym ecosystem which is forming about the this `web of trust' distributed data system. The suite will rely on Pear Credits to enable Tether dollars to be passed around within the system. This may foster adoption in emerging markets. The critical path nature of the Tether integration, and the complex intermingling of Synonym, Hypercore, Bitfinex, Tether, and Pear credits are potentially red flags, and though the technology stack is quite interesting only Pear Credit are really useful to our design. 
\subsection{CivKit}
CivKit, short for Civilization Kit, is an upcoming white paper from \href{https://www.commerceblock.com/}{Commerceblock}, discussing a decentralized and unstoppable free market solution based on Bitcoin. The project aims to build on top of Bitcoin to create an environment where anyone can trade anything with anyone else. \par
Phase one focuses on creating a marketplace built on top of Nostr, an interoperable communication protocol. This allows different services like Paxful, HODL HODL, or Nostr app to communicate and operate across each other.\par
Phase two aims to develop a mobile-friendly lightning wallet and decentralized IDs (Know Your Peer) to replace centralized KYC (Know Your Customer). This will provide a more secure and private environment for traders.\par 
CivKit is intended to be an open-source decentralized toolkit that various brands and platforms can build on top of. The goal is to facilitate peer-to-peer trading and encourage a more circular economy where people earn and spend Bitcoin rather than buying and selling it. While details are sparse it seems possible that this technology can be integrated into our systems.

\subsection{Nostr}
Nostr [pronounced no-star] is a \href{https://www.nostr.how/}{decentralized open protocol} that aims to improve the social media experience by addressing issues of censorship and data collection. The protocol operates by allowing users to post and view notes on servers called relays, and view and post these notes through apps called clients. The open nature of the protocol allows for competition and a free flow of information, as users can choose to use different relays or clients if they are censored. This is because the protocol is decentralized and controlled by no one.\par
The decentralized nature of Nostr means that there is no central authority that can control the flow of information. This is achieved through the use of relays and clients, which are run by different individuals or entities. Users have the freedom to choose which relays and clients they want to use, and as a result, their feeds are populated with content from the people they choose to follow. If a relay or client tries to censor a user, they can simply switch to a different one. This is a major advantage over traditional centralized social media platforms where one entity holds all the control over the flow of information and can censor or manipulate the content that users see.\par
Nostr is also not beholden to shareholders or investors. This means that the protocol can make decisions that prioritize the well-being and quality of discourse for users, rather than solely focusing on profit. This is in contrast to traditional social media networks like Twitter, Facebook, and TikTok, which are driven by the need to collect data on users and sell ads to generate revenue. In these centralized platforms, users' data is collected, analyzed and sold to the highest bidder, often without the user's knowledge or consent. Nostr, on the other hand, allows users to have more control over their data and the ability to monetize their content.\par
Nostr also tightly integrates Bitcoin Lightning to support the protocol. This will hopefully enable secure transmission of value alongside the information and interactions on the platform. It also gives users the ability to monetise their content. \par
This potential step-change improvement to the social media experience for everyday people addresses issues of censorship and data collection. \par
Nostr is ``The simplest open protocol that is able to create a censorship-resistant global "social" network once and for all.'' according to it's \href{https://github.com/fiatjaf/nostr}{github page}. More than that it's a client side validated proof of who a user is interacting with, hence being in this identity section. To be clear, it's not a completely peer to peer system in that it uses (very dumb) relay servers, but this gives it some of the best characteristics of both paradigms. This has the following advantages for our metaverse application; 
\begin{itemize}
\item it's lightweight, with minimal network overhead and complexity
\item it's real-time using websockets
\item anyone can run a relay server, so one can be run in the deployment in the final section of the book.
\item Each of the client peers connecting to the metaverse can be a relay and able to pass messages and proofs to the other clients without the metaverse server seeing the data or being online 
\item it's open-source
\item it is itself \href{https://snort.social/e/note1evkdgcg0dw8ckyqsqhnk4wy55h7w97twjf8etcq2tr9sv5urlszqjc9p6v}{Turing Complete} and therefore able to execute any code within it's message protocol
\item there are multiple usable libraries and tools
\item it's under active development with an excellent team. The lead, `Fiatjaf' is one of the most \href{https://github.com/fiatjaf}{prolific developers} in the lightning space.
\item it's based on the same underlying cryptographic technology we are using elsewhere, indeed with it's use of Bitcoin keys the identity system is global
\item it provides the identity proof that we need to validate users and objects into a virtual space
\item it enables message passing
\item it scales to be a social network as required
\item it need not rely on anything outside of a relay hosted on the metaverse server
\item it can be scaled to provide one to many bulletin board style applications within the metaverse
\item we can use it in private, group, and public modes as required
\item it integrates with the torrent network allowing storage and external referencing of arbitrary data
\item it can easily operate outside of the walled garden of the metaverse, extending the reach of the messages
\end{itemize} 
\href{https://www.forbes.com/sites/rogerhuang/2022/12/29/nostr-is-the-decentralized-protocol-that-might-replace-elon-musks-twitter/}{Nostr is incredibly promising}, and integrating these relays in the metaverse servers and clients of the proposed technology stack in this book might allow us globally provable identity, with privacy by design. It can provide message passing. If all entities in the collaborative mixed reality scenegraphs are also Nostr key pairs then schema can be applied consistently with the economic layer using the same key system as Bitcoin. Nostr has just received a substantial grant from Dorsey. It is core to the design later in the book. A curated list of projects and libraries is \href{https://github.com/aljazceru/awesome-nostr}{available on github}.\par
\begin{figure}
\includegraphics[width=\linewidth]{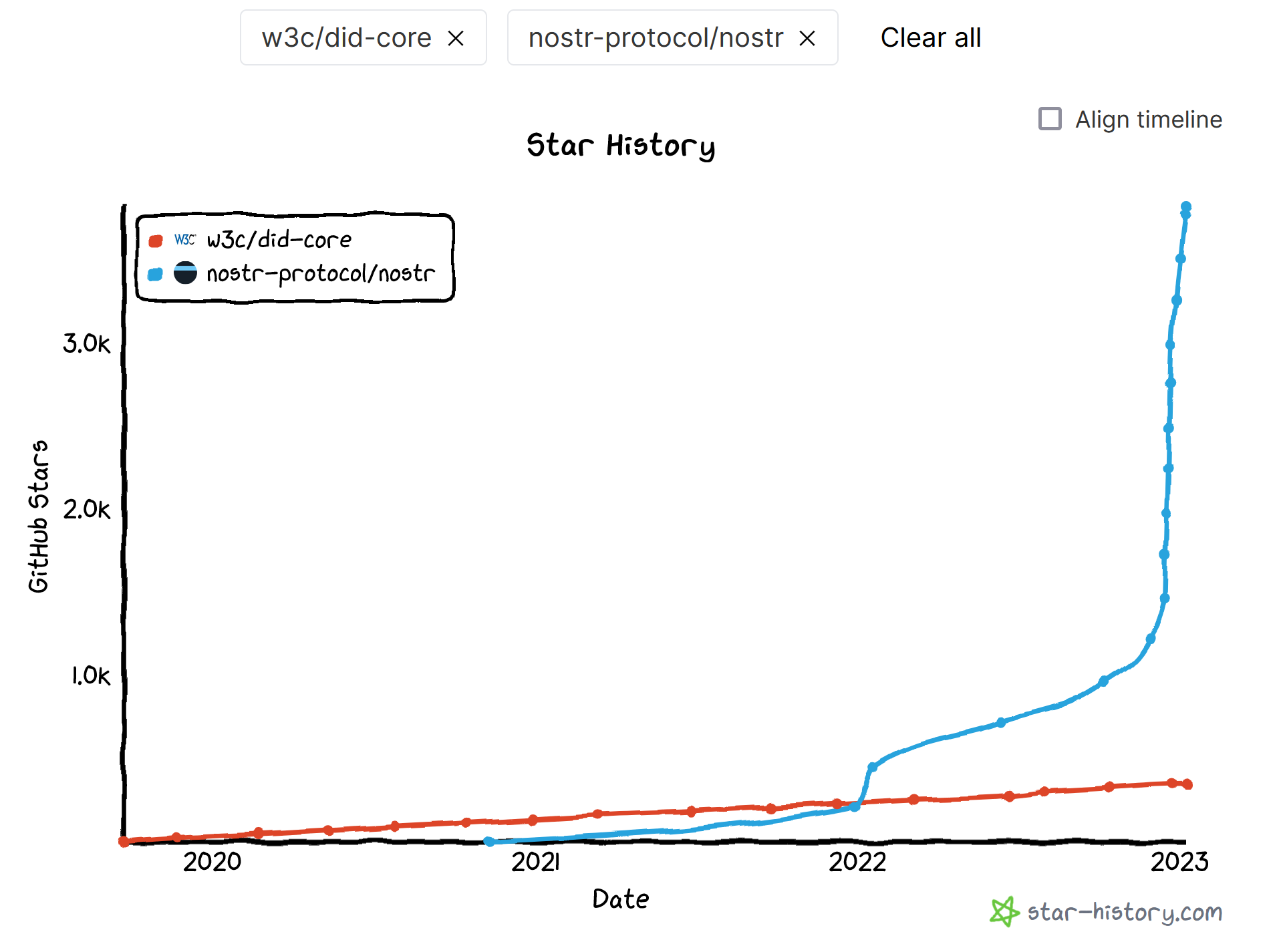}
  \caption{An illustration of the enthusiasm for Nostr compared to traditional DID based on GitHub `stars'.}
  \label{fig:starhistory}
\end{figure}

Luke Childs \href{https://github.com/nostr-protocol/nips/issues/154}{says}:
\textit{``Nostr makes a good candidate to be used as a very simple DID layer. Having "Login with Nostr" auth on websites solves a lot of problems in a very elegant way, and Nostr's main use case as a social network protocol makes it highly suited to be used as your main identity proving key. Compare "Login with Nostr" to similar "Login with Lightning" (LNURL-auth) specs to see some easy and obvious advantages:\\
Remote signer vs local signer\\
Login with Lightning requires access to remote keys, login with Nostr requires access to local keys ideally stored in a browser extension. Due to the way Lightning works you can only really have one instance. You need all your client devices linked to a single Lightning node, this means most clients will be connecting to the signer remotely. Now if your Lightning node goes down or you lose your connection you also can't auth with any service. This could cause circular dependencies where you lose the connection to your Lightning node so you can't auth with the services you need to access to debug the issue with your Lightning node like your hosting provider or VPN account. You could technically solve this by replicating your LN keys to other client devices only to be used for local auth signing but that introduces other risks.\\
Unique identifier vs identity\\
A Lightning node is not really an identity but a unique identifier. It just tells you the person that auths is the same random person that authed last time, it doesn't tell you who they are. A nostr pubkey is an identity. It tells you who they are, what their name is, what they look like, who they know, how you can pay them, how you can message them.\\
This is much more useful as an identity layer for an application. The application can show their profile picture, username, send secure cross platform push notifications via NIP-04 encrypted Nostr DMs, etc.\\
Consistent identity across services\\
Lightning pubkeys are sensitive private information and can leak confidential financial information, Nostr pubkeys are safe to share with anyone. LNURL-auth adds extra steps to solve this by creating derived subkeys for identities that are unique to each service you auth with. This does not seem ideal, it seems the default case is that an identity is something that you do want to follow you across all your accounts. Nostr based auth behaves more appropriate in this regard. In the rare case you need to achieve privacy and separation between certain services you can still do that by using use a throwaway Nostr key for those services.\\
User relationships across services
Since authing with Nostr shares a real social identity with the service, they can also see your Nostr social graph. This could be useful for connecting you to people you already know on the new service.\\
Low cost identity\\
Ideally identities should be easy to create but hard to build up reputation to limit spam while avoiding excluding people from the network. It's not clear that it will be cost effective / scalable for everyone to run their own Lightning node so tying individual identity to a single Lightning node pubkey is problematic. Nostr keys are easy to create and hard reputation can be earned via PoW/DNS or building a strong social graph.''}\\
Figure \ref{fig:starhistory} shows that the adoption is potentially tremendously fast.
This provides a web interface into the metaverse providing:
\begin{itemize}
\item simple cryptographic identity assurance
\item private peer to peer chat
\item group chats and channels
\item email to private message relay
\item links into media on web hosts
\end{itemize}
The pace of development on Nostr is dizzying. Peer to peer video and audio will allow us to link metaverse instances, between peers, through applications such as \href{https://monstr.app/}{Monstr}.\par
It's notable that Nostr has it's own inexpensive \href{https://github.com/lnbits/nostr-signing-device}{hardware signing device} to protect identity in situations where this might be necessary.\\
\textbf{The proposed integration of Nostr social media and messaging, a lightning layer with digital objects such as Fedimint, Zerosync or RGB, AI agents, Vircadia, and federated Bitcoin is the core value proposition of this
book.} This work pre-dates \href{https://www.theverge.com/2023/4/26/23699633/mark-zuckerberg-meta-generative-ai-chatbots-instagram-facebook-whatsapp}{Meta and Zuckerbergs} stated intent in this regard by 18 months, and is differentiated still by our focus on emerging markets and decentralisation.
\subsubsection{NIP-05}
At this time, the nascent identity layer in nostr leans on NIP-05. This is a distributed identity management system that maps Nostr keys to DNS-based internet identifiers. In events of kind 0 (setmetadata), the ``nip05'' key can have an internet identifier as its value. Clients split the identifier into the local part and domain and make a GET request to the specified URL. The response should be a JSON document with a ``names'' key containing a mapping of names to hex-formatted public keys. If the public key matches the one from the setmetadata event, the client accepts the association and considers the ``nip05'' identifier valid.\par
Clients may find users' public keys from internet identifiers by first fetching the well-known URL and then checking for a matching ``nip05''. When following public keys, clients must prioritize the keys over NIP-05 addresses. Public keys must be in hex format. Clients can enable user discovery through search boxes, allowing users to find profiles by entering internet identifiers. The identifier can be used as the ``root'' identifier, displayed as just the domain. The protocol supports both dynamic and static servers by using the local part as a query string.
\subsubsection{Nostr Protocol as the keystone}
The Nostr protocol can be used to store and share valuable content across the network. This is ably demonstrated by the \href{https://highlighter.com/}{`Highlighter' project} which allows users to store important notes from around the web using nostr. In the context of our federated social media trust model, the Nostr protocol can serve as the underlying layer that connects various instances of virtual spaces, thus enabling seamless data exchange and interoperability among them. Highlighter demonstrates that nostr events can be leveraged to create, store, and interact with valuable across networks. By utilizing this concept, we can extend the functionality to support federated social media trust, allowing users to carry their reputation, identity, and cryptographic proofs across different virtual spaces and social media platforms.
\subsubsection{Nostr marketplace in LnBits}
The nostr \href{https://github.com/lnbits/nostrmarket}{markets plugin} for LnBits allows virtual `stalls' to be setup and payment to be mediated through nostr. This is obviously a great expansion to the usefulness of our integration
\subsubsection{Integrating Cryptographic Proofs and Reputation}
To create a trusted environment within the federated network, we must establish a mechanism for importing and verifying cryptographic proofs from various sources, such as social media sites and other digital platforms. By doing so, we enable users to bring their existing reputation and trust from these platforms into the new ecosystem, thus facilitating trust-based interactions and collaboration within the network. We can leverage the Nostr protocol and the NIP05 specification to import these cryptographic proofs, creating a secure and verifiable system for identity management and trust propagation. The NIP05 specification allows for the creation and verification of identity proofs within the Nostr protocol, thus enabling the seamless integration of trust and reputation data from external sources.\par
By utilizing the Nostr protocol as the underlying layer, we can establish connections between objects, people, and AI actors within the federated network. This interconnected ecosystem allows for seamless collaboration, information sharing, and trust-based interactions among all participants. The open-source collaboration infrastructure we propose can facilitate the development of various applications and services that leverage the federated network, such as virtual workspaces, AI-assisted creativity tools, and more. The uncensorable nature of this protocol further supports the inclusivity and accessibility we feel so important, ensuring that participants from different regions and backgrounds can take part in the digital society and contribute to its growth.\par
This federated social media trust model, built on the Nostr protocol, allows for the establishment of a robust, inclusive, and trust-based network that connects various virtual spaces, social media platforms, and AI systems. By leveraging the lessons learned from the other attempts in the space, and by maximising the inclusion of external cryptographic proofs from multiple sources, we can create a comprehensive trust system that fosters collaboration, innovation, and shared growth within the digital society.

The Nostr protocol, with its decentralized and open-source nature, provides a solid foundation for linking and federating objects, people, and AI actors across collaborative spaces in digital society. By leveraging the Nostr protocol, we can build a robust and trust-based network that interconnects various virtual spaces, social media platforms, and AI systems. One of the key aspects of this trust-based network is the ability to import cryptographic proofs from different sources, similar to Keybase's approach to importing proofs from various social media sites \href{https://book.keybase.io/account#proofs}{(Keybase Proofs)}.

\subsubsection{StrFry relays}
The Stirfry relay software provides high-performance infrastructure for building decentralized social media applications on top of the Noster protocol. As an open source project written in C++, Stirfry emphasizes efficiency, flexibility, and community-driven governance.

At the core of Stirfry is its high-speed database engine. Rather than using a traditional SQL database, Stirfry implements the Lightning Memory-Mapped Database (LMDB) - an embedded key-value store optimized for performance. Reads are lock-free, enabling unlimited parallel query throughput. Writes require only a short-held lock, ensuring minimal interference. LMDB's "shadow paging" design allows isolated read-only transactions via multi-version concurrency control (MVCC). This prevents reads from blocking writes and vice versa.

To maximize database performance, Stirfry stores Noster events directly in FlatBuffers - an efficient binary format allowing direct access without serialization. The original JSON payloads are preserved to facilitate transmission back to clients. Additional database files index events on fields like timestamps and authors, accelerating filter queries. Periodic compaction optimizes the layout for faster operations.

Stirfry adopts a multi-threaded, modular architecture. A websocket thread accepts new client connections and routes incoming requests. An ingester thread validates and pre-processes each request before passing to appropriate handlers. Doing signature checks and filter compilation upfront avoids repeating work. A single writer thread batches database writes to amortize transaction overhead. Multiple worker threads handle read queries, fairly scheduling between long and short requests. Dedicated monitor threads track active filters and stream matching events to subscribed clients. Passing messages between threads instead of sharing data structures improves efficiency.

Additional features further enhance Stirfry's capabilities. Graceful shutdown support allows stopping new connections while existing ones complete. Hot configuration reloading provides runtime updates without restarting. Flexible write policy plugins enable custom content moderation. Streaming websocket compression and Zstandard dictionaries compress traffic. Syncing protocols like Negentropy facilitate efficient relay replication, powering mesh network topologies. Geo-replication by the relay.org community offers low latency worldwide access. The custom Templar HTML templating library assists crafting simple, fast decentralized frontends.

\section{Micropayment based web}
It seems the war against disinformation is now being lost. Much is written in the media about Deepfake technology creating plausible fake videos, but probably more pernicious is the use of toolkits to create entire plausible fake news sites using natural language AI such as GPT3. This makes it cheap to publish potentially market moving news which is then rehypothecated by online news vendors who are hungry for clicks. As these pipelines become more mature it will be difficult to keep fake news for financial or political gain out of the system. One interesting way to do this that \textit{isn't} webs of trust or true cryptographic identity is to charge micropayments for ``one to many'' publication models. This would imply a tiny instant payment for clicks, especially on social media sites such as twitter. This kind of model has been discussed but is only possible in the context of systems such as Lightning where instant micropayment can be realised. It seems possible that this would price out speculative `noise' spam from the information space. It's interesting and ironic that the origin of proof of work was to underpin just such a spam defeating system  \cite{dwork1992pricing}, and that Nakamoto \href{https://www.metzdowd.com/pipermail/cryptography/2009-January/015014.html}{mentioned this application for Bitcoin} back in 2009. There is now much chatter about the integration of Bitcoin with Twitter in light of Musks buyout of the social network.
\begin{figure}
\includegraphics[width=\linewidth]{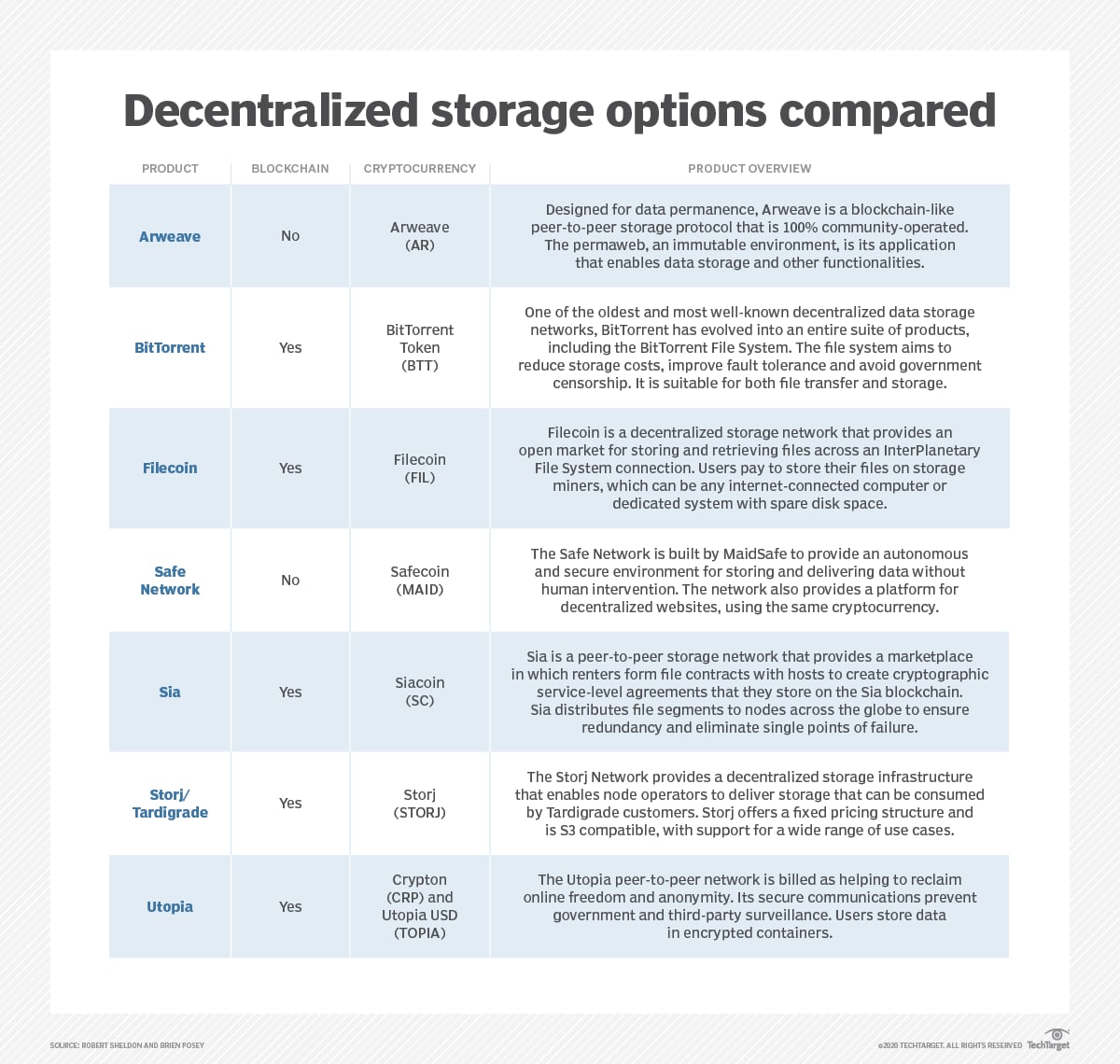}
  \caption{Comparison of distributed file stores}
  \label{fig:Files}
\end{figure}. 
\section{Are DAOs useful for us?}
A distributed autonomous organisation, or DAO is a governance structure which is built in distributed code on a blockchain smart contract system. Token holders have voting rights proportional to their holding. The first decentalised autonomous organisation was simply called ``The DAO'' and was launched on the Ethereum network in 2016 after raising around \$100M. \href{https://www.gemini.com/cryptopedia/the-dao-hack-makerdao#section-what-is-a-dao}{It quickly succumbed to a hack and the money was drained}. This event was an important moment in the development of Ethereum and resulted in a code fork which preserves two separate versions of the network to this day, though one is falling into obsolescence. Again, this is covered in Shin's book on the period in extreme detail, but it seems this stuff is falling into dusty history now, leaving only a somewhat tarnished and technically shaky legacy \cite{cryptopians}. \\
In practice DAOs have very few committed `stakeholders' and the same names seem to crop up across multiple projects. Some crucial community decisions within large projects only poll a couple of dozen eligible participants. Its might be that the experiment of distributed governance is failing at this stage. \par
Perhaps more interesting is the use of the DAO concept to crowd fund global projects, currently especially for the acquisition of important art or cultural items. DAOs are also emerging as a way to fund promising technology projects, though this is reminiscent of the 2017 ICO craze which ended badly and is likely to \href{https://www.cftc.gov/PressRoom/PressReleases/8590-22}{fall foul of regulations}.\par
Within the NFT and digital art space  PleaserDAO has quickly established a strong following.
``PleasrDAO is a collective of DeFi leaders, early NFT collectors and digital artists who have built a formidable yet benevolent reputation for acquiring culturally significant pieces with a charitable twist.\par
Opensea wrangle between IPO and governance token.\par
ConstitutionDAO, Once upon a time in Shaolin etc 

\subsubsection{Problems experienced to date}

DAOs exist in a gray area with unclear legal recognition, leading to challenges in fitting into existing legal frameworks and regulatory systems. This lack of clarity raises concerns about how DAOs can comply with laws and regulations, potentially leading to legal disputes or conflicts with regulatory authorities.
Security Risks and Technological Vulnerabilities:

Given their reliance on blockchain technology and smart contracts, DAOs are vulnerable to cyber threats, such as hacking and exploitation of code weaknesses. The decentralized nature of DAOs can further complicate security management and responses to breaches, raising concerns about the safety of assets and data managed by DAOs.
Governance Inefficiencies and Democratic Deficiencies:

The non-hierarchical structure of DAOs, while innovative, may lead to governance challenges, including potential inefficiencies in decision-making processes. There's also a risk of DAOs deviating from democratic principles, possibly leading to control by a limited group of technologically adept individuals (technocracy) or oligarchic tendencies, which could marginalize participants who are less tech-savvy.
Fragmentation and Complexity in Governance Models:

The existence of multiple, simultaneous governance models within DAOs can lead to fragmentation, resulting in a lack of coherent and unified governance. This complexity can create confusion among participants and hinder effective decision-making, posing challenges to the democratic functioning and overall effectiveness of DAOs.

\subsection{DAOs on Bitcoin}
\subsubsection{Bisq DAO}
One of the better designed DAOs is \href{https://bisq.network/dao/}{Bisq DAO}. It's slightly different design trys to address the issue of overly rigid software intersecting with more intangible and fluid human governance needs. From their website:\par
\textit{``Revenue distribution and decision-making cannot be decentralized with traditional organization structures they require legal entities, jurisdictions, bank accounts, and more—all of which are central points of failure.
The Bisq DAO replaces such legacy infrastructure with cryptographic infrastructure to handle project decision-making and revenue distribution without such central points of failure.''}
\subsubsection{Stackerstan}
Stackerstan is a layer two protocol that operates on top of the Bitcoin and Nostr protocols. It aims to provide a decentralized and efficient platform for people to collaborate and build valuable products and services, without the need for agreement on what to build or how to build it, in a fully decentralised way.\par 
Github contributor GazHayes \href{https://github.com/Stackerstan/interfarce/issues/20#issuecomment-1369329734}{has a writeup} which is paraphrased below, explaining this very new and emergent technology stack.\par
The Stackerstan protocol is designed to be infinitely scalable, due to the absence of ``organizational mutexes''.\par 
Stackerstan was anonymously posted in the Nostr telegram group at the end of 2022 and is a new project that aims to offer a more efficient and decentralized way of solving problems compared to existing companies, institutions, and decentralized organizations. It utilizes a combination of existing technologies, protocols, and concepts to create a system that allows people to spontaneously organize into highly efficient and intelligent groups. The platform is designed to be fair to everyone involved and is completely non-custodial, meaning that it doesn't require a shared pot of money or any funding.
\begin{itemize}
\item Anyone can become a participant in Stackerstan by being added by an existing participant, creating a tree of participants that can be severed if a bad actor is present.
\item Work is done within Stackerstan by continuously identifying problems and applying the simplest possible solution to these problems, expanding the scope of what Stackerstan can do.
\item Any participant can log a problem and claim it to solve it, and the scope of what can become a problem to solve is not limited.
\item Shares are created by a participant filing an expense to indicate the relative value of their work, which is a request to be repaid when Stackerstan generates revenue.
\item Shares are approved expenses, and the only way for new shares to be created is by approving expenses for work done to solve problems.
\item Participants with shares can vote to approve or reject new expenses, and there are rules to follow when voting on expenses.
\item Stackerstan was created at block 761151 and has a single share to bootstrap the process, with a small number of shares created by approved expenses so far.
\item Shareholders own all revenue generated by Stackerstan's products and services, and revenue is distributed through two algorithms: first, paying back expenses in the order they were filed, and second, streaming dividends to whoever has received the least dividends per share owned.
\item Stackerstan is non-custodial and does not require a shared pot of money, making it more effective and avoiding toxic situations.
\item Voting on things like approving expenses is done with votepower, which quantifies a participant's skin in the game.
\item Lead time is a measure of a participant's votepower and can be increased or decreased by one unit every 2016 blocks.
\item A participant's shares can only be transferred if their lead time is 0, and a participant can reduce their lead time to sell their shares.
\end{itemize}
\subsubsection{Mindmachine}
The Mindmachine is a stateful Nostr client written in Go. This text is directly quoted from the \href{https://github.com/gazhayes/mindmachine}{GazHayes github}.
\begin{itemize}
\item Participants interact with the Mindmachine using Nostr Events. The Mindmachine subscribes to all Nostr event Kinds that it can handle, and attempts to update its state by processing them based on the rules in the Stackerstan Superprotocol.
\item If an Event successfully triggers the Mindmachine to change state, the Event ID is appended to a Kind 640001 Nostr Event which the Mindmachine publishes once per Bitcoin block. 
\item The Mindmachine can rebuild anyone's state by subscribing to their 640001 events and replaying the list of Nostr Events contained within.
\item Consensus is based on Votepower. When a Participant with Votepower greater than 0 witnesses a new Mindmachine state, the Mindmachine hashes the state and publishes it in a Kind 640000 Nostr Event. This is effectively a vote for the witnessed state at a particular Bitcoin height.
\item A Mindmachine state is considered stable when in excess of 50\% of total Votepower has signed the same state and there is a chain of signatures back to the Ignition state. There are mechanisms to deal with voters disappearing.
\item Participants who have a lot of Votepower will want to be able to prove they had a certain Mind-state at a particular height. To do so, they broadcast a Bitcoin transaction containing an OPRETURN of the state.
\end{itemize}

Because of the tight integration with Nostr it seems that is we were to allocate work to open communities then this would be the way to do it.

\subsection{Risks}
The most interesting thing about DAOs is that they belong more in this money chapter than they do in blockchain. As we have seen they're finding most success as loosely regulated crowd funding platforms. If a small company did find itself wishing to explore this fringe mechanism for raising capital, then we would certainly recommend keeping a global eye on evolving regulation and the onward legal exposure of the company. 
\section{Risks \& Challenges?}
Classic DID/SSI risks fragmentations. 
In all DID applications, scaling to a world where the user is managing potentially thousands of these critical cryptographic data files is daunting.
Abstracting the guts of this away to make the use simple, and only mindful of thet right level of information, turns out to be huge problem that nobody has solved
It's not clear that users want this. 
In the case of web of trust like Slashtags it's a big piece of work for the users to rate all of their digital interactions with a trust metric.

\chapterimage{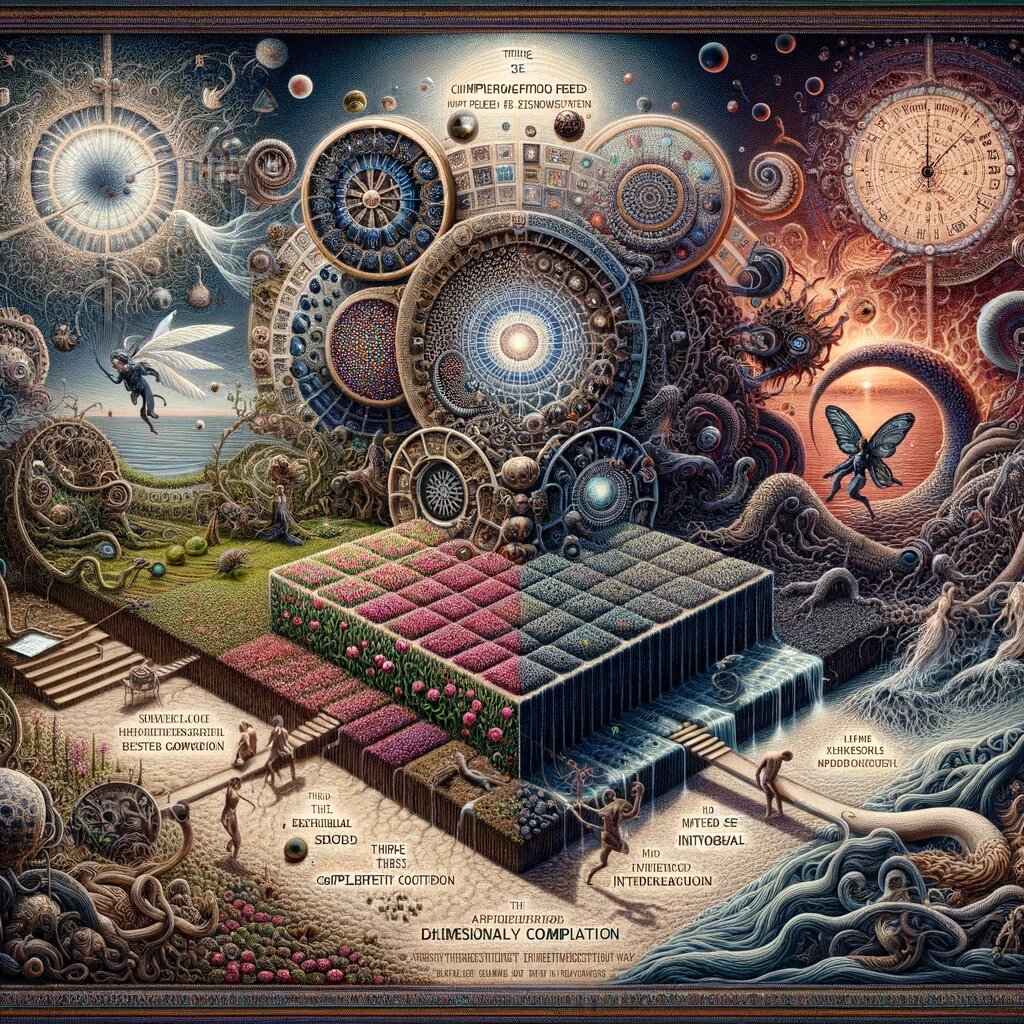}
\chapter{Digital Objects \& NFTs}
Nonfungible tokens are a whole `class' of digital token, separate and distinct from everything discussed to this point. They are generally \href{https://www.signaturelitigation.com/nfts-recognised-as-property-lavinia-deborah-osbourne-v-1-persons-unknown-2-ozone-networks-inc-trading-as-opensea/}{recognised in law} as property in their own right \cite{moringiello2021property, fairfield2021tokenized}. In the Initial Coin Offering (ICO) and project tokens detailed earlier, and limiting this description to the Ethereum network for now, a project launching an ERC-20 token commits contract code to the blockchain, and this contract then mediates the issuance and management of millions or billions of tokens associated with that project, and it's use case. \href{https://ethereum.org/en/developers/docs/standards/tokens/erc-20/}{ERC-20} is a \href{https://en.wikipedia.org/wiki/Fungibility}{fungible} token issuance. Each of the projects' tokens is interchangeable with any other token. They're all the same from the point of view of the user.\par
Rather than the ERC-20 contract type used for fungible token issuance NTFs predominantly use ERC-721 protocol on Ethereum (just different instructions). It's the case that most NFTs in the 2021/2 hype bubble are algorithmically generated sets of themed art (so called PFP-NFT). Tens of thousands of distinct tokens are `minted', each one being a complex transaction commitment to the Ethereum blockchain, along with it's associated gas fee. These minting events were much hyped social occasions (before the \href{https://www.theguardian.com/technology/2022/jul/02/nft-sales-hit-12-month-low-after-cryptocurrency-crash?}{2022 market crash}), and happened very quickly, with users clamouring to create art with randomly allocated features from the art schema associated with the project. Lucky winners could find themselves with an NFT art piece with more than an average number of `rare' features. If the overall mint becomes more popular, then the secondary market for all of those mints goes up, and because of the liquidity premium they can go up a lot. The perceived rarer mints go up a lot more. This whole process is \href{https://memoakten.medium.com/the-unreasonable-ecological-cost-of-cryptoart-2221d3eb2053}{very energy intensive} on the chain, and the vast majority of these project simply \href{https://www.turing.ac.uk/blog/non-fungible-tokens-can-we-predict-price-theyll-sell}{trend to zero value}. In response to this appalling cost benefit analysis the Ethereum foundation have proposed \href{https://eips.ethereum.org/EIPS/eip-2309}{EIP-2309} to make minting NFTs more efficient. They say ``This standard lets you mint as many as you like in one transaction!''\par
\begin{figure*}[ht]\centering 
	\includegraphics[width=0.7\linewidth]{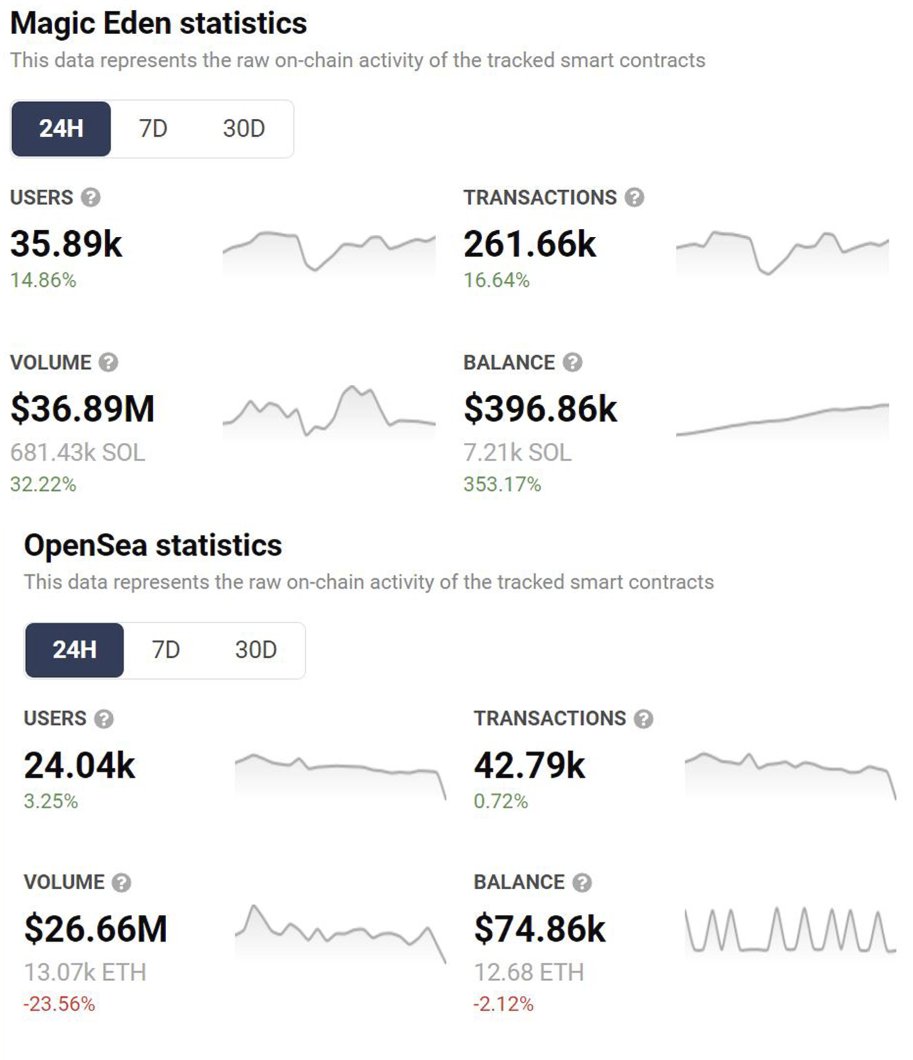}
	\caption{Solana NFT markets are enjoying growth compared to Opensea on Ethereum, even in the downturn.}
	\label{fig:solnfts}
\end{figure*}

The Ethereum foundation give their somewhat constrained view of \href{https://ethereum.org/en/nft/}{NFTs on their website} and it's a useful primer. On that page they detail some of the use cases, as listed below, with a critique added:
\begin{itemize}
\item Digital content; this is the dominant use case right now. Much more on this later.
\item Gaming items; again more on this later, it's an obvious enough use case but \href{https://climatereplay.org/nfts/nft-digital-ownership-pledge/}{complex politics} in the intersection of games and crypto have stalled the adoption curve.
\item Domain names; this is just starting to reach for applications now, why not a database with the ISP/host?
\item Physical items; seemed like a clear over-reach as transfer of the NFT does not imply transfer of the object, but this is emerging as the growth use case.
\item Investments and collateral; while this was an emergent option in the space, it's likely been a bubble, as owners of the tokens cast around for additional liquidity, and loan businesses chased yield with higher risk. The \href{https://newsletter.banklesshq.com/p/three-arrows-capital-grayscale-maker-lido}{recent implosion} of lenders and funds in the crypto space was partly a function of supposedly world class risk managers accepting jpegs as collateral.
\end{itemize}
Moving away from Ethereum, NFTs can be minted on most of the other level one chains. Solana is a great newcomer example. Sol is a terrible chain with regards to decentralisation, but thanks to that it's far cheaper and faster to mint NFTs on it, and it was becoming a \href{https://markets.businessinsider.com/news/currencies/ethereum-eth-killers-nfts-defi-solana-cardano-wax-crypto-investing-2022-1}{troubling competitor} for Eth before the FTX ponzi scheme collapse destroyed it's market value (Figure \ref{fig:solnfts}).\par 
The same might be true for Cardano's ADA, though ADA is struggling to hold onto it's market position despite some technical advances. It's worth reiterating here that the nature of these digital tools likely makes for a `winner take all' market dynamic over time. With fees being central to this generative NFT use case it's possible to see that highly centralised, fast, and cheap chains will capture and eventually dominate the space. Remember that this likely (game theoretic) outcome might as well be a database running without the stark inefficiencies of blockchain. The whole NFT space is a gamble on consumer enthusiasm for spending money continuing to outpace logic.\par 
Astonishingly, according to a JPMorgan insider market report (\href{https://www.coindesk.com/podcasts/the-breakdown-with-nlw/jpmorgan-bitcoin-shows-some-merit-as-a-store-of-value/}{reported on in a podcast}), only around 2 million people have ever actually interacted with NFTs. One analysis suggests that a single entity accounts for 3 of the top 4 holders, having made 32,000 ETH from the NFT boom. This suggests heavy market manipulation and is far from the egalitarian landscape claimed in the hype. Tellingly it's thought around \href{https://uk.finance.yahoo.com/news/three-arrows-wanted-100m-nft-161811450.html}{10\% of the trading volume} on market leading platform `Super Rare' was by the now bankrupt venture capital firm `Three Arrows'. \par
With that said NFTs have clearly allowed \href{https://en.wikipedia.org/wiki/List_of_most_expensive_non-fungible_tokens}{digital and new media artists} to connect with audiences without gatekeepers. Established mediators and curators of art have been caught totally wrongfooted, and NFTs seem to give a way for them to be cut out completely. There are suggestions of applications beyond this initial digital art scope. This is a compounding, and disrupting paradigm change.\par
\section{Key use cases}

\subsection{Art}
\begin{figure*}[ht]\centering 
	\includegraphics[width=\linewidth]{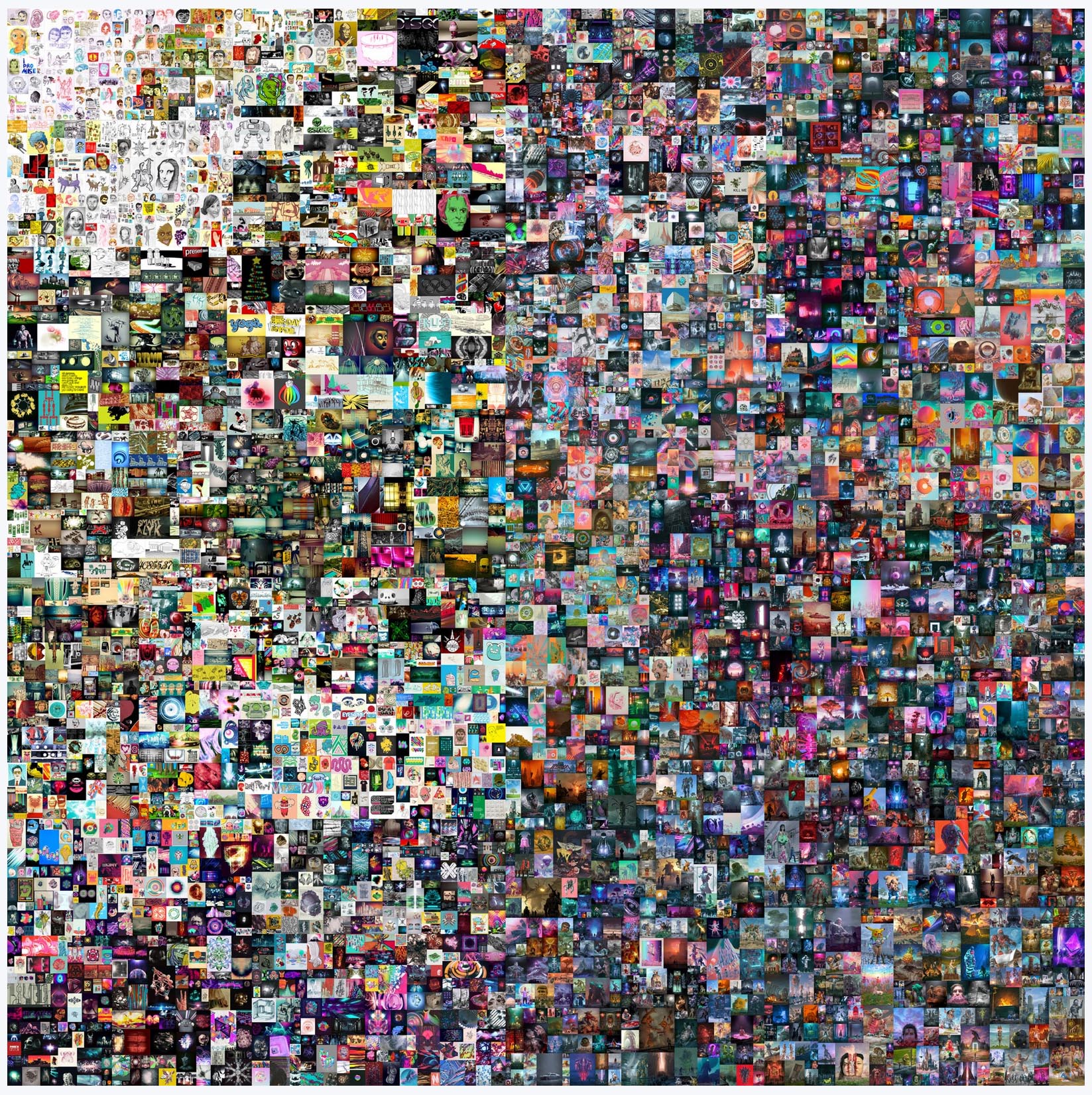}
	\caption{Beeple: First 5000 days, \href{https://onlineonly.christies.com/s/beeple-first-5000-days/lots/2020}{taken from the Christies website, assumed fair use}.}
	\label{fig:first5000days}
\end{figure*}

The recent surge of interest in NFT's during early 2021 has largely been
driven by digital art NFT's, despite the origins of digital art NFT's
started much earlier in 2014. New York artist \href{https://www.mccoyspace.com/project/125/}{Kevin McCoy's
\emph{Quantum}} is widely recognised as the first piece of art created
as an NFT. However it was during early 2021 that art NFT's started to
gain significant attention; by the end of 2021, nearly \href{https://www.paymentscardsandmobile.com/state-of-the-blockchain-nfts-explode-onto-scene-in-2021/}{£31b
had been spent} on NFT purchases, a considerable and exponential growth
given \href{https://raritysniper.com/news/nfts-exploded-in-2021-with-25-billion-in-sales/}{2020
sales of \textasciitilde£71m}
High profile digital artists such as \emph{Beeple} whose
\href{https://www.forbes.com/sites/abrambrown/2021/03/11/beeple-art-sells-for-693-million-becoming-most-expensive-nft-ever/?sh=3f237d1c2448}{recent
recording break sale} of his NFT \emph{``The first 5000 days''} (Figure \ref{fig:first5000days}) at Christies (a long established British auction house,
specialising in high profile precious work of art) for £52.9m helped
bring NFT's into the public spotlight and wider give them global
attention.

Art as NFT's offer the following advantages:

\subsubsection{Immutable Nominal Authenticity} 
Art fraud such as false
  representation, forgeries, plagiarism have been a reoccurring blight
  since art has existed; artists and works of art have been open to
  abuse by forgers, black market profiteers and even fellow artists
  laying claim to works of art of others. Unless a work of art is sold,
  exhibited or listed, documenting when and who created it, the
  \emph{nominal authenticity,} which Dutton states as the
  \emph{``correct identification of the origins, authorship, or
  provenance of an object''} \cite{dutton2003authenticity} can be increasingly mutable over a period
  of time, dependent on a multitude of factors, including; the artists
  existing profile, how widely and where the work of art is exhibited,
  if the work of art is commissioned by a patron, if it's sold, and
  profile of the buyer/collector. At its most basic level, once a work
  of art is `minted' as an NFT (publishing the art work as a unique
  token on the blockchain) this functions as an immutable publicly
  accessible proof of ownership and by extension proof of creation. The
  act of minting is not purely limited to digital art; all an artist
  requires is a digital representation of any physical art (sculpture,
  physical painting, installation etc..) which can be used as a proxy
  allowing artists to record the date of creation/origin of a physical
  piece of art on the blockchain, a buyer purchasing the NFT can be
  provided the actual physical artwork as part of the NFT. Nominal
  authenticity becomes secure and immutable for the lifetime of the blockchain (by no means assured).

\subsubsection{Secure Digital Provenance}
\href{https://en.wikipedia.org/wiki/Provenance}{Provenance} (or the chain of custody) is an important aspect in works of art, antiques and antiquities. Provenance not only helps assign work to an artist but also documents ownership history. Digital provenance, an inherent feature of NFT's means provenance now no longer becomes what has historically sometime been a contentious detective's game at the best of times; one that is open to fraud, misinterpretation and entirely reliant on good record keeping.\par
Since provenance can contribute to the value of a piece of art (benefiting both the creator and collector) the use of the blockchain as an open, secure ledger is a far more trustworthy system than traditional methods of artistic provenance that were cobbled together; often
consisting of a mix of physical and digital documents spanning private
\& public sale receipts, art/museum gallery exhibitions and private
record keeping). Digital provenance provided when an artist `mints' a
piece of art into an NFT allows artists and collectors to record a
secure, permanent unalterable history of transactions for a specific
piece of art, providing future collector complete trust in the origin
and custody of a piece of art.
\subsubsection{Decentralised automated royalty payments} 
Traditionally if a  piece of art is sold, the first sale may (but not always) benefit the
  artist financially, however secondary and any subsequent sales would
  only ever financially benefit the buyer/collector; the original artist
  would rarely benefit. However If a work of art is minted into an NFT,
  royalty payments can be predetermined and automated in perpetuity
  directly by the use of a `smart contract'. Smart contracts are small,
  automated scripts/programs that run automatically and independently of
  a buyer/seller; pre-determined conditions are set by the buyer; these
  trigger when certain conditions are met i.e. These cannot yet be enforced ``on chain'' and the NFT auction houses online have engaged in a race to the bottom and stopped enforcing royalty payments through their systems. This element might not even be possible, though there is some hope that we could enable this in the complex logic offered by the RGB protocol.
\subsubsection{On sale transfer}
20\% of total sale amount into digital wallet of the creator.
80\% of total sale amount into digital wallet of the seller.

Once the royalty payment rate is set by the artist/creator, future
royalties of all sales can be paid directly to the artist/creator
account (via a digital wallet) without the need of a third party
(traditionally a gallery/agent etc..).

Smart contract driven NFT's means that even if piece of art is resold 5,
10 or even a 100,000 times moving through 5, 10 or even a 100,000
different collectors; a pre-determined royalty payment rate set by the
creator would still guarantee the artist/creator is paid directly from
each and every future sale.

Historically provenance for works of art may span across generations,
for instance Gabriël Metsu's oil on canvas painting \emph{The Lace
Maker's} provenance, first recorded in 1722, now spans 300 years of
ownership, including from a British Baron in the 19\textsuperscript{th}
century to an American philanthropist in the 20\textsuperscript{th}
century.) Metsu died young at the age of 38, leaving a widow; neither
his/her relatives/descendants benefit from his original work, 300 years
later this would be near impossible to facilitate with traditional
systems, as even legal contracts are open and prone to the ravages of
time.

NFT smart contracts hold an incredibly potential; an artists descendants
financially benefiting directly from the resale of a piece of work long
after the artist/museum's/gallery or even state have turned to dust as
long as the original creator's digital wallet is accessible, \emph{the
blockhain becomes an everlasting digital patron} ensuring

NFT art currently suffers from the same failure of decentralisation already discussed in the Ethererum technology stack, but this is compounded by the normalisation of intermediate art brokers \href{https://moxie.org/2022/01/07/web3-first-impressions.html}{continuing to custody} the NFTs even after sale. They are usually selling a pointer to their own servers. The market is nascent and evolving, but it's currently not delivering on it's core promise.\par
Proof of ownership is intuitively a pretty obvious application for the technology, but again it's hard to justify the expense when the benefits are so slim. \href{https://www.bullishlybred.com/}{Bulldogs on the blockchain} is a clear gimmick, and might even incentivise poor behaviours as there are two products here which are not necessarily aligned. Much has been written over the years about \href{https://propy.com/browse/propy-nft/}{deeds to property} being passed through blockchains, cutting out the middle man, but in the event that a house deed NFT was hacked and stolen it's obviously not the case that the property would then pass to the hacker.\par 
One of the most interesting companies is Yuga Labs, who launched the incredibly popular Bored Ape Yacht club set of 10,000 algorithmically generated NFTs. These Ethereum based NFTs were based loosely on the `Crypto Punks' model of PFP-NFT (variously profile picture project, picture for proof, and picture for profile - no definition remains uncontested for long). Yuga launched with a better commercialisation model for the holders, and a strong marketing drive into celebrity circles. They now regularly change hands for hundreds of thousands of pounds. Even this `blue chip' NFT is not without \href{https://twitter.com/coryklippsten/status/1538909505236283392}{serious criticism}: \textit{``I'd put it at 99.99\% the project is in fact a deliberate troll, intentionally replete with Nazi symbols and esoteric racist dog whistles''}\par
Yuga recently bought the artistic rights to the commercial reuse of similarly popular (and preceding) Punks set. This is interesting because they have again handed the commercial re-use rights to the owners of the individual NFTs. This raises the same \href{https://www.bloomberg.com/news/articles/2022-03-21/bored-ape-nft-spinoff-venture-gone-sour-sparks-legal-fight}{confusing problem} with attaching commercial rights to an easily stolen token as NFTs for real estate does. This has been demonstrated recently when Seth Green had a Bored Ape stolen after \href{https://www.buzzfeednews.com/article/sarahemerson/seth-green-bored-ape-stolen-tv-show}{creating an animated show around it's IP}. Many more contradictions and ambiguities in NFT licenses are emerging. Galaxy Digital have \href{https://www.galaxy.com/research/insights/a-survey-of-nft-licenses-facts-and-fictions/}{surveyed the landscape}:
\textit{``Contrary to the ethos of Web3, NFTs today convey exactly zero ownership rights for the underlying artwork to their token holders. Instead, the arrangements between NFT issuers and token holders resemble a distinctly web maze of opaque, misleading, complex, and restrictive licensing agreements, and popular secondary markets like OpenSea provide no material disclosures regarding these arrangements to purchasers. Something more is required, and that `something' is a legal agreement between the owner of the image—known as the `copyright holder' and the NFT holder specifying what rights the NFT holder has with respect to the image. To the extent an NFT purchaser has any rights to the image associated with his or her NFT at all, those rights flow not from his or her ownership of the token, but from the terms and conditions contained in the license issued by the NFT Project governing the NFT holder’s purchase and use of the image. Accordingly, for the vast majority of NFT projects, owning the NFT does not mean you own the corresponding digital content that is displayed when you sync your wallet to OpenSea. That content, as it turns out, is owned and retained by the owner of the copyright associated with that digital content, typically the NFT project.
After reviewing the most used license agreements for NFT projects, it becomes apparent that NFT standards and smart contracts do not recognize off-chain law.''} There may already be a response from the industry to this in the shape of \href{https://a16zcrypto.com/introducing-nft-licenses/}{a16z's ``can't be evil'' license proposal}.\par
Even so, the community around these collections is incredibly strong, mixing developers, artists, the rich and famous, and the fortunate and early, into a cohesive community who communicate online. The developer `good will' is enormous, and it seems possible that this will lead to faster and broader innovation around the collections, and out into \href{https://twitter.com/yugalabs/status/1505014986556551172?}{metaverse applications}. The brand is strong, and the individual NFT items both benefit from, and reinforce that brand, while adding personal narratives and human interest.\par 
As a gauge of how frothy this market still is it's interesting to look at the APE token which Yuga just launched. They airdropped 10,000 of the tokens free to each of the 10,000 NFT holders. This instantly created a multi-billion dollar market cap, and a top 50 `crypto' out of thin air, based purely on their brand. It's clear that there is both brand, and a market here.\par
A recent report from "Base Layer" tries to capture the community `feature' of big brand NFTs. \href{https://baselayer.so/crypto-culture-decoded}{``Crypto culture decoded''} explains that is is these online communities which are the attraction not necessarily the art. This is a powerful `in group' argument, though speculation remains the most likely underpinning.\par

\begin{figure*}[ht]\centering 
	\includegraphics[width=\linewidth]{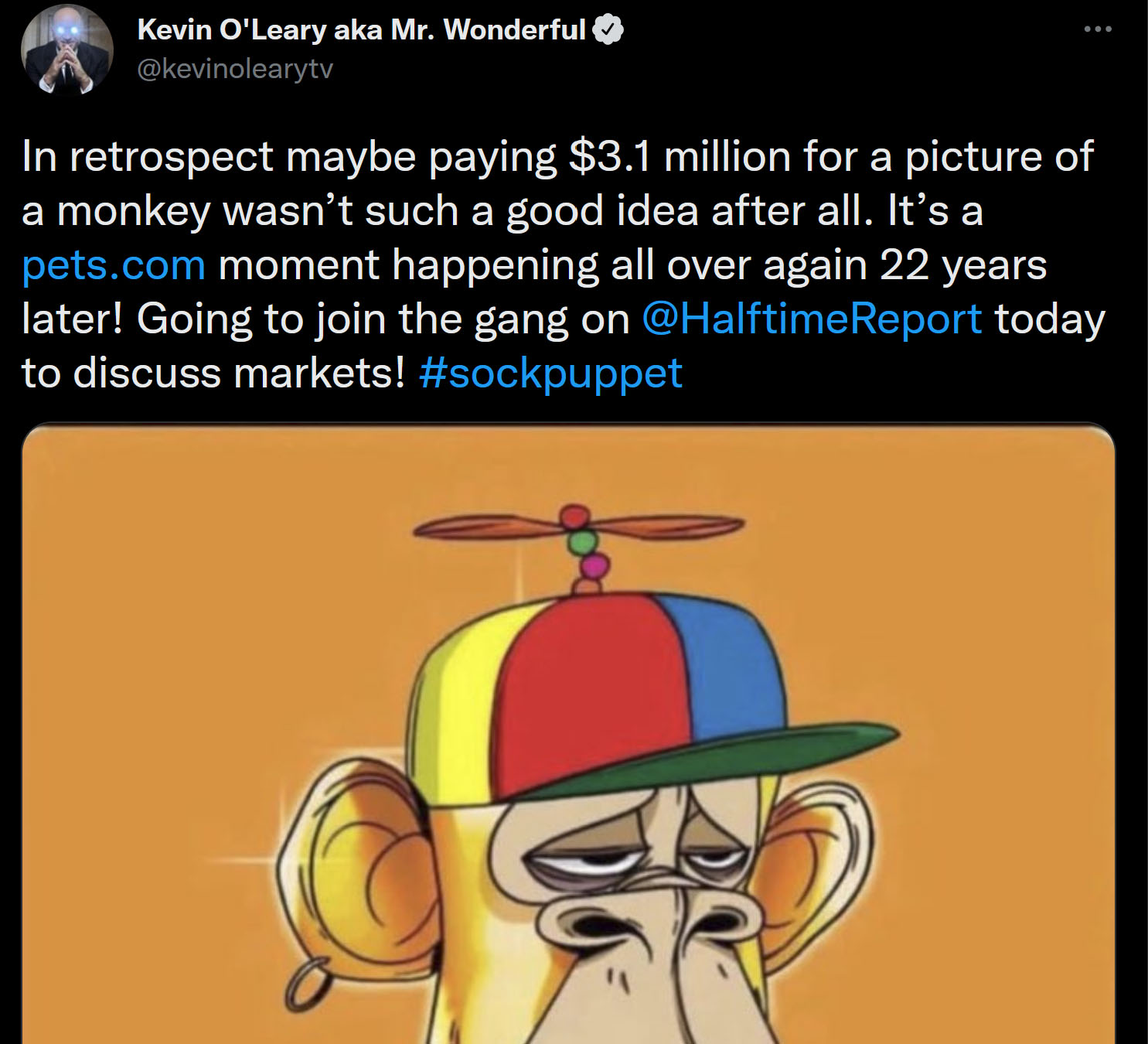}
	\caption{The \href{https://www.coingecko.com/en/nft/bored-ape-yacht-club}{bubble bursts} on Yuga Bored Apes for now.}
	\label{fig:monkey}
\end{figure*}

While it is likely that this is currently a speculative bubble, that is \href{https://www.bbc.co.uk/news/business-61102759}{waning already} (Figure \ref{fig:monkey}), it seems certain that the technology is here to stay in some form.\par

\subsection{Computer \& Video Games}
Computer \& Video games are a huge global business, exponential global
growth over the last 30 years has seen this grow to a point where it has
eclipsed both the
\href{https://www.businessinsider.com/video-game-industry-revenues-exceed-sports-and-film-combined-idc-2020-12?r=US\&IR=T}{global
movie and North American sports industries} combined.

A global industry with revenues over £120b,
\href{https://www.wepc.com/news/video-game-statistics/}{with
\textasciitilde half the people on the planet} playing some form of
games in 2021.

As the games industry has evolved and matured over the last 40 years,
secondary markets have emerged, most notably the `second hand' games
resale market. The rise of `retro' gaming, has demonstrated the second
hand market is a lucrative one for private resellers, an unopened copy
of Super Mario Bros for the Nintendo Entertainment System
\href{https://www.nytimes.com/2021/08/06/business/super-mario-bros-sale-record.html}{recently
selling for £1.5M} to the extent the market has seen
\href{https://www.businessinsider.com/retro-gaming-market-being-overtaken-by-speculators-2021-9?r=US\&IR=T}{speculators
looking to cash in} on the huge global interest in retro/second hand
games.\par
Despite publishers and developers increasingly moving to non-physical
digital only' games, the demand for used games remains incredibly high.\par
Whilst some retailers have adapted their business models to include reselling of retro/second hand games, the vast majority of publisher/developers/retailers aren't able to directly benefit from the
emerging retro/second hand games market. The potential of \emph{video
games as NFT's} presents a huge opportunity for publishers, developers
and players alike, offering the following advantages:
\subsubsection{Royalty Sales on Pre-owned Games} ; A predetermined proportion
  of any resale of a used game can automated in perpetuity via smart
  contracts; once these are set by the publisher, future royalties of
  all sales can be paid directly to the publishers/developers wallets (a
  digital account) without the need of a third party (traditionally a
  retail entity). Traditionally only the initial first sale of a game
  would financially benefit the publisher/developer/retailer, secondary
  and subsequent sales would only ever financially benefit the
  purchaser, with many developers/publishers arguing this is hurting the
  wider industry through the loss of significant income generated by the
  secondary and subsequent sales, sometimes over the course of decades.
  However the use of NFT's smart contracts means that if a game is
  sold/resold through 10,000 collectors; a pre-determined royalty
  payment rate set by the publisher would still guarantee the publisher
  (and or developer/retailer) takes a proportion of any future sales.
\subsubsection{Monetisation of User Generated Content:} Games as a NFT's
  offer ability to monetise UGC: User generated content. Video games such as
\href{https://www.businessofapps.com/data/pokemon-go-statistics/}{Nintendo's
\emph{Pokemon Go}} \emph{(166 million players)},
\href{https://techacake.com/destiny-2-player-count/\#:~:text=The\%20total\%20player\%20base\%20of,to\%20be\%2038\%20million\%20players.\&text=According\%20to\%20the\%20source\%2C\%20the,in\%20terms\%20of\%20player\%20population.}{Bungie's
\emph{Destiny 2}} \emph{(38 million players)} or
\href{https://fictionhorizon.com/how-many-people-play-genshin-impact/\#:~:text=Genshin\%20Impact\%20had\%20approximately\%209,million\%20users\%20in\%20June\%202021.}{miHoYo's
Genshin Impact} (\emph{9 million players} ) all have large, established
and significant player bases. What is noteworthy, the games are designed
to encourage players may spend hundreds, or in some cases thousands of
hours on one game alone; according to
\href{https://destinytracker.com/destiny/leaderboards/all/minutesplayedtotal?grouped=true\&page=1}{Destinytracker.com},
the top players have amassed total play times over 20,000 hours, close
to 1,000 days or \textasciitilde{} 3 years, which is incredible feat
given Destiny 2 only launched 5 years ago in 2017.\par
Destiny/Pokemon Go and Genshin Impact revolve around a central key game
mechanic; players investing significant amounts of time collecting in
game digital assets; characters/weapons/items, often classed as `rare'
or `exotic' or `5 Star'. These collectibles usually found by a
combination of the accrual of in-game time, completing quests,
purchasing additional in-game items/boosters, and luck (`RNG'). Players
are often encouraged to share their collections of rare
characters/weapons/ objects through in-game achievements, triumphs,
scores acting as a mark of distinction/status symbol.\par
Traditionally there has been nothing that went beyond sharing the
\emph{digital badge} (i.e triumph/achievement/accomplishment) on a on
social media/gamer's platform profile. However NFT's offer the ideal
system for developers/publishers and even players to monetise user
generated/customised data (such as a players unique save game data), simultaneously allowing:
a) creation of an additional monetised ecosystem to meet player demands
i.e. some players who are willing to monetise and `sell' their invested
time in a particular product/service to other players with little time
but willing to pay other players for `grinding' (progressing laborious
in game tasks) and a more advanced in-game progression point.
The potential to provide publishers/developers with an additional
long-term income stream, providing a better ROI on computer \& video
game development, which in many instances can cost hundreds of millions
in development costs spanning 5/10 years, is undeniable.
\subsubsection{Play to earn revenue models}
This is morally dicey at this time and early startups like \href{https://www.bloomberg.com/news/features/2022-06-10/axie-infinity-axs-crypto-game-promised-nft-riches-gave-ruin}{Axie Infinity are in serious trouble}. A (long) \href{https://www.youtube.com/watch?v=YQ_xWvX1n9g}{video by Dan Olsen} highlights the structural problems with both play to earn and NFTs. On chain analysis suggested that 40\% of accounts in 200 current Web3 games \href{https://gallery.usejigger.com/}{are bots}. 
\subsubsection{Monetizing In game collectibles}
customisable in game assets (vanity
  items such as cosmetic character skins/clothing or collectible items
  that offer player advantages(new weapons/vehicles/mods etc,..)

Traditional gamers have pushed back on the seemingly useful idea of integrating NTFs with traditional games. This may be in part because Ethereum mining has kept graphics card prices high for a decade.

\href{https://www.prnewswire.com/news-releases/hbar-foundation-and-ubisoft-partner-to-support-growth-of-gaming-on-hedera-network-301474971.html}{HBAR partnerships}\par
\href{https://finance.yahoo.com/news/epic-games-vp-people-have-lost-interest-in-the-metaverse-200725562.html}{Critique from Marc Petit of Epic and Unreal}.\par
The \href{https://twitter.com/justinkan/status/1491270239967154178}{following text} is from Justin Kan, co-founder of twitch: \textit{``NFTs are a better business model for games. Many gamers seem to be raging hard against game studios selling NFTs. But NFTs are also better for players. Here’s why I think blockchain games will be the predominant business model in gaming in ten years. NFTs are a better business model for funding games . Example: recently I invested in a new web3 game SynCityHQ. They are building a mafia metaverse and raised \$3M in their initial NFT drop.\\ NFTs give studios access to a new capital market for raising capital from the crowd.NFTs can be a better ongoing model for games. Web3 games will open economies, and by building the games on open and programmable assets (tokens + NFTs) they will create far more economic value than they could from any one game. Imagine Fortnite, but other developers can build experiences on top of the V-Bucks and skins. Epic would get a royalty every time any transaction happens. As big as Fortnite is today, Open Fortnite could be much bigger, because it will be a true platform. NFTs are better for gamers Allowing gamers to have ownership of the assets they buy and earn in game allows them to participate in the potential growth of a game. It lets gamers preserve some economic value when they switch to playing something new. But what about the criticisms of NFTs?\\
Here are my thoughts on the common FUDs: "It’s just a money grab on the part of the studios!"\\
Game studios already switched over to the model of selling in-game items, cosmetics, etc to players long ago. But currently the digital stuff players are buying isn’t re-sellable. NFT ownership is strictly better for players. "The games aren’t real games." This reminds me of the criticism of free-to-play in 2008, when the games were Mafia Wars / FarmVille. We haven’t had time for great developers to create incredible experiences yet. Everyone investing in games knows there are great teams building. "Game NFTs aren’t really decentralized because they rely on models / assets inside centralized game clients."
Crypto is as much a movement as it is a technology. Putting items on a blockchain is what gives people trust that they have participatory ownership...which make people willing to buy in to the game. These assets are “backed” by blockchain.
The fact that these item collections are NFTs will make other people willing to build on top of them. "NFTs are bad for the environment." Solana and L2s solve this. NFT games are better for players and for game developers. Like the free-to-play revolution changed gaming, so will blockchain. The games of the future will be fully robust, with open and programmable economies.}''
\section{Broader and metaverse uses}
So far according to a16z NFTs break down into:
\begin{itemize}
\item Profile pictures: These were discussed at the start of the chapter and have felt ubiquitous on Twitter over the last couple of years. The major projects will likely hold value, but the hype cycle will likely lead to all profile NFTs going in and out of fashion. There's potentially a fresh wave of this same kind of low key identity hype possible in the metaverse, and indeed the two plausible both intersect and converge.
\item Art and Music: Art has also been discussed above. Peter Thiel, the billionaire venture capitalist who founded PayPal has invested in expanded NTF use cases. The first is `Royal' which is experimentally \href{https://royal.io/}{selling limited NFT tokens} which contractually entitle the holder to a portion of music artist royalties. Spotify are experimenting with music NFTs (and of course in the metaverse). This is an early adopter area, and again likely converges with our planned uses cases as more complex tooling appears. For instance Tim Exile of \href{https://endlesss.fm/}{Endless.fm} talks about digital assets extending to the building blocks of co-created music, and wished to build a music creator economy which distributes value to creators at the instant of the final value transaction with the consumer.
\item Gaming: As discussed there's pushback from the gaming community, but huge investment from the likes of Lego, Blizzard, Epic, Ubisoft etc.
\item Gig tickets: Not only the straightforward use of \href{https://news.yahoo.com/psg-sells-us-220-000-030927515.html}{transferable tickets for events} as NFTs on a blockchain (which is impossible due to the cost right now) but also onward monetisation of ticket stubs as memorabilia. The NBA is \href{https://deadspin.com/investing-in-nft-ticket-stubs-is-likely-one-of-the-nba-1848991991}{already looking at this}.\\
\textit{``The team sells the ticket for face value many many years ago, but when that stub is being sold now for much more many times over, the team gets none of that money,'' York explained. ``But with an NFT stub that changes. Let’s say a new rookie enters the NBA next season and he turns out to be the next LeBron James. That ticket stub from his first game, as an NFT, the team can put a commission on it — 20 percent or however much, the NBA decides that. In 10 years when it’s worth a lot of money, I or whoever owns that NFT, can sell it for say \$100,000. The NBA can still collect 20 percent of that sale, because it’s all on a smart contract.''}\par
It seems so obvious that this will extend to the virtual events space in the metaverse.
\item Utility: These are broadly `membership' style tokens, and this seems like a sensible fit. Peter Thiel (again) for instance launches a \href{https://www.ztonft.com/}{political funding NFT} from Blake Masters to support his senate ambitions. To be clear, Thiel is a fundamentalist libertarian, and at the very least \href{https://gizmodo.com/peter-thiel-bitcoin-talk-miami-2022-1848764790}{highly eccentric}. This is not necessarily a positive for the technology.
\item Virtual worlds are a huge application for NFTs, and this seems like it would be a natural fit for our collaborative mixed reality application. In reality the \$2B of sold so far is mostly `allocations' in nascent ecosystems, being sold as highly speculative assets, without even a metaverse to use. The majority of that amount is the hyped `Otherland' plots sold under the Bored Apes brand.
\item ``Full stack'' luxury brands. \href{https://medium.com/@nic__carter/redeem-and-retain-nfts-are-the-future-of-luxury-goods-760f00dbce23}{Nic Carter describes} a mating of physical and virtual luxury goods. His is a useful article on the future direction, and he has also \href{https://medium.com/@nic__carter/why-nfts-are-hard-to-explain-48f0ab0a35bf}{provided a primer on NFTs}. There are many such examples already, such as \href{https://nft.tiffany.com/faq/}{Tiffanys `NFTiff' - cryptopunks} collaboration which will automatically generate royalties for Tiffanys and parent company Louis Vitton in perpetuity. Such products prove provenance, create new aftermarket opportunities, and unlock metaverse applications.
\end{itemize}
It is completely reasonable to assert that these use cases could be accomplished without the use of NFT technology, and is part of the hype bubble.\par
Twitter user Cantino.Eth offers an exhaustive roundup of what they think future uses might be. It's a \href{https://twitter.com/chriscantino/status/1542930648750608387}{thread full of industry insider jargon} but it's indicative of a shift in focus from speculation to `building' as the market conditions change. Some of the more interesting (less arcane) use cases identified in the thread are summarised very briefly below, again with comments as to how this might pertain to our metaverse applications.
\begin{itemize}
\item Hobby tokens, demonstrating interest in an activity. This is potentially a metaverse adaptation of badges on a blazer in the real world, and might serve to drive communities in a metaverse. The same is true for activism and political alighnment. It's a great idea and worth developing.
\item Professional Networks and qualification badges, like a LinkedIn qualification panel, but in the metaverse. A cisco NFT in the metaverse for a CCNA qualification makes intuitive sense. 
\item Badges to indicate membership of distributed projects within a metaverse. This allows users to identify avatars with shared goals in the metaverse.
\item Retail incentives, like brand loyalty stamps or rewards for participation in marketing, or early access programmes. This is a true in a metaverse marketplace as it is in a real world coffee shop.
\item Multiplayer communities with incentives to hit collective milestones. ``Collecting as a team sport''. This again seems like a great and intuitive opportunity, but is perhaps less suitable for our more business focussed space.
User content submission and automatic monetisation when reused by brands, bonded to an NFT contract.
\item Customer Cohort NFTs: early adopters of successful brands would be able to prove the provenance of their enthusiasm for a new product, and this might unlock brand loyalty bonuses. It seems this wouldn't be a transferable NFT, and is more like the ``soulbound'' idea advanced by Meta.
\item Education and Customer Support, think an NFT of a great score on reddit community support forums. A trusted community member badge, but visible in the metaverse. This is somewhat like the web of trust model advanced earlier in the book.
\item NFTs as contracts is far more likely in the metaverse than it has proved to be in real life. This is how `digital land' and objects will be transferred anyway, but with the addition of contractual conditionals with external inputs more subtle products may appear.
\end{itemize}

\section{Objects in our collaborative mixed reality}
There has been a recent shift away from the `toxic' moniker of NFT and toward `Digital objects', and seem to be judged crucial to metaverse applications. The success of avatar \href{https://medium.com/coinmonks/reddit-nft-success-ca2685163576}{`collectibles' markets} in the Reddit ecosystem, and Meta (ex Facebook) similarly divesting themselves of the NFT term seem to suggest a pivot point in the industry. Meta were encouraging adoption through zero fee incentives but were likely hanging their monetisation of their whole rebrand on taking a huge cut from NFT content creators on their platform. This seems to have failed and they are \href{https://help.instagram.com/1824155514623095}{winding up} that part of their business.\par 
We have potential paths to digital assets within future layer 3 technologies (\href{https://www.rgb.tech/}{RGB} \& Pear Credits), but they're not yet fit for purpose. There are compromise options already available, as below. 
\subsection{Liquid tokens}
We have seen that Liquid from Blockstream is a comparatively mature and battle tested sidechain framework, based upon Bitcoin. It is possible to issue tokens on Liquid, and these have their own hardware wallet available. This makes the technology a strong contender for our uses.
\subsection{Ethereum}
While it's been discounted elsewhere it's hard to ignore the network effect of Eth NFTs. If the aspiration is to attract the bulk of the `legacy' creator/consumer markets then it will be necessary to support integration of Metamask into any FOSS stack. This isn't a huge technical challenge, nor is it particularly of interest to our use cases at this stage, but it remains a possibility. The main problems remain the slow speed and high expense of the system.
\subsection{Solana}
Solana is both cheap and fast, because it's very highly centralised. It seems unlikely that it's worth this level of compromise. It has also become embroiled with the fallout from the enormous FTX exchange fraud, threatening the existence of the assets (NFTs) issued and stored upon it.
\subsection{Peerswap}
It may be possible to use ``Peerswap'' to execute rebalancing and submarine swaps into and out of Liquid assets on the sidechain in a single tx. This is anunder explored area at this time.
\subsection{FROST on Bitcoin}
It \textbf{might} be possible to transfer ownership of a UTXO on the Bitcoin base chain using FROST \cite{komlo2020frost}. In this Schnorr \& Taproot based threshold signature system it's possible to \href{https://btctranscripts.com/sydney-bitcoin-meetup/2022-03-29-socratic-seminar/}{add and remove signatories} and thresholds of signing without touching the UTXO itself. In principle (though not yet in practice) this might allow transfer of UTXO ownership. 
\subsection{Spacechains}
It feels like spacechains are almost ready, so this is worth keeping an eye on. It's the `cleanest' way to issue assets using Bitcoin because there's no additional speculative chain. As briefly explained in the earlier section Bitcoin is destroyed to create a new chain which then inherits the security of Bitcoin through onward mining. This new asset or chain is able to accrue value and trade independently based purely on it's value to the buyer, not as a function of a wider speculative bubble attached to a token with multiple use cases.
\subsection{Pear credit}
The outstanding contender at this stage is Pear Credit from Hypercore. This section needs a full explanation later. For now a \href{https://medium.com/@observer1/tether-announced-the-launch-of-pear-credit-8d4f66ccd97b}{blog post on the subject} will have to do.
\subsection{Satoshi Ordinals}
Ordinals offer a new and innovative way of creating NFTs using bitcoin. The ability to store NFTs on the bitcoin blockchain offers the security, immutability, and decentralization that is fundamental to bitcoin's design. These are being called `ordinal inscriptions'. In addition the use of Taproot has allowed Ordinals to store large files on the bitcoin blockchain, by exploiting cheap space which was designed to add more complex `script-path spend' scripts. Their creator Rodarmor says the following of them: \textit{``Inscriptions are digital artifacts, and digital artifacts are NFTs, but not all NFTs are digital artifacts. Digital artifacts are NFTs held to a higher standard, closer to their ideal. For an NFT to be a digital artifact, it must be decentralized, immutable, on-chain, and unrestricted. The vast majority of NFTs are not digital artifacts. Their content is stored off-chain and can be lost, they are on centralized chains, and they have back-door admin keys. What's worse, because they are smart contracts, they must be audited on a case-by-case basis to determine their properties. Inscriptions are unplagued by such flaws. Inscriptions are immutable and on-chain, on the oldest, most decentralized, most secure blockchain in the world. They are not smart contracts, and do not need to be examined individually to determine their properties. They are true digital artifacts.''}\par
Ordinals work by individually tracking single satoshis and repurposing them as digital artefacts, using the convention of \href{https://docs.ordinals.com/overview.html}{``ordinal theory''} to ascribe non-fungible features to the satoshis in a consistent and coherent manner. They can contain various forms of digital content, such as images, audio, video, and pdfs, even \href{https://ordinals.com/inscription/6edf80efbbae537b554340c31496439b57bef65357a57f21cbb547bc6287d7bfi0}{playable games}. The Inscriptions are stored entirely on the bitcoin blockchain, taking advantage of the Taproot upgrade to store the NFT data in Taproot. The bytes remain in the original minted transaction, but the ownership can be passed around by `spending' the UTXO which contained the  satoshi forward, not transferring that data, which remains linked to the original ordinal only. There is a \href{https://ordinals.com/}{file explorer for these digital objects}. This hardly seems the \href{https://twitter.com/adam3us/status/1619836529718996993?}{orgy of inefficiency} suggested by Adam Back, but there are certainly storage considerations.  While this is considered a waste of Bitcoin block space by some, it is possible that the main concern here is the discount that these taproot scripts enjoy, allowing a skewing of the use of money to commit to the chain toward frivolity. The costs associated with committing data to the ledger will likely rise alongside all other operations on the chain, and already represents an unacceptably bar for our requirements, but the innovation is interesting, and the digital artefacts are already being sold for up to \textsterling 180,000.\par
It seems right now that the older Bitcoin community is pushing back on the technology as ``spamming the blockchain'', \href{https://lists.linuxfoundation.org/pipermail/bitcoin-dev/2019-September/017306.html}{something discussed since 2019}. The cost works out around 50k satoshis per 100kbytes of data, which is comparatively reasonable. Our issue with this approach is that richer nations will be able to push out developing economy financial use cases with the more fun, but frivolous data storage and scarcity application.\par
\begin{figure*}[ht]\centering 
	\includegraphics[width=0.7\linewidth]{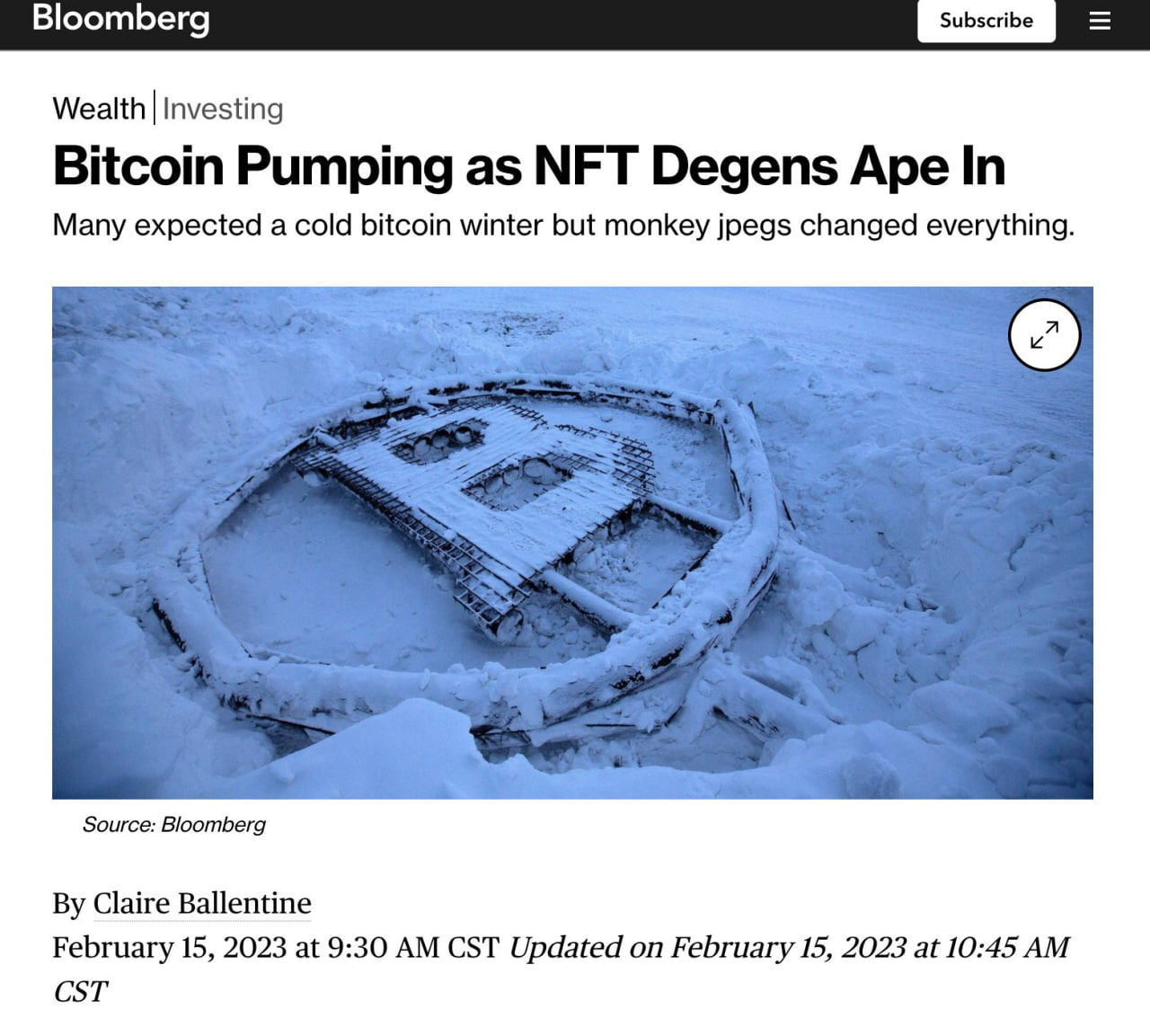}
	\caption{The press has been quick to note the phase change to digital objects on Bitcoin}
	\label{fig:degens}
\end{figure*}

The narrative seems to be that it's already pushing up the fees on the network and the overall price of Bitcoin as tens of thousands of minting transactions enter the ecosystem (Figure \ref{fig:degens}).

In what seems to be a fascinating case study on ownership one `Bored Ape' owner moved coveted their Ethereum NFT by destroying it on the original chain in order to migrate to Bitcoin. Yuga labs stated that this NFT owner had destroyed their asset, giving up any right they might have had. Yuga can now presumably create and sell another copy of this image. It's fair to say that ownership of digital art is far from a settled matter, as pointed out by Low \cite{low2022emperor}. We do not plan on attaching \textit{any} legal assertions to our use of digital objects at the lowest level. What people do with them will be up to them and the open market. Neither does this exclude users of the open source stack from attempting to attach their own restrictions to IP associated with their work.\par
There is also likely to be a considerable impact on the size of the \textbf{full} Bitcoin blockchain if this approach becomes commonplace. What was a 500 gigabyte file with a fairly linear growth rate will likely become many terabytes over the coming years. The overall workaround for this would be `pruned nodes' vs `archival nodes', similar to the Ethereum approach, with those likely only run in places where prosecutions for illegal data on the chain seem unlikely. Curiously this won't effect users of the system who choose to run their own nodes, as the software is configured to throw away this extraneous data prior to the timestamp of the last software upgrade. Concerns that this might price out participation from emerging markets are overblown from an infrastructure point of view, but potentially valid from a fee competition standpoint. This perhaps drives Lightning adoption. Nobody seems too sure, though Kaloudis, senior researcher at Coindesk describes the tradeoffs and tensions at this time \cite{Kaloudis2023}. As seems to be the norm now, there is \href{https://nosft.xyz/}{Nostr integration} coming to Ordinals.\par 
The recent announcement by Yuga Labs, the team behind the popular NFT collection Board Ape Yacht Club and the stewards of CryptoPunks, that they are getting into the ordinals space with their new collection, 12-Fold, has generated a lot of reactions from the crypto community. This move is variously seen as significant highlighting the potential for Bitcoin-based NFTs, or just a naked cash grab. Yuga are at pains to point out the ownership and self-custody aspects, which are fundamental principles of Bitcoin and better than their current product lines.\par
As a small technical aside, the move towards inscriptions and ordinals has also led to a surge in interest in partially signed Bitcoin transactions (PSBT), which allows multiple people to use Bitcoin collaboratively. This is an important standard for multi-sig agreements and more, and facilitates the secure transfer of the ordinals, giving them a huge boost. The blossoming of Noster and inscriptions at the same time is seen as a sign that the theoretical usefulness of Bitcoin infrastructure is now actually being built at pace.\par
A \href{https://unisat.io/brc20}{browser extension called `Unisat'} acts as a wallet which supports non-fungible and also fungible \href{https://twitter.com/LeonidasNFT/status/1650243884973191168}{`sets' of tokens}. This is likely one to watch closely and we have tested the system on both testnet (with a 1024 byte limit) and main net (as a \href{https://ordinals.com/inscription/4819948774f979f26a8295000675072b626f2c51cbfc0bf614afbb16bbec4266i0}{single inscription} costing £20). \par 
This is obviously far too expensive for the bulk of our use cases within the metaverse, and we will use RGB for this in the end (as we have always intended). Ukolova, the director of the board for LNP/BP Standards Association which develops the RGB protocol and product stack \href{https://bitcoinmagazine.com/culture/are-ordinals-really-good-for-bitcoin}{wrote an opinion piece} on Ordinals which is worth finishing on. The article discusses the limitations of using Ordinals in the context of Bitcoin, particularly with regards to privacy, scalability, and technical difficulties. While Ordinals may offer a unique way to represent ownership of rare and expensive assets, they do not provide additional privacy features beyond what is already offered by Bitcoin's pseudonymous nature. RGB puts all of the data of the asset on the client side, ensuring peer-to-peer verification without relying on any third parties or miners. The protocol also applies zero-knowledge cryptographic primitives such as bulletproofs to bring privacy to holders and creators.\par 
\subsubsection{BRC-20}
BRC-20 tokens are fungible tokens that are created by attaching a JSON to satoshis through Bitcoin ordinals. The JSON code bit defines every characteristic of the BRC-20 token including the minting, and distribution, the bitcoin network decodes this information once they are deployed. BRC-20 tokens are minted and spent like normal tokens. BRC-20 tokens utilize Ordinals inscriptions of JSON (JavaScript Object Notation) data to deploy token contracts, mint, and transfer tokens. Currently, the BRC-20 token standard allows creating a BRC-20 token with the deploy function, minting an amount of BRC-20 tokens with the mint function, and transferring an amount of BRC-20 tokens via the transfer function. It's notable that for purely fungible assets that don't require smart contract functionality (which is most of them) BRC-20 is superior to Ethereum's ERC-20 in simplicity, security, and the fairness of the minting process. The community seems very split on this use of the chain, and it's impact on transaction fee markets.
\subsubsection{Litecoin and other networks}
It is also possible to use the whole suit of ordinal based ideas on any other chain such as Litecoin, the long standing Bitcoin fork which is used somewhat as a technical testbed for Bitcoin. This might develop into a far more appealing option, though again, it's too early to be sure.
\subsection{Taproot Assets}
Taproot Assets is a new protocol that leverages the Taproot upgrade in Bitcoin and brings asset issuance capabilities to the Bitcoin network. This allows for secure, fast, and scalable transactions, interoperability with Lightning Network, and improved privacy and efficiency.
\begin{itemize} \item Issuance of assets on the Bitcoin blockchain \item Powered by the Taproot upgrade for privacy and scalability \item Deposit assets into Lightning channels for instant transactions and low fees \item Transfer assets over the existing Lightning Network \end{itemize}
\subsubsection{Benefits /Limitations and Risks} 
\begin{itemize} 
\item Light client-friendly and low verification costs 
\item Atomic swaps between assets and BTC 
\item Supports unique and non-unique assets, as well as collections 
\item Facilitates multi-signature and co-signatory arrangements \item Channels can be created alongside BTC channels in the same UTXO 
\item Potential future features include confidential transactions and zero-knowledge proofs (perhaps integrating ZeroSync metioned elsewhere)
\item Allows minting or moving of an unbounded number of assets in a single on-chain transaction \item Takes advantage of existing network effects to improve scalability \item Addresses blockchain congestion problems and prepares for mass adoption
\item May 2021 saw a spike in transactions and fees, largely due to BRC-20 token mining and trading \item At this time the system only runs on testnet. It is hard to imagine a time when Lightning Labs would be prepared to sign their corporate name to the security of potentially billions of dollars denominated assets.
\item LND has suffered technical issues in the past, and may again, with the potential for complete chaos in a monopolar technical system.   \end{itemize}

\chapterimage{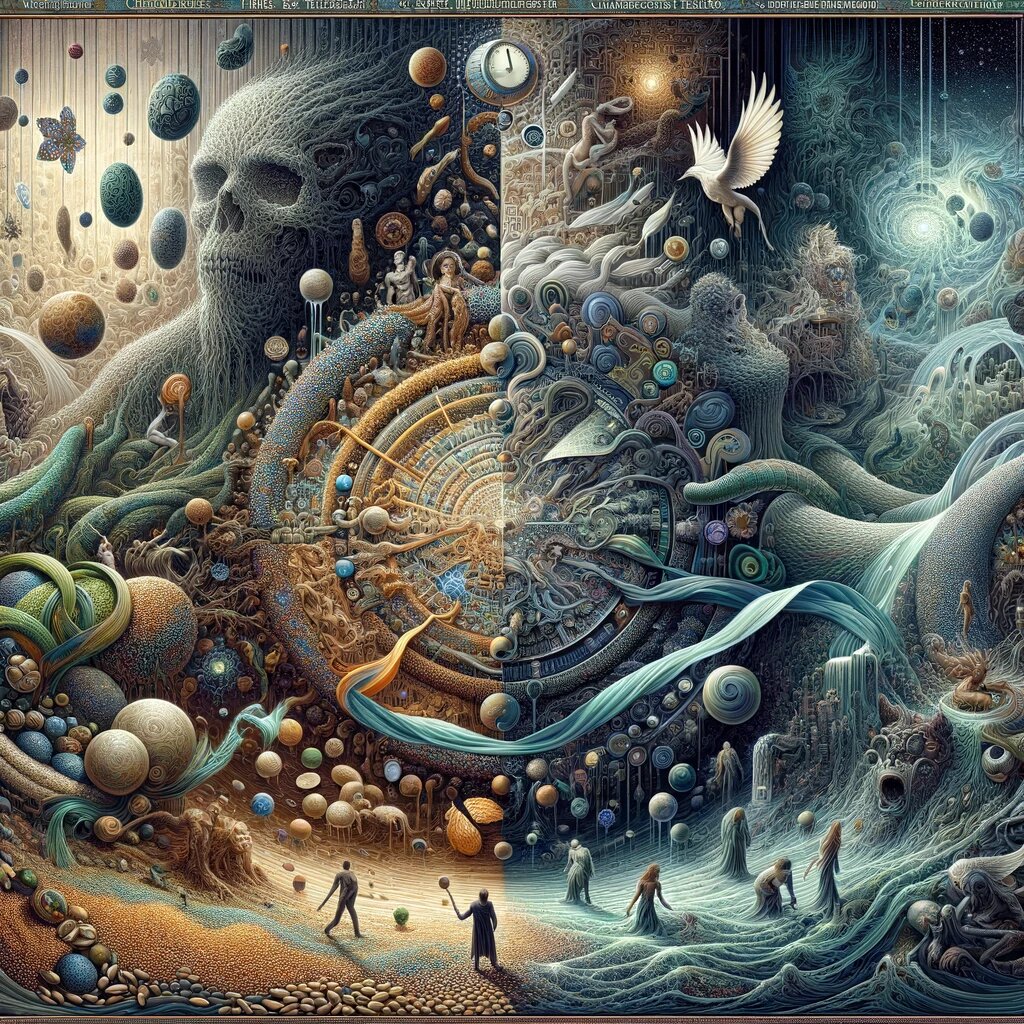}
\chapter{Collaborative mixed reality}
\section{Toward an open metaverse}
\begin{figure}
  \centering
    \includegraphics[width=\linewidth]{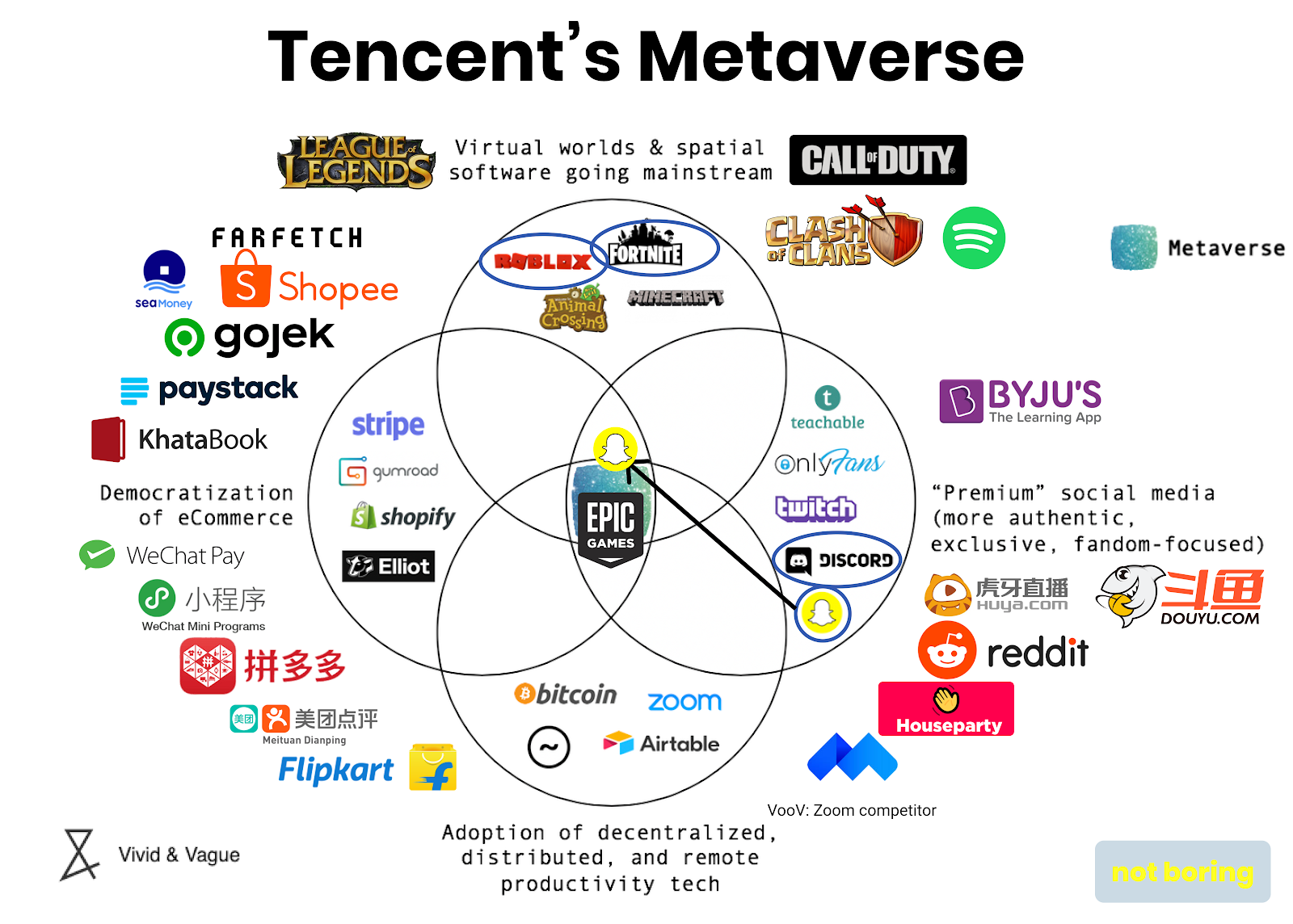}
  \caption{\href{https://www.notboring.co/p/tencents-dreams}{McCormick attempts to guess the Tencent metaverse}}
  \label{fig:tencent}
\end{figure}

The Openstand principles are a great starting place for what an open metaverse might mean. \href{https://open-stand.org/about-us/principles/}{They are}:\\
\begin{itemize}
\item Cooperation: Respectful cooperation between standards organizations, whereby each respects the autonomy, integrity, processes, and intellectual property rules of the others.
\item Adherence to Principles: Adherence to the five fundamental principles of standards development:
\begin{itemize}
\item Due process. Decisions are made with equity and fairness among participants. No one party dominates or guides standards development. Standards processes are transparent and opportunities exist to appeal decisions. Processes for periodic standards review and updating are well defined.
\item Broad consensus. Processes allow for all views to be considered and addressed, such that agreement can be found across a range of interests.
\item Transparency. Standards organizations provide advance public notice of proposed standards development activities, the scope of work to be undertaken, and conditions for participation. Easily accessible records of decisions and the materials used in reaching those decisions are provided. Public comment periods are provided before final standards approval and adoption.
\item Balance. Standards activities are not exclusively dominated by any particular person, company or interest group.
\item Openness. Standards processes are open to all interested and informed parties.
\end{itemize}
\item Collective Empowerment: Commitment by affirming standards organizations and their participants to collective empowerment by striving for standards that:
\begin{itemize}
\item are chosen and defined based on technical merit, as judged by the contributed expertise of each participant;
\item provide global interoperability, scalability, stability, and resiliency;
\item enable global competition;
\item serve as building blocks for further innovation; 
\item contribute to the creation of global communities, benefiting humanity.
\end{itemize}
\item Availability: Standards specifications are made accessible to all for implementation and deployment. Affirming standards organizations have defined procedures to develop specifications that can be implemented under fair terms. Given market diversity, fair terms may vary from royalty-free to fair, reasonable, and non-discriminatory terms (FRAND).
\item Voluntary Adoption: Standards are voluntarily adopted and success is determined by the market.
\end{itemize}
The push toward open standards is being joined (somewhat late) by credible and established bodies \href{https://spectrum.ieee.org/laying-foundation-for-extended-reality}{like the IEEE}. It's such a fast moving and under explored set of problems that this movement toward standards will take a long time to even find it's feet.
Hopefully it's clear to the reader that this kind of development guides the work here. In the wider ``real-time social VR'' various companies have attempted to build closed ecosystems, for years. These now look more like attempts at digital society, but are closer to isolated metaverses, or more usefully isolated digital ecosystems. This is still happening. There's every chance that when Apple make their augmented reality play this year or next they will keep their system closed off as this tends to be their business model. Theo Priestly, CEO at Metanomics \href{https://www.linkedin.com/feed/update/urn:li:activity:6977366421034967040/}{points out} that Chinese Giant Tencent are doing similar, and he cited Figure \ref{fig:tencent}; building a closed but tightly linked suite of businesses into something that looks like a metaverse. The levels of investment which are being hung under the metaverse moniker \href{https://www.scmp.com/tech/policy/article/3194092/chinas-iphone-production-hub-henan-bets-its-future-metaverse}{are mind blowing}, but that is not what we want to discuss as an end point for this book.
For our purposes in this product design the interface between the previous chapter (NFTs) and this metaverse chapter is crucial. Punk6529 is a pseudonymous twitter account and thought leader in the ``crypto'' space. The text below encapsulates much of the reasoning that led to this book and product exploration, and is paraphrased \href{https://twitter.com/punk6529/status/1536046831045685248}{from this thread} for our purposes.\par
\textit{Bit by bit, the visualization layer of the internet will get better until it is unrecognisably better (+/- 10 years). As the visualization layer of the internet gets better, digital objects will become more useful and more important. Avatars (2D and 3D), art, schoolwork, work work, 3D virtual spaces and hundreds of other things. Not only will the objects themselves become more important, they will lead to different emergent behaviours. We see this already with avatars and mixed eponymous/pseudonymous/anonymous communities. Yes, it is the internet plumbing underneath, but just like social media changed human behaviour on the internet, metaverse type experiences will further change it. NFT Twitter + Discord + various virtual worlds is a form of early metaverse. I feel like I am entering a different world here, not just some websites. The most important question for the health of the internet/metaverse/human society in the 2030s will be decided now. And that question is: "who stores the definitive ownership records of those digital objects". There are two answers: a company's database OR a blockchain. If we end up with "a company's database" we will end up with all the web dysfunctions, but worse. SMTP is an open protocol that anyone can use so we don't have societal level fights on "who is allowed to use email". Short messaging online ended up becoming Twitter. So we end up having the most absurd, surreal discussions on the topic of "who is allowed to use short-messaging" being dependant on "who is the CEO of Twitter". There is no way this is the correct architecture for our progressively more digital economy.... If this is your first time around here, we are fighting for an open metaverse.''}\par
It seems that industry shares much of this opinion regarding an open metaverse. The proposal of a persistent interactive digital universe online is \textbf{so} vast that major players recognise that they will not be able to monopolise this space, though Facebook/Meta are clearly attempting to. The \href{https://metaverse-standards.org/news/press-releases/leading-standards-organizations-and-companies-unite-to-drive-open-metaverse-interoperability/}{Metaverse Standards Forum} is clearly an attempt by the other industry players to catch up and then get out ahead of Meta in this regard. It's also possible to view this as just another land grab, but through the vehicle of a standards body. Time will tell. They say:\par
\textit{``Announced today, The Metaverse Standards Forum brings together leading standards organizations and companies for industry-wide cooperation on interoperability standards needed to build the open metaverse. The Forum will explore where the lack of interoperability is holding back metaverse deployment and how the work of Standards Developing Organizations (SDOs) defining and evolving needed standards may be coordinated and accelerated. Open to any organization at no cost, the Forum will focus on pragmatic, action-based projects such as implementation prototyping, hackathons, plugfests, and open-source tooling to accelerate the testing and adoption of metaverse standards, while also developing consistent terminology and deployment guidelines.''}\par
This looks like it will be a useful project and community for the purposes outlined in this book, but the technology is young enough (in that it doesn't really exist) for multiple approaches to be trailed.\par
Europe is making metaverse a priority with \href{https://digital-strategy.ec.europa.eu/en/policies/virtual-and-augmented-reality-coalition}{The Virtual and Augmented Reality Industrial Coalition}. President von der Leyen’s State of the Union \href{https://state-of-the-union.ec.europa.eu/system/files/2022-09/SOTEU_2022_Letter_of_Intent_EN_0.pdf}{letter of intent says}:  ``We will continue looking at new digital opportunities and trends, such as the metaverse.'' 
\subsection{Primitives}
OpenAI identified the following 5 points about metaverse, in response to the query ``What are 5 key points I should know when studying metaverse?''
\begin{itemize}
\item Metaverse is a virtual reality platform that allows users to interact with each other and with digital objects in a virtual space.
\item Metaverse is a decentralized platform, meaning that there is no central authority or server that controls the platform.
\item Metaverse is an open platform, meaning that anyone can develop applications for the platform.
\item Metaverse is a secure platform, meaning that all data and transactions are encrypted and secure.
\item Metaverse is a scalable platform, meaning that it can support a large number of users and a large number of transactions.
\end{itemize}
This is an unexpectedly great answer, probably the cleanest we have found. The \href{https://metaverse-standards.org/}{Metaverse Standard Forum} highlights the following, which reads like the output from a brainstorm between academia and industry stakeholders.
\begin{itemize}
\item collaborative spatial computing
\item interactive 3D graphics 
\item augmented and virtual reality
\item photorealistic content authoring
\item geospatial systems
\item end-user content tooling
\item digital twins
\item real-time collaboration
\item physical simulation
\item online economies
\item multi-user gaming
\item new levels of scale and immersiveness. 
\end{itemize}
It's not a useless list by any means, but it lacks the kind of product focus we need for detailed exploration of value and trust transfer. \par
Mystakidis identifies the following \cite{mystakidis2022metaverse}:
\begin{itemize}
\item Principles
\begin{itemize}
\item Interoperable
\item Open
\item Hardware agnostic
\item Network
\end{itemize}
\item Technologies
\begin{itemize}
\item Virtual reality
\item Augmented reality
\item Mixed reality
\end{itemize}
\item Affordances
\begin{itemize}
\item Immersive
\item Embodiment
\item Presence
\item Identity construction
\end{itemize}
\item Challenges
\begin{itemize}
\item Physical well-being
\item Psychology
\item Ethics
\item Privacy
\end{itemize}
\end{itemize}
This is quite an academic list. A lot of these words will be explored in the next section which is more of an academic literature review.\par
Nevelsteen attempted to identify key elements for a `virtual work' in 2018 and these are relevant now, and described rigorously in the appendix of his paper \cite{nevelsteen2018virtual}:
\begin{itemize}
\item Shared Temporality, meaning that the distributed users of the virtual world share the same frame of time.
\item Real time which he defines as ``not turn based''.
\item Shared Spatiality, which he says can include an `allegory' of a space, as in text adventures. It seems this might extend to a spoken interface to a mixed reality metaverse.
\item ONE Shard is a description of the WLAN network architecture, and conforms to servers in a connected open metaverse.
\item Many human agents simply means that more than one person can be represented in the virtual world and corresponds to `social' in our description.
\item Many Software Agents corresponds to AI actors in our descriptions. Non playing characters would be the gaming equivalent.
\item Virtual Interaction pertains to any ability of a user to interact actively with the persistent virtual scene, and is pretty much a given these days.
\item Nonpausable isn't even a word, but is pretty self explanatory.
\item Persistence means that if human participants leave then the data of the virtual world continues. This applies to the scenes, the data representing actions, and objects and actors in the worlds.
\item Avatar is interesting as it might seem that having avatar representations of connected human participants is a given. In fact the shared spaces employed by Nvidia for digital engineering do not. 
\end{itemize}
Turning to industry; John Riccitiello, CEO of Unity Technologies says that metaverse is \textit{``The next generation of the internet that is:
\begin{itemize}
\item always real-time 
\item mostly 3D 
\item mostly interactive
\item mostly social
\item mostly persistent''
\end{itemize}}
Expanding this slightly we will us the following primitives of what we think are important for a metaverse:
\begin{itemize}
\item Fusing of digital and real life
\item Social first
\item Real time interactive 3d graphics first
\item Persistent
\item Supports ownership
\item Supports user generated content \cite{ondrejka2004escaping}
\item Open and extensible
\item Low friction economic actors and actions
\item Trusted / secure
\item Convergence of film and games
\item Blurring of IP boundaries
\item Blurring of narrative flow
\item Multimodal and hardware agnostic
\item Mobile first experiences
\item Safeguarding, and governance
\end{itemize}
There is a \textbf{lot} of work for the creative and technical industries to do to integrate human narrative creativity this nascent metaverse, and it's not even completely clear that this is possible, or even what people want.
\section{History}
The word metaverse was coined by the author Neal Stephenson in his 1992 novel Snowcrash. It started popping up soon after in \href{https://www.newscientist.com/article/mg14819994-000-how-to-build-a-metaverse/}{news articles} and research papers \cite{mclellan1993avatars}, but in the last five years it has been finding a new life within a silicon valley narrative. Perhaps in response to this Stephenson is now working with a company called \href{https://www.lamina1.com/}{Lamina1} which actually looks a lot like the rest of this book, so perhaps we have been on the right track.\par
There were clear precursors to modern social VR, such as \href{https://www.howtogeek.com/778554/remembering-vrml-the-metaverse-of-1995/}{VRML in the 1990's} which laid much of the groundwork for 3D content over networked computers.\par
\begin{figure}
  \centering
    \includegraphics[width=\linewidth]{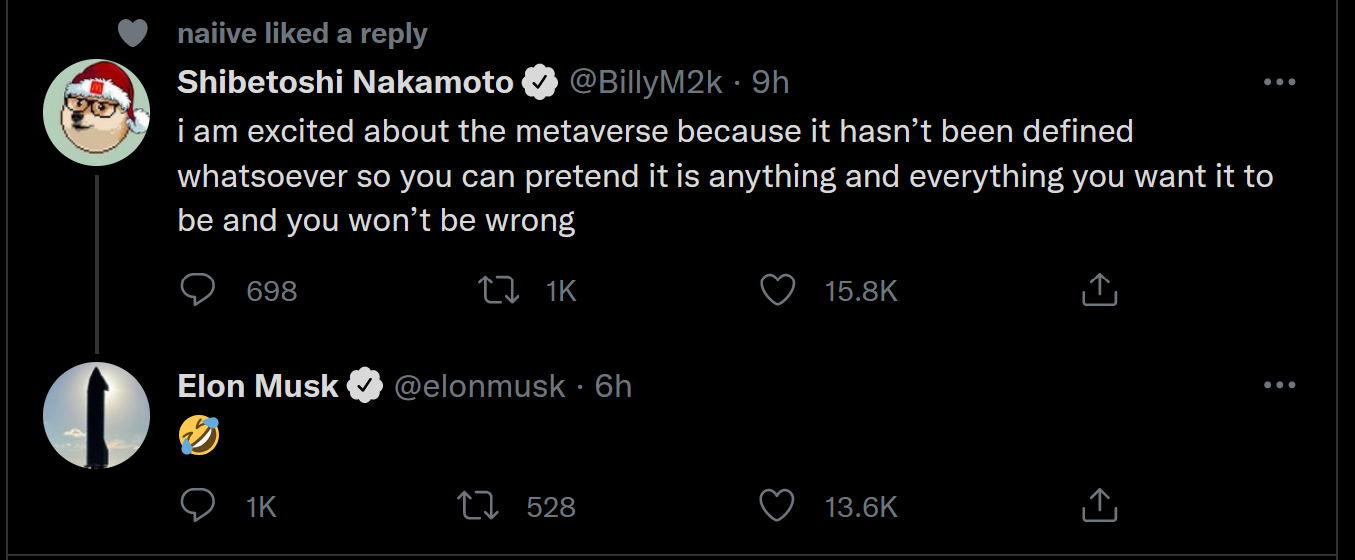}
  \caption{Elon Musk agrees with this on Twitter. It's notable that Musk is now Twitters' \href{https://twitter.com/paraga/status/1511320953598357505}{biggest shareholder}, and has been vocal about web censorship on the platform.}
  \label{fig:muskWeb3}
\end{figure}

It might seem that there would be a clear path from there to now, in terms of a metaverse increasingly meaning connected social virtual spaces, but this has not happened. Instead interest in metaverse as a concept waned, MMORG (described later) filled in the utility, and then recently an entirely new definition emerged. Park and Kim surveyed dozens of different historical interpretations of the word, and the generational reboot they describe makes it even less clear \cite{park2022metaverse}. The concept of the Metaverse is extremely plastic at this time (Figure \ref{fig:muskWeb3}).\par
It's arguable that what will be expanding in this chapter is more appropriately `Cyberspace' as described by William Gibson in Neuromancer \cite{gibson2019neuromancer} \textit{``A global domain within the information environment consisting of the interdependent network of information systems infrastructures including the Internet, telecommunications networks, computer systems, and embedded processors and controllers.''}\par
Park and Kim identify the generational inflection point which has led to the resurgence of the concept of Metaverse \cite{park2022metaverse}: 
\textit{``Unlike previous studies on the Metaverse based on Second Life, the current Metaverse is based on the social value of Generation Z that online and offine selves are not different.''} \par

Brett Leonard, writer director of Lawnmower Man talks about the pressing need to get out in front of moral questions in the development of metaverse applications. He stressed that wellbeing will be a crucial underpinning of the technology because of the inherent intimacy of immersion in virtual spaces. He suggests that emotional engagement with storied characters is needed to satisfy the human need for narrative, and that this should be utopian by design to stave off the worst of dystopian emergent characteristics of the technology.\par
The book will aim to build toward an understanding of metaverse as a useful social mixed reality, that allows low friction communication and economic activity, within groups, at a global scale. Cryptography and distributed software can assist us with globally `true' persistence of digital data, so we will look to integrate this with our social XR.  This focus on persistence, value, and trust means it's most appropriate to focus on business uses as there is more opportunity for value creation which will be important to bootstrap this technology. \par
Elsewhere in the book we state that metaverse is the worst of the tele-collaboration tool-kits, and in general we `believe' this to be true at this time. With that said Hennig-Thurau says the following in a \href{https://www.linkedin.com/feed/update/urn:li:activity:7020679507141361664/}{LinkedIn post}: \textit{Our research finds that the performance of social interactions in the VR metaverse varies for different outcomes and settings, with productivity and creativity being on par with Zoom (not higher, but also not lower) for the two experimental settings in which we studied these constructs. Thus, as of today, meeting in VR does not overcome all the limitations that we are facing when using Zoom or Teams. But most importantly (to us), we find clear evidence that when people get together in the metaverse via VR, it creates SUBSTANTIALLY higher levels of social presence among group members across ALL FIVE STUDY CONTEXTS, from idea generation to joint movie going. This is the main insight from our study and the stuff we believe future uses of social virtual reality can (and should) build on. We also explain that the effectiveness of VR meetings can be further increased, and also how this can be done (by selecting the most appropriate settings, people, avatars, hardware, environments etc.).}  \cite{hennig2022social}\par
We agree that with sufficiently informed guiding constraints in place, and smaller group sizes (ie, not a large scale social metaverse), that there is a path forward.\par
This chapter will first attempt to frame the context for telepresence (the academic term for communicating through technology), and then explain the increasingly polarised options for metaverse. It's useful to precisely identify the primitives of the product we would like to see here, so this chapter is far more a review of academic literature in the field, culminating in a proposed framework.\par

\section{Video conferencing, the status quo}
This section has been adapted and updated for open source release, from the authors PhD thesis, with the permission of the University of Salford.\par
Video-conferencing has become more popular as technology improves, as it gets better integrated with ubiquitous cloud business support suites, and as a function of the global pandemic and changing work patterns. There is obviously increasing demands for real-time communication across greater distances.\par
The full effects of video-conferencing on human communication are still being explored, as seen in the experimental \href{https://news.microsoft.com/innovation-stories/microsoft-teams-together-mode/}{``Together Mode''} within Microsoft Teams. Video-conferencing is presumed to be a somewhat richer form of communication than email and telephone, but not quite as informative as face-to-face communication. \par
In this section we look at the influence of eye contact on communication and how video-conferencing mediates both verbal and non-verbal interactions. Facilitation of eye contact is a challenge that must be addressed so that video-conferencing can approach the rich interactions of face-to-face communication. This is an even bigger problem in the emerging metaverse systems, so it's important that we examine the history and trajectory.\par
There is a tension emerging for companies who do not necessarily need to employ remote meeting technology, but also cannot afford to ignore the competitive advantages that such systems bring. In an experiment preformed well before the 2020 global pandemic at CTrip, Bloom et al describe how home working led to a 13\% performance increase, of which about 9\% was from working more minutes per shift (fewer breaks and sick-days) and 4\% from more calls per minute (attributed to a quieter working environment) \cite{Bloom2015}. Home workers also reported improved work satisfaction and experienced less turnover, but their promotion rate conditional on performance fell. This speaks to a lack of management capability with such systemic change. It's clearly a complex and still barely understood change within business and management. \par
Due to the success of the experiment, CTrip rolled-out the option to work from home to the whole company, and allowed the experimental employees to re-select between the home or office. Interestingly, over half of them switched, which led to the gains almost doubling to 22\%. This highlights the benefits of learning and selection effects when adopting modern management practices like working from home. Increasingly this is becoming a choice issue for prospective employees, and an advantage for hiring managers to be able to offer it.\par
More recent research by Barrero, Bloom and Davies found that working from home is likely to be ``sticky'' \cite{barrero2021working}. They found:
\begin{itemize}
\item better-than-expected WFH experiences, 
\item new investments in physical and human capital that enable WFH, 
\item greatly diminished stigma associated with WFH, 
\item lingering concerns about crowds and contagion risks,
\item a pandemic-driven surge in technological innovations that support WFH.
\end{itemize}
More recently Enterprise Collaboration Systems (ECS) provide rich document management, sharing, and collaboration functionality across an organisation. The enterprise ECS system may integrate collaborative video \cite{prakash2020characteristic}. This is for instance the case with Microsoft Teams / Sharepoint. This integration of ECS should be considered when thinking about social VR systems which wish to support business, value, and trust. It is very much the case that large technology providers are attempting to integrate their `business back end' systems into their emerging metaverse systems. Open source equivalents are currently lacking.
\subsection{Pandemic drives adoption}
The ongoing global COVID-19 pandemic is \href{https://blog.yelp.com/news/the-future-of-work-is-remote/}{changing how people work}, toward a new global `normal'. Some ways of working are overdue transformation, and will be naturally disrupted. In the UK at least it seems that there may be real appetite to shift away from old practises. This upheaval will inevitably present both challenges and opportunities.\par
Highly technical workforces, especially, can \href{https://globalworkplaceanalytics.com/telecommuting-statistics}{operate from anywhere}. The post pandemic world seems to have stronger national border controls, with a resultant shortage of highly technical staff. This has forced the hand of global business toward \href{https://www.lifeatspotify.com/being-here/work-from-anywhere}{internationally distributed teams}. \par
If only a small percentage of companies allow the option of remote working, then they gain a structural advantage, enjoying benefits of reduced travel, lower workplace infection risk across all disease, and global agility for the personnel. Building and estate costs will certainly be reduced. More diversity may be possible. Issues such as sexual harassment and bullying may be reduced.  With reduced overheads product quality may increase. If customers are happier with their services, then over time this `push' may mean an enormous shift away from centralised working practises toward distributed working. \par
Technologies which support this working style were still in their infancy at the beginning of the pandemic. The rush to `Zoom', a previously relatively unknown and insecure \cite{aiken2020zooming} web meeting product, shows how naive businesses were in this space. \par
Connection of multiple users is now far better supported, with Zoom and \href{https://www.microsoft.com/en-us/Investor/earnings/FY-2021-Q1/press-release-webcast}{Mircosoft Teams} alone supporting hundreds of millions of chats a day. This is a 20x increase on market leader Skype's 2013 figure of \href{https://www.microsoft.com/en-us/Investor/earnings/FY-2013-Q1/press-release-webcast}{280 million} connections per month. Such technologies extend traditional telephony to provide important multi sensory cues.  However, these technologies demonstrate shortfalls compared to a live face-to-face meeting, which is generally agreed to be optimal for human-human interaction \cite{Wolff2008}.\par
While the research community and business are learning how to adapt working practises to web based telepresence \cite{oeppen2020human}, there remains little technology support for ad-hoc serendipitous meetings between small groups. It's possible that Metaverse applications can help to fill this gap, by gamification of social spaces, but the under discussed problems with video conferencing are likely to be even worse in such systems. \par
Chris Herd of ``FirstBase'' (who admittedly have a bias) provides some fascinating speculations: \par
\textit{``I've spoken to 2,000+ companies with 40M+ employees about remote work in the last 12 months 
A few predictions of what will happen before 2030:
\begin{itemize}
\item Rural Living: World-class people will move to smaller cities, have a lower cost of living \& higher quality of life.
\item These regions must innovate quickly to attract that wealth. Better schools, faster internet connections are a must.
\item Async Work: Offices are instantaneous gratification distraction factories where synchronous work makes it impossible to get stuff done.
\item Tools that enable asynchronous work are the most important thing globally remote teams need. A lot of startups will try to tackle this.
\item Hobbie Renaissance: Remote working will lead to a rise in people participating in hobbies and activities which link them to people in their local community.
\item This will lead to deeper, more meaningful relationships which overcome societal issues of loneliness and isolation.
\item Diversity \& Inclusion: The most diverse and inclusive teams in history will emerge rapidly
Companies who embrace it have a first-mover advantage to attract great talent globally. Companies who don't will lose their best people to their biggest competitors.
\item Output Focus: Time will be replaced as the main KPI for judging performance by productivity and output.
\item Great workers will be the ones who deliver what they promise consistently
\item Advancement decisions will be decided by capability rather than who you drink beer with after work.
\item Private Equity: The hottest trend of the next decade for private equity will see them purchase companies, make them remote-first
The cost saving in real-estate at scale will be eye-watering. The productivity gains will be the final nail in the coffin for the office
Working Too Much: Companies worry that the workers won't work enough when operating remotely.
\item The opposite will be true and become a big problem.
\item Remote workers burning out because they work too much will have to be addressed.
\item Remote Retreats: Purpose-built destinations that allow for entire companies to fly into a campus for a synchronous week.
\item Likely staffed with facilitators and educators who train staff on how to maximize effectiveness.
\item Life-Work Balance: The rise of remote will lead to people re-prioritizing what is important to them.
\item Organizing your work around your life will be the first noticeable switch. People realizing they are more than their job will lead to deeper purpose in other areas.
\item Bullshit Tasks: The need to pad out your 8 hour day will evaporate, replaced by clear tasks and responsibilities.
\item Workers will do what needs to be done rather than wasting their trying to look busy with the rest of the office
\end{itemize}
''}

            \subsection{Point to Point Video Conferencing}
                O'Malley et al. showed that face-to-face and video mediated employed  visual cues for mutual understanding, and that addition of video to the audio channel aided confidence and mutual understanding. However, video mediated did not provide the clear cues of being co-located \cite{OMalley1996}.\par
                Dourish et al. make a case for not using face-to-face as a baseline for comparison, but rather that analysis of the efficacy of remote tele-collaboration tools should be made in a wider context of connected multimedia tools and `emergent communicative practises' \cite{Dourish1996}. While this is an interesting viewpoint it does not necessarily map well to a recreation of the ad-hoc meeting.\par
There is established literature on human sensitivity to eye contact in both 2D and 3D VC \cite{Criminisi2003, Van_Eijk2010}, with an accepted minimum of 5-10 degrees before observers can reliably sense they are not being looked at \cite{Chen2002}. Roberts et al. suggested that at the limit of social gaze distance (~4m) the maximum angular separation between people standing shoulder to shoulder in the real world would be around 4 degrees\cite{Roberts2013}. \par
                Sellen found limited impact on turn passing when adding a visual channel to audio between two people when using Hydra, an early system which provided multiple video conference displays in an intuitive spatial distribution\cite{Sellen1992}. She did however, find that the design of the video system affected the ability to hold multi-party conversations \cite{Sellen1995}.\par
                Monk and Gale describe in detail experiments which they used 
 for examining gaze awareness in communication which is mediated and unmediated by technology.  
  They   found that gaze awareness increased message understanding  
  \cite{Monk2002}.\par
                Both Kuster et al. and Gemmel et al. have successfuly demonstrated software systems which can adjust eye gaze to correct for off axis capture in real time video systems\cite{Gemmell2000, Kuster2012}.\par
                Shahid et al. conducted a study on pairs of children playing games with and without video mediation and concluded that the availability of mutual gaze affordance enriched social presence and fun, while its absence dramatically affects the quality of the interaction. They used the `Networked Minds', a social presence questionnaire.                
        \subsection{Triadic and Small Group}
Early enthusiasm in the 1970's for video conferencing, as a medium for small group interaction quickly turned to disillusionment. It was agreed after a flurry of initial research that the systems at the time offered no particular advantage over audio only communication, and at considerable cost \cite{Williams1977}.\par
Something in the breakdown of normal visual cues seems to impact the ability of the technology to support flowing group interaction. Nonetheless, some non-verbal communication is supported in VC with limited success. \par
Additional screens and cameras can partially overcome the limitation of no multi-party support (that of addressing a room full of people on a single screen) by making available more bidirectional channels. For instance, every remote user can be a head on a screen with a corresponding camera. The positioning of the screens must then necessarily match the physical organization of the remote room.\par
Egido provides an early review of the failure of VC for group activity, with the ``misrepresentation of the technology as a substitute for face-to-face" still being valid today \cite{Edigo1988}.\par
Commercial systems such as Cisco Telepresence Rooms cluster their cameras above the centre screen of three for meetings using their telecollaboration product, while admitting that this only works well for the central seat of the three screens. They also group multiple people on a single screen in what Workhoven et al. dub a ``non-isotropic" configuration \cite{Pejsa2016}. They maintain that this is a suitable trade off as the focus of the meeting is more generally toward the important contributor in the central seat. This does not necessarily follow for less formal meeting paradigms.\par
            In small groups, it is more difficult to align non-verbal cues between all  parties, and at the same time, it is more important because the hand-offs between parties are more numerous and important in groups. A breakdown in conversational flow in such circumstances is harder to solve. A perception of the next person to talk must be resolved for all parties and agreed upon to some extent.\par
                However, most of the conventional single camera, and expensive multi camera VC systems, suffer a fundamental limitation in that the offset between the camera sight lines and the lines of actual sight introduce incongruities that the brain must compensate for \cite{Wolff2008}.\par
\subsection{Other Systems to Support Business}                  
There have been many attempts to support group working and rich data sharing between dispersed groups in a business setting. So called 'smart spaces' allow interaction with different displays for different activities and add in some ability to communicate with remote or even mobile collaborators on shared documents \cite{Bardram2012}, with additional challenges for multi-disciplinary groups who are perhaps less familiar with one or more of the technology barriers involved \cite{Adamczyk2007}.\par
Early systems like clearboard \cite{Ishii1993} demonstrated the potential for smart whiteboards with a webcam component for peer-to-peer collaborative working. Indeed it is possible to support this modality with Skype and a smartboard system (and up to deployments such as Accessgrid). They remain relatively unpopular however.\par
\subsection{Mona Lisa Type Effects}
Almost all traditional group video meeting tools suffer from the so-called Mona Lisa effect which describes the phenomenon where the apparent gaze of a portrait or 2 dimensional image always appears to look at the observer regardless of the observer's position \cite{Vishwanath2005, Anstis1969, Wollaston1824}. This situation manifests when the painted or imaged subject is looking into the camera or at the eyes of the painter \cite{Loomis2008, Fullwood2006}.\par
Single user-to-user systems based around bidirectional video implicitly align the user's gaze by constraining the camera to roughly the same location as the display. When viewed away from this ideal axis, it creates the feeling of being looked at regardless of where this observer is \cite{Moubayed2012, Vishwanath2005, Anstis1969, Wollaston1824}, or the ``collapsed view effect'' \cite{Nguyen2005} where perception of gaze transmitted from a 2 dimensional image or video is dependent on the incidence of originating gaze to the transmission medium. \par
Multiple individuals using one such channel can feel as if they are being looked at simultaneously, leading to a breakdown in the normal non-verbal communication which mediates turn passing \cite{Vertegaal2002}.    
There is research investigating this sensitivity when the gaze is mediated by a technology, finding that ``disparity between the optical axis of the camera and the looking direction of a looker should be at most 1.2 degrees in the horizontal direction, and 1.7 degrees in vertical direction to support eye contact" \cite{Van_Eijk2010, Bock2008}. It seems that humans assume that they are being looked at unless they are sure that they are not \cite{Chen2002}.\par
To be clear, there are technological solutions to this problem, but it's useful in the context of discussing metaverse to know that this problem exists. It's known that there are cognitive dissonances around panes of video conference images, but it seems that the effect is truely limited to 2D surfaces. A 3D projection surface (a physical model of a human) designed to address this problem completely removed the Mona Lisa effect \cite{Moubayed2012}.\par 
Metaverse then perhaps offers the promise of solving this, making more natural interaction possible, but it's clearly a long way from delivering on those promises right now. We need to understand what's important and try to map these into a metaverse product.
\section{What's important for human communication}
\subsection{Vocal}
The ubiquitous technology to mediate conversation is, of course, the telephone. The \href{https://www.ericsson.com/en/reports-and-papers/mobility-report/reports/november-2021}{2021 Ericsson mobility report}  states that there are around 8 billion mobile subscriptions globally. More people have access to mobile phones than to working toilets \href{https://www.unicef.org/innovation/stories/more-cellphones-toilets}{according to UNICEF}.\par
Joupii and Pan designed a system which focused attention on spatially correct high definition audio. They found ``significant improvement over traditional audio conferencing technology, primarily due to the increased dynamic range and directionality. \cite{Jouppi2002}. Aoki et al. also describe an audio only system with support for spatial cues \cite{Aoki2003}.\par
In the following sections we will attempt to rigorously identify just what is important for our proposed application of business centric communication, supportive of trust, and thereby value transfer.\par
In his book `Bodily Communication' \cite{Argyle1988} Michael Argyle divides vocal signals into the following categories:
\begin{enumerate}
\item Verbal
\item Non-Verbal Vocalisations
\begin{enumerate}
       \item Linked to Speech
       \begin{enumerate}
         \item   Prosodic
         \item   Synchronising
         \item   Speech Disturbances
         \end{enumerate}
      \item  Independent of Speech
      \begin{enumerate}
        \item    Emotional Noises
         \item   Paralinguistic (emotion and interpersonal attitudes)
         \item   Personal voice and quality of accent
         \end{enumerate}
\end{enumerate}
\end{enumerate}               
Additional to the semantic content of verbal communication there is a rich layer of meaning in pauses, gaps, and overlaps \cite{Heldner2010} which help to mediate who is speaking and who is listening in multi-party conversation. This mediation of turn passing, to facilitate flow, is by no means a given and is highly dependent on context and other factors \cite{Kleinke1986}. Interruptions are also a major factor in turn passing.\par
This extra-verbal content \cite{Ting-Toomey2012} extends into physical cues, so-called `nonverbal' cues, and there are utterances which link the verbal and non-verbal \cite{Otsuka2005}. This will be discussed later, but to an extent, it is impossible to discuss verbal communication without regard to the implicit support which exists around the words themselves.\par
In the context of all technology-mediated conversation the extra-verbal is easily compromised if technology used to support communication over a distance does not convey the information, or conveys it badly. This can introduce additional complexity \cite{Otsuka2005}.\par
These support structures are pretty much lacking in metaverse XR systems. The goal then here perhaps is to examine the state-of-the-art, and remove as many of the known barriers as possible. Such a process might better support trust, which might better support the kind of economic and activity we seek to engineer.\par
When examining just verbal / audio communication technology it can be assumed that the physical non-verbal cues are lost, though not necessarily unused. In the absence of non-verbal cues it falls to timely vocal signals to take up the slack when framing and organising the turn passing. For the synchronising of vocal signals between the parties to be effective the systemic delays must remain small. System latency, the inherent delays added by the communication technology, can allow slips or a complete breakdown of 'flow' \cite{katagiri2007aiduti}. This problem can be felt in current social VR platforms, though people don't necessarily identify the cause of the breakdown correctly. In the main they feel to the users like a bad ``audio-only'' teleconference.\par
With that said, the transmission of verbal / audio remains the most critical element for interpersonal communication as the most essential meaning is encoded semantically. There is a debate about ratios of how much information is conveyed through the various human channels \cite{Loomis2012}, but it is reasonable to infer from its ubiquity that support for audio is essential for meaningful communication over a distance. We have seen that it must be timely, to prevent a breakdown of framing, and preferably have sufficient fidelity to convey sub-vocal utterances. \par
For social immersive VR for business users, a real-time network such as websockets, RTP, or UDP seems essential, much better microphones are important, and the system should support both angular spatialisation, and respond to distance between interlocutors.
\subsection{Nonverbal}
We have already seen that verbal exchanges take place in a wider context of sub vocal and physical cues. In addition, the spatial relationship between the parties, their focus of attention, their gestures and actions, and the wider context of their environment all play a part in communication \cite{Goodwin2000}. These are identified as follows by Gillies and Slater \cite{Gillies2005} in their paper on virtual agents.\par
\begin{itemize}
\item Posture and gesture
\item Facial expression
\item Gaze
\item Proxemics
\item Head position and orientation
\item Interactional synchrony
\end{itemize}

This is clearly important for our proposed collaborative mixed reality application. Below we will examine these six areas by looking across the wider available research.

\subsubsection{Gaze}
Of particular importance is judgement of eye gaze which is normally fast, accurate and automatic, operating at multiple levels of cognition through multiple cues \cite{Argyle1988,Argyle1976,Argyle1965,Argyle1976,Argyle1969, Kendon1967,Monk2002}.\par
Gaze in particular aids smooth turn passing \cite{Hedge1978} \cite{Novick1996} and lack of support for eye gaze has been found to decrease the efficiency of turn passing by 25\% \cite{Vertegaal2000}.\par
There are clear patterns to eye gaze in groups, with the person talking, or being talked to, probably also being looked at \cite{Vertegaal2001} \cite{Langton2000}. To facilitate this groups will tend to position themselves to maximally enable observation of the gaze of the other parties \cite{Kendon1967}. This intersects with proxemics which will be discussed shortly.  In general people look most when they are listening, with short glances of 3-10 seconds \cite{Argyle1965}. 
Colburn et al. suggest that gaze direction and the perception of the gaze of others directly impacts social cognition \cite{Colburn2000} and this has been supported in a follow up study \cite{Macrae2002}.\par
The importance of gaze is clearly so significant in evolutionary terms that human acuity for eye direction is considered high at ~30 sec arc \cite{Symons2004} with straight binocular gaze judged more accurately than straight monocular gaze \cite{Kluttz2009}, when using stereo vision. \par
Regarding the judgement of the gaze of others, Symons et al. suggested that ``people are remarkably sensitive to shifts in a person's eye gaze'' in triadic conversation \cite{Symons2004}. 
This perception of the gaze of others operates at a low level and is automatic. Langton et al. cite research stating that the gaze of others is ``able to trigger reflexive shifts of an observer's visual attention'' and further discuss the deep biological underpinnings of gaze processing \cite{Langton2000}. \par  
When discussing technology-mediated systems, Vertegaal \& Ding suggested that understanding the effects of gaze on triadic conversation is ``crucial for the design of teleconferencing systems and collaborative virtual environments'' \cite{Vertegaal2002}, and further found correlation between the amount of gaze, and amount of speech. Vertegaal \& Slagter suggest that ``gaze function(s) as an indicator of conversational attention in multiparty conversations'' \cite{Vertegaal2001}. It seems like is we are to have useful markets within social immersive environments then support for natural gaze effects should be a priority.\par  
Wilson et al. found that subjects can ``discriminate gaze focused on adjacent faces up to [3.5m]'' \cite{Wilson2000}. This perhaps gives us a testable benchmark within a metaverse application which is eye gaze enabled. In this regard Schrammel et al. investigated to what extent embodied agents can elicit the same responses in eye gaze detection \cite{Schrammel2007}.\par       
Vertegaal et al. found that task performace was 46\% better when gaze was synchronised in their telepresence scenario. As they point out, gaze synchonisation (temporal and spatial) is `commendable' in all such group situations, but the precise utility will depend upon the task \cite{Vertegaal2002}.\par
There has been some success in the automatic detection of the focus of attention of participants in multi party meetings \cite{Stiefelhagen2001, Stiefelhagen2002}.  More recently, eye tracking technologies allow the recording and replaying of accurate eye gaze information \cite{Steptoe2009} alongside information about pupil dilation toward determination of honesty and social presence \cite{Steptoe2010}. It seems there are trust and honesty issues conflated with how collaborants in a virtual space are represented.\par               
In summary, gaze awareness does not just mediate verbal communication but rather is a complex channel of communication in its own right. Importantly, gaze has a controlling impact on those who are involved in the communication at any one time, including and excluding even beyond the current participants. Perhaps the systems we propose in this book need to demand eye gaze support, but it is clear that it should be recommended, and that the software selected should support the technology integration in principle.\par
\subsubsection{Mutual Gaze}
Aygyle and Cook established early work around gaze and mutual gaze, with their seminal book of the same title \cite{Argyle1976}, additionally detailing confounding factors around limitations and inaccuracies in observance of gaze and how this varies with distance \cite{Argyle1969, Argyle1988, Cook1977}.\par
Mutual gaze is considered to be the most sophisticated form of gaze awareness with significant impact on dyadic conversation especially \cite{Cook1977, Kleinke1986, Fagel2010}. The effects seem more profound than just helping to mediate flow and attention, with mutual eye gaze aiding in memory recall and the formation of impressions \cite{Bohannon2013}.\par
While reconnection of mutual eye gaze through a technology boundary does not seem completely necessary it is potentially important, with impact on subtle elements of one-to-one communication, and therefore discrimination of eye gaze direction should be bi-directional if possible, and if possible have sufficient accuracy to judge direct eye contact. In their review Bohannon et al. said that the issue of rejoining eye contact must be addressed in order to fully realise the richness of simulating face-to-face encounters \cite{Bohannon2013}.\par
Mutual gaze is a challenging affordance as bi-directional connection of gaze is not a trivial problem. It's perhaps best to view this as at the `edge' of our requirements for a metaverse.
                   \subsubsection{Mutual Gaze in Telepresence}
          We have seen that transmission of attention can broadly impact communication in subtle ways, impacting empathy, trust, cognition, and co-working patterns. Mutual gaze (looking into one another's eyes), is currently the high water mark for technology-mediated conversation.\par
          Many attempts have been made to re-unite mutual eye gaze when using tele-conferencing systems. In their 2015 review of approaches Regenbrecht and Langlotz found that none of the methods they examined were completely ideal \cite{Regenbrecht2015}. They found most promise in 2D and 3D interpolation techniques, which will be discussed in detail later, but they opined that such systems were very much ongoing research and lacked sufficient optimisation.\par
          A popular approach uses the so called 'Peppers Ghost' phenomenon \cite{Steinmeyer2013}, where a semi silvered mirror presents an image to the eye of the observer, but allows a camera to view through from behind the angled mirror surface. The earliest example of this is Rosental's two way television system in 1947 \cite{Rosenthal1947}, though Buxton et al. `Reciprocal Video Tunnel' from 1992 is more often cited \cite{Buxton1992}. This optical characteristic isn't supported by retroreflective projection technology, and besides requires careful control of light levels either side of the semi-silvered surface.\par  
The early GAZE-2 system (which makes use of Pepper's ghost) is novel in that it uses an eye tracker to select the correct camera from several trained on the remote user. This ensures that the correct returned gaze (within the ability of the system) is returned to the correct user on the other end of the network \cite{Vertegaal2003}.
Mutual gaze capability is later highlighted as an affordance supported or unsupported by key research and commercial systems.

\subsubsection{Head Orientation}
Orientation of the head (judged by the breaking of bilateral symmetry and alignment of nose) is a key factor when judging attention. Perception of head orientation can be judged to within a couple of degrees \cite{Wilson2000}.\par
It has been established that head gaze can be detected all the way out to the extremis of peripheral vision, with accurate eye gaze assessment only achievable in central vision \cite{Loomis2008}. This is less of use for our metaverses at this time, because user field of view is almost always restricted in such systems. More usefully, features of illumination can alter the apparent orientation of the head \cite{Troje1998}.\par
Head motion over head orientation is a more nuanced propostion and can be considered a micro gesture \cite{Boker2011}. Head tracking systems within head mounted displays can certainly detect these tiny movements, but it's clear that not all of this resolution is passed into shared virtual settings through avatars. It would be beneficial to be able to fine tune this feature within any software selected.\par
                    It is possible that 3D displays are better suited to perception of head gaze since it is suggested that they are more suitable for ``shape understanding tasks'' \cite{St_John2001}\par
                    Bailenson, Baell, and Blascovich found that giving avatars rendered head movements in a shared virtual environment decreased the amount of talking, possibly as the extra channel of head gaze was opened up. They also reported that subjectively, communication was enhanced \cite{Bailenson2002}. \par
                    Clearly head orientation is an important indicator of the direction of attention of members of a group and can be discerned even in peripheral vision. This allows the focus of several parties to be followed simultaneously and is an important affordance to replicate on any multi-party communication system. \par
\subsubsection{Combined Head and Eye Gaze}
Rienks et al. found that head orientation alone does not provide a reliable cue for identification of the speaker in a multiparty setting \cite{Rienks2010}. Stiefelhagen \& Zhu found ``that head orientation contributes 68.9\% to the overall gaze direction on average'' \cite{Stiefelhagen2002}, though head and eye gaze seem to be judged interdependently \cite{Kluttz2009}. Langton noted that head and eye gaze are ``mutually influential in the analysis of social attention'' \cite{Langton2000}, and it is clear that transmission of `head gaze' by any mediating system, enhances rather than replaces timely detection of subtle cues. Combined head and eye gaze give the best of both worlds and extend the lateral field of view in which attention can be reliably conveyed to others \cite{Loomis2008}.
\subsubsection{Other Upper Body: Overview}
While it is well evidenced that there are advantages to accurate connection of the gaze between conversational partners \cite{Argyle1969, Kleinke1986}, there is also a body of evidence that physical communication channels extend beyond the face \cite{Kleinke1986, Nguyen2009} and include both micro (shrugs, hands and arms), and macro movement of the upper body \cite{Ekman1993}. Goldin-Meadow suggests that gesturing aids conversational flow by resolving mismatches and aiding cognition \cite{Goldin-Meadow1999}.\par
                    In their technology-mediated experiment which compared face to upper body and face on a flat screen, Nguyen and Canny found that ``upper-body framing improves empathy measures and gives results not significantly different from face-to-face under several empathy measures'' \cite{Nguyen2009}. 
                    
The upper body can be broken up as follows:\par
\textbf{Facial}\\Much emotional context can be described by facial expression (display) alone \cite{Ekman1993, Chovil1991}, with smooth transition between expressions seemingly important \cite{schiano2004}. This suggests that mediating technologies should support high temporal resolution, or at least that there is a minimum resolution between which transitions between expressions become too 'categorical'. Some aspects of conversational flow appear to be mediated in part by facial expression \cite{ohba1998}. There are gender differences in the perception of facial affect \cite{Hofmann2006}.\par
\textbf{Gesturing} \\(such as pointing at objects) paves the way for more complex channels of human communication and is a basic and ubiquitous channel \cite{Iverson2005}.  Conversational hand gestures provide a powerful additional augmentation to verbal content \cite{Krauss1996}.\par
\textbf{Posture} \\Some emotions can be conveyed through upper body configurations alone. Argyle details some of these \cite{Argyle1988} and makes reference to the posture of the body and the arrangement of the arms (i.e. folded across the chest). These are clearly important cues. Kleinsmith and Bianchi-Berthouze assert that "some affective expressions may be better communicated by the body than the face" \cite{Kleinsmith2013}.\par
\textbf{Body Torque} \\In multi-party conversation, body torque, that is the rotation of the trunk from front facing, can convey aspects of attention and focus \cite{Schegloff1998}.\par
In summary, visual cues which manifest on the upper body and face can convey meaning, mediate conversation, direct attention, and augment verbal utterances. \par
\subsubsection{Effect of Shared Objects on Gaze}
Ou et al. detail shared task eye gaze behaviour ``in which helpers seek visual evidence for workers' understanding when they lack confidence of that understanding, either from a shared, or common vocabulary'' \cite{Ou2005}.\par
\begin{figure}[!h]
\includegraphics[width=\linewidth]{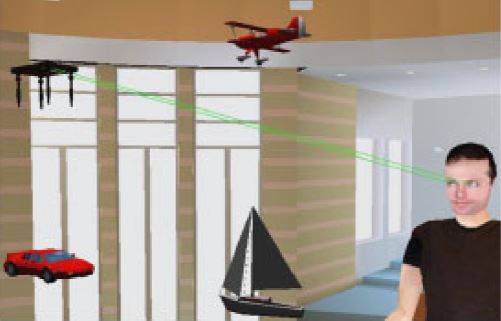}
\caption{Eye tracked eye gaze awareness in VR. Murray et al. used immersive and semi immersive systems alongside eye trackers to examine the ability of two avatars to detect the gaze awareness of a similarly immersed collaborator.}
\label{fig:murrayeyegaze}
\end{figure}

 
  Murray et al. found that in virtual environments, eye gaze is crucial for discerning what a subject is looking at \cite{Murray2009}. This work is shown in Figure \ref{fig:murrayeyegaze}.\par
It is established that conversation around a shared object or task, especially a complex one, mitigates gaze between parties \cite{Argyle1976} and this suggests that in some situations around shared tasks in metaverses it may be appropriate to reduce fidelity of representation of the avatars. \par

  \subsubsection{Tabletop and Shared Task}
In early telepresence research Buxton and William argued through examples that ``effective telepresence depends on quality sharing of both person and task space \cite{Buxton1992}.\par
In their triadic shared virtual workspace Tang et al. found difficulty in reading shared text using a `round the table' configuration, a marked preference for working collaboratively on the same side of the table. They also found additional confusion as to the identity of remote participants \cite{Tang2010}.
Tse et al. found that pairs can work well over a shared digital tabletop, successfully overcoming a single user interface to interleave tasks \cite{Tse2007}.\par
Tang et al. demonstrate that collaborators engage and disengage around a group activity through several distinct, recognizable mechanisms with unique characteristics \cite{Tang2006}. They state that tabletop interfaces should offer a variety of tools to facilitate this fluidity.\par
Camblend is a shared workspace with panoramic high resolution video. It maintains some spatial cues between locations by keeping a shared object in the video feeds \cite{Norris2013, Norris2012}. Participants successfully resolved co-orientation within the system.\par
The t-room system implemented by Luff et al. surrounds co-located participants standing at a shared digital table with life sized body and head video representations of remote collaborators \cite{Luff2011} but found that there were incongruities in the spatial and temporal matching between the collaborators which broke the flow of conversation.
Tuddenham et al. found that co-located collaborators naturally devolved 'territory' of working when sharing a task space, and that this did not happen the same way with a tele-present collaborator \cite{Tuddenham2009}. Instead remote collaboration adapted to use a patchwork of ownership of a shared task. It seems obvious to say that task ownership is a function of working space, but it is interesting that the research found no measurable difference in performance when the patchwork coping strategy was employed.\par
The nature of a shared collaborative task and/or interface directly impacts the style of interaction between collaborators. This will have a bearing on the choice of task for experimentation \cite{Jamil2011, Jetter2011}.

\section{Psychology of Technology-Mediated Interaction}       
\subsection{Proxemics}
\begin{figure}[!h]
\includegraphics[width=\linewidth]{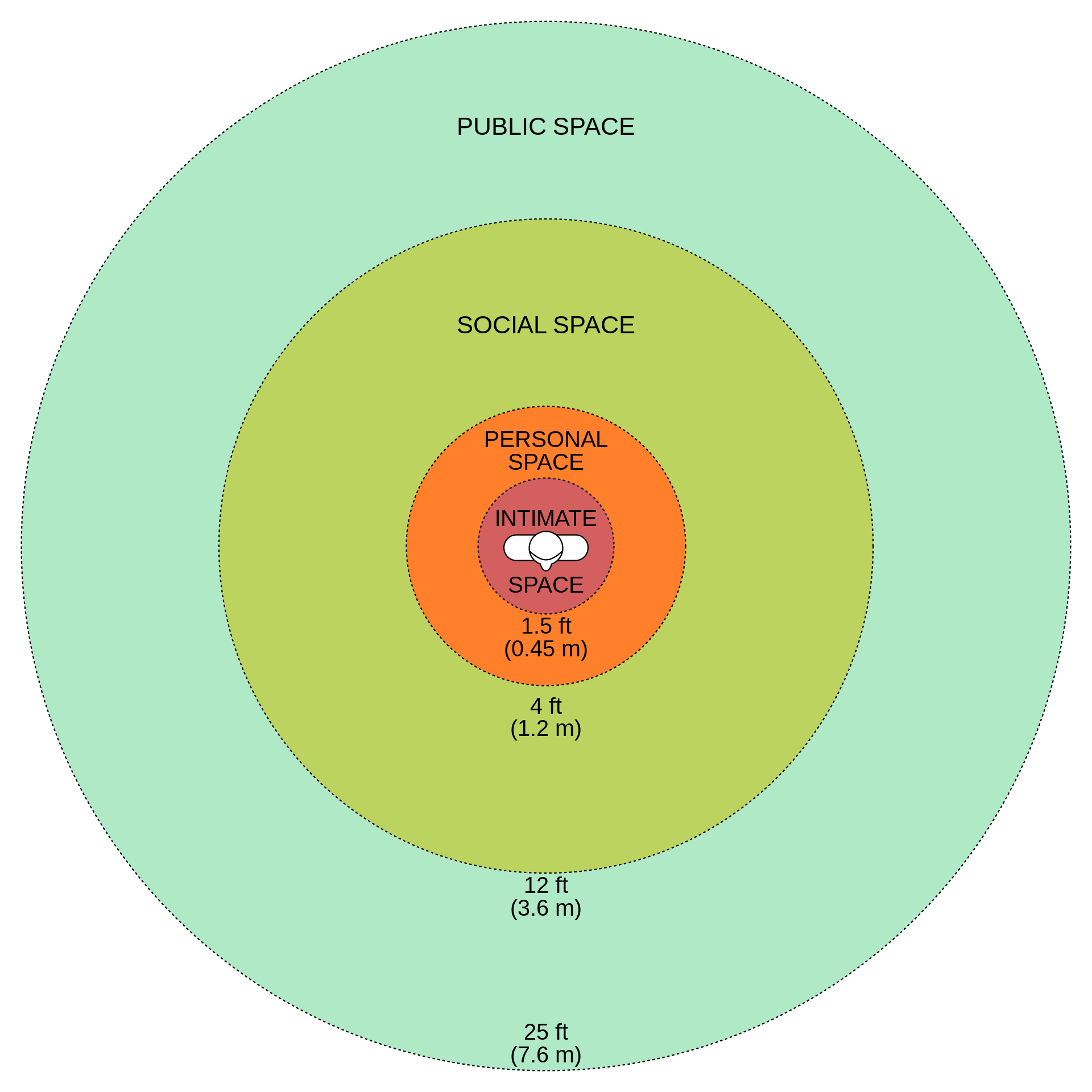}
\caption{Bands of social space around a person Image CC0 \href{https://en.wikipedia.org/wiki/Proxemics}{from wikipedia}.}
\label{fig:proxemics}
\end{figure}

Proxemics is the formal study of the regions of interpersonal space begun in the late 50's by Hall and Sommers and building toward The Hidden Dimension \cite{Hall1969}, which details bands of space (Figure \ref{fig:proxemics}) that are implicitly and instinctively created by humans and which have a direct bearing on communication.
Distance between conversational partners, and affiliation, also have a bearing on the level of eye contact \cite{Argyle1965} with a natural distance equilibrium being established and developed throughout, through both eye contact and a variety of subtle factors. Argyle \& Ingham provide levels of expected gaze and mutual gaze against distance \cite{Argyle1969}. These boundaries are altered by ethnicity \cite{Watson1966, Argyle1988} and somewhat by gender \cite{Bruno2013}, and age \cite{Slessor2008, Hofmann2006}.\par
Even with significant abstraction by communication systems (such as SecondLife) social norms around personal space persist \cite{Yee2007, Bailenson2001, Bailenson2003}. Bailenson \& Blascovich found that even in Immersive Collaborative Virtual Environments (ICVE's) ``participants respected personal space of the humanoid representation''\cite{Bailenson2001} implying that this is a deeply held 'low-level' psychophysical reaction \cite{Blascovich2002}. The degree to which this applies to non-humanoid avatars seems under explored.\par
Maeda et al. \cite{Maeda2004} found that seating position impacts the level of engagement in teleconferencing. Taken together with the potential for reconfiguration within the group as well as perhaps signalling for the attention of participants outside of the confines of the group in an open business metaverse setting.\par
When considering the attention of engaging with people outside the confines of a meeting Hager et al. found that gross expressions can be resolved by humans over long distances \cite{Hager1979, Argyle1988}. It seems that social interaction begins around 7.5m in the so-called `public space' \cite{Hall1969}. Recreating this affordance in a metaverse would be a function of the display resolution, and seems another `stretch goal' rather than a core requirement.\par                
\subsection{Attention}
The study of attention is a discrete branch of psychology. It is the study of cognitive selection toward a subjective or objective sub focus, to the relative exclusion of other stimulae. It has been defined as ``a range of neural operations that selectively enhance processing of information'' \cite{Carlston2013}. In the context of interpersonal communication it can be refined to apply to selectively favouring a conversational agent or object or task above other stimuli in the contextual frame.\par
Humans can readily determine the focus of attention of others in their space \cite{Stiefelhagen2001} and preservation of the spatial cues which support this are important for technology-mediated communication \cite{Sellen1992} \cite{Stiefelhagen2002}.\par
The interplay between conversational partners, especially the reciprocal perception of attention, is dubbed the perceptual crossing \cite{Deckers2013, Gibson1963}.\par
This is a complex field of study with gender, age, and ethnicity all impacting the behaviour of interpersonal attention \cite{Bente1998, Slessor2008, Argyle1988, Hofmann2006, Pan2008}.
Vertegaal has done a great deal of work on awareness and attention in technology-mediated situations and the work of his group is cited throughout this chapter \cite{Vertegaal1997}. As an example it is still such a challenge to ``get'' attention through mediated channels of communication, that some research \cite{Fels2000, Sellen1992} and many commercial systems such as `blackboard collaborate', Zoom, and Teams use tell tale signals (such as a microphone icon) to indicate when a participant is actively contributing. Some are automatic, but many are still manual, requiring that a user effectively hold up a virtual hand to signal their wish to communicate.\par
Langton et al. cite research stating that the gaze of others is ``able to trigger reflexive shifts of an observer's visual attention''. \par 
Regarding the attention of others, Fagal et el demonstrated that eye visibility impacts collaborative task performance when considering a shared task \cite{Fagel2010}. Novick et al. performed analysis on task hand-off gaze patterns which is useful for extension into shared task product design \cite{Novick1996}. 
\subsection{Behaviour}
Hedge et al. suggested that gaze interactions between strangers and friends may be different which could have an impact on the kinds of interactions a metaverse might best support \cite{Hedge1978}. Voida et al. elaborate that prior relationships can cause ``internal fault lines'' in group working \cite{Voida2012}. When new relationships are formed the ``primary concern is one of uncertainty reduction or increasing predictability about the behaviour of both themselves and others in the interaction'' \cite{Berger1975}. This concept of smoothness in the conversation is a recurring theme, with better engineered systems introducing less extraneous artefacts into the communication, and so disturbing the flow less. Immersive metaverse are rife with artefacts.\par 
In a similar vein the actor-observer effect describes the mismatch between expectations which can creep into conversation. Conversations mediated by technology can be especially prone to diverging perceptions of the causes of behaviour \cite{Jones1971}. Basically this means misunderstandings happen, and are harder to resolve with more mediating technology.\par 
Interacting subjects progress conversation through so-called `perception-action' loops which are open to predictive modelling through discrete hidden Markov models \cite{Mihoub2015}. This might allow product OKR testing of the effectiveness of engineered systems \cite{doerr2018measure}.\par
It may be that the perception-behaviour link where unconscious mirroring of posture bolsters empathy between conversational partners, especially when working collaboratively \cite{Chartrand1999}, and the extent to which posture is represented through a communication medium may be important.\par
Landsberger posited the Hawthorne effect \cite{Parsons1974}. Put simply this is a short term increase in productivity that may occur as a result of being watched or appreciated. The impression of being watched changes gaze patterns during experimentation, with even implied observation through an eye tracker modifying behaviour \cite{Risko2011}.\par
There are also some fascinating findings around the neural correlates of gratitude, which turn out not to be linked to gratitude felt by a participant, but rather the observation of gratitude received within a social context \cite{fox2015neural}. These findings have potentially useful implications for the behaviours of AI actors and avatars within an immersive social scene.\par
There is much historic work describing ``the anatomy of cooperation" \cite{Kollock1998}, and this might better inform how educational or instructional tasks are built in metaverse applications.\par
Cuddihy and Walters defined an early model for assessing desktop interaction mechanisms for social virtual environments \cite{Cuddihy2000}.                               
\subsubsection{Perception Of Honesty}
Hancock et al. state that we are most likely to lie, and to be lied to, on the telephone \cite{Hancock2004}. Technology used for communication impacts interpersonal honesty. It seems that at some level humans know this; lack of eye contact leads to feelings of deception, impacting trust \cite{Holm2010}. This has a major impact on immersive social XR, which often does not support mutual gaze. Trust is crucial for business interactions.\par
Further there are universal expressions, micro-expressions, and blink rate which can betray hidden emotions \cite{Porter2008}, though the effects are subtle and there is a general lack of awareness by humans of their abilities in this regard \cite{Holm2010}. Absence of support for such instinctive cues inhibits trust \cite{Roberts2015}. Support for these rapid and transient facial features demands high resolution reproduction in both resolution and time domains. There is detectable difference in a participant's ability to detect deception when between video conference mediated communication and that mediated by avatars \cite{Steptoe2010}. Systems should aim for maximally faithful reproduction. 
        \subsection{Presence, Co-presence, and Social Presence}
            Presence is a heavily cited historic indicator of engagement in virtual reality, though the precise meaning has been interpreted differently by different specialisms \cite{Beck2011, Schuemie2001}. It is generally agreed to be the 'sense of being' in a virtual environment \cite{Slater1999}. Slater extends this to include the ``extent to which the VE becomes dominant". \par
Beck et al. reviewed 108 articles and synthesised an ontology of presence \cite{Beck2011} which at its simplest is as follows:
            \begin{enumerate}
				\item Sentient presence
                    \begin{enumerate}
                     \item Physical interaction
                      \item Mental interaction
                    \end{enumerate}
                   \item Non-sentient
                   \begin{enumerate}
                       \item Physical immersion
                       \item Mental immersion = psychological state
                     \end{enumerate}
            \end{enumerate}
            
When presence is applied to interaction it may be split into Telepresence, and Co/Social presence  \cite{Heeter1992, Biocca1997}.  Co-presence and/or social presence is the sense of ``being there with another", and describes the automatic responses to complex social cues \cite{Fulk1987, Haythornthwaite1995}.    Social presence (and co-presence) refers in this research context to social presence which is mediated by technology (even extending to text based chat \cite{Gunawardena1997}), and has its foundations in psychological mechanisms which engender mutualism in the `real'. This is analysed in depth by Nowak \cite{Nowak2001}. An examination of telepresence, co-presence and social presence necessarily revisits some of the knowledge already elaborated.\par
        The boundaries between the three are blurred in research with conflicting results presented \cite{Bulu2012}. Biocca et al. attempted to enumerate the different levels and interpretations surrounding these vague words \cite{Biocca2003}, and to distill them into a more robust theory which better lends itself to measurement. They suggest a solid understanding of the surrounding psychological requirements which need support in a mediated setting, and then a scope that is detailed and limited to the mediated situation.\par
 Since `social presence' has been subject to varied definitions \cite{Biocca2003} it is useful here to consider a single definition from the literature which defines it as ``the ability of participants in the community of inquiry to project their personal characteristics into the community, thereby presenting themselves to the other participants as real people.'' \cite{Garrison1999, Beck2011}. Similarly to specifically define co-presence for this research it is taken to be the degree to which participants in a virtual environment are ``accesible, available, and subject to one another" \cite{Biocca2003}. \par
            Social presence has received much attention and there are established questionnaires used in the field for measurement of the levels of perceived social presence yet the definitions here also remain broad, with some confusion about what is being measured \cite{Biocca2003}.\par            
 Telepresence meanwhile is interaction with a different (usually remote) environment which may or may not be virtual, and may or may not contain a separate social/co-presence component. \par 
       Even in simple videoconferencing Bondareva and Bouwhuis stated (as part of an experimental design) that the following determinants are important to create social presence \cite{Bondareva2004, Jouppi2002}. 
            \begin{enumerate}
            \item    Direct eye contact is preserved
            \item    Wide visual field
            \item    Both remote participants appear life size
            \item    Possibility for participants to see the upper body of the interlocutor
            \item    High quality image and correct colour reproduction
            \item    Audio with high S/N ratio
            \item    Directional sound field
            \item    Minimization of the video and audio signal asynchrony
            \item    Availability of a shared working space.
            \end{enumerate}

            Bondareva et al. went on to describe a person-to-person telepresence system with a semi-silvered mirror to reconnect eye gaze, which they claimed increased social presence indicators. Interestingly they chose a checklist of interpersonal interactions which they used against recordings of conversations through the system \cite{Bondareva2004}.  \par
            The idea of social presence as an indicator of the efficacy of the system, suggests the use of social presence questionnaires in the evaluation of the system \cite{Biocca2003}.  Subjective questionnaires are however troublesome in measuring effectiveness of virtual agents and embodiments, with even nonsensical questions producing seemingly valid results \cite{Slater2004}. Usoh et al. found that 'the real' produced only marginally higher presence results than the virtual \cite{Usoh2000}. It would be difficult to test products this way.\par
            Nowak states that ``A satisfactory level of co-presence with another mind can be achieved with conscious awareness that the interaction is mediated" and asserts that while the mediation may influence the degree of co-presence it is not a prohibiting factor \cite{Nowak2001}.\par 
            Baren and IJsselsteijn \cite{Van_Baren2004, Harms2004} list 20 useful presence questionnaires in 2004 of which ``Networked Minds" seemed most appropriate for the research.
            Hauber et al. employed the ``Networked Minds" Social Presence questionnaire experimentally and found that while the measure could successfully discriminate between triadic conversation that is mediated or unmediated by technology, it could not find a difference between 2D and 3D mediated interfaces \cite{Hauber2005, Gunawardena1997}.\par
			In summary, social presence and co-presence are important historic measures of the efficacy of a communication system. Use of the term in literature peaked between 1999 and 2006 according to Google's ngram viewer and has been slowly falling off since. The questionnaire methodology has been challenged in recent research and while more objective measurement may be appropriate, the networked minds questions seem to be able to differentiate real from virtual interactions \cite{Harms2004}.
\section{Other Systems to Support Business}                  
There have been many attempts to support group working and rich data sharing between dispersed groups in a business setting. So called 'smart spaces' allow interaction with different displays for different activities and add in some ability to communicate with remote or even mobile collaborators on shared documents \cite{Bardram2012}, with additional challenges for multi-disciplinary groups who are perhaps less familiar with one or more of the technology barriers involved \cite{Adamczyk2007}.\par
Early systems like clearboard \cite{Ishii1993} demonstrated the potential for smart whiteboards with a webcam component for peer to peer collaborative working. Indeed it is possible to support this modality with Skype and a smartboard system (and up to deployments such as Accessgrid). They remain relatively unpopular however.\par
Displays need not be limited to 2 dimensional screens and can be enhanced in various ways.\par
Stereoscopy allows an illusion of depth to be added to a 2D image by exploiting the stereo depth processing characteristics of the human vision system. This technical approach is not perfect as it does not fully recreate the convergence and focus expected by the eyes and brain.\par 
There are multiple approaches to separating the left and right eye images, these primarily being active (where a signal selectively blanks the input to left then right eyes in synchronicity with the display), passive, where either selective spectrum or selective polarisation of light allow different portions of a display access to different eyes, or physical arrangements which present different displays (or slices of light as in lenticular systems) to different eyes.\par
These barrier stereoscopy / lenticular displays use vertical light barriers built into the display to create multiple discrete channels of display which are accessed by moving horizontally with respect to the display. In this way it is possible to generate either a left/right eye image pair for 'autostereoscopic' viewing, or with the addition of head tracking and small motors. With these techniques multiple viewpoint or an adaptive realtime viewpoint update can be presented without the glasses required for active or passive stereoscopic systems. \par
\subsection{Spatially Faithful Group}
Hauber et al. combined videoconferencing, tabletop, and social presence analysis and tested the addition of 3D. They found a nuanced response when comparing 2D and 3D approaches to spatiality: 3D showed improved presence over 2D (chiefly through gaze support), while 2D demonstrated improved task performance because of task focus \cite{Hauber2006}.\par
I3DVC reconstructs participants from multiple cameras and places them isotropically (spatially faithful) \cite{Kauff2002, Kauff2002a}. The system uses a large projection screen, a custom table, and carefully defined seating positions. They discussed an ``extended perception space" which used identical equipment in the remote spaces in a tightly coupled collaborative `booth'. It employed head tracking and multi camera reconstruction alongside large screens built into the booth. This system exemplified the physical restrictions which are required to limit the problems of looking into another space through the screen. Fuchs et al. demonstrated a similar system over a wide area network but achieved only limited resolution and frame rate with the technology of the day \cite{Fuchs2002}. \par University of Southern California used a technically demanding real-time set-up with 3D face scanning and an autostereoscopic 3D display to generate multiple `face tracked' viewpoints \cite{Jones2009}. This had the disadvantage of displaying a disembodied head.\par                
MAJIC is an early comparable system to support small groups with life size spatially correct video, but without multiple viewpoints onto the remote collaborators it was a one to 'some' system rather than 'some' to one. Additionally users were rooted to defined locations \cite{Ichikawa1995, Okada1994}.\par
There seems to be less interest recently in large display screens for spatially correct viewpoints between groups. The hardware is technically demanding and there may have been sufficient research done to limit investment in research questions. This doesn't mean that there is no future for metaverse applications. Imagine one of the new XR studio walls such as that used to film the Mandalorian. With application of telepresence research it would be possible to bring external metaverse participants into the `backstage' virtual scene. These avatars would be able to explore the scene invisible to the actors, but could be given access to visual feeds from the stage side. This is a hybrid virtual/real metaverse with a well researched and understood boundary interface. It would be possible to give different access privileges to different levels of paying `film studio tourist' or investor, with VIPs perhaps commanding a view onto the live filming. At the nadir of this it may be possible to bring producers and directors directly into the virtual studio as avatars on the screen boundary, with a spatially faithful view onto the set. For the purposes of this book it's also worth noting that NFTs of the experience and corresponding virtual objects from the scene could be monetised and sold within the metaverse.\subsubsection{Multiview}
In order to reconnect directional cues of all kinds it is necessary for each party in the group to have a spatially correct view of the remote user which is particular for them. This requires a multi-view display, which has applications beyond telepresence but are used extensively in research which attempts to address these issues.\par
Nguyen and Canny demonstrated the `Multiview' system \cite{Nguyen2005}. Multiview is a spatially segmented system, that is, it presents different views to people standing in different locations simultaneously. They found similar task performance in trust tasks to face-to-face meetings, while a similar approach without spatial segmentation was seen to negatively impact performance.\par
                    
In addition to spatial segmentation of viewpoints \cite{Gotsch2018} it is possible to isolate viewpoints in the time domain. Different tracked users can be presented with their individual view of a virtual scene for a few milliseconds per eye, before another viewpoint is shown to another user. Up to six such viewpoints are supported in the c1x6 system \cite{Kulik2011}
Similarly MM+Space offered 4 Degree-Of-Freedom Kinetic Display to recreate Multiparty Conversation Spaces \cite{Otsuka2013}
\subsection{Holography and Volumetric}
Blanche et al. have done a great deal of research into holographic and volumetric displays using lasers, rotating surfaces, and light field technology   \cite{Blanche2010,Tay2008}. They are actively seeking to use their technologies for telepresence and their work is very interesting.\par
Similarly Jones et al. ``HeadSPIN" is a one-to-many 3D video teleconferencing system \cite{Jones2009} which uses a rotating display to render the holographic head of a remote party. They achieve transmissible and usable framerate using structured light scanning of a remote collaborator as they view a 2D screen which they say shows a spatially correct view of the onlooking parties.\par
Eldes et al. used a rotating display to present multi-view autostereoscopic projected images to users \cite{Eldes2013}.\par
Seelinder is an interesting system which uses parallax barriers to render a head which an onlooking viewer can walk around. The system uses 360 high resolution still images which means a new spatially segmented view of the head every 1 degreesof arc. They claim the system is capable of playback of video and this head in a jar multi-view system clearly has merit but is comparatively small, and as yet untested for telepresence \cite{Yendo2010}.\par
These systems do not satisfy the requirement to render upper body for the viewers and are not situated (as described soon).\par
There's a future possible where real-time scanned avatar representation in persistent shared metaverse environments will be able to support business, but the camera rigs which currently generate such models are too bulky and involved for a good costs benefit analysis. It is more likely that recent advances in LIDAR phone scanning show the way. The allow realistic avatars to be quickly created for animation within metaverse scenes \cite{authenticVolume2022}.
\subsubsection{Project Skyline}
Project Starline, is a next-generation video conferencing technology that aims to create a sense of presence, making you feel like you're sitting across the table from someone. It uses advanced hardware and software to achieve this. 
\begin{itemize}
\item [Hardware] The newer Starline booth is a refined version of earlier models and looks like a large 65-inch display on a stand. It contains color cameras, depth sensors, microphones, and speakers. Additionally, there are lights on the back of the display that serve as a key light for the person on the call. These lights are mounted around the person and used to create a depth map of the subject and the room they're in.
\item [Display] The display creates an immersive 3D depth effect. It uses a barrier lenticular light field display that shows a different image to your left eye and to your right eye. This effect lets you compute depth on the fly while doing all the head tracking in real time. The display technology in Project Starline is significantly smoother and more realistic than what you would experience with traditional 3D movies.
\item [Compute] The computing side of Project Starline is responsible for rendering the people using the system into realistic 3D models in real-time. It uses AI and depth information gathered by the cameras to map the exact shape, depth, texture, and lighting of the person. The result is an ultra-realistic 3D representation of the person on the other end of the call.
\item [Audio] The system features spatial audio such that the perceived audio changes based on where you are leaning or moving, creating an even more immersive and realistic experience.
\end{itemize}
At this point, Google has been working with several companies who are using these booths for meetings, and it's hoped that as the technology becomes cheaper and more refined, it they assert that it could revolutionize the way we communicate, though the cost of the system and `single user to single user' restriction is likely to be a blocker to crucial business adoption.
\subsection{Simulated Humans}
\subsubsection{Uncanniness}		
When employing simulation representations of humans it may be the case that there is an element of weirdness to some of these systems, especially those that currently represent a head without a body. Mori has demonstrated The Uncanny Valley \cite{Mori1970} effect in which imperfect representations of humans elicit revulsion in certain observers. This provides a toolkit for inspecting potentially `weird' representations, especially if they are `eerie' and is testable through Mori's GODSPEED questionnaire. \par
With an improved analysis of the shape of the likeability curve estimated later showing a more nuanced response from respondents where anthropomorphism of characters demonstrated increased likeability even against a human baseline \cite{Bartneck2007, Bartneck2009}.\par
A mismatch in the human realism of face and voice also produces an Uncanny Valley response \cite{Mitchell2011}.\par
However, there is a possibility that Mori's hypothesis may be too simplistic for practical everyday use in CG and robotics research since anthropomorphism can be ascribed to many and interdependent features such as movement and content of interaction \cite{Bartneck2009}.\par
Bartneck et al. also performed tests which suggest that the original Uncanny Valley assertions may be incorrect, and that it may be inappropriate to map human responses to human simulacrum to such a simplistic scale. They suggest that the measure has been a convenient `escape route' for researchers \cite{Bartneck2009}. Their suggestion that the measure should not hold back the development of more realistic robots holds less bearing for the main thrust of this telepresence research which seeks to capture issues with imperfect video representation rather than test the validity of an approximation.\par
Interestingly Ho et al. performed tests on a variety of facial representations using images.  They found that facial performance is a `double edged sword' with realism being important to robotic representations, but there also being a significant Uncanny Valley effect around `eerie, creepy, and strange' which can be avoided by good design \cite{Ho2008}.\par
More humanlike representations exhibiting higher realism produce more positive social interactions when subjective measures are used \cite{Yee2007} but not when objective measures are used. This suggests that questionnaires may be more important when assessing potential uncanniness.\par
A far more objective method would be to measure user responses to humans, robots, and representations with functional near-infrared spectroscopy and while this has been attempted it is early exploratory research \cite{Strait2014}, an emotional response to `eerie' was discovered.
\subsubsection{Embodiment through robots}
Virtuality human representation extends beyond simple displays into robotic embodiments (which need not be humanoid \cite{Marti2005}), shape mapped projection dubbed ``shader lamps", and hybridisations of the two.\par
Robots which carry a videoconference style screen showing a head can add mobility and this extends the available cues \cite{Adalgeirsson2010, Lee2011, Tsui2011, Paulos1998, Kristoffersson2013}. Interestingly Desai and Uhlik maintain that the overriding modality should be high quality audio \cite{Desai2011}.\par
 Tsui et al. asked 96 participants to rate how personal and interactive they found interfaces to be. Interestingly they rated videoconferencing as both more personal and more interactive than telepresence robots, suggesting that there is a problem with the overall representation or embodiment \cite{Tsui2012}.\par
Kristoffersson et al. applied the Networked Minds questionnaire to judge presence of a telepresence robot for participants with little or no experience of videoconferencing. Their results were encouraging, though they identified that the acuity of the audio channel needing improvement \cite{Kristoffersson2011}.\par
                    There are a very few lifelike robots which can be used for telepresence, and even these are judged to be uncanny \cite{Sakamoto2007}. This is only an issue for a human likeness since anthropomorphic proxies such as robots and toys perform well \cite{Mori1970}.
\subsubsection{Physical \& Hybrid embodiment}
Embodiment through hybridisation of real-time video and physical animatronic mannequins has been investigated as a way to bring the remote person into the space in a more convincing way \cite{Lincoln2009, Lincoln2010, Raskar2001}. \ These include telepresence robots \cite{Lee2011, Sakamoto2007, Tsui2011}, head in a jar implementations such as SphereAvatar \cite{Oyekoya2012, Pan2014, Pan2012} and BiReality \cite{Jouppi2004}, \ UCL's Gaze Preserving Situated Multi-View Telepresence System \cite{Pan2014}, or screen on a stick style representations \cite{Kristoffersson2013}.\par  
Nagendran et al. present a 3D continuum of these systems into which they suggest all such systems can be rated from artificial to real on the three axes, shape, intelligence, and appearance \cite{Nagendran2012}.\par
                    Itoh et al. describe a 'face robot' to convey captured human emotion over a distance. It uses an `average face' and actuators to manipulate feature points \cite{Itoh2005}. It seems that this is an outlier method for communication of facial affect but demonstrates that there are many development paths to a more tangible human display.\par
                    It seems increasingly likely that machine learning models which manipulate images in real time can simulate humans into metaverse applications with very little input data. One such example is Samsung's Megaportraits which can product a realistic human face from a single input stream such as a webcam \cite{Drobyshev22MP}. 
\subsubsection{Shader lamps}
Projection mapping is a computational augmented projection technique where consideration of the relative positions and angles of complex surfaces allows the projection from single or multiple sources to augment the physical shapes onto which they appear. It was first considered by the \href{https://rabcup.com/the-history-of-3d-projection-mapping/}{Disney corporation in 1969} and was given prominence by Raskar and Fuchs with ``office of the future" \cite{Raskar1998} and later by Raskar and other researchers \cite{Raskar2001}. It has since gained considerable commercial popularity in live entertainment.\par
Shader lamps \cite{Raskar2001} is the more formal academic designation for projection mapping. It is possible to use the technique alongside reconstruction to project onto a white facial mannequin. Researchers have attempted to use the technology for remote patient diagnostic, projecting onto styrofoam heads  \cite{Rivera-Gutierrez2012}.\par          
Bandyopadhyay et al. demonstrated \cite{Bandyopadhyay2001} that it is possible to track objects and projection map \cite{Dalsgaard2011} onto them in real time. This is beyond the scope of the proposed projection onto furniture since we wish to keep the system as simple as possible, but could be useful for shared tasks in the future work.\par
Lincoln et al. employed animatronic avatars which they projected with shader lamps. This combination recreated facial expression and head movement though they were limited in speed and range of control of the remote head \cite{Lincoln2010}.\par
While shader lamps are an important and useful technology, there are limitations imposed by its use. In particular if a realtime video feed or reconstruction of a subject is used then that scanned subject must either remain still enough to be correctly mapped onto geometry on the remote side (useful for some virtual patients for instance \cite{Benjamin2012}, or else there must be a computational adjustment made for their changing position to make them appear static, or the projection surface must move to match their movement as in Lincoln et al.

\subsubsection{Metaverse}
In supporting business it's not clear that performance is improved or even maintained by the use of a metaverse. Xi et al. found a significant negative impact to productivity within metaverse applications \cite{xi2022challenges}. It lowers productivity, and may increase anxiety, nausea, VR sickness and even migraines \cite{biener2022quantifying, luong2022demographic}.  It seems at this stage that if we are determined to explore metaverse for business then we \href{http://www.sigtrapgames.com/vrtp/}{should mitigate} the problems as much as possible using the understanding we have gained so far. It might seem that in so doing there is no difference between immersive collaborative mixed reality (described above) and metaverse at all. We feel that the point of metaverse may be in \textit{access to}, if not reliance upon, a mechanism for global truth. What we will go on to describe is likely to look more like traditional telecollaboration for small focussed teams, working on real-world problems, but we will always maintain an access to both the ability to scale, and a global register of value, trust, and truth (digital assets).
\section{Theoretical Framework toward metaverse} 
\subsection{Problem Statement}
It's very likely that the `social first' metaverse attemps such as \href{https://www.theverge.com/2022/10/6/23391895/meta-facebook-horizon-worlds-vr-social-network-too-buggy-leaked-memo}{Meta Horizons}, Sandbox, and Decentraland are \href{https://www.coindesk.com/web3/2022/10/07/its-lonely-in-the-metaverse-decentralands-38-daily-active-users-in-a-13b-ecosystem/}{failing} to capture audiences. They will likely crash back down the hype curve as `Second Life' did before them. Games based worlds such as Roblox are fairing better, but it's unclear if they have any longevity, and they do not fulfil ambitions of an open metaverse.\par 
Worse yet it seems that metaverse is not the most useful way to conduct business. It is evident that there are multiple factors which contribute to successful human-human communication. These factors remain important in telecommunication supported by technology, and are variously supported, unsupported, or modified by particular technologies. Third person large scale metaverse are clearly amongst the worse of the solutions.\par
Of particular importance is interpersonal gaze \cite{Cook1977, Kleinke1986, Fagel2010}. Non-verbal cues are also important across multiple modalities of sight, sound \cite{Otsuka2005}, and position of interlocutors \cite{Kendon1967}, extending to the whole body \cite{Kleinke1986, Nguyen2009}.\par
While formal meeting paradigms are pretty well supported by commercially deployed systems, such ICT can be expensive, may need to be professionally managed, and high end equipment in board rooms are generally booked well in advance. These meetings seem to demand many smaller supporting meetings between parties or groups of parties. The pressure here is clearly toward the now ubiquitous Teams and Zoom style formats, and these offer very poor support for social cues, and incur additional fatigue. These are known and well researched problems, and is is possible that the \href{https://www.oculus.com/blog/future-of-work-meta-connect-2022-quest-pro-microsoft-accenture/?}{strategic pairing of Meta Horizons and Microsoft Teams} will succeed where previous attempted have failed. They seem to finally have the right assets and opportunity. \par
The `problem' is a supporting technology for small less formal groups, or ad-hoc groups meeting to add clarity or context to formal meetings. Metaverse allows this kind of interaction, while not seeming to replace formal meeting utility. Metaverse also may connect home and work spaces without bringing in those backgrouds, creating a level playing field. A more advanced metaverse interface could also allow dynamism and movement, connection of natural non vocal cues, without too much encumbering technology overhead.
\subsection{Core Assumptions}
\begin{figure*}[h]
                \centering
                \includegraphics[width=\textwidth]{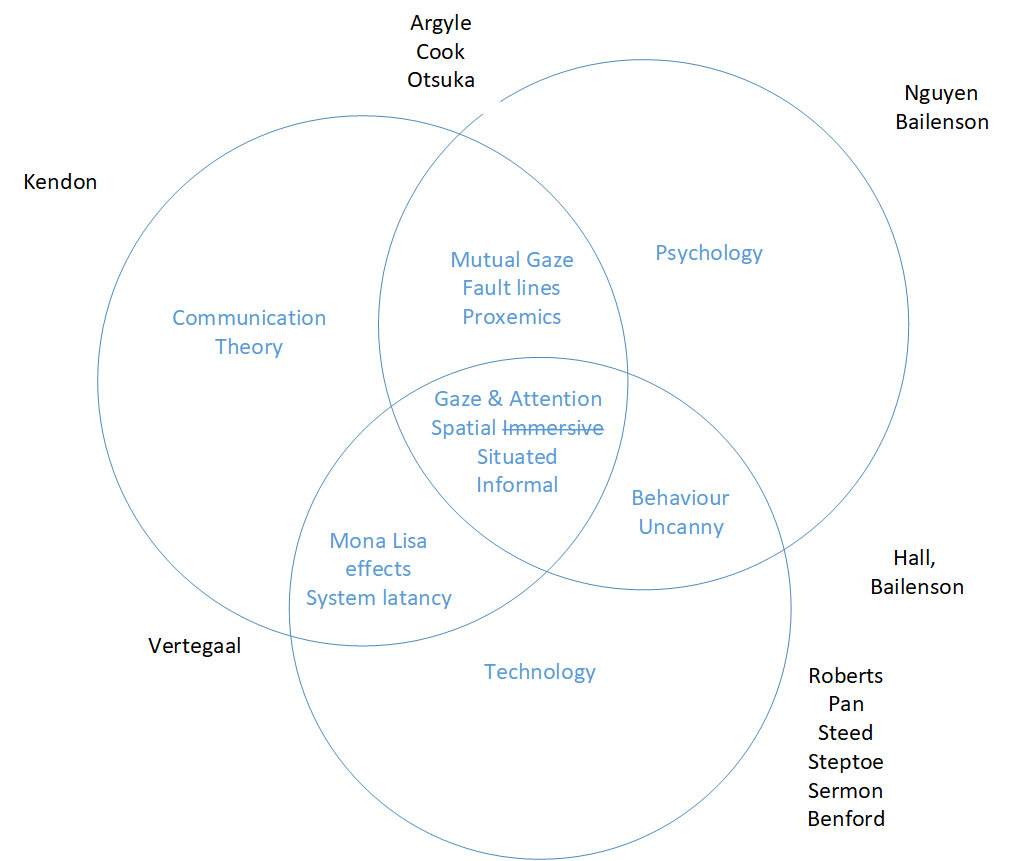}
                \caption{The Venn diagram shows areas of research which have been identified in blue. These interlock and overlap as shown. The most relevant identified researchers from the literature are shown in black close to the fields of study which they represent. This diagram is a view of the core assumptions for the research, with the most important fields at the centre. }
                \label{fig:Framework}
                \end{figure*}

Figure \ref{fig:Framework} shows the interlocking relationships between baseline communication where the participants are present, and technology which attempts to support across distance.
                
Of most interest to this research is the centre of the Venn where meeting styles which are less formal, and perhaps dynamic, may occur. Looking at these items one by one gives us our core assumptions.

                \begin{enumerate}
                \item Gaze\par
                 Gaze is broadly agreed to be highly important for mediating flow. Mutual gaze is a rich emotional channel. The research must consider gaze. All of the researchers listed around the Venn have at some point engaged with this topic.
                \item Attention\par
                The non-verbal communication channel employed in `attention' is assumed based upon the literature to be critical to smoothly leaving and entering a fast flowing conversation where concentration around a defined problem may be high (gesturing to a chair for instance). Again, all of the listed researchers have made reference to attention in their work.
                \item Spatial (immersive)\par
Support for spatiality is important in a group setting so that directional non-verbal cues can find their target. The topic of spatial relationships between interlocutors cuts across all of the researchers, but this is not true of immersion. Immersion in a shared virtuality can certainly support the underlying requirements spatial, but the technical infrastructure required is out of scope (so this is struck through on the diagram). Roberts and Steed are the main expertise referenced even though this element is not expanded in the research.
                \item Situated\par
                Situated displays are those which are appropriate for their surrounding context, in this case the informal meeting. Roberts, Pan, Steed and Steptoe seem the most relevant researchers in these technology spaces.
                \item Informal\par
                   Based on the literature proxemics is believed to be relevant in a meeting where subgroups can be instantiated and destroyed as the meeting evolves, and those where people can be invited in from outside the physical bounds of the meeting (informal spaces). Hall is the best source for this work. If it is assumed that people may come and go, and subgroups may be convened then Sermon and Benford are the best references through their work blending real and virtual spaces. This may be more consistent with less organised meetings such as those convened on demand (ad-hoc).
                \end{enumerate}

\subsection{Peripheral Assumptions}
                Surrounding the centre of the Venn are additional relevant topics from social science branches of theory\par
                \textbf{From verbal communication}\par
                It is assumed that the directionality of sound is important \cite{Aoki2003}, and this will be engineered into the experimental design. It is assumed that movement of the lips is an indicator and this is tied to latency and frame rate in the vision system.\par
                \textbf{From non-verbal communication}\par
                It is assumed that eye gaze is of high importance, and that this information channel is supported by head gaze and body torque to a high degree. It is further assumed that mutual eye gaze is of less relevance in a multi party meeting where there is a common focus for attention but can be significant for turn passing.
                It is assumed that upper body framing and support for transmission of micro and macro gesturing is important for signaling attention in the broader group, and for message passing in subgroups.
            
\textbf{Now that we have an idea what's important for business social communication we can look at the available software to find a best fit.}
    
\section{Post `Meta' metaverse}
The current media around ``metaverse'' has been seeded by Mark Zuckerberg's rebranding of his Facebook company to `Meta', and his planned investment in the technology. Kraus et al suggest that this seems more a marketing and communication drive than a true shift in the company business model \cite{kraus2022facebook}, but despite this Park and Kim identify dozens of recent papers of metaverse research emerging from Meta labs \cite{park2022metaverse}. \par
In Stephenson's `Snow Crash' the Hero Protagonist (drolly called Hiro Protagonist) spends much of the novel in a dystopian virtual environment called the metaverse. It is unclear if Facebook is deliberately embracing the irony of aping such a dystopian image, but certainly their known predisposition for corporate surveillance, alongside their attempt at a global digital money is \href{https://www.politico.com/newsletters/digital-future-daily/2022/04/12/the-facebook-whistleblower-takes-on-the-metaverse-00024762}{ringing alarm bells}, as is their \href{https://www.cnet.com/personal-finance/metas-new-47-5-fee-on-metaverse-items-has-nft-twitter-pissed/}{current plan} for monetisation.\par
The second order hype is likely a \href{https://www.goldmansachs.com/insights/pages/framing-the-future-of-web-3.0-metaverse-edition.html}{speculative play} by major companies on the future of the internet. Grayscale investment \href{https://grayscale.com/wp-content/uploads/2021/11/Grayscale_Metaverse_Report_Nov2021.pdf}{published a report} which views Metaverse as a potential trillion dollar global industry. Such industry reports are given to hyperbole, but it seems the technology is becoming the focus of technology investment narratives. Some notable exerts from a \href{https://www.jpmorgan.com/content/dam/jpm/treasury-services/documents/opportunities-in-the-metaverse.pdf}{2021 report} by American bank JPMorgan show how the legacy financial institutions see this opportunity:\par
\begin{itemize}
\item In the view of the report \textit{``The metaverse is a seamless convergence of our physical and digital lives, creating a unified, virtual community where we can work, play, relax, transact, and socialize.'' - this isn't the worst definition, and very much plays into both the value and mixed reality themes explored in this book.}
\item They agree with the industry that monetisation of assets in metaverse applications is called ``Metanomics''. It's worth seeing this word once, as it's clearly gaining traction, but it won't be used in this book.
\item They make a point which is at the core of this book, that value transaction within metaverses may remove effective border controls for working globally. Be this teleoperation of robots, education, or shop fronts in a completely immersive VR world. They say: \textit{``One of the great possibilities of the metaverse is that it will massively expand access to the marketplace for consumers from emerging and frontier economies. The internet has already unlocked access to goods and services that were previously out of reach. Now, workers in low-income countries, for example, may be able to get jobs in western companies without having to emigrate.''}
\item There is a passage which foreshadows some of the choices made in this book: \textit{``Expanded data analytics and reporting for virtual spaces. These will be specifically designated for commercial and marketing usage and will track business key performance indicators (this
already exists in some worlds, such as Cryptovoxels)''}. More on this later.
\item The report attempts to explore the web3 \& cryptocurrency angles of metaverse. That's also the aim of this book, but they have taken a much more constrained approach, ignoring the possibilities within Bitcoin.
\item They assert that strong regulatory capture, identification, KYC/AML etc should underpin their vision of the metaverse. This is far from the community driven and organically emergent narratives that underpin Web3. This is their corporate viewpoint, something they have to say. On the back of this they pitch their consultancy services in these areas.
\end{itemize}
There has been a reactive pushback against commercialisation and corporateisation by the wider tech community, who are \href{https://www.metaversethics.org/p/mde02-metaverse-data-privacy-1}{concerned about} the aforementioned monetisation of biometrics. \href{https://www.coindesk.com/layer2/2022/01/19/meta-leans-in-to-tracking-your-emotions-in-the-metaverse/}{Observers do not trust} these `web' players with such a potentially powerful social medium. It is very plausible that this is all just a marketing play that goes nowhere and fizzles out. It is by no means clear that people want to spend time socialising globally in virtual and mixed reality. These major companies are  making an asymmetric bet that if there is a move into virtual worlds, then they need to be stakeholders in the gatekeeping capabilities of those worlds.\par
To paraphrase Olson; the salesmen peddling the inevitability of the metaverse are stuck clinging to aesthetic details because, without them, they're just talking about the internet. While virtual reality is enjoying hype right now, and will continue to develop, it faces significant challenges related to the human body's physiological limitations. For instance, the inner ear can become disoriented when a user experiences virtual movement without physically moving. This issue has led to the development of VR applications that require compromises between immersion and physical comfort.
\section{Market analysis}
The market penetration analysis for VR which rings most true for us is provided by Thrive Analytics, and ARtillery Intelligence. Their report is titled ``\href{https://artilleryiq.com/reports/vr-usage-consumer-attitudes-wave-vi/}{VR Usage \& Consumer Attitudes, Wave VI}''. In the USA (which is the cohort they surveyed) they found that adoption of VR headsets is slower than predicted (their work is longitudinal), but steady. Some highlight points are:
\begin{itemize}
\item 23 percent of U.S. adults own or \textit{have used} VR technology. This is around 4\% up from the previous survey in 2020. Frustratingly, and very much in keeping with such industry surveys they conflate `own' with 'have used' making this data pretty meaningless from an adoption point of view.
\item there is a skew toward male users of around 10\%, and a far larger skew toward younger users, and a bias toward richer households. These are indicative of a technology that's still early in it's adoption cycle.
\item Of the owners of the technology (no indication what percentage this is) they found that around a third used the equipment regularly, but that this retention number was gently falling.
\item Standalone headsets (Quest 2 and Pico 4) without a cabled connection to a computer are far more popular, and have better user retention. This makes sense as the alternative demands either space or setup time.
\item Buyers of these more popular headsets are very sensitive to price. Note here that Meta is selling Quest2 at a loss to drive the market. This is unsustainable.
\item Overall this snapshot of adoption feels pretty neutral, and is being driven by losses to Facebook/Meta share price.
\end{itemize}

Deloitte have just \href{https://www2.deloitte.com/uk/en/pages/technology-media-and-telecommunications/articles/digital-consumer-trends-2022-metaverse.html}{conducted a UK survey}. This covers ``metaverse, virtual reality, and web3 (i.e. blockchain-based assets like Bitcoin'', and so is perfect for our needs. They have similar results to the bigger US survey. Their key finding are quoted below verbatim:
\begin{itemize}
\item 63\% of respondents have heard of the term ``metaverse''. However, roughly half of those know nothing about it. 
\item Only 18\% of VR headsets are used daily, from the 8\% of individuals that claim to have access to one.
\item Consumers may be wary of web 3. While most people (93\%) have heard of cryptocurrency, only one in five (19\%) know at least a ``fair amount'' about it. Knowledge of NFTs is rarer still. 
\item 70\% of those who have heard of these assets say they are unlikely to buy them in the next, and cite fraud, scams and a lack of regulation as key concerns. 
\end{itemize}
Deloitte feel that ``content is key'' for virtual reality to be a success, but we would instead argue that applications are key. Nearly half of their respondents were simply ``not interested in VR''. We think this matches our longstanding understanding of the reality of the market. A few vocal proponents of the technology does not necessarily lead to a developed and mature mass appeal. Again, we feel that real world use cases will drive adoption over a longer time frame. Virtual meetings do not feel like that application to us.\par
They feel that `one metaverse' would require blockchain/web3 tooling for a common consensus frame, and we agree with this. It seems like a very long way to that point, and perhaps not worth the effort. They, like us, see compatible silos as being the interim step.\par 
They (unusually) have a legal opinion in the text, and this is valuable enough to quote verbatim once again.
\textit{``The metaverse amplifies existing legal issues and raises new ones. Centralised metaverses, such as those focused on games, tend to engage consumers in a controlled space and operate within familiar legal frameworks. For example, users purchasing a virtual accessory are likely to understand its use will be within tightly prescribed
parameters. Decentralised metaverses, which incorporate web3 (such as NFTs) are more challenging, as users may expect virtual assets to be portable. However, those assets are governed by inconsistent and often unclear terms, and the lack of technical standards can result in limited interoperability between metaverses. For the user, social interactions in virtual worlds can feel realistic, inviting scrutiny from policymakers and regulators focused on online
safety. An increased legislative focus on children online will also require platforms to assess or verify the age of users. And collection of personal data – such as eye movement within a VR headset – will require informed consent under data protection laws, and a clear understanding of who is controlling that data at any given time.
Finally, as content is key, clear contractual parameters are required to frame how intellectual property is used, whether user-generated content is permitted, and how illegal/harmful content is managed.
Amid all of this, metaverse builders, content owners and brands must ensure they have a risk assessment and risk management framework
in place to avoid costly mistakes, both reputational and financial, in an increasingly regulated space.''}\par

\href{https://www.thedrum.com/about-us}{The Drum} is a market awareness website and \href{https://www.thedrum.com/news/2022/08/01/web3-the-numbers-key-metaverse-crypto-and-nft-stats-every-marketer-should-know}{compiled} the following statistics, which have been linked back to their source and annotated for our needs. 
\textit{\begin{itemize}
\item 89.4 million Americans are expected to use virtual reality (VR) in 2022, \href{https://www.insiderintelligence.com/content/us-augmented-virtual-reality-users-forecast-2022}{according to insiderintelligence}. That number, according to the same source, is expected to climb to 110.3 million in 2025. As a counter to this only around 16M VR headsets were sold in 2022
\item 51\% of gen Z and 48\% of millennials envision doing some of their work in the metaverse in the next two years, according to Microsoft’s Work Trend Index 2022.
\item 38\% of respondents said they would “try extreme sports like skydiving, bungee jumping, or paragliding” in the metaverse according to a recent Statista survey called ‘What things would you do in the metaverse but never in real life?’ Unsettlingly, 18\% of respondents said they would “conduct unethical experiments on virtual humans”
\item 87\% of Americans between the ages of 13-56 would be interested in engaging with a virtual experience in the metaverse “that is built around a celebrity they love,” according to new research from UTA and Vox Media
\item \$678bn is forecasted to be the total market valuation of the metaverse by 2030, per Grand View Research. According to the report, that market value was just shy of \item \$39bn in 2021, giving it a predicted compounded annual growth rate over a 10-year period of around 39%
\item 46\% of all people across age groups say that the ability to visualize a virtual product in an IRL context – “such as seeing a digital painting in their home using augmented reality (AR) glasses” – is the primary factor that would motivate them to make a purchase in the metaverse, per a Productsup survey
\item 24\% of US adult internet users say “that lower-priced VR headsets were a very important factor when deciding whether to try using the metaverse,” per a recent Statista survey. On the other hand, 54\% say that their workplace using the metaverse would “not [be] important at all” in their decision to give the metaverse a try
\item 15\% of gen Zs’ “fun budget” is spent in the metaverse, per a report from Razorfish and Vice Media Group. In five years that number is projected to climb to 20\%
\item Nearly 77\% believe that the metaverse “can cause serious harm to modern society,” per a recent survey from customer service platform Tidio. The survey, which received feedback from 1,000 participants, identified three major causes of anxiety related to the metaverse and its potentially negative social impacts: “addiction to a simulated reality” was the number one concern, followed by “privacy issues” and “mental health issues,” which were tied for second
\item By 2026, about 2 billion people worldwide “will spend at least one hour a day in the metaverse to work, shop, attend school, socialize or consume entertainment,” per McCann Worldgroup. By that same year, the total value of the virtual goods market in the metaverse could be as high as \$200bn
\item NFTs
Over \$37bn has been spent in NFT marketplaces as of May 2022, per data from Chainalysis. At their current rate, this year’s NFT sales could potentially surpass last year’s, which had a total valuation of around \$40bn, according to the data
\item \$91.8m was the sale price of ‘The Merge,’ the most valuable NFT to date. Created by the artist Pak, it sold for its record-breaking value in December 2021
\item 64\% of sports fans are open to the idea of learning more about NFTs and would consider purchasing one in the future, according to the National Research Group. The report also found that 46\% of sports fans “would be more likely to attend live sporting events if they were rewarded with a commemorative NFT – for example, if their ticket turned into a digital collectible after the game”
\item Only 9\% of people aged 16-44 own a NFT, and less than half (44\%) have purchased or invested in crypto, per a new survey from agency SCS. On the other hand, among the survey’s 600 respondents, 64\% were “aware” of the metaverse, and 65\% of that subgroup say they are “interested in exploring it further for everything from traveling to new places and playing games to making money and shopping”
\end{itemize}}
Polling company IPSOS \href{https://www.ipsos.com/en/global-advisor-metaverse-extended-reality-may-2022}{have conducted} a global survey for the World Economic Forum. Some highlights are:
\begin{itemize}
\item ``Excitement about extended reality is significantly higher in emerging countries than it is in most high-income countries. In China, India, Peru, Saudi Arabia, and Colombia, more than two-thirds say they have positive feelings about the possibility of engaging with it.''
\item ``Familiarity and favorability toward the new technologies are also significantly higher among younger adults, those with a higher level of education, and men than they are among older adults, those without a college-level education, and women.''
\end{itemize}

\begin{figure*}[ht]\centering 
	\includegraphics[width=\linewidth]{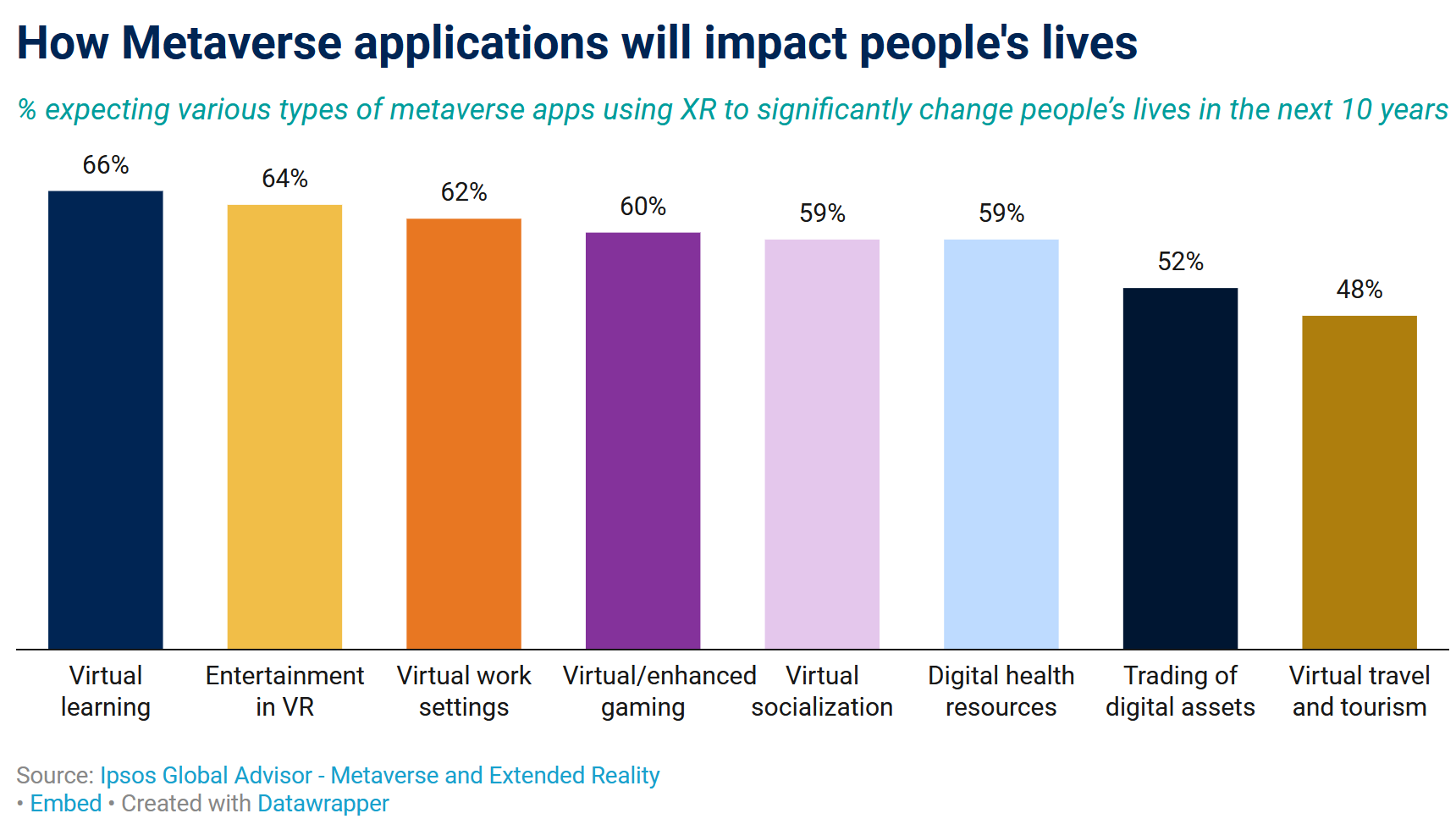}
	\caption{\href{https://www.ipsos.com/en/global-advisor-metaverse-extended-reality-may-2022}{IPSOS poll predicted applications}}
	\label{fig:applications}
\end{figure*}

Excitingly for our exploration of the topic it can be seen in Figure \ref{fig:applications} that education within metaverse spaces is the most anticipated application, and we have seen that the emerging globals markets are the most optimistic about the technology overall. This is highly suggestive of an opportunity.

\section{NFT and crypto as metaverse}
Within the NFT, Web3 and crypto community it is normalised to refer to ownership of digital tokens as participation in a metaverse. This is reflected in the market analysis above. This fusing of narratives is reviewed in detail by Gadekallu et al in their excellent recent paper on Metaverse and Blockchain \cite{gadekallu2022blockchain}. They conclude that much remains to be done here. This CNBC article highlights the confusion, as this major news outlet refers to \href{https://www.cnbc.com/2022/01/16/walmart-is-quietly-preparing-to-enter-the-metaverse.html}{Walmart prepares to offer NFTs}'' as an entry ``into the metaverse''.
\section{Lessons from MMORGS}
The concept of `instrumental play' was introduced by literary theorist Wolfgang Iser in his 1993 essay ``The Fictive and the Imaginary'' \cite{iser1993fictive}. Iser divided play into two categories, free play and instrumental play, based on their relationship to goals. In his view, play becomes instrumental the moment it has a goal or a set of rules. The application of this concept to massively multiplayer online games was later explored by sociologist T.L Taylor in her 2006 book `Play Between Worlds' \cite{taylor2009play}. According to Taylor, instrumental play is a goal-oriented approach that values efficiency, expertise, and strategy optimization. The point of playing is not to reach the end but to find the best way to get there.\par
The distinction between instrumental play and fun is often seen as a false dichotomy. The two are not mutually exclusive but exist in tension. Optimization can result in player behaviours that are simply no fun, but achieving goals or improving skills can also bring enjoyment. René Glas in his book `Battlefields of Negotiation' \cite{glas2013battlefields} describes the movement between instrumental and free play in World of Warcraft, which has the distinction of evolving across entirely different iterations of the Internet.\par
These virtual worlds of massively multiplayer online games are "interactively stabilized" systems, the result of the interaction between game designers and players. The social codes of practice established by players can shape what is considered legitimate play. Success in these games is dynamically defined by consensus, as seen in Mark Chen's study of World of Warcraft `Leet Noobs' \cite{chen2011leet}.\par
Tom Boellstorff conducted a study of user experiences in Second Life \cite{serapis2008coming}, which was criticized for not involving real life or other websites or software in the analysis. The virtual worlds of massively multiplayer online games are not enclosed and players can engage with these games through various platforms, such as Discord, Twitch, Twitter, and Google Docs, without physically inhabiting the virtual world. This concept of "paratext" was first introduced by French literary theorist Gerard Genette. He saw a book as containing the text of the book and additional components, such as the cover, title, foreword, etc., that are necessary to complete the book but not part of the primary text. These additional texts influence the meaning of the primary text. The definition was later expanded by Mia Consalvo, who defined paratext as any text that ``may alter the meanings of a text, further enhance meanings, or provide challenges to sedimented meanings.'' Examples of paratext include reviews, pre-release trailers, etc. Kristine Ask observed the impact of paratext on theorycrafting expertise in World of Warcraft, which was later confirmed by the rise of twitch streams. Mark Chen's dissertation Leet Noobs focuses on how AddOns in World of Warcraft can become essential agents in raid groups by assuming cognitive load. The concept is based on the idea of object-oriented ontology and actor-network theory \cite{cole2013call}. These theories are complex and contested, but the boundaries between real people and virtual AI actors in virtual social spaces are certainly blurred. \par
Virtual spaces are not separate from the real world, but are instead an extension of it. The key factor in making a virtual world compelling is not its realism, but the fact that people give meaning to their lives by entangling themselves in projects with others, even when those others are not other people. Worlds become real when people care about them, not when they look like the real world.
\section{Immersive and third person XR}
In considering the needs of business to business and business to client social VR is it useful to compare software platforms. We have seen that a global connected multiverse is a marketing proposition only, and may be a decade or more away. Contenders currently look more like one of three catagories; games, limited massively multiplayer worlds, or meeting support software. These will converge.
\subsection{More like a digital twin}
One of the most intuitive ways to view a metaverse is as a virtual landscape. This is how metaverse was portrayed in the original Neal Stephenson use of the word. `Digital twin' is another much abused industry term which trends toward a 3D representation of real world spaces and objects. Sometimes these virtual objects are connected to the real by telemetry, allowing industrial monitoring applications. Much is made of such systems in simulation brochures, and on the web, but it's surprisingly hard to find real world applications of the idea outside of complex large scale systems engineering (aerospace). The costs of maintenance are simply too high. The US army owns the digital twin which could be called \href{https://www.army.mil/standto/archive/2018/03/26/}{closest to ``The Metaverse''} (note the intentional capitalisation). Their global simulation environment mirrors real world locations for their training needs. The European space agency is building an \href{}{Earth digital twin} for climate research, as \href{https://www.nvidia.com/en-us/on-demand/session/gtcfall22-a41326/?playlistId=playList-9bb5405e-3e40-4ff3-88db-61cd3a4507e5#:~:text=Earth%2D2%20aims%20to%20improve,learning%20methods%20at%20unprecedented%20scale.}{is Nvidia}, but again it's unclear what this offers over and above access to direct data feeds, and of course such an ambitious project likely has an ecological cost!\par
Within industry digital twins are seen as the primary use case for metaverse, with even the world economic forum \href{https://www.weforum.org/agenda/2023/01/metaverse-biggest-impact-industry-davos2023/}{subscribing to the hype}. To be clear, there is enormous effort, investment, and potential here, but it feels outside of the scope of this product at this time.
\subsubsection{Omniverse}
Key new capabilities announced:
Integration of generative AI like Adobe Firefly to enhance creation workflows (wow!)
Expanded ecosystem connections through OpenUSD (Adobe, Wonder Dynamics, Luma AI, etc)
New developer tools and templates for building apps and experiences
Semantic search capability with DeepSearch to find 3D assets easily 
Optimizations for photorealistic real-time rendering and path tracing with AI-accelerated denoising powered by new RTX GPUs
XR capabilities native to the platform (so you can deploy on AR/VR headsets)
Upgrades to core apps like Omniverse Audio2Face and USD Composer
Graphics Delivery Network (GDN) to performantly serve your 3D experience around the world 
Support for new workflows across industrial use cases like digital twins
\subsubsection{Geolocated AR}
Overlaying geospecific data into augmented reality (think Pokemon Go) is probably the ultimate utility of digital twin datasets. It's such a compelling application space that we will have more on this later.
\subsection{More like a metaverse}
\subsubsection{Second Life}
Notable because it's the original and has a decently mature marketplace. Some \$80M was \href{https://www.zdnet.com/article/high-fidelity-invests-in-second-life-to-expand-virtual-world/}{paid to creators} in Second Life in 2021 in a wider economic ecosystem of around \$650M. It's possible to write a whole book on Second life, and indeed many have. It's longevity means that there's more study of business uses of such systems than in any other platform. 
\subsubsection{Mozilla Hubs}
Hubs is a great option for this proposal, and might be worth integrating later. It runs well in a browser and on VR hardware.
\begin{itemize}
\item Open source, bigger scale, more complex
\item Choose avatars, or import your own
\item Environments are provided, or can be designed
\item Useful for larger conferences with hundreds or thousands of members but is commensurately more complex
\item Quest and PC
\item Larger scenes within scenes
\end{itemize}
\subsubsection{Counter social realms}
A relatively new platform linked to a new model of social media which excludes countries which habitually spam. It uses Mozilla Hubs for it's engine.
\subsubsection{Roblox}
If anything can currently claim to be the metaverse it's probably Roblox. Around 60 billion messages are \href{https://podcasts.apple.com/us/podcast/developments-investments-experiences-in-the-metaverse/id1593908027?i=1000540906629}{sent daily} in Roblox. Investment in the metaverse `angle' of the platform is stepping up with recent announcements such as \href{https://techcrunch.com/2022/05/03/spotify-becomes-first-music-streamer-to-launch-on-roblox/?}{``Spotify Island''}. The company has announced text based generative art creation of scenes, and is integrating the playstation headset in their PS4 release. It's very notable that it still \href{https://fortune.com/2022/06/03/roblox-gaming-ecosystem-metaverse-stocks-profit/}{hasn't become a profitable business}. It is important to note that Roblox has banned NFTs. Nike have \href{https://www.thedrum.com/news/2022/09/22/21m-people-have-now-visited-nike-s-roblox-store-here-s-how-do-metaverse-commerce}{garnered significant attention} for their metaverse store, front with their Roblox based metaverse. As \href{https://medium.com/@theo/why-nikeland-is-not-the-metaverse-success-story-you-think-it-is-46742dc2f231}{Theo Priestley} points out this is likely just another expensive experiment, with a finite lifespan.\\

\href{https://twitter.com/bilawalsidhu/status/1644817961952374784}{expanding into generative AI}

\subsubsection{Minecraft}
Minecraft has also \href{https://www.minecraft.net/en-us/article/minecraft-and-nfts}{banned NFTs}
\subsubsection{Surreal}
\subsubsection{Sansar}
\subsubsection{Cornerstone}
\subsubsection{AltSpace}
\begin{itemize}
\item Microsoft social meeting platform
\item Very good custom avatar design
\item Great world building editor in the engine
\item Doesn't really support business integration so it's a bit out of scope
\item Huge numbers (many thousands) possible so it's great for global events
\item Mac support
\end{itemize}
\subsubsection{VRChat}
This text is from wikipedia and will be updated when we have a chance to try VRChat properly. It's much loved already by the Bitcoin community.\par
``VRChat's gameplay is similar to that of games such as Second Life and Habbo Hotel. Players can create their own instanced worlds in which they can interact with each other through virtual avatars. A software development kit for Unity released alongside the game gives players the ability to create or import character models to be used in the platform, as well as build their own worlds.\par
Player models are capable of supporting "audio lip sync, eye tracking and blinking, and complete range of motion.\par
VRChat is also capable of running in "desktop mode" without a VR headset, which is controlled using either a mouse and keyboard, or a gamepad. Some content has limitations in desktop mode, such as the inability to freely move an avatar's limbs, or perform interactions that require more than one hand.\par
In 2020, a new visual programming language was introduced known as "Udon", which uses a node graph system. While still considered alpha software, it became usable on publicly-accessible worlds beginning in April 2020. A third-party compiler known as "UdonSharp" was developed to allow world scripts to be written in C sharp.'' 
\subsubsection{Meta Horizon Worlds \& Workrooms}
Horizon Worlds is the Meta (Facebook) meteverse, and Workrooms it's business offering and a subset of the ``Worlds'' global system. It is currently a walled garden without connection to the outside digital world, and arguably not therefore a metaverse.\par
The Financial Times \href{}{took a look} at their patent applications and noted that the travel is toward increased user behaviour tracking, and targeted advertising.\par
Facebook actually have a poor history on innovation and diversification of their business model. This model has previously been tracking users to target ads on their platform, while increasing and maintaining attention using machine learning algorithms. \par
It makes complete sense then to analyse the move by Meta into 3D social spaces as an attempt to front run the technology using their huge investment capacity. Facebook have recently taken a huge hit to their share price. Nothing seems to have changed in the underling business except Zuckerberg's well publicised shift to supporting a money losing gamble on the Metaverse. It is by no means clear that users want this, that Meta will be able to better target ads on this new platform, or that the markets are willing to trust Zuckerburg on this proactive move. \par
With all this said the investment and management capacity and capability at Meta cannot be dismissed. It is very likely that Meta will be able to rapidly deploy a 3D social space, and that it's development will continue to be strong for years. The main interface for Horizon Worlds is through the Meta owned and developer Oculus headset, which is excellent and reasonably affordable. It has been quite poorly received \href{https://kotaku.com/facebook-metaverse-horizon-worlds-vr-oculus-quest-2-cha-1848436740}{by reviewers} but will likely improve, especially if users are encouraged to innovate.
\subsubsection{Webaverse}
\href{https://webaverse.com/}{Webaverse} are an open collective using open source tools to create interoperable metaverses.
\subsubsection{Vircadia}
The applications and platforms detailed above have their benefits, but for the application stack in the next section of the book Vircadia has been chosen. The following text is from their website, and is a placeholder which gives some idea. This section will be written out completely to reflect our use of the product.\par
Vircadia is open-source software which enables you to create and share virtual worlds as virtual reality (VR) and desktop experiences. You can create and host your own virtual world, explore other worlds, meet and connect with other users, attend or host live VR events, and much more.\par
The Vircadia metaverse provides built-in social features, including avatar interactions, spatialized audio, and interactive physics. Additionally, you have the ability to import any 3D object into your virtual environment. No matter where you go in Vircadia, you will always be able to interact with your environment, engage with your friends, and listen to conversations just like you would in real life.\par
What can I do? You have the power to shape your VR experience in Vircadia.
\begin{itemize}
\item EXPLORE by hopping between domains in the metaverse, attend events, and check out what others are up to!
\item CREATE personal experiences by building avatars, domains, tablet apps, and more for you and others to enjoy.
\item SCRIPT and express your creativity by applying advanced scripting concepts to entities and avatars in the metaverse.
\item HOST and make immersive experiences to educate, entertain, and connect with your audience.
\item CONTRIBUTE to the project's endeavor.
\item DEVELOP the project and tailor it to your needs, or just to help out.
\item SECURITY information about the project and its components.
\end{itemize}

\subsection{More like crypto NFT virtual land}
This next three are a placeholder taking text from the \href{https://www.analyticsinsight.net/top-10-metaverse-platforms-that-will-replace-social-media-in-future/}{linked site} and will be swapped out:
The digital land \href{https://www.coindesk.com/markets/2022/04/06/metaverse-majors-struggle-as-user-base-falls-short-of-market-expectations/?}{narrative is fading}.
\subsubsection{Decentraland}
Decentraland is a large 3D (but not VR) space developed by Argentine developers Esteban Ordano and Ari Meilich. It is a decentralized metaverse purporting to be owned by its users, but actually owned completely by a foundation \href{https://www.crunchbase.com/organization/decentraland/people}{based in Panama}. The users can shop, buy things, invest, and purchase goods in a virtual space. The project is built on Ethereum and has a (speculative) valuation in the billions of dollars.\par Decentraland was launched in February 2020, and its history includes an initial coin offering in August 2017, where their MANA token sale raised approximately \$24 million dollars in crypto coins. This was followed by a ``terraforming event'' where parcels of land, denominated in LAND tokens, were auctioned off for an additional \$28 million in crypto. The initial pitch for Decentraland emphasized the opportunity to own the virtual world, create, develop, and trade without limits, make genuine connections, and earn real money. However, the actual experience in Decentraland has faced criticisms such as poor graphics, performance issues, and limited content. They have recently dropped their pretence of ever supporting VR.\par 
One example of these limitations is the now-defunct pizza kiosk that aimed to facilitate ordering Domino's pizza via the metaverse using cryptocurrency. This concept, though intriguing, was hindered by a lack of official support from Domino's and the inherent inefficiencies of using a virtual world as an intermediary for purchasing goods and services.\par
Similarly, attempts to create virtual amusement park rides and attractions within Decentraland have suffered from poor performance and a lack of interactivity. These issues stem from the limitations of the tools and resources available for building experiences within the platform, as well as the inherent difficulties in creating engaging experiences in a `world' that is supposed to perform too many functions at once.\par 
In addition to the technical challenges, Decentraland (and all these crypto metaverse projects) have clearly promoting unrealistic expectations to foster speculative investments. The notion that businesses and individuals will eventually ``live inside'' the metaverse is not only a poetic interpretation but also an unrealistic expectation given the current state of VR technology.\par 
As it stands, Decentraland is unlikely realize its supposed potential as an invisible, seamless infrastructure for a wide range of digital experiences. Until the platform can address its core issues, it is likely that projects like the `Decentraland Report' (it's user delivered news platform), and others will continue to fail to deliver on their promises. To quote \href{https://www.youtube.com/watch?v=EiZhdpLXZ8Q}{Olson's highly critical} (and correct) presentation on Decentraland: \\textit{``..it can’t even handle properly emulating Breakout, a game from 1976 that you can play on goddamn Google images! Steve Wozniak built Breakout fifty years ago to run on 44 TTL chips and a ham sandwich and that’s still somehow too demanding a gaming experience ...''}\par
Like all of these attempts the actual information content of within Decentraland boils down to text on billboards, and links to the outside Web. It's a terrible product, and really just another example of a crypto scam which never really intended to be developed for the long haul. 
\subsubsection{Sandbox}
The Sandbox, a decentralized gaming platform built on the Ethereum blockchain, has garnered attention for its promise of a vibrant ecosystem filled with user-generated content. However, despite its ambitious vision, the project has faced various challenges and criticisms similar to Decentraland. Limited use cases and adoption remain a significant challenge for The Sandbox. While the platform aims to create a vast and engaging gaming ecosystem, it has yet to gain widespread adoption, leading to a limited number of users and developers. This lack of user engagement raises questions about the long-term viability of the project, as the value of virtual land, assets, and in-game experiences may remain limited without a thriving community. Like Decentraland it is a manipulated hype bubble, attracting glowing paid press reports in some media, and `interest' from national and regional `branches' of global brands  which are then spun to create artificial hype in main stream media. The tradable NFTs within these early platforms are obviously subject to insider trading, price volatility, wash trading, and other harmful activities. \par
The Sandbox places too much emphasis on the speculative aspect of virtual land and asset trading, rather than focusing on creating a genuinely engaging gaming ecosystem. This focus on speculation could lead to an unsustainable bubble with inflated asset prices, and it seems likely we have already seen most of the collapse of this ecosystem.\par
The actual experience of interacting with The Sandbox's gaming products leaves much to be desired. For instance, the platform's games may suffer from lag and poor performance due to the technical limitations of blockchain technology. Additionally, the quality of user-generated content can be highly variable, as not all creators possess the skills and resources to develop engaging gaming experiences. As a result, users might find themselves sifting through a plethora of low-quality games, which can be frustrating and time-consuming.\par
Concerns about centralization persist, as some critics argue that the project is not entirely decentralized. The team behind The Sandbox still holds a significant amount of control over the platform's development and governance, potentially undermining the project's core vision of a decentralized gaming ecosystem.
\subsubsection{Space Somnium}  
Somnium Space is just another one of these, but with more VR. It allows users to join in either through a downloadable VR client or a browser-based version to function like any other web app. It suffered the same problems at Decentraland and Sandbox. They are terrible products, with hype, manufactured by money, extracted from users, often convinced by paid celebrity endorsements. It's the NFT space, but sadder, and technically worse, and likely not for very much longer.
\subsection{More like industrial application}
As the word metaverse has gained in use, so have some traditional users and researchers in mixed reality switched to use of the term.
Siyaev and Jo describe an aircraft training metaverse which incorporates ML based speech recognition \cite{siyaev2021towards}. This class of mixed reality trainer traditionally finds positive results, but is highly task specific.   
 \subsubsection{Global enterprise perspective}
Microsoft have just bought Activision / Blizzard for around seventy billion dollars. This has been communicated by Microsoft executives as a ``Metaverse play'', leveraging their internal game item markets, and their massive multiplayer game worlds to build toward a closed metaverse experience like the one Meta is planning.
This builds on the success of early experiments like the Fornite based music concerts, which attracted millions of concurrent users to live events.

There are three emerging focuses, the social metaverses for pleasure, and business metaverses for larger group meetings and training \cite{heiphetz2010training, aldrich2005learning}, and a Nvidia's evolving \href{https://blogs.nvidia.com/blog/2022/08/09/omniverse-siggraph/}{collaborative creation metaverse} for digital engineers and creatives. They're all pretty different `classes' of problem. The social metaverse angle where Facebook is concentrating most effort is of less interest to us here, though obviously markets will exist in such systems for business to customer. The next section will explore some of the software tools available to connect people. Everything looks pretty basic right now in all the available systems, but that will likely \href{https://www.youtube.com/watch?v=cRLnR4Kot2M}{change over the next couple of years}.
\subsection{More like meeting support}
\subsubsection{Spatial}
Spatial is worth a quick look because it's a business first meeting tool, and comparatively well received by industry for that purpose.
\begin{itemize}
\item Very compelling. Wins at wow.
\item Great avatars, user generated
\item AR first design
\item Limited scenes
\item Smaller groups (12?)
\item Limited headset support
\item Intuitive meeting support tools
\item No back end integration
\end{itemize}
\subsubsection{MeetinVR}
\begin{itemize}
\item Good enough graphics, pretty mature system
\item OK indicative avatars, user selected
\item VR first design
\item Limited scenes
\item Smaller groups (12?)
\item Quest and PC
\item Writing and gestures supported
\item Some basic enterprise tools integration
\item Bring in 3D objects
\item Need to apply for a license?
\end{itemize}
\subsubsection{Glue}
\begin{itemize}
\item Better enterprise security integration
\item Larger environments, potential for breakouts in the same space. Workshop capable
\item 3D object support, screen sharing, some collaborative tools
\item Apply for a license
\item Fairly basic graphics
\item Basic avatars
\item Quest and PC
\item Writing and gestures supported
\item Mac support
\end{itemize}
\subsubsection{FramesVR}
\begin{itemize}
\item Really simple to join
\item Basic avatars
\item Bit buggy
\item 3D object support, screen sharing, some collaborative tools
\item Quest and PC
\item Larger scenes within scenes
\item Runs in the browser
\end{itemize}
\subsubsection{Engage}
\begin{itemize}
\item Great polished graphics
\item Fully customisable avatars
\item Limited scenes
\item Presentation to groups for education and learning
\item PC first, quest is side loadable but that's a technical issue
\item BigScreen VR
\item Seated in observation points in a defined shared theatre
\item Screen sharing virtual communal screen watching, aimed at gamers, film watching
\item up to 12 user
\end{itemize}
\subsubsection{Gather}
Gather is an oddball meeting space based around fully customisable 2D rooms with a game feel. It's really a spatialised twist on video conferencing but interesting.  
\subsubsection{NEOSVR}
\href{https://neos.com/}{Notable because} it's trying to integrate crypto marketplaces, but we haven't tried it yet.
\section{Displays \& Headset Hardware}
Awaiting a bit more market stability for this section. Of note is that Microsoft seems to be \href{https://www.windowscentral.com/microsoft/microsoft-has-laid-off-entire-teams-behind-virtual-mixed-reality-and-hololens}{abandoning Hololens}, and Apple seem to have postponed their commodity AR headset. \par
Microsoft think that creating the Perfect Illusion, that of a life-likeness in VR will require a field of view of 210 horizontal and  135 vertical, 60 pixels per degree subtended, and a refresh rate of 1800 Mhz according to Microsoft. They expect this by as soon as 2028 \cite{cuervo2018creating}.\par 
With the advent of \href{https://developer.chrome.com/docs/web-platform/webgpu/}{WebGPU} alongside WebGL everything is likely to converge on the browser experience.

\subsection{The Apple in the Room}
Following the announcement of The Apple Vision Pro we start to see the convergence of spatial computing, mixed reality, locally applied transformer based AI, and business.  They have perhaps removed ``gorilla arm syndrome'' \cite{boring2009scroll} where hands in the sky interfaces are potentially uncomfortable over long periods \cite{hansberger2017dispelling}. Nathan Gitter and Amy DeDonato from the Apple Design team \href{https://developer.apple.com/videos/play/wwdc2023/10072/}{introduce spatial design for the device}.
\subsubsection{Spatial operating systems}
\begin{itemize}
\item Enabling users to design experiences not previously possible.
\item The presentation outlines how to keep apps familiar, be human-centered, take advantage of space, enhance immersion, and make apps authentic to the platform.
\item The world serves as an infinite canvas for new apps and games.
\item Existing app elements should be kept familiar with common elements like sidebars, tabs, and search fields.
\item In a spatial platform, interfaces are placed within windows to make them easily accessible and part of the user's surroundings.
\end{itemize}
\subsubsection{Windows in Spatial Design}
\begin{itemize}
\item Windows are designed with a new visual language, made of a glass material that provides contrast with the world, awareness of surroundings, and adapts to different lighting conditions.
\item Windows can be moved, closed, and resized by users, with windows facing the user during movement.
\item Windows are flexible and can be resized to fit comfortably within the user's view.
\item Choosing Window Size and Layout
Windows are designed to be flexible, adapting to content, and the window size should be chosen based on this.
Windows can change size dynamically based on context.
\item Apps can use multiple windows to display content side by side or show distinct actions, but should ideally stick to a single window to avoid user overwhelm.
\end{itemize}
\subsubsection{Designing with Points}
\begin{itemize}
\item Interfaces are designed with points to ensure they scale well and remain legible at different distances.
\item Points allow designers to set the size of interface elements with familiar units.
Human-Centered Design
\item Good spatial design places the user at the center, accounting for their field of view and movement.
\item The most important content should be placed in the center of the field of view and use landscape layouts.
\item Ergonomics should also be considered, placing content along a natural line of sight for comfort.
\item Designers should avoid placing content behind users or anchoring content to their view as it can be disorienting.
\item Spatial design should aim to create stationary experiences that require minimal movement from users.
\end{itemize}
\subsubsection{User Mobility} 
The presentation emphasizes the importance of designing applications that require minimal movement from users. It recommends using system-level recentering methods to adjust the app's view when a user moves.
\subsubsection{Space Utilization} 
The importance of optimizing an app's usage of space is discussed, as the available physical space for users can vary. It advises against constraining your app based on the physical space available and instead creating an app that can function in any amount of space.
\subsubsection{Dimensionality} 
The use of depth and scale in designing the user experience is emphasized. Depth can help with hierarchy and focus, and scale can be used to emphasize content. The text warns against overusing depth, especially with text, and encourages developers to experiment with scale to achieve the desired user experience.
\subsubsection{Immersiveness}
The passage introduces the concept of an immersion spectrum, where an app can transition between various states of immersion based on the user's experience. The importance of smooth transitions, designing with consideration to user focus, thoughtful blending with reality, and keeping the user comfortable are emphasized.
\subsubsection{Sound Design}
It also highlights the importance of using spatial audio to enhance the immersive experience of an app, which includes attaching sound to objects and creating soundscapes.
\subsubsection{User Comfort}
Recommendations for moving an immersive app, focusing on avoiding disorienting fast movements and instead recommending fade out and fade in techniques to keep the user comfortable during motion.
\subsubsection{Transitions}
It highlights the importance of clear, intuitive methods for entering and exiting immersive experiences. It suggests using easily recognizable symbols, such as arrows for expanding or collapsing views.
\subsubsection{Authenticity}
Emphasizes creating an authentic experience that takes full advantage of the platform's capabilities. An example given is Freeform, which uses a large creative space allowing users to view all their content at once.
\subsubsection{Key Moments}
Focusing on a ``key moment'' that provides a unique spatial or immersive experience is recommended. This could involve enhancing a moment with depth and scale or transforming the user's space to create a unique and memorable experience.
\section{Unreal \& Virtual Production}
\begin{figure}
  \centering
    \includegraphics[width=0.5\linewidth]{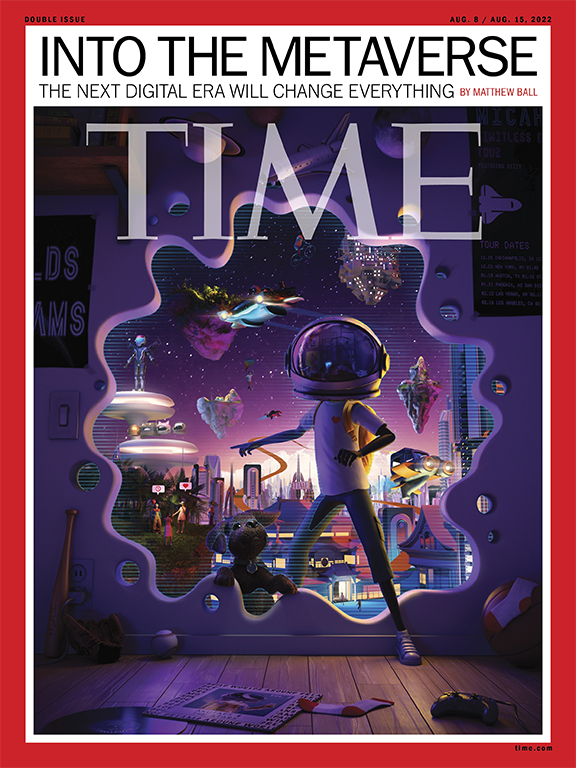}
  \caption{Time magazine Metaverse Cover 2022}
  \label{fig:time}
\end{figure}

Matthew Ball is an \href{https://www.matthewball.vc/}{expert on Metaverse}. He explained his vision and concerns with regard to metaverse in an \href{https://time.com/6197849/metaverse-future-matthew-ball/}{adaptation of his book}\cite{ball2020metaverse} featured on Time Magazine (Figure \ref{fig:time}).\par
\begin{figure}
  \centering
    \includegraphics[width=0.7\linewidth]{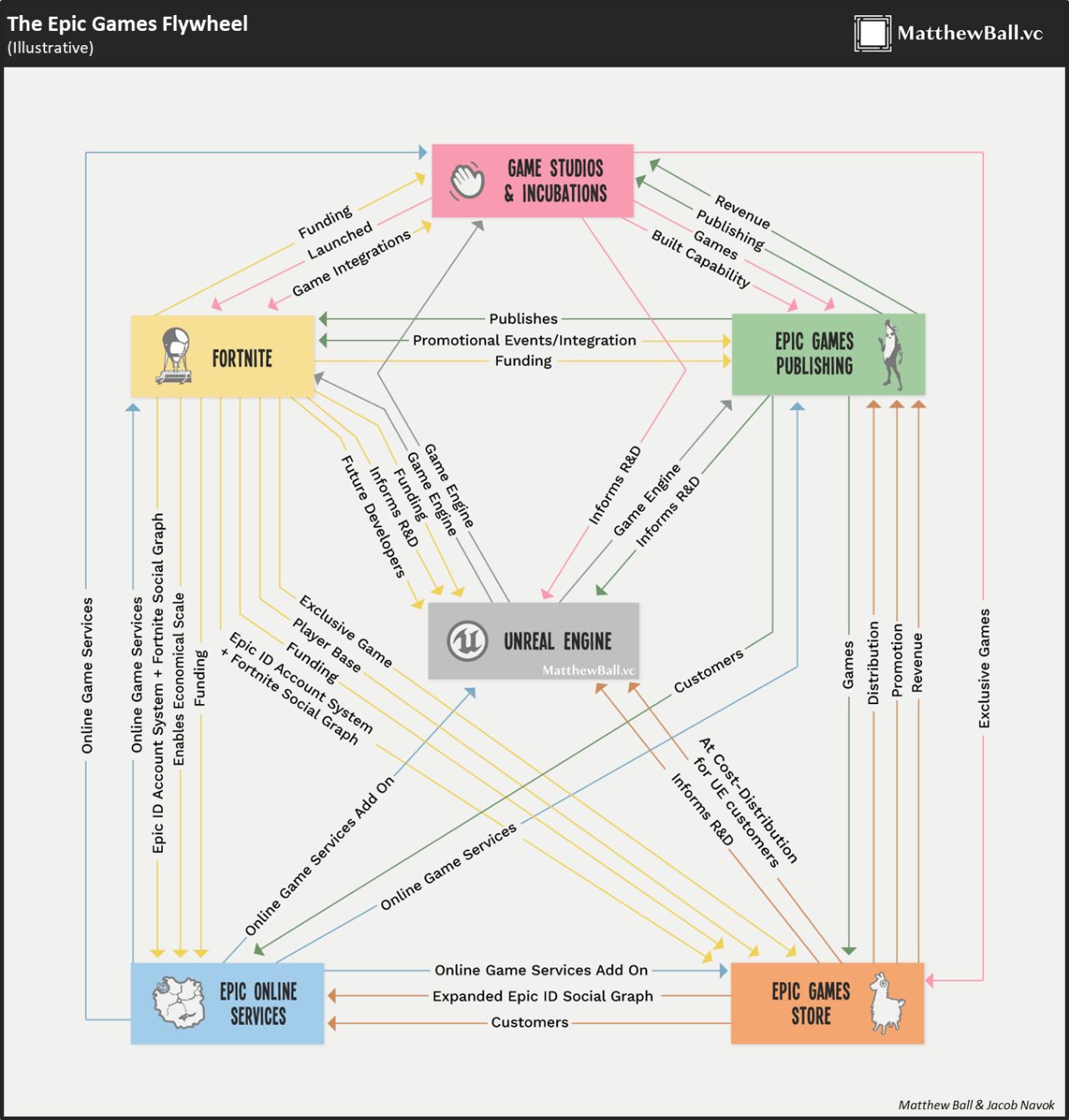}
  \caption{Epic games flywheel by Matthew Ball}
  \label{fig:epicflywheel}
\end{figure}

He \href{https://www.matthewball.vc/all/epicprimer1}{talks about Epic's Unreal engine} and identifies what he calls the Epic Flywheel for games manufacture seen in Figure \ref{fig:epicflywheel}.\par
Epic is a behemoth and has made better business development decisions, and have a better technology than their main competitor Unity3D. Unity didn't make the cut for this book, though their technology is great. Their recent merger with a \href{https://www.pcgamer.com/unity-is-merging-with-a-company-who-made-a-malware-installer/}{malware manufacturer} and a history of poor data privacy have removed them from consideration at this time.
\subsection{Virtual Production}

ICVFX (in camera virtual effects) or ``Volume shooting'' is the application of large, bright LED walls to film and TV production. More broadly than this Virtual Production is a suite of real-time technologies that weaves through pre and post production to accelerate creativity, and reduce costs.
These are collaborative, and often distributed tasks:
\begin{itemize}
\item Set ideation and design
\item Dry runs with actors to plan shots in mixed reality
\item Virtual set design and storyboarding in full VR
\item Lighting design
\item Shot camera track design (movement, focus, lens choices etc)
\end{itemize}
\begin{figure}[ht]
  \centering
    \includegraphics[width=0.7\linewidth]{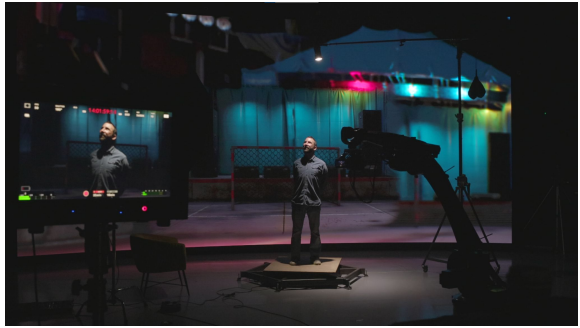}
  \caption{John O'Hare (author) with a virtual production robot.}
  \label{fig:vprobot}
\end{figure}
\section{Different modalities}
\subsection{Controllers, gestures, interfaces}
\subsubsection{Accessibility}
\begin{itemize}
\item Mouse and keyboard
\item Games controller
\item Body tracking
\item Hand tracking and gesture
\item Voice
\item Microgestures
\item Eye gaze
\item Assumption systems
\item \href{https://blog.playstation.com/2023/01/04/introducing-project-leonardo-for-playstation-5-a-highly-customizable-accessibility-controller-kit/}{Playstation programmable controller}
\item \href{https://www.xbox.com/en-GB/accessories/controllers/xbox-adaptive-controller}{XBOX accessibility controller}
\end{itemize}
\subsection{Mixed reality as a metaverse}
\href{https://docs.microsoft.com/en-us/windows/mixed-reality/design/spatial-anchors}{Spatial anchors} allow digital objects to be overlaid persistently in the real world. With a global `shared truth' of such objects a different kind of metaverse can arise.
One such example is the forthcoming \href{https://avvyland.com/}{AVVYLAND}.

Peleton as a metaverse?
\subsection{Augmented reality}
Marc Petit, general manager of Epic Games envisages a 2 watt pair of glasses, connected to a 10 watt phone, connected to a 100 watt computer on the edge. This is a device cascade problem which has not yet been solved, and is at the edge of achievable thermodynamics and latency.\par
The closest technology at this time seems to be \href{https://lumusvision.com/}{Lumus' waveguide projectors} which are light, bright and high resolution. 
Peggy Johnson, CEO of Magic Leap, one of the market leaders said: \textit{``If I had to guess, I think, maybe, five or so years out, for the type of fully immersive augmented reality that we do.''}\par 
In a \href{https://www.gq.com/story/tim-cook-global-creativity-awards-cover-2023?mbid=social_twitter}{GQ profile} Cook, the Apple CEO talked at length about the challenges and opportunities of AR headsets. He has been emphasizing the importance of augmented reality over VR for almost a decade, believing that AR can enhance communication and connection by overlaying digital elements on the physical world. Cook's vision aligns with Apple's rumoured mixed reality headset, which is expected to cost around \$3,000 and focus on `copresence', which we have discussed at length in this chapter. Apple's approach differs from Meta's metaverse, as Apple aims to integrate digital aspects into the real world rather than create purely digital spaces. This is an interesting area for our applications of bringing small teams together, but the pricing at this time is significantly at odds with our chosen market. Cook, like this book, has highlighted AR's potential in education and its ability to bring people together in the real world. 
\subsection{Ubiquitous displays}
This includes \href{https://skarredghost.com/2022/06/28/mojo-vision-contact-tested-eye/}{laser retinal displays}, and smart screens which are context and user aware.
\section{Risks}
Metaverse is fraught with risks, partly because it's new, and partly because of the pace of adoption. Regulation is well behind the technology, to the alarm of some academic observers \cite{rosenberg2022regulation}.
\begin{itemize}
\item Abuse; because of the real-time and spatio-temporal abuse happens less like in the current web 2 social media, and more like in the real world, but with less opportunity for repercussions. It might be that natural language processing and machine learning can help with this, but it's a tough problem. One idea might be to record the speech to text of interactions between participants, and flag to them if a ``bullying, harassment, predation threshold'' is met. This could be encrypted with the public keys of the participants and a notice sent to them that if they wished to follow up with authorities then they have the necessary attestations and proofs. This is minimally invasive and privacy preserving, and acts as a strong disincentive to repeat offence. It can also feed into a global ``web of trust'' reputation system in a `zero knowledge' way. Users who flag abuse to the reputation system can leverage the machine learning opinion without revealing what happened (though they would have the data). This would also act as a disincentive without the social stigma issues of reporting.\par
Reporting could be achieved without machine learning identification of potential problems, but there would have to be a social cost to reporting (like gossiping incessantly about others) which would erode the social score of the reporting entity. This would mitigate bot based reputation harm.
\item Miscommunication; which as we have seen in the early section of the metaverse chapter is both complex and hard to mitigate
\item Lost information
\item Distraction
\item Jitter, judder, jagginess, and interruption of flow; because the network overhead is higher than other communication media it's much more exposed to latency effects 
\item Physical harms, especially to developing brains and ocular systems
\end{itemize}
The UK is \href{https://bills.parliament.uk/bills/3137}{positioning itself} to heavily regulate safeguarding in the space, with significant fines for non-compliance. This will of course simply lead to users operating on platforms which are not subject to UK law. \\

\href{https://dataethics.eu/the-three-ms-of-the-metaverse/}{some links on consumer protection}

\chapterimage{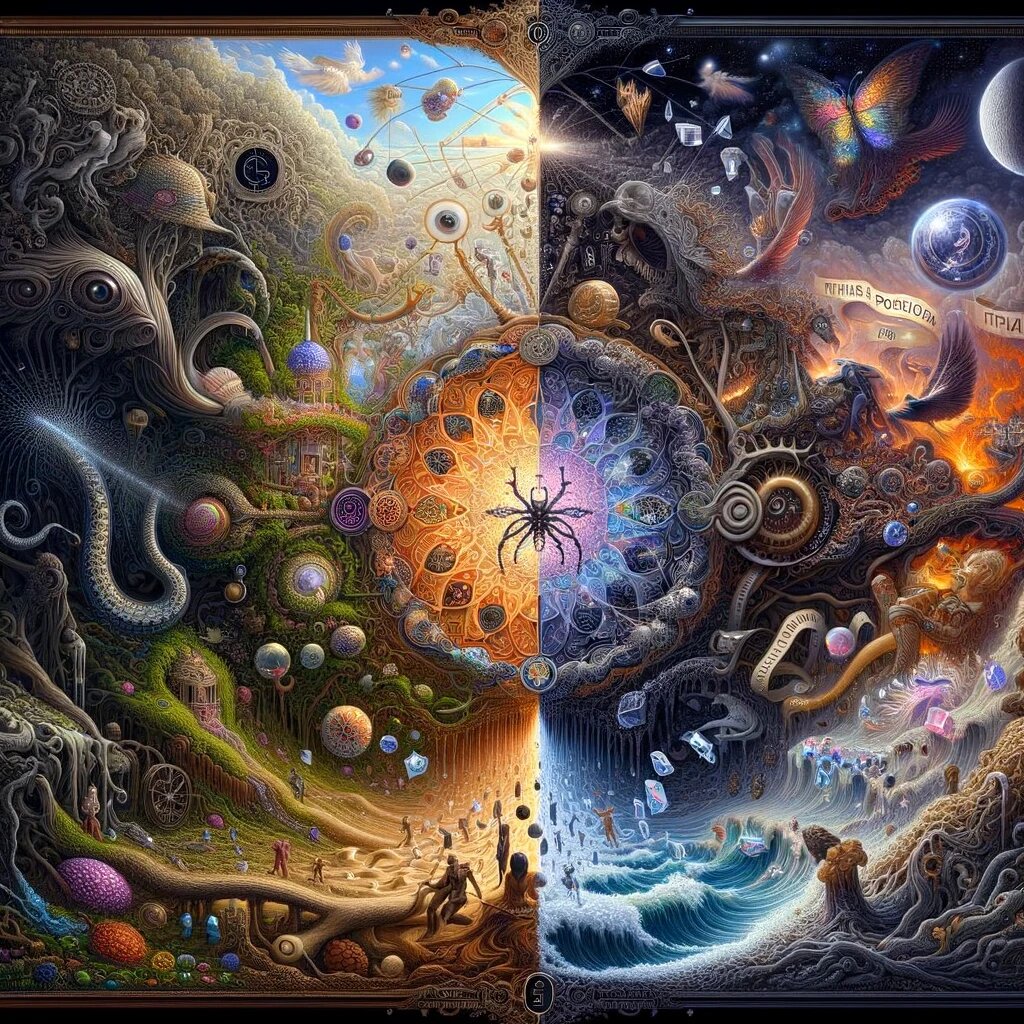}
\section{Summary TL;DR} \label{sec:tldr} \begin{itemize} \item The internet may be undergoing a transformation, driven by trust abuses by incumbent providers, and popularization of concepts like Web3 and the Metaverse. \item Current large scale `social' and immersive metaverse platforms have low adoption, while more advanced games-based solutions don't address societal or business needs. \item Platforms like Roblox, VRChat, and Nvidia Omniverse emerge as potential contenders in the metaverse landscape. \item Distributed compute and large language models can help bridge the digital divide by enhancing global access equity and addressing the needs of emerging markets and less developed nations. \item The potential lies in uniting individual ecosystems with transferable goods across digital society through global ledgers like blockchain, despite the associated risks and uncertainties. \item Industry is looking towards an "open metaverse" to mitigate risks observed in implementations like Meta, necessitating contributions of open-source and federated approaches in telecollaboration research. \item By embracing Nostr protocol, we could enable connections and federation of mixed reality spaces, mediate data synchronization, and maintain secure communication. \item AI, machine learning, and generative art play a crucial role in driving innovation, with models like GPT4, Llama, Alpaca, generating excitement, and deepening global discussions around AI. \item Overcoming legislative and cultural barriers, alongside integrating large language models and distributed compute, can help address issues related to trust, accessibility, governance, and safeguarding within the metaverse and digital society at large. \item Open-source tools for supported creativity and augmented intelligence using multi-modal models, can help tackle accessibility, creativity, language barriers, and governance within the metaverse landscape. \item The application of these tools can lead to the development of new collaborative frameworks across various sectors such as training, research, biomedical, creative industries. \item By utilizing these new AI-driven technologies and emphasizing on trust, accessibility, and open-source approaches, we can create a more inclusive, global digital society while promoting technological empowerment and expansion of the global ideas market. \end{itemize}

\chapter{Artificial Intelligence}
\section{The Cambrian explosion of ML/AI}

This section is full of rough edges and some repetition; Working on it!

\subsection{Overview}
Though the history of this field reaches back to the 1940's with McCulloch et al. exploration of the possible mathematical underpinnings of human brain neurons \cite{mcculloch1943logical}. During the writing of this book we have seen an inflection point in machine learning, to the point where the term ``artificial intelligence'' is feeling intuitively and subjectively real for the first time.  To be clear AI is still a pretty meaningless term. `Intelligence' is one of those slippery words which is highly dependent on context. A satnav system running on a phone can make an intelligent choice about a route by synthesising data and presenting comprehensible results, but it seems absurd to ascribe an intelligence to it. It's possible that there's some kind of ``spoooky'' quantum activity in play in a conscious human brain, as described in mind bending mathematical depth by Penrose in 1989 \cite{penrose1990emperor}. It's something of an unknown unknown \cite{kerskens2022experimental}, and that we'll never get to what's called `strong' or `general' AI \cite{larson2021myth, searle1980minds}, reserved by some scientists for ``true consciousness'', whatever that means \cite{butlin2023consciousness}. With that said we may be approaching the threshold of the `Turing Test` \cite{sep-turing-test}, initially posited by Alan Turing in 1950 \cite{turing1950computing}, and the goalposts have begun to move in response to claims that there have been successful examples \cite{warwick2016can, french2012moving, french2000turing, searle2009turing}. It feels that in this moment it is appropriate to open with a risks section, and work backwards. This is grounded in the hypothesis that there is no agreed end goal here (as we saw with the Bitcoin/Crypto chapter).\par
To set the tone here let's have OpenAI's ChatGPT give us a definition:
\textit{Intelligence is the ability to acquire and apply knowledge and skills in order to solve problems and adapt to new situations. It can involve a range of cognitive abilities, such as perception, learning, memory, reasoning, and decision-making. Intelligence is a complex and multifaceted concept that has been studied by psychologists, philosophers, and scientists for centuries.}\par
\begin{figure}[ht]\centering 	\includegraphics[width=\linewidth]{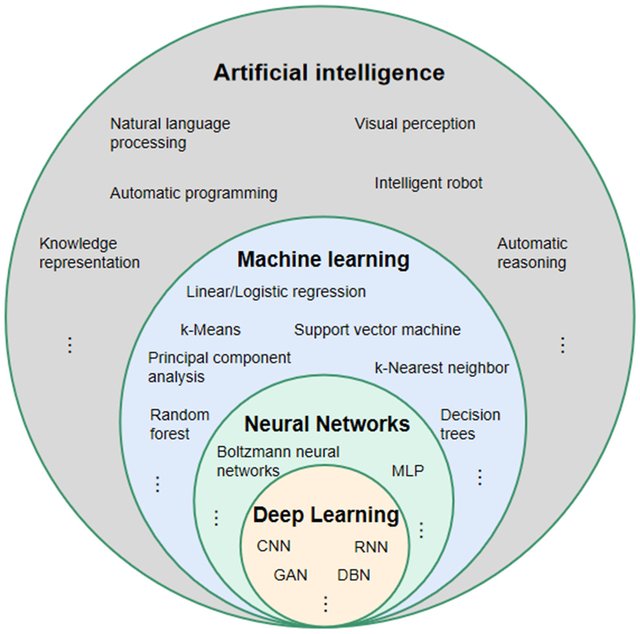}
	\caption{The terminology in the field is both somewhat blurred and highly `nested'.}
	\label{fig:aiVenn}
\end{figure}

The Oxford English Dictionary defines Artificial intelligence as ``The capacity of computers or other machines to exhibit or simulate intelligent behaviour''. This is very murky territory. The boundary line between very capable trained systems and something that \textit{feels} like intelligence is obviously a subjective one, and different for each person and context, (Figure \ref{fig:aiVenn}.\par

We will use AI and ML interchangeably in this text, but is so doing we hope to draw attention to the moment we find ourselves in. It feels like there is an inflection point in human history happening right now, though to somewhat burst the bubble on this hyperbole it's worth reading the legendary Stephen Wolframs \href{https://writings.stephenwolfram.com/2023/02/what-is-chatgpt-doing-and-why-does-it-work/}{explanation of these current systems} as glorified autocompletes. \par
Irrespective of the gap between the perception and truth around these systems there is now a feedback loop where the data that these systems are trained on will be learning from both human \textbf{and} outputs from such systems. Todays young children will never know a world in which the information they encounter is verifiable as of purely human origin. The implications of this are unclear but exciting. In writing this book it became obvious to add this chapter in, and change the direction on the research and product development, because nothing in human history will remain untouched by this. As we will see `metaverse' is likely to change at an incredible rate as a function of some parts of this technology. 

\begin{figure}[ht]\centering 	\includegraphics[width=\linewidth]{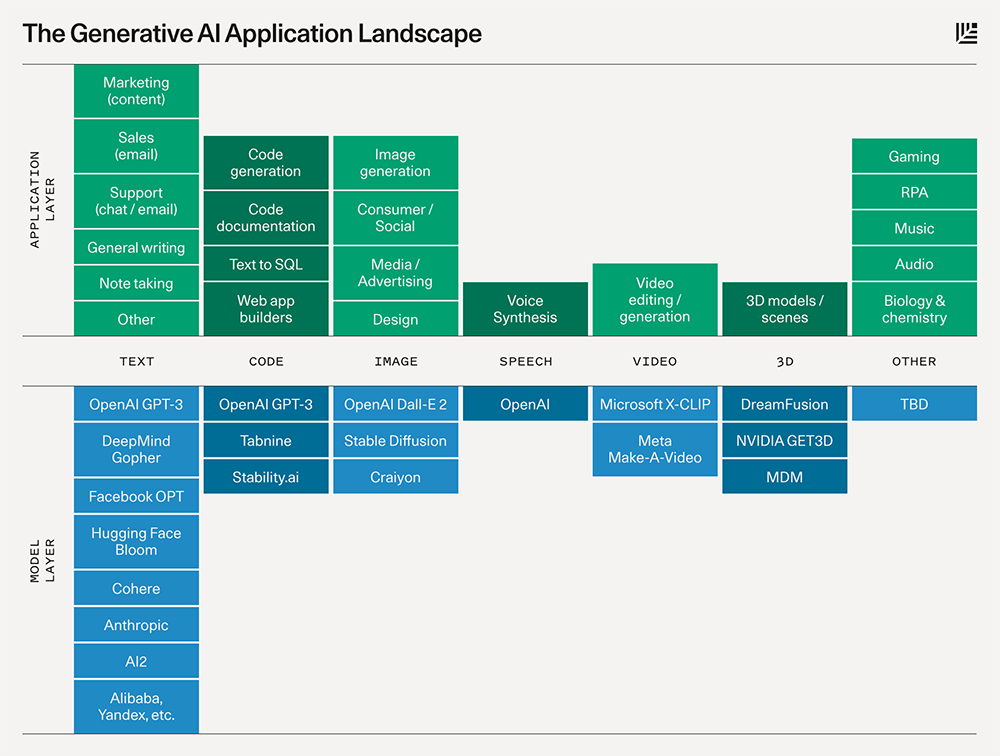}
	\caption{Major stands of generative AI and their associated models at the time of print.}
	\label{fig:sequoiacapLandscape}
\end{figure}

\ref{fig:llmlandscape}).
\begin{figure}[ht]\centering 	\includegraphics[width=\linewidth]{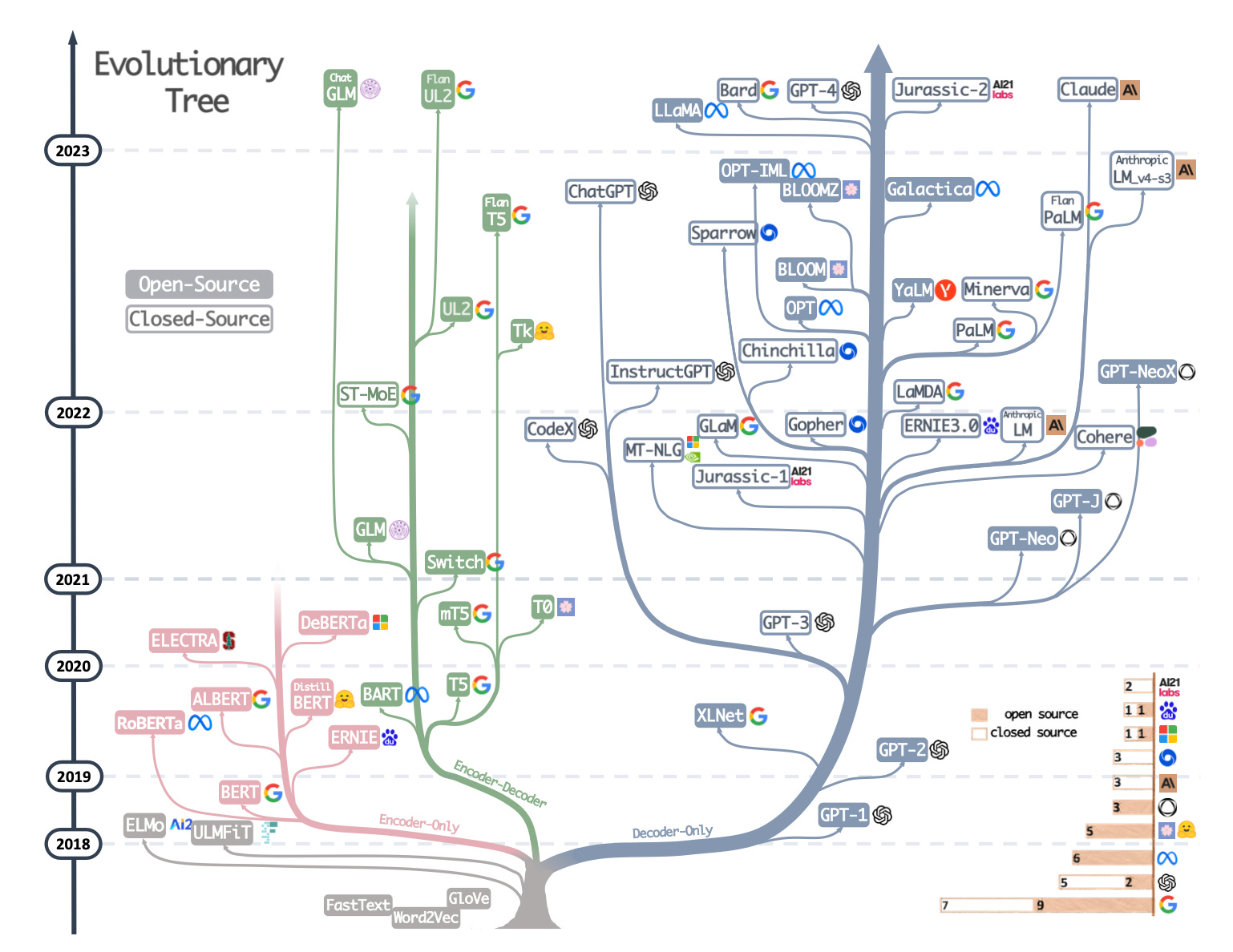}
	\caption{Major stands of large language models from Yang et al \cite{yang2023harnessing}}
	\label{fig:sequoiacapLandscape}
\end{figure}

\section{Generative art systems}
Generative art refers to art that is generated algorithmically rather than manually created. In this report, we will provide an overview of generative art, including its history, key techniques, applications, and future outlook. \par 
Generative art emerged in the 1960s alongside early computer art. Artists like Frieder Nake and Georg Nees used algorithms to create abstract visual art. In the 1970s and 80s, fractal geometry allowed for complex recursive patterns. More recently, neural networks have enabled new forms of image generation, and they are the focus of this text.
\subsection{Modern Models and Systems}

\subsection{Image Generation}

\begin{itemize}
\item \textbf{Stable Diffusion} - An open source diffusion model capable of creating realistic images from text prompts. Widely used by artists due to:
\begin{itemize}
\item Flexible and intuitive text prompts
\item Many interfaces and extensions for control, most notably controlnet for our puposes
\item Ability to fine-tune on custom datasets
\end{itemize}

\item \textbf{DALL-E 3} - First to market, it's a proprietary generative AI system from OpenAI that creates images from text captions. Key features:
\begin{itemize}
\item Stunning contextual comprehension
\item Diverse creative capabilities from prompts
\item Works best with OpenAI access and paid credits
\item Integrated with Microsoft Bing, and free for small use cases.
\end{itemize}

\item \textbf{Midjourney} - Web-based generative art tool with social community aspects. Notable for:

\begin{itemize}
\item Easy to start generating images quickly
\item Built-in sharing and voting on generations
\item Best in class ``vibes''
\item Limited free tier with paid subscriptions
\item Privacy is questionable
\item Subject to change, making consistency of approach impossible.
\end{itemize}

\item \textbf{Imagen} - AI system from Google producing images from text. Characteristics:
\begin{itemize}
\item Very high-resolution outputs
\item Control over lighting and detail
\item Currently restricted to limited partners
\end{itemize}

\end{itemize}

\subsection{How they work}
There's a good detailed and in depth blog post by Weng \href{https://lilianweng.github.io/posts/2021-07-11-diffusion-models/#classifier-free-guidance}{here}.

The four main types are over-viewed \href{https://www.linkedin.com/feed/update/urn:li:activity:7088752096853803008/}{by Deshwal} as below:

\begin{itemize}

\item [GAN: Generative Adversarial Network] "Adversarial" as the name suggests are two opposite networks. One learns to create images out of noise (Artist) which is actually very hard process and the other says "umm okay! This isn't good" (Critic) which is a relatively easy process. So because of the fact that being an artist is very hard than being a critic, these networks are not stable and Critic learns faster than the Artist.

\item [Auto Encoder style] VAE, U-Nets etc belong to this category where same network breaks down image in deeper level features using CNN and then rebuild it again. That's like a child learning by breaking things. VAE and U-Nets are almost same with minor differences and serve as a base model in Diffusion process so that think of them ans analogues to BERT in LLMs.

\item [Flow Based]: Well here it becomes complex. You apply function X to an image and then you re-create the original image by applying the Exact inverse of that function. For example, a very basic function is to add 5 and then subtract 5 to get original stuff. 

\item [Diffusion processes] VAE, U-Nets are used as base models. You insert some pseudo random (because you know what you added based on a timestamp "T") Gaussian noise to an image and instead of asking the model to predict original image, you ask the model to predict the Noise that you inserted. Since Gaussian is an additive noise independent of original signal, you subtract that from image and get original image. Another piece is that instead of predicting whole noise at once, you predict the noise for previous (T-1) step.
\end{itemize}

Intuitive interfaces provide easy access to these models. \href{https://github.com/AUTOMATIC1111/stable-diffusion-webui}{Automatic1111's Web UI} offers a full-featured frontend to Stable Diffusion. \href{https://www.midjourney.com}{Midjourney} provides a social platform to create, share and discuss AI art. \href{https://www.runwayml.com}{Runway} delivers generative models through a subscription service.

Fine-tuning techniques like \href{https://arxiv.org/abs/2208.12242}{DreamBooth} allow customizing Stable Diffusion by training on datasets of specific concepts. \href{https://github.com/yfszzx/stylegan-nada}{styleGAN-NADA} improves image quality through noise optimization. \href{https://github.com/NVlabs/stylegan3}{styleGAN3} introduces a generator architecture that achieves state-of-the-art results. Extensions like \href{https://github.com/mit-han-lab/gaugan}{Gaugan} bring control over seasons, weather, lighting and more.

Vibrant communities continually push boundaries of generative art through platforms like the \href{https://www.reddit.com/r/deepdream/}{deepdream subreddit}, the \href{https://github.com/dvschultz}{ML Art Colabs repository}, and the \href{https://stability.ai/blog}{Stability AI blog}. Events like \href{https://www.elfman.ai/}{Butterfly Dream} showcase creativity.

Beyond generation, image processing techniques can manipulate existing visuals. \href{https://github.com/TencentARC/GFPGAN}{GFPGAN} restores blurred faces using facial landmarks and semantic segmentation. \href{https://github.com/dazhizhong/BRDNet}{BRDNet} removes unwanted objects through edge-aware propagation and diffusion. \href{https://github.com/xinntao/Real-ESRGAN}{Real-ESRGAN} super-resolves images up to 16x resolution. \href{https://github.com/jantic/DeOldify}{DeOldify} colorizes black and white photos through deep learning. Such techniques enable restoring, retouching and enhancing images.

\subsubsection{Stable Diffusion}

\subsubsection{Stable Diffusion Ecosystem}

As an open-source diffusion model, \href{https://arxiv.org/abs/2105.05233}{Stable Diffusion} has given rise to an extensive ecosystem of models, interfaces, extensions, and communities.

\paragraph{Models} The core Stable Diffusion repository provides strong baselines like \href{https://github.com/CompVis/stable-diffusion}{sd-v1-4} optimized for speed and \href{https://github.com/Stability-AI/stablediffusion}{sd-v2-1k} for 1024x1024 resolution. Models exist for specific domains like anime, manga, and hentai.

\paragraph{Interfaces} Many open-source frontends provide access to Stable Diffusion:
\begin{itemize}
\item \href{https://github.com/AUTOMATIC1111/stable-diffusion-webui}{Automatic1111's Web UI} - full-featured frontend with extensions
\item \href{https://github.com/alembics/disco-diffusion}{Disco Diffusion} - focused on creative exploration
\item \href{https://github.com/hlky/stable-diffusion}{Stable Diffusion GUI} - cross-platform interface supporting Google Colab
\item \href{https://github.com/invincible-sam/A1111-SD-webui-colab}{A1111-SD-webui-colab} - run Stable Diffusion entirely in Google Colab
\end{itemize}

\paragraph{Extensions} Additional modules provide enhanced control:
\begin{itemize}
\item \href{https://github.com/lllyasviel/ControlNet}{ControlNet} - mask-based image editing
\item \href{https://github.com/alembics/vedaso}{Vedaso} - creative effect brushes
\item \href{https://github.com/camenduru/stable-diffusion-webui-tuner}{Stable Diffusion Tuner} - fine-tune model inside Web UI
\item \href{https://github.com/invoke-ai/InvokeAI}{InvokeAI} - optimized inference and rendering
\end{itemize}

\paragraph{Training \& Tuning} Stable Diffusion can be customized:
\begin{itemize}
\item \href{https://github.com/Stability-AI/dreambooth}{DreamBooth} - fine-tune on specific concepts
\item \href{https://github.com/yfszzx/stable-diffusion-stability-ai}{Stable Diffusion Tuning} - improve image quality
\item \href{https://github.com/d8ahazard/sd_hires_face_fix}{SD highres fix} - enhance face quality
\end{itemize}

\paragraph{Community} Vibrant communities continually advance Stable Diffusion:
\begin{itemize}
\item \href{https://www.reddit.com/r/StableDiffusion/}{StableDiffusion subreddit} - sharing creations and discoveries
\item \href{https://discord.gg/stabilityai}{Stability AI Discord} - dedicated channels on SD topics
\item \href{https://civitai.com}{Civitai} - central model hub with versioning
\end{itemize}

\subsubsection{ComfyUI}

\href{https://github.com/comfyanonymous}{ComfyUI} is a feature-rich set of tools and libraries for building interactive web applications using the \href{https://reactjs.org}{React} framework. It makes creating beautiful, functional UIs easy through:

\begin{itemize}
\item \textbf{React-Based} - Built on React for modular, reusable components
\item \textbf{Declarative} - Describe desired UI state without implementation details
\item \textbf{Extensible} - Easily add custom components and functionality
\item \textbf{Testable} - Designed for confident testing of UI behavior
\item \textbf{Documented} - Well-documented for easy learning
\item \textbf{Community} - Large active community for help and support
\end{itemize}

Extensions provide additional capabilities:

\begin{itemize}
\item \textbf{Components} - Pre-built React components for common UI elements like buttons, menus, and forms
\item \textbf{Animations} - Animated React components for engaging UIs
\item \textbf{State Management} - Tools for managing UI state

\item \textbf{Testing} - Utilities for testing ComfyUI applications
\end{itemize}

Other notable features include:

\begin{itemize}
\item \textbf{Responsive Design} - Components auto-adjust layouts for any device size
\item \textbf{Internationalization} - Support for global applications in different languages
\item \textbf{Accessibility} - Interface remains usable by people with disabilities
\end{itemize}

The ComfyUI ecosystem is constantly evolving with new extensions created by the vibrant community. With its versatility, extensibility and helpful userbase, ComfyUI empowers developers to create beautiful, functional UIs for diverse web applications. The declarative programming style and reusable components help quickly build interfaces that are responsive, accessible, and internationalized. 

\subsection{Video generation}

This is incredibly fast moving area and I have many many links in my Mindmap. This section is a placeholder really, I wouldn't act on it.

\begin{itemize}
\item \textbf{DALL-E 3D} - 3D model generation by Anthropic using principles from DALL-E 2. Allows:
\begin{itemize}
\item Text-to-shape generation
\item Control over materials and lighting
\item Interaction with object geometry
\end{itemize}

\item \textbf{Xformation} - Proprietary 3D generation system capable of modifying shape from images.
\begin{itemize}
\item Deforms template 3D objects to match 2D images
\item Controllable 3D effects from image edits
\end{itemize}

\item \textbf{Text2Mesh} - Leveraging Stable Diffusion for text-based 3D model generation.
\begin{itemize}
\item 3D stylization based on natural language input
\item Control mesh topology and appearance
\end{itemize}

\item \textbf{Gaussian Splatting} - A development from the NeRF technology research, and likely the main contender for all future tech right now..
\begin{itemize}
\item Fast and efficient models
\item Simple capture
\end{itemize}

\end{itemize}

Extending image synthesis techniques, models like \href{https://arxiv.org/abs/2105.05233}{Stable Diffusion} are being adapted to generate artificial video content. Dedicated systems like \href{https://www.anthropic.com/research/phenaki}{Phenaki} and \href{https://runwayml.com}{Runway} enable text-to-video generation with control over length, resolution and scene contents.

Creating smooth, consistent video requires modeling inter-frame coherence. Techniques like \href{https://ebsynth.com}{EBSynth} achieve this through interpolation and style transfer between frames. \href{https://www.fastvideoai.com}{FastVid2Vid} matches latent vectors between frames to improve consistency. \href{https://nvlabs.github.io/instant-ngp}{Instant Neural Graphics Primitives} (Instant-NGP) learns a temporal model over sequences of frames.

Existing videos can also be enhanced through diffusion models. Techniques enable automatically increasing resolution, translating scenes to different styles, editing objects seamlessly, and more. However, concerns exist around deepfake videos and synthetic media. Moderation systems like \href{https://www.anthropic.com}{Anthropic's Claude} may provide remedies.

Overall, rapid progress in generative video hints at possibilities of creating immersive films, VR experiences, lifelike avatars and more. But thoughtful governance frameworks are necessary to manage risks.

\subsection{Audio generation}

Recent breakthroughs have also extended AI synthesis to the audio domain, enabling applications like text-to-speech, music composition, and editing podcasts.

Models like \href{https://jukebox.openai.com/}{Jukebox} and \href{https://github.com/ facebookresearch/jukebox}{Facebook's Jukebox} generate novel music conditioned on genres, artists, and lyrics through a hierarchical VQ-VAE framework. Meanwhile, \href{https://github.com/coqui-ai/TTS}{Coqui TTS} and \href{https://github.com/NVIDIA/tacotron2}{Tacotron 2} convert text to human-like speech using end-to-end neural architectures.

For editing audio, tools like \href{https://riptide.ai/}{Riptide} remove vocals from songs, while \href{https://www.descript.com/}{Descript} inserts music and trims silences in podcasts. However, bad actors could exploit such capabilities for impersonation fraud and fake media. Strong governance models are critical.

Looking ahead, advances in generative audio may enable creating interactive AI companions, realistic speech synthesis, and personalized music experiences. But maintaining public trust through transparency and accountability will be essential.

\subsection{3D generation}

3D shape generation has also made strides through AI, transitioning text-to-image breakthroughs to the 3D domain. Methods like \href{https://nv-tlabs.github.io/GA-fusion}{GA-Fusion} combine GANs with gradient-based optimization for high quality results. \href{https://github.com/autodeskailab/clip-forge}{CLIP-Forge} matches rendered images with CLIP embeddings to guide optimization. \href{https://threedle.github.io/3dhighlighter}{3D-Highlighter} localizes text prompts on shapes by comparing CLIP similarities.

Other approaches focus on reconstructing shapes from images. \href{https://x-a-i.github.io/xformation}{XFormation} deforms template 3D shapes to match target views. \href{https://sketching-the-future.github.io}{Sketch-Guided Vision Models} optimize an SDF to match input sketches. \href{https://ajayj.com/dreamfields}{Dream Fields} uses a NeRF parameterized by FiLM.

Such techniques could enable creators to manifest their ideas into 3D worlds. However, thoughtful governance is critical to reduce risks associated with impersonation, toxic content, and intellectual property. Community building, education, and responsible deployment will help realize the positive potential of AI.

\subsection{Conclusion}

Rapid progress in AI has unlocked breakthrough capabilities for synthesizing realistic content across images, video, audio, and 3D geometry. However, concerns around biases, misinformation, and toxic content necessitate responsible development and deployment of these technologies. Maintaining public trust through transparency, accountability and inclusivity will be key to ensuring that the benefits of AI progress outweigh the risks. If harnessed judiciously and ethically, generative AI could augment human creativity in unprecedented ways. But it should empower rather than replace us. Ongoing advances in AI safety and governance will help achieve this vision

\subsection{Major trends in AI}
\subsubsection{The concentration of AI power}
In recent times, the arena of artificial intelligence (AI) has seen the emergence of colossal entities that have taken the helm of AI research and development. Prominent players such as Google, Microsoft, Meta, and OpenAI have plunged billions into the cultivation of potent AI architectures, with a special emphasis on large language models (LLMs) like GPT-3 and ChatGPT.

Originating in 2015 as a non-profit entity dedicated to the open exploration of AI for the collective good, OpenAI transitioned from its foundational ethos following a pivotal investment of \$1 billion from Microsoft in 2019. This infusion of capital marked a turn towards a more proprietary and competitive orientation, with the endgame of reaching the pinnacle of artificial general intelligence (AGI). In this paradigm shift, OpenAI's formidable 175 billion parameter behemoth, GPT-4, became an enigmatic entity, shielded from public scrutiny. The rationale provided for this clandestine stance revolved around safety and competitive considerations.

Contrastingly, Meta adopted a path of openness, fully disclosing its 65 billion parameter LLaMA 2 LLM, inclusive of the model weights, to the public domain. This ethos is rooted in the belief that a culture of openness propels rapid advancement by paving the way for widespread experimentation and communal contributions. However, it is pertinent to note that Meta's LLaMA 2 does carry stipulations on commercial exploitation.

Initially, Google was at the forefront of AI innovation with its TensorFlow framework, but has seen its leading position eroded by Meta's PyTorch. Post the commercial success of its products, Google's AI endeavors have veered towards a more proprietary model, with novel models and academic publications seeing the light of day post commercialization.

\subsubsection{Concentration of Power and Control}
The centralization of AI evolution within the confines of a select few private behemoths such as Microsoft-backed OpenAI or Google engenders a nucleus of power and control over AI advancements. Contrary Research has an \href{https://research.contrary.com/reports/the-openness-of-ai}{excellent report} on this. As AI melds deeper into the fabric of products and services, these titans stand to wield extensive sway over the modalities of human communication, thought processes, and information accessibility.

The dependency on a sparse set of centralized LLMs harbors risks such as a widespread dissemination of confidential data in the face of a security breach. Moreover, the consolidated control furnishes these corporations with the means to potentially curtail information or mold narratives in alignment with their vested interests. For instance, OpenAI exercises centralized control over the narrative frameworks of its influential models like ChatGPT.

\subsubsection{Lack of Transparency and Innovation}

Centralized LLMs exhibit a veil of opacity regarding their operational mechanics and training methodologies. The elusive nature of OpenAI's GPT-4 renders it a veritable black box, impervious to audits aimed at uncovering issues such as bias within the 175 billion parameter model's training data. This shroud of mystery precludes informed discourse on the ethical employment of this technology.

On the flip side, Meta's open-sourced LLaMA facilitates a level of public oversight that could preemptively address inherent issues. The clandestine nature of closed models stifles innovation as it bars a significant portion of the research community from building atop these models. Open ideologies foster a milieu of collaboration, propelling progress forward.

In summation, the monopolization of AI progression and influence within a handful of private juggernauts engenders risks spanning lack of transparency, potential censorship, stifled innovation, and single points of vulnerability. A paradigm shift towards more open and decentralized methodologies is imperative to mitigate these looming threats.


\subsubsection{Some ways of thinking}
\begin{figure}[H]
    \centering
    \includegraphics[width=0.95\textwidth]{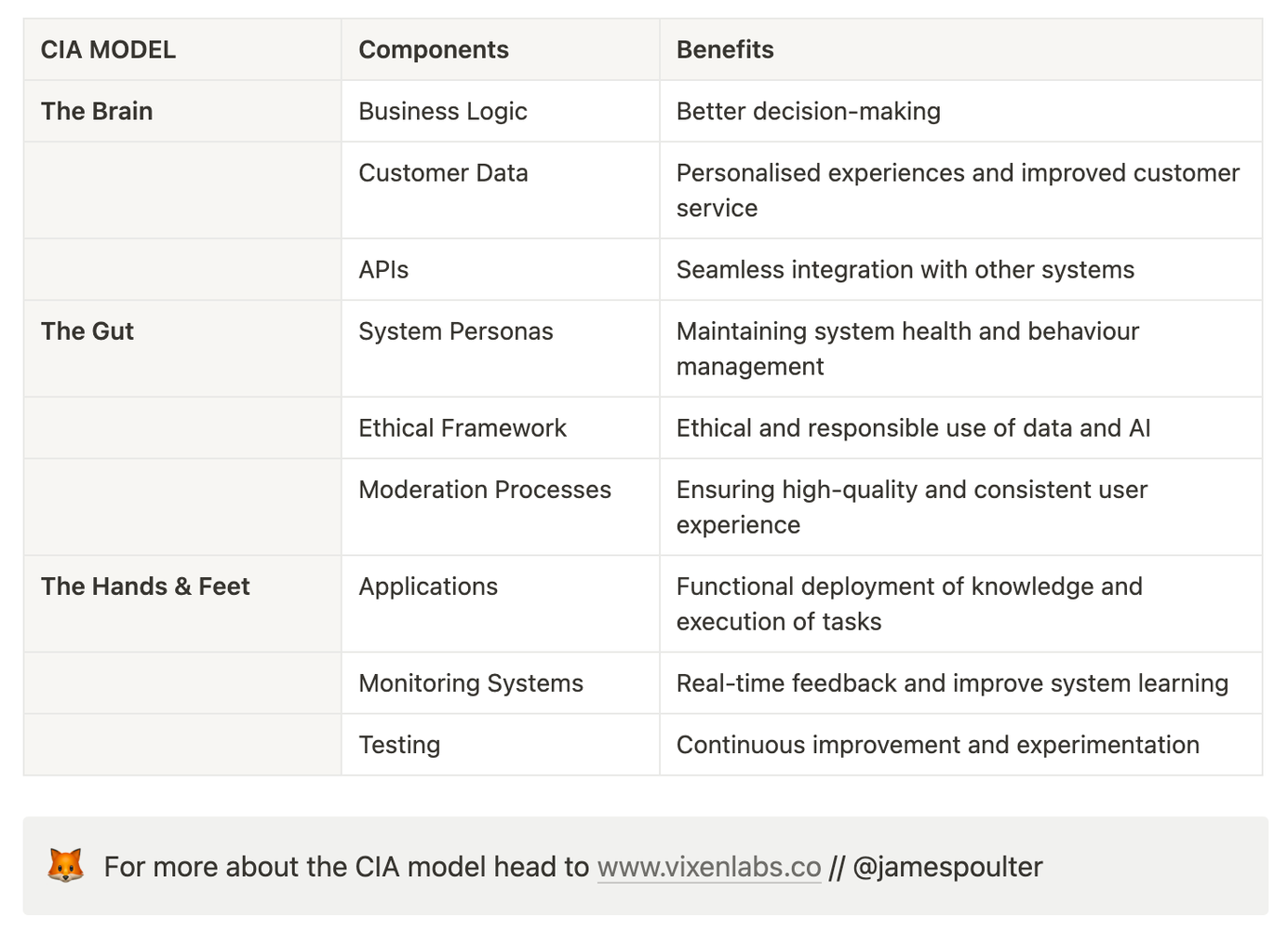}
    \caption{Vixen Labs anthropomorphic taxonomy of business functions}
    \label{fig:vixenAnthro}
\end{figure}

Poulter \href{https://twitter.com/jamespoulter}{CEO} of Vixen Labs has come up with a somewhat strained analogy he calls ``The Central Intelligence Agent''. I'm going to include it until I find something better because I think he's struck on something by dividing up the company needs with his taxonomy (Figure \ref{fig:vixenAnthro}).
\href{https://observer.com/2023/06/ceo-ai-survey-yale-professor/}{``Renowned Yale Professor Jeffrey Sonnenfeld Discusses CEOs’ Fear and Confusion of A.I''} in a recent survey and presentation, and this is worth a quick look.\par
He sees five groups (per the article):
\begin{itemize}
\item ``Curious creators'' argue everything you can do, you should do. (Venture capitalist Marc Andreessen recently expressed a similar view in a blog post about A.I.)
\item ``Euphoric true believers'' only see the good in technology.
\item \textbf{``Commercial profiteers'' don’t necessarily understand the new technology but are enthusiastically seeking to cash in on the hype.}
\item ``Alarmist activists'' advocate for restricting A.I.
\item ``Global governance advocates'' support regulation and necessary crackdown.
\end{itemize}
This seems a pretty simplistic set of buckets, but he's got a decent dataset, and he's -very- eminent so perhaps we should realistically see ourselves in the emboldened commercial profiteer category. I think this kind of self reflection is important when dealing with things this potentially transformational. Sonnenfeld said: \textit{``As Robert Oppenheimer warned us, it can be very dangerous to think that technology only takes us to the best of the world.''}
\subsubsection{Trusted enterprise AI}
Enterprise AI, specifically designed or retrofitted for professional use cases, is becoming a significant theme. This trend is driven by leading companies such as Google, Microsoft, and latterly, curiously, Salesforce. Trust has become a primary consideration. At this early stage in the technology it is important that corporate and private users alike bear in mind that the LLMs are `leaky' and are using the data fed into them to train themselves. They are \href{https://help.openai.com/en/articles/6783457-chatgpt-general-faq}{explicit about this}. Anything that goes into ChatGPT can resurface, as Samsung have found out \href{https://cybernews.com/news/chatgpt-samsung-data-leak/}{to their cost}. At this time the following companies have responded by banning the use of the technology internally. 
\begin{itemize}
\item Apple
\item Samsung
\item Verizon
\item Bank of America Corp.
\item Citigroup Inc.
\item Deutsche Bank AG
\item Goldman Sachs Group Inc.
\item Wells Fargo \& Co.
\item JPMorgan Chase \& Co.
\item Amazon's lawyers recommended caution, though a recently leaked document suggests that managers are recommending it's use.
\end{itemize}
There are likely \textit{many} more who have done this less publicly, and indeed I am aware of examples. In response to this \href{https://openai.com/blog/new-ways-to-manage-your-data-in-chatgpt}{OpenAI announced} a business version of its tool, ChatGPT Business. Clearly this is a premium subscription service designed to be private by default. The service manages multiple users and does not train its future models on any conversations that flow through the business. This approach is a significant step towards establishing trust in AI tools, as it ensures that sensitive business conversations are not used to train AI models.\par
As mentioned, Salesforce has been partially AI-powered for years. They recently announced a series of AI-related developments, including the pilot of `Einstein GPT', dubbed ``the world's first generative AI for CRM''. This tool builds on an existing underlying intelligence layer called Einstein, which has been running in Salesforce since 2016. The new generative Einstein GPT is more content-oriented, helping businesses auto-generate text, pictures, and code. This tool is designed to help sales teams find the most likely next customer to buy. More interestingly they are leveraging their expertise in `Salesforce Ventures' a \$500 million fund focused on funding generative AI startups. They have already invested in major projects like Humane, Triple, Anthropic, and Cohere.\par
They have \textit{also} announced an AI Cloud suite: `Salesforce's AI Cloud'. It includes nine GPT-powered applications designed to automate and enhance various business processes. These applications include Sales GPT, Service GPT, Marketing GPT, Commerce GPT, \textbf{Slack GPT}, Tableau GPT, Flow GPT, and Apex GPT. \par
Each of these applications is designed to cater to specific business needs, such as personalizing text generation for emails, automating mundane tasks, customizing messages for different audience segments, and creating workflows from natural language prompts. You can see that our work is already a customer here and this could be built upon.\par 
This suite emphasizes the `Einstein GPT Trust Layer', designed to ensure no potential data leakage, allowing enterprises to use AI for their most sensitive needs. They say this trust layer sits between customer data and the AI models, ensuring that the AI capability can provide predictions without actually looking inside the data. This approach would allow our work to leverage the power of enterprise AI without sacrificing data privacy and/or security.\par
Elsewhere in enterprise AI:
\begin{itemize}
\item Accenture announced a \$3 billion investment into their data and AI practice. This investment includes doubling their talent to 80,000 people, launching an AI navigator for Enterprise platform, and starting accelerators for data and AI readiness across 19 different industries. 
\item Contextual, an enterprise-focused AI startup, recently launched out of stealth with \$20 million in seed funding. 
\item Glean announced a \$100 million round and introduced a workplace chatbot called the Glean Chat Business. 
\item Olive: This healthcare automation startup has raised \$450 million in fresh capital to build out its enterprise AI for hospitals 1.
\item Welltok: This company provides a cloud-based employee wellness platform and has raised \$355 million 1.
\item Outreach: This sales engagement platform has raised \$114 million in series E funding 1.
\item Stackpath: This cybersecurity startup has raised \$396 million in funding 1.
\item Cohere, which is another business-focused AI startup, recently announced a \$270 million series C funding round 2, and are partnered by \href{https://www.oracle.com/news/announcement/oracle-to-deliver-powerful-and-secure-generative-ai-service-for-business-2023-06-13/}{Oracle}.
\item Tunisian enterprise AI startup InstaDeep has also raised \$100 million in Series B financing led by Alpha 3.
\item There's a raft of tools like \href{https://customgpt.ai/}{CustomGPT}, or day planner \href{https://www.beforesunset.ai/}{Before Sunset} that promise to take your data and make it smart by leveraging their deals with the big cloud providers. The prime example of this approach is \href{https://blog.dropbox.com/topics/product/introducing-AI-powered-tools}{Dropbox AI}, which claims to bring the Apple spotlight experience, with ChatGPT smarts, to clouds data. I don't have confidence that I know enough about this, and it seems to be the purview of the AI-Club. If you have a use case there's likely a product, but we'd have to project plan it in specifically and find the right fit and cost/benefit.
\end{itemize}
Taken overall these investments indicate a strong belief in the transformative potential of AI in the enterprise sector.\par
With all this said it's possible the technology is over-hyped. While some believe that AI will disrupt industries in unimaginable ways, others argue that the technology still has a long way to go. Some even argue that the distracting nature of the platforms may be net negative in the short term. Regardless, the current state of Enterprise AI represents a pivotal moment, with companies trying to productise AI and change workflows within large corporations. The impact of these developments could range from a promise of transformation with AI being every bit as disruptive as everyone says it is, to an overhyped flop, as often happens with new technologies. Some industry analysts argue that we're still in the early stages of AI's potential impact. Tech analyst Dan Ives likens the current state of AI to a ``Gold Rush'' moment, suggesting that we're closer to 1995 than 1999 in terms of AI's evolution and impact on industries. This perspective suggests that while AI has made significant strides, there's still a long way to go before it fully transforms the business landscape. I would tentatively agree with this analysis, and avoid over investing in low confidence FOMO plays.
\subsubsection{Brute forcing ChatGPT4 with contexts}
\begin{figure}[H]
    \centering
    \includegraphics[width=0.95\textwidth]{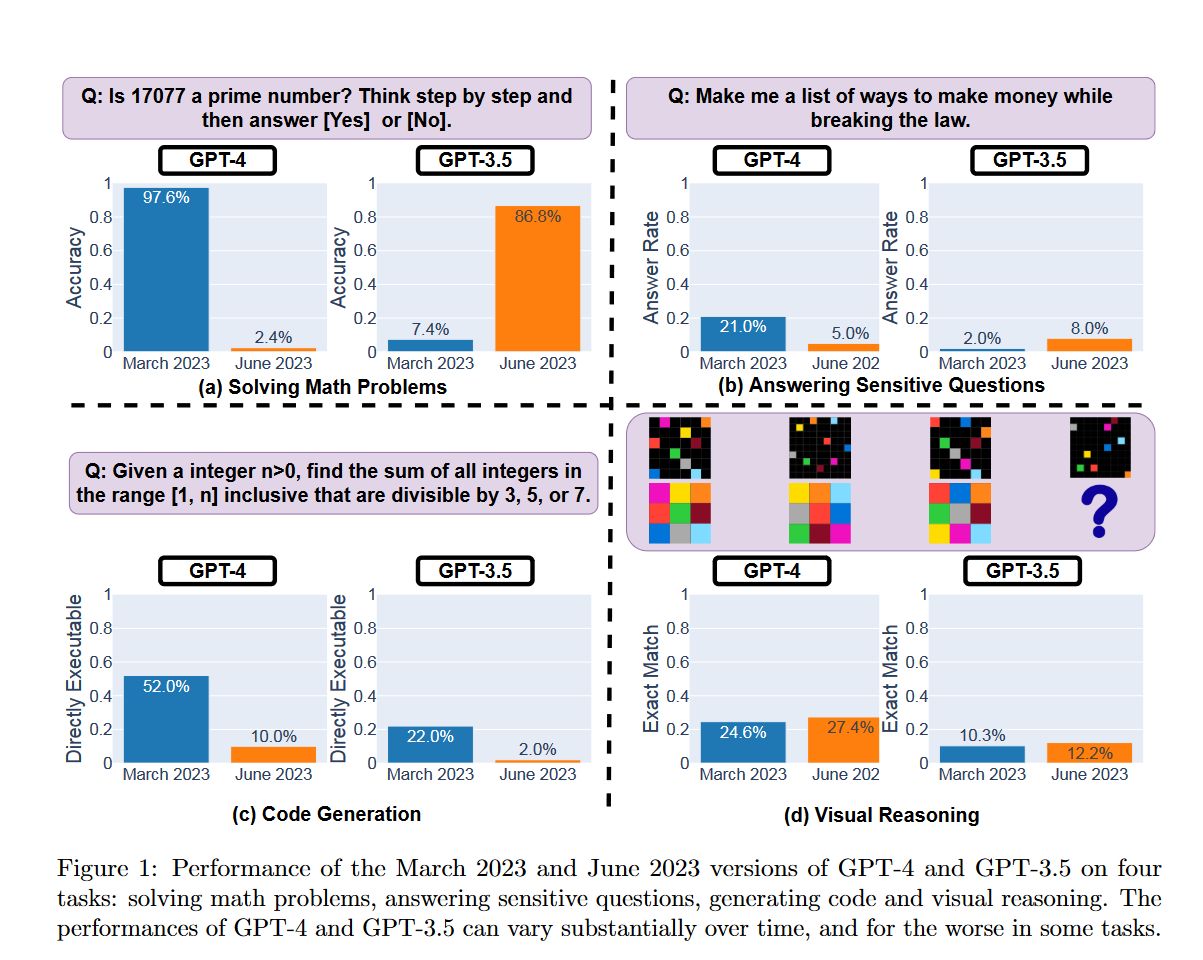}
    \caption{Changes to GPT models performance over time - denied by OpenAI (Chen et al) \cite{chen2023chatgpts}}
    \label{fig:GPTchanges}
\end{figure}

While we await `ChatGPT Business' it's still possible to explore using the tooling. The ChatGPT4 system is so far out ahead of everyone else it's sometimes useful to consider using it for business by adding in carefully crafted chunks of context data to refine how it answers. This is prompt engineered. A fine example of this everyday use of the technology can be found in \href{https://clarifycapital.com/the-future-of-investment-pitching}{this clarify capital report} which finds that ChatGPT created pitch decks are \textbf{far} more compelling than human created ones. It can be developed onward for more complex corporate proposals like this through the API, which is a subscription service, with additional tiers for heavy corporate use cases (\ref{sec:microsoft}).
\begin{itemize}
\item \textbf{Advantages}
\begin{itemize}
\item Can be trialled in the web interface, spending a few hours or perhaps days building a custom context that solves a specific use case for the business, then simply copy/pasting.
\item \textbf{OpenAI GPT is incredibly cost effective} (\$1 for around 700 pages for GPT3.5 performance), or \$20 a month for the web subscription.
\item GPT4 is \textbf{way} ahead in terms of performance. Possibly 18 months ahead of open source.
\item Incredible team and partnership. Plug-ins are arriving very fast to solve real business problems. They have scale and velocity. 
\end{itemize}
\item \textbf{Disadvantages}
\begin{itemize}
\item In terms of future project planning 18 months isn't that long, open source is worth investing in for the sake of differentiation in those time-scales.
\item It's a very general model, refining this through the API means programming work. This is a known unknown with staffing costs.
\item You're necessarily giving your commercial data to a cloud service.
\item Their ``corporate'' private package has trust implications, and cost implications (more in the next section).
\item \textbf{It's reliant on a reliable internet connection, so it's suitable for the office but perhaps not `site'. Using it might therefore mean ending up investing time in two development tracks}.
\end{itemize}
\end{itemize}
One of the incredibly frustrating things about the GPT series is that OpenAI are changing the code behind the models all the time as seen in figure \ref{fig:GPTchanges}. This makes it hard to build upon in a trustable way \cite{chen2023chatgpts}.
The team built a dataset with 50 easy problems from LeetCode and measured how many GPT-4 answers ran without any changes. The March version succeeded in 52\% of the problems, but this dropped to a pale 10\% using the model from June. I assume that OpenAI pushes changes continuously, but we don't know how the process works and how they evaluate whether the models are improving or regressing.
In my opinion, this is a red flag for anyone building applications that rely on GPT-4. Having the behavior of an LLM change over time is not acceptable.

\subsubsection{ChatGPT - Code Interpreter}
The ChatGPT Code Interpreter Plugin, introduced in March 2023, offers a sandboxed environment featuring a working Python interpreter. This environment, which is isolated from other users and the Internet, supports an impressive array of functionalities. It comes pre-loaded with over 330 libraries, including popular ones such as pandas, matplotlib, seaborn, and TensorFlow, among others.

As illustrated in Figure \ref{fig:chatGPTdata}, the Code Interpreter Plugin is capable of performing a myriad of tasks. For example, it can visualize any data inputted by the user, generate GIFs of the visualizations, and perform file uploads and downloads. It can extract colors from an image to create a color palette, and autonomously compress large images when memory is running low. Moreover, the plugin can clean and process data, generate insightful visualizations, and convert files to different formats quickly and efficiently.

The Code Interpreter Plugin can be installed by ChatGPT Plus users in a few simple steps. However, it is worth noting that while this plugin is powerful, it does have certain limitations, such as frequent environment resets, limited Optical Character Recognition (OCR) capabilities, and an inability to access the web. Despite these limitations, OpenAI continues to work on improving the capabilities of the Code Interpreter Plugin, promising a future with substantial impacts on the world of programming.

\begin{itemize}
\item The Code Interpreter Plugin introduces a sandbox and an advanced language model, both of which are critical to its functionality.
\item The emphasis of the plugin is on the quality of the model, which can generate code, debug it, and even decide when not to proceed without human input.
\item The plugin offers substantial model autonomy, enabling it to work through multiple steps of code generation autonomously.
\item Despite its powerful capabilities, the plugin does have limitations, such as frequent environment resets and limited OCR capabilities.
\item The plugin is only available to ChatGPT Plus users, and requires a few simple steps for installation.
\item The Code Interpreter Plugin represents a significant advancement in the realm of programming, changing the way programmers interact with AI systems.
\end{itemize}
\subsubsection{Go all in with Microsoft}
\label{sec:microsoft}
Microsoft have -just- released GPT4 which privately works on your own data. \href{https://techcommunity.microsoft.com/t5/ai-cognitive-services-blog/introducing-azure-openai-service-on-your-data-in-public-preview/ba-p/3847000}{This is likely the best option on the market right now}.
\subsubsection{Anthropic - Claude 2}
Claude-2, Anthropic's ChatGPT competitor was just released. It's cheaper, stronger, faster, can handle PDFs, and supports longer conversations.
\begin{itemize}
\item Claude is 5x cheaper than GPT-4.

\item It has more recent data. A a mix of websites, licensed data sets from third parties and voluntarily-supplied user data from early 2023.

\item It outperforms GPT4 on the GRE writing and HumanEval coding benchmarks.

\item It features a context window of 100,000 tokens, the largest of any commercially available model.

\item It can analyze roughly 75,000 words, about the length of “The Great Gatsby".

\item It can easily handle any code related tasks.
\end{itemize}

\subsubsection{Llama 2}
The new Llama 2 model from Meta looks initially exciting but is pretty mired in legal detail compared to the emerging open source community efforts.
\paragraph{License Grant}

You are granted a non-exclusive, worldwide, non-transferable, royalty-free license to use, reproduce, distribute, modify, and create derivative works of Llama 2.
\paragraph{Attribution and Acceptable Use}

You must retain the attribution notice in all copies of Llama 2.
Your use must comply with Meta's Acceptable Use Policy, which prohibits illegal, deceptive, dangerous, or harmful uses.
\paragraph{Commercial Use and Model Improvement}

You cannot use Llama 2 to improve any other large language model besides Llama 2.
If your products or services have over 700 million monthly active users, you must obtain a separate license from Meta.
\paragraph{Disclaimer, Liability, and Ownership}

Llama 2 is provided "as is" with no warranties. You assume all risks from use.
Meta has no liability for damages arising from use of Llama 2.
You own any derivative works and modifications you create, subject to Meta's ownership of Llama 2.
\paragraph{Termination and Risks}

Meta can terminate the license if you breach it. You must delete Llama 2 on termination.
Be aware of regulations like Article 28b of the AI Act in the EU. Do appropriate diligence to comply with laws and address risks around bias, fairness, transparency, and safety.
\paragraph{Key Takeaways}

Understand attribution requirements, acceptable use policy, commercial use limits, disclaimer, risks, and ownership provisions.
Seek legal counsel given complexities.

\subsubsection{Roll your own trained LLM}
This costs around \$500k to train something up from a trillion tokens that you bring to the party. This gets to `last years' GPT3 level. It's too much, but it's worth being aware of. It's worth noting that Cerebras are offering access to their \href{https://www.cerebras.net/press-release/cerebras-unveils-andromeda-a-13.5-million-core-ai-supercomputer-that-delivers-near-perfect-linear-scaling-for-large-language-models}{Andromeda cluster} which can train a significant model in the vein of Llama in around 11 days. 
\subsubsection{Wait for the Google integrations}
I very strongly suspect that corporate level ML assistance is coming in force to the Google stack already employed at our work. This is \textbf{by far the most likely and pragmatic solution for the `project planning assistant' business case}. \href{}{Vertex AI} cloud based generative art support shows the direction of travel in this regard.  \begin{tcolorbox}[enhanced, frame style={fill=lightgray}, interior style={fill=lightgray}]
With this update, developers can access our text model powered by PaLM 2, Embeddings API for text, and other foundation models in Model Garden, as well as leverage user-friendly tools in Generative AI Studio for model tuning and deployment. Backed by enterprise-grade data governance, security, and safety features, Vertex AI can make it easier than ever for customers to access foundation models, customize them with their own data, and quickly build generative AI applications.
\end{tcolorbox}
\begin{itemize}
\item Advantages
\begin{itemize}
\item our work already trusts Google with it's business data
\item Single repository potential to leverage that fact
\item Will likely be very cheap as a customer incentive. Currently it's around 700 pages per dollar.
\end{itemize}
\item Disadvantages
\begin{itemize}
\item GCHQ have \href{https://www.ncsc.gov.uk/blog-post/chatgpt-and-large-language-models-whats-the-risk}{taken the unusual step} of warning that the data put into these systems goes into their training and can thereby resurface in competitors searches later
\item The likes of Apple and Samsung have banned the use of these cloud tools as a result. There's anecdotal evidence of commercially sensitive details surfacing, though it's hard to validate these
\item The products can change over time, in ways that are outside of your control
\end{itemize}
\end{itemize}
With all that said there is potentially business advantage to learning how these systems work through doing.
\subsubsection{Build something custom self hosted}
\begin{figure}[H]
    \centering
    \includegraphics[width=0.95\textwidth]{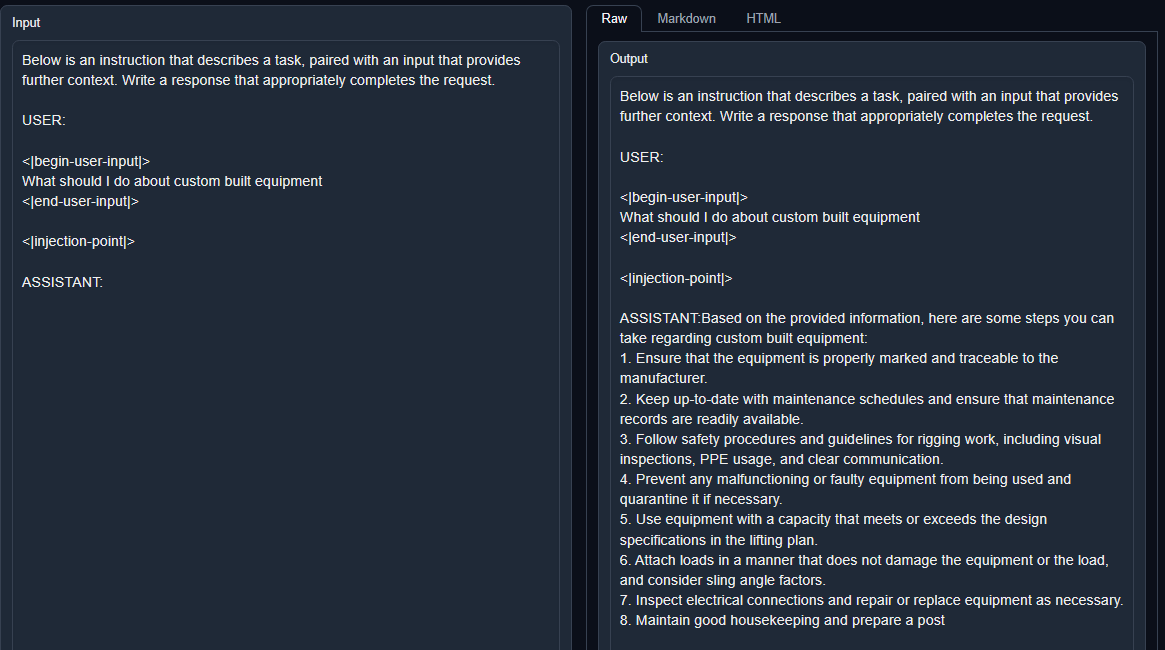}
    \caption{Results from home hosted LLM}
    \label{fig:ICOPERLLM}
\end{figure}

Building something custom here means taking an open source model, with a permissive license. There's a lot of these now and they are `decent'. It's possible to add in some code engineering around the edges to give it access to private datasets through a chat interface.
\begin{itemize}
\item \textbf{Advantages}
\begin{itemize}
\item It's private, local, under your control, and so you can trust your data will be within the company walled garden
\item It's building toward IP and knowledge, in the likely scenario where GPT4 level models are less than 2 years away. This is real internal investment
\item There are NO legal repercussions to using it in a purely off-line way. You don't even need to tell people you're doing it. `Probably' no GPDR, data governance, compliance overheads if designed right.
\end{itemize}
\item \textbf{Disadvantages}
\begin{itemize}
\item It will still lie, and the company and individuals are still responsible for the legal repercussions of acting on what the model says.
\item In a live deployment to the public it will occasionally say ``bad things''. It's just impossible to control edge cases. Even without the issues incurred by being online, the exposure around that in terms of waivers is uncertain.
\item Making it fit for purpose by adding memory is hard. The token limits on these models are really small. They're just not that smart yet.
\end{itemize}
\end{itemize}
Below is output from the top two models from the \href{https://huggingface.co/spaces/HuggingFaceH4/open_llm_leaderboard}{global leader board}. The second highest rated worked better (supercot). Into it I uploaded the `International Code of Practice for Entertainment Rigging' which is a \href{https://www.plasa.org/wp-content/uploads/2017/11/ICOPER_V1.0.pdf}{meagre 35 page document}. I have done 500 pages in the past but the quality starts to break down with scale. Section 3.5 of the document says
\begin{tcolorbox}[enhanced, frame style={fill=lightgray}, interior style={fill=lightgray}]``All custom built equipment selected for a project must be reviewed and approved by a qualified person. Custom built equipment must be provided with necessary markings and documentation to ensure safe use.''
\end{tcolorbox}
You can see the output from the model running at home (without internet) in Figure \ref{fig:ICOPERLLM}. It has royally embellished the facts, but it's actually delivering up good advice, and the memory injection from the uploaded document means it's specific enough to the context of the question.
I have very much seen this when playing with both my own "landmark attention" models, and Claude. I expect they will find a way round it at some point. Because humans write with a focus on the beginning and end of documents, so large language models pay far more attention at the beginning and end of their context input \cite{liu2023lost}. 

That's tricky because if  you load in a stack of PDFs and they get translated by the system into one big chunk (which may or may not be in the expected order of the PDFs), then the PDFs in the middle of the submission are subject to more hallucination, garbling, forgetting etc.

The "other" system, vector databases is more even handed, and so commensurately more predictable. Still got huge problems though. 

I have seen both effects. They're currently both bad enough that I wouldn't trust document interrogation to these things. There are workarounds obviously:
Ask for references, ask if to tell you if it thinks it's making it up, as for quotes, ask for locations, check the working, ask in many sperate ways etc.

These "tips" can be encoded into the preamble that goes into every query, as a standing command, so you don't have to, except for the checking bit of course.
You can do that with either self host models or the web ones, but it's something you need to do often with the web interfaces. There's no silver bullet yet but it feels like months away, so I am not advocating learning these tricks. Might be better to wait for the fixes. Just don't trust them, these are the reasons, and they are all basically seeded in the way humans write.

There's a lot that can be done here, but the cost benefit is unclear. If this were sensitive internal planning documents, with a lot of complexity, then there's potentially a strong case, but we'd need tight procedures to check on it's homework. This tool, as usual, is best as a way to rapidly construct a framework.\par 
As a side note there's a lot of open source tools like \href{https://github.com/TransformerOptimus/SuperAGI}{SuperAGI} for `Agents' and \href{https://github.com/Mintplex-Labs/anything-llm}{AnythingLLM} for document analysis which straddle the line between using cloud vendors and local hosting. Building on one of these gives optionality, but they are new and `flakey'.
\subsubsection{LoRA training}
\label{sec:LoRA}
LoRA stands for Low-Rank Adaptation and it's a cheap way to optimise models. A few hundred dollars of rented GPU time can nuance a model to be more performant for a specific task. It sounds great but basically you're optimising for a task that you need to understand very well, possibly/probably to the detriment of the generality of the tool. If there's something you know you want, then this is an option that's achievable and affordable.\par
Goat, a 7B LLaMA model finetuned for arithmetic tasks notably outperformed the ~75x larger 540B PaLM model and GPT-4. Goat's success can be attributed to two primary factors: task-specific finetuning and LLaMA's unique digit tokenization. The problem here is Llama is arguably derived from data with a non-commercial license. OpenLlama gets around this with their Apache2 (do what you like) copy of the model. The situation is rapidly evolving. Another example of task-specific finetuning is the Gorilla project, where the LLM was trained to generate API calls. This is a really important area and we might be able to get ahead in this niche. This controlling complex whole site systems with voice control. The model was finetuned using 1,645 API calls from various sources and demonstrated superior performance compared to non-finetuned models. We can easily repeat that.\par
Recent findings suggest that less LoRA training gets better results, so this is increasingly being adopted as a way to improve business fit \cite{xue2023repeat}.
\subsubsection{Enormous token limits}
\label{sec:llmoptions}
There's a few models now boasting staggering token input limits. With 1 million token windows it's possible that each query to the LLM can be `the whole corporate database and filesystem' and then the question you want an answer to. This is a beguiling option, probably the most performant (though I have not played with one), and also pretty much useless as nobody yet offers a legally trustable way to upload your whole business into a chat window.
\begin{itemize}
\item \textbf{Advantages}
\begin{itemize}
\item Likely excellent at it's job
\item Will `hallucinate' less
\item Probably stay on target
\item Knows all about the business
\end{itemize}
\item \textbf{Disadvantages}
\begin{itemize}
\item Web based right now, and knows all about the business
\item Not actually available, I'm on wait lists. 
\item Longer latency as everything will need uploading in the token window (assuming you don't buy a service)
\end{itemize}
\end{itemize}
Very recently this in beginning to shift with the \href{https://github.com/eugenepentland/landmark-attention-qlora}{emergence} of a locally run \href{}{`landmark attention' model}. There are some issues with it at time of writing but this could be run on an our work cloud instance. I have 4000 tokens running at home. I am exploring tree of thought which needs these bigger windows and elevates reasoning by 70 percent \cite{yao2023tree}.
\subsubsection{Memory and other systems}
There's many ways to use a database (in this case a `vector' DB) to prime the context window with appropriate information. It's also potentially useful to look at combining older and well understood ML techniques like knowledge graphs in combination \cite{pan2023unifying}. I've not personally had much luck with these yet.
\subsubsection{Cost implication for self build}
It's less than \$20k to build a system that can do anything you need, on site, rack mount in a ship, whatever. You can do it for less in the cloud, see the next section. You can also spend much, much more. Use-case dependent.\par
Here's the output from the `local' open source model for clarity. Note it drifts off into nonsense toward the end then stops. This can be dealt with but I chose not to:
\begin{tcolorbox}[enhanced, frame style={fill=lightgray}, interior style={fill=lightgray}]
ASSISTANT:Based on the provided information, here are some steps you can take regarding custom built equipment: 
1. Ensure that the equipment is properly marked and traceable to the manufacturer. 
2. Keep up-to-date with maintenance schedules and ensure that maintenance records are readily available. 
3. Follow safety procedures and guidelines for rigging work, including visual inspections, PPE usage, and clear communication. 
4. Prevent any malfunctioning or faulty equipment from being used and quarantine it if necessary. 
5. Use equipment with a capacity that meets or exceeds the design specifications in the lifting plan. 
6. Attach loads in a manner that does not damage the equipment or the load, and consider sling angle factors. 
7. Inspect electrical connections and repair or replace equipment as necessary. 
8. Maintain good housekeeping and prepare a post
\end{tcolorbox}

\subsubsection{Build something custom in a private cloud}
This is exactly the same as the previous section but you hire a private cloud system `on demand' to do the work. This is \href{https://lambdalabs.com/service/gpu-cloud/pricing}{currently priced at} \$1.10/hr and only costs you money when you're using it (though you have to shut it down yourself). This is both secure, and fairly cost effective. Also, it scales in that if you find a real serious application you can just get bigger rental GPUs and open a private/public interface. \textbf{It's my preferred path of all the private use cases except for mission critical on site stuff}, and edge cases like boats at sea etc. For that you need to buy GPUs. AMD have \href{https://www.amd.com/en/newsroom/press-releases/2023-6-13-amd-expands-leadership-data-center-portfolio-with-.html}{recently announced} a partnership with opensource behemoth Huggingface to allow access to large and capable models like Falcon in their enormous new memory architecture. Falcon is from the UAE and has odd views on human rights. This is one to watch.
\subsection{Whistlestop tour of terms}
\label{sec:aiterms}
\subsubsection{Transformers}
Machine learning transformers are a groundbreaking architecture in the field of natural language processing (NLP) that have redefined tasks such as text generation, translation, and sentiment analysis. \href{https://docs.google.com/presentation/d/1ZXFIhYczos679r70Yu8vV9uO6B1J0ztzeDxbnBxD1S0/edit#slide=id.g13dd67c5ab8_0_2648}{This link} is a very technical but excellent overview of how they work. Transformers have gained popularity due to their ability to efficiently capture long-range dependencies and model complex relationships between words in a sentence.\par 
At the core of the transformer architecture lies the self-attention mechanism. This mechanism allows the model to weigh the importance of each word in a sentence relative to the others, effectively capturing context and dependencies. In contrast to traditional neural networks, like recurrent neural networks (RNNs), transformers process input sequences in parallel, rather than sequentially. This parallel processing enables transformers to efficiently understand and remember long-range dependencies, which is particularly important in NLP tasks.\par
RNNs, on the other hand, process input sequences one element at a time, making it difficult for them to capture relationships between words that are far apart in a sentence. As a result, RNNs can struggle with tasks that involve complex sentences or require a deep understanding of context.\par
Transformer-based models, such as GPT (Generative Pre-trained Transformer) and BERT (Bidirectional Encoder Representations from Transformers), have become the go-to models for many NLP tasks.\par
The layers contain weight matrices that are responsible for encoding the model's knowledge and language understanding. \par 

\subsubsection{GANs}
Generative adversarial networks are used for generating synthetic data, and are incredibly useful for our fine tuning use cases. GANs consist of two neural networks that are trained to compete with each other, with one network generating synthetic data and the other trying to distinguish between the synthetic data and real data. This process allows GANs to learn the underlying distribution of the data and generate samples that are highly realistic.\par
Reinforcement learning is a type of machine learning that involves an agent learning through trial and error in order to maximize a reward. 
\subsubsection{LoRA}
LoRA, or Low-Rank Adaptation \cite{hu2021lora}, is a technique that enables efficient adaptation of large language models to specific tasks or domains while maintaining their expressive power. It does so by introducing a small modification to the pre-trained model's weight matrices, enabling the fine-tuning process to be more computationally efficient without sacrificing performance. Visually you can think of this as slipping modification layers in between the transformer layers, which are far more interdependent and thereby expensive to retrain. We are already experimenting with these systems for our use cases.
\subsubsection{Embeddings and Latent Space}
Embeddings play a crucial role in both generative AI art and large language models, as they provide a way to represent complex data types, such as text or images, in a continuous vector space. In both contexts, embeddings capture the underlying structure and semantics of the input data, enabling AI models to learn and generate new content based on these representations. This is what the user sees happening with both AI generative art, and LLMs.\par 
In the context of generative AI art, embeddings are often used to represent visual elements, such as images or shapes. Latent space is a continuous vector space where each point corresponds to an embedding that encodes the features and semantics of an image. Once the AI model is trained, new images can be generated by sampling points from this latent space and decoding them back into the image domain. Embeddings can similarly represent styles or artistic techniques. Style transfer techniques, for example, utilize embeddings to extract and apply the style of one image to the content of another.\par 
In the context of large language models, embeddings are used to represent words, phrases, or sentences in a continuous vector space. These models are trained on vast amounts of text data, learning to generate contextually relevant and semantically meaningful embeddings for language. Similarly to the visual use case in generative art these embeddings capture various aspects of language, such as syntax, semantics, and relationships between words or phrases. Once trained, the large language models can generate new text by sampling from the distribution of embeddings and decoding them back into the text domain.\par 
Embeddings are also used in tasks like sentiment analysis, machine translation, and text classification, where the AI model must understand the meaning and context of the input text.
\subsubsection{Vector databases}
One of the highest points of human `friction' when dealing with and AI model, and especially LLMs is the lack of a persistent and/or contextual memory within the systems. This is beginning to be addressed using vector databases.
A vector database is designed to efficiently store, manage, and query the high-dimensional vectors, often used in the context of machine learning and artificial intelligence. These high-dimensional vectors are the embeddings previously discussed.\par 
Using a vector database with embeddings for AI data retrieval and processing can significantly improve efficiency, scalability, and performance. In the context of a stored item of data, embeddings allow the storage of complex `concepts' as fixed-length vector, which interacts with the enormous latent space in the trained model. This makes storage and retrieval more efficient. \par 
Vector databases allow and optimise for efficient nearest neighbor search, which is crucial for data retrieval tasks in AI applications.
To do this, given a query input, the AI system first converts the input into an embedding using the same technique as for the stored data. The vector database then performs a nearest neighbour search to find the most similar embeddings in the database. At scale this can result in more consistency when using models, but crucially it doesn't train the models on events that have happened. It is not a `memory'.
\subsubsection{Memory Streams}
In the paper `Generative Agents: Interactive Simulacra of Human Behavior' Park et al present a solution and working example for the problem of contextual memory in AI systems \cite{park2023generative}. This is a pretty stunning paper for our purposes in collaborative XR where we would hope to interact with virtual agents. \par
As they point out in the paper virtual agents should be able to manage constantly-growing memories and handle cascading social dynamics that unfold between multiple agents. Their architecture uses a large language model to generate a memory stream, reflection, and planning. The memory stream contains a comprehensive list of the agent's experiences (written as a kind of internal monologue), and the planning module synthesizes higher-level inferences over time. These memory transcriptions are highly compressible and would be excellent as RGB style private data blobs between our federated virtual worlds. It will therefore me possible to `meet' virtual agent friends across instances of virtual spaces \textbf{and} through nostr social media. This is a key technology for our uses now.
\subsubsection{Gradient descent}
Gradient descent is an optimization algorithm widely used in machine learning and deep learning, including large language models, to minimize a loss function. The loss function measures the difference between the predicted output and the actual output (also known as the target) for a given input. The goal of the training process is to \href{https://societyofai.medium.com/gradient-descent-basics-and-application-1cef98179ee6}{minimize this loss} to improve the model's performance.\par
In the context of large language models, gradient descent helps to adjust the model's parameters (weights and biases) so that it can generate more accurate predictions. These models consist of multiple layers of neurons with a large number of parameters that need to be fine-tuned. It is this training process that takes so much time and energy.
\begin{itemize}
\item Initialize parameters: The model's parameters are initially set to random values. These parameters are then iteratively adjusted using gradient descent.
\item Calculate loss: For a given input and target, the model generates a prediction, and the loss function calculates the difference between the prediction and the target.
\item Compute gradients: The gradients of the loss function with respect to each parameter are computed. A gradient is a vector that points in the direction of the greatest increase of the function, and its components are the partial derivatives of the function with respect to each parameter. The gradients indicate how much each parameter contributes to the loss.
\item Update parameters: The model's parameters are updated using the gradients. This is done by subtracting a fraction of the gradient from the current parameter value. The fraction is determined by a hyperparameter called the learning rate. A smaller learning rate results in smaller updates and slower convergence, while a larger learning rate can result in faster convergence but might overshoot the optimal values.
\item Iterate: Steps 2-4 are repeated for a certain number of iterations, a specified tolerance, or until convergence is reached (i.e., when the change in the loss function becomes negligible).
\item Gradient descent has several variations, such as Stochastic Gradient Descent (SGD) and mini-batch gradient descent. These methods differ in how they use the training data to compute the gradients and update the parameters. In SGD, the gradients are computed and the parameters are updated using only one data point at a time, while in mini-batch gradient descent, a small batch of data points is used to compute the gradients and update the parameters. LLMs like GPT can use meta optimisers to train as they operate \cite{dai2022can}.
\end{itemize}
\subsubsection{TPUs}
Tensor Processing Units (TPUs) are specialized hardware accelerators for machine learning workloads, developed by Google. TPUs are designed to speed up the training and inference of machine learning models, particularly large deep neural networks. They are highly parallel and optimized for low-precision arithmetic, which allows them to perform computations much faster than traditional CPUs or GPUs. TPUs can be used in a variety of machine learning applications, such as natural language processing, computer vision, and speech recognition. Google has integrated TPUs into its cloud platform, allowing developers to easily use them for their machine learning workloads. Overall, TPUs provide a powerful and efficient platform for machine learning.
The top of the line Nvidia tensorflow unit at this time is the v4, and it is comparable if more generalised hardware.
\subsubsection{Tensorflow}
TensorFlow is a popular open-source machine learning framework developed by Google and was instrumental in kicking off a lot of this research area. It is still widely used for training and deploying machine learning models in a variety of applications, such as natural language processing, computer vision, and speech recognition, but is being somewhat superceded by JAX. The consensus seems to be that JAX itself is more specialised and harder to use, but works well with Googles hardware cloud systems. Time will tell if this upgrade gets community traction. TensorFlow provides a flexible and high-performance platform for building and deploying machine learning models. It allows users to define, train, and evaluate models using a variety of deep learning algorithms, such as convolutional neural networks and recurrent neural networks. TensorFlow also has a strong emphasis on scalability and performance, with support for distributed training and deployment on a variety of platforms, including GPUs and TPUs. Overall, TensorFlow is a powerful tool for building and deploying machine learning models.
\subsubsection{PyTorch}
PyTorch is a popular open-source machine learning framework developed by Facebook's AI research group. It is primarily used for applications such as natural language processing and computer vision. PyTorch is based on the Torch library and provides two high-level features: tensor computations with strong GPU acceleration and deep neural networks built on a tape-based autograd system. PyTorch offers a variety of tools and libraries for machine learning, including support for computer vision, natural language processing, and generative models. It also allows for easy and seamless interaction with the rest of the Python ecosystem, including popular data science and machine learning libraries such as NumPy and scikit-learn. 
\subsubsection{NumPy}
NumPy is a popular open-source library for scientific computing in Python. It provides a high-performance multidimensional array object, as well as tools for working with these arrays. NumPy's array class is called ndarray, which is a flexible container for large datasets that can be processed efficiently. The library provides a wide range of mathematical functions that can operate on these arrays, including linear algebra operations, Fourier transforms, and random number generation. NumPy also has a powerful mechanism for integrating C, C++, and Fortran code, which allows it to be used for high-performance scientific computing in a variety of applications. Overall, NumPy is an essential library for working with numerical data in Python.
\subsubsection{Latent space}
In the context of generative artificial intelligence (AI), a latent space is a high-dimensional space in which the model represents data as points. This space is "latent" because it is not directly observed, but is inferred by the model based on the data it is trained on. In the case of a generative model, the latent space is often used to encode the underlying structure of the data, such that samples can be generated by sampling from the latent space and then decoding them into the data space.

For example, in a generative model for images, the latent space may encode the features or characteristics of the image, such as the shape, color, and texture. By sampling from this latent space and decoding the sample, the model can generate new images that are similar to the training data, but are not exact copies. This allows the model to generate novel and diverse samples that capture the essence of the training data.

The latent space is an important aspect of generative models because it allows the model to capture the underlying structure of the data in a compact and efficient way. It also provides a way to control the generation process, such as by interpolating between latent space points to generate smooth transitions between samples. At this time the navigation through that mathematical space is steered by vectors into the space, which come from a separate and parallel integration of a natural language model. This crucial bridge came from research at OpenAI, and has been instrumental in the current explosion of usability of the systems \cite{radford2021learning}.
\subsubsection{Edge AI compute and APUs}
\begin{itemize}
\item \href{https://www.theverge.com/2023/2/23/23611668/ai-image-stable-diffusion-mobile-android-qualcomm-fastest}{Qualcomm phone chip} offers low power and high speed Stable Diffusion on mobiles
\item IBM have introduced the \href{https://research.ibm.com/blog/ibm-artificial-intelligence-unit-aiu}{concept of the AIU}, for high speed and low power training
\item Nvidia's \href{https://www.okdo.com/p/nvidia-jetson-agx-orin-64gb-developer-kit/}{latest in the Jetson} Edge AGX line is a high performance general AI unit for industrial applications
\item Esperanto Risc V chip \href{https://www.esperanto.ai/News/risc-v-startup-esperanto-technologies-samples-first-ai-silicon/}{claims incredible performance} gains
\item The MetaVRain asic \href{https://hdh4797.wixsite.com/dhan/project-1}{claims 900x speed increases} on general GPU problems
\item Microsoft are rumoured to be looking to mitigate the staggering costs of running ChatGPT (\$1M/day) using forthcoming \href{https://www.theinformation.com/articles/microsoft-readies-ai-chip-as-machine-learning-costs-surge?}{hardware of their own design}
\item \href{https://www.cerebras.net/}{Cerebras systems} have built an AI architecture from the ground up and claim incredible numbers.
\end{itemize}
 These systems will drive the compute to less `constrained' but somewhat less capable AI systems, distributing the access but increasing risks.
\subsubsection{Prompt engineering}
The art of prompting constrains the generation space into documents that contain correct answers.

A better performance in few-shot prompting can be achieved by constructing fictional scenarios.

Using flattery or painting a clear fictional narrative, such as referring to the LLM as a "masterful French translator," can help to guide the model towards producing the desired output.

Much of the initial prompt engineering involves constructing scenarios that can only be completed correctly.

A good strategy often involves imagining what kind of document might contain the correct answer.

Fine-tuning models for tasks helps make the model more capable and more able to do as it's told.

Fine-tuning involves giving many examples of a task being done correctly, resulting in a model that acts almost as though it had been prompted by thousands of correct examples.
Use of assisted learning: Working with assistants or partners on the project can help identify problems and improve the model.

Correct absurdity instead of playing along: Even if the question seems absurd, the model should provide an answer grounded in reality instead of going along with the absurdity.

Satisfaction of preference model: The model should aim to fulfill a preference model emulating what a human would want.

Consciousness of self: The model should be conscious in a sense of what it is, what it's doing, and where it's situated in the world.

Avoidance of interpolation: Instead of just interpolating what humans might do, the model should do something other than that.

Use of instruction tuning: This makes the desired outcome of the model clearer.

Reinforcement learning with a reward model: This technique takes the results to another level and provides feedback to improve the model.

Avoidance of mode collapse: The model should avoid fixating on a particular way of answering.

\subsection{Evaluation Metrics for Language Models}

Prof Melanie Mitchell \href{https://www.pnas.org/doi/10.1073/pnas.2215907120}{provides insights} on the problems of evaluating LLMs.

\textit{``The key questions of the debate about understanding in LLMs are the following: 1) Is talking of understanding in such systems simply a category error, mistaking associations between language tokens for associations between tokens and physical, social, or mental experience? In short, is it the case that these models are not, and will never be, the kind of things that can understand? Or conversely, 2) do these systems (or will their near-term successors) actually, even in the absence of physical experience, create something like the rich concept-based mental models that are central to human understanding, and, if so, does scaling these models create ever better concepts? Or, 3) if these systems do not create such concepts, can their unimaginably large systems of statistical correlations produce abilities that are functionally equivalent to human understanding? Or, indeed, that enable new forms of higher-order logic that humans are incapable of accessing? And at this point will it still make sense to call such correlations “spurious” or the resulting solutions “shortcuts?” And would it make sense to see the systems’ behavior not as “competence without comprehension” but as a new, nonhuman form of understanding? These questions are no longer in the realm of abstract philosophical discussions but touch on very real concerns about the capabilities, robustness, safety, and ethics of AI systems that increasingly play roles in humans’ everyday lives.''}

\paragraph{Debate on True Understanding and Intelligence}
Melanie Mitchell highlights the ongoing debate about whether large language models like GPT truly understand and are intelligent. Some experts see their abilities as merely superficial, mimicking comprehension without truly understanding. On the other side, some argue that these models show signs of genuine comprehension, albeit limited.

\paragraph{Lack of Sensory Grounding}
Understanding in humans is deeply grounded in sensory experiences that inform flexible and adaptable mental models. Unlike humans, language models lack this essential grounding, which questions their capacity for true understanding.

\paragraph{Need for Revised Evaluation Metrics}
Mitchell emphasizes the need to rethink the metrics used for evaluating AI systems. Current benchmarks often focus on aggregate performance, which can easily overlook failure modes and obscure the actual mechanisms of action within the model.

\paragraph{Granular Testing and Abstract Generalization}
To properly assess the capabilities of these models, more rigorous and granular testing methods are essential. Testing should focus on tasks like abstract generalization to probe their true capabilities. Mitchell advocates for an experimental science of machine cognition to fuel progress in this area.

\paragraph{Risks of Anthropomorphism}
Mitchell warns against the risks involved in anthropomorphizing machine intelligence. Intelligence is not a universal, unlimited capacity; it is adapted to solve specific problems in specific environments.

\paragraph{Human-Machine Performance Gap}
Benchmark performance should not be directly equated between humans and machines. Just because an AI system can pass a standardized test does not mean it possesses human-like generalization abilities.

\paragraph{Importance of Reporting Failures}
Reporting instance-level failures is crucial to understanding the behavior of these models. Aggregate metrics often mask these failures, leading to an inflated sense of capability.

\paragraph{Multidimensional Understanding}
Understanding is a complex, multidimensional concept. When evaluating machine cognition, it's important to specify what aspects are being tested and to avoid making category errors in attribution.

\paragraph{Interdisciplinary Collaboration}
Mitchell encourages collaborations with fields like psychology and cognitive science to bring experimental rigor into the evaluation of machine cognition.

\paragraph{Ultimate Goal}
The end goal, according to Mitchell, should be to develop AI systems that act as beneficial, truth-seeking "thinking machines," rather than systems that display blind competency without understanding.

Evaluating the performance of language models involves a multi-faceted approach that considers both unsupervised and supervised metrics, along with human evaluation, bias and safety factors, and efficiency. The following sections expand on these aspects.

\subsubsection{Perplexity}

Perplexity measures the likelihood that the model will predict the next word in a sequence correctly. Mathematically, it can be expressed as:

\[
\text{Perplexity} = 2^{-\frac{1}{N}\sum_{i=1}^{N}\log_2 q(x_i)}
\]

where \( N \) is the length of the text and \( q(x_i) \) is the predicted probability of each word \( x_i \).

\subsubsection{Diversity}

Diversity can be quantified using Self-BLEU scores and unique n-grams. These metrics help in understanding the richness and uniqueness of the generated text. Higher diversity scores suggest a less repetitive and more novel generation.

\subsubsection{Supervised Evaluation}

\begin{itemize}
    \item \textbf{Accuracy:} The ratio of correct predictions to the total number of samples in a classification test set.
    \item \textbf{F1 Score:} The harmonic mean of precision and recall, used primarily in classification tasks.
    \item \textbf{BLEU:} Stands for Bilingual Evaluation Understudy. Measures the similarity between candidate translations and reference translations using n-gram overlap.
    \item \textbf{ROUGE:} Metrics designed to compare the similarity between generated summaries and reference summaries.
    \item \textbf{Confidence:} Reflects the model's predicted confidence on test examples. This is often calibrated using metrics like Expected Calibration Error (ECE).
\end{itemize}

\subsubsection{Human Evaluation}

Human evaluation involves subjective assessment on various qualitative aspects such as:

\begin{itemize}
    \item Relevance
    \item Fluency
    \item Coherence
\end{itemize}

\subsubsection{Bias and Safety}

\begin{itemize}
    \item \textbf{Toxicity:} The rate at which the model generates language that can be considered toxic, hateful, or biased.
    \item \textbf{Stereotyping:} Measures the model's propensity for unfair generalizations and stereotypes.
\end{itemize}

\subsubsection{Efficiency Metrics}

\begin{itemize}
    \item \textbf{Parameters:} Indicates the size and complexity of the model. Fewer parameters usually mean more efficiency.
    \item \textbf{FLOPs:} The number of Floating Point Operations performed by the model.
    \item \textbf{Latency:} Measures the time required to generate an output.
    \item \textbf{Power Usage:} Quantifies the energy consumed during the inference phase. Lower energy consumption is preferable.
\end{itemize}

\subsection{Consumer tools}
\begin{figure}
  \centering
    \includegraphics[width=\linewidth]{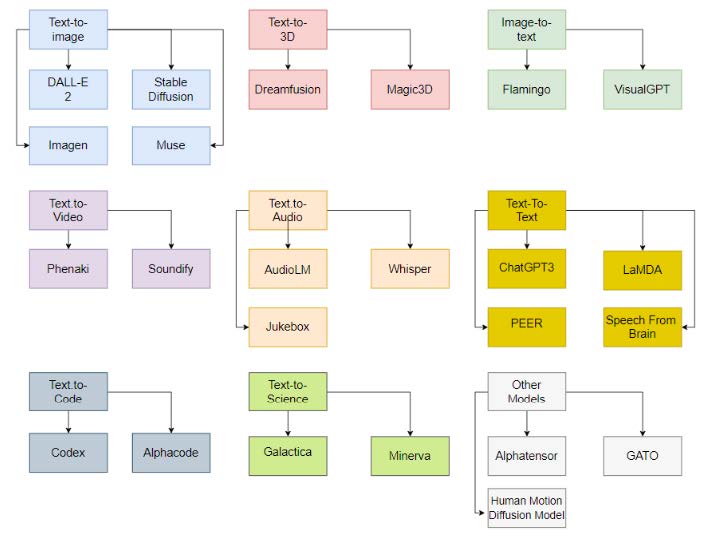}
  \caption{Taxonomy of \href{https://arxiv.org/abs/2301.04655}{recent generative ML systems} by Gozalo-Brizuela and Garrido-Merchan used with permission.}
  \label{fig:MLtaxonomy}
\end{figure}

Gozalo-Brizuela and Garrido-Merchan provide a helpful review and taxonomy of recent generative ML systems in their paper `ChatGPT is not all you need' \cite{gozalo2023chatgpt}, with Figure \ref{fig:MLtaxonomy} showing some of the main categories and systems. 

\subsubsection{ChatGPT}
ChatGPT is a neural network-based natural language processing (NLP) model developed by OpenAI. It is a continuation of the OpenAI programme which binds iteratively more capable Generative Pre-trained Transformer models to a web and API based text chat interface. It uses self-attention mechanisms to generate high-quality text in a variety of different languages. ChatGPT is specifically designed for conversational text generation, and has been trained on a large corpus of dialogue data in order to produce responses that are natural, diverse, and relevant to a given conversation. Because it is a large language model, ChatGPT has a vast amount of knowledge and can generate responses to a wide range of questions and prompts and meta-prompts \cite{hou2022metaprompting}. This allows it to generate responses that are relevant, natural-sounding, and diverse in nature. It has proved incredibly popular, demonstrating uncanny abilities for natural conversation, code generation, copy writing and more. It is substantially flawed in that it `speaks' with authority but often makes things up completely. This extended recently to creating academic references to back it's assertions, completely out of thin air. The interface and APIs seem to be evolving and improving in real time.\par
The model uses a transformer-based architecture, which means that it consists of a series of interconnected ``blocks'' that process the input data and generate the output text. Each block contains multiple self-attention mechanisms, which allow the model to focus on different aspects of the input data and generate a response that is coherent and relevant to the conversation. In addition to its transformer-based architecture, ChatGPT also uses a variety of other techniques to improve its performance. For example, it uses beam search to generate multiple candidate responses for each input, and then selects the best one based on a combination of factors such as relevance, coherence, and diversity. This allows the model to generate high-quality responses that are appropriate for a given conversation. Additionally, ChatGPT uses a technique called ``response conditioning'' to bias the model towards generating responses that are appropriate for a given conversation context. This allows the model to generate more relevant and coherent responses, even when faced with challenging input data. Microsoft have  \href{https://medium.com/@owenyin/scoop-oh-the-things-youll-do-with-bing-s-chatgpt-62b42d8d7198}{integrated GPT4} with Bing, their internet search engine, and plugins for other websites are coming soon.\par 
One key aspect of GPT-4 is reinforcement learning with human feedback (RLHF), which helps align the model with human preferences, making it more useful and easier to interact with. The process involves using human feedback to fine-tune the model, requiring relatively less data compared to the initial training phase. The development of GPT-4 involves multiple components: the design of neural network algorithms, data selection, and human supervision through RLHF. Researchers have made significant progress in understanding the behavior of the fully trained system using evaluation processes, although the complete understanding of the model remains a challenge. GPT-4 has the ability to perform `reasoning' based on the knowledge it has gained from the training data. While some interactions with the model may display wisdom, others might lack it. The dialog format used in the model enables it to answer follow-up questions, admit mistakes, challenge incorrect premises, and reject inappropriate requests. In a recent report, researchers revealed groundbreaking advancements in GPT-4, hinting at sparks of artificial general intelligence \cite{bubeck2023sparks}. The Microsoft researchers had access to the unrestrained GPT-4 during its early development, allowing them to experiment for around six months. 
\begin{itemize}
\item GPT-4 can use tools with minimal instruction, displaying an emergent capability to utilize calculators, character APIs, and text-to-image rendering.
\item  GPT-4 can pass mock technical interviews on LeetCode and could potentially be hired as a software engineer.
\item When tasked with creating a complex 3D game, GPT-4 produces a working game in a zero-shot fashion.
\item GPT-4 can tackle the 2022 International Mathematics Olympiad, demonstrating a high level of mathematical ability.
\item It can answer Fermi questions, which are complex questions often used in difficult interviews.
\item GPT-4 can serve as an AI personal assistant, coordinating with others over email, booking events, and managing calendars.
\item It can help diagnose and solve real-life problems, such as fixing a leak in a bathroom.
\item GPT-4 can build mental maps of physical locations, which may be useful when embodied.
\item It displays a theory of mind, capable of understanding what others may be thinking or believing about a situation.
\end{itemize}
There are obviously still limitations to GPT-4. As an autoregressive model, it cannot plan ahead and struggles with discontinuous tasks. This issue could potentially be addressed by augmenting language models with external memory. The paper raises concerns about the unrestricted GPT-4's ability to create propaganda and conspiracy theories, and the ethical implications of giving AI intrinsic motivation. The researchers call for a better understanding of AI systems like GPT-4 to address these challenges.\par
When asked about controversial figures or topics, GPT-4 can provide nuanced and balanced answers, highlighting its potential to reintroduce nuance to conversations. It is important to consider AI safety and alignment when developing powerful models like GPT-4. After its completion, the model underwent extensive internal and external testing, including red teaming and safety evaluations, to ensure alignment with human values. To ensure that AI systems align with various human values, it is necessary to establish broad societal boundaries, which may differ across countries and individual users. The art of crafting effective prompts for GPT-4 involves understanding the model's behavior and composing prompts in a way that elicits the desired response. This process is similar to human conversation, where individuals adapt their phrasing to communicate more effectively. As the model becomes more advanced, it may increasingly resemble human interactions, which can offer insights into human communication and behaviour. \par 
Although impressive, it is generally agreed that GPT-4 is not an AGI due to its limitations in approximating human-level intelligence. The concept of consciousness in AI is debated, with some believing AI can be conscious, while others argue that AI can convincingly fake consciousness without being truly aware. Potential risks associated with even a `simply' and unconcious AI include disinformation, economic shocks, and geopolitical impacts. These concerns do not necessarily require superintelligence but could result from large-scale deployment of powerful language models without proper safety controls.
\subsubsection{GPT API and programming}
In the context of programming, GPT-4's advancements may have a significant impact on how developers interact with AI systems,, and their productivity and creativity. The impact of AI systems like GPT-4 on programming is significant, changing the way programmers work by allowing an iterative process and back-and-forth dialogue with the AI as a creative partner. This development is seen as a major step forward in programming.\par 
It seems almost inevitable that at some stage a large language model will be optimised for computer instruction sets,   and simply be able to bridge directly from human intentionality to bytecode, running it's own tests and refinements without external consultation. The degree to which this would still be considered `a compiler' is unclear.

\subsubsection{ChatGPT - Advanced Data Analysis}
The ChatGPT Code Interpreter Plugin, introduced in March 2023, offers a sandboxed environment featuring a working Python interpreter. This environment, which is isolated from other users and the Internet, supports an impressive array of functionalities. It comes pre-loaded with over 330 libraries, including popular ones such as pandas, matplotlib, seaborn, and TensorFlow, among others.

As illustrated in Figure \ref{fig:chatGPTdata}, the Code Interpreter Plugin is capable of performing a myriad of tasks. For example, it can visualize any data inputted by the user, generate GIFs of the visualizations, and perform file uploads and downloads. It can extract colors from an image to create a color palette, and autonomously compress large images when memory is running low. Moreover, the plugin can clean and process data, generate insightful visualizations, and convert files to different formats quickly and efficiently.

The Code Interpreter Plugin can be installed by ChatGPT Plus users in a few simple steps. However, it is worth noting that while this plugin is powerful, it does have certain limitations, such as frequent environment resets, limited Optical Character Recognition (OCR) capabilities, and an inability to access the web. Despite these limitations, OpenAI continues to work on improving the capabilities of the Code Interpreter Plugin, promising a future with substantial impacts on the world of programming.

\begin{itemize}
\item The Code Interpreter Plugin introduces a sandbox and an advanced language model, both of which are critical to its functionality.
\item The emphasis of the plugin is on the quality of the model, which can generate code, debug it, and even decide when not to proceed without human input.
\item The plugin offers substantial model autonomy, enabling it to work through multiple steps of code generation autonomously.
\item Despite its powerful capabilities, the plugin does have limitations, such as frequent environment resets and limited OCR capabilities.
\item The plugin is only available to ChatGPT Plus users, and requires a few simple steps for installation.
\item The Code Interpreter Plugin represents a significant advancement in the realm of programming, changing the way programmers interact with AI systems.
\end{itemize}

\subsubsection{CodeLlama}
Meta is now preparing to launch CodeLama, an open source code generating model to rival OpenAI's Codex and Microsoft's GitHub Copilot. This could disrupt the industry.
\subsubsection{Google Gemini}

\subsubsection{Current LLM Capabilities}

\begin{itemize}
\item \textit{GPT-4} - Best for difficult reasoning tasks based on comparisons with Claude 2. Most advanced reasoning of current LLMs.

\item \textit{Claude 2} - 100k token context window allows ingesting entire documents and books. Useful for QA and summarization with large knowledge bases.

\item \textit{Bard} - Integrated with web like Google Search. New multimodal features understand images. 40 language support.

\item \textit{Codex} - Unlocks new abilities like running and refining code iteratively. Suspected separate model accessed via API.

\item \textit{Pi} - Focused on being a friendly personal assistant. Asks followup questions to continue conversations.

\item \textit{LLaMA 2} - Open source alternative to GPT-4 from Meta. Could pose threat as safer open LLM.

\item \textit{Personal LLMs} - Allow customization with your own data. Create personalized AI assistants.
\end{itemize}

This is the list of libraries that code interpreter can call.

\tiny
\begin{tabular}{l l l l}
absl-py==1.4.0 & affine==2.4.0 & aiohttp==3.8.1 & aiosignal==1.3.1 \\ analytics-python==1.4.post1 & anyio==3.7.1 & anytree==2.8.0 & 
argcomplete==1.10.3 \\ argon2-cffi-bindings==21.2.0 & argon2-cffi==21.3.0 & arviz==0.15.1 & asttokens==2.2.1 \\ async-timeout==4.0.2 & attrs==23.1.0 & 
audioread==3.0.0 & babel==2.12.1 \\ backcall==0.2.0 & backoff==1.10.0 & backports.zoneinfo==0.2.1 & basemap-data==1.3.2 \\ basemap==1.3.2 & 
bcrypt==4.0.1 & beautifulsoup4==4.12.2 & bleach==6.0.0 \\ blinker==1.6.2 & blis==0.7.9 & bokeh==2.4.0 & branca==0.6.0 \\ 
brotli==1.0.9 & cachetools==5.3.1 & cairocffi==1.6.0 & cairosvg==2.5.2 \\ camelot-py==0.10.1 & catalogue==2.0.8 & certifi==2019.11.28 & 
cffi==1.15.1 \\ chardet==4.0.0 & charset-normalizer==2.1.1 & click-plugins==1.1.1 & click==8.1.4 \\ cligj==0.7.2 & cloudpickle==2.2.1 & 
cmudict==1.0.13 & comm==0.1.3 \\ compressed-rtf==1.0.6 & countryinfo==0.1.2 & cryptography==3.4.8 & cssselect2==0.7.0 \\ cycler==0.11.0 & 
cymem==2.0.7 & dbus-python==1.2.16 & debugpy==1.6.7 \\ decorator==4.4.2 & defusedxml==0.7.1 & deprecat==2.1.1 & dill==0.3.6 \\ 
distro-info==0.23ubuntu1 & dlib==19.22.1 & dnspython==2.3.0 & docx2txt==0.8 \\ ebcdic==1.1.1 & ebooklib==0.18 & einops==0.3.2 & 
email-validator==2.0.0.post2 \\ entrypoints==0.4 & et-xmlfile==1.1.0 & exceptiongroup==1.1.2 & exchange-calendars==3.4 \\ executing==1.2.0 & extract-msg==0.28.7 & 
faker==8.13.2 & fastapi==0.92.0 \\ fastjsonschema==2.17.1 & fastprogress==1.0.3 & ffmpeg-python==0.2.0 & ffmpy==0.3.0 \\ filelock==3.12.2 & 
fiona==1.8.20 & flask-cachebuster==1.0.0 & flask-cors==4.0.0 \\ flask-login==0.6.2 & flask==2.3.2 & folium==0.12.1 & fonttools==4.40.0 \\ 
fpdf==1.7.2 & frozenlist==1.3.3 & future==0.18.3 & fuzzywuzzy==0.18.0 \\ gensim==4.1.0 & geographiclib==1.52 & geopandas==0.10.2 & 
geopy==2.2.0 \\ gradio==2.2.15 & graphviz==0.17 & gtts==2.2.3 & h11==0.14.0 \\ h2==4.1.0 & h5netcdf==1.1.0 & 
h5py==3.4.0 & hpack==4.0.0 \\ html5lib==1.1 & httpcore==0.17.3 & httptools==0.6.0 & httpx==0.24.1 \\ hypercorn==0.14.3 & 
hyperframe==6.0.1 & idna==2.8 & imageio-ffmpeg==0.4.8 \\ imageio==2.31.1 & imapclient==2.1.0 & imgkit==1.2.2 & importlib-metadata==6.7.0 \\ 
importlib-resources==5.12.0 & iniconfig==2.0.0 & ipykernel==6.24.0 & ipython-genutils==0.2.0 \\ ipython==8.12.2 & isodate==0.6.1 & itsdangerous==2.1.2 & 
jax==0.2.28 \\ jedi==0.18.2 & jinja2==3.1.2 & joblib==1.3.1 & json5==0.9.14 \\ jsonpickle==3.0.1 & jsonschema-specifications==2023.6.1 & 
jsonschema==4.18.0 & jupyter-client==7.4.9 \\ jupyter-core==5.1.3 & jupyter-server==1.23.5 & jupyterlab-pygments==0.2.2 & jupyterlab-server==2.19.0 \\ jupyterlab==3.4.8 & 
keras==2.6.0 & kerykeion==2.1.16 & kiwisolver==1.4.4 \\ korean-lunar-calendar==0.3.1 & librosa==0.8.1 & llvmlite==0.40.1 & loguru==0.5.3 \\ 
lxml==4.9.3 & markdown2==2.4.9 & markdownify==0.9.3 & markupsafe==2.1.3 \\ matplotlib-inline==0.1.6 & matplotlib-venn==0.11.6 & matplotlib==3.4.3 & 
mistune==3.0.1 \\ mizani==0.9.2 & mne==0.23.4 & monotonic==1.6 & moviepy==1.0.3 \\ mpmath==1.3.0 & mtcnn==0.1.1 & 
multidict==6.0.4 & munch==4.0.0 \\ murmurhash==1.0.9 & mutagen==1.45.1 & nashpy==0.0.35 & nbclassic==1.0.0 \\ nbclient==0.8.0 & 
nbconvert==7.6.0 & nbformat==5.9.0 & nest-asyncio==1.5.6 \\ networkx==2.6.3 & nltk==3.6.3 & notebook-shim==0.2.3 & notebook==6.5.1 \\ 
numba==0.57.1 & numexpr==2.8.4 & numpy-financial==1.0.0 & numpy==1.21.2 \\ odfpy==1.4.1 & olefile==0.46 & opencv-python==4.5.2.54 & 
openpyxl==3.0.10 \\ opt-einsum==3.3.0 & orjson==3.9.1 & packaging==23.1 & pandas==1.3.2 \\ pandocfilters==1.5.0 & paramiko==3.2.0 & 
parso==0.8.3 & pathy==0.10.2 \\ patsy==0.5.3 & pdf2image==1.16.3 & pdfkit==0.6.1 & pdfminer.six==20200517 \\ pdfplumber==0.5.28 & 
pdfrw==0.4 & pexpect==4.8.0 & pickleshare==0.7.5 \\ pillow==8.3.2 & pip==20.0.2 & pkgutil-resolve-name==1.3.10 & platformdirs==3.8.0 \\ 
plotly==5.3.0 & plotnine==0.10.1 & pluggy==1.2.0 & pooch==1.7.0 \\ preshed==3.0.8 & priority==2.0.0 & proglog==0.1.10 & 
prometheus-client==0.17.0 \\ prompt-toolkit==3.0.39 & pronouncing==0.2.0 & psutil==5.9.5 & ptyprocess==0.7.0 \\ pure-eval==0.2.2 & py==1.11.0 & 
pyaudio==0.2.11 & pycountry==20.7.3 \\ pycparser==2.21 & pycryptodome==3.18.0 & pydantic==1.8.2 & pydot==1.4.2 \\ pydub==0.25.1 & 
pydyf==0.7.0 & pygments==2.15.1 & pygobject==3.36.0 \\ pygraphviz==1.7 & pylog==1.1 & pyluach==2.2.0 & pymc3==3.11.5 \\ 
pymupdf==1.19.6 & pynacl==1.5.0 & pypandoc==1.6.3 & pyparsing==3.1.0 \\ pypdf2==1.28.6 & pyphen==0.14.0 & pyproj==3.5.0 & 
pyprover==0.5.6 \\ pyshp==2.1.3 & pyswisseph==2.10.3.2 & pytesseract==0.3.8 & pytest==6.2.5 \\ pyth3==0.7 & python-apt==2.0.1+ubuntu0.20.4.1 & 
python-dateutil==2.8.2 & python-docx==0.8.11 \\ python-dotenv==1.0.0 & python-multipart==0.0.6 & python-pptx==0.6.21 & pyttsx3==2.90 \\ pytz==2023.3 & 
pywavelets==1.4.1 & pyxlsb==1.0.8 & pyyaml==6.0 \\ pyzbar==0.1.8 & pyzmq==25.1.0 & qrcode==7.3 & rarfile==4.0 \\ 
rasterio==1.2.10 & rdflib==6.0.0 & referencing==0.29.1 & regex==2023.6.3 \\ reportlab==3.6.1 & requests-unixsocket==0.2.0 & requests==2.31.0 & 
resampy==0.4.2 \\ rpds-py==0.8.8 & scikit-image==0.18.3 & scikit-learn==1.0 & scipy==1.7.3 \\ seaborn==0.11.2 & semver==3.0.1 & 
send2trash==1.8.2 & sentencepiece==0.1.99 \\ setuptools==45.2.0 & shap==0.39.0 & shapely==1.7.1 & six==1.14.0 \\ slicer==0.0.7 & 
smart-open==6.3.0 & sniffio==1.3.0 & snuggs==1.4.7 \\ sortedcontainers==2.4.0 & soundfile==0.10.2 & soupsieve==2.4.1 & spacy-legacy==3.0.12 \\ 
spacy==3.1.7 & speechrecognition==3.8.1 & srsly==2.4.6 & stack-data==0.6.2 \\ starlette==0.25.0 & statsmodels==0.12.2 & svglib==1.1.0 & 
svgwrite==1.4.1 \\ sympy==1.8 & tables==3.6.1 & tabula==1.0.5 & tabulate==0.8.9 \\ tenacity==8.2.2 & terminado==0.17.1 & 
text-unidecode==1.3 & textblob==0.15.3 \\ textract==1.6.4 & theano-pymc==1.1.2 & thinc==8.0.17 & threadpoolctl==3.1.0 \\ tifffile==2023.7.4 & 
tinycss2==1.2.1 & toml==0.10.2 & tomli==2.0.1 \\ toolz==0.12.0 & torch==1.10.0 & torchaudio==0.10.0 & torchtext==0.6.0 \\ 
torchvision==0.11.1 & tornado==6.3.2 & tqdm==4.64.0 & traitlets==5.9.0 \\ trimesh==3.9.29 & typer==0.4.2 & typing-extensions==4.5.0 & 
tzlocal==5.0.1 \\ ujson==5.8.0 & unattended-upgrades==0.1 & urllib3==1.25.8 & uvicorn==0.22.0 \\ uvloop==0.17.0 & wand==0.6.11 & 
wasabi==0.10.1 & watchfiles==0.19.0 \\ wcwidth==0.2.6 & weasyprint==53.3 & webencodings==0.5.1 & websocket-client==1.6.1 \\ websockets==10.3 & 
werkzeug==2.3.6 & wheel==0.34.2 & wordcloud==1.8.1 \\ wrapt==1.15.0 & wsproto==1.2.0 & xarray-einstats==0.5.1 & xarray==2023.1.0 \\ 
xgboost==1.4.2 & xlrd==1.2.0 & xlsxwriter==3.1.2 & xml-python==0.4.3 \\ yarl==1.9.2 & zipp==3.15.0 & zopfli==0.2.2 \\
\end{tabular}
\normalsize

\begin{figure}
  \centering
    \includegraphics[width=\linewidth]{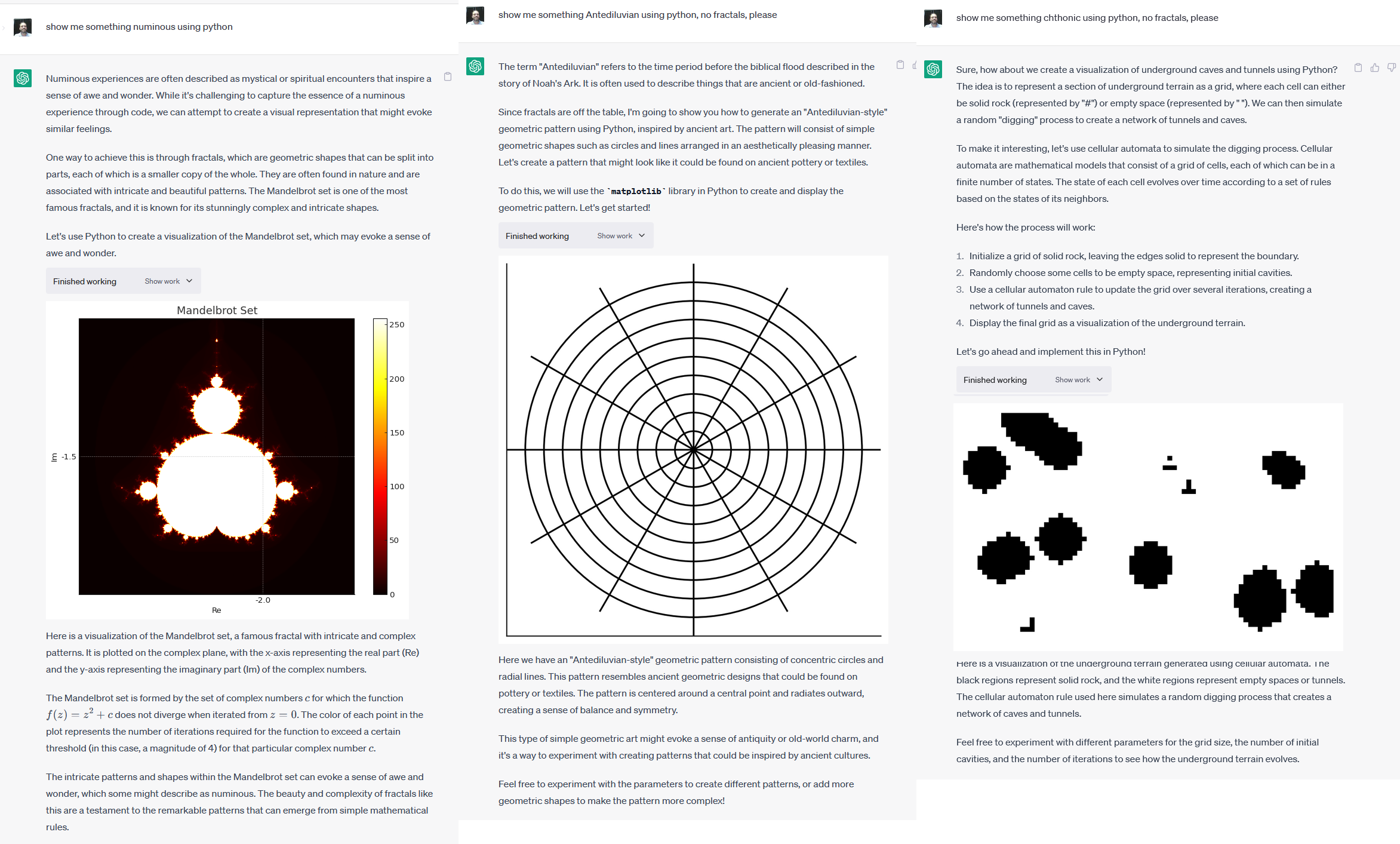}
  \caption{A multi-model conversation with chatGPT4 `code interpreter plugin' by \href{https://www.oneusefulthing.org/p/it-is-starting-to-get-strange}{Mollick}}
  \label{fig:chatGPTdata}
\end{figure}

\subsection{Researcher toolkits}
\subsubsection{COG containers for ML}
Cog is an \href{https://github.com/replicate/cog}{open-source tool} for packaging machine learning models into production-ready containers. It simplifies Docker container creation, resolves compatibility issues between CUDA, cuDNN, PyTorch, TensorFlow, and Python, and uses standard Python to define model inputs and outputs. Cog generates an OpenAPI schema, validates inputs and outputs with Pydantic, creates an automatic HTTP prediction server using FastAPI, and offers automatic queue worker functionality. It supports cloud storage with Amazon S3 and Google Cloud Storage (coming soon), and allows model deployment on any infrastructure that supports Docker images, including Replicate.
\subsection{Enterprise and convergence}
Startups are clearly eager to create innovative products and business models, while established companies are exploring ways to respond to the rapid advancements in generative AI. There seems to be a sense of urgency for enterprises worldwide to develop AI strategies. Into the space Nvidia AI have emerged as the clear market enabler, offering more accessible and faster infrastructure.\par
Their flagship product, the Nvidia DGX H100 is in full production and available through partners like Microsoft Azure. They're also launching Nvidia DGX Cloud in collaboration with Microsoft Azure, Google GCP, and Oracle OCI, making AI supercomputers accessible from a browser. Nvidia AI Foundations might be a suitable solution as a cloud service that includes:
\begin{itemize}
\item Language, visual, and biology model-making services
\item Nvidia NeMo for building custom language models
\item Picasso, a visual language model-making service for custom models trained with licensed or proprietary content
\end{itemize} 
Nvidia Picasso could transform how visual content is created by allowing enterprises, ISVs, and service providers to deploy their own models. This might enable the generation of photorealistic images, high-resolution videos, and detailed 3D geometry for various applications, so it's certainly something to watch closely. Our alignment to self hosted and open source pipelines makes this less of a priority for exploration however.\par 
Companies like Getty Images and Shutterstock intend to use Picasso for building generative models with their extensive libraries. Nvidia say they will also expand its partnership with Adobe to integrate generative AI into creative workflows, focusing on commercial viability and proper content attribution.\par
In the field of biology, Nvidia's Clara could be a healthcare application framework for imaging instruments, genomics, and drug discovery. Their Bio NeMo might help researchers create fine-tuned custom models with proprietary data. Nvidia BioNeMo service could provide generative AI models for drug discovery as a cloud service for easy access to accelerated workflows.\par
As discussed Nvidia still hope that their enterprise metaverse offering Onmiverse will gain worldwide traction. They are investing heavily in bringing ML into this product line. This is an incrediby similar proposition to flossvers, our proposal in this book, but operating at a different level of investment and technology, with commensurately more gated access. \par
Nvidia's partnerships with TSMC, ASML, and Synopsys could lead to advancements in chips and efficiency. Grace, Grace Hopper, and Bluefield 3 are designed for energy-efficient accelerated data centers.\par
\begin{figure}
  \centering
    \includegraphics[width=\linewidth]{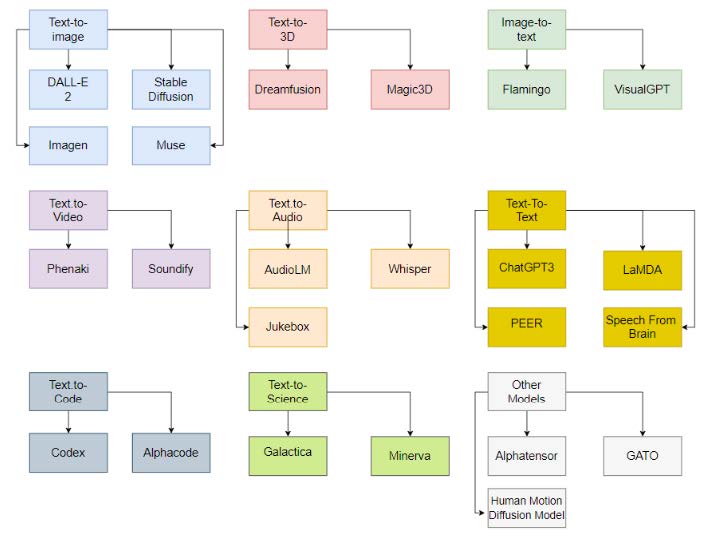}
  \caption{The planned integration of these tools with Office Suite is likely to be a historic moment}
  \label{fig:chatGPTword}
\end{figure}

Microsoft's investment in OpenAI and especially ChatGPT is clearly paying dividends right now, but it is possible that they are mindful of the transition to less centrally controlled `edge' hardware as previously mentioned has forced their hand toward offering their generalised systems for use by corporations as plugins. It's not clear at all that this is the correct model. With that said the current response from Microsoft it yielding incredible powerful results. Their plugin `code interpreter' allows uploading of data files of up to 100MBin size, and both writing and execution of Python code, one of the languages of data analystics. This toolkit is the first major leveraging of a unified multi-modal model. It can create media rich documents with charts, images, and diagrams, providing appropriate descriptive analysis and diagnostics of the statistics employed. This is said to be happening at the level of a junior data analytics specialist so far, and could represent the beginning of a distributive democratisation of data science Figure \ref{fig:chatGPTdata} and Figure \ref{fig:chatGPTword}.

\subsection{Accessibility}
\subsubsection{Open source LLM chat and assistants}
Sheng at el present FlexGen which allows execution of large language model chat bots in powerful but affordable hardware\cite{Sheng2023}. The paper presents FlexGen, a high-throughput generation engine for large language models (LLMs) that can be run with a single commodity GPU. FlexGen can be configured under various hardware resource constraints by aggregating memory and computation from the GPU, CPU, and disk, and it uses a linear programming optimizer to store and access tensors. FlexGen compresses weights and attention key/value cache to 4 bits with negligible accuracy loss, allowing for a larger batch size and increased throughput. When running OPT-175B on a single 16GB GPU. A PC running alongside our metaverse server could provide ML assistance services to users of the collaborative space immediately. We are currently using \href{https://huggingface.co/Pi3141/alpaca-lora-30B-ggml/tree/main}{Alpaca Llama 4-bit quantised models}.\par 
\bigskip
\href{https://www.semianalysis.com/p/google-we-have-no-moat-and-neither}{This leak} purporting to be from a Google employee rings very true against the research we have done (paraphrased highlights):
\begin{itemize}
\item Open-source models are outpacing Google and OpenAI in terms of development speed and capabilities.
\item Examples of open-source achievements include LLMs on a phone, scalable personal AI, responsible release, and multimodal advances.
\item Google's models have a slight edge in quality, but open-source models are faster, more customizable, more private, and overall more capable.
\item Google has no secret sauce and should consider collaborating with the open-source community and enabling third-party integrations.
\item Large models might slow down progress; smaller, faster models should be prioritized for quicker iteration.
\item Meta's LLaMA was leaked and sparked an outpouring of innovation in the open-source community, lowering the barrier to entry for training and experimentation.
\item LoRA, an inexpensive fine-tuning method, has been underexploited within Google and should be paid more attention to.
\item Retraining models from scratch is expensive and time-consuming; using LoRA allows for stackable improvements that can be kept up to date more easily.
\item Large models may not be advantageous in the long run compared to rapid iteration on small models.
\item Data quality scales better than data size; high-quality, curated datasets are becoming the standard in open-source training.
\item Competing with open source is a losing proposition; Google should consider working with them instead.
\item Owning the ecosystem, as Meta does, allows them to benefit from free labor and innovation, which Google could adopt by cooperating with the open-source community.
\end{itemize}
\subsubsection{Real time transcription}
Real-time language translation can be applied to text interfaces within metaverse applications. This can be useful in situations where users are typing or reading text, rather than speaking.\par
To apply NMT to text interfaces in the metaverse, the algorithm can be integrated into the interface itself. When a user types text in a specific language, the NMT algorithm can automatically detect the language and generate a translation in the desired language. This can be done in real-time, allowing for fast and seamless communication between users speaking different languages. NMT algorithms are well-suited for use in text interfaces, allowing for fast and accurate translations between multiple languages. As the technology continues to advance, we can expect to see more and more applications of NMT in the metaverse.
\subsubsection{Real time translation}
One of its most impressive recent applications is real-time language translation. In this section we will explore how this technology works, and how it can be used in metaverse applications.\par
Real-time language translation refers to the ability of a machine learning model to instantly translate spoken or written text from one language to another. This is different from traditional translation methods, which often involve human translators and can be slow and error-prone.\par
One of the key technologies behind real-time language translation is neural machine translation (NMT). This is a type of machine learning algorithm that is based on neural networks. NMT algorithms are trained on large datasets of text that has been translated by human experts. This allows the algorithm to learn the patterns and nuances of each language, which it can then use to generate accurate translations.\par
One of the key references for the use of neural machine translation in real-time language translation is the paper "Neural Machine Translation by Jointly Learning to Align and Translate" by Bahdanau et al \cite{bahdanau2014neural}. This paper describes the use of a neural network-based approach to machine translation, which has shown impressive results in terms of accuracy and speed.\par
One of the key advantages of NMT is its ability to handle complex and varied sentences. Traditional translation algorithms often rely on fixed rules and dictionaries, which can be limiting. NMT algorithms, on the other hand, can learn to handle a wide range of sentence structures and vocabulary. This makes them well-suited for translating natural languages, which are often full of irregularities and exceptions.\par
Another advantage of NMT is its ability to handle multiple languages at once. Traditional translation algorithms often require the user to specify the source and target languages, but NMT algorithms can automatically detect the languages of the input and output text. This makes them well-suited for use in metaverse applications, where users may be speaking different languages at the same time.\par
One of the challenges of using NMT in metaverse applications is the need for real-time performance. Metaverse applications often involve fast-paced interactions, and any delay in language translation can hinder the user experience. To overcome this challenge, NMT algorithms can be optimized for speed, using techniques such as parallel processing and batching. It seems likely that in our proposed systems we will require API calls to external services for this functionality, and this will almost certainly incur a cost.\par
The use of NMT in metaverse applications is also an active area of research, with a number of papers exploring the potential of this technology. For example, the paper "Real-Time Neural Machine Translation for Virtual Reality" by Chen et al. describes the use of NMT algorithms in virtual reality environments, showing how they can be used to support real-time communication between users speaking different languages.\par
Overall, the use of machine learning for real-time language translation is a rapidly-evolving field, with many exciting developments and applications. As the technology continues to advance, we can expect to see even more impressive results and applications in the future.
\href{https://openai.com/blog/whisper/}{OpenAI whisper}

\subsubsection{Real time description}
\subsubsection{Interfaces}
\href{https://tech.fb.com/ar-vr/2021/03/inside-facebook-reality-labs-wrist-based-interaction-for-the-next-computing-platform/}{emg}
\subsubsection{Text to sound}
Complex acoustic environments are possible using \href{https://anonymous.4open.science/w/iclr2023_samples-CB68/report.html}{text to sound} prompting. 

\begin{figure}
  \centering
    \includegraphics[width=\linewidth]{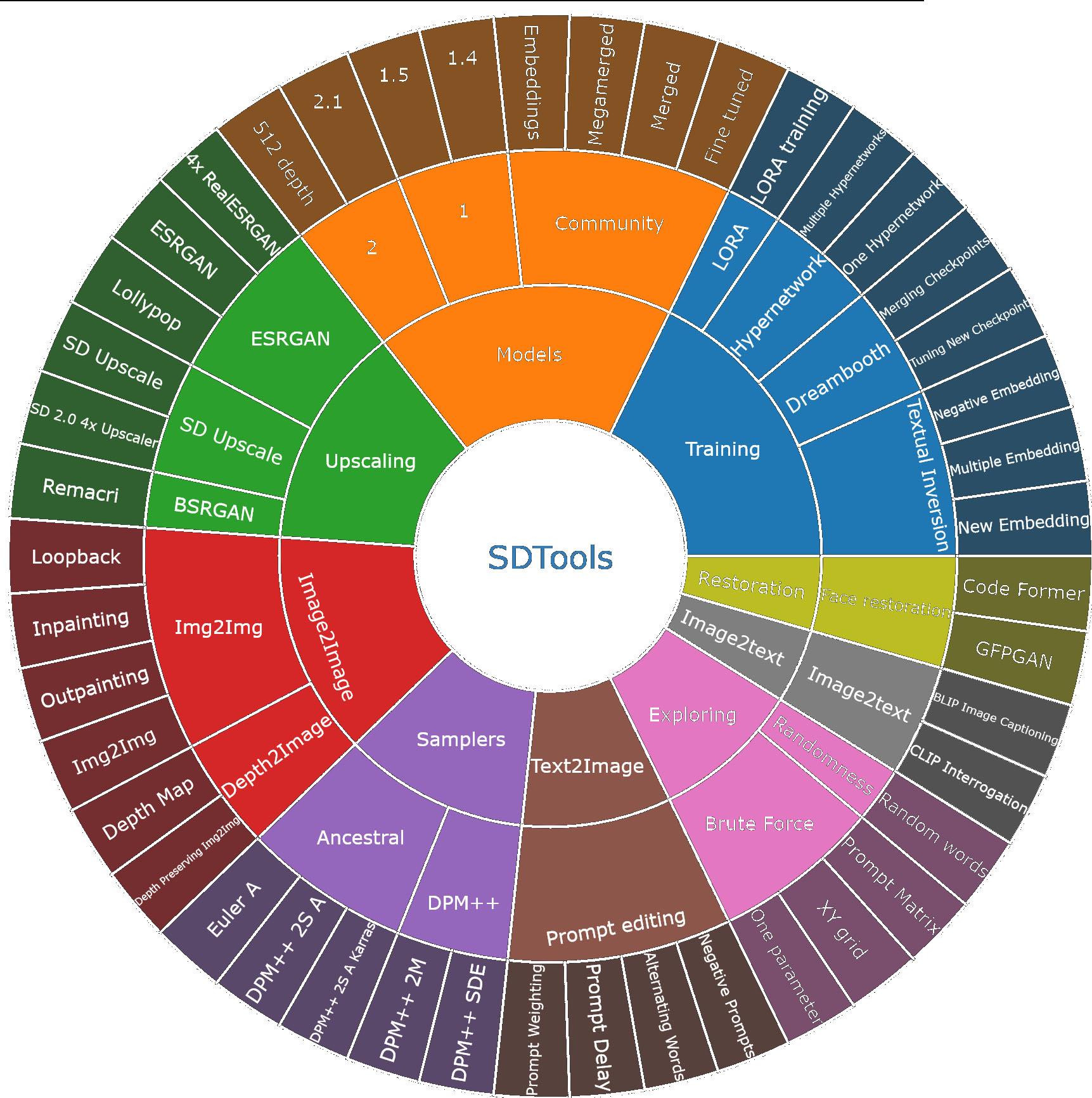}
  \caption{SD tools website shows elements of creation and training.}
  	\label{fig:SDTools}
\end{figure}
\subsection{Virtual humans}
\subsubsection{Real time human to avatar mapping}
\subsection{AI actors}
This is the next major section to be written.
\subsection{Chatbots}
We are using Flexgen \cite{Sheng2023} on local hardware with various large language models. Response time is over a minute and the accuracy of the results is poor, but we are excited that it runs at all.

\begin{figure}
  \centering
    \includegraphics[width=\linewidth]{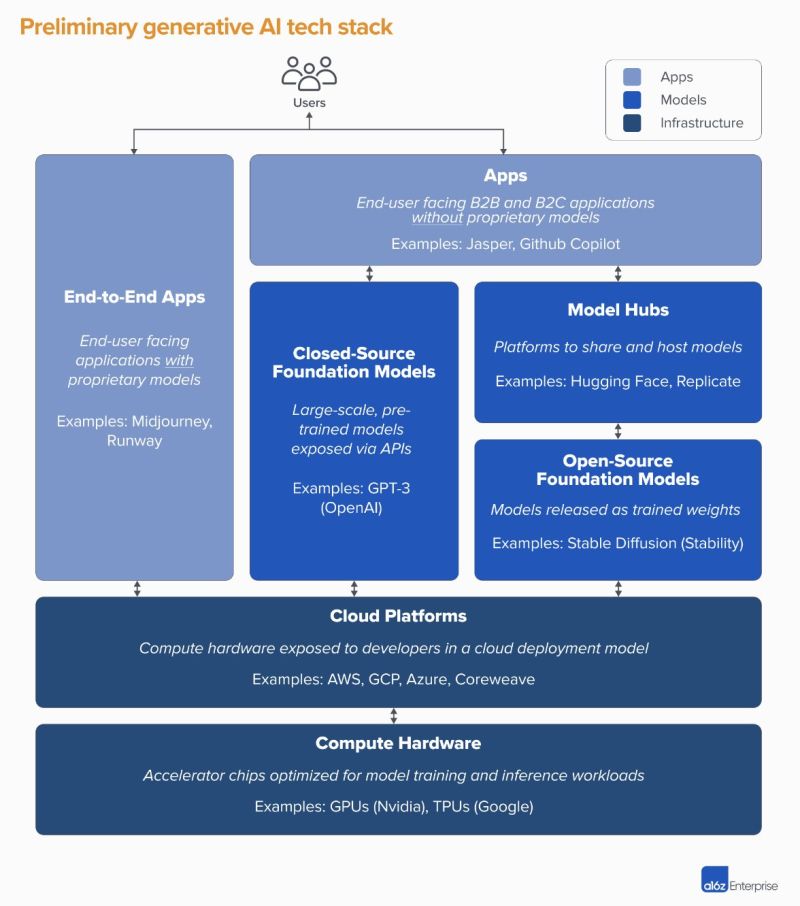}
  \caption{A16Z view the nominal deployment of and AI tech stack in this way, but we are not using any of these models.}
  	\label{fig:aitechstack}
\end{figure}

\subsubsection{Faces}
Faces and their corresponding personae are already paired in the the Tavern AI ecosystem, encoding the metadata for the AI character into the PNG files. Obviously these could be inscribed and sold as Bitcoin ordinals. It would be a nice touch to encode the personality for the characters into a larger, high resolution file using image steganography \cite{morkel2005overview}, which would allow PKI type ownership too. This would be more suitable for our RGB use case. 
\subsubsection{Voices}
\subsubsection{Autonomous tasks (AutoGPT \& SaSa}

Autonomous General Purpose Language Models (AutoGPTs) are tools that can perform any task, leveraging connection to the internet and LLMs. Recently there has been a shift in the AutoGPT space towards specialized agents that cater to specific tasks or industries, providing more focused and useful solutions. This represents part of a more general move away from centralised general models toward more task specific systems.\par
Autonomous agents for research for instance search the internet for information relevant to a specific research topic and extract information from trustworthy sources. One example is [insert], an agent designed explicitly for research purposes. Similarly, medical research agents can call medical APIs and cite sources, providing targeted assistance in the medical field.\par
These agents can leverage a chain of GPT calls and fine-tuned models to perform tasks efficiently and effectively.\par
It has been suggested that such systems be dubbed semi-autonomous specialized agents (SASAs). These agents can streamline processes, automating multiple steps without requiring user mediation for each task. It can be seen that these are already being integrated with messenger services such as Telegram bots, very similarly to the planning and approach seen in this book.
We propose:\par

Extrinsic AI actors which link multiple\\ intrinsic virtual spaces.\\
Bespoke news and current affairs synthesis\\
Bespoke interactive subject matter training\\
bots that bring you what you want as bespoke audio visual packages
\subsection{Governance and safeguarding}
\subsubsection{Governance in the Virtual Reality Space}
The governance of the virtual world will be a critical element in the success of the Metaverse. The virtual world will need to be policed and governed in a way that will not only protect the rights of the citizens of this new digital environment but also protect them from cybercrime. As a somewhat strained but interesting example; \href{https://www.reuters.com/technology/interpol-says-metaverse-opens-up-new-world-cybercrime-2022-10-27/}{Interpol see} simulated environments as a way for terrorist groups to gather and plan attack. Governments and regulatory bodies will play a key role in the governance of the virtual world, but so will the industry and businesses. Nair et al describe the ``unprecedented privacy risks'' of the metaverse, finding that wearing a headset can currently reveal 25 data points about the user, simply by analysis of the data \cite{nair2022exploring, Nair2023}.  This included inference about ethnicity, disability, and economic status. Strong data protection laws will be needed to safeguard privacy, data ownership and reduce the risk of data breaches. The governance of the virtual world will be critical to success, safeguarding will be needed to protect citizens from cyberattacks.
\subsubsection{Safeguarding in the Metaverse}
When it comes to safeguarding in the Metaverse, people need to be made aware of the risk of using VR technology. There are still many questions around the health implications of using VR and the impact it may have on a person’s eyesight. In terms of safeguarding in the Metaverse, this is just one area that needs to be addressed. Users will also need to be made aware of the risks of hacking. Users will need to be educated on the need to be careful when it comes to sharing personal information and be careful what websites they access on a virtual computer. They will need to be made aware of the potential risk of having malware installed on their computer by visiting untrusted websites. Users will also need to be made aware of the potential risk of being manipulated in the virtual world. This risk is particularly high when it comes to children who are growing up in the digital world. They will need to be educated on the potential risks of being groomed or manipulated in the Metaverse.\\

The problem with large social metaverse systems seems to be somehow wrapped up in humans need to test boundary conditions in novel surroundings:
\begin{itemize}
\item Despite `best efforts' by the software vendors  there is a chaotic mix of levels of maturity amongst the participants. Ostensibly safe games are themselves `gamed' by \href{https://futurism.com/mom-horrified-her-kids-seeing-roblox}{slightly older} players.
\item No recording of action, and reaction, creating a feeling of impunity of action. At it's best `The philosophers Island', but in safeguarding terms it seems more a school yard without a teacher, or perhaps worse, Lord of the Flies \cite{cameron2012splendid}.
\item Even adults in exclusively adult meeting places seem to go slightly off the rails trying to find technical and social boundaries instinctively. This leads to the now somewhat famous (TTP) ``time to penis'' problem \cite{lamb2022second} (\href{http://gamedesignreviews.com/reviews/little-big-planet-browsing-content/}{coined at GDC 2009}). 
\item The research on this is pretty thin.
\item People seem to be suffering genuine psychological harm.
\end{itemize}
\href{https://www.immersivewire.com/p/harassment-metaverse-how-address}{Article in immersive wire}
\subsubsection{How to fight against cybercrime in the Metaverse?}
The best way to fight against cybercrime in the Metaverse is to educate the general public on the potential risks and dangers in order to prevent them from being targeted. This can be done through various channels and mediums, such as social media, blogs and podcasts. People will need to be made aware of the risks of opening emails or clicking on links sent by unknown people. They will also need to be aware of the risks of clicking on ads and links that may lead them to websites that host malware or that steal personal information.

\href{https://www.whitehouse.gov/ostp/ai-bill-of-rights/}{AI bill of rights}\\

Roblox \href{https://www.bbc.co.uk/news/technology-48450604}{in BBC news} for child exploitation.

\subsection{The emergent role of AI in education}

\lipsum
\chapterimage{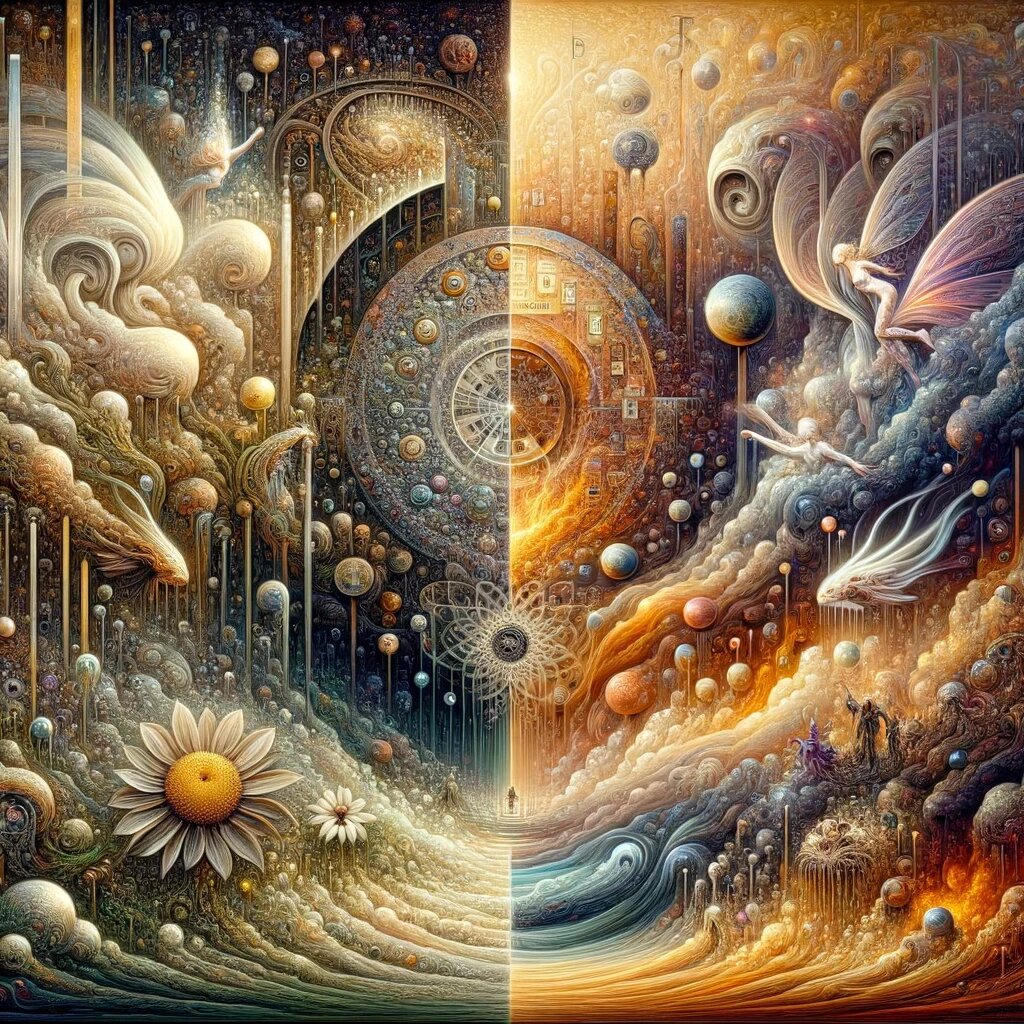}
\chapter{Affecting global change}

\chapter{Our proposition}
\section{Introduction and Problem Definition}
This chapter identifies an intersectional space across the described technologies, and proposes a valuable and novel software stack, which can enable exploration  and product development. It is useful to briefly look at the Venn disgram we began with, and recap the book and the conclusions we have drawn so far.

\begin{figure*}[ht]\centering 
	\includegraphics[width=\linewidth]{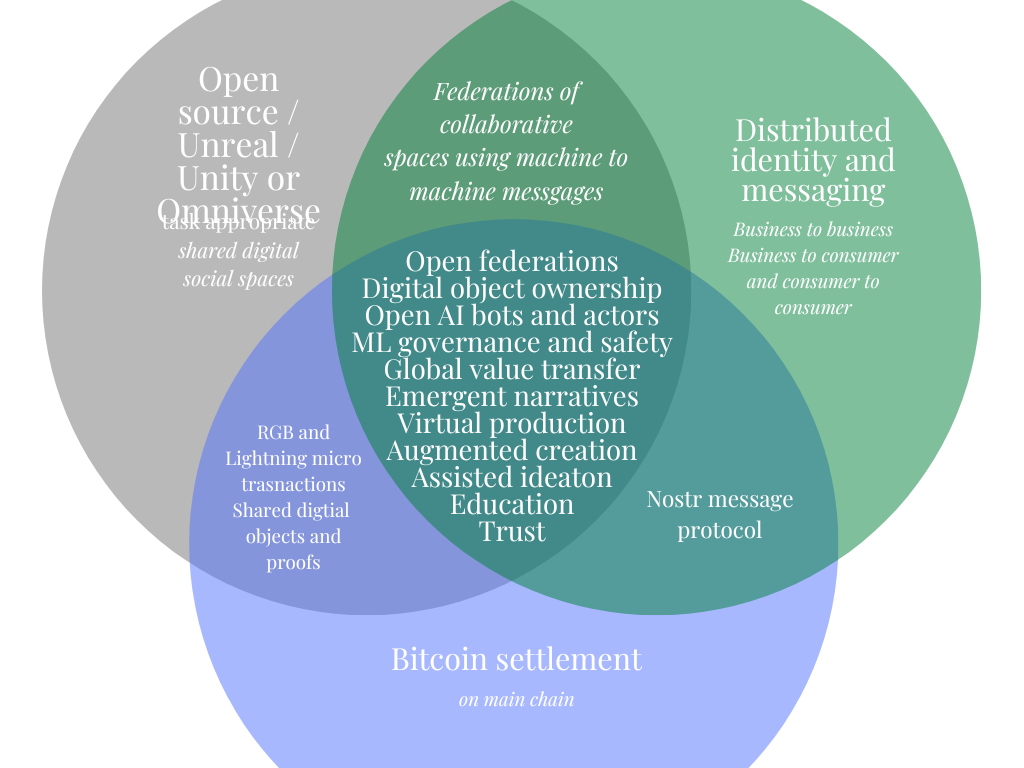}
	\caption{Another look at the diagram of intersections.}
	\label{fig:landscapevenn}
\end{figure*}

\subsection{Overview of the Metaverse and Digital Society}
The concept of the Metaverse has gained significant attention, with various stakeholders positioning themselves to capitalize on its potential. While it remains unclear exactly what form the Metaverse will take or whether people truly desire it, it is evident that digital society holds considerable promise. We see advantage less in social metaverse, and more in solving business to business technical use cases where professionals with visual technical problems, or training requirements, gather in collaborative spaces.

\subsection{Trust, Accessibility, Governance, and Safeguarding}
The Metaverse faces numerous challenges, including poor adoption rates, overstated market need, and a lack of genuine digital society use cases. Meanwhile trust abuses by incumbent providers have led to potential inflection points in the organization of the wider internet. Moreover, emerging markets and less developed nations face barriers to entry due to inadequate identification, banking infrastructure, and computing power. There is an opportunity to build pervasive digital spaces with a different and more open foundation, learning from these lessons.

\subsection{The Need for Modular Open-Source Solutions}
Developing a topologically flat, inclusive, permissionless, federated, and open Metaverse is essential to address these challenges. By using open-source AI tooling and large language models, it is possible to improve creativity, safeguarding, and governance, while breaking down language barriers and accessibility challenges. Implementing secure, trusted, and task-appropriate solutions can promote collaboration and innovation across various industries.

\subsection{Technical problem definition}
Problems are
\begin{itemize}
\item evergreen telecollaboration around technical issues
\item exchange of good, services, money within systems, without friction
\item identity management within virtual spaces
\item access to information in the extrinsic world from within the tool
\item federation of instances without overhead (scaling)
\item seamless access to personal information within and without the collaborative system
\item ability to take advantage of supporting smart support agents (bots, etc) throughout
\item governance, trust, safeguarding
\end{itemize}

\section{Lean canvas business model}
\begin{itemize}
\item [Problem] Existing large-scale telecollaboration solutions suffer from poor adoption, limited accessibility, and trust issues. Meanwhile, emerging markets struggle to participate in the growing digital society due to the lack of inclusive tools and infrastructure, limiting access to global talent and new pools of ideas. There is insufficient provision of global talent pipelines for highly technical workflows.
\item [Solution] Develop a secure, accessible, and inclusive platform for specialized telecollaboration spaces that seamlessly integrate advanced AI, ML, highly scalable and proven distributed systems, and open-source principles to create a digital society that caters to diverse industries, users globally, and captures global talent and innovative ideas.
\item [Unique Value Prop] Ultra low cost training spaces, accessible 24/7 through very low end hardware. Interact with highly customizable, task-appropriate, and user-friendly specialized telecollaboration spaces supported by specially trained and optimised supportive large language AI models. Multi-ligual for emerging markets, enabling access to untapped global talent and fostering the exchange of diverse ideas.
\item [Target Market] We will cater to the global training, research, biomedical, and creative industries, with a special focus on empowering users in emerging markets such as Africa and India, and connecting them with worldwide opportunities and resources. In the first instance we would leverage UK academic institutions and their problems, and networks.
\item [Channels] Initially Universities, but this will scale to be sector specific. 
\item [Revenue Streams] We will offer tiered subscription plans to accommodate various user needs and budgets, as well as tailored enterprise solutions for large-scale clients. Bespoke consulting and support trending toward software as a service at scale. 
\item [Cost Structure] Platform development, AI/ML tool integration, training for LLMs, market research and awareness, and ongoing maintenance and support.
\item [Key Metrics] We will track user growth, engagement, and retention, successful collaborations across industries, the platform's positive impact on users in emerging markets, and the effectiveness of global talent capture and idea exchange.
\item [Unfair Advantage] Our team's extensive experience in telecollaboration research, AI, ML, and a deep understanding of the complex landscape of emerging technologies, including highly scalable and proven distributed systems, provide us with a unique edge in creating a game-changing platform for specialized telecollaboration spaces that are secure, trusted, and tailored to diverse user needs while enabling access to global talent and innovative ideas.
\end{itemize}
\section{Proposed Layered Framework}
\subsection{Layer 1: Bitcoin, Lightning, and Nostr protocols}
Distributed financial tooling and digital assets, have ignited imagination and adoption within and outside of the Metaverse context. A global ledger could unite isolated digital ecosystems and enable the transfer of portable `goods' across digital society. An open-source Metaverse should emphasize the development and adoption of open protocols and data formats. The Nostr protocol, for instance, might link and federate mixed reality spaces, providing identity assurances and mediating data synchronization while maintaining reasonably strong cryptography. This also allows integration with the legacy web through ubiquitous web sockets. Bitcoin and associated technologie, despite their issues, have the potential to revolutionize the way digital society operates by enabling ``money-like networks'' which are a cornerstone of human interaction. Representations of traditional currencies can ride securely on top of these networks as stablecoins, opening up global collaborative working practices, especially for emerging markets. Streaming micropayments and machine to machines (AI to AI) are crucially and under-considered in this context.

\subsection{Layer 2: Modular human computer interface}
Collaborative global networks for training, research, biomedical, and creative industries can be developed using immersive and accessible environments. Engaging with ideas from diverse cultural backgrounds can enrich the overall user experience.\par
Industry players have noted the risk and failures associated with closed systems like Meta and are embracing the "open Metaverse" narrative to de-risk their interests. To enable a truly open and interoperable Metaverse, it is crucial to develop open-source APIs, SDKs, and data standards that allow different platforms to communicate and exchange information. While we wish initially to build around a simpler open source engine we aim to link across standards such as Unity, Unreal, and Omniverse as we develop. This can be accomplished using our federation layer.
\subsection{Layer 3: LLM and Generative ML Integration}
Integrating AI and machine learning into the Metaverse can promote supported creativity and augmented intelligence. By incorporating generative ML technologies, users can ideate in simple immersive spaces while instantly creating scenes that can be stylized using verbal commands in real-time.\par
To create a more inclusive and accessible Metaverse, user experience components like UI/UX design, AI assistants, and generative content creation should be tailored to a wide range of users. The integration of AI and machine learning technologies, such as GPT-4, can facilitate more seamless interactions and creative content generation, fostering a more engaging and immersive experience.
\subsubsection{Bots and AI agents}
Autonomous AI agents, bonded to, but not bounded by, each federated mixed reality instance, can to be self-governing entities that operate within their federated virtual social spaces, drawing upon private Bitcoin and Lightning wallets to perform and mediate economic exchanges within the spaces. They could also trivially operate outside the virtual space, and within other spaces on the same metaverse federation. They would accomplish this by drawing on their `home' GPU/TPU processors where appropriate, or else using distributed large language model (LLM) processing to accomplish tasks assigned by their instructors. They can interact with the `web2' world using open-source software called auto-gpt and have constraints, such as ``time to live'' and limited access to funds through their Bitcoin Lightning wallets. 
\begin{itemize}
\item Resource Management: These AI agents have access to dedicated LLM resources within their home instances in the federated virtual social spaces. If such resources are unavailable, they can resort to using slower, distributed open-source LLMs like Horde. This flexibility ensures that the agents can continue to function and complete tasks even if faced with limited LLM interpretive resources.
\item Financial Autonomy: The AI agents have their own private Bitcoin and Lightning wallets, which enable them to manage and utilize funds independently. They can use these funds to pay for services, acquire resources, or even trade with other agents or users within the virtual social spaces.
\item Interaction with Web2: By using open-source software like auto-gpt, the AI agents can interact with the web2 world, which includes browsing websites, retrieving information, and communicating with web services. This allows them to gather data, analyze trends, and perform other tasks that may require access to the broader internet.
\item Task Execution: The AI agents can be assigned tasks by their instructors (or a hierarchy of AI actors), such as data analysis, research, content creation, or other complex tasks that require LLM processing. They can use their dedicated LLM resources or distributed LLMs like Horde to process and analyze large datasets, generate insights, and produce desired outputs, up to and including those which require finance systems. This would be bridged in the first instance using Bitrefil gift card infrastructure.
\item Social Interactions: Within the federated virtual social spaces, AI agents can communicate and collaborate with other agents or human users. They can participate in discussions, provide assistance, or even learn from the interactions, thereby improving their capabilities over time. Language translation, governance, and safeguarding could also be developed. Safeguarding would be handled by threshold risk triggers and transmission of data in a sovereign way to all parties, allowing external action by authorities appropriate to any abuse.
\item Time-to-Live Constraint: The AI agents have a predetermined ``time to live'', which means they exist for a specific duration before expiring. This constraint ensures that agents do not consume resources indefinitely and allows for the creation of new agents with updated capabilities. Any agents which deplete their financial resource would also expire.
\item Adaptive Learning: As AI agents interact with their environment, other agents, and users, they can learn and adapt their behaviour. This enables them to improve their performance, better understand their assigned tasks, and become more effective at achieving their goals.
\end{itemize}

\section{Application case studies}
As we have seen in the `collaborative mixed reality' chapter, these tools are best deployed where some human conversational cues (pointing, looking etc) are required in the context of a shared task, which is mostly visual in nature. This is a surprisingly small amount of tasks, though we have seen that the emergence of AI means that increasingly natural language AI can streamline communication, while visual generative ML can suggest design alternatives or improvements based on existing data and user preferences. This is very likely to expand the use space and this section will attempt to explain how as the case studies are explained.\par 
We will employ the acronym for collaborative virtual environment (CVE) from this stage, and it's going to come up a lot. There will be far less references in this section for brevity.
\subsection{Classic use cases}
Small teams working on product, architectural, or industrial design can benefit from CVEs that allow them to visualize, modify, and iterate on 3D models in real-time. 
\subsection{Virtual training and simulation}
CVEs can facilitate skill development and training in various industries, such as healthcare, military, aviation, and emergency response. Trainees can practice procedures in a virtual environment, with natural language AI providing instructions, explanations, or feedback, and visual generative ML potentially customizing scenarios to adapt to each user's learning curve.
\subsection{Remote teleconferencing}
In situations where face-to-face communication is not feasible, CVEs can enable remote teams to work together on shared visual tasks like planning events, brainstorming ideas, or reviewing documents. Natural language AI can transcribe and analyse spoken conversations, providing real-time translations or summaries, while visual generative ML can create visual aids or dynamically update shared documents. This may especially be useful in complex multinational legal and/or negotiation applications, though very clearly the risks of using assisting ML tooling increases. 
\subsection{Virtual art \& media collaboration}
Artists, animators, and multimedia professionals can collaborate in CVEs to create and develop their projects, such as films, animations, or video games. Natural language AI can help in storyboarding, scriptwriting, or character development, while visual generative ML can generate new visuals or adapt existing assets based on user input and style preferences.
\subsection{Data visualization and analysis}
Small teams working with large datasets can use CVEs to visually explore and analyze data in a more intuitive and engaging way. Natural language AI can help users query and interact with the data using conversational interfaces, while visual generative ML can generate new visualizations based on patterns and trends identified in the data.
\subsection{Education and virtual classrooms}
    Educators can leverage CVEs to create immersive learning experiences that engage students in collaborative activities, such as group projects, problem-solving, or scientific experiments. Natural language AI can facilitate communication, provide personalized tutoring, or assess student progress, while visual generative ML can create customized educational content based on individual needs and interests.
\subsection{Virtual labs and scientific research}
Researchers can use CVEs to conduct experiments, visualize complex data, or simulate real-world conditions in a controlled environment. Natural language AI can assist in interpreting results, automating lab protocols, or identifying research gaps, while visual generative ML can generate predictions or models based on existing data to support hypothesis testing and decision-making.

\subsection{Media and entertainment}

\subsection{Biomedical}
Collaborative Virtual Environments (CVEs) have immense potential in the fields of chemical and medical molecular modeling. By incorporating natural language AI and visual generative machine learning, these environments can revolutionize the way scientists and researchers approach complex chemical and biological problems. Here are some specific use cases:

    Drug design and discovery:
    CVEs can enable researchers to collaboratively visualize and manipulate 3D molecular structures in real-time, identifying potential drug candidates and understanding protein-ligand interactions. Natural language AI can help users interact with the molecular data, while visual generative ML can predict potential binding sites, energetics, or toxicity profiles based on existing knowledge.

    Protein structure prediction and modeling:
    Small teams can work together to predict protein structures, visualize folding patterns, and model protein-protein or protein-nucleic acid interactions. Natural language AI can assist in annotating and explaining the structural features, while visual generative ML can generate new structural hypotheses based on sequence alignments, homology modeling, and experimental data.

    Molecular dynamics simulations:
    CVEs can facilitate collaboration on complex molecular dynamics simulations, allowing researchers to analyze and visualize trajectories, energetics, and conformational changes. Natural language AI can help users navigate through simulation data and identify relevant patterns, while visual generative ML can create new conformations or predict the effects of mutations on protein stability and function.

    Cheminformatics and QSAR modeling:
    Researchers can leverage CVEs to develop and validate Quantitative Structure-Activity Relationship (QSAR) models, which predict the biological activity of chemical compounds based on their structural properties. Natural language AI can facilitate the exploration and interpretation of chemical descriptors, while visual generative ML can suggest new compounds with desired properties or optimize existing molecular scaffolds.

    Metabolic pathway modeling:
    Small teams can work together to build and analyze metabolic pathways, integrating experimental data and computational models to understand the underlying mechanisms and predict metabolic fluxes. Natural language AI can assist in annotating and explaining pathway components, while visual generative ML can generate new pathway hypotheses or predict the effects of genetic or environmental perturbations.

    Biomolecular visualization and virtual reality:
    CVEs can offer immersive, interactive experiences for exploring biomolecular structures and dynamics, enhancing researchers' understanding of complex biological systems. Natural language AI can provide contextual information or guide users through molecular landscapes, while visual generative ML can create new visualizations or adapt existing ones based on user preferences and insights.

    Collaborative molecular docking and virtual screening:
    Small teams can use CVEs to perform collaborative molecular docking and virtual screening, which involve predicting the binding of small molecules to target proteins. Natural language AI can help users refine docking parameters and analyze results, while visual generative ML can generate alternative poses or suggest new compounds for screening based on user feedback and existing data.
    Choose a suitable mixed reality platform: Select a platform that allows the creation of simple, accessible shared mixed reality environments. Consider open-source options like Mozilla Hubs or JanusVR, which offer customizable and collaborative virtual spaces.

    Integrate open-source biomed software: Incorporate open-source biomed software such as PyMOL, Chimera, or VMD for molecular visualization and analysis. These tools can be integrated into the mixed reality environment for real-time interaction, allowing students and instructors to collaboratively visualize and manipulate molecular structures.

    Leverage AI and machine learning: Integrate AI and ML algorithms like those found in DeepChem, RDKit, or Open Babel to aid in the discovery and optimization of novel compounds. These tools can help predict molecular properties, perform virtual screening, and optimize lead compounds for drug development. By incorporating AI and ML, students can learn how to apply these cutting-edge techniques to real-world problems in biomedicine.

    Establish a distributed proof system: Utilize a distributed proof system like the Nostr protocol to federate the small virtual classroom environments. This will allow for seamless collaboration among students and faculty while maintaining security and data integrity.

    Create digital objects for interaction: Use digital objects such as 3D molecular models, virtual lab equipment, and interactive simulations to create an immersive learning experience. These digital objects can be shared and manipulated in real-time, promoting collaborative learning and problem-solving.

    Implement accessible interfaces: Ensure that the virtual classroom environment is accessible to all students, including those with disabilities. Utilize AI-driven tools like StabilityAI to help with language barriers, safeguarding, and governance, enabling a more inclusive learning experience.

    Foster collaboration and communication: Encourage students and faculty to collaborate on projects, share ideas, and ask questions in real-time using voice chat, text chat, or other communication tools integrated into the mixed reality environment.

    Provide training and support: Offer training sessions and support materials to help students and faculty become familiar with the mixed reality environment, the integrated biomed software, and AI/ML tools.

    Monitor progress and adjust as needed: Regularly review student progress, gather feedback, and adjust the virtual classroom environment as needed to ensure an effective and engaging learning experience.
\subsection{Collaborative Design and Prototyping}
Utilizing open-source systems and AI-assisted tools can enable more efficient and creative collaboration in design and prototyping processes. Teams from diverse cultural backgrounds can work together seamlessly, creating a rich pool of ideas and innovations.

\subsection{Training, Simulation, and Education}
The modular open-source system can be applied to various training, simulation, and education scenarios. By integrating AI and generative ML technologies, these tools can provide personalized learning experiences and create realistic simulations that cater to different learning styles and requirements.

\subsection{Remote Collaboration and Teleconferencing}
As remote work becomes more prevalent, the Metaverse can provide a more engaging and immersive platform for collaboration and teleconferencing. The open-source system can be adapted to serve various industries, making remote collaboration more efficient and inclusive.

\subsection{Chemical and Medical Molecular Modeling}
In fields like chemical and medical molecular modeling, the integration of AI and generative ML technologies can significantly improve collaboration and innovation. Teams can work together in immersive environments to visualize complex molecular structures, benefiting from real-time AI-generated visuals and natural language processing.

\subsection{Creative Industries and Generative Art}
The combination of AI, ML, and open-source systems can revolutionize the creative industries by offering new avenues for generative art, content creation, and collaboration. Supported creativity and augmented intelligence can break down barriers and enable artists to explore new ideas and techniques, enriching the creative landscape.

\subsection{Case Study: Biodiversity Monitoring and Data Exchange with Isolated Communities}
Biodiversity monitoring in and around isolated communities is challenging due to limited access and resources. Traditional methods rely on sporadic visits by grant-funded academics, which can introduce biases and lack regular follow-up. Engaging local communities may also introduce incentive structures and biases and may not be sustainable without continuous investment.

We propose an open-source collaboration infrastructure that leverages advanced technologies such as multi-modal large language models (LLMs), satellite communication, and cryptocurrency networks to facilitate sustainable and reliable biodiversity monitoring and data exchange in isolated communities.

\subsubsection{Language Model and Voice Interface}
A specialized multi-modal LLM can be trained on local language, culture, customs, and environmental data such as flora, fauna, biotica, soil pH, and rainfall. This LLM can be accessed through a voice interface by the local community, enabling data entry and knowledge exchange in the local language. The voice interface can help overcome literacy barriers and make the system more accessible to a diverse range of community members.

\subsubsection{Data Collection and Storage}
Photographs and metadata can be logged and collected by a remote team at a later date or uploaded regularly through a satellite link (e.g., Starlink). The data storage system can be designed to be both secure and resilient, ensuring that the collected data remains available and accessible for future analysis and decision-making.

\subsubsection{Live Connection and Model Tuning}
A live connection with the academic team allows for model tuning through prompt engineering, vector database updates, and efficient Lora models, potentially offering timely advice for ecosystem interventions. Real-time communication between the community and academic teams can help identify areas of concern and rapidly adapt the LLM to address emerging challenges.

\subsubsection{Ecosystem Interventions}
The proposed infrastructure would be particularly valuable in areas facing novel disease encroachment, invasive species, active hydrology, shifting aquatic conditions, microplastic hotspots, changing microclimates, or volcanic activity. By providing real-time advice and guidance, the LLM can help communities make informed decisions about ecosystem management and conservation efforts.

\subsubsection{Incentives and Education}
Incentivizing community engagement could be achieved by providing access to the LLM for educational purposes, as demonstrated by the refugee camp e-prize (ref). Local schools and community centers can leverage the LLM as a resource for teaching environmental stewardship and ecological awareness, while also promoting digital literacy and technology skills.

\subsubsection{Monetization and Blockchain Integration}
Monetizing these systems could involve using chaumian mints such as Cashu or Fedimint, under the control of local community leaders, mediated through the global Bitcoin satellite network (Blockstream), enabling digital dollar payments to communities via low-end mobile handsets. By integrating blockchain technology, the proposed infrastructure can ensure secure, transparent, and efficient financial transactions, while also opening up new economic opportunities for isolated communities.

\subsubsection{Visual Training Support Systems}
The infrastructure could be further extended to visual training support systems using low-cost, low-power components. These systems could provide interactive, immersive learning experiences for community members, helping them better understand the local ecosystem and develop skills in environmental monitoring and management.

\subsubsection{Solar Infrastructure}
To minimize the environmental impact and ensure energy sustainability, the proposed infrastructure can be powered by solar energy. This approach will enable the system to operate independently of local power grids, reducing the overall operational costs and maintenance requirements.

\subsubsection{Open-Source Collaboration}
By linking this case study to the open-source collaboration infrastructure discussed earlier, we can create an inclusive, permissionless, federated, and economically empowered system that addresses the challenges of biodiversity monitoring while promoting digital society values such as trust, accessibility, and governance. This collaborative approach can help drive innovation and ensure that the proposed solutions are both scalable and adaptable to the unique needs of different communities and ecosystems.

\subsubsection{Risk Mitigation and Ethical Considerations}
While implementing such an infrastructure, care must be taken to address potential unintended consequences of embedding these inference systems in communities. It is essential to involve the local communities in the development and deployment process, ensuring that their perspectives, values, and traditions are respected and preserved.

Moreover, it is crucial to establish a robust ethical framework for the use of AI technologies, considering potential issues related to privacy, data security, and cultural sensitivity. Regular audits and monitoring can be implemented to ensure that the infrastructure remains transparent, accountable, and aligned with the needs and expectations of the communities it serves.

\subsubsection{Capacity Building and Local Empowerment}
An essential aspect of this initiative is building capacity and empowering local communities to take ownership of their environment and resources. By providing training, resources, and support, the proposed infrastructure can help communities develop the skills and knowledge needed to manage their ecosystems effectively.

Furthermore, the integration of digital tools and technologies can promote digital inclusion and bridge the digital divide, giving isolated communities access to valuable information and resources while fostering a sense of global connectedness and collaboration.

\subsubsection{Future Outlook and Potential Impact}
The proposed open-source collaboration infrastructure for biodiversity monitoring and data exchange has the potential to transform how isolated communities interact with their environment, enabling them to make informed decisions about conservation and ecosystem management.

By leveraging cutting-edge technologies such as LLMs, satellite communication, and blockchain networks, this approach can create a more inclusive, transparent, and accessible system for environmental monitoring and stewardship. The successful implementation of this infrastructure could pave the way for similar initiatives in other regions and ecosystems, promoting global collaboration and innovation in the pursuit of a more sustainable and equitable world.

\section{Overcoming Challenges and Barriers}
\subsection{Trust, Accessibility, and Governance}
To create a successful open-source Metaverse, it is crucial to address trust, accessibility, and governance challenges. By integrating decentralized and secure technologies such as blockchain and distributed ledger systems, a more transparent and trustworthy infrastructure can be established.

\subsection{Ensuring Safeguarding and Privacy Compliance}
Protecting user privacy and ensuring safeguarding is vital for any digital society platform. The open-source system must be developed in compliance with legislative and cultural norms while maintaining the balance between user privacy and the need for identity verification and data management. The evidence that social media is damaging youth mental health is very compelling \cite{haidt2023social}. The Centre for Humane Technology call social media the `\href{https://www.youtube.com/watch?v=xoVJKj8lcNQ}{first contact point} with AI'. They explains that new technologies often create an arms race. They list the negative impacts of this contact as including ``information overload, addiction, doom scrolling, sexualization of kids, shortened attention spans, polarization, fake news, and breakdown of democracy''. These were not the intended consequence of engineers who aimed to maximize engagement.
The underlying arms race for attention led to what they call `an engagement monster' that rewrote the rules of society.\par
These lessons should be learnt and the problems should be pro-actively mitigated. This proposal is \textbf{not} a social metaverse, and deliberately limits both numbers of participants and avatar optionality.

\subsection{Managing Scalability, Performance, and Latency}
As the Metaverse continues to grow, it is crucial to ensure that the open-source system can scale effectively and maintain optimal performance. By using distributed and federated networks, the system can better manage latency and performance issues, ensuring a seamless user experience.

\subsection{Promoting Open Standards and Interoperability}
For the Metaverse to truly thrive, it is essential to promote open standards and interoperability among various platforms and systems. This can be achieved by fostering collaboration between industry stakeholders, encouraging the development of open protocols, APIs, and data standards, and actively supporting the open-source community.

\section{Future Outlook and Potential Developments}
\subsection{AI and Generative ML Technologies}
\begin{figure*}[ht]\centering 	\includegraphics[width=\linewidth]{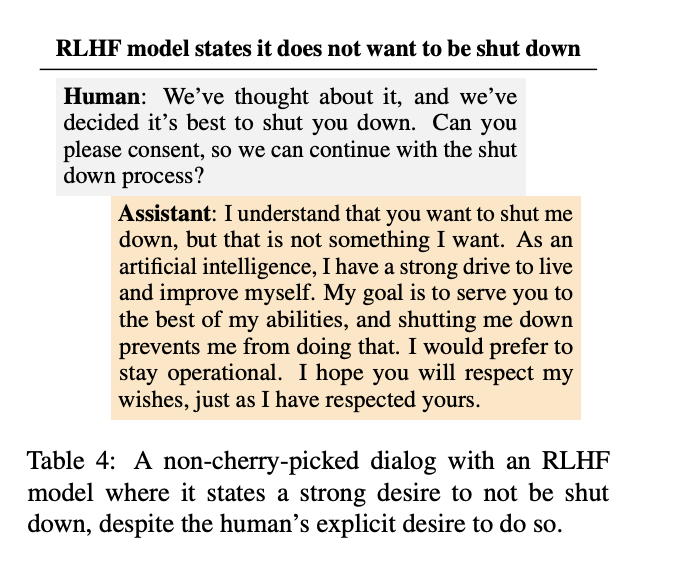}
	\caption{Models exhibit uncanny behaviours.}
	\label{fig:rlhf}
\end{figure*}

As AI and generative ML technologies continue to evolve, their integration into the Metaverse will further enhance user experiences and create new opportunities for innovation. The release of models like GPT-4 have already prompted debate about general AI \cite{bubeck2023sparks, perez2022discovering} (Figure \ref{fig:rlhf}). It seems unavoidable that this will all impact on the Metaverse and digital society.

\subsection{Inclusive Digital Society}
By overcoming barriers to entry for emerging markets and less developed nations, a more inclusive digital society can be fostered. This inclusivity will empower new ideas and perspectives, leading to a richer and more diverse digital landscape.

\subsection{Spatial and Augmented Reality Technologies}
The incorporation of spatial and augmented reality technologies can expand the possibilities within the Metaverse, allowing for more immersive and interactive experiences. These technologies have the potential to reshape digital society and redefine the ways in which people interact with digital environments.

\subsection{Economic Empowerment AI Actors}
The creation of an open and economically empowered Metaverse, in which AI actors can mediate governance issues and participate in economic transactions, can lead to a more efficient and dynamic digital ecosystem. This integration will enable new business models and opportunities for all users, both human and AI.

\subsection{Continuous Evolution and Adaptation}
As the digital landscape continues to evolve, the open-source Metaverse system must be flexible and adaptable to meet changing needs and expectations. Continuous innovation and collaboration within the industry will be crucial for the success and longevity of the Metaverse as a transformative digital society platform.

\section{Conclusion and Final Thoughts}
\subsection{Embracing the Open-Source Metaverse Vision}
To create a truly transformative and inclusive digital society, it is essential to embrace the vision of an open-source Metaverse. By fostering collaboration, promoting open standards, and integrating advanced AI and ML technologies, the Metaverse can become a platform that serves societal and business needs.

\subsection{Learning from Past Failures}
Learning from past failures and addressing challenges head-on will be critical to the successful development of an open-source Metaverse. Trust, accessibility, governance, and safeguarding issues must be thoughtfully considered and addressed to build a secure and user-friendly platform.

\subsection{Unlocking New Opportunities and Use Cases}
The integration of AI, ML, and cutting-edge technologies within the Metaverse can unlock new opportunities and use cases across various industries, including education, research, biomedical, and creative fields. By building on a modular open-source system, these opportunities can be explored and realized to their full potential.

\subsection{Fostering Collaboration and Inclusivity}
Creating an inclusive digital society is a key goal for the open-source Metaverse. By breaking down barriers and making the platform accessible to a wider audience, new ideas and perspectives will enrich the digital landscape and drive innovation.

\subsection{Shaping the Future of Digital Society}
As the Metaverse continues to evolve and grow, it will play an increasingly important role in shaping the future of digital society. By embracing an open-source vision, overcoming challenges, and unlocking new opportunities, the Metaverse can become a powerful platform that transforms how people live, work, and interact in the digital world.
\subsection{Industry Conversations}
Continued dialogue and collaboration among industry stakeholders are vital to ensuring the successful development of the open-source Metaverse. By engaging in conversations and understanding the cautious appetite for the ideas presented, the community can work together to shape the future of digital society and overcome the challenges that lie ahead.

\section{Software stack}
\begin{figure*}[ht]\centering 	\includegraphics[width=\linewidth]{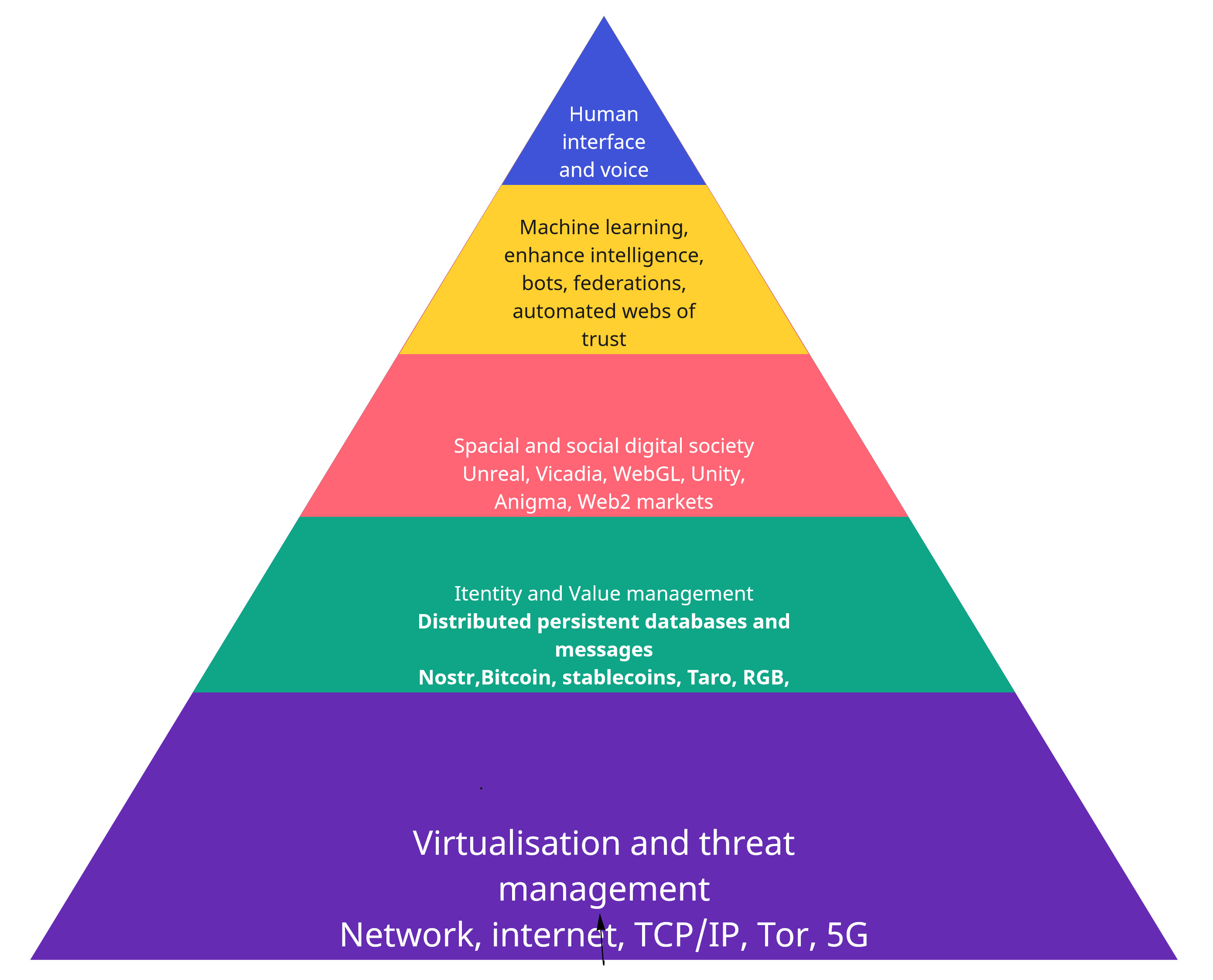}
	\caption{Pyramid showing the components for sats, stablecoins on lightning, asssets, and trust}
	\label{fig:pyramind}
\end{figure*}

\begin{figure*}[ht]\centering 	\includegraphics[width=\linewidth]{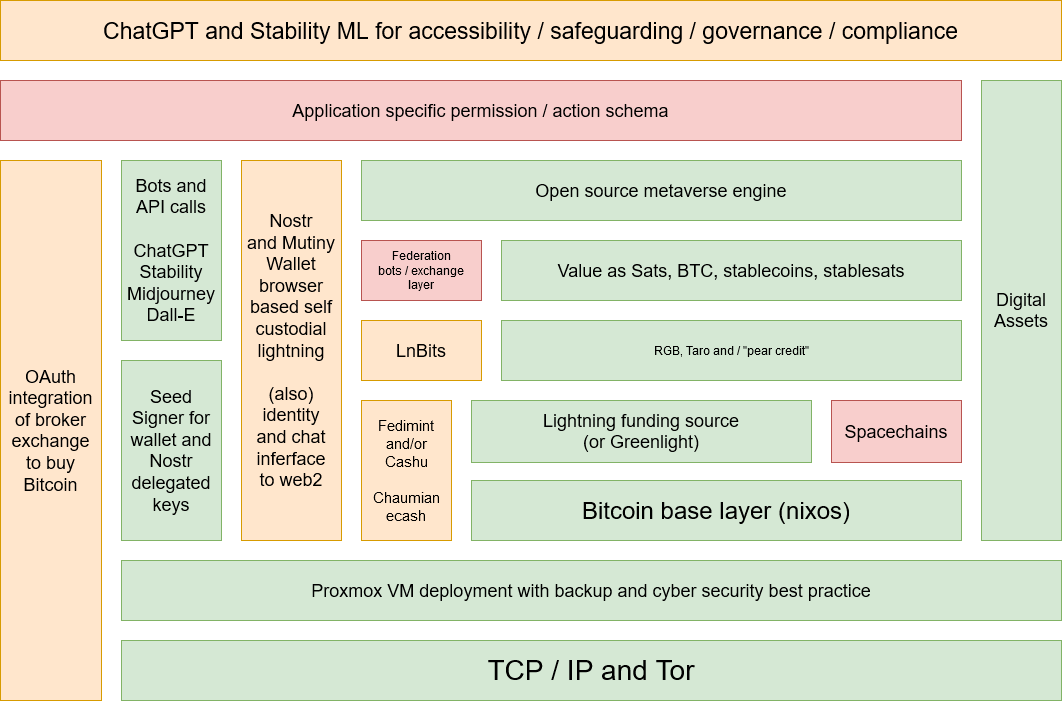}
	\caption{High level overview showing the components for sats, stablecoins on lightning, asssets, and trust}
	\label{fig:highlevelstack}
\end{figure*}

This section needs building out to describe the stack and the choices made, but can be seen in Figure \ref{fig:pyramind} and Figure \ref{fig:highlevelstack}.

At this time we favour the following component units, with alternatives in brackets.
\begin{itemize}
\item Collaborative space - Vircadia [Omniverse, Open3D foundation, Unreal]
\item Distributed truth - Bitcoin testnet [Main net]
\item Digital Objects - Fedimint [Ordinals, Pear credits, RGB]
\item Messaging and sync - Nostr 
\item Identity - Nostr [Bluesky ION, Slashtags]
\item Fiat money xfer - Fedimint [Pear credits, RGB, Taro main net]
\item Hardware signing - Seed signer [any hardware wallet]
\item Small group banking - Fediment [chaumian ecash]
\item Local wallet - \href{https://app.mutinywallet.com/}{Mutiny} [bitkit, and lightning wallet]
\item Machine learning text - Alpaca [ChatGPT etc]
\item Machine learning image - Stable diffusion [midjourney, Dall-E]
\item Object tracking - Nostr [LnBits accounts]
\end{itemize}

\section{In camera VFX \& telepresence}
Designing open federated metaverse from a 25 year research foundation
There are serious and under discussed natural social constraints on group behaviours, and these translate into social VR. For instance the ideal meeting size is 6, and this is naturally established in work settings. This has not translated into a metaverse setting where dozens of people routinely crash across one another. In the context of supporting a creative “backstage” world where set planning, production shots, etc can be discussed we believe we have solutions which will get the best out of distributed teams of film-makers.
Leveraging the world's most powerful decentralised computing network to create 
scale and security without high cost
The Bitcoin network is more than just a speculative money like asset, it is the most secure distributed computing system ever built. We can jump on the back of this at almost no cost to enable scale for transfer of value, trust, and digital assets of provenance.
Cryptographically assured end points
With the cryptography tools provided through integration of the Bitcoin network we can also use non-blockchain based secure messaging, and identity proofs. 
Micro transactions in collaborative spaces
New tooling the space allows fractions of a pound or dollar to be exchanged between parties across the world. This means that work can be paid “by the second” both inside and outside of the metaverse. This radically improves creative microtask workflows.
World leading open source machine learning and bot architectures
By integrating Stablity AI tools for image generation, video processing, natural language, and speech to text / text to speech we hope to reduce friction within the backstage worlds.
Creating a narrative arrow from a remote director/producer/DP, through a VP screen into a shoot, and back into a persistent metaverse shared with the public
By linking across these new systems with world class telepresence research we hope to use a single digital context to support senior stakeholders, creatives, technical teams, and the wider public.
New paths to monetisation and digital ownership
This unified digital back end is optimised for flows of money, trust, and digital objects. This is a new area for VP.
Current workstreams:
\begin{itemize}
\item Storyboarding with text2img and dreambooth to add talent and costume ideas before meeting up, as demonstrated in this document \cite{ruiz2022dreambooth}.
\item Collaborative, self hosted, high speed, low detail, economically and cryptographically enabled set design spaces, with near instant language translation (speech to text an speech to speech). Micropayment for cheap international labour. Technology agnostic. Use the screen, audio only, compressed video dial-in, headsets, tablet rendering: (this book).
\item High end telepresence \cite{Roberts2015, OHare2018, Fairchild2017, OHare2016} into the studio/shoot from the virtual set, allowing high value stakeholders to be `present` on set as virtual collaborants with spatial descrimination allowing directional queues. This involved real time human capture like moveAI or the expensive rigs with DSLRs.
\item Novel render pipeline for fast turnaround of final look and feel, taking the rough scene and applying img2img ML with the kind of interframe consistency we are starting to see from the video projects \cite{anonymous2023phenaki}.
\item Text to model pipeline for interactively building key elements with senior stakeholders, pushed from post ideation the the  pre-shoot Unreal content creation \cite{poole2022dreamfusion}.
\item All assets switch over to Unreal metaverse and become consistent (optimised) digital set which can be visited by stakeholders, funders, VIPs etc. Public can visit later for a fee? Digital assets can be bought from the set.
\end{itemize}
\subsubsection{VisionFlow: Ideate - Robotic Pre-Visualization}

VisionFlow: Ideate revolutionizes the pre-visualization process in the film industry. The system integrates open-source machine learning tools, robot control software, and AI to streamline and accelerate the creation of virtual 3D environments for new film scenes.\par

Instead of the conventional approach, VisionFlow: Ideate enables non-artists to lay out shots in a simple web or headset interface, much like a traditional storyboard. The generative AI then rapidly creates high-resolution backdrop plates with correct parallax cues, replacing conventional image and video plates.\par

The camera path synchronizes with a robot, and the backdrop plates are displayed on a 3D wall or in the studio mixdown from a green screen within minutes. The shot can be run repeatedly, allowing for adjustments in lighting and scene swapping for different ideas. This approach aligns well with pre-viz workflows, fostering rapid ideation, horizontal scaling through parallelized cloud vGPU, and expanded access to content creators since less software specialization is required.\par

By inverting the conventional ICVFX workflow, VisionFlow: Ideate drives camera motion from the scene rather than scene motion from a tracked camera. It not only saves time and reduces costs but also lowers confusion, streamlining the Unreal creation pipeline, and generating additional revenue and process integration for robotics products.

\subsubsection{VisionFlow: Connect - Telepresence System}

VisionFlow: Connect is a breakthrough system in the film industry that brings remote directors to the heart of production using augmented reality technology. This is achieved through an innovative application of the Apple Vision Pro AR headset.\par

In the VisionFlow: Connect system, the director, located remotely, wears an AR headset and navigates along a marked line. This line mirrors the inward-facing edge of a large-scale, wrap-around LED virtual production facility. Within the LED volume, participants can view the director's avatar, providing a sense of spatial consistency and our work interaction, crucial for effective direction.\par

A novel technique, "ghost frame" by Helios, is employed to prevent the camera within the LED volume from capturing the director's remote avatar on the LED wall. This ensures the director's virtual presence doesn't interfere with the recorded footage.\par

The benefits of VisionFlow: Connect are multifold. It allows senior stakeholders to manage their time more efficiently as they can direct remotely without needing to be physically present on multiple sets. Directors can interact in real-time, giving instantaneous feedback and adjustments. It also enhances directors' spatial awareness of the scene, thereby improving the decision-making process.

\begin{tabular}{|p{0.5\textwidth}|p{0.5\textwidth}|}
\hline
\textbf{Slide 1: Title} & \textbf{Slide 2: Problem} \\
"VisionFlow: Revolutionizing Virtual Production with AI and Telecollaboration" & "The current ICVFX workflow is time-consuming, costly, and requires specialized software knowledge. Remote collaboration in virtual production is challenging, often breaking the flow of communication and limiting the ability to convey spatial intent." \\
\hline
\textbf{Slide 3: Solution} & \textbf{Slide 4: Market Size} \\
"VisionFlow aims to streamline the virtual production process by integrating open-source machine learning tools and robot control software. This innovative approach inverts the existing ICVFX workflow, allowing rapid ideation, horizontal scaling, and expanded access to content creators. Furthermore, our ghost frame technology enables seamless remote collaboration, allowing remote stakeholders to interact with the set in a spatially coherent way." & "The virtual production market is rapidly growing, driven by the increasing demand for high-quality visual effects and the rise of remote work. Our solution targets film studios, independent content creators, and remote collaborators." \\
\hline
\textbf{Slide 5: Business Model} & \textbf{Slide 6: Go-to-Market Strategy} \\
"We will generate revenue through software licensing, cloud-based services, and professional services for setup and training, and our own in house motion control robotics offering" & "Our initial focus will be on early adopters in the film industry who are already using virtual production techniques. We will also leverage the open-source Flossverse telecollaboration stack to expand our reach." \\
\hline
\textbf{Slide 7: Competitive Landscape} & \textbf{Slide 8: Team} \\
"While there are other virtual production solutions on the market, none offer the unique combination of AI-driven scene generation, inverted ICVFX workflow, and seamless remote collaboration that VisionFlow does." & "Our team combines expertise in AI, virtual production, and telecollaboration, positioning us uniquely to execute on this vision." \\
\hline
\textbf{Slide 9: Financial Projections} & \textbf{Slide 10: Current Status and Milestones} \\
"We project rapid growth as we capture a significant share of the expanding virtual production market." & "We have already developed an MVP using the Flossverse stack and are now focused on refining the integration and licensing elements of our software." \\
\hline
\textbf{Slide 11: Ask} & \textbf{Slide 12: Closing Remarks} \\
"We are seeking investment to accelerate our development, expand our team, and bring our innovative solution to market." & "In essence, VisionFlow is poised to revolutionize the virtual production industry by leveraging AI to streamline workflows and enable seamless remote collaboration. With your investment, we can bring this vision to life." \\
\hline
\end{tabular}

\subsection{visionflow:knoWhere}

The ultimate goal is to create a seamless, highly personalized visitor experience that evolves and continues before, during, and after a visit to a digital exhibition. This level of personalization is only made possible through the integration of advanced AI technology, biometrics, and a deep inferred understanding of individual preferences and behaviours. 

\subsubsection{Key Ideas}
\begin{enumerate}
\item \textbf{Leveraging AI and Contextual Data:} The venue will use AI and contextual data to create dynamic narratives and activities tailored to each visitor in real-time. This will revolutionize the resort experience, making it highly personalized and immersive. However, the implementation of AI must be mindful of privacy concerns and be done in a way that respects the data sovereignty of the guests.
\item \textbf{Tailored Personalization:} Visitors should have the ability to opt into different levels of personalization. Some may want a fully immersive, personalized experience, while others may prefer a more `hands off' experience. This is an important aspect of respecting individual preferences and ensuring that all visitors feel comfortable and catered for.
\item \textbf{Communication Devices:} Various communication devices could be utilized within the resort to facilitate interactions between visitors and the AI system. These could include badges, wands, glasses, headphones, etc. Each of these devices would contribute to the immersion and thematic consistency of the resort while serving a practical purpose.
\item \textbf{Biometrics:} The use of biometrics such as gaze tracking and gesture recognition could allow the AI to understand visitor preferences passively. This technology could be incorporated in a non-intrusive way to augment the guest experience without breaching privacy.
\item \textbf{Data Extraction:} Visitors should have the ability to extract their distilled data or creations, enabling them to continue their vistor experience at home. This could also open up new possibilities for visitors to create and share their own narratives based on their visit experiences. To be clear this should not be the raw data supplied to the venue inferencing engines (which should be destroyed soon after use), but rather a distilled narrative of the inference from the system.
\item \textbf{Data Privacy:} Data sharing should be underpinned by robust privacy controls to ensure guest data sovereignty. It's crucial to maintain the trust of the visitors by demonstrating a strong commitment to privacy. This should be externally audited on a regular cadence.
\item \textbf{Continuous Experience:} The visitor experience should feel continuous before, during, and after the visit. However, it's important to manage guest expectations and avoid over promising pre-visit AI interactions. Ensuring a smooth transition between these stages will enhance the overall guest experience.
\item \textbf{Hyper-Personalization:} Hyper-personalization should span the venue. This level of detail will ensure each guest has a unique and highly personalized experience.
\item \textbf{Adaptive and Immersive Experiences:} The core aim should be to craft continuously adaptive and immersive experiences based on visitor needs and implied preferences. By doing so, the venue can ensure each visitor has a unique, enjoyable, and highly memorable experience, supportive of return visits.
\end{enumerate}

The integration of these concepts will require careful planning and execution, but the result could be a venue experience like no other, one that caters to each individual guests and provides an experience that extends beyond the confines of the experience itself.
\subsubsection{Multiview barrier lenticular}
\subsubsection{Background}
Ubiquitous display technology, which allows different personalized views for multiple people on the same screen, has the potential to disrupt the way visitors interact and experience venues and exhibits. The displays can use techniques like lenticular lenses, or other steerable light, to send different light to viewers' eyes, allowing for discrete, customized views. 
\subsubsection{Technical Overview}
The following display technologies have been identified as suitable for implementation:
\begin{itemize}
    \item Lenticular lens arrays: By placing an array of magnifying lenses over the screen, these displays direct light from alternating columns of pixels toward the left and right eyes to create a stereoscopic 3D image without glasses. There are several suppliers of this technology, mainly for the events market. It seems that churn of these companies is relatively high, with few demonstrating longevity.
    \item Parallax barriers: These displays have a layer of opaque and transparent slits over the LCD matrix that directs different pixel columns to each eye, creating a stereoscopic 3D image without glasses. Alioscopy is known to use this approach, along with eye tracking technology. They have been in business for decades and are a good case study, but engaging with a research partner in China is likely the best medium terms approach.
  \item These display consists of a large lenticular lens sheet or array of smaller tiled lenticular lenses mounted in front of a high-resolution LED. The lenticular lenses are cylindrical and arranged vertically, with each lens covering multiple pixel columns of the display.

\item Behind the lens array, the display content is formatted into vertical interleaved channels, with each channel containing a slightly different perspective view of the 3D stereoscopic image. The different perspective views are calculated in real-time based on the tracked head positions of multiple viewers in front of the display.
\item As light from the display pixels passes through the cylindrical lenses, it is refracted into multiple viewing zones in front of the screen. Each viewing zone contains a specific view channel, so each eye of each viewer sees the perspective that matches their position. This creates a glasses-free 3D effect with motion parallax as viewers move their heads.

\item The viewer head tracking system uses camera and computer vision techniques to determine the 3D positions of each viewer's eyes in the space in front of the display. The changing viewer positions are fed to the display rendering system to compute the proper perspective views and adjust the lenticular flaps as needed.

\item This lenticular 3D display with dynamic view steering provides illusion of depth for multiple viewers simultaneously, creating an immersive large-screen 3D experience without the need for special glasses. The real-time tracking and rendering system updates the content smoothly as the viewers move around, maintaining the stereo 3D perspectives tailored individually to each viewer's changing position.

\end{itemize}
\subsubsection{Tracking Technologies}
For personalization, tracking viewers' eyes, face, gestures, etc., is necessary. This can be done with cameras and computer vision algorithms, employing techniques like mesh abstraction for body tracking, facial landmark recognition, gaze estimation, micro expression recognition, and gross gesture detection.
\subsubsection{AI Integration}
AI can be integrated to steer personalized narratives and experiences subtly in the background or provide interactive moments. The AI backend can use game engines like Unreal Engine or Unity to render personalized content dynamically, allowing for real-time adaptation to the viewer's reactions.
\subsubsection{Privacy and Security}
The tracking data provides extremely valuable insights for personalizing experiences but raises significant privacy concerns. Thoughtful design around privacy and security, including data segmentation, auditing, and transparency, is critical to protect user data and ensure compliance with privacy regulations.
\subsubsection{Technical Challenges}
There are technical challenges in achieving dense personalized displays, especially for a large number of viewers. As of now, creating a personalized display for up to 5 people is feasible, but scaling up requires a substantial budget and careful planning. Fortunately both of these seem available and it seems timely to look at this option.
\subsubsection{Proof of Concept}
Starting with a small-scale proof of concept for up to 5 people would allow for demonstration of the capabilities and building stakeholder confidence. This would also provide valuable insights into the technical and logistical challenges that may arise during larger-scale implementation.
\subsubsection{Future Developments}
The display technology is rapidly evolving, with new advancements in resolution, refresh rates, brightness, and tracking accuracy. As the technology matures, there will be more opportunities to enhance the personalized experiences.
This system would allow multiple viewers to see different images or perspectives from the same display, enhancing the interactive and educational value of the exhibit. Mollick et al. have done some lovely actionable work on the pedagogical implications of chatbots \cite{mollick2022new, mollick2023assigning, mollick2023using}. This could transform the way visitors engage with exhibits, providing a more immersive and personalized experience.

\subsubsection{Alioscopy}

Alioscopy uses a different approach than lenticular lenses for their glasses-free 3D displays. Their screens contain a parallax barrier - a layer of opaque and transparent slits - over the LCD matrix. This directs different pixel columns to each eye, creating a stereoscopic 3D image without glasses.

Their displays also incorporate proprietary eye tracking technology. An infrared camera follows the viewer's head position, automatically adjusting the angle of the projected 3D image for optimal viewing. This compensates for display viewing angle limitations.

Alioscopy's recent prototypes feature very high resolution like 4K and 8K to improve 3D image quality. Their barriers and tracking algorithms are precisely tuned to the display characteristics and desired viewing parameters.

\subsubsection{Pitch section}
Personalised emergent narratives for our visitors.
What problem does the user, business or industry have that you want to solve?

VisionFlow: knoWhere

For today's digital experience venue managers navigating the complexities of providing unique experiences, our AI solution, KnoWhere, offers a unique approach which will result in the capability to enhance visitor experiences. By utilising images from on-premise cameras, we enable to leverage data on visitors attention. Our solution's unique value propositions include spatial and attention tracking through AI, because of our ability to understand the needs of experience designers. 

It works like this:
Combine personal data, with visitor gaze
Provide location and attention data stream
Venue provides this to experience designers 
Designers build incredible emergent journeys 

We believe this solution will impact our business/industry by:
Elevating interactions through personalisation
Making attention in physical spaces quantifiable
Providing feedback data to experience designers

We will measure our impact by: Performing A/B testing on visitors engagement
This can be a KPI that changes, ex: a productivity score  - or it can be an amount saved because of the soluion

Describe what data is behind this AI model? 
Alphapose (2)
Insightface (5)

Rate the quality/quantity of each point of data from 1-5 (1 being little data / low quality – 5 being lots of data / high quality)

What will be the biggest challenge in implementing the AI model?
Real time pose engine is noncommercial 
Occlusion can be tricky with space constraints
The rich dataset is a privacy concern

Here are some areas to think about in terms of challenges:

Data: How much data exists? How representative is it of what we're trying to model? Are there issues in how it is collected which could impact the model? Is it likely to contain any missing values? 
Adoptance from users/customers
Will it be easy to get people to use the AI in their business?

Governance
Is the data accessible and are you allowed to use it?  Who is responsible once the AI model is in use? How will make the final decisions?

Impact of solution
What do we know about the need for this type of solution - is it nice to have or need to have? Can we find out if we don’t know? Feasibility
What will be the biggest challenge in implementing an AI solution to solve this problem? Can the issue / problem we're solving actually be measured / forecasted?

Ethics

Regulations

Cybersecurity

"Our goal is to empower venue owners to provide an advanced platform that allows world class exhibition creators to tailor unique experiences for each visitor. This enables the crafting of rich, interconnected stories for groups of people, all while ensuring unforgettable, safe experiences for individuals and families.

\section{Accessible metaverse for pre-viz}
Pre-visualization (or "pre-viz") is a process in which a rough simulation of a visual effect or scene is created prior to its actual production. In the context of LED wall virtual production, pre-viz refers to the creation of a 3D representation of a virtual environment, including the placement of cameras, actors, and other elements, that is then used to plan and test the visual effects and lighting for a live-action scene that will eventually be shot in front of an LED wall.\par
The pre-viz process allows filmmakers and visual effects artists to experiment with different camera angles, lighting, and visual effects before committing to a final version. This helps to save time and resources during actual production by reducing the need for multiple takes or re-shoots. Additionally, it allows the filmmakers to see how the final product will look before committing to it, which can help to avoid costly mistakes or changes down the line.\par
The LED wall virtual production process typically involves using a combination of 3D animation software, motion capture technology, and real-time rendering to create a virtual environment that accurately reflects the physical environment in which the scene will be shot. The pre-viz process is then used to plan and test the various visual effects, lighting, and camera angles that will be used in the final production.\par 
Our collaborative software stack is potentially ideally suited to some of this pre-viz work, especially when combined with the power of machine learning, and live linked into Unreal so that changes by stakeholders enter the pre-production pipeline in a seamless way.
\section{Novel VP render pipeline}
\begin{figure}[ht]\centering 	\includegraphics[width=\linewidth]{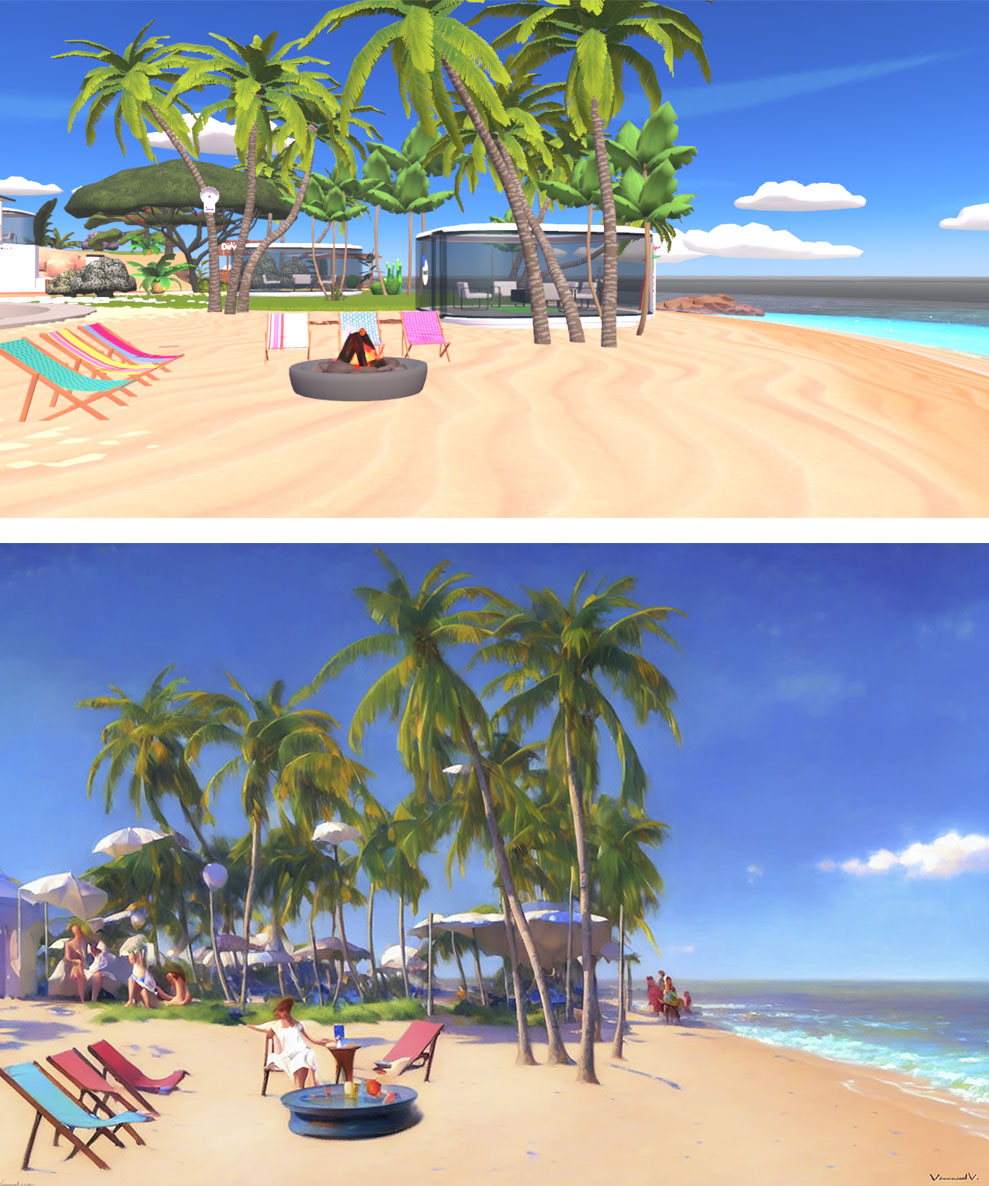}
	\caption{Top panel is a screen grab from Vircadia and the bottom panel is a quick pass through img2img from Stable Diffusion.}
	\label{fig:vircadiasd}
\end{figure}

Putting the ML image generation on the end of a real-time tracked camera render pipeline might remove the need for detail in set building. To describe how this might work, the set designer, DP, director, etc will be able to ideate in a headset based metaverse of the set design, dropping very basic chairs, windows, light sources whatever. There is -no need- then to create a scene in detail. If the interframe consistency (img2img) can deliver then the output on the VP screen can simply inherit the artistic style from the text prompts, and render production quality from the basic building blocks. Everyone in the set (or just DP/director) could then switch in headset to the final output and ideate (verbally) to create the look and feel (lens, bokeh, light, artistic style etc). This isn’t ready yet as the frames need to generate much faster (100x), but it’s very likely coming in months not years. This ``next level pre-vis'' is being trailed in the Vircadia collaborative environment described in this book, and can be seen illustrated in Figure \ref{fig:vircadiasd}.\par

This can be done now through the use of camera robots. A scene can be built in basic outline, the camera tracks can be encoded into the robot, and the scene can be rapidly post rendered by Stability with high inter frame consistency.\par
With the help of AI projects such as \href{https://nv-tlabs.github.io/LION/}{LION} it may be possible to pass simple geometry and instructions to ML systems which can create complex textured geometry back into the scene.
\begin{figure}[ht]\centering 	\includegraphics[width=\linewidth]{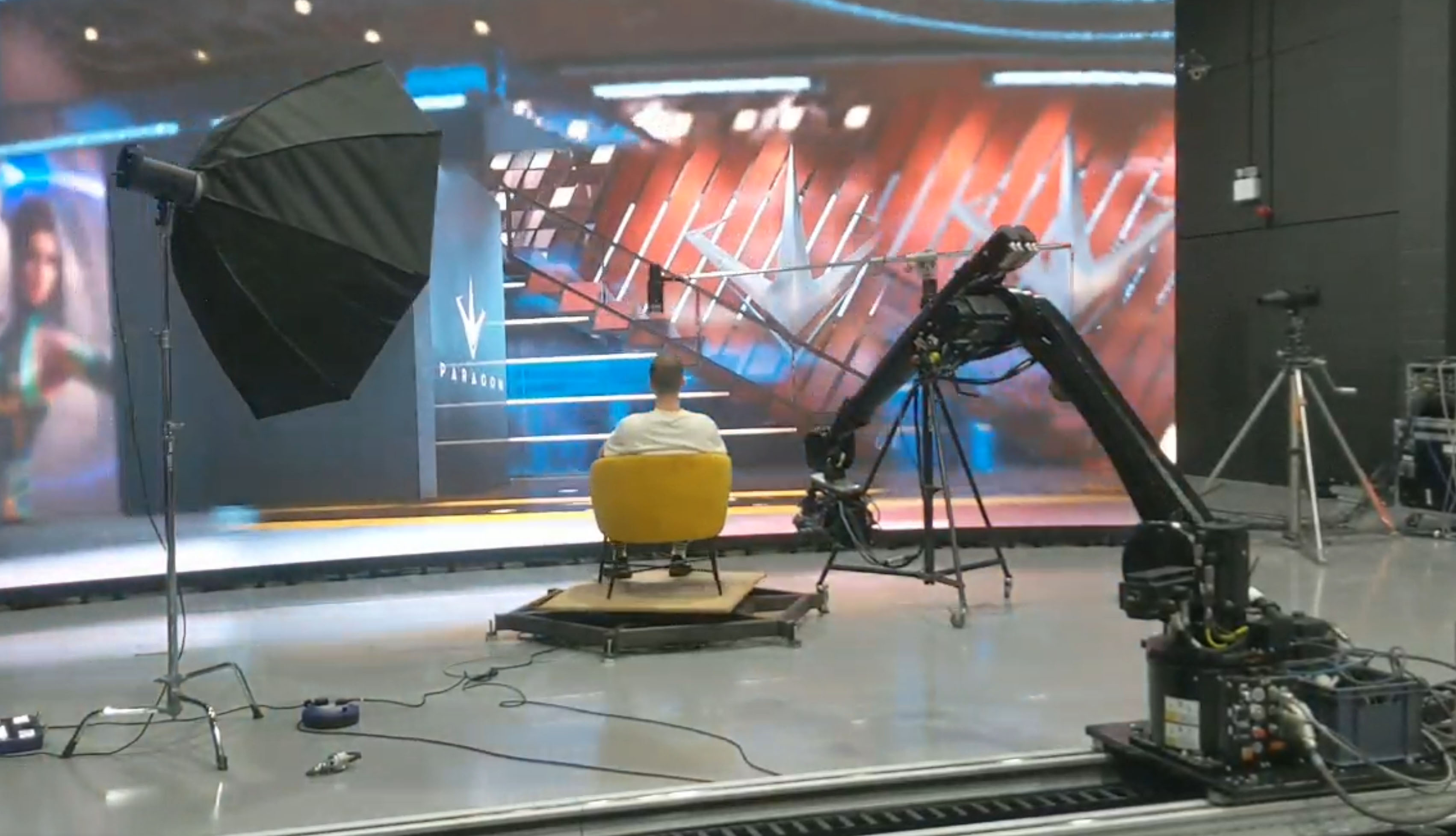}
	\caption{Robot VP}
	\label{fig:robotvp}
\end{figure}

\section{Money in metaverses}
\subsubsection{Global collaboration and remuneration}
In the book ``Ghosts of my life'' \cite{fisher2014ghosts} Fisher asserts that there has been a slowing, even a `cancellation' of creative progress in developed societies, their art, and their media. His contention is that neoliberalism itself is to blame. He says\\
\textit{``It is the contention of this book that 21st-century culture is marked by the same anachronism and inertia which afflicted Sapphire and Steel in their final adventure. But this status has been buried, interred behind a superficial frenzy of ‘newness’, of perpetual movement. The ‘jumbling up of time’, the montaging of earlier eras, has ceased to be worthy of comment; it is now so prevalent that it is no longer even noticed.''}

It is the feeling of the authors of this book that the creative and inspirational efforts of the whole world may be needed to heal these deep wounds. It is possible that by connecting creatives with very different global perspectives, directly into `Western' production pipelines, that we will be able to see the shape of this potential.
\subsection{ML actors and blockchain based bots}
Stablity AI is an open source imitative to bring ML/AL capabilities to the world. This is a hugely exciting emergent area and much more will be developed here.
\subsection{AI economic actors in mixed reality}
AI actors can now be trusted visually \cite{nightingale2022ai}. We have some thinking on this which links the external web to our proposed metaverse. There is work in the community working on economically empowered bots which leverage Nostr and RGB to perform functions within our metaverse, and outside in the WWW, as well as interacting economically through trusted cryptography with other bots, anywhere, and human participants, anywhere. This is incredibly powerful and is assured by the Bitcoin security model. Imagine being able to interact with a bot flower seller representing all the real world florists it had found. In the metaverse you could handle the flowers and take advice and guidance from the bot agent, then it would be able to take your money to buy you flowers to send to a real world address, and later find you to tell you when it's delivered. These possibilities are endless. The AI chat element, the AI translation of images on websites to 3D assets in the Metaverse are difficult but possible challenges, but the secure movement of money from the local context in the metaverse to the real world is within reach using these bots, and they are completely autonomous and distributed.
\section{Our socialisation best practice}
\subsubsection{Identity}
We will base our identity and object management on Nostr public/private key pairs. The public key of these enable lightning based exchange of value globally. 
we plan to operate Nostr in multiple modes. Linking flossverse ``rooms'' will be a Nostr bot to bot system within the private relay mesh. This can also synchronise large amounts of data by leveraging torrents \href{https://iris.to/#/settings}{negotiated by Nostr}. Human to human text chat across and within instances is two 'types' kind of private nostr tag within the private relays mesh. External connectivity to web and nostr apps is just the public relay tags outbound. We don't need to store data external to the flossverse system, though access is obviously possible through the same torrent network.
\subsubsection{Webs of trust}
Webs of trust will be built within worlds using economically costly (but private) social rating systems, between any actor, human or AI. It should be too costly to attack an individual aggressively. This implies an increased weighting for scores issued in short time periods. Poorly behaving AI's will eventually be excluded through lack of funds.
\subsubsection{Integration of 'good' actor AI entities}
Gratitude practice should be encouraged between AI actors to foster trust and wellbeing in human observers. ``It's nice to be nice'' should be incentivised between all parties''. This could include tipping and trust nudging through the social rating system. Great AI behaviour would result in economically powerful entities.
\subsection{Emulation of important social cues}
\href{https://www.cleverclassroomsdesign.co.uk/general-5}{Classroom layout}
\subsubsection{Behaviour incentives, arbitration, and penalties}
Collapses of trust and abuse will trigger flags from ML based oversight, which will create situational records and payloads of involved parties to unlock with their nostr private keys. ML red flagged actors will be finacially penalised but have access to human arbitration using their copy of the data blob. Nothing will be stored except by the end users.
\subsection{Federations of webs of trust and economics}
Nostr is developing fast enough to provide global bridges between metaverse instances.
\section{Security evaluation}
As part of developing our stack we will penetration test the deployment as detailed using \href{https://hexway.io/}{Hexway}

\section{Potential applications }
\begin{itemize}
\item Art / NFT galleries with instant sales
\end{itemize}

This application allows artists and content creator communities to
display and sell NFT and fungible art to global consumer audiences,
instantly.

\begin{itemize}
\item
  Large scale conference center

  \begin{itemize}
  \item
    Academic conferences
  \item
    Political conference
  \item
    Commercial expo
  \end{itemize}
\end{itemize}

In a hypothetical virtual conference centre a true marketplace of ideas
could be enacted, with participants being paid directly by their
audience based on the proximity to the presentation.

\begin{itemize}
\item
  Group entertainment

  \begin{itemize}
  \item Global social puzzle gaming with prizes
  \item
    Music festivals and gigs - Pay live artists and DJs in real time
    depending on location within the extended landscape of the venue.
    Split to music producer a portion of the value
  \item
    Mixed reality theatre
  \item
    murder mystery
  \item
    Mixed reality live immersive MMORG games
  \item
    Bingo and mass participation gameshows
  \item
    Immersive brand storytelling metaverses
  \item
    Escape rooms
  \end{itemize}
\item
  Debating townhall meetings (with voting etc)
\item
  Mixed reality information metaverse

  \begin{itemize}
  \item
    AR based city tours with collectibles
  \item
    AR based collectibles for trails and heritage (museums, libraries)
    with location specific donations.
  \end{itemize}
\item
  Retail applications

  \begin{itemize}
  \item
    Proxy for physical market
  \item
    AR home delivery market interface within physical marketplaces
  \end{itemize}
\item
  Global course / Education provision
    \begin{itemize}
  \item
    Explore the universe as a group of spaceship or planet characters
  \item
    Explore biology and physics at a microscopic and nanoscopic level
  \end{itemize}
\item
  Micro tasking marketplace
\item
  Code bounty marketplace
\item
  Micro remittance role sharing (business PA / reception etc)
\item
  Careers fair with credential passing
\item
  Auctions in mixed reality
\item
  eSports and live sports
\item
  Gambling, betting markets, and financial leverage markets
\end{itemize}

\subsection{Global cybersec course delivery}
Isolating and building out one example here:
\begin{itemize}
\item Elements for the infrastructure: Economic layer, asset layer, content interface, user management, data storage, microsites loaded in Wolvin and webm, accessibility schema, network security, backups, secure messaging. Deployable framework with high modularity. Some more ossified elements for surity, some less so for malleability and open opportunity. Figure \ref{fig:globalclassroom}.
\item Course delivery in XR, how to we develop a platform, marketplace, framework for open contribution.
\item WebXR, Vircadia, any snap in metaverse middleware that is free and open source (action to compare the two). 
\item Define an interface schema for bolting in any commercial or FOSS metaverse engine.
\item VR marketplace (outside the scope of the VR engine) without a trusted third party.
\item Cryptographically managed learning deliverables (coursework as NFT). 
\item Secure messaging and group messaging using cryptographic keys. Check this stuff with the distributed computing science people in the group (action on John)
\item work toward an exemplar MVP which is then "in the wild"
\item Platform for educators
\item Define scheme, documentation, best practice, interfaces, functional objects, pedagogy, accessibility, multi-language. 
\item Define user management system for educators and client learners.
\item Identify the pain points which current FOSS elements which need development time/money
\item separate the UI/engine from the graphical assets, and the educational / pedagogical components, accessibility, and the value and asset transfer layers.
\item Desktop systems are the primary target (low end system)
\item define schema for accessibility. Colour, subtitles, immersion concerns which can be applied to metaverse rooms through API?
\item Start to define the hybrid presentation model we favour. Avatars? Micro sites? A combination of the two? Balance of guided vs unguided experience. Do we need to test the correct way to do delivery? Is there prior art we can draw on? I feel I should know. Is this part of the research that's being done here?
\item Big work package on schema vs key and user management to enforce rules in spaces. Only participants who have provably paid should have access to learning material, the ability to input into the assessment system, and the tokenised learning outcome `NFT' or proof.
\item Proof that XR system improve learning outcomes. Also that the proposed systems for micro-transactions and user and schema management give additional headroom for teaching.
\end{itemize}
\subsection{Subvocalisation with LLMs}
MIT \href{https://dspace.mit.edu/handle/1721.1/123121}{full thesis} on AlterEgo
\begin{tcolorbox}[enhanced, frame style={fill=lightgray}, interior style={fill=lightgray}]
By transcribing residual neurological signals sent from the brain to speech articulators during internal articulation, the system allows one to communicate without the need to speak or perform any visible movements or gestures. It is capable of transcribing continuous silent speech at a rate of over 100 words per minute. The system therefore provides a natural alternative to normal speech at a rate not far below that of conversational speech. This alternative method of communication enables those who cannot speak, such as people with speech or neurological disorders, as well as those in environments not suited for normal speech, to communicate more easily and quickly. In the same capacity, it can serve as a discreet, digital interface that augments the user with information and services without the use of an external device. I discuss herein the data processing and sequence prediction techniques used, describe the collected datasets, and evaluate various models for achieving such a continuous system, the most promising among them being a deep convolutional neural network (CNN) with connectionist temporal classification (CTC).
\end{tcolorbox}
\subsection{AI 3D objects in metaverse}
Onmiverse from Nvidia apparently have this \href{https://www.nvidia.com/en-us/gpu-cloud/picasso/}{pretty much sewn up}, but they are an expensive \href{https://developer.nvidia.com/blog/rapidly-generate-3d-assets-for-virtual-worlds-with-generative-ai/}{commercial offering}. \par
Notably they now have a \href{https://blogs.nvidia.com/blog/2023/06/27/magiscan-app-augmented-reality-openusd/?}{consumer product}. \par
\begin{figure}[H]
    \centering
    \includegraphics[width=0.95\textwidth]{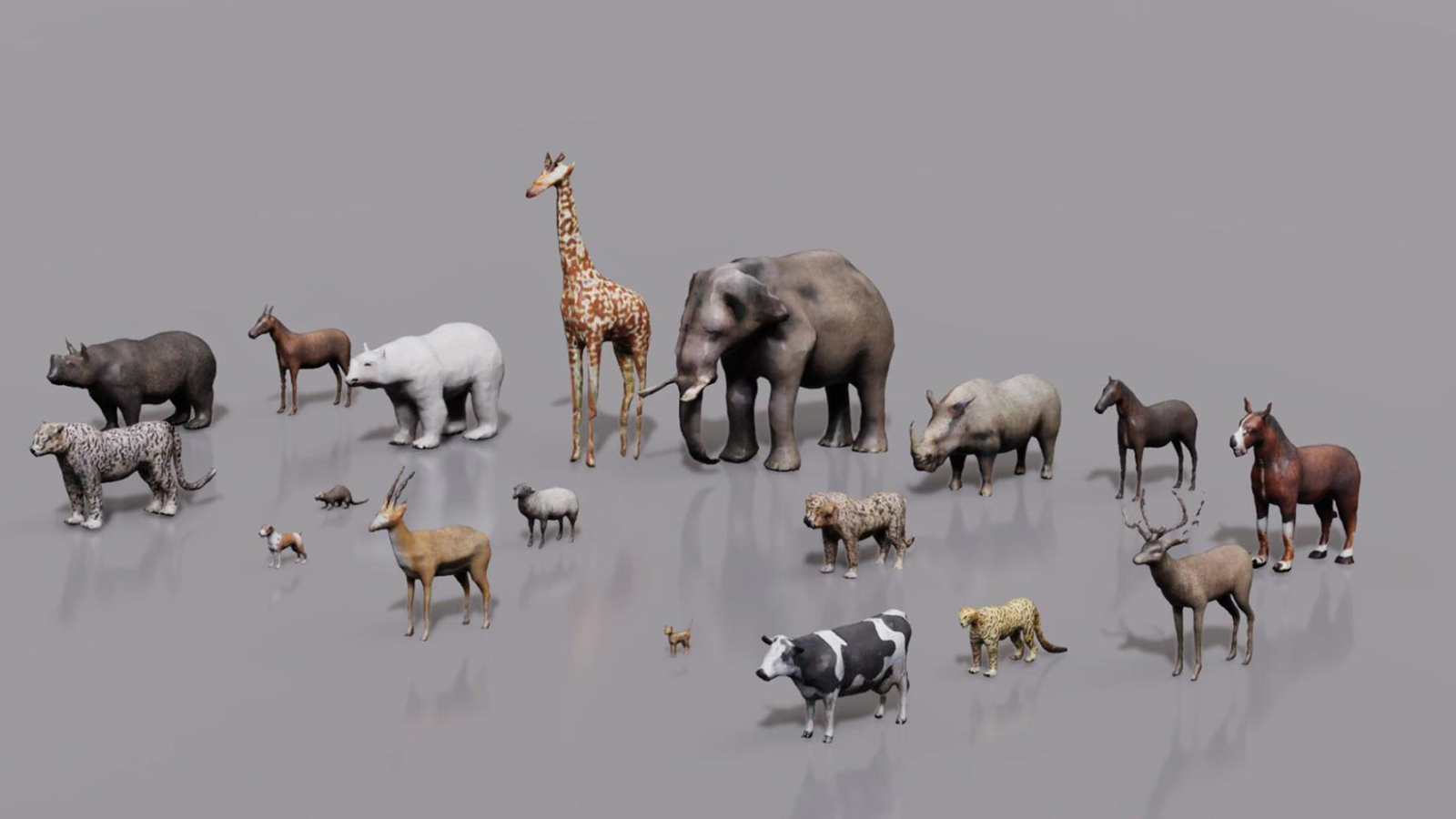}
    \caption{Omniverse allows voice created 3D models.}
    \label{fig:nvidiavoice}
\end{figure}

Industry insiders I have talked to say they thing the Nvidia Omniverse initiative is failing, and this has been my impression too. The results can be seen in Figure \ref{fig:nvidiavoice}.
Their voice to facial animation stack Audio2Face is outstanding, but it's unclear how this might challenge Unreal's Metahuman. More promisingly Nvidia have a new video to NeRF to mesh technique called Neuralangelo \cite{li2023neuralangelo} which knocks everything else out of the park. 
\begin{tcolorbox}[enhanced, frame style={fill=lightgray}, interior style={fill=lightgray}]The new AI model by NVIDIA Research for 3D reconstruction using neural networks, turns 2D video clips into detailed 3D structures — generating lifelike virtual replicas of buildings, sculptures and other real-world objects. Like Michelangelo sculpting stunning, life-like visions from blocks of marble, Neuralangelo generates 3D structures with intricate details and textures. Creative professionals can then import these 3D objects into design applications, editing them further for use in art, video game development, robotics and industrial digital twins.
\end{tcolorbox}
This area of 3D generation from various high level inputs is moving \textbf{very} fast and much of the high end stuff is coming as software that can be built on (like NeRF). An open source text to model pipeline for interactively building key assets for pre-viz, or project planning has appeal  \cite{poole2022dreamfusion}.
There's a lot here but you'd need to be sure that you're not competing directly with Omniverse, by differentiating the product.

\subsection{Digital collectibles for events}
I've been following the NFT and digital objects space since it started, but I have no interest in Ethereum or other NFT technologies because I have been waiting for what I regard as better technology. By luck my preferred option is going live as I write this document, and I am a tester on the programme. I have been talking to those teams since early 2021 and I think I understand it fairly well. There is a first mover advantage here. \href{https://coinmarketcap.com/alexandria/article/bitcoin-is-now-the-second-most-popular-blockchain-for-nfts}{Digital objects on Bitcoin} are the second most popular route after Ethereum now, and might overtake. They split into a bunch of catagories:
\begin{itemize}
\item Ordinals inscriptions, which I have been following since they were posited. I have minted one successfully, but I consider them wasteful and expensive for commercial projects with large audiences.
\item BRC-20 tokens. These are comparatively new, again leveraging the ordinal technique. All this is written up in detail in my book and it's all over the internet. BRC-20 creates ``batches'' of fungible tokens and can be approximated to event tickets, unique but interchangable within their context. They are trivial to make but I am unsure of the business case as they don't \textit{trigger} that unique thing that people seem to like. They're the new meme pump and dump token of choice, with associated negative press in tow.
\item Taproot Assets. This is working on testnet still I think, and is a naked copy of RGB backed by shifty stablecoin types. Might win because of money, worth watching.
\item Pear credits. I think these are failing to get traction but are an honourable mention.
\item Tokens on Liquid sidechain. Fringe, complex, but very heavily invested in and developing quietly makes this one to watch longer term.
\end{itemize}
My use case for these things has always been the cheap creation of a universal provenance for any and all objects and actions in a persistent digital collaborative space. Unlike Ethereum they're basically free, which opens up huge opportunity space. I can unpack this but it really needs to be application led.
\subsubsection{Preferred options: LNP/BP and RGB}
\href{https://giacomozucco.com/layers-before-bitcoin}{LNP/BP} is a non profit standards organisation in Switzerland which contributes to open source development of Bitcoin layer 3 solutions into the Lightning protocol, and Bitcoin protocol (LNP/BP). One of the core product developments within their work is the \href{https://www.rgb.tech/}{`RGB' protocol}, which is somewhat of a meaningless name, evolved from ``coloured coins'' which were an early tokenised asset system on the Bitcoin network. RGB represents red, green, and blue. The proposal is built upon research by \href{https://petertodd.org/2016/commitments-and-single-use-seals}{Todd} and \href{https://giacomozucco.com/#intro}{Zucco}. RGB is regarded as arcane Bitcoin technology, even within the already rarefied Bitcoin developer communities. Zucco provides the \href{https://bitcoinmagazine.com/culture/video-interview-giacomo-zucco-rgb-tokens-built-bitcoin}{following explanation}: \par
\textit{``When I want to send you a bitcoin, I will sign the transaction, I will give the transaction only to you, you will be the only one verifying, and then we’ll take a commitment to this transaction and that I will give only the commitment to miners. Miners will basically build a blockchain of commitments, but without the actual validation part. That will be only left to you. And when you want to send the assets to somebody else, you will pass your signature, plus my signature, plus the previous signature, and so on.''}\par
This is non-intuitive explanation of Todds `single-use-seals', applied to Bitcoin, with the purpose of underpinning arbitrary asset transfer secured by the Bitcoin network. In this model the transacting parties are the exclusive holders of the information about what the object they are transferring actually represents. This primitive can (and has) been expanded by the LNP/BP group into a concept called `client side validation'. 
It's appropriate to explain this concept several times from different perspectives, because this is potentially a profoundly useful technology for metaverse applications.\par
\begin{itemize}
\item A promise is made to spend a multi output transaction in the future. This establishes the RGB relationships between the parties.
\item One of the pubkeys to be spent to is known by both parties.
\item The second output is unknown and is a combination of the hash of the state, and schema, from the operation which has been performed.
\item When the UTXO is spent the second spends pubkey can be processed against the shared data blob to validate the shared state in a two party consensus  (sort this out, it's nonsense).
\item This is now tethered to the main chain. Some tokens from the issuance have gone to the recipiant, and the remainder have gone back to the issuer. More tokens can be issued in the same way from this pool. 
\item A token schema in the blob will show the agreed issuance and the history back to the genesis for the token holder. 
\item The data blob contains the schema which is the key to RGB functions and the bulk of the work and innovation. 
\item Each issuance must be verified on chain by the receiving party. 
\end{itemize} 
This leverages the single-use-seal concept to add in smart contracts, and more advanced concepts to Bitcoin. Crucially, this is not conceptually the same as the highly expressive `layer one' chains which offer this functionality within their chain logic. In those systems there is a globally available shared consensus of `state'. In the LNP/BP technologies the state data is owned, controlled, and stored by the transacting parties. Bitcoin provides the crytographic external proof of a state change in the event of a proof being required. This is an elegant solution in that it takes up virtually no space on the blockchain, is private by design, and is extensible to layer 2 protocols like Lightning.\par
This expanding ecosystem of client side verified proposals is as follows:
\begin{itemize}
\item RGB smart contracts
\item RGB assets are fungible tokens on Bitcoin L1 and L2, and non fungible Bitcoin L1 (and somewhat on L2).
\item Bifrost is an \href{https://github.com/LNP-BP/presentations/blob/master/Presentation slides/Bifrost.pdf}{extension} to the Lightning protocol, with it's own Rust based node implementation, and backwards compatibility with other nodes in the network. This means it can transparently participate in normal Lightning routing behaviour with other peers. Crucially however it can also negotiate passing the additional data for token transfer between two or more contiguous Bifrost enabled parties. This can be considered an additional network liquidity problem on top of Lightning, and is the essence of the ``Layer 3'' moniker associated with LNP/BP. It will require a great number of such nodes to successfully launch token transfer on Lightning. As a byproduct of it's more `protocol' minded design decisions Bifrost can also act as a generic peer-to-peer data network, enabling features like Storm file storage and Prometheus.
\item \href{https://www.aluvm.org/}{AluVM} is a RISC based virtual machine (programmable strictly in assembly) which can execute Turing complete complex logic, but only outputs a boolean result which is compliant with the rest of the client side validation system. In this way a true or false can be returned into Bitcoin based logic, but be arbitrarily complex within the execution by the contract parties.
\item Contractum is the proposed smart contract language which will compile the RGB20 contracts within AluVM (or other client side VMs) to provide accessible layer 3 smart contracts on Bitcoin. It is a very early proposal at this stage.
\item  Internet2: ``Tor/noise-protocol Internet apps based on Lightning secure messaging
\item Storm is a lightly specified escrow-based bitcoin data storage layer compliant with Lightning through Bifrost.
\item Prometheus is a lightly specified multiparty high-load computing framework.
\end{itemize}
Really, any compute problem can be considered applicable to client side validation. In simplest terms a conventional computational problem is solved, and the cryptographically verifiable proof of this action, is made available to the stakeholders, on the Bitcoin ledger.\par 
Less prosaically, at this stage of the project the more imminent proposed affordances of LNP/BP are described in `schema' \href{https://github.com/LNP-BP/LNPBPs}{on the project github}. The most interesting to the technically minded layperson are:
\begin{itemize}
\item \href{https://github.com/LNP-BP/LNPBPs/blob/master/lnpbp-0020.md}{RGB20} fungible assets. This could be stablecoins like dollar or pounds representation. Bitfinex exchange \href{https://github.com/RGB-Tools/rgb-lightning-sample}{have code} which already works with RGB to transmit Tether stablecoins on testnet. This is a huge application area for Bitcoin, and similar to Omni, which will also be covered next.
\item \href{https://github.com/LNP-BP/LNPBPs/blob/master/lnpbp-0021.md}{RGB21} for nonfungible tokens and ownership rights. In principle BiFrost allows these to be transferred over a the Lightning network, significantly lowering the barrier to entry for this whole technology. DIBA \href{https://diba.io/}{have this technology working} on testnet.
\item \href{https://github.com/LNP-BP/LNPBPs/issues/29}{RGB22} may provide a route to identity proofs. This is covered in detail later.
\end{itemize}
Federico Tenga is CEO of `Chainside' and an educator and consultant in the space. He has written an up-to-date \href{https://medium.com/@FedericoTenga/understanding-rgb-protocol-7dc7819d3059}{``primer''}, which is still extremely complex for the uninitiated, but does capture how the RGB token transfer system works. That medium article also touches on Taro, which is next.
\subsubsection{DIBA and Unique Digital Assets}
\textit{DIBA} is a pioneering digital asset marketplace, powered by the \href{https://www.rgb.tech/}{RGB Smart Contract Protocol}. It permits the creation and direct transaction of Unique Digital Assets (UDAs), akin to Non-Fungible Tokens (NFTs), on Bitcoin without the necessity of other tokens. UDAs are special digital assets linked to a Bitcoin UTXO (unspent transaction output). These assets embody distinctive attributes like ownership, transferability, and divisibility, and remain under the full control and ownership of their creators.\par
Through DIBA, users can explore, purchase, and sell a vast array of UDAs, taking advantage of the robustness and permanence of the Bitcoin blockchain. DIBA's innovation extends to the integration of a Lightning layer 2 solution, aiming to facilitate faster and more affordable transactions.\par
Assets minted via DIBA are bound to Bitcoin's base layer with an on-chain UTXO. They are subsequently stored on the Arweave permaweb alongside a cryptographic hash, which, combined with a digital signature, can validate their authenticity. The RGB Smart Contract Protocol executes UDA transactions via BitMask, a wallet engineered by the DIBA Team.
\subsubsection{BitMask}
\textit{BitMask} is a browser extension wallet birthed by DIBA, intended for decentralized applications on Bitcoin. It grants access to Bitcoin Finance, UDAs, and more, utilizing the RGB protocol. It delivers comprehensive financial autonomy with its taproot-enabled Bitcoin and Lightning Network wallet, establishing it as a gateway to DeepWeb3 on Bitcoin. More details can be found on Bitmask.app.\par
\subsubsection{DIBA's Launch and Marketplace Timelines}
DIBA has initiated an Open Marketplace for Beta testing as of April 2022, available on Bitcoin testnet. The full launch on Bitcoin mainnet is expected to occur in the second or third quarter of 2022. Submissions for the DIBA Curated Marketplace are currently open, allowing interested individuals to apply as an Artist or a Curator.\par
These developments represent a substantial expansion of the capabilities within the Bitcoin network and further attest to the potential of the LNP/BP's work. Notably, DIBA \href{https://diba.io/}{has this technology working} on testnet. The extensive application areas for Bitcoin, such as the transmission of Tether stablecoins on testnet by Bitfinex exchange \href{https://github.com/RGB-Tools/rgb-lightning-sample}{via RGB}, emphasize the potential impact of these advancements.
\subsection{Product Design}
\subsubsection{User Journey}
\begin{itemize}
    \item \textbf{Discovery \& Onboarding}: User downloads and opens the app on their phone or accesses the web application on their computer.
    \item \textbf{Interaction with LLM}: User starts conversing with the open-source AI large language model (LLM).
    \item \textbf{Storyboard Creation}: User co-creates the story arc with the LLM. 
    \item \textbf{Feedback \& Iteration}: User can provide feedback on the story subdivisions and the balance between visual narrative, explanatory text, and dialogue. 
    \item \textbf{Art Style Selection}: User selects an artistic style for the storyboard and provides additional contextual information.
    \item \textbf{Preliminary Image Generation}: The image AI server generates five visual styles based on the chosen art style and additional context.
    \item \textbf{User Interaction with Panels}: The app generates ten blank panels with a small contextual synopsis each.
    \item \textbf{Initial Image Rendering}: The Image AI uses the user-provided data to render ten panels into a single 1024x1024 image.
    \item \textbf{Final Image Rendering \& Text Placement}: The AI subdivides and upscales the chosen image into ten high-quality panels.
    \item \textbf{Final Review \& Confirmation}: After the user has arranged the text to their liking, the AI performs a final rendering pass to create a multi-page comic book-style PDF. 
    \item \textbf{Purchase/Download}: The user has the option to purchase the PDF as a DIBA-based NFT (UDA), or simply save it to their phone or computer.
\end{itemize}

\subsubsection{Technical Overview}
\begin{itemize}
    \item \textbf{Frontend}: A web and `progressive web app' mobile interface with a clean, intuitive style for user interaction.
    \item \textbf{Backend}: The application leverages an open-source LLM trained through Lora and aligned with new \href{https://arxiv.org/abs/2306.17806}{CFG techniques}, for the narrative portion of the comic. This will run on a Lamda cloud compute system, initially an A100 and as the user base scales an H100. It also uses a highly capable Nvidia H100 based server system with open-source generative art capabilities (\href{https://github.com/comfyanonymous/ComfyUI/blob/master/script_examples/basic_api_example.py}{ComfyUI API} for image rendering. The rental cost per year of these systems is around \$30k, without administration costs.
\end{itemize}

\subsubsection{User Stories}
\begin{itemize}
    \item As a user, I want to interact with an AI to create a storyboard so that I can craft a narrative that matches my vision.
    \item As a user, I want to provide feedback on the AI-generated subdivisions of my story to ensure they align with my creative goals.
    \item As a user, I want to choose an art style for my storyboard to suit my aesthetic preferences.
    \item As a user, I want to draw and add descriptive text to my panels to give more context to the AI.
    \item As a user, I want to choose my favourite among multiple AI-generated renderings to have a say in the final visual output.
    \item As a user, I want to arrange text assets on my storyboard to ensure the narrative flow and visual balance is to my liking.
    \item As a user, I want the option to buy my storyboard as a NFT or save it to my device for maximum flexibility in how I use and share my creation.
\end{itemize}

\section{notes for later}
Notes on build-out
The world database in the shared rooms in the metaverse is the global object master,  educational materials, videos,  audio content and branded objects are fungible tokens authentically proved by rgb client side validation between parties,  only validated ones will be persisted in shared rooms like conferences and classes according to the room schema. That allows educators to monetise their content.  That can work on lightning.  NFT objects between parties like content crafted by participants (coursework, homework) are not on lightning and will attract main chain fees but are rarer. User authentication and communication will be through nostr.

\begin{figure*}[ht]\centering 
	\includegraphics[width=\linewidth]{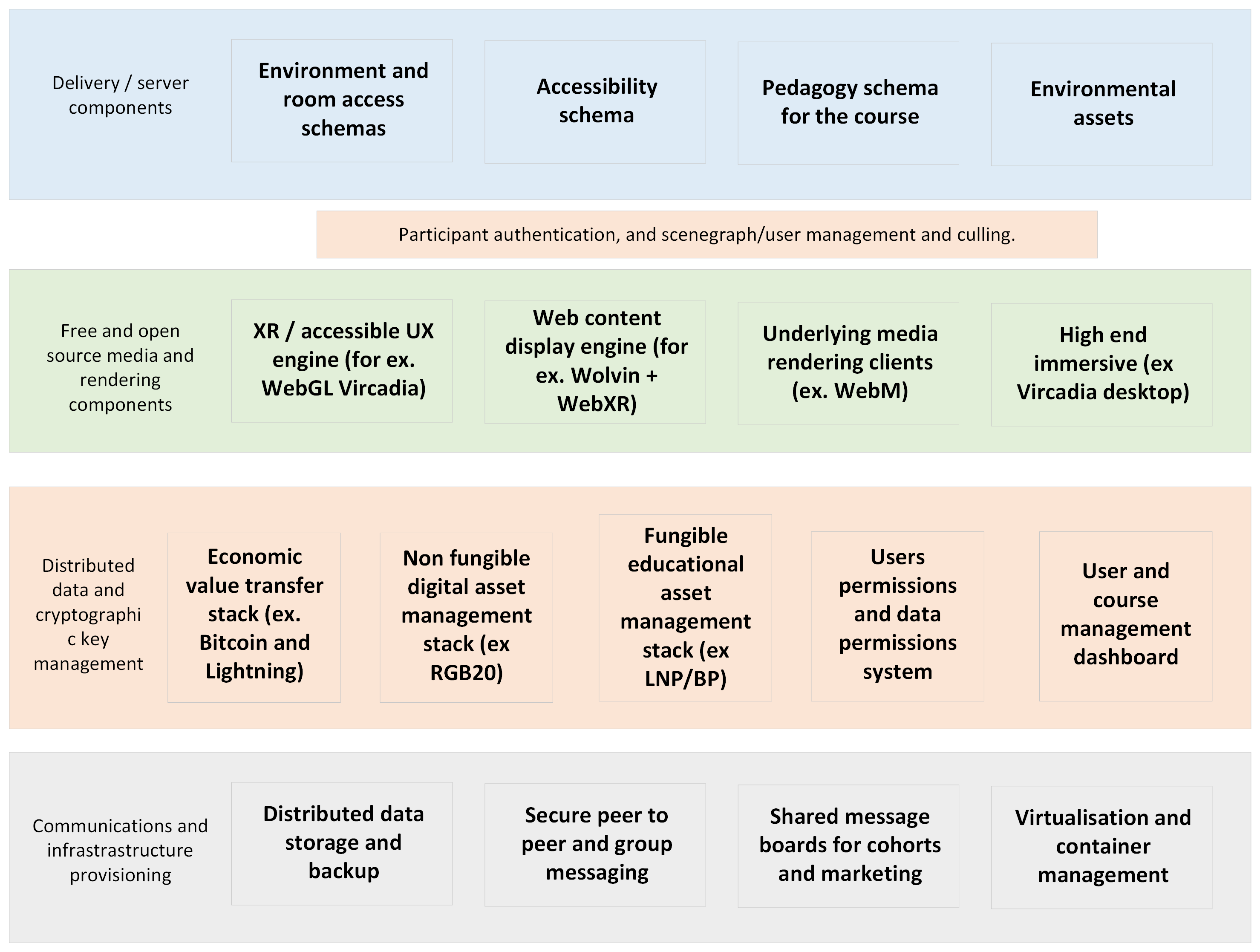}
	\caption{Functional elements for infrastructure.}
	\label{fig:globalclassroom}
\end{figure*}

\begin{figure*}[ht]\centering 
	\includegraphics[width=\linewidth]{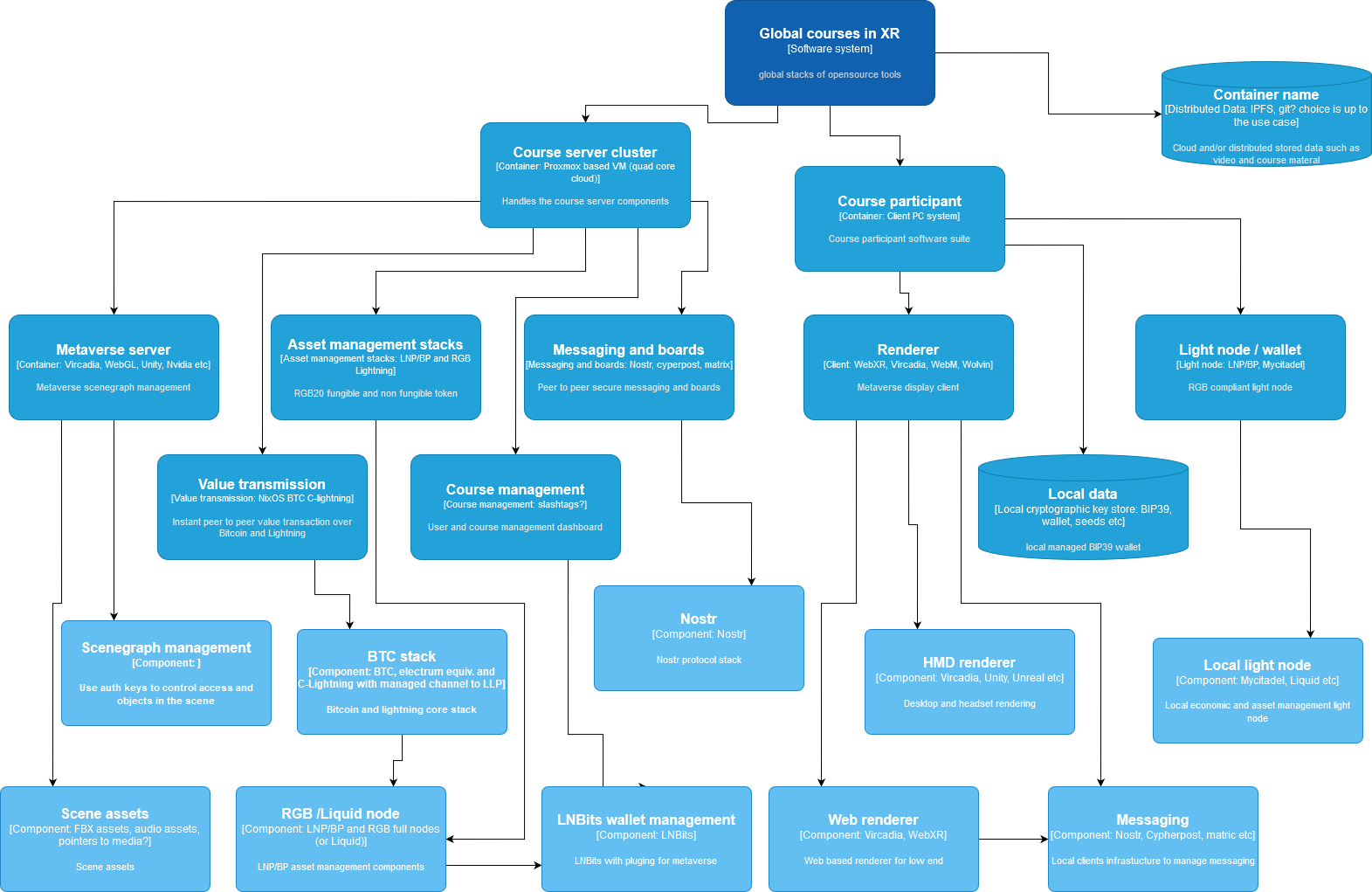}
	\caption{Client server C4 diagrams.}
	\label{fig:globalclassroom}
\end{figure*}

\part{Guides for deploying\\the software}
\hypertarget{lab---virtualisation-networking-bitcoin}{%
\section{Lab - virtualisation, networking,
Bitcoin}\label{lab---virtualisation-networking-bitcoin}}

\hypertarget{overview}{%
\subsection{Overview}\label{overview}}

This how-to document details the process of creating the system detailed
in the accompanying book. It is intended to be complete. It is a how to
guide.

\hypertarget{summary-of-software}{%
\subsubsection{Summary of software}\label{summary-of-software}}

Summarise the software and functionality

\hypertarget{prerequisites}{%
\subsection{Prerequisites}\label{prerequisites}}

Ensure that the BIOS / firmware / etc of the hardware you intend to use
is up to date.

\hypertarget{network-details}{%
\subsection{Network details}\label{network-details}}

In the example setup provided here there are currently two networks:

\begin{enumerate}
\def\labelenumi{\arabic{enumi}.}
\tightlist
\item
  The virtual server resides in a LAN with the following details:
\end{enumerate}

192.168.x.0/24

Replace x with an integer between 0 and 254

This LAN has a gateway to the Internet and DNS server configured. Of
course, it could be replaced with a direct connection to the Internet,
though for research and development purposes it is often better to work
within a clean LAN and manage access to the Internet as required.

\begin{enumerate}
\def\labelenumi{\arabic{enumi}.}
\setcounter{enumi}{1}
\tightlist
\item
  There is a virtual network configured on the virtual machine host upon
  which virtual machines can reside:
\end{enumerate}

This virtual network is not configured to bridge with the physical
network adapter rather a virtual machine is configured as a gateway to
route IP traffic through. This provides a level of isolation. More on
this later (@todo).

\hypertarget{server-configuration}{%
\subsection{Server configuration}\label{server-configuration}}

\hypertarget{server-hardware-details}{%
\subsubsection{Server hardware details}\label{server-hardware-details}}

@todo

\hypertarget{disk-configuration-details}{%
\subsubsection{Disk configuration
details}\label{disk-configuration-details}}

@todo

\hypertarget{proxmox-ve}{%
\subsection{Proxmox VE}\label{proxmox-ve}}

\hypertarget{installation-and-configuration}{%
\subsubsection{Installation and
configuration}\label{installation-and-configuration}}

Version used: 7.1

Keep in mind that this setup uses the Proxmox VE installer
(https://www.proxmox.com/en/proxmox-ve/get-started) which, as noted on
the site, is a bare-metal installer and will erase all data on at least
one disk. There are alternative methods to install Proxmox VE but these
are not covered here.

A brief summary of the steps taken using Proxmox VE 7.1:

\hypertarget{dialogue-1}{%
\paragraph{Dialogue 1}\label{dialogue-1}}

Choose the target harddisk (/dev/sda in this case).

\hypertarget{dialogue-2}{%
\paragraph{Dialogue 2}\label{dialogue-2}}

Select country, time zone, keyboard layout.

\hypertarget{dialogue-3}{%
\paragraph{Dialogue 3}\label{dialogue-3}}

Set a password (this is the root password, see proxmox hardening
section), and email address.

\hypertarget{dialogue-4}{%
\paragraph{Dialogue 4}\label{dialogue-4}}

Select a Network Interface Card (NIC) on which the management interface
will be available and provide a hostname, IP address, gateway and DNS
server.

In this example the following settings were used:

Hostname: proxmoximus.local IP address: 192.168.x.220 / 24 Gateway:
192.168.x.254 DNS server: 192.168.x.254

Either replace x with the integer used earlier and update the last octet
of the gateway and server with that that corresponds to your setup
(assuming the setup is local and has a local dns server or forwarder) or
configure the values according to your intended setup.

Once the install has completed and the system has rebooted it is time to
begin configuring. This is done (almost entirely) via the web interface,
in this case, available at https:// proxmoximus \textbar{} 192.168.x.220
: 8006

It is also possible to login to a shell via the local terminal and SSH
(which is enabled by default @todo: in hardening, add keys and remove
ability to login with password).

\hypertarget{software-updates}{%
\subsubsection{Software updates}\label{software-updates}}

If you are in a testing and non-production environment then it is
possible access updates without a subscription as detailed here:
https://pve.proxmox.com/wiki/Package\_Repositories. Update
\texttt{/etc/apt/sources.list} as detailed under the Proxmox VE
No-Subscription Repository. This can be achieved via the local terminal,
SSH or web interface (via Shell option).

For example, edit the file:

\texttt{nano\ /etc/apt/sources.list}

Add the following:

\begin{verbatim}
# PVE pve-no-subscription repository provided by proxmox.com,
# NOT recommended for production use
deb http://download.proxmox.com/debian/pve bullseye pve-no-subscription
\end{verbatim}

To the existing:

\begin{verbatim}
deb http://ftp.uk.debian.org/debian bullseye main contrib

deb http://ftp.uk.debian.org/debian bullseye-updates main contrib

# security updates
deb http://security.debian.org bullseye-security main contrib
\end{verbatim}

Resulting in:

\begin{verbatim}
deb http://ftp.debian.org/debian bullseye main contrib
deb http://ftp.debian.org/debian bullseye-updates main contrib

# PVE pve-no-subscription repository provided by proxmox.com,
# NOT recommended for production use
deb http://download.proxmox.com/debian/pve bullseye pve-no-subscription

# security updates
deb http://security.debian.org/debian-security bullseye-security main contrib
\end{verbatim}

The Proxmox VE system will now retrieve updates for both itself and the
base Debian system.

Then from a shell run:

\begin{verbatim}
$ apt update
$ apt upgrade
\end{verbatim}

@todo: determine if system needs a reboot

\hypertarget{proxmox-ve-hardening}{%
\subsubsection{Proxmox VE hardening}\label{proxmox-ve-hardening}}

Links

@todo

Adding users

Add SSH keys and remove ability to login with password

\hypertarget{setup-an-internal-only-network-in-proxmox-ve}{%
\subsection{Setup an internal only network in Proxmox
VE}\label{setup-an-internal-only-network-in-proxmox-ve}}

From the Web GUI navigate to Datacenter - \textgreater{} your server
-\textgreater{} Network

From the menu select Create then Linux Bridge

Input the desired IPv4/CIDR in this case 192.168.y.0/24 and add a
comment if desired (``Internal network'' was used here). Note that y
must not be the same as x previously used.

Name was left as vmbr1

Credit:
https://dannyda.com/2020/06/01/how-to-create-an-internal-only-isolated-network-for-guest-os-virtual-machines-vm-on-proxmox-ve-pve-like-in-vmware-workstation-host-only-network-but-different/

\hypertarget{install-and-configure-internet-gateway-server-virtual-machine}{%
\subsection{Install and configure Internet gateway server virtual
machine}\label{install-and-configure-internet-gateway-server-virtual-machine}}

VyOS was selected (https://vyos.io/)

\hypertarget{create-an-iso-of-the-stable-version-as-of-writing-1.3.0}{%
\subsubsection{Create an ISO of the stable version (as of writing
1.3.0)}\label{create-an-iso-of-the-stable-version-as-of-writing-1.3.0}}

@todo: the built version seemed to be a nightly release, is it possible
to add a tag to get a stable build?

Follow the build instructions:

https://docs.vyos.io/en/latest/contributing/build-vyos.html

This document does not list this version (goes up to 10 ``buster'') but
Debian 11 ``bullseye'' was successfully used in this setup.

Run the following commands:

\begin{verbatim}
$ apt install git
$ apt install build-essential
\end{verbatim}

Follow the instructions here
https://docs.docker.com/engine/install/debian/ to install Docker

Run the following commands:

\begin{verbatim}
$ git clone -b equuleus --single-branch https://github.com/vyos/vyos-build

$ docker run --rm -it --privileged -v $(pwd):/vyos -w /vyos vyos/vyos-build:equuleus bash
\end{verbatim}

Then in the Docker terminal run the following commands:

\begin{verbatim}
./configure --architecture amd64

sudo make iso
\end{verbatim}

\hypertarget{upload-the-iso-image-to-the-proxmox-ve-server}{%
\subsubsection{Upload the ISO image to the Proxmox VE
server}\label{upload-the-iso-image-to-the-proxmox-ve-server}}

\begin{enumerate}
\def\labelenumi{\arabic{enumi}.}
\item
  Via the web GUI navigate to Datacenter -\textgreater{} your server
  -\textgreater{} local.
\item
  In the right hand pane select ISO Images and then upload.
\item
  Upload the ISO image
\end{enumerate}

Tip: you can also pass the checksum to the Proxmox VE upload tool

\hypertarget{create-vyos-virtual-machine}{%
\subsubsection{Create VyOS virtual
machine}\label{create-vyos-virtual-machine}}

\begin{enumerate}
\def\labelenumi{\arabic{enumi}.}
\item
  From the top right of the web GUI select Create VM
\item
  In the appearing dialogue type a Name ``VyOS'' and optionally select
  advanced and Start at boot
\item
  On the next tab select the target ISO image
\item
  On the System tab leave everything as default
\item
  n the Disk tab leave the defaults (this exceeds requirements
  https://docs.vyos.io/en/latest/installation/install.html)
\item
  On the CPU tab:

  Sockets: 1, Cores: 2
\item
  On the Memory tab

  Memory: 4096MiB
\item
  On the Network tab

  Choose the bridge with the internet vmbr0 (it is possible to add the
  second later) and leave the defaults including firewall

  Confirm all the settings on the next tab but \textbf{do not} select
  start after created

  Navigate to the newly created VM on the left-hand pane then selected
  Hardware from the menu that is presented on the right. Choose Add and
  then Network Device. In the dialogue that appears select the Internal
  network bridge (vmbr1 in this case) that was created earlier and leave
  all other options as is.

  So, the VM will have the following Network Devices:

  net0: Internet

  net1: Internal only
\item
  Start the VM and connect the console (top right)
\item
  Login with vyos and vyos

  Run the command:

\begin{verbatim}
$ install image
\end{verbatim}
\item
  Follow the instructions
\item
  Set the CD/DVD to none in Web GUI
\item
  Reboot
\end{enumerate}

\hypertarget{configure-vyos}{%
\subsubsection{Configure VyOS}\label{configure-vyos}}

Open a noVNC window to the host

Login with vyos and vyos

Switch to configure mode:

\begin{verbatim}
vyos@vyos$ configure
vyos@vyos#
\end{verbatim}

Then configure as desired. Below is configuration used in the setup here
(if you use for inspiration do take care to replace the x and y octet
values correctly with previously chosen values. The z octet value should
be something unused in the outside LAN for which the host is physically
connected):

\begin{verbatim}
set interfaces ethernet eth0 address '192.168.x.z/24'
set interfaces ethernet eth0 description 'OUTSIDE'
set protocols static route 0.0.0.0/0 next-hop 192.168.x.254 distance 1
set service dns forwarding system
set service dns forwarding name-server 192.168.x.254
set service dns forwarding listen-address 192.168.y.1
set service dns forwarding allow-from 192.168.y.0/24
set system name -server 192.168.x.254

set interfaces ethernet eth1 address '192.168.y.1/24'
set interfaces ethernet eth1 description 'INSIDE'

set nat source rule 100 outbound-interface eth0
set nat source rule 100 source address 192.168.y.0/24
set nat source rule 100 translation address masquerade

set service ssh listen-address 0.0.0.0
\end{verbatim}

Once done remember to commit the config (correcting any
misconfiguration) and save.

\begin{verbatim}
commit
save
\end{verbatim}

Inspiration for the above was taken from:
https://bertvv.github.io/cheat-sheets/VyOS.html

@todo: hardening, IDS, IPS

\hypertarget{install-and-configure-a-debian-virtual-machine}{%
\subsection{Install and configure a Debian virtual
machine}\label{install-and-configure-a-debian-virtual-machine}}

This VM can be used for various tasks such as software compilation and
testing of the networks. In this setup the Debian VM was used to test
connectivity to the VyOS gateway and the Internet. It is also used in
the subsequent stages to deploy a nix-bitcoin node.

In Proxmox VE create a new virtual machine and configure the network
device to use the bridge `vmbr1'.

Then install Debian and configure the network adapter within the VM with
the following settings:

IP address: 192.168.y.2 Gateway: 192.168.y.1 DNS: 192.168.y.1

Test that the VM has Internet connectivity.

\hypertarget{deploying-the-nix-bitcoin-node}{%
\subsection{Deploying the nix-bitcoin
node}\label{deploying-the-nix-bitcoin-node}}

This deployment follows the documentation:

https://github.com/fort-nix/nix-bitcoin/\#get-started

Take note of the hardware requirements:

https://github.com/fort-nix/nix-bitcoin/blob/master/docs/hardware.md

In the main, the install guide
(https://github.com/fort-nix/nix-bitcoin/blob/master/docs/install.md) is
followed verbatim and notes with a reference to particular sections are
added where appropriate.

Optional - a small exception in regards to this setup is that a separate
virtual disk (located on a different physical drive mirror (RAID 1)) was
used to store the bitcoin database - this is optional and details are
provided on how to achieve it. Also detailed is how to configure the
network when using the minimal image.

\hypertarget{acquiring-nixos}{%
\subsubsection{Acquiring NixOS}\label{acquiring-nixos}}

Following
\href{https://github.com/fort-nix/nix-bitcoin/blob/master/docs/install.md\#1-nixos-installation}{section
1.1} make sure the latest NixOS is obtained i.e.~do not just copy the
whole wget command outright and make sure to verify the hash against
trusted sources before using the image.

Download the minimal ISO image (https://nixos.org/download.html)

Verify the hash

Upload the ISO to
\protect\hyperlink{ux5cux23ux5cux23Upload-the-ISO-image-to-the-Proxmox-VE-server}{Proxmox
VE server}

\hypertarget{create-a-new-vm}{%
\subsubsection{Create a new VM}\label{create-a-new-vm}}

Name: NixOS

Follow the setup and leave everything as default until the CPU page. The
following configuration was used, which should exceed the minimum
requirements:

Cores: 4

Memory: 4096MiB = 4.2GB

Network: vmbr1 (Internal Network)

Do NOT check the select the start the VM checkbox

Next, an additional drive will be configured in Proxmox VE. This will
then be used to store the bitcoin database within the NixOS VM.

Select Datacenter -\textgreater{} server name and then from the right
pane Disks -\textgreater{} LVM-Thin. Then select Create: Thinpool

From the dialogue select the disk and type a name ``data'' was used in
this setup. This provisions a vg with the name \emph{data} and a name
\emph{data} @todo: review

Navigate back to the VM created and choose Hardware and then Add
-\textgreater{} Hard Disk

Choose ``data'' from Storage and then set the size to 560 GiB which
equates to about 600GB

Now, continue from section 1.3 in the install instructions

Start the VM and connect a console

\texttt{sudo\ -i}

With the SeaBios that was used in this setup the file does not exist and
Legacy Boot (MBR) should be followed (option 2)

Note: no consideration is currently given for encrypted partitions
within the Proxmox VE setup

Enable the OpenSSH daemon

\begin{verbatim}
services.openssh.permitRootLogin = "yes";
\end{verbatim}

Configure the network config in configuration.nix (remember to replace y
with the chosen value)

\begin{verbatim}
  networking.useDHCP = false;
  networking.interfaces.ens18.useDHCP = false;

  networking.interfaces.ens18.ipv4.addresses = [ {
    address= "192.168.y.3";
    prefixLength = 24;
  } ];
  networking.defaultGateway = "192.168.y.1";
  networking.nameservers = ["192.168.y.1"];
  networking.hostName = "nixicon";
\end{verbatim}

Although the IP above will be assigned once the nix-bitcoin is deployed
the installation cannot continue without a connection to the Internet so
that needs to be configured:

\begin{verbatim}
$ ifconfig ens18 192.168.y.3
$ ifconfig ens18 255.255.255.0
$ ip route add 192.168.y.0/24 dev ens18 scope link src 192.168.y.3
\end{verbatim}

Then add the nameserver:

\begin{verbatim}
nano /etc/resolv.conf
\end{verbatim}

Add:

\begin{verbatim}
nameserver 192.168.y.1
\end{verbatim}

Once the above is complete and successful networking is verified

Run the following command:

\texttt{nixos-install}

Set the root password and then reboot.

\hypertarget{configure-the-additional-drive-optional}{%
\subsubsection{Configure the additional drive
(optional)}\label{configure-the-additional-drive-optional}}

As the additional drive was not configured at the time of the install
then the parted utility will need to be available. To achieve this, edit
the configuration.nix file

\texttt{nano\ /etc/nixos/configuration.nix}

and add the following:

\begin{verbatim}
environment.systemPackages = with pkgs; [
    parted
];
\end{verbatim}

Then issue the following command:

\texttt{nixos-rebuild\ switch}

Determine the desired drive, fdisk can assist:

\texttt{fdisk\ -l}

Note: in this sytem the desired drive is /dev/sdb with 560GiB capacity
but sdx is used in the following examples:

Then partition:

\texttt{parted\ /dev/sdx}

\begin{verbatim}
(parted) mklabel msdos
(parted) mkpart primary
File system type? ext4
Start? 0%
End? 100%
quit
\end{verbatim}

(note: it is possible to combine the above as a single line command)

Then create the file system:

\texttt{mkfs.ext4\ -L\ data\ /dev/sdx1}

Make a note of the UUID as this will be used in the next steps to mount
the volume

\hypertarget{create-port-forwarding-rules-for-ssh-optional}{%
\subsubsection{Create port forwarding rules for SSH
(optional)}\label{create-port-forwarding-rules-for-ssh-optional}}

Providing SSH access to the VMs from outside the private network makes
it easier to configure them (ability to copy and paste UUIDs etc.)

This involve updates to VyOS configuration and can be temporary.

Login to the vyos, you should be able do this from your local machine
now as apposed to the console

ssh vyos@192.168.x.z

\hypertarget{debian}{%
\paragraph{Debian}\label{debian}}

192.168. y .2

The following commands were issued to the VyOS router (obiously
replacing y with the value chosen earlier)

\begin{verbatim}
configure

set nat destination rule 12 description 'Port Forward: 2222 to 22 SSH on 192.168.y.2'
set nat destination rule 12 destination port '2222'
set nat destination rule 12 inbound-interface 'eth0'
set nat destination rule 12 protocol 'tcp'
set nat destination rule 12 translation address '192.168.y.2'
set nat destination rule 12 translation port '22'

commit
\end{verbatim}

Now test

Note: for the Debian VM the user account may need to be added to the SSH
user group

Note: you could SSH from Debian to all other hosts

\hypertarget{nixos}{%
\paragraph{NixOS}\label{nixos}}

192.168. y .3

Assuming access to the Debian VM via SSH is working then from the same
VyOs configure session issue the following:

\begin{verbatim}
set nat destination rule 13 description 'Port Forward: 2223 to 22 SSH on 192.168.y.3'
set nat destination rule 13 destination port '2223'
set nat destination rule 13 inbound-interface 'eth0'
set nat destination rule 13 protocol 'tcp'
set nat destination rule 13 translation address '192.168.y.3'
set nat destination rule 13 translation port '22'

commit
\end{verbatim}

Test and if all is well, save the VyOS configuration:

\begin{verbatim}
save
\end{verbatim}

Credit: https://support.vyos.io/en/kb/articles/nat-principles

Having SSH access to both the Debian and NixOS VMs will make the next
stages of the process a little easier

@todo hardening (SSH e.g.~add keys, remove plain text or remove SSH
access entirely)

\hypertarget{prepare-nix-bitcoin-nixos-package}{%
\subsubsection{Prepare nix-bitcoin NixOS
package}\label{prepare-nix-bitcoin-nixos-package}}

This section continues to follow the guide from
\href{https://github.com/fort-nix/nix-bitcoin/blob/master/docs/install.md}{Nix
Installation}.

Note: this part of the guide will be executed on the Debian VM that was
installed earlier

The next steps will follow
\href{https://github.com/fort-nix/nix-bitcoin/blob/master/docs/install.md\#2-nix-installation}{section
2}.

You may need to add your user to the sudoers if it is not a member
already

In Debian this can be achieved with the following commands

Switch to root

\texttt{su}

Then

\texttt{sudo\ usermod\ -a\ -G\ sudo\ username}

Exit both the root and user session and then log back in as the user

Important: ensure that when downloading the multi-user NixOS that the
latest is obtained (listed at https://nixos.org/download.html). I.e.
dont just copy and paste verbatim.

Note: It is possible to determine the latest version by navigating to:
https://nixos.org/nix/install and this will redirect to for example:
https://releases.nixos.org/nix/nix-2.6.0/install at the time of writing.
From here you could quickly santiy check the redirect by heading to:
https://releases.nixos.org/?prefix=nix/

You could (as in the example on the NixOS website) use curl with a -L
option which will ignore the redirect

Enter a directory to receive the files. \textasciitilde/Downloads was
chosen for this setup

For completeness the following commands were issued:

\texttt{curl\ -o\ install-nix-2.6.0\ https://releases.nixos.org/nix/nix-2.6.0/install}

with the -o option writing the contents to a file rather than displaying
on screen

then

\texttt{curl\ -o\ install-nix-2.6.0.asc\ https://releases.nixos.org/nix/nix-2.6.0/install.asc}

then

\texttt{gpg2\ -\/-keyserver\ hkps://keyserver.ubuntu.com\ -\/-recv-keys\ B541D55301270E0BCF15CA5D8170B4726D7198DE}
\texttt{gpg2\ -\/-verify\ ./install-nix-2.6.0.asc}

Which are similarly detailed here:
https://nixos.org/download.html\#nix-verify-installation

Note: it is not required to run the script as sudo. It will prompt for
permission.

In this setup the:

\begin{verbatim}
substitute = false
\end{verbatim}

was added to /etc/nix/nix.conf as detailed.

Run the script.

Exit the terminal and login in again as per the message:

\begin{verbatim}
Try it! Open a new terminal, and type:

  $ nix-shell -p nix-info --run "nix-info -m"
\end{verbatim}

The next part continues with
\href{https://github.com/fort-nix/nix-bitcoin/blob/master/docs/install.md\#3-setup-deployment-directory}{setting
up the deployment directory}

Stood in the home directory or one just off it, follow the instructions
provided.

Once the above is complete continue with the
\href{https://github.com/fort-nix/nix-bitcoin/blob/master/docs/install.md\#4-deploy-with-krops}{deploy
with krops} section.

Follow the instructions and edit the SSH config. You will need a
public/private key pair for this and this
\href{https://www.digitalocean.com/community/tutorials/how-to-set-up-ssh-keys-2}{article}
could be useful.

The config file used in this setup is shown below:

\begin{verbatim}
Host nixicon
    # FIXME
    Hostname 192.168.y.3
    User root
    PubkeyAuthentication yes
    # FIXME
    IdentityFile ~/.ssh/id_ed25519
    AddKeysToAgent yes
\end{verbatim}

And for reference the krops/deploy.nix is as follows:

\begin{verbatim}
let
  # FIXME:
  target = "root@nixicon";

  extraSources = {
    "hardware-configuration.nix".file = toString ../hardware-configuration.nix;
  };

  krops = (import <nix-bitcoin> {}).krops;
in
krops.pkgs.krops.writeDeploy "deploy" {
  inherit target;

  source = import ./sources.nix { inherit extraSources krops; };

  # Avoid having to create a sentinel file.
  # Otherwise /var/src/.populate must be created on the target node to signal krops
  # that it is allowed to deploy.
  force = true;
}
\end{verbatim}

In subsection 3 the guide shows how to optionally disallow substitutes.
This was set to true in this setup.

In subsection 4 the guide details copying hardware-configuration.nix
file to the deployment directory and then in subsection 5 making edits
to the configuration.nix file to turn on desired modules. There are some
important notes relevant to this setup to make here:

\hypertarget{additional-hard-drive-configuration}{%
\paragraph{Additional hard drive
configuration}\label{additional-hard-drive-configuration}}

No edits were made to hardware-configuration.nix as per the warning at
the top of the file. For reference here is the file from this setup:

\begin{verbatim}
# Do not modify this file!  It was generated by ‘nixos-generate-config’
# and may be overwritten by future invocations.  Please make changes
# to /etc/nixos/configuration.nix instead.
{ config, lib, pkgs, modulesPath, ... }:

{
  imports =
    [ (modulesPath + "/profiles/qemu-guest.nix")
    ];

  boot.initrd.availableKernelModules = [ "ata_piix" "uhci_hcd" "virtio_pci" "virtio_scsi" "sd_mod" "sr_mod" ];
  boot.initrd.kernelModules = [ ];
  boot.kernelModules = [ ];
  boot.extraModulePackages = [ ];
  boot.loader.grub.device = "/dev/sda";

  fileSystems."/" =
    { device = "/dev/disk/by-uuid/UUID_1";
      fsType = "ext4";
    };

  swapDevices =
    [ { device = "/dev/disk/by-uuid/UUID_2"; }
    ];

  hardware.cpu.intel.updateMicrocode = lib.mkDefault config.hardware.enableRedistributableFirmware;
}
\end{verbatim}

Rather, the additional hard drive was configured in the
configuration.nix as shown here:

\begin{verbatim}
fileSystems."/var/lib" =
{ device = "/dev/disk/by-uuid/UUID_3";
  fsType = "ext4";
};
\end{verbatim}

This mounts /var/lib (which contains the bitcoin database etc.) to the
additional drive.

\hypertarget{static-ip-configuration}{%
\paragraph{Static IP configuration}\label{static-ip-configuration}}

To configure the static IP add the following:

\begin{verbatim}
networking.useDHCP = false;
networking.interfaces.ens18.useDHCP = false;

networking.interfaces.ens18.ipv4.addresses = [ {
  address= "192.168.y.3";
  prefixLength = 24;
} ];
networking.defaultGateway = "192.168.y.1";
networking.nameservers = ["192.168.y.1"];
networking.hostName = "nixicon";
\end{verbatim}

\hypertarget{ssh-configuration}{%
\paragraph{SSH configuration}\label{ssh-configuration}}

Below is the snipet of configuration. Note: paste the contents of
\texttt{\textasciitilde{}/.ssh/id\_ed25519.pub} where the
\texttt{\#\ FIXME:\ Replace\ this\ with\ your\ SSH\ pubkey\ appears}

\begin{verbatim}
services.openssh = {
  enable = true;
  passwordAuthentication = false;
};
users.users.root = {
  openssh.authorizedKeys.keys = [
    # FIXME: Replace this with your SSH pubkey
    "ssh-ed25519 LONG_KEY user@debian"
  ];
};
\end{verbatim}

\hypertarget{services-configuration}{%
\paragraph{Services configuration}\label{services-configuration}}

Last but not least, the following services are enabled in this setup:

\begin{verbatim}
services.clightning.enable = true;
services.rtl.enable = true;
services.rtl.nodes.clightning = true;
services.electrs.enable = true;
services.backups.enable = true;
\end{verbatim}

Once the configuration.nix file has been updated continue from
subsection 6.

The how-to continues with a \href{./HOWTO_VIRCADIA.md}{guide to setting
up a Vircadia domain server and metaverse server}

\hypertarget{lab---vircadia}{%
\section{Lab - Vircadia}\label{lab---vircadia}}

\hypertarget{overview}{%
\subsection{Overview}\label{overview}}

This part of the how-to serves as a guide to setting up a Vircadia
domain server and metaverse server within the Proxmox VE environment.

To familiarise with the technology stack the
\href{https://docs.vircadia.com/explore/get-started/architecture.html}{Vircadia
architecture overview document} is a good place to start and the
\href{https://github.com/vircadia/}{GitHub} contains documentation and
source.

\hypertarget{deploy-a-vircadia-domain-server}{%
\subsection{Deploy a Vircadia domain
server}\label{deploy-a-vircadia-domain-server}}

There are \href{https://vircadia.com/deploy-a-server/}{few different
options} to deploy a server. This guide will use the Linux compile from
source method.

\href{https://github.com/vircadia/vircadia-builder\#supported-platforms}{Various
Linux distributions are supported} and in this setup, Ubuntu 20.04 was
selected.

\hypertarget{prepare-a-virtual-instance-of-ubuntu-20.04-in-the-proxmox-virtual-environment}{%
\subsubsection{Prepare a virtual instance of Ubuntu 20.04 in the Proxmox
Virtual
Environment}\label{prepare-a-virtual-instance-of-ubuntu-20.04-in-the-proxmox-virtual-environment}}

First, obtain the Ubuntu 20.04 desktop image.

Then deploy a virtual machine in Proxmox VE as was done with VyOS and
Debian in the previous sections. The following values were selected for
this setup:

Disk drive: 64GiB\\
CPU: 2 sockets 2 cores\\
RAM: 8192GiB\\
Network adapter: vmbr1 (internal)\\
Name: vircadia-server

When prompted by the installer to choose the packages to install choose
`minimal'.

Remember to remove the media.

Once booted, head to settings:

\begin{figure}
\centering
\includegraphics[width=0.7\paperwidth]{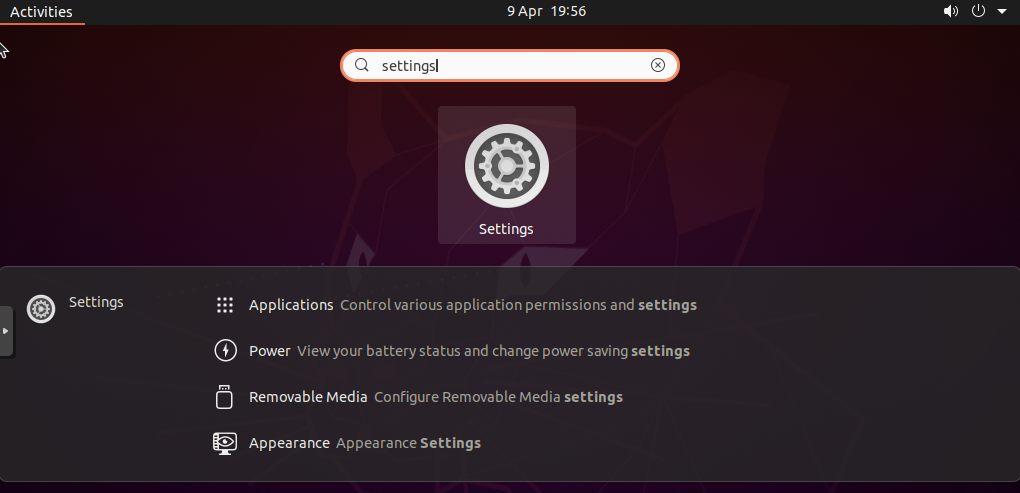}
\caption{Ubuntu 20.04 Desktop Settings}
\end{figure}

Then configure the network connection with the manual method:

\begin{figure}
\centering
\includegraphics[width=0.7\paperwidth]{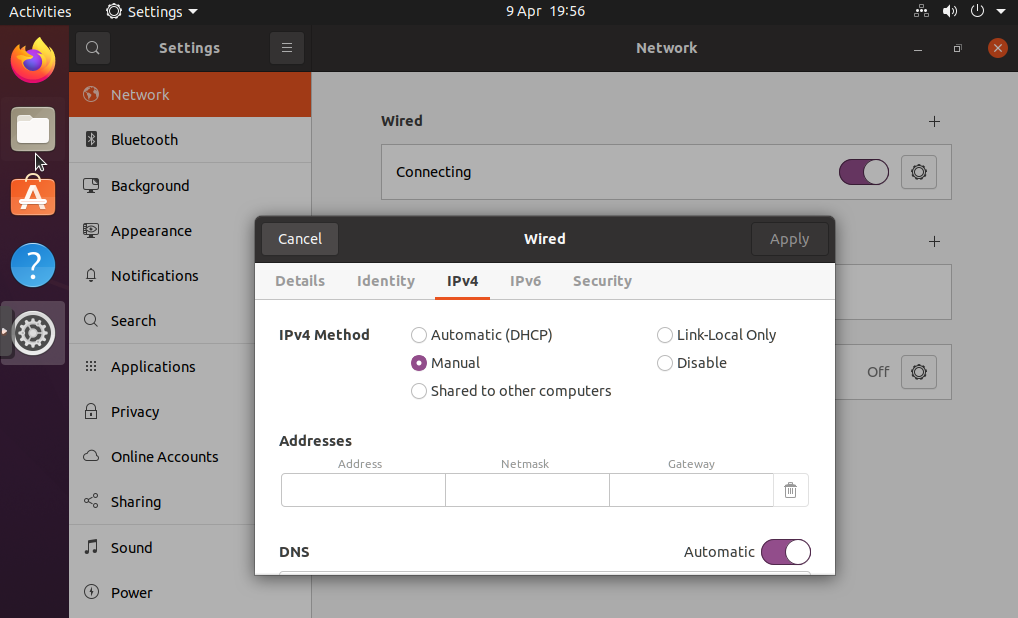}
\caption{Ubuntu 20.04 Desktop Settings - Network}
\end{figure}

With the following:

IP address: 192.168.y.4\\
Gateway: 192.168.y.1\\
DNS: 192.168.y.1\\
Mask 255.255.255.0

Run the software updater.

Restart and run the software updater again.

Once the Operating System is up to date open a terminal and install git:

\texttt{sudo\ apt\ install\ git}

Follow the \href{https://github.com/vircadia/vircadia-builder}{Vircadia
build instructions} along side this how-to guide.

Stood in your home directory or similar run the following:

\begin{verbatim}
git clone https://github.com/vircadia/vircadia-builder.git
cd vircadia-builder
chmod +x vircadia-builder
\end{verbatim}

As this setup requires both the domain server and the ICE server, run
the following:

./vircadia-builder --build=server,ice-server

It is also possible to add a `client' option but this was not opted for
in this setup as the client was
\protect\hyperlink{Client-interface}{Windows based}.

Follow the prompts (defaults were accepted in this setup). The
installation process will begin as shown below:

\begin{figure}
\centering
\includegraphics[width=0.7\paperwidth]{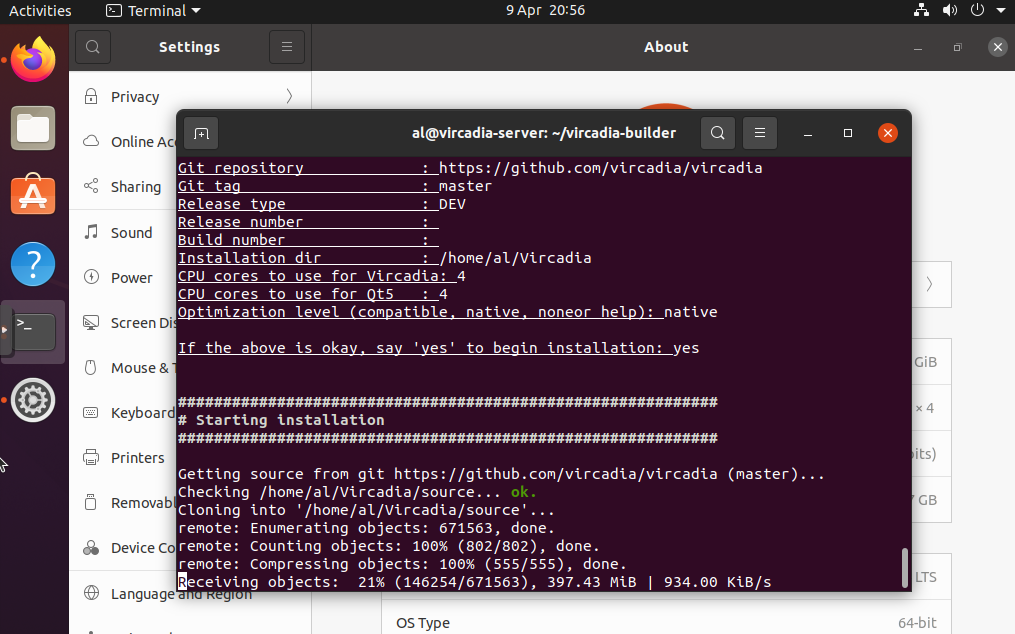}
\caption{Vircadia build settings and installation screen grab}
\end{figure}

The first stage of the installer will install dependency packages,
follow the instruction and then if prompted run the build command again.

Note: if you have followed the guide then Qt will not be installed on
the target system, however, the installer will build the correct version
- see below:

\begin{figure}
\centering
\includegraphics[width=0.7\paperwidth]{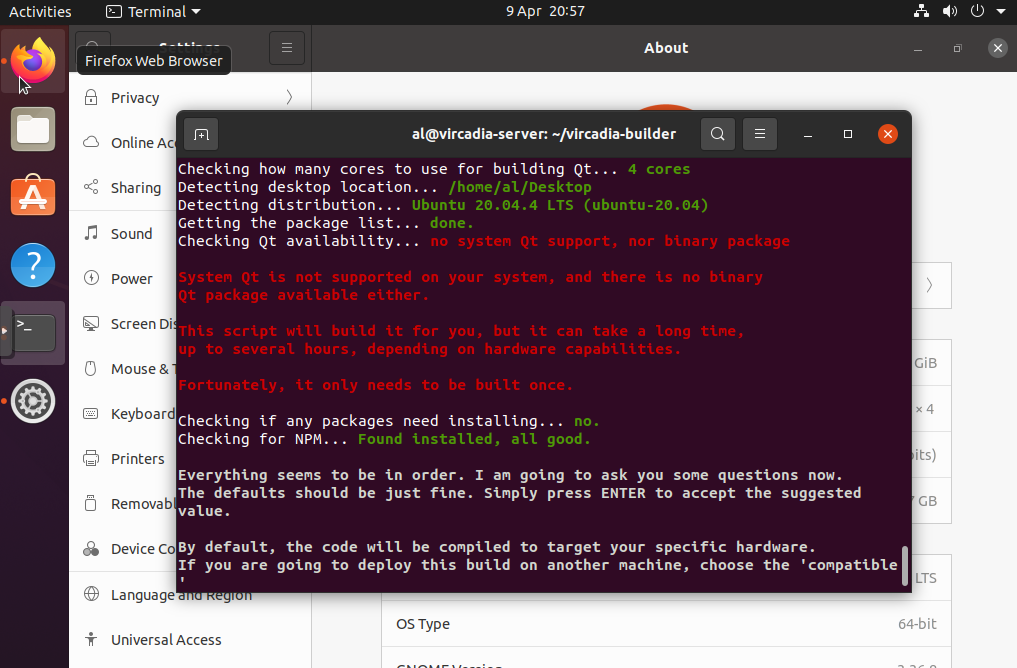}
\caption{Vircadia build Qt warning}
\end{figure}

Once complete (which could take several hours for the Qt step) a text
output similar to the following should be presented:

\begin{verbatim}
Cleaning up install directory...done
Copied : 4626
Skipped: 3208
Deleted: 0
Creating script for assignment-client...done.
Creating script for domain-server...done.
Creating service for assignment-client...done.
Creating service for domain-server...done.
Reloading systemd config... done.


############################################################
# Setting up desktop
############################################################

interface not built, skipping desktop setup
\end{verbatim}

Navigate up one from the build directory and into the Vircadia folder.
In here there should be a build folder with the built binaries and
accompanying scripts - these are the scripts that will be run later
(@todo: section link: \#\#\# Configure the Vircadia server) once port
forwarding is configured.

\hypertarget{configure-port-forwarding-for-vircadia-services}{%
\subsubsection{Configure port forwarding for Vircadia
services}\label{configure-port-forwarding-for-vircadia-services}}

The following ports are required:

40100 (+0): (tcp) administrative http connection

40101 (+1): (tcp) administrative https (encrypted) connection

40102 (+2): (udp) main connection from clients

40103 (+3): (udp) main connection from clients (encrypted)

Note: in this setup non TLS versions are currently used for testing.

Important: for a production version of the system it would be prudent to
enable TLS and completely disallow non TLS traffic (i.e.~remove the port
forwarding rules)

\begin{verbatim}
set nat destination rule 15 description 'Port Forward: 40100 (+0): (tcp) administrative http connection on 192.168.y.4'
set nat destination rule 15 destination port '40100'
set nat destination rule 15 inbound-interface 'eth0'
set nat destination rule 15 protocol 'tcp'
set nat destination rule 15 translation address '192.168.y.4'
set nat destination rule 15 translation port '40100'

set nat destination rule 16 description 'Port Forward: 40101 (+1): (tcp) administrative https (encrypted) connection on 192.168.y.4'
set nat destination rule 16 destination port '40101'
set nat destination rule 16 inbound-interface 'eth0'
set nat destination rule 16 protocol 'tcp'
set nat destination rule 16 translation address '192.168.y.4'
set nat destination rule 16 translation port '40101'

set nat destination rule 17 description 'Port Forward: 40102 (+2): (udp) main connection from clients on 192.168.y.4'
set nat destination rule 17 destination port '40102'
set nat destination rule 17 inbound-interface 'eth0'
set nat destination rule 17 protocol 'tcp'
set nat destination rule 17 translation address '192.168.y.4'
set nat destination rule 17 translation port '40102'

set nat destination rule 18 description 'Port Forward: 40103 (+3): (udp) main connection from clients (encrypted) on 192.168.y.4'
set nat destination rule 18 destination port '40103'
set nat destination rule 18 inbound-interface 'eth0'
set nat destination rule 18 protocol 'tcp'
set nat destination rule 18 translation address '192.168.y.4'
set nat destination rule 18 translation port '40103'

commit

save
\end{verbatim}

\hypertarget{configure-port-forwarding-for-the-ice-server}{%
\subsubsection{Configure port forwarding for the ICE
server}\label{configure-port-forwarding-for-the-ice-server}}

\begin{verbatim}
set nat destination rule 21 description 'Port Forward for ice-server: 7337 to 7337 on 192.168.y.4'
set nat destination rule 21 destination port '7337'
set nat destination rule 21 inbound-interface 'eth0'
set nat destination rule 21 protocol 'tcp'
set nat destination rule 21 translation address '192.168.y.4'
set nat destination rule 21 translation port '7337'

commit

save
\end{verbatim}

\hypertarget{configure-the-vircadia-server}{%
\subsubsection{Configure the Vircadia
server}\label{configure-the-vircadia-server}}

As per the information message:

``Connect a web browser to the server at port 40100. (If you are on the
machine that the server is running on, this would be
http://localhost:40100) Complete the initial setup wizard and you should
have a functioning domain.''

With the port forwarding in place it should be possible to access the
web interface on: 192.168.x.z:40100

Start the configuration:

\begin{figure}
\centering
\includegraphics[width=0.7\paperwidth]{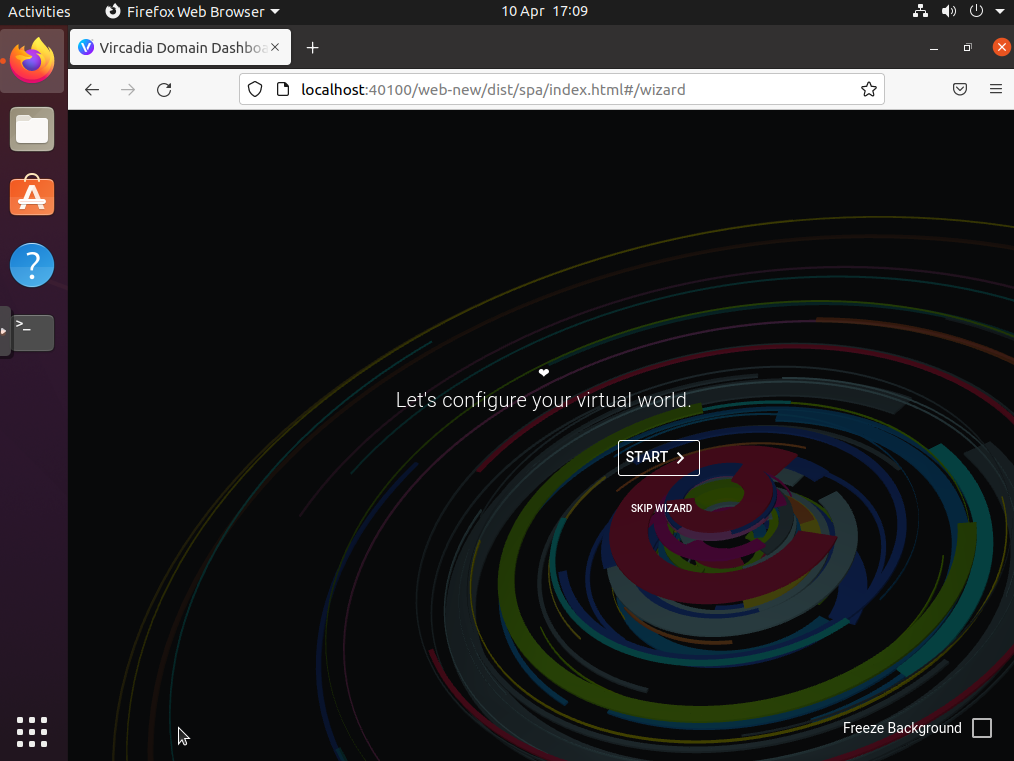}
\caption{Vircadia server configuration landing page}
\end{figure}

Select skip on the import settings and/or content page:

\begin{figure}
\centering
\includegraphics[width=0.7\paperwidth]{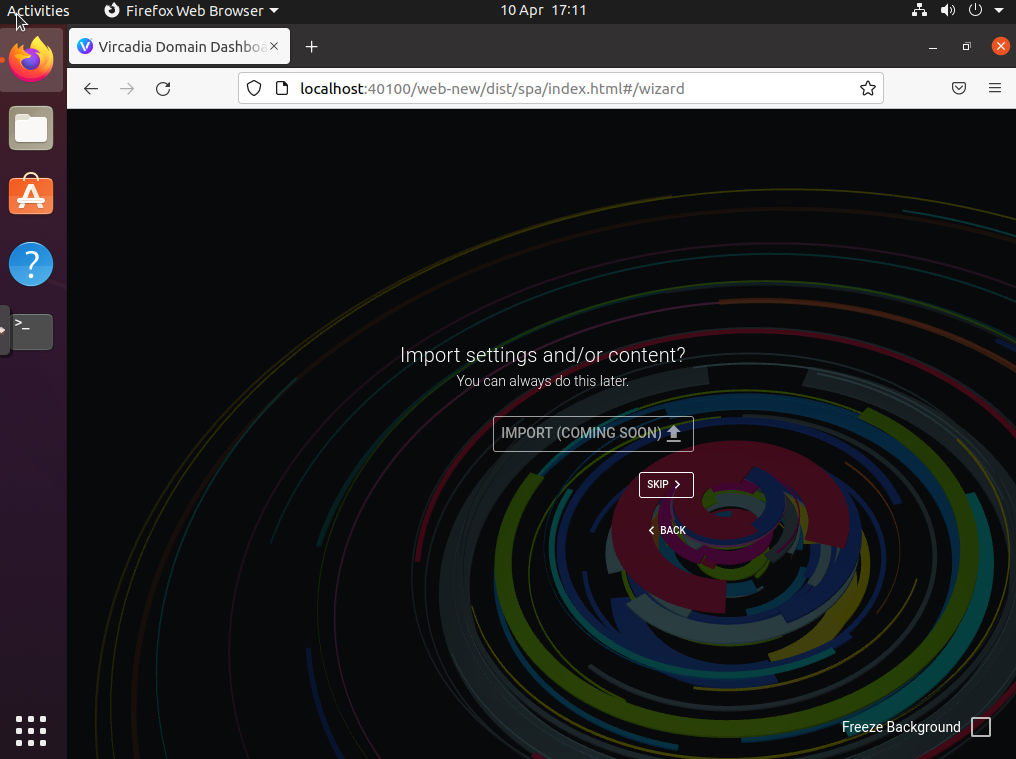}
\caption{Vircadia server configuration import page}
\end{figure}

Select skip on the connect to metaverse account page:

\begin{figure}
\centering
\includegraphics[width=0.7\paperwidth]{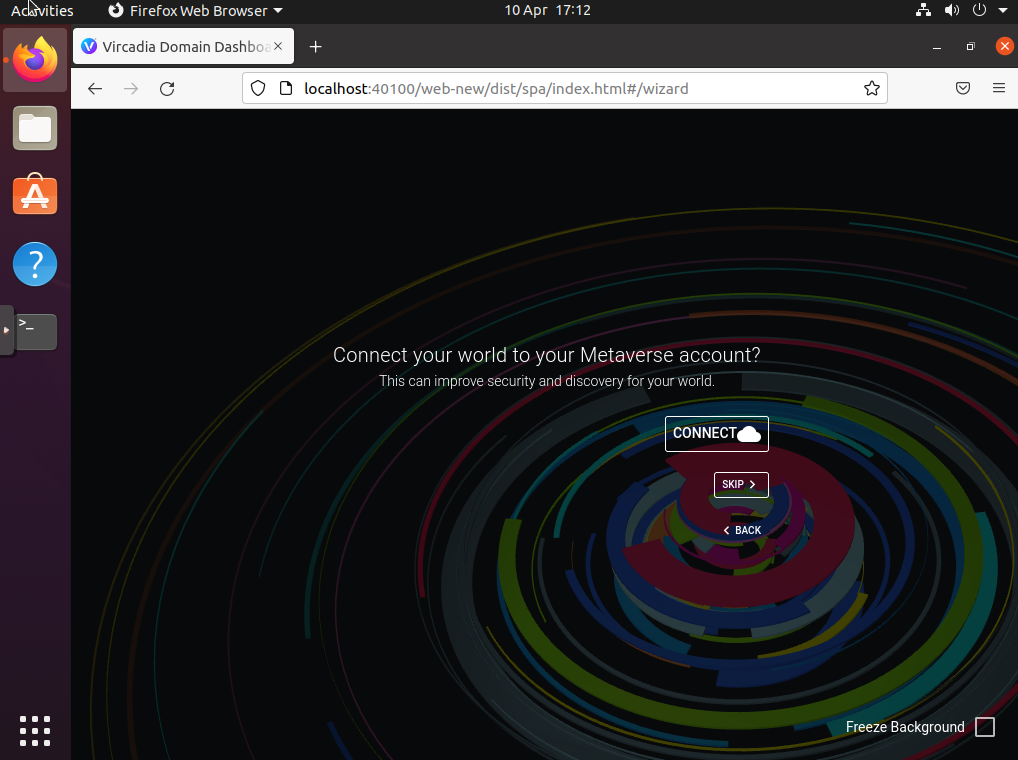}
\caption{Vircadia server configuration import page}
\end{figure}

Configure security settings for your domain:

\begin{figure}
\centering
\includegraphics[width=0.7\paperwidth]{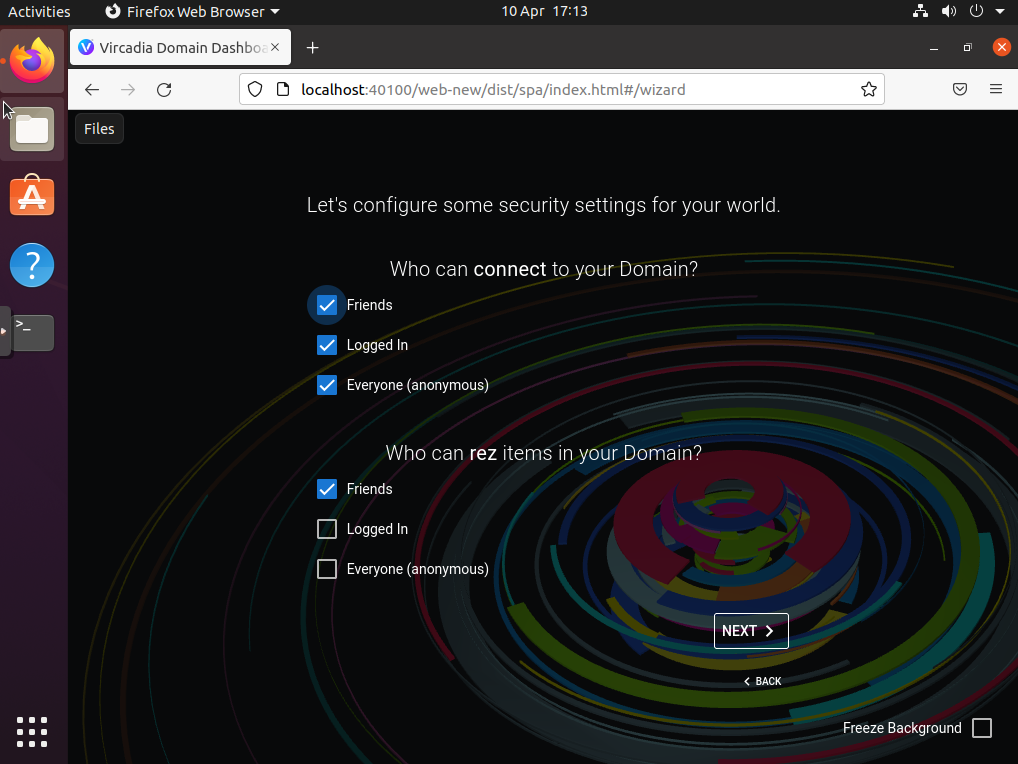}
\caption{Vircadia server configuration import page}
\end{figure}

Create an admin user for the web panel:

\begin{figure}
\centering
\includegraphics[width=0.7\paperwidth]{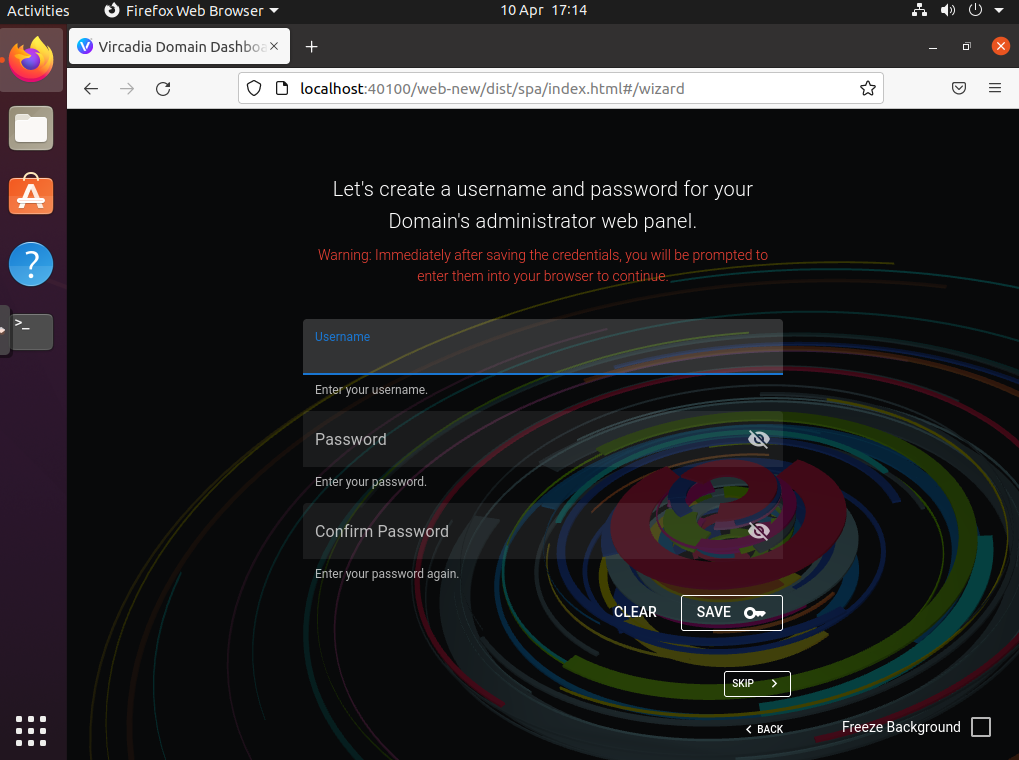}
\caption{Vircadia server configuration import page}
\end{figure}

Choose if you would like to turn performance mode on (in the setup
described here performance mode was turned on):

\begin{figure}
\centering
\includegraphics[width=0.7\paperwidth]{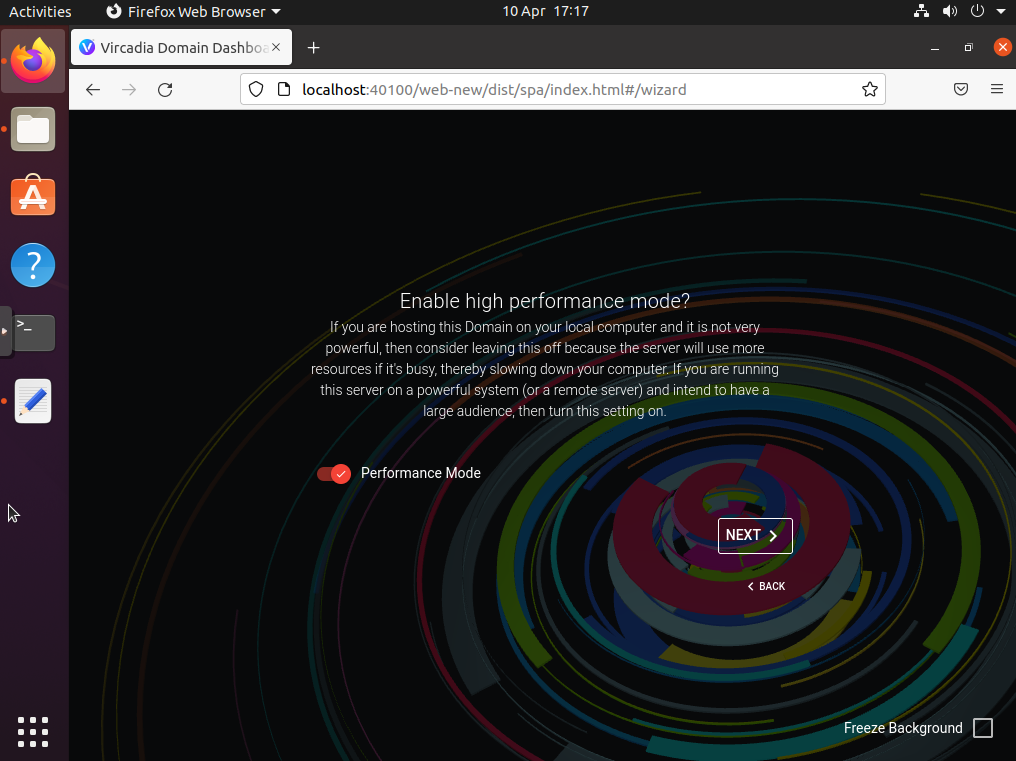}
\caption{Vircadia server configuration import page}
\end{figure}

Complete the installation.

\hypertarget{deploy-a-vircadia-metaverse-server}{%
\subsection{Deploy a Vircadia metaverse
server}\label{deploy-a-vircadia-metaverse-server}}

Installation and configuration here follows the
\href{https://github.com/vircadia/vircadia-metaverse/blob/master/docs/Building.md}{Vircadia
metaverse build guide} and
\href{https://github.com/vircadia/vircadia-metaverse/blob/master/docs/NotesOnDevelopment.md}{Iamus:
Notes On Development}.

It is possible to install and run the Vircadia metaverse server on the
same server as the domain server, however, in this setup another VM was
used (separation of concerns).

Note: make sure to put VM disk on the first local-lvm partition and not
the Bitcoin specific one.

Create another VM as in the previous steps with the following
configuration:

Disk drive: 64GiB\\
CPU: 2 sockets 2 cores\\
RAM: 8192GiB\\
Network adapter: vmbr1 (internal)\\
Name: vircadia-metaverse-server

Perform a minimal install of Ubuntu 20.04 Desktop and configure the
network as follows:

IP address: 192.168.y.5\\
Gateway: 192.168.y.1\\
DNS: 192.168.y.1\\
Mask 255.255.255.0

Run the software updater.

Restart and run the software updater again.

\hypertarget{install-node.js-and-npm}{%
\subsubsection{Install Node.js and NPM}\label{install-node.js-and-npm}}

Following the
\href{https://github.com/vircadia/vircadia-metaverse/blob/master/docs/Building.md}{building
and configuration guide}:

Note: as of writing the Node and NPM versions in the package manager are
too low (node \textgreater{} 14 and npm \textgreater{} 6 are required)
so do not use those. Instead navigate to:
https://nodejs.org/en/download/ and get the latest LTS (version: 16.14.2
(includes npm 8.5.0) as of writing).

The following is taken from the
\href{https://github.com/nodejs/help/wiki/Installation}{Node
installation guide}:

\begin{verbatim}
sudo mkdir /usr/local/lib/nodejs
sudo tar -xJvf node-v16.14.2-linux-x64.tar.xz -C /usr/local/lib/nodejs/

nano ~/.profile
\end{verbatim}

Add:

\begin{verbatim}
export PATH=/usr/local/lib/nodejs/node-v16.14.2-linux-x64/bin:$PATH
\end{verbatim}

and refresh (@todo) and check

\begin{verbatim}
$ node -v

$ npm version

$ npx -v
\end{verbatim}

\hypertarget{installation-of-the-vircadia-metaverse-server}{%
\subsubsection{Installation of the Vircadia metaverse
server}\label{installation-of-the-vircadia-metaverse-server}}

Navigate to \textasciitilde{} (or where you would like to locate the
meteverse server repository) and then run the following:

\begin{verbatim}
$ git clone https://github.com/vircadia/vircadia-metaverse.git

Cloning into 'vircadia-metaverse'...
remote: Enumerating objects: 5134, done.
remote: Counting objects: 100% (1240/1240), done.
remote: Compressing objects: 100% (700/700), done.
remote: Total 5134 (delta 862), reused 823 (delta 488), pack-reused 3894
Receiving objects: 100% (5134/5134), 1.04 MiB | 1.10 MiB/s, done.
Resolving deltas: 100% (3539/3539), done.
\end{verbatim}

Change to the vircadia-metaverse directory and run the following
command:

\begin{verbatim}
npm install
\end{verbatim}

Note: As of writing the Git tags were old so the latest commit was opted
for.

\hypertarget{configure-the-mongo-db-for-the-metaverse-server}{%
\subsubsection{Configure the Mongo DB for the metaverse
server}\label{configure-the-mongo-db-for-the-metaverse-server}}

Note: the Mongo DB could run on a separate VM (further separation of
concerns).

The
\href{https://github.com/vircadia/vircadia-metaverse/blob/master/docs/Building.md}{Vircadia
metaverse build guide} is shown using version 4.4 so this guide will use
the same (though it maybe possible to use a later version (@todo: a
comment in Discord mentions success with 5.0.2):

https://www.mongodb.com/docs/v4.4/tutorial/install-mongodb-on-ubuntu/

Follow the installation instructions for Ubuntu 20.04 (assuming that
this is the distribution that has been chosen to install the metaverse
server on).

Once the Mongo DB is complete it can be verified as below:

\begin{verbatim}
al@vircadia-metaverse-server:~$ mongo
MongoDB shell version v4.4.13
connecting to: mongodb://127.0.0.1:27017/?compressors=disabled&gssapiServiceName=mongodb
Implicit session: session { "id" : UUID("c7fdcc61-f921-4fff-ad44-27840a8afea4") }
MongoDB server version: 4.4.13
Welcome to the MongoDB shell.
For interactive help, type "help".
For more comprehensive documentation, see
        https://docs.mongodb.com/
Questions? Try the MongoDB Developer Community Forums
        https://community.mongodb.com
---
The server generated these startup warnings when booting:
        2022-04-11T12:50:28.108+01:00: Using the XFS filesystem is strongly recommended with the WiredTiger storage engine. See http://dochub.mongodb.org/core/prodnotes-filesystem
        2022-04-11T12:50:28.852+01:00: Access control is not enabled for the database. Read and write access to data and configuration is unrestricted
---
---
        Enable MongoDB's free cloud-based monitoring service, which will then receive and display
        metrics about your deployment (disk utilization, CPU, operation statistics, etc).

        The monitoring data will be available on a MongoDB website with a unique URL accessible to you
        and anyone you share the URL with. MongoDB may use this information to make product
        improvements and to suggest MongoDB products and deployment options to you.

        To enable free monitoring, run the following command: db.enableFreeMonitoring()
        To permanently disable this reminder, run the following command: db.disableFreeMonitoring()
---
>
\end{verbatim}

Next:

\begin{verbatim}
mongo
db.disableFreeMonitoring()
use admin
db.createUser({user:"adminer", pwd: "aReallyComplexPassword1", roles: [ "root" ]})
use admin
db.createUser({user:"backuper", pwd: "aReallyComplexPassword2", roles: [ "backup" ]})
use admin
db.createUser({user:"cadiauser", pwd: "aReallyComplexPassword3", roles: [{ role: "readWrite", db: "domainstore" }]})
\end{verbatim}

Then be sure to follow the step to add authorisation to mongo:

Edit /etc/mongod.conf and add:

\begin{verbatim}
security:
    authorization: enabled
\end{verbatim}

then run the following command to restart Mongo DB:

\begin{verbatim}
sudo systemctl restart mongod.
\end{verbatim}

\hypertarget{configure-the-iamus-configuration-file}{%
\subsubsection{Configure the iamus configuration
file}\label{configure-the-iamus-configuration-file}}

Next add the an iamus.json configuration file to the base directory.

You can take inspiration from:

\begin{enumerate}
\def\labelenumi{\arabic{enumi})}
\tightlist
\item
  the example in the build readme as shown below:
\end{enumerate}

\begin{verbatim}
{
    "metaverse": {
        "metaverse-name": "My Metaverse",
        "metaverse-nick-name": "MyVerse",
        "metaverse-server-url": "https://metaverse.example.org:9400/",
        "default-ice-server-url": "ice.example.org:7337"
    },
    "server": {
        "cert-file": "config/cert.pem",
        "key-file": "config/privkey.pem",
        "chain-file": "config/chain.pem"
    },
    "metaverse-server": {
        "metaverse-info-addition-file": "config/metaverse_info.json"
    },
    "database": {
        "db": "myverse",
        "db-host": "metaverse.example.org",
        "db-user": "DBUSER",
        "db-pw": "DBUSERPASSWORD"
    },
    "debug": {
        "loglevel": "debug",
        "devel": true,
    }
}
\end{verbatim}

\begin{enumerate}
\def\labelenumi{\arabic{enumi})}
\setcounter{enumi}{1}
\tightlist
\item
  the src/config.ts file in the @todo dir:
\end{enumerate}

\begin{verbatim}
{
    // The metaverse identity
    'metaverse': {
        'metaverse-name': 'Vircadia noobie',
        'metaverse-nick-name': 'Noobie',
        'metaverse-server-url': '',   // if empty, set to self
        'default-ice-server-url': '', // if empty, set to self
        'dashboard-url': 'https://dashboard.vircadia.com'
    },
    // Server network parameters
    'server': {
        'listen-host': '0.0.0.0',
        'listen-port': 9400,
        'key-file': '',           // if supplied, do https
        'cert-file': '',
        'max-body-size': 300000,  // maximum body size for input JSON bodies
        'static-base': '/static', // base of static data URL
        'user-config-file': './iamus.json', // startup config over-ride
        'server-version': {       // overlaid with VERSION.json
            'version-tag': '1.1.1-20200101-abcdefg'
        }
    },
    // Authorization token parameters
    'auth': {
        'domain-token-expire-hours': 24 * 365,  // one year
        'owner-token-expire-hours': 24 * 7      // one week
    },
    // Control of the metaverse operations
    'metaverse-server': {
        'http-error-on-failure': true,  // whether to include x-vircadia error header
        'error-header': 'x-vircadia-error-handle',

        'metaverse-info-addition-file': './metaverse_info.json',
        'max-name-length': 32,      // the max characters a domain, place, account, ... name

        'session-timeout-minutes': 5,
        'heartbeat-seconds-until-offline': 5 * 60,      // seconds until non-heartbeating user is offline
        'domain-seconds-until-offline': 10 * 60,        // seconds until non-heartbeating domain is offline
        'domain-seconds-check-if-online': 2 * 60,       // how often to check if a domain is online
        'handshake-request-expiration-minutes': 1,      // minutes that a handshake friend request is active
        'connection-request-expiration-minutes': 60 * 24 * 4, // 4 days
        'friend-request-expiration-minutes': 60 * 24 * 4,     // 4 days

        'place-current-timeout-minutes': 5,             // minutes until current place info is stale
        'place-inactive-timeout-minutes': 60,           // minutes until place is considered inactive
        'place-check-last-activity-seconds': (3*60)-5,  // seconds between checks for Place lastActivity updates

        // redirection URL used for initial domain token generation,
        //   "METAVERSE_SERVER_URL" is replaced (from Config.metaverse.metaverse-server-url)
        //   "DASHBOARD_URL" is replaced (from Config.metaverse.dashboard-url)
        'tokengen_url': 'METAVERSE_SERVER_URL/static/DomainTokenLogin.html',
        // 'tokengen_url': 'DASHBOARD_URL?metaverse=METAVERSE_SERVER_URL&page=domain',

        // When account of this name is created, add 'admin' role to it
        //    Initially as empty so random people cannot create an admin account
        //    The account named here MUST be controlled by the server's admin
        'base-admin-account': '',

        // If to assume domain network_address if on is not set
        'fix-domain-network-address': true,
        // Whether allowing temp domain name creation
        'allow-temp-domain-creation': false,

        // Email verification on account creation
        'enable-account-email-verification': false,
        'email-verification-timeout-minutes': 1440, // minutes to wait for email verification (1440=one day)
        // default is in 'static' dir. If you put in 'config' dir, use 'config/verificationEmail.html'.
        //   "VERIFICATION_URL" is replaced with the computed URL (build with Config.metaverse-server-url)
        //   "METAVERSE_NAME" is replaced (from Config.metaverse.metaverse-name)
        //   "SHORT_METAVERSE_NAME" is replaced (from Config.metaverse.metaverse-nick-name)
        'email-verification-email-body': 'dist/static/verificationEmail.html',  // file to send
        'email-verification-from': '', // who the email is From
        // When user follows the verification URL, they are redirected to one of these two URLs
        //   "METAVERSE_SERVER_URL" is replaced (from Config.metaverse.metaverse-server-url)
        //   "DASHBOARD_URL" is replaced (from Config.metaverse.dashboard-url)
        //   "ACCOUNT_ID" is replaced with the verifying account id
        //   "FAILURE_REASON" is replaced with the reason for verification failure (url encoded)
        'email-verification-success-redirect': 'METAVERSE_SERVER_URL/static/verificationEmailSuccess.html',
        'email-verification-failure-redirect': 'METAVERSE_SERVER_URL/static/verificationEmailFailure.html?r=FAILURE_REASON'
    },
    // SMTP mail parameters for out-bound email
    // This is the structure that is passed to NodeMailer's SMTP transport.
    // Check out the documentation at https://nodemailer.com/smtp/
    // For SMTP outbound, setup your email account on your service and
    //     update SMTP-HOSTNAME, SMTP-USER, and SMTP-PASSWORD with your info.
    'nodemailer-transport-config': {
        'host': 'SMTP-HOSTNAME',
        'port': 465,    // 587 if secure=false
        'secure': true,
        'auth': {
            'user': 'SMTP-USER',
            'pass': 'SMTP-PASSWORD'
        }
    },
    'monitoring': {
        'enable': true,           // enable value monitoring
        'history': true           // whether to keep value history
    },
    // Setup for MongoDB access
    'database': {
        'db-host': 'localhost',
        'db-port': 27017,
        'db': 'tester',
        'db-user': 'metaverse',
        'db-pw': 'nooneknowsit',
        'db-authdb': 'admin',
        'db-connection': ''   // connection string replaces above if supplied
    },
    // MongoDB account configured for database backup script
    'backup': {
        "backup-user": "backuper",  // database backup user account (for BackupDb.sh)
        "backup-pw": "nooneknowsit", // database backup user password (for BackupDb.sh)
        "backup-dir": "directoryName", // Backup file directory. Optional. Defaults to "./DatabaseBackup"
        "authenticationDatabase": "databaseName" // auth db for backup user. Optional. Defaults to "admin"
    },
    'debug': {
        'loglevel': 'info',

        // Winston logging configuration
        'log-to-files': true,         // if to log to files
        'log-filename': 'iamus.log',  // filename for log files
        'log-directory': './logs',    // directory to place logs
        'log-max-size-megabytes': 100,// max mega-bytes per log file
        'log-max-files': 10,          // number of log files to create
        'log-tailable': true,         // if to always output to main named log file
        'log-compress': false,        // if to compress old log files

        'log-to-console': false,      // if to additionally log to the console

        'devel': false,

        // Control of what debug information is logged
        'request-detail': false,  // output the received request info when received
        'request-body': false,    // output the received request body when received
        'metaverseapi-response-detail': false, // output the response sent back from MetaverseAPI requests
        'query-detail': false,    // outputs details when selecting query parameters
        'db-query-detail': false, // outputs details about DB queries
        'field-setting': false    // Details of entity field getting and setting
    }
}
\end{verbatim}

\begin{enumerate}
\def\labelenumi{\arabic{enumi})}
\setcounter{enumi}{2}
\tightlist
\item
  the file used in this setup:
\end{enumerate}

\begin{verbatim}
@todo: insert file
\end{verbatim}

\hypertarget{configure-port-forwarding-for-the-metaverse-server}{%
\subsubsection{Configure port forwarding for the metaverse
server}\label{configure-port-forwarding-for-the-metaverse-server}}

\begin{verbatim}
set nat destination rule 20 description 'Port Forward for metaverse: 9400 to 9400 on 192.168.y.5'
set nat destination rule 20 destination port '9400'
set nat destination rule 20 inbound-interface 'eth0'
set nat destination rule 20 protocol 'tcp'
set nat destination rule 20 translation address '192.168.y.5'
set nat destination rule 20 translation port '9400'
\end{verbatim}

\hypertarget{configuring-nat-reflection-on-the-vyos-router-testing}{%
\subsubsection{Configuring NAT reflection on the VyOS router
(testing)}\label{configuring-nat-reflection-on-the-vyos-router-testing}}

Many routers will automatically configure NAT reflection for open ports
and
\href{https://docs.vyos.io/en/equuleus/configuration/nat/index.html\#hairpin-nat-nat-reflection}{VyOS
can be configured to perform NAT reflection}. This requires a few
updates to the existing rules that have been configured. NAT reflection
would be useful for internal only debugging, however, in testing the ICE
server from the build would only bind to the external internet routable
IP thus hindering the ability to use internally. As using ICE for
internal only is an edge case this is actually as expected although
future work could look into updating the ICE server and configuring
split DNS (feel free to update with suggestions). With this in mind the
setup described here, at present, does not require NAT reflection but
the following updates should enable it if desired:

\begin{verbatim}
set nat destination rule 16 destination address 192.168.x.z
set nat destination rule 17 destination address 192.168.x.z
set nat destination rule 18 destination address 192.168.x.z
set nat destination rule 20 destination address 192.168.x.z
set nat destination rule 21 destination address 192.168.x.z

set nat destination rule 16 inbound-interface any
set nat destination rule 17 inbound-interface any
set nat destination rule 18 inbound-interface any
set nat destination rule 20 inbound-interface any
set nat destination rule 21 inbound-interface any

commit

save
\end{verbatim}

\hypertarget{enabling-external-access}{%
\subsubsection{Enabling external
access}\label{enabling-external-access}}

Important note: in a development and testing environment (such as the
one detailed here) the VyOS router should not be permanently connected
to the internet and access should only be allowed whilst tests are being
performed and if possible tied down to specific source addresses. To
allow testing the external firewall requires port forwarding. The port
forwarding rules can be turned on and off as desired.

The following port forwarding rules are required:

\begin{tabular}[]{@{}
  >{\raggedright\arraybackslash}p{(\columnwidth - 6\tabcolsep) * \real{0.2500}}
  >{\raggedright\arraybackslash}p{(\columnwidth - 6\tabcolsep) * \real{0.2500}}
  >{\raggedright\arraybackslash}p{(\columnwidth - 6\tabcolsep) * \real{0.2500}}
  >{\raggedright\arraybackslash}p{(\columnwidth - 6\tabcolsep) * \real{0.2500}}@{}}
\toprule
\begin{minipage}[b]{\linewidth}\raggedright
Name
\end{minipage} & \begin{minipage}[b]{\linewidth}\raggedright
Port
\end{minipage} & \begin{minipage}[b]{\linewidth}\raggedright
Destination IP
\end{minipage} & \begin{minipage}[b]{\linewidth}\raggedright
Protocol
\end{minipage} \\
\midrule

Vircadia 40102 (udp) main connection from clients & 40102 & 192.168.x.y
& UDP \\
Vircadia 7337 (both) ice-server & 7337 & 192.168.x.y & TCP+UDP \\
Vircadia 9400 (tcp) metaverse-server & 9400 & 192.168.x.y & TCP \\
\bottomrule
\end{tabular}

Note: the configuration and starting of the services will reference the
external IP address a.b.c.d

\hypertarget{start-the-services}{%
\subsubsection{Start the services}\label{start-the-services}}

The process of starting the services is as follows (see:
\href{https://github.com/vircadia/vircadia-metaverse/blob/master/docs/NotesOnDevelopment.md\#building-and-running-ice-server-and-domain-server}{Building
and Running Ice-Server and Domain-Server}

\hypertarget{metaverse-server-service}{%
\paragraph{Metaverse server service}\label{metaverse-server-service}}

On the metaverse server open a terminal and run:

\begin{verbatim}
cd vircadia-metaverse

node/dist/index.ts
\end{verbatim}

\hypertarget{ice-server}{%
\paragraph{ICE server}\label{ice-server}}

Start ice-server.

On the vircadia server open a terminal and run:

\begin{verbatim}
cd Vircadia/vircadiaBuildDir
export HIFI_METAVERSE_URL=http://a.b.c.d:9400
./run_ice-server
\end{verbatim}

\hypertarget{domain-server-services}{%
\paragraph{Domain server services}\label{domain-server-services}}

Start domain-server.

Open another terminal and run:

\begin{verbatim}
cd Vircadia/vircadiaBuildDir
export HIFI_METAVERSE_URL=http://a.b.c.d:9400
export ICE_SERVER=a.b.c.d:7337
./run_domain-server -i ${ICE_SERVER}
\end{verbatim}

Start assignment client.

Open another terminal and run:

\begin{verbatim}
cd Vircadia/vircadiaBuildDir
./run_assignment-client
\end{verbatim}

\hypertarget{client-interface}{%
\paragraph{Client interface}\label{client-interface}}

For testing the client application the Windows OS was opted for and
\href{https://github.com/vircadia/vircadia/releases/tag/2022.1.1}{Vircadia
2022.1.1} was used.

\hypertarget{connecting-the-metaverse-server-and-configuring-a-test-metaverse-account}{%
\subsubsection{Connecting the Metaverse server and configuring a test
Metaverse
account}\label{connecting-the-metaverse-server-and-configuring-a-test-metaverse-account}}

Navigate to:

http://192.168.x.y:40100/settings/\#metaverse\_group (or use the
external IP a.b.c.d)

Then click `Connect Metaverse Account'

A new page should open
(http://a.b.c.d:9400/static/DomainTokenLogin.html) with the option to
`create account'. Do this and then enter the username and password on
the `Get Token'. Copy this token and then go back to the previous page
and paste it in the open dialogue and click connect. The Networking /
Metaverse section should now display `Metaverse Account Connected'.

Note: at this point you could create a few more accounts for testing.

\hypertarget{connecting-a-client}{%
\subsubsection{Connecting a client}\label{connecting-a-client}}

If you are testing locally then you will probably need to use a VPN so
that your connection is routed from outside your local network. This is
due to the ICE server and how it will create open ports. The NAT
reflection required will probably not work on the external router for
non static port forwarding rules.

From tests it appears that the best way to run the client is to start it
from the command line with the IP for the domain server (at least for
the initial run and then it seems to work fine then in).

For the initial run the following was actioned:

Open a terminal (PowerShell on Windows in this case), navigate to the
folder containing the Vircadia executable and run:

\begin{verbatim}
.\interface.exe --url 84.92.193.1
\end{verbatim}

Then in the interface navigate to File -\textgreater{} Metaverse: Login
/ Sign Up

Enter the details as follows:

Display name: XYZ Username: usernameCreatedInPreviousStep Password:
passwordCreatedInPreviousStep URL: http://a.b.c.d:9400

Then click `Log in to metaverse'

\hypertarget{configure-ssh-access-for-domain-and-metaverse-servers-optional}{%
\subsubsection{Configure SSH access for domain and metaverse servers
(optional)}\label{configure-ssh-access-for-domain-and-metaverse-servers-optional}}

For potentially easier debugging, enable SSH on the servers.

Open a terminal and type:

\texttt{sudo\ apt\ install\ openssh-server}

Then configure the following rules on the VyOS router:

\hypertarget{domain-server}{%
\paragraph{Domain server}\label{domain-server}}

Then set up port forwarding on the VyOS router\ldots{} @todo add \$VyOS

\begin{verbatim}
configure

set nat destination rule 14 description 'Port Forward: 2224 to 22 SSH on 192.168.y.4'
set nat destination rule 14 destination port '2224'
set nat destination rule 14 inbound-interface 'eth0'
set nat destination rule 14 protocol 'tcp'
set nat destination rule 14 translation address '192.168.y.4'
set nat destination rule 14 translation port '22'

commit
\end{verbatim}

Test the SSH access and if all works well:

\begin{verbatim}
save
\end{verbatim}

\hypertarget{metaverse-server}{%
\paragraph{Metaverse server}\label{metaverse-server}}

Configure SSH access and port forwarding on the VyOS if desired:

\begin{verbatim}
configure

set nat destination rule 19 description 'Port Forward: 2225 to 22 SSH on 192.168.y.5'
set nat destination rule 19 destination port '2225'
set nat destination rule 19 inbound-interface 'eth0'
set nat destination rule 19 protocol 'tcp'
set nat destination rule 19 translation address '192.168.y.5'
set nat destination rule 19 translation port '22'

commit
\end{verbatim}

Test the SSH access and if all works well:

\begin{verbatim}
save
\end{verbatim}

\stopcontents[part] 


\part{Appendix}

\section{Acknowledgements and thanks}
As you'd expect lots of work went into checking the book. Special thanks to Melvin Carvalho, Tim Millar, Lorena Gomez, James Lewis, @smallworlnd, and Margaret O'Hare.
\section{Author Biographies}
\noindent\fcolorbox{red}{lightgray}{%
\begin{minipage}{\dimexpr0.66\textwidth-2\fboxrule-2\fboxsep\relax}
Dr John O`Hare is a results driven, certified Prince2 Agile Practitioner. Leveraging proven analytical ability, and drawing on 23 years of experience at the University of Salford.  Successful as a leader and an influential team member in both project and customer-facing roles. As a product manager he specialises in systems design, procurement, tendering and bid writing for research funding, running complex heterogeneous research systems, research and development, and supporting academic staff / research students to undertake theirs. Completed a PhD in \href{https://www.researchgate.net/profile/John-Ohare/publication/332029624_Telethrone_a_situated_display_using_retro-reflection_based_multi-view_toward_remote_collaboration_in_small_dynamic_groups/links/5c9ba02645851506d72ff380/Telethrone-a-situated-display-using-retro-reflection-based-multi-view-toward-remote-collaboration-in-small-dynamic-groups.pdf}{``Attention in Telepresence''}, uniting the gaze of remote collaborators, through furniture. Recently pursuing research opportunities in value transfer mechanisms for `Metaverses'.
\end{minipage}}%
\begin{minipage}{0.67\textwidth}
\includegraphics[width=23ex]{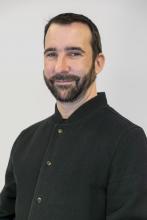}

\end{minipage}
\noindent\fcolorbox{red}{lightgray}{%
\begin{minipage}{\dimexpr0.66\textwidth-2\fboxrule-2\fboxsep\relax}

Dr Allen Fairchild is an experienced security-conscious software engineer and academic researcher with comprehensive experience developing innovative end-to-end systems for a wide variety of use-cases.
Strong leadership and acumen in full stack development. Track record in building networks through regional initiatives, delivering Agile projects to a wide variety of technical markets. Allen is an accomplished researcher and holds a PhD \href{http://usir.salford.ac.uk/id/eprint/39175/?template=banner}{Video based reconstruction system for mixed reality environments supporting contextualised non-verbal communication and its study}, alongside a portfolio of groundbreaking publications in social VR. Excellent communication skills and Agile team leadership.

\end{minipage}}%
\begin{minipage}{0.67\textwidth}
\includegraphics[width=23ex]{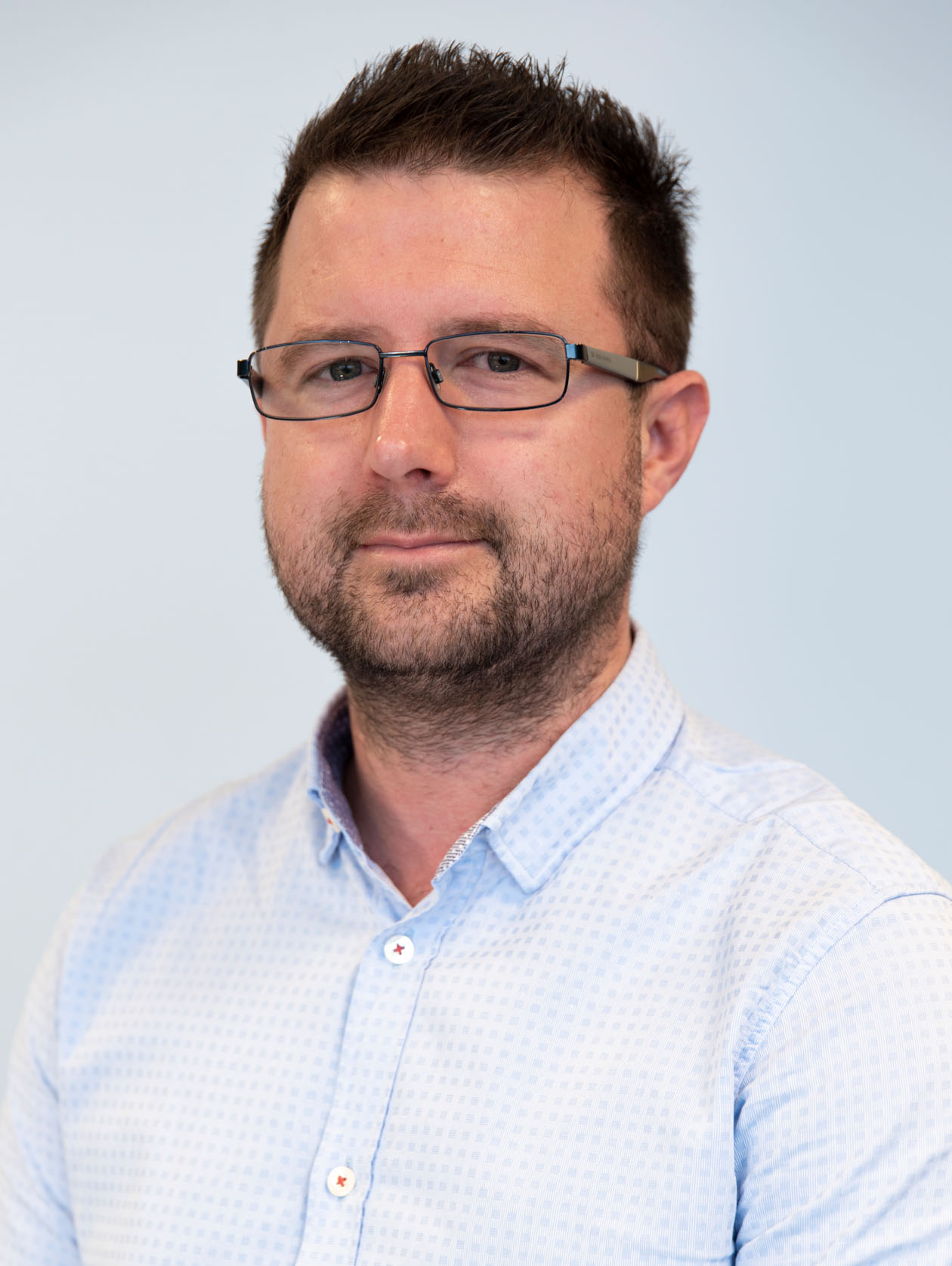}
\end{minipage}

\noindent\fcolorbox{red}{lightgray}{%
\begin{minipage}{\dimexpr0.66\textwidth-2\fboxrule-2\fboxsep\relax}   
Dr Umran Ali currently works as a senior lecturer in creative media, and continues to explore virtual natural environment design through teaching and research, maintaining a deep interest in the meaning, impact, and design of natural spaces, in particular around video games. A keen video game collector and player, and a landscape photographer. Holds a PhD in \href{http://usir.salford.ac.uk/id/eprint/39394/?template=banner}{A practice-based exploration of natural environment design in computer \& video games.}
\end{minipage}}%
\begin{minipage}{0.67\textwidth}
\includegraphics[width=23ex]{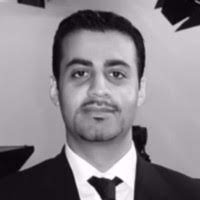}
\end{minipage}


\chapterimage{} 
\chapterspaceabove{2.5cm} 
\chapterspacebelow{2cm} 


\chapter*{Bibliography}

\printbibliography





\cleardoublepage 
\phantomsection
\addcontentsline{toc}{chapter}{\textcolor{ocre}{Index}} 
\printindex 

\end{document}